\documentclass[11pt]{article}
\usepackage[margin=.8in,left=.8in]{geometry}
\usepackage{amsthm}
\usepackage{mathtools}
\usepackage{color}
\usepackage{xcolor}
\usepackage{stmaryrd}
\usepackage{rotating}

\usepackage{adjustbox}

\usepackage{hyperref}
\hypersetup{
  colorlinks = true,
  allcolors=.           
}

\usepackage[all]{xy}

\usepackage{xfrac}

\usepackage{tikz}
\usetikzlibrary{3d,shadings}
\usetikzlibrary{arrows.meta}
\usetikzlibrary{decorations.pathmorphing}
\usepackage{smartdiagram}

\newcommand{\boldpi}{\mbox{$\pi$\hspace{-6.5pt}$\pi$}}

\newcommand{\Truncation}[1]{
  \pi_{#1}
}

\newcommand{\ConstrainedMaps}[3]{
  \mathrm{Map}^{\mathrm{\ast}/}
  #1(
    #2
    ,\,
    #3
  #1)
}

\DeclareMathAlphabet{\mathpzc}{OT1}{pzc}{m}{it} 

\newcommand\mathscr[1]{\scalebox{1.1}{$\mathpzc{#1}$}}

\newcommand\degree[1]{{}^{{#1}}}

\usepackage[titletoc]{appendix}

\usepackage[safe]{tipa} 

\definecolor{darkblue}{rgb}{0.05,0.25,0.65}
\definecolor{greenii}{RGB}{20,140,10}
\definecolor{lightgray}{rgb}{0.9,0.9,0.9}
\definecolor{orangeii}{RGB}{200,100,5}
\definecolor{darkyellow}{rgb}{.91,.91,0}

\usepackage{multirow}

\usepackage{stmaryrd}    

\usepackage{enumerate} 

\usepackage{amssymb,amsmath}

\newcounter{sqindex}

\usepackage{mathptmx}
\usepackage{amsmath}
\usepackage{graphicx}

\newcommand{\themap}{f}

\newcommand{\PointedHomotopyTypes}{
  \mathrm{Ho}
  \big(
    \mathrm{Spaces}^{\ast/}_{\mathrm{Qu}}
  \big)
}

\newcommand{\Spaces}{
  \mathrm{Spaces}
}

\newcommand{\PointedSpaces}{
  \mathrm{Spaces}^{\ast/}
}

\newcommand{\HoSpaces}{
  \mathrm{Ho}\big(
    \Spaces
  \big)
}

\newcommand{\HoPointedSpaces}{
  \mathrm{Ho}\big(
    \PointedSpaces
  \big)
}

\newcommand{\LCHSpaces}{
  \mathrm{LCHSpaces}^{\mathrm{prop}}
}

\newcommand{\mapsdown}{\rotatebox[origin=c]{-90}{$\mapsto$}}

\usepackage[new]{old-arrows}   

\newdir{> }{{}*!/10pt/@{>}}

\usepackage{amssymb}

\DeclareRobustCommand{\rchi}{{\mathpalette\irchi\relax}}
\newcommand{\irchi}[2]{\raisebox{\depth}{$#1\chi$}} 

\makeatletter
\newif\if@sup
\newtoks\@sups
\def\append@sup#1{\edef\act{\noexpand\@sups={\the\@sups #1}}\act}%
\def\reset@sup{\@supfalse\@sups={}}%
\def\mk@scripts#1#2{\if #2/ \if@sup ^{\the\@sups}\fi \else%
  \ifx #1_ \if@sup ^{\the\@sups}\reset@sup \fi {}_{#2}%
  \else \append@sup#2 \@suptrue \fi%
  \expandafter\mk@scripts\fi}
\def\tensor#1#2{\reset@sup#1\mk@scripts#2_/}
\def\multiscripts#1#2#3{\reset@sup{}\mk@scripts#1_/#2%
  \reset@sup\mk@scripts#3_/}
\makeatother

\makeatletter
\newbox\slashbox \setbox\slashbox=\hbox{$/$}
\def\itex@pslash#1{\setbox\@tempboxa=\hbox{$#1$}
  \@tempdima=0.5\wd\slashbox \advance\@tempdima 0.5\wd\@tempboxa
  \copy\slashbox \kern-\@tempdima \box\@tempboxa}
\def\slash{\protect\itex@pslash}
\makeatother

\def\clap#1{\hbox to 0pt{\hss#1\hss}}
\def\mathllap{\mathpalette\mathllapinternal}
\def\mathrlap{\mathpalette\mathrlapinternal}
\def\mathclap{\mathpalette\mathclapinternal}
\def\mathllapinternal#1#2{\llap{$\mathsurround=0pt#1{#2}$}}
\def\mathrlapinternal#1#2{\rlap{$\mathsurround=0pt#1{#2}$}}
\def\mathclapinternal#1#2{\clap{$\mathsurround=0pt#1{#2}$}}

\let\oldroot\root
\def\root#1#2{\oldroot #1 \of{#2}}
\renewcommand{\sqrt}[2][]{\oldroot #1 \of{#2}}

\DeclareSymbolFont{symbolsC}{U}{txsyc}{m}{n}
\SetSymbolFont{symbolsC}{bold}{U}{txsyc}{bx}{n}
\DeclareFontSubstitution{U}{txsyc}{m}{n}

\DeclareSymbolFont{stmry}{U}{stmry}{m}{n}
\SetSymbolFont{stmry}{bold}{U}{stmry}{b}{n}

\DeclareFontFamily{OMX}{MnSymbolE}{}
\DeclareSymbolFont{mnomx}{OMX}{MnSymbolE}{m}{n}
\SetSymbolFont{mnomx}{bold}{OMX}{MnSymbolE}{b}{n}
\DeclareFontShape{OMX}{MnSymbolE}{m}{n}{
    <-6>  MnSymbolE5
   <6-7>  MnSymbolE6
   <7-8>  MnSymbolE7
   <8-9>  MnSymbolE8
   <9-10> MnSymbolE9
  <10-12> MnSymbolE10
  <12->   MnSymbolE12}{}

\usepackage{cleveref}

\crefformat{section}{\S#2#1#3} 
\crefformat{subsection}{\S#2#1#3}
\crefformat{subsubsection}{\S#2#1#3}

\theoremstyle{italics}
\newtheorem{theorem}{Theorem}[section]
\newtheorem{lemma}[theorem]{Lemma}
\newtheorem{prop}[theorem]{Proposition}

\theoremstyle{definition}
\newtheorem{defn}[theorem]{Definition}
\newtheorem{notation}[theorem]{Notation}
\newtheorem{example}[theorem]{Example}
\newtheorem{examples}[theorem]{Examples}
\newtheorem{remark}[theorem]{Remark}

\usepackage{amsfonts}
\usepackage{colortbl}

\newcommand{\underoverset}[3]{\underset{#1}{\overset{#2}{#3}}}

\renewcommand{\emph}{\textit}

\begin{document}

\title{
  M/F-Theory as $M\!f$-Theory
}

\author{
  Hisham Sati${}^{\ast \dagger}$
  \;\;
  and
  \;\;
  Urs Schreiber${}^{\ast}$
}

\maketitle

\begin{abstract}
  \noindent
  In the quest for mathematical foundations of M-theory,
  the {\it Hypothesis H}
  that fluxes are quantized in Cohomotopy theory,
  implies,
  on flat but possibly singular spacetimes,
  that M-brane charges
  locally organize into equivariant homotopy groups of spheres.
  Here we show how this leads to a correspondence
  between phenomena conjectured in M-theory
  and fundamental mathematical concepts/results in
  stable homotopy,
  generalized cohomology
  and Cobordism theory $M\!f$:

  - stems of homotopy groups
  correspond to
  charges of probe $p$-branes near black $b$-branes;

  - stabilization within a stem
  is
  the boundary-bulk transition;

  - the Adams d-invariant
  measures
  $G_4$-flux;

  - trivialization of the d-invariant
  corresponds to
  $H_3$-flux;

  - refined Toda brackets measure $H_3$-flux;


  - the refined Adams e-invariant sees the $H_3$-charge lattice;

  - vanishing Adams e-invariant
  implies
  consistent global $C_3$-fields;

  - Conner-Floyd's e-invariant
  is the $H_3$-flux seen in the Green-Schwarz mechanism;

  -
  the Hopf invariant
  is the
  M2-brane Page charge ($\widetilde G_7$-flux);

  -
  the Pontrjagin-Thom theorem
  associates the
  polarized brane worldvolumes sourcing all these charges.

  \noindent
  In particular, spontaneous K3-reductions
  with 24 branes
  are singled out from first principles:

  - Cobordism in the third stable stem
  witnesses
  spontaneous KK-compactification on K3-surfaces;

  - the order of the third stable stem
  implies the 24 NS5/D7-branes in M/F-theory on K3.

  \noindent
  Finally, complex-oriented cohomology emerges
  from Hypothesis H,
  connecting it to all
  previous proposals for brane charge quantization
  in the chromatic tower:
  K-theory, elliptic cohomology, etc.:

  - quaternionic orientations
  correspond to
  unit $H_3$-fluxes near M2-branes;

  - complex orientations
  lift these unit $H_3$-fluxes to heterotic
  M-theory with heterotic line bundles.

  \noindent
  In fact, we find
  quaternionic/complex Ravenel-orientations
  bounded in dimension;
  and we find the bound to be 10,
  as befits spacetime dimension 10+1.
\end{abstract}

\vfill

\hrule
\vspace{5pt}

{
\footnotesize
\noindent
\def\arraystretch{1}
\tabcolsep=0pt
\begin{tabular}{ll}
${}^*$\,
&
Mathematics, Division of Science; and
\\
&
Center for Quantum and Topological Systems,
\\
&
NYUAD Research Institute,
\\
&
New York University Abu Dhabi, UAE.
\end{tabular}
\hfill
\adjustbox{raise=-15pt}{
\includegraphics[width=3cm]{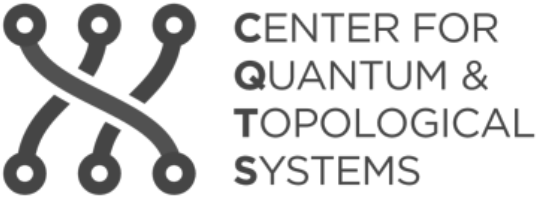}
}

\vspace{1mm}
\noindent ${}^\dagger$The Courant Institute for Mathematical Sciences, NYU, NY

\vspace{.2cm}

\noindent
The authors acknowledge the support by {\it Tamkeen} under the
{\it NYU Abu Dhabi Research Institute grant} {\tt CG008}.
}

\medskip

\newpage

\tableofcontents

\vspace{2cm}

\begin{center}

\begin{tikzpicture}

  \draw[densely dotted, line width=.33]
    (-3,0) arc (180:0:3 and .7);

  \begin{scope}[rotate=(-90), line width=.33]
  \draw[densely dotted]
    (-3,0) arc (180:0:3 and .7);
  \end{scope}

  \begin{scope}[rotate=(6)]

  \shade[ball color=gray!60, opacity=2]
    (-3:3)
      -- (3:3)
      .. controls (4:1.1) and (11:1.2) ..
      (15-3:3)
      -- (15+3:3)
      .. controls (15+4:1.2) and (15+11:1.2)
      .. (15+15-3:3)
      -- (2*15+3:3)
      .. controls (2*15+4:1.2) and (2*15+11:1.2)
      .. (2*15+15-3:3)
      -- (3*15+3:3)
      .. controls (3*15+4:1.2) and (3*15+11:1.2)
      .. (3*15+15-3:3)
      -- (4*15+3:3)
      .. controls (4*15+4:1.2) and (4*15+11:1.2)
      .. (4*15+15-3:3)
      -- (5*15+3:3) .. controls
         (5*15+4:1.2) and
         (5*15+11:1.2)
      .. (5*15+15-3:3)
      -- (6*15+3:3) .. controls
         (6*15+4:1.2) and
         (6*15+11:1.2)
      .. (6*15+15-3:3)
      -- (7*15+3:3) .. controls
         (7*15+4:1.2) and
         (7*15+11:1.2)
      .. (7*15+15-3:3)
      -- (8*15+3:3) .. controls
         (8*15+4:1.2) and
         (8*15+11:1.2)
      .. (8*15+15-3:3)
      -- (9*15+3:3) .. controls
         (9*15+4:1.2) and
         (9*15+11:1.2)
      .. (9*15+15-3:3)
      -- (10*15+3:3) .. controls
         (10*15+4:1.2) and
         (10*15+11:1.2)
      .. (10*15+15-3:3)
      -- (11*15+3:3) .. controls
         (11*15+4:1.2) and
         (11*15+11:1.2)
      .. (11*15+15-3:3)
      -- (12*15+3:3) .. controls
         (12*15+4:1.2) and
         (12*15+11:1.2)
      .. (12*15+15-3:3)
      -- (13*15+3:3) .. controls
         (13*15+4:1.2) and
         (13*15+11:1.2)
      .. (13*15+15-3:3)
      -- (14*15+3:3) .. controls
         (14*15+4:1.2) and
         (14*15+11:1.2)
      .. (14*15+15-3:3)
      -- (15*15+3:3) .. controls
         (15*15+4:1.2) and
         (15*15+11:1.2)
      .. (15*15+15-3:3)
      -- (16*15+3:3) .. controls
         (16*15+4:1.2) and
         (16*15+11:1.2)
      .. (16*15+15-3:3)
      -- (17*15+3:3) .. controls
         (17*15+4:1.2) and
         (17*15+11:1.2)
      .. (17*15+15-3:3)
      -- (18*15+3:3) .. controls
         (18*15+4:1.2) and
         (18*15+11:1.2)
      .. (18*15+15-3:3)
      -- (19*15+3:3) .. controls
         (19*15+4:1.2) and
         (19*15+11:1.2)
      .. (19*15+15-3:3)
      -- (20*15+3:3) .. controls
         (20*15+4:1.2) and
         (20*15+11:1.2)
      .. (20*15+15-3:3)
      -- (21*15+3:3) .. controls
         (21*15+4:1.2) and
         (21*15+11:1.2)
      .. (21*15+15-3:3)
      -- (22*15+3:3) .. controls
         (22*15+4:1.2) and
         (22*15+11:1.2)
      .. (22*15+15-3:3)
      -- (23*15+3:3) .. controls
         (23*15+4:1.2) and
         (23*15+11:1.2)
      .. (23*15+15-3:3)
      ;

 \foreach \n in {1, ..., 3}
 {
   \draw
     (\n*15:3.25) node {\scalebox{.8}{\color{darkblue}$S^3$}};
 }

 \draw
   (-.3,.3)
   node
   {

    ${\scalebox{.6}{\bf\color{darkblue} K3}}
    \mathrlap{\raisebox{0pt}{\scalebox{.6}{$\setminus ( 24 \cdot D^4 )$}}}$
   };

  \end{scope}

  \begin{scope}
    \clip
      (-134.9-0.2:0.96) rectangle (-134.9+0.2:0.1);
    \draw[densely dotted]
      (-134.9-0.2:4.3) to (-134.9-0.2:.4);
    \draw[densely dotted]
      (-134.9-0.0:4.3) to (-134.9+0.0:.4);
    \draw[densely dotted]
      (-134.9+0.2:4.3) to (-134.9+0.2:.4);
  \end{scope}
  \begin{scope}[shift={(-134.9-0.2:.4)}]
    \draw[densely dotted, line width=1.5pt]
      (-120:.1) arc (-120:360-150:.1);
  \end{scope}

  \begin{scope}[rotate=(180)]
    \shade[ball color=green!40, opacity=.6] (0,0) circle (3.005);
  \end{scope}

  \begin{scope}[rotate=(-90)]
  \draw[line width=.33]
    (-3,0) arc (180:360:3 and .7);
  \end{scope}

  \begin{scope}[rotate=(6)]

  \foreach \n in {1, ..., 24}
  {
    \draw[line width=.33, fill=lightgray]
     (\n*15-3.1:3)
       arc (\n*15-3.1:\n*15+2.9:3)
     to[bend right=70] (\n*15-2.9:3);
  }

   \end{scope}

  \draw[line width=.33]
    (-3,0) arc (180:360:3 and .7);

  \draw
    (-134.9:0.97)+(.23,-.2)
    node
    {\scalebox{.8}{\color{darkblue}$\mathbf{S}^7$}};

\end{tikzpicture}

\end{center}

\medskip

\newpage

\section{Introduction}
\label{theIntroduction}

\noindent {\bf Charges and Generalized cohomology.}
There is a close link between
{\it non-perturbative physics}
(
e.g. \cite{BakulevShirkov10})
and
{\it algebraic topology} (e.g. \cite{May99}):
Pure algebraic topology may be understood as the study of
homotopy types of spaces through functorial assignment
of their algebraic invariants;
but seen through the lens of mathematical physics,
this is all about identifying {\it charges} carried
by solitonic charged objects
in spacetimes (black holes, black $p$-branes)
as {\it cohomology classes}
in some generalized cohomology theory evaluated on
the ambient spacetime:

\vspace{-2mm}
\begin{center}
{\small
\begin{tabular}{|p{13.5em}|c|p{13.5em}|}
  \hline
  \begin{minipage}[center]{5.2cm}
    \bf
    $\mathclap{\phantom{\vert^{\vert}}}$
    Non-perturbative physics
  \end{minipage}
  &
  \begin{minipage}[center]{4.6cm}
    \begin{center}
    $\longleftrightarrow$
    \end{center}
  \end{minipage}
  &
  \begin{minipage}[center]{5.2cm}
    \bf
    $\mathclap{\phantom{\vert^{\vert}}}$
    Algebraic Topology
  \end{minipage}
  \\
  \hline
  \hline
  \begin{minipage}[center]{5.2cm}
   $\mathclap{\phantom{\vert^{\vert}}}$
   Charges carried by
   \\
   $\mathclap{\phantom{\vert_{\vert}}}$
   microscopic $p$-branes
  \end{minipage}
  &
  \begin{minipage}[center]{4.6cm}
    \begin{center}
      $
      \mathclap{\phantom{\int_{\vert_{\vert_{\vert}}}}}
      \Big[
      \;\;\,
      \xymatrix{
        \overset{
          \mathclap{
          \;\;
          \raisebox{3pt}{
            \tiny
            \color{darkblue}
            \bf
            \begin{tabular}{c}
              ambient
              \\
              spacetime
            \end{tabular}
          }
          }
        }{
          X
        }
        \ar[rr]^-{
          \mbox{
            \tiny
            \color{greenii}
            \bf fields
          }
        }
        &&
        \overset{
          \mathclap{
          \raisebox{3pt}{
            \tiny
            \color{darkblue}
            \bf
            \begin{tabular}{c}
              classifying
              \\
              space
            \end{tabular}
          }
          }
        }{
          A
        }
      }
      \;\;
      \Big]
      $
    \end{center}
  \end{minipage}
  &
  \begin{minipage}[center]{5.2cm}
    $\mathclap{\phantom{\vert^{\vert}}}$
    Generalized cohomology
    \\
    $\mathclap{\phantom{\vert_{\vert}}}$
    of external spacetime
  \end{minipage}
  \\
  \hline
\end{tabular}
}
\end{center}

\noindent
\vspace{-4mm}

\noindent
{\it Field strengths} or {\it flux densities} in physics are
collections of differential forms $F_i$ on spacetime, satisfying
differential relations: {\it Bianchi identities}.
The corresponding rational charges are reflected
in the periods of these differential forms,
expressing the total flux through any closed hypersurfaces.
On these fields, $\,$a {\it charge quantization law}$\,$
(see

\hspace{-.95cm}
\begin{tabular}{ll}
\label{ChargeQuantizationLaw}
\begin{minipage}[left]{7.1cm}
\cite{Freed00}\cite{Sati10b}\cite{FSS20c})
is a non-abelian cohomology theory
$\widetilde A(-)$ \cite[\S 2]{FSS20c}\cite[p. 6]{SS20b}
with the requirement
that true fields are cocycles in
(differential) $A$-cohomology
whose flux densities are just their images under the
{\it character map} \cite[\S 4]{FSS20c}\cite[\S 3.4]{SS20c}.
\end{minipage}

&

\hspace{2mm}
\raisebox{15pt}{
$
  \xymatrix@R=3pt{
    \overset{
      \mathclap{
      \raisebox{3pt}{
        \tiny
        \color{darkblue}
        \bf
        \begin{tabular}{c}
          $A$-cohomology of
          \\
          spacetime $X$
        \end{tabular}
      }
      }
    }{
      \widetilde A
      (X)
    }
    \ar[rr]^-{
       \overset{
         \raisebox{3pt}{
           \tiny
           \color{greenii}
           \bf
           character map
         }
       }{
         \mathrm{ch}_A
       }
    }
    \ar@{}[d]|-{
      \mathclap{
      \raisebox{-3pt}{
        \tiny
        \color{orangeii}
        \bf
        \begin{tabular}{c}
          charge quantization law
        \end{tabular}
      }
      }
    }
    &&
    {
    \overset{
      \mathclap{
      \raisebox{3pt}{
        \tiny
        \color{darkblue}
        \bf
        \begin{tabular}{c}
          de Rham cohomology with coefficients
          \\
          in Whitehead $L_\infty$-algebra
        \end{tabular}
      }
      }
    }{
      H_{\mathrm{dR}}
      \big(
        X;
        \mathfrak{l}A
      \big)
    }
    }
    \\
    \big[
      X
      \xrightarrow{
        \overset{
          \mathclap{
          \raisebox{3pt}{
            \tiny
            \color{darkblue}
            \bf
            $A$-cocycle
          }
          }
        }{
          c
        }
      }
      A
    \big]
    \ar@{}[rr]|-{
      \longmapsto
    }^-{
        \mathclap{
        \raisebox{3pt}{
          \tiny
          \color{greenii}
          \bf
          \begin{tabular}{c}
            classical
            \\
            approximation
          \end{tabular}
        }
        }
    }
    &&
    \Big\{
    \big(
      \overset{
        \mathclap{
        \raisebox{3pt}{
          \tiny
          \color{darkblue}
          \bf
          \begin{tabular}{c}
            field strenghts/
            \\
            flux densities
          \end{tabular}
        }
        }
      }{
        F_i
          \in
        \Omega^\bullet_{\mathrm{dR}}(X)
      }
    \big)_{i}
    \,\Big\vert\,
      \overset{
        \mathclap{
        \raisebox{3pt}{
          \tiny
          \color{darkblue}
          \bf
          \begin{tabular}{c}
            Bianchi identities/
            \\
            charge conservation laws
          \end{tabular}
        }
        }
      }{
        \big(
          d\, F_i \,=\, P_i\big( \{F_j\} \big)_j
        \big)_i
      }
    \Big.
    \Big\}
  }
$
}

\end{tabular}

\medskip
\noindent
Since charge quantization thus identifies
generalized cohomology theories
with microscopic properties of physical objects,
definitions/theorems in algebraic topology translate
to formulations of putative physical theories,
non-perturbatively.

\vspace{.2cm}

\label{ExampleMagneticChargeInOrdinaryCohomology}
\hspace{-.9cm}
\begin{tabular}{ll}

\begin{minipage}[left]{6cm}
{\bf Example: Magnetic charge in ordinary cohomology.}
The archetypical example is {\it Dirac's charge quantization}
\cite{Dirac31} (review in \cite{Alvarez85}\cite[\S 2]{Freed00}),
which is the observation that the quantum nature of
electrons requires the magnetic charge carried by a
magnetic monopole (say, a charged black hole)
to be identified with a class in ordinary integral degree-2 cohomology
of the spacetime surrounding the monopole.

\end{minipage}

&

\label{VortexStrings}
\hspace{-.4cm}

\raisebox{-68pt}{
\scalebox{.57}{
\begin{tikzpicture}[decoration=snake]

  \begin{scope}[shift={(0,1)}]

  \begin{scope}[shift={(0,-1)}]

  \draw (6,4.1)
    node
    {
      \scalebox{1.5}{
      $
        \underset{
               \mbox{
                 \color{greenii}
                 \tiny
                 \bf
                 \begin{tabular}{c}
                   electromagnetic field
                   \\
                   sourced by monopole
                 \end{tabular}
               }
        }{
        \xrightarrow{
          \scalebox{.7}{
          $
          \phantom{AAAAAAAAAA}
          c
          \phantom{AAAAAAAAAA}
          $
          }
        }
        }
      $
      }
     };

  \draw (0,4)
   node
     {
       \scalebox{1.5}{
       $
         \overset{
           \underset{
             \mbox{
               \color{darkblue}
               \tiny
               \bf
               \begin{tabular}{c}
                 spacetime around
                 magnetic monopole
               \end{tabular}
             }
           }{
             \scalebox{.8}{$
             \mathllap{X :=}\;
             \mathbb{R}^{3,1}
             \setminus
             \mathbb{R}^{0,1}
             \;\;\simeq\;\;
             \mathbb{R}^{0,1}
             \times
             \mathbb{R}_{\mathrm{rad}}
             \times
             S^2
             $}
           }
         }{
           \overbrace{\phantom{\mbox{\hspace{5cm}}}}
         }
      $
      }
     };

  \draw (12,4)
   node
     {
       \scalebox{1.5}{
       $
         \overset{
           \underset{
             \mbox{
               \tiny
               \color{darkblue}
               \bf
               \begin{tabular}{c}
                 classifying space for
                 ordinary cohomology
               \end{tabular}
             }
           }{
             \scalebox{.8}{$
               B \mathrm{U}(1)
               \;\simeq\;
               \mathbb{C}P^\infty
             $}
           }
         }{
           \overbrace{\phantom{\mbox{\hspace{5cm}}}}
         }
       $
       }
     };

  \end{scope}

  \begin{scope}[scale=1.5]

  \begin{scope}[shift={(0,-.6)}]

  \shade[ball color=gray!40, opacity=.6] (0,0) circle (2);

  \draw[fill=white, fill opacity=.50]
    (0,0) circle (.2);
  \draw (0,-.37)
    node
    {
      \scalebox{.78}{
      \color{orangeii}
      \bf
      monopole
      }
    };
  \draw (-.2,0) arc (180:360:.2 and .06);
  \draw[densely dotted] (.2,0) arc (0:180:.2 and .06);

  \end{scope}

  \end{scope}

  \end{scope}

  \begin{scope}[shift={(12,0)}]

  \begin{scope}[scale=1]

  \shade[ball color=gray!40, opacity=.6] (0,0) circle (2);
  \draw (-2,0) arc (180:360:2 and 0.6);
  \draw[dashed] (2,0) arc (0:180:2 and 0.6);

  \draw[decorate] (0,0) circle   (3);
  \draw (-114:1.5) node
    {
      \large
      \color{blue}
      $\mathbb{C}P^1$
    };
  \draw (-50:3.8)
    node
    {
      \scalebox{1.5}{
      \tiny
      \color{darkblue}
      \bf
      \begin{tabular}{c}
        higher cells
      \end{tabular}
      }
    };

  \end{scope}

  \end{scope}

  \begin{scope}
  \clip
    (9,2)
    rectangle
    (10,-2);
  \draw[line width=6, white]
    (-.7,1) to[bend left=20] (10.4,1);
  \draw[line width=6, white]
    (-.7,-1) to[bend right=10] (10.4,-1);
  \end{scope}

  \draw[line width=1.2, olive, arrows={[scale=1.3]|->[scale=2.9]}]
    (-.7,1) to[bend left=20] (10.4,1);
  \draw[line width=1.2, olive, arrows={[scale=1.3]|->[scale=2.9]}]
    (-.7,-1) to[bend right=10] (10.4,-1);

\draw
  (6,-3.2)
  node
  {
    \scalebox{1.3}{$
  [ c ]
  \;\in\;
  \underset{
    \mathclap{
    \!\!\!\!\!\!\!\!\!\!\!\!\!\!
    \!\!\!\!\!\!\!\!\!\!\!\!\!\!
    \mbox{
      \tiny
      \color{darkblue}
      \bf
      \begin{tabular}{c}
        charge = homotopy class
      \end{tabular}
    }
    }
  }{
  \big\{
    X
    \longrightarrow
    B \mathrm{U}(1)
  \big\}_{
    \scalebox{.7}{$
      \!\!\big/\mathrm{hmpty}
    $}
  }
  }
  \;\simeq\;\;
  \underset{
    \mathclap{
    \mbox{
      \tiny
      \color{darkblue}
      \bf
      \begin{tabular}{c}
        \\
        charge
        \\
        lattice
      \end{tabular}
    }
    }
  }{
    \mathbb{Z}
  }
  $}
  };

\end{tikzpicture}
}
}

\\
\cline{2-2}

\begin{minipage}[left]{6cm}

\noindent $\mathclap{\phantom{\vert^{\vert^{\vert^{\vert}}}}}$
While magnetic monopoles remain
hypothetical, the same mechanism governs
magnetic flux quantization in superconductors
(e.g. \cite[\S 2]{Chapman00}), which is
experimentally observed. The difference here
is that, instead of removing the worldline of a
singular point source from
spacetime, fields are required to {\it vanish at infinity}
along some directions -- here: along a plane perpendicular to the superconductor,
e.g. \cite[\S IV.B]{AGZ98}.
The result is integer numbers of unit flux tubes: vortex strings \cite{NielsenOlesen73}
(review in \cite{Tong09}).

\end{minipage}

&

\raisebox{-97pt}{
\scalebox{.75}{
\begin{tikzpicture}[decoration=snake]

  \shade[right color=lightgray, left color=white]
    (3,-3)
      --
    (-1,-1)
      --
        (-1.21,1)
      --
    (2.3,3);

  \draw[dashed]
    (3,-3)
      --
    (-1,-1)
      --
    (-1.21,1)
      --
    (2.3,3)
      --
    (3,-3);

  \begin{scope}[rotate=(+8)]
  \draw[dashed]
    (1.5,-1)
    ellipse
    (.2 and .37);
  \draw
   (1.5,-1)
   to node[above]{
     \;\;\;\;\;\;\;\;\;\;\;\;\;
     \rotatebox[origin=c]{7}{
     \scalebox{.7}{
     \color{orangeii}
     \bf
     flux tube
     }
     }
   }
   (-2.2,-1);
  \draw
   (1.5+1.2,-1)
   to
   (4,-1);
  \end{scope}

  \begin{scope}[shift={(-.2,1.4)}, scale=(.96)]
  \begin{scope}[rotate=(+8)]
  \draw[dashed]
    (1.5,-1)
    ellipse
    (.2 and .37);
  \draw
   (1.5,-1)
   to
   (-2.3,-1);
  \draw
   (1.5+1.35,-1)
   to
   (4.1,-1);
  \end{scope}
  \end{scope}

  \begin{scope}[shift={(-1,.5)}, scale=(.7)]
  \begin{scope}[rotate=(+8)]
  \draw[dashed]
    (1.5,-1)
    ellipse
    (.2 and .32);
  \draw
   (1.5,-1)
   to
   (-1.8,-1);
  \end{scope}
  \end{scope}

  \begin{scope}[shift={(9.6,.6)}, scale=(.9)]

  \begin{scope}[scale=1]

  \shade[ball color=gray!40, opacity=.6] (0,0) circle (2);
  \draw (-2,0) arc (180:360:2 and 0.6);
  \draw[dashed] (2,0) arc (0:180:2 and 0.6);

  \draw[decorate] (0,0) circle   (3);
  \draw (-114:1.5) node
    {
      \large
      \color{blue}
      $\mathbb{C}P^1$
    };
  \draw (+90:2.2) node
    {
      \scalebox{.8}{
        $0$
      }
    };
  \draw (-90:2.2) node
    {
      $\infty$
    };

  \draw (-50:3.7)
    node
    {
      \scalebox{1.2}{
      \tiny
      \color{darkblue}
      \bf
      \begin{tabular}{c}
        higher cells
      \end{tabular}
      }
    };

  \end{scope}

  \end{scope}

  \begin{scope}
  \clip
    (2,0)
    rectangle
    (8,4);
  \draw[line width=3.5, white]
    (3.3,.9) to[bend left=20] (9.5,2.38);
  \end{scope}
  \draw[line width=1, olive, arrows={[scale=1.2]|->[scale=2]}]
    (3.3,.9) to[bend left=20] (9.5,2.38);

  \begin{scope}
  \clip
    (2,0)
    rectangle
    (8,-4);
  \draw[line width=3.5, white]
    (2.92,-2.2) to[bend left=7] (9.5,-1.2);
  \end{scope}
  \draw[line width=1, olive, arrows={[scale=1.2]|->[scale=2]}]
    (2.92,-2.2) to[bend left=7] (9.5,-1.2);

 \draw
   (1,4)
   node
   {
     $
       \overset{
         \;\;\;
         \scalebox{1.1}{$
         \underset{
           \raisebox{3pt}{
             \tiny
             \color{darkblue}
             \bf
             \begin{tabular}{c}
               spacetime seen by fields
               vanishing at transversal infinity
             \end{tabular}
           }
         }{
           \scalebox{.8}{$
           X \,:=\,
           \mathbb{R}^{1,1}
           \times
           \mathbb{R}^2_{\mathrm{cpt}}
           \,\simeq\,
           \mathbb{R}^{1,1}
           \times
           S^2
           $}
         }
         $}
       }{
         \overbrace{
           \phantom{------------------}
         }
       }
     $
   };

 \draw
   (5.4,4.3)
   node
   {
     \scalebox{1.5}{
     \xymatrix@C=30pt{
      \ar[rr]^-{\scalebox{.55}{$c$}}_-{
        \scalebox{.7}{
          \tiny
          \color{greenii}
          \bf
          \begin{tabular}{c}
            magnetic flux
            \\
            through transversal plane
          \end{tabular}
        }
      }
      &&
     }
     }
   };

 \draw
   (9.6,4.1)
   node
   {
     $
       \overset{
         \;
         \scalebox{1.1}{$
           \underset{
             \raisebox{3pt}{
               \tiny
               \color{darkblue}
               \bf
               classifying space for ordinary cohomology
             }
           }{
           \scalebox{.8}{$
           B \mathrm{U}(1)
           \,=\,
           \mathbb{C}P^\infty
           $}
           }
           \mathclap{\phantom{\big(\mathbb{R}^2 \big)_{\mathrm{cpt}}}}
         $}
       }{
       \overbrace{
         \phantom{------------------}
       }
       }
     $
   };

\draw
  (6.3,-3.4)
  node
  {
    \scalebox{1}{$
  [ c ]
  \;\in\;
  \underset{
    \mathclap{
    \!\!\!\!\!\!\!\!\!\!\!\!\!\!
    \!\!\!\!\!\!\!\!\!\!\!\!\!\!
    \mbox{
      \tiny
      \color{darkblue}
      \bf
      \begin{tabular}{c}
        total flux = homotopy class
      \end{tabular}
    }
    }
  }{
  \big\{
    X
    \longrightarrow
    B \mathrm{U}(1)
  \big\}_{
    \scalebox{.7}{$
      \!\!\big/\mathrm{hmpty}
    $}
  }
  }
  \;\simeq\;\;
  \underset{
    \mathclap{
    \mbox{
      \tiny
      \color{darkblue}
      \bf
      \begin{tabular}{c}
        \\
        charge
        \\
        lattice
      \end{tabular}
    }
    }
  }{
    \mathbb{Z}
  }
  $}
  };

\end{tikzpicture}
}
}

\end{tabular}

\vspace{-.0cm}

\noindent Notice that this
{\it charge/flux quantization reveals the proper nature of
the electromagnetic field},
as it reflects the physical reality
of the ``vector potential'', hence of the photon's gauge field,
not seen in the old Maxwell theory.

\medskip

\noindent {\bf Example: Nuclear and gravitational charge in 1st non-abelian cohomology.}
Analogous comments apply to the nuclear force fields,
and the Atiyah-Hitchin quantization law of their
monopole solutions \cite{AtiyahHitchin88}.
In fact, analogous comments also apply to the field of
gravity, as reflected by gravitational instantons
(review in \cite[\S 10.2]{EguchiGilkeyHanson80}).
In these cases, the cohomology theory in question is
{\it non-abelian} (see \cite[\S 2, \S 4.2]{FSS20c} for pointers),
represented by the classifying spaces
$B G$ of the respective structure/gauge groups $G$, namely  $S\mathrm{U}(n)$ and
$\mathrm{Spin}(n)$.

\medskip

\label{OpenProblemOfUnifiedFundamentalTheory}
\noindent {\bf Open problem: Charge quantization in a unified fundamental theory.}
While charge quantization in gauge theories and in
gravity is thus well understood separately,
their expected unification in a ``theory of everything''
such as string theory
is as famous as it remains incomplete, certainly in
the putative non-perturbative completion
(envisioned in the ``second superstring revolution'' \cite{Schwarz96})
to M/F-theory (e.g. \cite{Duff96}\cite{Duff99}\cite{BBS06})
whose actual (namely: mathematical)
formulation remains a wide open problem
(e.g., \cite[\S 6]{Duff96}\cite[p. 6]{Duff98}\cite[p. 330]{Duff99}\cite[\S 12]{Moore14}\cite[$@$21:15]{Witten19}\cite[$@$17:14]{Duff19}).
This is the problem with which we are concerned here.

\medskip

\noindent {\bf Example: D-Brane charge in K-theory?}
\label{DBraneChargeInKTheory}
A prominent example of identifying a physical theory
by determining its charge quantization law in generalized cohomology
is the conjecture that the charges of D-branes seen in string theory
are quantized in topological K-cohomology \cite{MinasianMoore97}
(see also \cite{Witten01}\cite{Evslin06}\cite{EvslinSati06}\cite{Fredenhagen08}\cite{GS19c}).
This

\vspace{1mm}
\hspace{-.9cm}
\begin{tabular}{ll}

\begin{minipage}[left]{12cm}
conjecture is
largely based on the suggestion \cite[\S 3]{Witten98}
that, in turn, {\it Sen's conjecture}
about tachyon condensation
of open superstring states stretched between D-branes/anti-branes
\cite{Sen98} (review in \cite{Sen05})
should imply the defining equivalence relation of topological K-theory
on the Chan-Paton vector bundles carried by D-branes.
\\
This suggestion still remains to be checked
(see \cite[p. 35]{Erler13}\cite[p. 112]{Erler19}).
Even assuming this conjecture,
open problems remain within the K-theory hypothesis itself:
\end{minipage}

&

\hspace{.4cm}

\raisebox{-2pt}{
\begin{minipage}[left]{5.6cm}

\scalebox{.94}{
\begin{tikzpicture}
\def\reduce{.4}
\begin{scope}[shift={(-.8,0)}]
  \draw[fill = black]
    (-.05,1.5-\reduce) rectangle (+.05,-1.5+\reduce);
  \draw[fill=white]
    (0,0) circle (.23);
  \draw[fill=lightgray, fill opacity=.6]
    (0,0) circle (.23);
  \begin{scope}
    \clip
      (0,0) circle (.22);
    \draw (0,0)
      node
      {
        \color{blue}
        $\mathcal{V}$
      };
  \end{scope}
\end{scope}

\begin{scope}[shift={(-0,0)}]
  \draw[fill = black]
    (-.05,1.5-\reduce) rectangle (+.05,-1.5+\reduce);
  \draw[fill=white]
    (0,0) circle (.23);
  \draw[fill=lightgray, fill opacity=.6]
    (0,0) circle (.23);
  \begin{scope}
    \clip
      (0,0) circle (.22);
    \draw (0,0)
      node
      {
        \color{blue}
        $\mathcal{W}$
      };
  \end{scope}
\end{scope}

\begin{scope}[shift={(+.8,0)}]
  \draw[fill=lightgray, fill opacity=.6]
    (0,0) circle (.23);
  \draw[fill=white, draw opacity=0]
    (-.05,1.5-\reduce) rectangle (+.05,-1.5+\reduce);
  \draw
    (-.05,1.5-\reduce) to (-.05,-1.5+\reduce);
  \draw
    (+.05,1.5-\reduce) to (+.05,-1.5+\reduce);
  \draw[fill=white, draw opacity=0]
    (0,0) circle (.23);
  \draw[fill=lightgray, draw opacity=0, fill opacity=.6]
    (0,0) circle (.23);
  \begin{scope}
    \clip
      (0,0) circle (.22);
    \draw (0,0)
      node
      {
        \color{blue}
        \raisebox{1pt}{
          $\overline{\mathcal{W}}$
        }
      };
  \end{scope}
\end{scope}

  \draw
    (2,0)
    node
    {
      \scalebox{2.4}{
        $\rightleftharpoons$
      }
    };

\begin{scope}[shift={(+3.2,0)}]
  \draw[fill = black]
    (-.05,1.5-\reduce) rectangle (+.05,-1.5+\reduce);
  \draw[fill=white]
    (0,0) circle (.23);
  \draw[fill=lightgray, fill opacity=.6]
    (0,0) circle (.23);
  \begin{scope}
    \clip
      (0,0) circle (.22);
    \draw (0,0)
      node
      {
        \color{blue}
        $\mathcal{V}$
      };
  \end{scope}
\end{scope}

 \draw[white,line width=1.4]
   (-1, 1.32-\reduce) to (3.4, 1.32-\reduce);
 \draw[white,line width=1.4]
   (-1, 1.1-\reduce) to (3.4, 1.1-\reduce);
 \draw[white,line width=1.4]
   (-1, .89-\reduce) to (3.4, .89-\reduce);

 \draw[white,line width=1.4]
   (-1, -1.32+\reduce) to (3.4, -1.32+\reduce);
 \draw[white,line width=1.4]
   (-1, -1.1+\reduce) to (3.4, -1.1+\reduce);
 \draw[white,line width=1.4]
   (-1, -.89+\reduce) to (3.4, -.89+\reduce);

  \draw
    (2,-.6)
    node
    {
      \tiny
      \color{greenii}
      \bf
      \def\arraystretch{.75}
      \begin{tabular}{c}
        tachyon
        \\
        condensation
      \end{tabular}
    };

\draw
  (-.4,-1.8+\reduce)
  node
  {
    \tiny
    \color{darkblue}
    \bf
    D-branes
  };

\draw
  (+.8,-1.8+\reduce)
  node
  {
    \tiny
    \color{darkblue}
    \bf
    \def\arraystretch{.75}
    \begin{tabular}{c}
      anti-
      \\
      D-brane
    \end{tabular}
  };

\end{tikzpicture}
}

\end{minipage}
}

\end{tabular}

\vspace{3pt}

\noindent {\bf (a)} In type IIB, it is incompatible with S-duality
(see \cite{KrizSati05a}\cite[\S 8.3]{Evslin06}).

\noindent {\bf (b)} In type IIA, it requires with the D8-brane also a D(-2)-brane,
which remain mysterious (e.g. \cite{AJTZ10}).

\noindent {\bf (c)} K-theory seems to predict physically spurious charges
(e.g. \cite[\S 4.5.2]{dBDHKMMS02}\cite[p. 1]{FredenhagenQuella05}),
such as irrational charges
(\cite[(2.8)]{BachasDouglasSchweigert00}; for more discussions and recent progress see \cite{BSS19}).

\noindent {\bf (d)}
The lift of K-theoretic charge quantization
to M-theory has remained elusive
(e.g. \cite[\S 4.6.5]{dBDHKMMS02}).

\medskip

\noindent {\bf Example: M-Brane charge in modified ordinary cohomology?}
\label{MBraneChargeInModifiedOrdinaryCohomology}
It has been tradition to describe the expected
M-brane charge quantization via ordinary cohomology
equipped with an incremental sequence of constraints
(e.g. \cite{Witten97a}\cite{Witten97b}\cite{DMW00}\cite{DFM03}).
From the
obstruction-theoretic perspective of
\cite{KrizSati04}\cite{KSpin}\cite{Sati10b}
this looks just like the first steps of
lifting through a character map
(p. \pageref{ChargeQuantizationLaw}),
via an Atiyah-Hirzebruch spectral sequence
of a generalized cohomology for M-theory,
without that cohomology theory having been determined yet.

\vspace{.1cm}

\hspace{-.9cm}
\begin{tabular}{ll}

\begin{minipage}[left]{13.2cm}
As a result, the list of conditions to be imposed
on ordinary cohomology seems open ended
(see \cite[Table 1]{FSS19b}\cite{GS21}),
with every new phenomenon argued for in the string physics literature
being another potential candidate to add to the list;
such as non-abelian DBI effects (review in \cite{Myers03})
including M-brane polarizations,
seen only through more recent developments in M2-brane theory
(review in \cite[\S 6]{BLMP13}).

This tradition is hence a perpetual {\it bottom-up approach} to formulating M-theory.
\\
In contrast,
following \cite{Sati10b}\cite{Sati14}\cite{FSS19b}\cite{FSS19c}\cite{SS19a}\cite{SS19b},
we consider here
a {\it top-down approach}:
{\it postulating} a generalized cohomology theory
for charge quantization in M-theory  as advocated early on in \cite{Sa05a}\cite{Sa05b}\cite{Sa06},
and then rigorously {\it deriving} its implications for physics.
\end{minipage}

&

\hspace{.5mm}

\raisebox{2pt}{
\begin{minipage}[center]{4.1cm}
\begin{center}
\hspace{-.7cm}
\scalebox{.9}{
$
  \xymatrix@R=35pt{
    \fbox{
      \tiny
      {\begin{tabular}{c}
        Generalized
        \\
        cohomology
      \end{tabular}}
    }
    \ar@{<..}@<-15pt>[dd]|-{
      \mbox{
        \tiny
        {\begin{tabular}{c}
          \color{darkblue}
          \bf
          Bottom-up approach:
          \\
          incrementally
          \\
          impose constraints
          \\
          suggested by
          \\
          physics arguments
        \end{tabular}}
      }
      \;\;\;\;\;\;\;\;\;\;\;\;\;\;\;\;\;\;\;\;\;
    }
    \ar@{->}@<+15pt>[dd]|-{
      \;\;\;\;\;\;\;\;\;\;\;\;\;\;\;\;\;\;\;\;\;
      \mbox{
        \tiny
        \begin{tabular}{c}
          \color{darkblue}
          \bf
          Top-down approach:
          \\
          systematically
          \\
          derive constraints
          \\
          rigorously obtained by
          \\
          mathematical analysis
        \end{tabular}
      }
    }
    \\
    \\
    \fbox{
      \tiny
      \begin{tabular}{c}
        Ordinary
        \\
        cohomology
      \end{tabular}
    }
  }
$
}
\end{center}
\end{minipage}
}

\end{tabular}

\medskip


\newpage

\noindent
{\bf M-brane charge in complex-oriented cohomology?}
\label{MBraneChargeInComplexOrientedCohomology}
Several arguments suggest
\cite{KrizSati04}\cite{KrizSati05a}\cite{KrizSati05b}\cite{Sati06b} \cite{Sati10b}
that charge quantization in M/F-theory
involves {\it elliptic cohomology}.
Interestingly,  all these candidates --

\vspace{.1cm}

\hspace{-.9cm}
\begin{tabular}{ll}

\begin{minipage}[left]{7.8cm}

(0) ordinary cohomology, (1) complex K-theory, (2) elliptic cohomology
--
are {\it complex-orientable} cohomology theories
(see \cref{Orientation}). As such, they are
the first three stages in the {\it chromatic tower}
of cohomology theories (see \cite{Ravenel86}\cite{Lurie10}),
whose culmination is
complex {\it Cobordism cohomology} $M \mathrm{U}$
(e.g. \cite{Kochman96}).
Moreover, complex Cobordism stands out in that
the initial map
$\mathbb{S} \simeq M \mathrm{Fr} \xrightarrow{\;} M \mathrm{U}$
\eqref{dInvariantAndBoardman}
that it receives from
{\it framed Cobordism cohomology}
(see Example \ref{ExamplesOfMultiplicativeCohomologyTheories}
for pointers)
is a co-cover, meaning that
computations in $M \mathrm{U}$-theory
co-descend to reveal $M\mathrm{Fr}$-theory;
this is the statement of the
{\it Adams-Novikov spectral sequence}
(\cite[Thm. 3.1, 3.3]{Novikov67}\cite[Thm. 3.1]{Ravenel78},
see \cite{Wilson13} for modern exposition).
One may wonder:

\vspace{2mm}
{\it Is M/F-theory charge quantized in $M\mathrm{Fr}$-theory?}

\end{minipage}

&

\hspace{-.4cm}

\begin{minipage}[left]{9cm}

$$
  \xymatrix@R=5pt@C=8pt{
    \mathllap{
      \mbox{
        \tiny
        \color{darkblue}
        \bf
        \begin{tabular}{c}
          stable
          \\
          Cohomotopy
        \end{tabular}
      }
      \!\!
    }
    \mathbb{S}
    \ar[drrrr]
      ^-{
        \;
        \mathclap{\phantom{\vert^{\vert}}}
        \mbox{
          \tiny
          \color{greenii}
          \bf
          Adams-Novikov co-cover
        }
        \mathclap{\phantom{\vert_{\vert}}}
        \;
      }
    \ar@{=}[d]
    \\
    \underset{
      \mathclap{
      \raisebox{-6pt}{
        \tiny
        \color{darkblue}
        \bf
        \begin{tabular}{c}
          framed
          \\
          Cobordism
        \end{tabular}
      }
    }
    }{
      M\mathrm{Fr}
    }
    \ar[rr]
    &&
    \underset{
      \mathclap{
      \raisebox{-3pt}{
        \tiny
        \color{darkblue}
        \bf
        \begin{tabular}{c}
          quaternionic
          \\
          Cobordism
        \end{tabular}
      }
      }
    }{
      M\mathrm{Sp}
    }
    \ar[rr]
    &&
    M\mathrm{U}
    \mathrlap{
      \;\;\;
      \mbox{
        \tiny
        \color{darkblue}
        \bf
        complex Cobordism
      }
    }
    \\
    && &&
    \mbox{$\vdots$}
    \mathrlap{
      \;\;\;\;
      \mbox{
        \tiny
        \color{darkblue}
        \bf
        \begin{tabular}{l}
          Morava K-theories/
          \\
          CY-cohomology
        \end{tabular}
      }
    }
     \ar@{}@<-24pt>[ddd]
       |-{
         \rotatebox{90}{
           \tiny
           \color{orangeii}
           - - -
           {\bf
           chromatic tower
           }
           - - -
         }
       }
    \\
    && &&
    \mathrm{E}\ell\ell
    \mathrlap{
      \;\;\;\;
      \mbox{
        \tiny
        \color{darkblue}
        \bf
        elliptic cohomology
      }
    }
    \\
    &&
    K \mathrm{O}
    &&
    K \mathrm{U}
    \mathrlap{
      \;
      \;\;\;
      \mbox{
        \tiny
        \color{darkblue}
        \bf
        K-theory
      }
    }
    \\
    && &&
    H^{\mathrm{ev}} \! R
    \mathrlap{
      \;\;\;
      \mbox{
        \tiny
        \color{darkblue}
        \bf
        ordinary cohomology
      }
    }
    \\
    &&
    \mbox{
      \tiny
      \color{orangeii}
      \bf
      \begin{tabular}{c}
        quaternionic
        \\
        oriented
      \end{tabular}
    }
    \ar@{}[rr]|-{ \supset }
    &&
    \mbox{
      \tiny
      \color{orangeii}
      \bf
      \begin{tabular}{c}
        complex
        \\
        oriented
      \end{tabular}
    }
  }
$$

\end{minipage}

\end{tabular}

\medskip

\vspace{.1cm}

\hypertarget{HypothesisH}{}
\noindent {\bf Hypothesis H: M-brane charge in Cohomotopy theory.}
Given that M-theory is supposedly the pinnacle of
fundamental physics,
we suspect
that M-brane charge quantization
corresponds not to random but rather to the
deep phenomena seen in algebraic topology.
This idea has been the guiding light for
investigations into M-theory in
\cite{Sati10b}\cite{Sati11a}\cite{Sati11b}\cite{Sati14}.
But the most fundamental of all
cohomology theories is
Borsuk-Spanier {\it Cohomotopy theory} $\pi^\bullet$
\eqref{CohomotopyInIntroduction}
whose stabilization is
the Whitehead-generalized cohomology theory
{\it stable Cohomotopy} $\mathbb{S}^\bullet$ \eqref{StableCohomotopy},
which the Pontrjagin-Thom construction identifies
\eqref{BulkInteractionsAsStabilization}
with framed Cobordism cohomology $M\mathrm{Fr}^\bullet$.

\medskip

\noindent Our {\it Hypothesis H}
posits that
{\it M-brane charge is quantized in 4-Cohomotopy}
$\pi^4$
\cite[\S 2.5]{Sati13}\cite{FSS15b}\cite{FSS16a}
(exposition in \cite[\S 7, 12.2]{FSS19a});
specifically in tangentially twisted 4-Cohomotopy
\cite{FSS19b}\cite{FSS19c}, which
on flat orbifolds means \cite[Thm. 5.16]{SS20b}
tangentially $RO$-graded equivariant Cohomotopy
\cite{HSS18}\cite{SS19a}\cite{BSS19}.

\noindent Evidence for Hypothesis H comes from
analyzing the image of twisted Cohomotopy...

\vspace{-.1cm}
\begin{itemize}

\vspace{-.2cm}
\item [{\bf (a)}]
... in  ordinary rational cohomology
\cite[\S 7]{FSS19a}\cite{FSS19b}\cite{FSS19c}\cite{HSS18}\cite{BMSS19}\cite{SS20a}\cite{FSS20b};
\\
\phantom{...}
under the cohomotopical character map
\cite[\S 5.3]{FSS20c}\cite{SS20c};

\vspace{-.25cm}
\item [{\bf (b)}]
... in ordinary integral cohomology and differential cohomology  \cite{GS21}, via the Postnikov tower;

\vspace{-.25cm}
\item[{\bf (c)}]
... in K-theory \cite{SS19a}\cite{BSS19},
under the Hurewicz-Boardman homomorphism;

\vspace{-.25cm}
\item [{\bf (d)}]
... in equivariant Cobordism  \cite{SS19a},
under the Pontrjagin construction;

\vspace{-.25cm}
\item [{\bf (e)}]
... in configuration spaces \cite{SS19b}, under the May-Segal theorem.

\end{itemize}

\medskip

\hypertarget{MFTheory}{}
\noindent
{\bf M/F-Brane charge in $M\mathrm{Fr}$-Theory.}
Here we explain how \hyperlink{HypothesisH}{Hypothesis H},
specialized
\eqref{ChargeQuantizationIn4CohomotopyOnHomotopicallyFlatSpacetime}
to homotopically flat spacetimes
(Remark \ref{FramedSpacetimes}, hence to the situation where the
tangential twisting of Cohomolotopy disappears),
connects expected physics of M/F-theory
with the conceptual heart of the mathematical field of
{\it stable homotopy theory and generalized cohomology}
\cite{Adams74}. Since this field may be advertized
(e.g. \cite[p. xv]{Ravenel86}\cite[\S 1]{MahowaldRavenel87}\cite[p. 1]{Wilson13}\cite[\S 1]{IsaksenWangXu20})
as revolving all around the
stable homotopy of spheres $\mathbb{S}_{-\bullet}$,
hence the stable Cohomotopy $\mathbb{S}^{\bullet}$,
and so equivalently the framed Cobordism cohomology $M\mathrm{Fr}^\bullet$,
it makes sense to refer to it as
{\it $M\mathrm{Fr}$-theory}, for short.
We find that the correspondence that we lay out in
\cref{TheDictionary} (summarized in \cref{Conclusions})

\vspace{-2mm}
$$
  \xymatrix{
    \underset{
      \mathclap{
      \raisebox{-3pt}{
        \tiny
        \color{darkblue}
        \bf
        \begin{tabular}{c}
          fundamental physics
        \end{tabular}
      }
      }
    }{
      \mbox{M/F-Theory}
    }
    \ar@{<->}[rr]
      ^-{
        \mbox{
          \tiny
          \color{greenii}
          \bf
          Hypothesis H
        }
      }
    &{\phantom{AAA}}&
    \underset{
      \mathclap{
      \raisebox{-3pt}{
        \tiny
        \color{darkblue}
        \bf
        \begin{tabular}{c}
          algebraic topology
        \end{tabular}
      }
      }
    }{
      \mbox{$M\mathrm{Fr}$-Theory}
    }
  }
$$
not only supports the Hypothesis that M-brane charge
is quantized in Cohomotopy theory, by showing that
theorems in $M\mathrm{Fr}$-theory rigorously imply
effects expected in the physics of M/F-theory, but also
illuminates $M\mathrm{Fr}$-theory by providing coherent physics intuition
for several of its fundamental definitions and theorems
(discussed in \cref{BordismAndStableHomotopyForMTheory}).

\newpage

\section{The correspondence}
\label{TheDictionary}

We develop a correspondence (summarized in \cref{Conclusions})
that translates, under \hyperlink{HypothesisH}{Hypothesis H},
fundamental concepts expected in M-theory on
flat (Rem. \ref{FramedSpacetimes})
and fluxless (Rem. \ref{FluxlessBackgrounds})
backgrounds
to fundamental concepts of generalized cohomology and of stable homotopy theory
(\hyperlink{MFTheory}{$M\mathrm{Fr}$-theory}).
This section is expository, focusing on conceptual identifications.
All definitions and proofs that we appeal to are
treated rigorously in \cref{BordismAndStableHomotopyForMTheory}.

\medskip

Before entering the discussion below in \cref{FullMBraneChargeAndUnstableCohomotopyTheory},
here are some comments on our running assumption of
flat and fluxless backgrounds:

\medskip

\begin{remark}[Focus on cohomologically fluxless backgrounds]
  \label{FluxlessBackgrounds}
  From \cref{M5ThreeFlux} on we focus on
  backgrounds for which the 4-flux $G_4$ is trivialized
  in cohomology
  (in {\it generalized} cohomology, that is, see \eqref{FlatDifferential4Cohomotopy}),
  its trivialization being the
  twisted $H_3$-flux, or equivalently the sum of the $C_3$-field
  with a closed $H_3$-flux \eqref{TrivializationOfZTorsorOfKU3Fluxes}.
  Seen in ordinary rational cohomology
  $H\mathbb{R}$, the $G_4$-flux is represented as a closed
  differential form, the {\it flux density},
  whose vanishing in cohomology means that the background is {\it fluxless}
  in that the integrated 4-flux through any closed 4-manifold
  inside spacetime vanishes. Beware that the literature
  sometimes conflates ``vanishing flux''
  (i.e. flux vanishing in cohomology)
  with ``vanishing flux density''.
  The distinction is fully resolved in enhancement of the
  present considerations to {\it differential} cohomology
  (as in \cite[\S 4.3]{FSS20c}),
  which we hope to discuss elsewhere
  (but see Remark \ref{H3FluxesAsExtraordinaryFlatDifferentialCohomotopy}).
\end{remark}

\begin{remark}[Focus on homotopically flat spacetimes]
  \label{FramedSpacetimes}
  We focus on
  spacetime topologies with trivializable tangent bundles,
  hence admitting a framing, as in \cite{Sati13}\cite{Sati14}.
  For such
  ``homotopically-flat'' spacetimes,
  Hypothesis H says
  \eqref{ChargeQuantizationIn4CohomotopyOnHomotopicallyFlatSpacetime}
  that M-brane charge is measured in
  plain Cohomotopy
  (in contrast to tangentially twisted Cohomotopy, since the
  tangential twist is given by the class of the tangent bundle)
  on which we wish to concentrate for focus of presentation
  (details in \cref{MBraneWorldvolumesAndThePontrjaginConstruction}).

  \noindent
  {\bf (a)}
  Typical examples of homotopically flat spacetimes of interest here are
  product manifolds of Minkowski spacetimes $\mathbb{R}^{d,1}$ with unit spheres in
  a normed real division algebra, i.e. with
  0-spheres $S^0 \,=\, \ast \sqcup \ast$, circles $S^1$,
  3-spheres $S^3$ and 7-spheres $S^7$ (\cite[p. 21]{Adams60}).

  \noindent
  {\bf (b)}
  Restriction to homotopically flat spacetimes is of course a strong constraint,
  which is however alleviated by understanding
  it in the generality of {\it equivariant} homotopy theory
  (specifically: equivariant Cohomotopy \cite{SS19a}\cite[Def. 5.28]{SS20b})
  where flat spacetimes include flat orbifolds (see pointers in \cite[\S 1]{SS20b}):
  Such ``flat'' (often: ``Euclidean'')
  orbifolds are really curved spaces for which
  all curvature is concentrated in singularities.
  While we do not discuss this here, the point is that
  all the algebraic topology that we do discuss has
  fairly immediate equivariant generalization
  applicable to this case.

  \noindent
  {\bf (c)}
  While restrictive, the special case is of central interest:
  Much of the particle physics model building in string theory
  (review in \cite{IbanezUranga12})
  utilizes flat orbifold spacetimes (review in \cite{Reffert06}).
  Moreover, flat and locally fluxless
  orbifold spacetimes are, as ``universal spacetimes'' \cite{CGHP08},
  among the few known classes of
  exact M-theory backgrounds, i.e. those satisfying supergravity
  equations of motion subject to the full tower of M-theoretic
  higher curvature corrections
  (whatever these may be, see e.g. \cite{CGNT05}).

  \noindent
  {\bf (d)}
  Since we consider M-brane charge {\it vanishing at infinity}
  (see \cref{MBraneWorldvolumesAndThePontrjaginConstruction}),
  we may think of non-compact flat orbifolds
  as local charts inside a global curved spacetime orbifold with
  M-brane charge localized inside them.
  This way our considerations here provide
  the local building blocks
  for a fully general discussion of $E$-cohomological M-brane charge
  on globally curved spacetimes
  (to be given elsewhere).

\noindent
{\bf (e)}
Finally, while flat Minkowski spacetimes themselves are homotopically trivial,
it is this constraint that charges vanish at infinity
which makes them appear to their charge cohomology theory
as {\it effective} spheres with non-trivial topology.
\end{remark}

We close these introductory remarks by further amplification of this last point:

\newpage

\begin{remark}[Generalized charge cohomology vanishing at infinity]
It is a familiar fact in (higher) gauge theory,
recalled in a moment, that the
(non-abelian and generalized) cohomology
of flat Euclidean $n$-space, subject to the constraint that
{\it fluxes vanish at infinity}, is equivalent to the unconstrained cohomology
of the $n$-sphere \eqref{IncarnationsOfCompactlySupportedCohomologyOnEuclideanSpace}:
This fact governs the usual classification of
gauge instantons in 4d \eqref{TopologicalClassificationOfBPSTInstantons}
(and of Skyrmions in 3d)
as well as the traditionally conjectured classification of D-brane charges in K-theory
\eqref{DBraneChargeForFlatTransversalSpaceSeenInKTheory}.
While this fact is classical and widely used, it may be worthwhile
to recall some of its aspects, applications and pertinent references.
The discussion in the main text
(notably in \cref{MBraneWorldvolumesAndThePontrjaginConstruction})
is a direct application of this classical fact; indeed it is the direct analog
of \eqref{DBraneChargeForFlatTransversalSpaceSeenInKTheory},
with K-theory replaced by Cohomotopy.

\medskip

Despite the prominent role that $n$-spheres play in the discussion
of finite-flux configurations on flat Euclidean space,
it is important to beware that the topological $n$-sphere
in question is just a mathematical tool for reasoning about
fields on flat space subject to asymptotic constraints,
and not to be conflated with a curved spacetime.
In particular, there is no non-trivial contribution of the field
of gravity in this discussion, which means that the charge cohomology theories
in question are not twisted by non-trivial (generalized) tangent bundles.
One way to bring out the actual mechanism at work is:

\noindent
{\bf (1)}
to understand the
$n$-sphere as being just the homeomorphism type of
the ``one-point compactification'' of Euclidean $n$-space
(\cite{Aleksandrov24}\cite[p. 150]{Kelly55} review in \cite{Cutler20}), hence of
Euclidean $n$-space with a formal ``point at infinity'' adjoined
\eqref{OnePointCompactificationAndFunctionsOnIt}:

\vspace{-.9cm}
\begin{equation}
  \label{OnePointCompactificationInIntroduction}
  \overset{
    \raisebox{5pt}{
      \tiny
      \color{darkblue}
      \bf
      \begin{tabular}{c}
        flat Euclidean $n$-space
        \\
        with formal ``point-at-infinity
      \end{tabular}
    }
  }{
    \mathbb{R}^n_{\mathrm{cpt}}
  }
  \;\;
  \underset{
    \mathclap{
    \raisebox{-3pt}{
      \tiny
      \color{greenii}
      \bf
      homeomorphism
    }
    }
  }{
    \simeq
  }
  \;\;
  \overset{
    {
      \raisebox{5pt}{
        \tiny
        \color{darkblue}
        \bf
        \begin{tabular}{c}
          topological
          $n$-sphere
        \end{tabular}
      }
    }
  }{
    S^n
  }
\hspace{1cm}
\raisebox{-1.2cm}{
\begin{tikzpicture}
\draw (0,0) circle (1.3);
\draw (140:1.5) node {$S^n$};
\draw (-3.2,-1.3)
  to
  node[above, very near start]{$\mathbb{R}^n$}
  (3.2,-1.3);
\draw[dotted] (3.2,-1.3) to (3.8,-1.3);
\draw[dotted] (-3.2,-1.3) to (-3.8,-1.3);
\draw[densely dashed]
  (2.3,-1.3)
  to
  (0,1.3)
  ;
\draw[fill=black] (2.3,-1.3) circle (.05);
\draw[draw=gray, fill=gray] (1.29,-.152) circle (.05);
\draw[densely dashed]
  (0,1.3)
  to
  (1.2,-1.3)
  ;
\draw[draw=gray, fill=gray] (.99,-.84) circle (.05);
\draw[fill=black] (1.2,-1.3) circle (.05);
\draw[fill=white] (0,1.3) circle (.05);
\draw (0,1.3+.2) node {$\infty$};
\end{tikzpicture}
}
\end{equation}
This ``point at infinity'' is a mathematical tool for conceptualizing asymptotic
boundary conditions, {\it not} a point in spacetime. It may transparently be understood
as the pole $\infty \,\in\, S^n$ from which a stereographic projection conformally identifies $\mathbb{R}^n$ with the complement $S^n \setminus \{\infty\} \,\simeq\, \mathbb{R}^n$.

\noindent
{\bf (2)}
to notice that the
{\it cohomology with compact support}
of flat $\mathbb{R}^n$ is equivalent
to the reduced cohomology of its one-point compactification,
hence to pointed-homotopy classes
$\Truncation{0}\ConstrainedMaps{}{-}{-}$
of continuous maps from this formal $S^n$ to the given classifying space
\eqref{NonabelianCohomologyInIntroduction}:
\begin{equation}
  \label{IncarnationsOfCompactlySupportedCohomologyOnEuclideanSpace}
  \overset{
    {
    \raisebox{6pt}{
      \tiny
      \color{darkblue}
      \bf
      \begin{tabular}{c}
        $A$-cohomology
        \\
        with compact support
        \\
        of flat Euclidean space
      \end{tabular}
    }
    }
  }{
    H^0_{\mathrm{cpt}}
    \big(
      \mathbb{R}^n
      ;\,
      A
    \big)
  }
  \;\;
  \simeq
  \;\;
  \overset{
    {
    \raisebox{6pt}{
     \tiny
     \color{darkblue}
     \bf
     \begin{tabular}{c}
       pointed homotopy classes
       \\
       from one-point compactification
       \\
       to the classifyiong space
     \end{tabular}
     }
    }
  }{
    \Truncation{0}
    \ConstrainedMaps{\big}
      { \mathbb{R}^n_{\mathrm{cpt}} }
      { A }
  }
  \;\;
  \simeq
  \;\;
  \overset{
    \raisebox{6pt}{
      \tiny
      \color{darkblue}
      \bf
      \begin{tabular}{c}
        pointed homotopy classes
        \\
        from $n$-sphere
        \\
        to the classifying space
      \end{tabular}
    }
  }{
    \Truncation{0}
    \ConstrainedMaps{\big}
      { S^n }
      { A }
  }
  \;\;
  \simeq
  \;\;
  \overset{
    {
    \raisebox{6pt}{
      \tiny
      \color{darkblue}
      \bf
      \begin{tabular}{c}
        reduced $A$-cohomology
        \\
        of the $n$-sphere
      \end{tabular}
    }
    }
  }{
    \widetilde{A}(S^n)
  }
  \;\;
  \simeq
  \;\;
  \overset{
    {
    \raisebox{6pt}{
      \tiny
      \color{darkblue}
      \bf
      \begin{tabular}{c}
        $n$th homotopy group
        \\
        of classifying space
      \end{tabular}
    }
    }
  }{
    \pi_n(A)
    \,.
  }
\end{equation}
This follows since every continuous function $c \,:\, S^n \to A$ which vanishes at $\infty$
(in that it takes $\infty$ to the given base-point of the classifying space $A$)
is naturally homotopic to a function $c \circ \phi$ which vanishes in an open neighbourhood
$D^n_\infty \subset S^n$ of $\infty$, hence whose support is in the compact subset $S^n \setminus D^4$. Here $\phi : S^n \to S^n$ may be taken to
homeomorphically identify $S^n \setminus \overline{D^n_\infty}$ with $S^n \setminus \{\infty\}$
while sending $D^n_\infty \to \{\infty\}$. Such $\phi$ is homotopic to the identity on
$S^n$, as may readily be checked by explicit coordinate expressions, but also
follows on the general grounds of the Hopf degree theorem
(e.g. \cite[\S IX, Cor. 5.8]{Kosinski93}).

For instance, K-theory with compact support
on flat Euclidean spaces
is {\it defined} this way
\eqref{IncarnationsOfCompactlySupportedCohomologyOnEuclideanSpace}
as
\begin{equation}
  \label{CompactlySupportedKTheory}
  \overset{
    \raisebox{6pt}{
      \tiny
      \color{darkblue}
      \bf
      \begin{tabular}{c}
        compactly-supported K-theory
        \\
        of flat Euclidean space
      \end{tabular}
    }
  }{
    K_{\mathrm{cpt}}(\mathbb{R}^n)
  }
    \,:=\,
  \overset{
    \raisebox{6pt}{
      \tiny
      \color{darkblue}
      \bf
      \begin{tabular}{c}
        reduced K-theory of
        \\
        one-point compactification
      \end{tabular}
    }
  }{
    \widetilde{K}(\mathbb{R}^n_{\mathrm{cpt}})
  }
    \;=\;
  \overset{
    \raisebox{6pt}{
      \tiny
      \color{darkblue}
      \bf
      \begin{tabular}{c}
        reduced K-theory
        \\
        of $n$-sphere
      \end{tabular}
    }
  }{
    \widetilde{K}(S^n)
  }
\end{equation}
(e.g. \cite[\S II, Ex. 4.4]{Karoubi78}\cite[(2.20) and below (2.58)]{OlsenSzabo99Constructing})
and the analogous statement holds for any generalized cohomology theory
(e.g. \cite[p. 28]{MiskinValette03}).
The left hand side of \eqref{CompactlySupportedKTheory}
makes manifest that we are dealing
with flat spacetime, even though the equivalent re-expression on the right involves
an $n$-sphere domain.
\end{remark}

\medskip

\noindent
{\bf Classical Example: BPST instanton charge in non-abelian cohomology.}
When the coefficients in \eqref{IncarnationsOfCompactlySupportedCohomologyOnEuclideanSpace}
are
$A \,=\, B \mathrm{SU}(2)$, the classifying space for
$\mathrm{SU}(2)$-gauge bundles, then
\eqref{IncarnationsOfCompactlySupportedCohomologyOnEuclideanSpace}
recovers the traditional classification
(e.g. \cite[\S 8.2]{EguchiGilkeyHanson80})
of multi-center BPST-instantons on flat 4-space:
\begin{equation}
  \label{TopologicalClassificationOfBPSTInstantons}
  \overset{
    \mathclap{
    \raisebox{5pt}{
      \tiny
      \color{darkblue}
      \bf
      \begin{tabular}{c}
        non-abelian cohomology
        \\
        with compact support
        \\
        of flat Euclidean 4-space
      \end{tabular}
    }
    }
  }{
  H^1_{\mathrm{cpt}}
  \big(
    \mathbb{R}^4
    ;\,
    \mathrm{SU}(2)
  \big)
  }
  \;\;
  =
  \;\;
  H^0_{\mathrm{cpt}}
  \big(
    \mathbb{R}^4
    ;\,
    B \mathrm{SU}(2)
  \big)
  \;\;
  \simeq
  \;\;
  \overset{
    \raisebox{6pt}{
      \tiny
      \color{darkblue}
      \bf
      \begin{tabular}{c}
         pointed-homotopy classes
         \\
         from the $n$-sphere
         \\
         to the classifying space
      \end{tabular}
    }
  }{
    \Truncation{0}
    \ConstrainedMaps{}
      { S^4 }
      { B \mathrm{SU}(2) }
  }
  \;\;
  \simeq
  \;\;
  \pi_4
  \big(
    B \mathrm{SU}(2)
  \big)
  \;\;
  \simeq
  \;\;
  \pi_3
  \big(
    \mathrm{SU}(2)
  \big)
  \;\;
  \simeq
  \;\;
  \overset{
    \mathclap{
    \raisebox{6pt}{
      \tiny
      \color{darkblue}
      \bf
      \begin{tabular}{c}
        set of
        multi-center BPST
        \\
        instanton numbers
      \end{tabular}
    }
    }
  }{
    \mathbb{Z}
  }
  \,.
\end{equation}
Notice that this concerns instantons in Yang-Mills theory on flat space(-time) without
any coupling to gravity, hence without any spacetime curvature.

In contrast,
if we would consider a curved spacetime manifold of the form $\mathbb{R}^{d,1} \times S^n$
(say) with non-trivial metric on $S^n$ (either round or squashed), then we would have
to consider charges of Einstein-Yang-Mills theory instead of plain Yang-Mills theory,
and the plain classifying space $B \mathrm{SU}(2)$ would be generalized to
some suitable twisted differential classifying stack.
This situation of twisted differential cohomology of curved spacetimes
can of course
be discussed, too (see \cite{FSS20c}) but is disregarded in the present article
for sake of focus.

\medskip

\noindent
{\bf Traditional Example: D-Brane charge in non-twisted K-theory.}
When the coefficients in \eqref{IncarnationsOfCompactlySupportedCohomologyOnEuclideanSpace}
are
$A \,=\, \mathrm{KU}_0 \,=\, \mathbb{Z} \times B \mathrm{U}$,
the classifying space for complex K-theory, then
\eqref{IncarnationsOfCompactlySupportedCohomologyOnEuclideanSpace}
reproduces
the traditionally conjectured classification of D-brane charge in type IIB string theory
for flat transversal space,
seen in compactly supported K-theory of the transversal $\mathbb{R}^{9-p}$,
understood
as the reduced K-theory of its one-point compactification
(e.g. \cite[\S II, Ex. 4.4]{Karoubi78}\cite[(2.20) and below (2.58)]{OlsenSzabo99Constructing}):
\begin{equation}
  \label{DBraneChargeForFlatTransversalSpaceSeenInKTheory}
  \overset{
    \mathclap{
    \raisebox{6pt}{
      \tiny
      \color{darkblue}
      \bf
      \begin{tabular}{c}
        compactly supported K-theory
        \\
        of flat transversal space
      \end{tabular}
    }
    }
  }{
    \mathrm{K}_{\mathrm{cpt}}
    \big(
      \mathbb{R}^{9-p}
    \big)
  }
  \;\;\;\;
  =
  \;\;\;\;
  \overset{
    \mathclap{
    \raisebox{6pt}{
      \tiny
      \color{darkblue}
      \bf
      \begin{tabular}{c}
        reduced K-theory of
        \\
        transversal compactification
      \end{tabular}
    }
    }
  }{
  \widetilde
  {\mathrm{K}}
  \big(
    \mathbb{R}^{9-p}_{\mathrm{cpt}}
  \big)
  }
  \;\;
  \simeq
  \;\;
  \overset{
    \mathclap{
    \raisebox{6pt}{
      \tiny
      \color{darkblue}
      \bf
      \begin{tabular}{c}
        reduced K-theory
        \\
        of $9-p$-sphere
      \end{tabular}
    }
    }
  }{
    \widetilde
    {\mathrm{K}}
    \big(
      S^{9-p}
    \big)
  }
  \;\;
  \simeq
  \;\;
  \Truncation{0}
  \ConstrainedMaps{\big}
    { S^{9-p} }
    { B \mathrm{U} }
  \;\;
  \simeq
  \;\;
  \pi_{9-p}
  \big(
    B \mathrm{U}
  \big)
  \;\;
  \simeq
  \;\;
  \overset{
    \raisebox{6pt}{
      \tiny
      \color{darkblue}
      \bf
      \begin{tabular}{c}
        set of
        \\
        D$p$-brane charges
      \end{tabular}
    }
  }{
    \left\{
    \!\!\!
    \begin{array}{ll}
      \mathbb{Z} & \mbox{for $p$ odd}
      \\
      0 & \mbox{otherwise}.
    \end{array}
    \right.
  }
\end{equation}
This is made explicit, for instance, in
\cite[\S 2.1]{Gukov99}\cite[\S 2.1]{BergmanGimonHorava99}\cite[\S 3]{OlsenSzabo99Constructing}\cite[p. 25]{Schwarz01},
following \cite[\S 4.1]{Witten98}\cite[\S 2]{Horava98}.
The same kind of expressions hold, more generally, for D-brane charge on
orbi-orientifolds, seen in equivariant KR-theory;
this is made explicit, for instance, in
\cite[\S 3]{Gukov99}\cite[\S 3]{GaberdielStefanski99}\cite[p. 3]{GarciaCompeanEtAl08}.

The formulation below in \cref{FullMBraneChargeAndUnstableCohomotopyTheory} of
M-brane charge for flat transversal space seen in
Cohomotopy vanishing at transversal infinity is the direct and evident analogue of
this traditional situation \eqref{DBraneChargeForFlatTransversalSpaceSeenInKTheory},
for K-theory replaced by Cohomotopy.

\medskip

\noindent
{\bf In contrast: Brane charge in twisted generalized cohomology.}
Again, notice that the traditional classification of D-branes on spacetimes that
are not flat but
already have themselves the homotopy type of an $n$-sphere
is in general different: The non-trivial
background metric (field of gravity) on the sphere
will generically come with a non-trivial $B$-field, and that
changes the charge quantization law from plain K-theory to twisted K-theory.
(The archetypical and well-studied example of this case
is
the classification of D-branes for the
WZW-model on $\mathrm{SU}(2) \,\simeq\, S^3$ with its B-field proportional to the
volume class of the 3-sphere, e.g.
\cite{FredenhagenSchomerus00}\cite{Braun03}\cite[\S 1]{GaberdielGannon04},
review in \cite[\S 3.1]{Evslin06}.)
This, too, is closely analogous to the situation for Cohomotopy instead of K-theory,
under {\it Hypothesis H}: If spacetime itself is a spherical fibration,
then Hypothesis H predicts charge quantization in tangentially twisted Cohomotopy,
see \cite{SS20a}.

While, in either case, brane charges
in  twisted generalized cohomology (either twisted K-theory or twisted Cohomotopy)
on spherical spacetimes
is an important subject, we disregard it in the present article for sake of focus.
The point to amplify here is that, nevertheless,
the compactly-supported cohomology of flat transversal space which we do consider
happens toth an expression in terms of non-twisted cohomology of spheres
-- which is a well-known mathematical phenomenon, familiar from
traditional discussion of D-brane charge in K-theory
\eqref{DBraneChargeForFlatTransversalSpaceSeenInKTheory}.

\medskip

\newpage

\subsection{M-Brane charge and Borsuk-Spanier Cohomotopy}
\label{FullMBraneChargeAndUnstableCohomotopyTheory}

\noindent
  {\bf Higher gauge fields and homotopy \& cohomology.}

\vspace{2pt}

\makeatletter

\hspace{-.9cm}
\begin{tabular}{ll}

\begin{minipage}[left]{6.5cm}

\noindent
  A {\it non-abelian higher gauge field}
  species \cite{SSS09}\cite{dcct}
  is a reduced\footnotemark\global\let\saved@Href@A\Hy@footnote@currentHref
  {\it non-abelian cohomology theory},
namely
\cite[\S 2]{FSS20c} \cite[p. 6]{SS20b}
(see \cref{BorsukSpanierCohomotopy})
a contravariant functor $\widetilde A(-)$ sending
pointed topological spaces $X$
to the sets of homotopy-classes of their pointed
continuous functions
into a given pointed space $A$,
the {\it classifying space} for $A$-cohomology:\footnotemark\global\let\saved@Href@B\Hy@footnote@currentHref

\end{minipage}
&

\hspace{-.1cm}

{
\begin{minipage}[left]{10cm}

\scalebox{.88}{
\def\arraystretch{1.3}
\begin{tabular}{|l|l|l|}
  \hline
  {\bf
    Physical field
  }
  &
  $A$
  &
  {\bf Cohomology theory}
  \\
  \hline
  \hline
  Electromagnetic
  &
  $B \mathrm{U}(1)$
  &
  Ordinary abelian cohomology
  \\
  \cline{1-3}
  Nuclear
    &
    $B \mathrm{SU}(n)$
    &
  Ordinary non-abelian cohomology
  \\
  \cline{1-3}
  Gravity
    &
    $B \mathrm{Spin}(n)$
    &
  Ordinary non-abelian cohomology
  \\
  \hline
  \hline
  Neveu-Schwarz
    &
    $B^2 \mathrm{U}(1)$
    &
  Ordinary abelian cohomology
  \\
  \cline{1-3}
  Ramond-Ramond
    &
    $K\mathrm{U}$
    &
  Topological K-theory {\it \small (Hypothesis K)}
  \\
  \hline
  \hline
  C-field
  &
  $S^4 \simeq B (\Omega S^4)$
  &
  Cohomotopy {\it \small (Hypothesis H)}
  \\
  \cline{1-3}
\end{tabular}
}
\end{minipage}
}

\end{tabular}

  \addtocounter{footnote}{-1}
  \let\Hy@footnote@currentHref\saved@Href@A
  \footnotetext{
    We consider reduced cohomology throughout,
    since this captures the crucial physics concept
    of fields {\it vanishing at infinity}
    (\cref{MBraneWorldvolumesAndThePontrjaginConstruction}).
    Notice that reduced cohomology $\widetilde A(-)$
    {\it subsumes} non-reduced cohomology $A(-)$, as
    $A(X) = {\widetilde A}(X_+)$.
  }

 \stepcounter{footnote}

 \let\Hy@footnote@currentHref\saved@Href@B
 \footnotetext{
   Strictly speaking, bare $A$-cohomology
   gives only the ``topological sector''
   or ``instanton sector'' of the higher gauge field;
   while the full field content is in
   {\it differential} $A$-cohomology \cite[\S 4.3]{FSS20c}.
   Here we disregard the differential refinement
   just for focus of the exposition;
   we come back to this elsewhere.
   But see also Remark \ref{H3FluxesAsExtraordinaryFlatDifferentialCohomotopy}
   below.
 }

\makeatother

\begin{equation}
  \label{NonabelianCohomologyInIntroduction}
  \hspace{1cm}
  \underset{
    \mathclap{
    \raisebox{-6pt}{
      \tiny
      \color{darkblue}
      \bf
      \begin{tabular}{c}
        = pointed
        topological space
      \end{tabular}
    }
    }
  }{
    \overset{
      \mathclap{
      \raisebox{6pt}{
        \tiny
        \color{darkblue}
        \bf
        \begin{tabular}{c}
          spacetime with
          \\
          ``point at infinity''
        \end{tabular}
      }
      }
    }{
      X
    }
  }
  \qquad \quad
    \longmapsto
  \quad
  \underset{
    \mathclap{
    \raisebox{-3pt}{
      \tiny
      \color{darkblue}
      \bf
      \begin{tabular}{c}
        =
        non-abelian reduced
        $A$-cohomology of $X$
      \end{tabular}
    }
    }
  }{
    \overset{
      \mathclap{
      \raisebox{3pt}{
        \tiny
        \color{darkblue}
        \bf
        \begin{tabular}{c}
          gauge-equivalence classes
          of $A$-fields on $X$
          \\
          vanishing at infinity
        \\
        \end{tabular}
      }
      }
    }{
    \widetilde A(X)
    \;\;
      :=
    \;\;
      \pi_0
      \mathrm{Maps}^{\ast/}
      \!\!
      \big(
        X,\; A
      \big)
    }
  }
  \;\;
    =
  \;\;
  \left\{
   \phantom{\mbox{\tiny\bf (e.g. spacetime)}}
   \;\;
  \xymatrix{
    \mathllap{
      \mbox{
        \tiny
        \color{darkblue}
        \bf
        \begin{tabular}{c}
          spacetimes =
          \\
          domain space
        \end{tabular}
      }
    }
    \!\!\!\!
    X
    \ar@/^2pc/[rr]
      |-{
        \mathclap{\phantom{\vert}}
        \;c\;
      }
      ^-{
        \mathclap{
          \tiny
          \color{greenii}
          \bf
          \begin{tabular}{c}
  \bf          field configuration =
            \\
  \bf          pointed map / cocycle
          \end{tabular}
        }
      }
      _-{\ }="s"
    \ar@/_2pc/[rr]
      |-{
        \mathclap{\phantom{\vert}}
        \;c'\;
      }
      _-{
        \mathclap{
          \tiny
          \color{greenii}
          \bf
          \begin{tabular}{c}
 \bf           field configuration =
            \\
\bf            pointed map/cocycle
          \end{tabular}
        }
      }
      ^-{\ }="t"
    &&
    A
    \mathrlap{
      \!\!\!\!\!
      \mbox{
        \tiny
        \color{darkblue}
        \bf
        \begin{tabular}{c}
          classifying space
          \\
          for $A$-cohomology
        \end{tabular}
      }
    }
    \ar@{=>}
      |-{
        \mbox{
          \tiny
          \color{orangeii}
          \bf
          $\mathclap{\phantom{\vert^{\vert}}}$
          \begin{tabular}{c}
            gauge transf.
            \\
            = homotopy/
            \\
            coboundary
          \end{tabular}
          $\mathclap{\phantom{\vert_{\vert}}}$
        }
      }
      "s"; "t"
  }
   \phantom{\mbox{\tiny\bf for $A$-cohomology}}
  \right\}_{
    \!\!\!\!\!
      \Big/
    \!\!\!\!\!\!\!\!\!\!
    \mbox{
      \tiny\color{orangeii}
      \bf
      \begin{tabular}{c}
        gauge equiv.
        \\
        = homotopy
      \end{tabular}
    }
  }
\end{equation}

\noindent
Familiar examples include:

\vspace{0cm}
\begin{itemize}

\vspace{-.2cm}
\item
For $A = K(R,n)$ an Eilenberg-MacLane space,
this construction \eqref{NonabelianCohomologyInIntroduction}
is ordinary abelian cohomology  as computed by
singular- or {\v C}ech-cochains (see \cite[Ex. 2.2]{FSS20c}
for pointers);
specifically as computed by (PL-)differential forms in the case
that $R = \mathbb{R}$ (see \cite[\S 3, Ex. 4.9]{FSS20c} for pointers).
Notice that $K(\mathbb{Z}, n+1) \simeq B^{n}\mathrm{U}(1)$
is the classifying space for circle $(n-1)$-gerbes
(see \cite[Ex. 2.12]{FSS20c} for pointers).

For  $n = 1$ this models the electromagnetic field
(the ``vector potential $A$-field'', e.g. \cite[\S 10.5]{Nakahara03}),
while for $n = 2$ this models
the Kalb-Ramond $B$-field \cite{Gawedzki86}\cite{FreedWitten99}\cite{CJM04}.
The case for $n = 3$ is often taken as a
first approximation to a model for the supergravity $C$-field
(see p. \pageref{MBraneChargeInModifiedOrdinaryCohomology}).
The point of our discussion here is to improve on this
approximation for the $C$-field.

\vspace{-.2cm}
\item
For $A = B G$ the classifying space of a compact topological group $G$
(e.g. the infinite Grassmannian $\mathrm{Gr}_n \simeq B \mathrm{O}(n)$),
the definition
\eqref{NonabelianCohomologyInIntroduction} reduces to
ordinary non-abelian cohomology in the sense  of  Chern-Weil theory
(see \cite[Ex. 2.2, 2.3]{FSS20c} for pointers).

For $G$
in the Whitehead tower of $\mathrm{SU}(n)$ this
models the generic non-abelian/nuclear force gauge field
(e.g. \cite[\S 10.5]{Nakahara03});
while for $G$
in the Whitehead tower of $\mathrm{SO}(n)$
this models gravity (see \cite[Fig. 1,2]{SSS09}):
\end{itemize}

\vspace{-.4cm}
\begin{equation}
  \label{OrdinaryCohomologyInIntroduction}
  \overset{
    \mathclap{
    \raisebox{6pt}{
      \tiny
      \color{darkblue}
      \bf
      \begin{tabular}{c}
        ordinary cohomology
        \\
        (in particular: abelian)
      \end{tabular}
    }
    }
  }{
  \widetilde H^{n}
  \big(
    X;
    \,
    R
  \big)
  }
  \;\simeq\;
  \pi_0
  \mathrm{Maps}^{\ast/}
  \!\!
  \big(
    X
    \,,\,
    \overset{
    \mathclap{
    \raisebox{6pt}{
      \tiny
      \color{greenii}
      \bf
      \begin{tabular}{c}
        Eilenberg-MacLane
        \\
        space
      \end{tabular}
    }
    }
    }{
      K(R,n)
    }
  \big)
  \,,
  \phantom{AAAAAAA}
  \overset{
    \mathclap{
    \raisebox{6pt}{
      \tiny
      \color{darkblue}
      \bf
      \begin{tabular}{c}
        non-abelian cohomology
        \\
        (as in Chern-Weil theory)
      \end{tabular}
    }
    }
  }{
    \widetilde H^1(X;\,G)
  }
  \quad \simeq\;
  \pi_0
  \mathrm{Maps}^{\ast/}
  \!\!
  \big(
    X
    \,,\,
    \overset{
    \mathclap{
    \raisebox{6pt}{
      \tiny
      \color{greenii}
      \bf
      \begin{tabular}{c}
        classifying space
        \\
        of compact group
      \end{tabular}
    }
    }
    }{
      B G
    }
  \big)
  \,.
\end{equation}
In generalization of the ordinary abelian case
on the left of \eqref{OrdinaryCohomologyInIntroduction},
if $A = E^n$ is a space in a {\it spectrum} of pointed spaces
(``$\Omega$-spectrum'')\footnote{
  Beware that we write the indices on the component
  spaces of a spectrum \eqref{AnOmegaSpectrum}
  as {\it  super}scripts,
  instead of the more conventional subscripts.
  This is to harmonize the index placement in
  common formulas such as the suspended unit maps
  $\Sigma^n(1^E) \; :\; S^n \xrightarrow{\;} E^n$
  in \eqref{UnitMorphismOfSpectra} below.
  Our {\it sub}script is reserved for the
  coefficient groups \eqref{WhiteheadGeneralizedCoohomology},
  matching the expected index placement for
  stable homotopy groups of spheres
  $
    \pi_n^s
      \,=\,
    \mathbb{S}_n
  $ and cobordism rings
  $
    \Omega_n^f
      \,=\,
    (M\!f)_n
  $.
 }
\begin{equation}
  \label{AnOmegaSpectrum}
  \overset{
    \mathclap{
    \raisebox{4pt}{
      \tiny
      \color{darkblue}
      \bf
      \begin{tabular}{c}
        spectrum
        \\
        (of spaces)
      \end{tabular}
    }
    }
  }{
    E
  }
  \;\;=\;\;
  \Big\{
    \overset{
      \mathclap{
      \raisebox{4pt}{
        \tiny
        \color{darkblue}
        \bf
        graded component spaces (pointed)
      }
      }
    }{
      E^n
      \,:=\,
      \Omega^{\infty - n } E
      \,:=\,
      \Omega^{\infty} \Sigma^n E
    }
    \,
    ,
    \,\;
    \xymatrix@C=34pt{
      E^n
      \ar[rrr]
        _-{ \simeq }_-{}
        ^-{
          \mathclap{
          \mbox{
            \tiny
            \color{greenii}
            \bf
            \begin{tabular}{c}
                classifying maps for
                \\
                suspension isomorphism
            \end{tabular}
          }
          }
        }
      &&&
      \overset{
        \mathclap{
        \raisebox{3pt}{
          \tiny
          \color{darkblue}
          \bf
          \begin{tabular}{c}
            loop spaces
          \end{tabular}
        }
        }
      }{
        \Omega E^{n+1}
      }
    }
  \Big\}_{n \in \mathbb{N}}
\end{equation}
then definition \eqref{NonabelianCohomologyInIntroduction}
reduces (\cite[Ex. 2.13]{FSS20c}) to that of
{\it Whitehead's generalized cohomology}\footnote{Whitehead's
generalized cohomology is traditionally just called
{\it generalized cohomology}, for short.
Since this becomes ambiguous as we consider yet more general
non-abelian cohomology \eqref{NonabelianCohomologyInIntroduction},
we re-instantiate Whitehead's name for definiteness.}
(\cite{Whitehead62}\cite{Adams74})
\begin{equation}
  \label{WhiteheadGeneralizedCoohomology}
  \overset{
    \mathclap{
    \raisebox{6pt}{
      \tiny
      \color{darkblue}
      \bf
      \begin{tabular}{c}
        Whitehead-generalized
        \\
        reduced  cohomology
      \end{tabular}
    }
    }
  }{
    \widetilde E^n(X)
  }
  \;\simeq\;
  \pi_0 \mathrm{Maps}^{\ast/}
  \!\!
  \big(
    X
    \,,\,
    \overset{
    \mathclap{
    \raisebox{6pt}{
      \tiny
      \color{greenii}
      \bf
      \begin{tabular}{c}
        $n$th-space in
        \\
        a spectrum
      \end{tabular}
    }
    }
    }{
      E^n
    }
  \big)
  \,,
  \;\;\;
  \mbox{in particular (Ex. \ref{StableHomotopyGroupsAndCohomologyOfSpheres})}:
  \;\;
  \underset{
    \mathclap{
    \raisebox{-3pt}{
      \tiny
      \color{darkblue}
      \bf
      graded coefficient group
    }
    }
  }{
    E_\bullet
    \;:=\;
    \pi_\bullet(E)
  }
  \;:=\;
  \underset{
    \raisebox{-3pt}{
      \tiny
      \color{darkblue}
      \bf
      reduced $E$-cohomology of spheres
    }
  }{
    \widetilde E^k(S^{k+\bullet})
    \;\simeq\;
    \pi_{\bullet + k}\big(E^{k}\big)
  }
  \,.
\end{equation}
such as complex K-cohomology
(conjectured to model the RR-field, see p. \pageref{DBraneChargeInKTheory})
or complex cobordism cohomology
(see Examples \ref{ExamplesOfMultiplicativeCohomologyTheories}):
\begin{equation}
  \overset{
    \mathclap{
    \raisebox{6pt}{
      \tiny
      \color{darkblue}
      \bf
      \begin{tabular}{c}
        complex top.
        \\
        K-theory
      \end{tabular}
    }
    }
  }{
    \widetilde {{K\mathrm{U}}}{}^0(X)
  }
  \;\simeq\;
  \pi_0 \mathrm{Maps}^{\ast/}
  \!\!
  \big(
    X
    \,,\,
    \overset{
    \mathclap{
    \raisebox{6pt}{
      \tiny
      \color{greenii}
      \bf
      \begin{tabular}{c}
        stable unitary
        \\
        classifying space
      \end{tabular}
    }
    }
    }{
      B \mathrm{U}
    }
  \big)
  \,,
  \phantom{AAAAA}
  \overset{
    \mathclap{
    \raisebox{6pt}{
      \tiny
      \color{darkblue}
      \bf
      \begin{tabular}{c}
        complex
        \\
        cobordism
      \end{tabular}
    }
    }
  }{
    \widetilde {\mathrm{MU}}{}^0(X)
  }
  \;\simeq\;
  \pi_0 \mathrm{Maps}^{\ast/}
  \!\!
  \Big(
    X
    \,,\,
    \overset{
    \mathclap{
    \raisebox{6pt}{
      \tiny
      \color{greenii}
      \bf
      \begin{tabular}{c}
        Thom space of cplx. vector bdl.
        \\
        over classifying space
      \end{tabular}
    }
    }
    }{
    \mathrm{Th}
    \big(
      \mathcal{V}_{B \mathrm{U}}
    \big)
    }
  \Big)
  \,.
\end{equation}
But in full non-abelian generalization of the second case in
\eqref{OrdinaryCohomologyInIntroduction}, every
$\infty$-group $\mathcal{G}$ -- namely every based loop group
(see \cite[Prop. 2.8]{FSS20c} for pointers)
with its homotopy-coherent
group operations by concatenation of loops --
defines a generalized non-abelian cohomology theory,
such as that classifying non-abelian $G$-gerbes
(see \cite[Ex. 2.6]{FSS20c} for pointers):
\begin{equation}
  \label{HigherNonAbelianCohomology}
  \left.
  \begin{aligned}
    &
    A \,\simeq\, B \mathcal{G}
    \\
    \Leftrightarrow
    \;
    &
    \mathcal{G} \,\simeq\, \Omega A
  \end{aligned}
  \right\}
  \;\;\;\;\;
  \overset{
    \mathclap{
    \raisebox{6pt}{
      \tiny
      \color{darkblue}
      \bf
      \begin{tabular}{c}
        general
        \\
        non-abelian cohomology
      \end{tabular}
    }
    }
  }{
    \widetilde H^1
    \big(
      X;
      \,
      \mathcal{G}
    \big)
  }
  \;\simeq\;
  \pi_0 \mathrm{Maps}^{\ast/}\!\!
  \big(
    X
    \,,\,
    \overset{
    \mathclap{
    \raisebox{6pt}{
      \tiny
      \color{greenii}
      \bf
      \begin{tabular}{c}
        any
        \\
        connected space
      \end{tabular}
    }
    }
    }{
      B \mathcal{G}
    }
  \big)
  \,,
  \phantom{AA}
  \mbox{e.g.:}
  \phantom{A}
  \overset{
    \mathclap{
    \raisebox{6pt}{
      \tiny
      \color{darkblue}
      \bf
      \begin{tabular}{c}
        non-abelian cohomology
        \\
        (as in Giraud-Breen theory)
      \end{tabular}
    }
    }
  }{
  \widetilde H^1
  \big(
    X;
    \,
    \mathrm{Aut}
    (
      B G
    )
  \big)
  }
  \;\simeq\;
  \pi_0 \mathrm{Maps}^{\ast/}\!\!
  \big(
    X
    \,,\,
    \overset{
    \mathclap{
    \raisebox{6pt}{
      \tiny
      \color{greenii}
      \bf
      \begin{tabular}{c}
        automorphism 2-group
        \\
        of $G$-classifying space
      \end{tabular}
    }
    }
    }{
    \mathrm{Aut}
    (
      B G
    )
    }
  \big)
  \,.
\end{equation}

\medskip

\noindent
{\bf The M-Theory C-field and Borsuk-Spanier Cohomotopy.}
The most fundamental example of general non-abelian
cohomology \eqref{NonabelianCohomologyInIntroduction}
is unstable {\it Cohomotopy}
\cite{Borsuk36}\cite{Pontrjagin38}\cite{Spanier49}\cite{Peterson56},
whose classifying spaces are the $n$-spheres,
$S^n \,\simeq\, B \big(\Omega S^n\big)$,
hence whose higher gauge group \eqref{HigherNonAbelianCohomology}
is the loop $\infty$-group
$\mathcal{G} = \Omega S^n$ of $n$-spheres:
\vspace{-2mm}
\begin{equation}
  \label{CohomotopyInIntroduction}
  \overset{
    \mathclap{
    \raisebox{6pt}{
      \tiny
      \color{darkblue}
      \bf
      \begin{tabular}{c}
        Cohomotopy
      \end{tabular}
    }
    }
  }{
    \widetilde \pi^{\, n}
    (X)
  }
  \;:=\;
  \pi_0
  \mathrm{Maps}^{\ast/}
  \!\!
  \big(
    X
    \,,\,
    \overset{
    \mathclap{
    \raisebox{6pt}{
      \tiny
      \color{greenii}
      \bf
      \begin{tabular}{c}
        $n$-sphere
      \end{tabular}
    }
    }
    }{
      S^n
    }
  \big)
  \,,
  \phantom{AAAAAA}
  \xymatrix@C=5em{
    \overset{
      \mathclap{
      \raisebox{6pt}{
        \tiny
        \color{darkblue}
         \begin{tabular}{c} \bf
          Cohomotopy
        \end{tabular}
      }
      }
    }{
      \widetilde \pi^{\, n}(-)
    }
    \ar[rr]^-{
     (
        \!\!
        \raisebox{1pt}{
          \scalebox{.7}{$
            S^n \xrightarrow{g} A
          $}
        }
        \!\!
      )
      _\ast
    }_-{
    \raisebox{-3pt}{
      \tiny
      \color{greenii}
      \bf
      \begin{tabular}{c}
        non-abelian cohomology operation
        \\
        induced by any $[g] \in \pi_n(A)$
      \end{tabular}
    }
    }
    &&
    \overset{
      \mathclap{
      \raisebox{6pt}{
        \tiny
        \color{darkblue}
        \bf
        \begin{tabular}{c}
          any $A$-cohomology
        \end{tabular}
      }
      }
    }{
      \widetilde A(-)
    }
  }
\end{equation}
Over a smooth manifold, such as
$X =  (M^d)^{\scalebox{.6}{cpt}} \!\!\wedge \mathbb{R}^{p,1}_+ $
\eqref{OnePointCompactificationAndFunctionsOnIt},
the {\it rationalization} of Cohomotopy in even degrees
\eqref{CohomotopyInIntroduction}
\begin{equation}
  \label{RationalCohomotopy}
  \mathclap{
  \xymatrix@R=7pt@C=13pt{
    \overset{
      \mathrlap{
        \raisebox{3pt}{
          \tiny
          \color{darkblue}
          \bf
          Cohomotopy
        }
      }
    }{
    {\widetilde \pi}{}^{\,2k}
    \big(
      M^d_+
    \big)
    }
    \ar@{=}[r]
    \ar@/_1pc/[drrrr]
      _-{
        \mbox{
          \tiny
          \color{greenii}
          \bf
          cohomotopical character map
        }
        \mathclap{\phantom{\vert_{\vert}}}
      }
      _>>>>>>>>>>>{
        \mathrm{ch}_{\pi^{2k}}
      }
    &
    \mathrm{Maps}^{\ast/\!\!}
    \big(
      X,S^n
    \big)
    \ar[r]
      ^-{
        \mathclap{
        \mbox{
          \tiny
          \color{greenii}
          \bf
          \begin{tabular}{c}
            rationalization
            \\
            \phantom{A}
          \end{tabular}
        }
        }
      }
    &
    \mathrm{Maps}^{\ast/\!\!}
    \big(
      X, L_{\mathbb{R}}S^n
    \big)
    \ar@{=}[r]
    &
    \overset{
      \mathclap{
      \raisebox{3pt}{
        \tiny
        \color{darkblue}
        \bf
        \begin{tabular}{c}
          rational
          \\
          Cohomotopy
        \end{tabular}
      }
      }
    }{
      {\widetilde \pi}{}^{\,2k}_{\mathbb{R}}
      \big(
        M^d_+
      \big)
    }
    \ar@{=}[r]
      ^-{
        \mbox{
          \tiny
          \color{greenii}
          \bf
          \begin{tabular}{c}
            non-abelian
            \\
            de Rham theorem            \\
            \phantom{A}
          \end{tabular}
        }
      }
    &
    \overset{
      \raisebox{3pt}{
        \tiny
        \color{darkblue}
        \bf
        \begin{tabular}{c}
          non-abelian
          \\
          de Rham cohomology
        \end{tabular}
      }
    }{
      H_{\mathrm{dR}}
      \big(
        M^d, \mathfrak{l}S^n
      \big)
    }
    \ar@{=}[d]
    \\
    &&&&
    \left\{ \footnotesize
      \!\!\!\!
      {\begin{array}{c}
        G_{2k},
        \\
        2 G_{4k-1}
      \end{array}}
      \!\!\!\!\!\!
      \in
      \Omega^\bullet_{\mathrm{dR}}(M^d)
      \left\vert
      {\begin{array}{rl}
        d\, G_{2k \phantom{-1\;}}
        &
        \!\!\!\!\!\!
        = 0
        \\
        d\, 2G_{4k-1}
        &
        \!\!\!\!\!\!
        = - G_{2k} \wedge G_{2k}
      \end{array}}
      \right.
      \!\!\!\!
    \right\}
  }
  }
\end{equation}
is equivalent \cite[Prop. 2.5]{FSS19b},
via a cohomotopical character map \cite[Ex. 5.23]{FSS20c},
to concordance classes of pairs of differential forms
$(G_{2k}, G_{4k-1})$ that satisfy
the  differential relations known from
the C-field 4-flux form and its Hodge dual
\cite[(5.11b)]{DAuriaFre82}\cite[(III.8.53)]{CDF91}
(review in \cite[(3.23)]{MiemicSchnakenburg06})
in  11-dimensional supergravity \cite{CremmerJuliaScherk78},
with their characteristic quadratic dependence.

\medskip
Noticing the direct analogy with how the image of
the twisted K-theoretic Chern character produces
differential forms $(H_3, \{F_{2k}\}_k)$ satisfying the
differential relations $d F_{2k+2} = H_3 \wedge F_{2k}$
(see \cite[\S 5.1]{FSS20c})
known from the RR-field flux forms $F_{2k \leq 5}$ and their
Hodge duals $F_{2k \geq 5}$,
the form of the cohomotopical character \eqref{RationalCohomotopy}
is the first indication
\cite[\S 2.5]{Sati13}\cite{FSS16a}\cite[\S 7]{FSS19a}
that the C-field wants to be charge-quantized
in Cohomotopy theory.

\medskip
In the special case of interest here,
this \hyperlink{HypothesisH}{\it Hypothesis H} reads,
in more detail, as follows:

\newpage

\hypertarget{HypothesisHOnHomotopicallyFlatSpacetimes}{}
\noindent
{\bf Hypothesis H on homotopically flat spacetimes:}
{\it
If $X$ is (the pointed topological space underlying)
a homotopically flat 11-dimensional spacetime
(Remark \ref{FramedSpacetimes})
equipped with a point at infinity
\eqref{OnePointCompactificationAndFunctionsOnIt},
then the C-field on $X$ is charge-quantized in
reduced Borsuk-Spanier 4-Cohomotopy \eqref{CohomotopyInIntroduction},
in that (the topological sector of)
a full C-field configuration is a
$[c] \in {\widetilde \pi}{}^4(X)$
whose corresponding 4- and dual 7-flux density
are the image
$\big[G_4(c),G_7(c)\big] \,=\, \mathrm{ch}_{\pi^4}(c)$
under the cohomotopical character map \eqref{RationalCohomotopy}:}
\begin{equation}
  \label{ChargeQuantizationIn4CohomotopyOnHomotopicallyFlatSpacetime}
  \xymatrix@R=-12pt{
    \overset{
      \mathclap{
      \raisebox{3pt}{
        \tiny
        \color{darkblue}
        \bf
        \begin{tabular}{c}
        Borsuk-Spanier
        \\
        4-Cohomotopy
        \end{tabular}
      }
      }
    }{
      \pi^4(X)
    }
    \ar[rr]
      _-{
        \mathrm{ch}_\pi
      }
      ^-{
        \mbox{
          \tiny
          \color{greenii}
          \bf
          cohomotopical character map
        }
      }
    &
    {\phantom{AAAAAAA}}
    &
    \overset{
      \mathclap{
      \raisebox{3pt}{
        \tiny
        \color{darkblue}
        \bf
        \begin{tabular}{c}
          non-abelian
          \\
          de Rham cohomology
        \end{tabular}
      }
      }
    }{
      H_{\mathrm{dR}}
      \big(
        X;
        \mathfrak{l}S^4
      \big)
    }
    \\
    \underset{
      \mathclap{
      \raisebox{-3pt}{
        \tiny
        \color{darkblue}
        \bf
        \begin{tabular}{c}
          C-field charge, quantized
          \\
          in Cohomotopy theory
        \end{tabular}
      }
      }
    }{
      [c]
    }
    \ar@{}[rr]
      |-{ \longmapsto }
    &&
    \underset{
      \mathclap{
      \raisebox{-3pt}{
        \tiny
        \color{darkblue}
        \bf
        \begin{tabular}{c}
          C-field 4-flux density
          \\
          and its 7-form dual
        \end{tabular}
      }
      }
    }{
      \big[
        G_4(c), G_7(c)
      \big]
    }
  }
\end{equation}

\medskip

\noindent
{\bf The abelianized C-field and Stable Cohomotopy.}
Non-abelian cohomology \eqref{NonabelianCohomologyInIntroduction}
satisfies, in general, fewer conditions than
abelian (Whitehead-generalized) cohomology \eqref{WhiteheadGeneralizedCoohomology},
notably it may lack gradings and suspension isomorphisms \eqref{AnOmegaSpectrum}.
This means that non-abelian cohomology is {\it richer}
than abelian Whitehead-generalized cohomology, just as non-abelian groups
are richer than abelian groups).

\medskip
But there are non-abelian cohomology operations
(\cite[Def. 2.17]{FSS20c})
which approximate non-abelian cohomology by abelian cohomology;
a famous example is the Chern-Weil homomorphism
(see \cite[\S 4.2]{FSS20c} for pointers).
Indeed, every non-abelian cohomology theory $A$ \eqref{NonabelianCohomologyInIntroduction}
has a universal approximation
by a Whitehead-generalized
abelian cohomology theory \eqref{WhiteheadGeneralizedCoohomology},
namely that represented by the suspension spectrum $\Sigma^\infty A$ of $A$.
(\cite[Ex. 2.24]{FSS20c}); a famous example
appears in Snaith's theorem \cite{Snaith79}\cite{Mathew}
(and its variants \cite{GepnerSnaith08})
which says that complex K-theory theory is close to being the
stabilization of the non-abelian cohomology theory classified by
the pointed space $\mathbb{C}P^\infty_+$.

\medskip
The abelianization of Borsuk-Spanier Cohomotopy \eqref{CohomotopyInIntroduction}
is {\it stable Cohomotopy} \cite[p. 204]{Adams74}\cite{Segal74}\cite{Rognes04},
the Whitehead-generalized cohomology
theory \eqref{WhiteheadGeneralizedCoohomology}
whose classifying spectrum \eqref{AnOmegaSpectrum}
is the sphere spectrum $\mathbb{S}$ with
$\mathbb{S}^n = \Omega^\infty S^{\infty + n}$.
Stable Cohomotopy, in turn, has canonical images
in any multiplicative abelian cohomology theory, via the
Hurewicz-Boardman homomorphism
\cite[\S II.6]{Adams74}\cite{Hunton95}\cite{Arlettaz04},
coming from the fact that $\mathbb{S}$ is the initial ring
homotopy-commutative spectrum in analogy to how the integers
$\mathbb{Z}$ are the initial commutative ring,
see \eqref{UnitCohomologyOperations} and \eqref{dInvariantAndBoardman} below.

\medskip
Therefore, the universal cohomology operation from unstable/non-abelian
to stable/abelian Cohomotopy initiates, under
\hyperlink{HypothesisHOnHomotopicallyFlatSpacetimes}{Hypothesis H},
a web of abelian approximations to
cohomotopical C-field charge, as measured by a web of
Whitehead-generalized cohomology theories, a small part of
which looks as follows:
\begin{equation}
  \label{StableCohomotopy}
  \hspace{-6mm}
  \mathclap{
  \raisebox{29pt}{
  \xymatrix@R=-4pt@C=9.5pt{
    &&&&
    &
    \widetilde {M \mathrm{Sp}}{}^n(X)
    \ar[r]
    &
    \widetilde{K \mathrm{O}}{}^n(X)
    \ar[r]
    &
    H^{4(n+\bullet)}(X\!;\mathbb{Z})
    \\
    \overset{
      \mathclap{
      \raisebox{3pt}{
        \tiny
        \color{darkblue}
        \bf
        Cohomotopy
      }
      }
    }{
      {\widetilde \pi}{}^n
      \big(
        X
      \big)
    }
    \!
    \ar@{=}[r]
    &
    \!
    \pi_0
    \mathrm{Maps}^{\ast/\!\!}
    \big(
      X
      \!
      ,
      S^n
    \big)
    \!
    \ar[r]
      _-{
        \mbox{
          \tiny
          \color{greenii}
          \bf
          \begin{tabular}{c}
            \phantom{a}
            \\
            stabilization/
            \\
            abelianization
          \end{tabular}
        }
      }
    &
    \pi_0
    \mathrm{Maps}^{\ast/\!\!}
    \big(
      X
      \!
      ,
      \mathbb{S}^n
    \big)
    \ar@{=}[r]
    \ar@/_.8pc/[drrr]
      _-{
        \mbox{
          \tiny
          \color{greenii}
          \bf
          {\begin{tabular}{c}
            Hurewicz-Boardman
            \\
            homomorphism
          \end{tabular}}
        }
      }
    &
    \overset{
      \mathclap{
      \raisebox{3pt}{
        \tiny
        \color{darkblue}
        \bf
        \begin{tabular}{c}
          stable
          \\
          Cohomotopy
        \end{tabular}
      }
      }
    }{
      \widetilde {\mathbb{S}}{}^{n}
      \big(
        X
      \big)
    }
    \!
    \ar@{=}[r]
    &
    \overset{
      \mathllap{
        \raisebox{3pt}{
          \tiny
          \color{darkblue}
          \bf
          \begin{tabular}{c}
            framed
            \\
            Cobordism
          \end{tabular}
        }
      }
    }{
      \widetilde {M \mathrm{Fr}}{}^n(X)
    }
    \ar[dr]
    \ar[ur]
    &
    \mathclap{
    \mbox{
      \tiny
      \color{darkblue}
      \bf
      Cobordism
    }
    }
    &
    \mathclap{
    \mbox{
      \tiny
      \color{darkblue}
      \bf
      K-theory
    }
    }
    &
    \mathclap{
    \mbox{
      \tiny
      \color{darkblue}
      \bf
      \begin{tabular}{c}
        ordinary
        \\
        cohomology
      \end{tabular}
    }
    }
    \\
    &&&&
    &
    {\widetilde {M \mathrm{U}}}{}^n(X)
    \ar[r]
      _-{
        \mbox{
          \tiny
          \color{greenii}
          \bf
          \begin{tabular}{c}
            {\phantom{a}}
            \\
            orientation
          \end{tabular}
        }
      }
    &
    {\widetilde {K \mathrm{U}}}{}^n(X)
    \ar[r]
      _-{
        \mbox{
          \tiny
          \color{greenii}
          \bf
          \begin{tabular}{c}
            Chern
            \\
            character
          \end{tabular}
        }
      }
    &
    H^{2(n+\bullet)}(X\!;\mathbb{Z})
  }
  }
  }
\end{equation}
Previously we had discussed
\hyperlink{HypothesisH}{\it Hypothesis H}

\noindent {\bf (a)}
in  the (twisted and equivariant) rational approximation \eqref{RationalCohomotopy}
in much detail
\cite{FSS19b}\cite{FSS19c}\cite{FSS20b}\cite{SS20a} \cite{SS20c},
and

\noindent {\bf (b)}
in the (equivariant) stable approximation \eqref{StableCohomotopy}
in codimension-4 \cite{SS19a}\cite{BSS19}.

\medskip
\noindent
We now set out to discuss the abelian/stable approximation \eqref{StableCohomotopy}
to cohomotopical M-brane charge more systematically.

\medskip

\newpage

\subsection{M-Brane worldvolumes and Pontrjagin-Thom collapse}
\label{MBraneWorldvolumesAndThePontrjaginConstruction}

\hspace{-.4cm}
\begin{tabular}{ll}

\begin{minipage}[left]{10.5cm}

\noindent
{\bf Brane interactions and bordisms.}
At the foundations of perturbative string theory
is, famously, the postulate that the  multitude  of
{\it interactions} of fundamental particles
--  traditionally encoded  by  Feynman graphs with labeled
interaction vertices  -- are secretly just (conformal)
{\it bordisms} between strings \cite{Segal04}\cite{MooreSegal06}\cite{StolzTeichner11}.

\end{minipage}

&

\begin{minipage}[left]{6cm}

 \includegraphics[width=1\textwidth]{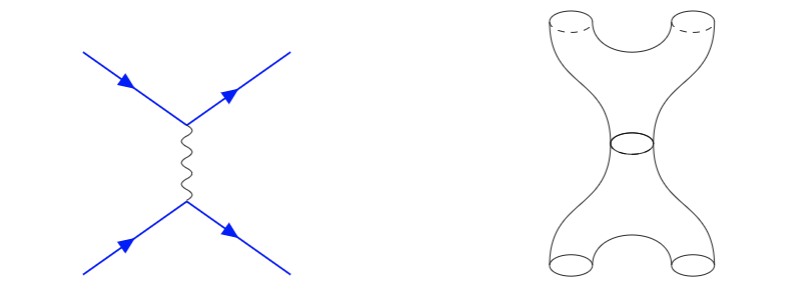}

\end{minipage}

\end{tabular}

\vspace{2mm}

While an analogous perturbation theory for
fundamental higher dimensional
$p$-branes is not expected to exist, it is
expected that the interactions of
{\it probe $p$-branes}
(solitonic branes so light/decoupled
as not to disrupt the topology of spacetime)
is similarly given by
$(p+1)$-dimensional bordisms between them,
thought of as the time-evolution of $p$-dimensional brane
volumes merging and splitting, much like soap bubbles do.
This expectation is reflected in attempts
\cite{Witten10}\cite{Freed12}\cite[p. 32]{Freed13}\cite{FreedTeleman14}\cite{BenZvi14}\cite{Mueller20}
to conceive of at least part of the 5-brane's worldvolume field theory
as a functor on a bordism category.
Field theories regarded as such functorial bordism representations
are well-understood in the ``topological sector''
\cite{Atiah89}\cite{Lurie08}\cite{Freed13}
and are plausible in more generality,
where, however, precise definitions are tricky
and remain largely elusive.

\medskip

\noindent
{\bf Conserved brane charge and Cobordism cohomology.}
But as we turn attention away from
attempts to formulate a na{\"i}ve worldvolume dynamics of $p$-branes,
and instead look to the
{\it generalized cohomology theory of their charges in spacetime},
the assumption that brane interactions
is encoded by bordisms leads to a robust
conclusion:

If total brane charge is to be preserved by brane interactions,
then {\it brane charge is a bordism invariant} of the
brane's $p$-dimensional shape within spacetime.
The finest such invariant, hence the universal brane charge,
is the {\it cobordism class} itself,
which is an element in
(unstable) degree-$(d-p)$ {\it Cobordism cohomology} of spacetime:

\vspace{-5.5mm}

\begin{center}
\hypertarget{TableInteractions}{}
\begin{tabular}{|l|l|}
  \hline
    $\mathclap{\phantom{\vert^{\vert^{\vert}}}}$
    {\bf
    Expected aspect of brane physics}
    $\mathclap{\phantom{\vert_{\vert}}}$
  &
    {\bf
    Formulation in differential topology}
  \\
  \hline
  \hline
    $\mathclap{\phantom{\vert^{\vert^{\vert}}}}$
    Codimension-$(n = d-p)$
    probe $p$-branes
    in space $M^d$
    $\mathclap{\phantom{\vert_{\vert}}}$
  &
    Codimension-$n$
    submanifolds
    of manifold $M^d$
  \\
  \hline
    $\mathclap{\phantom{\vert^{\vert^{\vert}}}}$
    $p$-brane interactions
    in spacetime $X^{d,1}$
    $\mathclap{\phantom{\vert_{\vert}}}$
  &
    Codimension-$n$ bordisms in $M^d \times [0,1]$
  \\
  \hline
  $\mathclap{\phantom{\vert^{\vert^{\vert}}}}$
  (Anti-)brane charge carried by single $p$-brane
    $\mathclap{\phantom{\vert_{\vert}}}$
  &
  $f$-structure on normal bundle
  \\
  \hline
    $\mathclap{\phantom{\vert^{\vert^{\vert}}}}$
    Total brane charge
    preserved by brane interactions
    $\mathclap{\phantom{\vert_{\vert}}}$
  &
    Function of cobordism classes
    of submanifolds
 \\
 \hline
   $\mathclap{\phantom{\vert^{\vert^{\vert}}}}$
   Finest conserved charge of
   codimension-$n$ branes in $M^d$
   $\mathclap{\phantom{\vert_{\vert}}}$ \!\!\!\!\!
 &
 Cobordism class in unstable Cobordism cohomology
 \\
 \hline
 $\mathclap{\phantom{\vert^{\vert^{\vert}}}}$
 Brane charge seen in bulk spacetime
 $\mathclap{\phantom{\vert_{\vert}}}$
 &
 Class in stable Cobordism cohomology $M\! f^n(M^d)$
 \\
 \hline
\end{tabular}
\end{center}

\vspace{-2mm}

\noindent {\footnotesize {\bf Table I.} If brane interactions are
given by bordisms, then {\bf (a)}
individual brane/anti-brane charge is reflected
in $f$-structure on the brane's normal bundle and
{\bf (b)} conserved total brane charges are $f$-bordism invariants.
The finest such is the $f$-cobordism class itself,
which reflects brane charge quantized in $f$-Cobordism cohomology.
This is unstable (non-abelian) cohomology
for branes constrained to asymptotic spacetime boundaries, but
approaches stable (abelian) Cobordism cohomology $M \!f$ as
interaction processes are allowed to probe into higher bulk dimensions.}

\vspace{4mm}
\hspace{-.9cm}
\begin{tabular}{ll}
\label{AntiBranesAndNormalStructure}
\begin{minipage}[left]{11.6cm}
\noindent
{\bf Anti-Branes and normal $f$-structure.}
Here an {\it anti-brane} is supposed to be a brane with opposite orientation
relative to the ambient spacetime, and such that there are
interaction processes of brane/anti-brane pair creation/annihilation.
With brane interactions perceived as bordisms,
this is naturally reflected by bordisms with
{\it $f$-structure} on their normal bundle,
named after fibrations
$\xymatrix@C=18pt{B G\! \ar[r]|<<<{\,f\,} & B \mathrm{O}}$
factoring the bundle's classifying map.

Initial  among all $f$-structures is framing structure
$\xymatrix@C=9pt{B \mathrm{Fr} \simeq E \mathrm{O}\! \ar[r] & B \mathrm{O}}$;
where the two possible choices of normal structure
on branes of the simple Cartesian shape
$\mathbb{R}^p \subset \mathbb{R}^d$ reflect the
unit brane charge and its opposite anti-brane charge,
while the normally framed
cup- or cap-shaped bordisms
witness their pairwise (dis-)appearance
into/out of the vacuum.

Hence where the refinement of point particles to $p$-branes
serves to {\it geometrize} both their interactions and their
charges, both of these aspects are naturally unified in
Cobordism cohomology theory.

\end{minipage}

&

\begin{minipage}[left]{6cm}

\begin{tikzpicture}

  \draw[fill=lightgray, draw opacity=0, fill opacity=.4]
    (-3,-3) rectangle (0,3);

  \draw[line width=5, lightgray]
    (0,3) to (0,-3);

  \begin{scope}

  \clip (-2.2,-2.2) rectangle (0,2.2);

  \draw[line width=2]
    (0,0) circle (2);

  \end{scope}

  \draw[fill=black]
    (0,2) circle (.1);

  \draw[fill=white]
    (0,-2) circle (.1);

  \draw[->, orangeii]
    (90:2) to  (90:2.5);

  \draw[->, orangeii]
    (90+30:2) to  (90+30:2.5);

  \draw[->, orangeii]
    (90+60:2) to  (90+60:2.5);

  \draw[->, orangeii]
    (90+90:2) to  (90+90:2.5);

  \draw[->, orangeii]
    (-90:2) to  (-90:2.5);

  \draw[->, orangeii]
    (-90-30:2) to  (-90-30:2.5);

  \draw[->, orangeii]
    (-90-60:2) to  (-90-60:2.5);

  \draw
    (0, 2.4)
    node
    {
      \rlap{
      \scalebox{.5}{
      \color{orangeii}
      \bf
      \begin{tabular}{c}
        normal framing
        in space
      \end{tabular}
      }
      }
    };

  \draw
    (0, 2)
    node
    {
      \rlap{
      \scalebox{.5}{
      \color{darkblue}
      \bf
      \begin{tabular}{c}
        brane
      \end{tabular}
      }
      }
    };

  \draw
    (0, -2.4)
    node
    {
      \rlap{
      \scalebox{.5}{
      \color{orangeii}
      \bf
      \begin{tabular}{c}
        opposite normal framing
      \end{tabular}
      }
      }
    };

  \draw
    (0, -2)
    node
    {
      \rlap{
      \scalebox{.5}{
      \color{darkblue}
      \bf
      \begin{tabular}{c}
        anti-brane
      \end{tabular}
      }
      }
    };

  \draw
    (-3.2, -1.6)
    node
    {
      \rlap{
      \scalebox{.5}{
      \color{orangeii}
      \bf
      \begin{tabular}{c}
        normal framing
        \\
        in spacetime
      \end{tabular}
      }
      }
    };

  \draw
    (-3.1, +2.8)
    node
    {
      \rlap{
      \scalebox{.5}{
      \color{darkblue}
      \bf
      \begin{tabular}{c}
        spacetime
      \end{tabular}
      }
      }
    };

  \draw
    (-1,0)
    node
    {
      $
        {
          \xleftrightharpoons{
          \mbox{
            \tiny
            \color{greenii}
            annihilation
          }
          }
        }
      $
    };

  \draw
    (-1,-.38)
    node
    {
      \raisebox{10pt}{
        \tiny
        \color{greenii}
        pair creation
      }
    };

  \draw
    (0,+1)
    node
    {
      \rotatebox[origin=c]{-90}{
      \scalebox{.5}{
      \color{darkblue}
      \bf
        space
      }
      }
    };

\end{tikzpicture}

\end{minipage}

\end{tabular}

\medskip

\noindent
\label{BraneChargeInKTheoryAndMultiplicativeGenera}
{\bf Brane charge in K-theory and multiplicative genera.}
Regarded through the prism of stable homotopy theory,
framed Cobordism splits up into a
chromatic spectrum of spectra, a small part of which looks as
follows:
$$
  \xymatrix@R=2pt{
    &
    \mbox{
      \tiny
      \color{darkblue}
      \bf
      framed
    }
    &
    \mbox{
      \tiny
      \color{darkblue}
      \bf
      \begin{tabular}{c}
        quaternionic {\color{black}/}
        \\
        symplectic
      \end{tabular}
    }
    &
    \mbox{
      \tiny
      \color{darkblue}
      \bf
      \begin{tabular}{c}
        complex {\color{black}/}
        \\
        unitary
      \end{tabular}
    }
    &
    \mbox{
      \tiny
      \color{darkblue}
      \bf
      \begin{tabular}{c}
        real {\color{black}/}
        \\
        orthogonal
      \end{tabular}
    }
    \\
    &
    &
    &
    M \mathrm{SU}
    \ar[r]
    \ar[dd]
    &
    M \mathrm{SO}
    \ar[dd]
    \\
    \mathllap{
      \raisebox{2pt}{
        \tiny
        \color{darkblue}
        \bf
        Cohomotopy
      }
      \;\;
    }
    \mathbb{S}
    \ar@{=}[r]
    &
    M \mathrm{Fr}
    \ar[r]
    &
    M \mathrm{Sp}
    \ar[ur]
    \ar[dr]
    \ar[ddd]
      |>>>>>>>{
        {\phantom{A \atop A}}
      }
    &
    &
    &
    \mbox{
      \tiny
      \color{darkblue}
      \bf
      Cobordism
    }
    \\
    &
    &&
    M \mathrm{U}
    \ar[dd]
      |<<<<<{
        {\phantom{A \atop A}}
      }
    \ar[r]
    &
    M \mathrm{O}
    \\
    &
    {\phantom{A \atop A }}
    \ar@<+4pt>@{..}[rrrr]
    &&&&
    \\
    &
    &
    K \mathrm{O}
    \ar[r]
    &
    K \mathrm{U}
    &
    &
    \mbox{
      \tiny
      \color{darkblue}
      \bf
      K-theory
    }
  }
$$
This reveals that brane charge quantization in $M \! f$-theory
is finer than but closely related to charge quantization in
K-theory: The Conner-Floyd isomorphism
(\cite[Thm. 10.1]{ConnerFloyd66}\cite[Thm. 6.35]{TamakiKono06}) says
that complex K-theory ${K\mathrm{U}}$ is the quotient of complex Cobordism cohomology $M\mathrm{U}$,
obtained by identifying
all submanifolds that do not ``wrap compact dimensions''
(see \eqref{PontrjaginIsomorphismForNonCompactManifolds} below) with the
index of their Dirac operator, i.e. with their Todd class.
An analogous statement holds for ${K\mathrm{O}}$
and quaternionic Cobordism.

\medskip
Therefore, and in view of unresolved issues
(e.g. \cite[\S 4.5.2, 4.6.5]{dBDHKMMS02}\cite[\S 8]{Evslin06}\cite[\S 1]{FredenhagenQuella05})
surrounding the
popular but still conjectural statement that D-brane charge is
quantized in K-theory, it is worthwhile to examine
M-brane charge quantization in $M\!f$-Cobordism theory.

\medskip

\noindent
{\bf Brane charge and Pontrjagin-Thom collapse.}
The above motivation of Cobordism cohomology as the
natural home for $p$-brane charge,
summarized in \hyperlink{TableInteractions}{Table I},
is in itself only a plausibility argument,
just as is the traditional motivation (\cite[\S 3]{Witten98})
of K-theory as the natural home for D-brane charge.
However, this physically plausible conclusion
is rigorously {\it implied by \hyperlink{HypothesisHOnHomotopicallyFlatSpacetimes}{Hypothesis H}}, and hence
is supported by and adds to the other
evidence for that Hypothesis:
This implication is the statement of
{\it Pontrjagin's isomorphism},
which says that the operation of
assigning to a normally framed closed submanifold
its {\it asymptotic directed distance} function
(traditionally known as the {\it Pontrjagin-Thom collapse construction})

\vspace{-.2cm}
\begin{center}
{\hypertarget{FigureD}{}}
\begin{tikzpicture}[scale=0.69]
  \node
    (X) at (-4.5,6)
    {
      \small
      $M^d$
    };
  \node (sphere) at (6,6)
    {\small
      $
        \mathbb{R}^n_{\scalebox{.6}{$\mathrm{cpt}$}}
          =
        S^n
    $};

  \draw (X)+(-2.65,-.4)
    node
    {
      $
        \xymatrix@C=26pt{
          \underset{
            \mathclap{
            \raisebox{-5pt}{
              \tiny
              \color{darkblue}
              \bf
              \begin{tabular}{c}
                closed submanifold,
                \\
                \color{orangeii}
                normally framed
              \end{tabular}
            }
            }
          }{
            \Sigma^{d-n}
          }
          \;\;
          \ar@{^{(}->}[rr]
          &&
        }
      $
    };

  \draw[->] (X) to node[above] {\footnotesize $c$} (sphere);

  \node at (-4.5,5.4)  {\tiny \color{darkblue} \bf manifold};
  \node at (-4.5,4.4)  {$\overbrace{\phantom{--------------------}}$};

  \node at (6,5.3)
    {
      \tiny \color{darkblue} \bf
      \begin{tabular}{c}
        Cohomotopy classifying space
        \\
        ($n$-sphere)
      \end{tabular}
    };
  \node at (6,4.4)  {$\overbrace{\phantom{--------------}}$};

  \node at (.04,5.7)
    {
      \tiny
      \color{greenii}
      \bf
      \begin{tabular}{c}
        \\
        directed asymptotic distance from $\Sigma$
        \\
        {\color{black} $\;\simeq\;$}
        cocycle representing Cohomotopy charge of $\Sigma$
      \end{tabular}
    };

  \begin{scope}[shift={(-6,-1.5)}]
    \clip (-2.9,-2.9) rectangle (5.9,5.9);
    \draw[step=3, dotted] (-3,-2) grid (6,6);
    \draw[very thick] (-4,1.3) .. controls (-1,-3.2) and (2.3,6.6)  .. (7,4.2);
    \begin{scope}[shift={(0,.9)}]
      \draw[dashed] (-4,1.3) .. controls (-1,-3.2) and (2.3,6.6)  .. (7,4.2);
    \end{scope}
    \begin{scope}[shift={(0,-.9)}]
      \draw[dashed] (-4,1.3) .. controls (-1,-3.2) and (2.3,6.6)  .. (7,4.2);
    \end{scope}
    \begin{scope}[shift={(0,.45)}]
      \draw[dashed, thick] (-4,1.3) .. controls (-1,-3.2) and (2.3,6.6)  .. (7,4.2);
    \end{scope}
    \begin{scope}[shift={(0,-.45)}]
      \draw[dashed, thick] (-4,1.3) .. controls (-1,-3.2) and (2.3,6.6)  .. (7,4.2);
    \end{scope}
  \end{scope}
  \begin{scope}[shift={(4,0)}]
    \draw (2,0) circle (2);
    \node at (+.6,0)
      {{\tiny $0$}
        \raisebox{.0cm}{
          $
          \mathrlap{
          \!\!\!\!\!\!\!\!\!\!
          \mbox{ \bf
          \tiny \color{darkblue}
            \begin{tabular}{c}
              regular
              \\
              value
            \end{tabular}
          }}
          $
        }
      };
    \node (zero) at (0,0) {$-$};
    \node (infinity) at (4,0) {\colorbox{white}{$\infty$}};

   \fill[black] (2,0) ++(40+180:2) node (minusepsilon)
     {\begin{turn}{-45} $)$  \end{turn}};
   \fill[black] (2,0) ++(180-40:2) node (epsilon)
     {\begin{turn}{45} $)$ \end{turn}};
   \fill[black] (2.3,0.25) ++(40+180:2) node
     { \tiny $-\epsilon$ };
   \fill[black] (2.3,-0.25) ++(-40-180:2) node
     { \tiny $+\epsilon$ };
  \end{scope}

  \draw[|->, thin, brown] (-5.1-.25,.95-.25)
    to[bend right=6.7]
    (epsilon);
  \draw[|->, thin, brown] (-5.1+.25,.05+.25)
    to[bend right=6.7]
    (minusepsilon);
  \draw[|->, thin, brown] (-5.1,.5)
    to[bend right=6.7]
    node
      {
        \colorbox{white}{
          \tiny
          \color{greenii}
          \bf
          constant on $0$ at $\Sigma$
        }
      }
    node
    {
      \tiny
      \color{greenii}
      {\phantom{
      \colorbox{white}{\bf
        codimension $n$ submanifold
      }
      }}
      $\mathrlap{
        \;\;\;\;\;\;\;\;
        \raisebox{-47pt}{
        \begin{turn}{90}
          \colorbox{white}{
            \begin{tabular}{c}
              \tiny \bf directed distance
              \\
              \tiny \bf near $\Sigma$
            \end{tabular}
          }
        \end{turn}
        }
      }$
    }
    (zero);

  \draw[|->, thin, olive]
    (-5.1-.5,1.4-.45)
    to[bend left=26]
    (infinity);
  \draw[|->, thin, olive] (-4.9,3.2) to[bend left=26] (infinity);
  \draw[|->, thin, olive] (-4.7,-2.7) to[bend right=30] (infinity);
  \draw[|->, thin, olive] (-5.1+.5,-.4+.45)
    to[bend right=33] node[below]
    {\colorbox{white}{
    \tiny
    \color{greenii}
    \bf
    \begin{tabular}{c}
      constant on $\infty$
      far away from $\Sigma$
      \\
    \end{tabular}
    \hspace{-.34cm}
    }}
    (infinity);

  \draw
    (-9.2,-1.2)
    node
    {
      \scalebox{.8}{
        $\Sigma$
      }
    };

  \begin{scope}[shift={(-7+.59,-.97+.41)}]
    \draw[->, orangeii]
      (0,0)
      to
      (124:.3);
  \end{scope}

  \begin{scope}[shift={(-7,-.97)}]
    \draw[->, orangeii]
      (0,0)
      to
      (118:.33);
  \end{scope}

  \begin{scope}[shift={(-7-.6,-.97-.28)}]
    \draw[->, orangeii]
      (0,0)
      to
      (107:.35);
  \end{scope}

  \begin{scope}[shift={(-7-1.3,-.97-.4)}]
    \draw[->, orangeii]
      (0,0)
      to
      (88:.38);
  \end{scope}

\end{tikzpicture}
\end{center}
\vspace{-9mm}
\noindent {\bf \footnotesize Figure D --  The Pontrjagin construction.}
{\footnotesize
The {\it charge in Cohomotopy} of a manifold $M^d$,
sourced by a normally framed closed submanifold $\Sigma^{d-n}$,
is the homotopy class of the
function that assigns directed asymptotic distance from
$\Sigma$, measured along its normal framing.}

\vspace{.2cm}

\noindent
{\it identifies} framed Cobordism with Cohomotopy,
as non-abelian cohomology theories,
over {\it closed manifolds} $M^d$:
\begin{equation}
  \label{PontrjaginEquivalenceBetweenUnstableFramedCobordismAndCohomotopy}
 \hspace{-3mm}
  \xymatrix@C=9pt{
    \overset{
      \mathclap{
      \raisebox{6pt}{
        \tiny
        \color{darkblue}
        \bf
        \begin{tabular}{c}
          framed unstable
          \\
          $n$-Cobordism of $M^d$
        \end{tabular}
      }
      }
    }{
      \mathrm{Cob}_{\mathrm{Fr}}^n\big( M^d \big)
    }
    \ar@{}[r]|-{ := }
    \ar@/^2pc/[rrrr]|-{
      \mbox{
        \tiny
        \color{greenii}
        \bf
        assign Cohomotopy charge
      }
    }
    \ar@/_2pc/@<-4pt>@{<-}[rrrr]|-{
      \mbox{
        \tiny
        \color{greenii}
        \bf
        reconstruct submanifold
        from its charge
      }
    }
    &
    \mathrm{NFramedSubmflds}_{d-n}
    \big(
      M^d
    \big)_{\!\!/\mathrm{brdsm}}
    \ar@<+8pt>[rr]|-{
      \mbox{
        \tiny
        \color{greenii}
        \bf
        directed asymptotic distance
      }
    }
    \ar@{}[rr]|-{ \simeq }
    \ar@<-8pt>@{<-}[rr]|-{
      \mbox{
        \tiny
        \color{greenii}
        \bf
        pre-image of regular value 0
      }
    }
    &
    {\phantom{AAAAAAAAAAAAA}}
    &
   \;\; \mathrm{Maps}
    \big(
      M^d
      \!,\,
      S^n
    \big)_{\!\!/{\mathrm{hmtpy}}}
    \ar@{}[r]|-{ =: }
    &
    \overset{
      \mathclap{
      \raisebox{6pt}{
        \tiny
        \color{darkblue}
        \bf
        \begin{tabular}{c}
          unstable
          \\
          $n$-Cohomotopy of $M^d$
        \end{tabular}
      }
      }
    }{
      \pi^n
      \big(
        \underset{
          \mathclap{
          \raisebox{-6pt}{
            \tiny
            \bf
            \begin{tabular}{c}
              \color{purple}
              closed
              \\
              \color{darkblue}
              manifold
            \end{tabular}
          }
          }
        }{
          M^d
        }
      \big).
    }
  }
\end{equation}

\vspace{.2cm}

\noindent
{\bf Charges vanishing at infinity and reduced cohomology.}
In the case that $M^d \,=\, \mathbb{R}^d$ is contractible
(and hence, in particular, not closed), the homotopy-invariance
of Cohomotopy theory immediately implies that {\it all Cohomotopy charges
vanish} on such $M^d$, even though there are non-trivial cobordism classes
of normally framed submanifolds in $\mathbb{R}^d$.
But translating the trivialization of their Cohomotopy charge
back through the reconstruction map \eqref{PontrjaginEquivalenceBetweenUnstableFramedCobordismAndCohomotopy}
shows that
this is a result of these submanifolds
being allowed to ``escape to infinity'',
carrying their charges away with them.
Hence the sensible notion of brane charge
quantized in any pointed cohomology theory $A$ is
that which is constrained not to escape to
infinity, hence to {\it vanish at infinity}:

A
{\it charge vanishes at infinity}
in a non-abelian cohomology theory $A(-)$ \eqref{NonabelianCohomologyInIntroduction}
with a zero-element
$0_A \in A$ in its classifying space,
if it is represented by a cocycle map
$X \xrightarrow{\;c\;} A$
that extends with value $0_A$ to the
{\it one-point compactification} of $X$ \eqref{OnePointCompactificationInIntroduction}:

\begin{equation}
  \label{OnePointCompactificationAndFunctionsOnIt}
  \mathclap{
  \scalebox{.85}{$
  \overset{
    \mathclap{
    \raisebox{3pt}{
      \tiny
      \color{darkblue}
      \bf
      \begin{tabular}{c}
      one-point-
      \\
      compactification
      \end{tabular}
    }
    }
  }{
    M_{\scalebox{.6}{$\mathrm{cpt}$}}
  }
  \;\coloneqq\;
  \left(
    \begin{aligned}
    &
    M \sqcup \{\infty_{{}_X}\},\;
    \mbox{topologized such that}:
    \\
    &
    \!\!
    \begin{aligned}
      &
      \mbox{
        $M \hookrightarrow M \sqcup \{\infty_{{}_X}\}$
        is open embedding
      }
      \\
      &
      \mbox{
        $O_\infty$
        is open nbrhd
      }
      \!\Leftrightarrow\!
      \mbox{
        $
        M \!\setminus\! O_\infty
        $
        is cmpct
      }
    \end{aligned}
    \end{aligned}
    \!\!
  \right)
  \;\;\mbox{i.e.:}\;\;\;
  \arraycolsep=2pt
  \begin{array}{lcl}
    \mbox{ map $X \xrightarrow{\;c\;} A$... }
    &
    \leftrightarrow
    &
    \mbox{
      extension
      $M_{\scalebox{.6}{cpt}} \xrightarrow{\;\widetilde c\;} A$
      ...
    }
    \\
    \mbox{...vanishes at infinity}
    &\Leftrightarrow&
    \mbox{...equals $0_A$ at $\infty_{{}_M}$ }
    \\
    \mbox{...compactly supported}
    &\Leftrightarrow&
    \mbox{...equals $0_A$ on nbrhd of $\infty_{{}_M}$}
  \end{array}
  $
  }
  }
\end{equation}
The {\it reduced cohomology} $\widetilde A(-)$
of $X \coloneqq M_{\scalebox{.6}{cpt}} \wedge \mathbb{R}^{p,1}_+$
is that given by cocycles and coboundaries which vanish at infinity:
\begin{equation}
  \label{ReducedCohomotopy}
  \overset{
    \mathclap{
    \raisebox{3pt}{
      \tiny
      \color{darkblue}
      \bf
      \begin{tabular}{c}
        reduced
        \\
        $A$-cohomology
      \end{tabular}
    }
    }
  }{
    {\widetilde A}^n
    (
      X
    )
  }
  \;:=\;
  \pi_0
  \overset{
    \mathclap{
    \raisebox{3pt}{
      \tiny
      \color{darkblue}
      \bf
      \begin{tabular}{c}
        maps that take
        \\
        $\infty \in X$ to $0 \in A$
      \end{tabular}
    }
    }
  }{
    \mathrm{Maps}^{\ast/\!}
  }
  (
    X
    ,
    A
  )
  \,,
  \phantom{AAAA}
  \overset{
    \mathclap{
    \raisebox{3pt}{
      \tiny
      \color{darkblue}
      \bf
      \begin{tabular}{c}
        reduced
        \\
        Cohomotopy
      \end{tabular}
    }
    }
  }{
    {\widetilde \pi}^{\, n}
    (
      X
    )
  }
  \;:=\;
  \pi_0
  \overset{
    \mathclap{
    \raisebox{3pt}{
      \tiny
      \color{darkblue}
      \bf
      \begin{tabular}{c}
        maps that take
        \\
        $\infty \in X$
        to
        $0 \in S^n$
        \rlap{
          defined as
          $\infty \in \mathbb{R}^n_{\scalebox{.6}{cpt}}$
        }
      \end{tabular}
    }
    }
  }{
    \mathrm{Maps}^{\ast/\!}
  }
  \big(
    X
    ,
    \mathbb{R}^n_{\scalebox{.6}{$\mathrm{cpt}$}}
  \big)
  \,.
\end{equation}

\medskip

\noindent
{\bf Wrapped branes and the Pontrjagin theorem.}
In terms of reduced Cohomotopy \eqref{ReducedCohomotopy},
the Pontrjagin isomorphism \eqref{PontrjaginEquivalenceBetweenUnstableFramedCobordismAndCohomotopy}
takes the following form, now valid for possibly non-compact
manifolds $M^d$ (see Prop. \ref{PontrjaginIsomorphism}, Rem. \ref{AttributionsForPontrjaginTheorem}):
\begin{equation}
  \label{PontrjaginIsomorphismForNonCompactManifolds}
  \xymatrix@C=5em{
    \underset{
      \!\!\!\!\!\!\!\!\!
      \!\!\!\!\!\!\!
      \raisebox{-16pt}{
      \scalebox{.9}{
      $
  \def\arraystretch{.3}
  \begin{array}{c}
    \mathllap{
      \raisebox{2pt}{
      \scalebox{1.2}{
        \tiny
        \color{darkblue}
        \bf
        branes wrapped on:
      }
      }
    }
    \Sigma^{\mathrlap{d-n}}
    \\
    \times
    \\
    \mathllap{
      \raisebox{2pt}{
      \scalebox{1.2}{
        \tiny
        \color{darkblue}
        \bf
        extended along:
      }
      }
    }
    \mathbb{R}^{\mathrlap{p,1}}
  \end{array}
  \;\;\;\subset\!\!
  \def\arraystretch{.3}
  \begin{array}{c}
    M^{\mathrlap{d}}
    \\
    \times
    \\
    \mathbb{R}^{\mathrlap{p,1}}
  \end{array}
      $}
      }
    }{
    \overset{
      \mathclap{
      \raisebox{6pt}{
        \tiny
        \color{darkblue}
        \bf
        \begin{tabular}{c}
          framed unstable
          \\
          $n$-Cobordism of $M^d$
        \end{tabular}
      }
      }
    }{
      \mathrm{Cob}^n_{\mathrm{Fr}}
      \big(
        M^d
      \big)
    }
    }
    \quad
    \ar@/^.9pc/[rr]
      ^-{
        \;
        \mbox{
          \tiny
          \color{greenii}
          \bf
          assign Cohomotopy charge
        }
        \;
      }
    \ar@{<-}@/_.9pc/[rr]
      _-{
        \;\;\;\;
        \mbox{
          \tiny
          \color{greenii}
          \bf
          find worldvolume of given charge
        }
        \;
      }
    \ar@{}[rr]|-{\simeq}
    &
    {\phantom{AAAAAAAAA}}
    &
    \qquad
    \overset{
      \mathclap{
      \raisebox{8pt}{
        \tiny
        \color{darkblue}
        \bf
        \begin{tabular}{c}
          reduced
          \\
          $n$-Cohomotopy
        \end{tabular}
      }
      }
    }{
    {
      \underset{
        \mathclap{
        \raisebox{-14pt}{
          \tiny
          \color{darkblue}
          \bf
            \begin{tabular}{c}
              charge
              \\
              vanishes
              \\
              at infinity
            \end{tabular}
          }
        }
      }{
        \widetilde \pi}^{\, n}
      }
    }
    \Big(
      \underset{
        \mathclap{
        \raisebox{-10pt}{
          \tiny
          \color{darkblue}
          \bf
          \begin{tabular}{c}
            along
            these...
          \end{tabular}
        }
       }
      }{
          M^d_{\scalebox{.6}{$\mathrm{cpt}$}}
      }
      \wedge
      \underset{
        \mathclap{
        \raisebox{-14pt}{
          \tiny
          \color{darkblue}
          \bf
          \begin{tabular}{c}
            ...but not
            \\
            along these
            \rlap{
              dimensions
            }
          \end{tabular}
        }
       }
      }{
        \mathbb{R}^{p,1}_+
      }
    \Big).
  }
\end{equation}
On the right of \eqref{PontrjaginIsomorphismForNonCompactManifolds}
we have included a contractible factor, for conceptual completeness:
By the homotopy-invariance of Cohomotopy this does not
affect the nature of the charges, which we may interpret as saying
that:

\noindent
{\it The Cohomotopy charge of a brane filling
a flat space(-time) factor $\mathbb{R}^{p,1}$
and wrapped on a closed submanifold $\Sigma$
inside the remaining spatial dimensions $M^d$
depends only on this compact factor} (with its normal framing).

\medskip

We turn to the archetypical class of examples of this
situation:

\medskip

\noindent {\bf Probe branes near black brane horizons
 and homotopy groups of spheres.}
With the above formulation we may formally grasp the
spacetime topologies supporting charges of
{\it probe $p$-branes} near horizons of
{\it solitonic $b$-branes} --
generalizing the classical case
(from p. \pageref{ExampleMagneticChargeInOrdinaryCohomology})
of
magnetic monopoles (solitonic 0-branes)
and magnetic flux lines (probe 1-branes):

\noindent
{\bf (a)}
The {\it near horizon topology} of
(infinitely extended, coincident)
{\it solitonic $b$-branes} in an 11-dimensional spacetime is
the complement $\mathbb{R}^{10,1}  \setminus  \mathbb{R}^{b,1}$
of their (geometrically singular)
worldvolume inside an ambient
spacetime chart;
and
the corresponding {\it asymptotic boundary of spacetime}
is, topologically, the subspace
$\simeq  \, \mathbb{R}^{b,1}  \times  S^{9 -b} $ of any fixed radius $r$
(geometrically a limit $r \to 0$) from the singular locus.

\noindent
{\bf (b)} The {\it effective spacetime topology}
for (infinitely extended, not necessarily coincident)
{\it probe $p$-branes}
localized
(i.e. with  charges vanishing at infinity)
near the horizon of such
solitonic $p$-branes
is the one-point compactification
\eqref{OnePointCompactificationAndFunctionsOnIt}
of their transverse space
$\mathbb{R}^{p-b} \times S^{9-p}$ inside this near-horizon domain:

\vspace{-.4cm}
\begin{equation}
  \label{SolitonicBraneSpacetime}
  \mathclap{
  \hspace{-7mm}
  \xymatrix@C=25pt{
  \overset{
    \mathclap{
    \raisebox{6pt}{
      \tiny
      \color{darkblue}
      \bf
      \begin{tabular}{c}
        ambient
        \\
        spacetime
      \end{tabular}
    }
    }
  }{
    \mathbb{R}^{10,1}
  }
  \setminus
  \underset{
    \mathclap{
    \raisebox{-6pt}{
      \tiny
      \color{darkblue}
      \bf
      \begin{tabular}{c}
        singular $b$-brane
        \\
        worldvolume
      \end{tabular}
    }
    }
  }{
    \mathbb{R}^{b,1}
  }
  \;\simeq\;
  \overset{
    \mathclap{
    \raisebox{+6pt}{
      \tiny
      \color{darkblue}
      \bf
      \begin{tabular}{c}
        asymptotic
        \\
        worldvolume
      \end{tabular}
    }
    }
  }{
    \mathbb{R}^{b,1}
  }
  \!\times
  \underset{
    \mathclap{
    \raisebox{-6pt}{
      \tiny
      \color{darkblue}
      \bf
      \begin{tabular}{c}
        sphere around
        \\
        $b$-brane
      \end{tabular}
    }
    }
  }{
    S^{9-b}
  }
  \!\times\!
  \overset{
    \mathclap{
    \raisebox{+6pt}{
      \tiny
      \color{darkblue}
      \bf
      \begin{tabular}{c}
        radial
        \\
        distance
      \end{tabular}
    }
    }
  }{
    \mathbb{R}_{\mathrm{rad}}
  }
  \;
  \ar@{~>}[r]
    _-{
      \mbox{
        \tiny
        \color{greenii}
        \bf
        \begin{tabular}{c}
          pass to
          \\
          asymptotic
          \\
          boundary
        \end{tabular}
      }
    }
  &
  \;
  \overset{
    \mathclap{
    \raisebox{+6pt}{
      \tiny
      \color{orangeii}
      \bf
      \begin{tabular}{c}
        asymptotic boundary
        \\
        of $b$-brane spacetime
      \end{tabular}
    }
    }
  }{
    \mathbb{R}^{b,1} \times S^{9-b}
  }
  \;\simeq\;
  \underset{
    \mathclap{
    \raisebox{-6pt}{
      \tiny
      \color{darkblue}
      \bf
      {\begin{tabular}{c}
        probe brane
        \\
        worldvolume
      \end{tabular}}
    }
    }
  }{
    \mathbb{R}^{p,1}
  }
  \!\times\!
  \overset{
    \mathclap{
    \raisebox{+6pt}{
      \tiny
      \color{darkblue}
      \bf
      \begin{tabular}{c}
        transversal space to
        \\
        probe $p$-branes
        near black $b$-brane
      \end{tabular}
    }
    }
  }{
    \mathbb{R}^{b-p}
    \!\times\!
    S^{9 - b}
  }
  \ar@{~>}[r]
    _-{
      \mbox{
        \tiny
        \color{greenii}
        \bf
        \begin{tabular}{c}
          localize
          \\
          probe $p$-branes
          \\
          near black $b$-brane
        \end{tabular}
      }
    }
  &
  \;
  \overset{
    \mathclap{
    \raisebox{4pt}{
      \tiny
      \color{orangeii}
      \bf
      \begin{tabular}{c}
        asymptotic boundary
        of $b$-brane spacetime
        \\
        seen by localized $p$-brane probes
      \end{tabular}
    }
    }
  }{
  \mathbb{R}^{p,1}
  \!\wedge\!
  \underset{
    \mathclap{
    \raisebox{-4pt}{
      \tiny
      \color{orangeii}
      \bf
      \begin{tabular}{c}
        space transversal to $p$-branes
        \\
        including transversal $\infty$
      \end{tabular}
    }
    }
  }{
  \big(
    \mathbb{R}^{b-p}
      \!\times\!
    S^{9-b}
  \big)_{\mathrm{cpt}}.
  }
  }
  }
}
\end{equation}
\vspace{-.4cm}

\noindent
\label{InterpretingProbeBraneWorldvolumes}
Now, since the
reduced 4-Cohomotopy charge \eqref{ReducedCohomotopy}
of the transversal space
is identified \eqref{PontrjaginIsomorphismForNonCompactManifolds}
under the Pontrjagin isomorphism
\eqref{PontrjaginEquivalenceBetweenUnstableFramedCobordismAndCohomotopy}
with
$(9 - 4 = 5)$-brane worldvolumes
extended along $\mathbb{R}^{p,1}$ and wrapped
on (cobordism classes of normally framed) submanifolds $\Sigma$
of the transversal asymptotic boundary \eqref{SolitonicBraneSpacetime}:
\vspace{-2mm}
\begin{equation}
  \label{CohomotopyChargeOnBlackBraneAsymptoticBoundary}
  \hspace{-6mm}
  \xymatrix@R=-13pt{
  \underset{
    \mathclap{
    \raisebox{-3pt}{
      \tiny
      \begin{tabular}{c}
        \color{darkblue}
        \bf
        charge lattice
        (under \hyperlink{HypothesisHOnHomotopicallyFlatSpacetimes}{\it Hypothesis H}) of
        \\
        \color{darkblue}
        \bf
        probe $p$-branes
        at $b$-brane horizons
        \\
      \end{tabular}
    }
    }
  }{
  \mbox{
    $
      \big(
        \mathrm{M}p @ \mathrm{M}b
      \big)
      \!
    $
    BndrCharges
  }
  }
  \!\!
  \coloneqq
  {\widetilde \pi}^4
  \Big(\!
    \mathbb{R}^{p,1}_+
      \wedge
    \big(
      \mathbb{R}^{b - p}
        \times
      S^{9 - b}
    \big)_{\mathrm{cpt}}
 \! \Big)
  \ar[rr]^-{ \simeq }
  &
  {\phantom{AAAAA}}
  &
  \big\{\!
    \mathbb{R}^{p,1}
  \! \big\}
\!   \times
  \mathrm{Cob}^4_{\mathrm{Fr}}
    \big(
      \mathbb{R}^{b - p}
       \! \times \!
      S^{9 - b}
    \big)
    \\
  \hspace{5.5cm}   [c]
    {
      \left(
      \!\!\!
      \scalebox{.6}{
        \begin{tabular}{l}
          representative $c$
          \\
          factoring through $\mathrm{pr}_2$
          \\
          and regular at $0 \in S^4$
        \end{tabular}
      }
      \!\!\!
      \right)
    }
    \ar@{|->}[rr]^-{
            \mbox{
        \tiny
        \color{greenii}
        \bf
        \begin{tabular}{c}
          associate with Cohomotopy charge
          \\
          the wrapped brane sourcing it
        \end{tabular}
      }
    }
    &&
    \big[
      \mathbb{R}^{p,1}
        \times
      \Sigma
    \big]
    \;:=\;
    \big[
      c^{-1}(\{0\})
    \big]
  }
\end{equation}


\vspace{-.15cm}

\hspace{-.9cm}
\begin{tabular}{ll}

\begin{minipage}[left]{6.5cm}

we interpret, under
\hyperlink{HypothesisHOnHomotopicallyFlatSpacetimes}{\it Hypothesis H}, the product manifold
$$
  \mathbb{R}^{p,1} \times \Sigma
  \;\subset\;
  \mathbb{R}^{p,1}
    \times
  \big(
    \mathbb{R}^{b-p} \times S^{9-b}
  \big)
$$
of dimension
$$
  \begin{aligned}
  \mathrm{dim}
  \big(
    \mathbb{R}^{p,1} \times \Sigma
  \big)
  &
  =
  (p + 1) +  \big( (9 - p) - 4\big)
  \\
  & = \, 5 + 1
  \end{aligned}
$$
as the probe 5-brane worldvolume which carries
the given Cohomotopy charge $[c]$,
and hence interpret $\Sigma$ as
the ``brane polarization''
(in the sense of \cite[\S 4]{Myers03})
of the probe $p$-brane worldvolume,
puffing it up to a 5-brane worldvolume,
due to the background 5-brane charge $[c]$.
\end{minipage}

&

\hspace{1mm}
\raisebox{-140pt}{
\scalebox{.8}{
\begin{tikzpicture}

\draw[gray, line width=1.7, dashed]
  (0,+4.3)
  ellipse
  (6 and .9);

\draw
  (1.7,4.7)
  node
  {
    {
    \scalebox{.8}{
      \color{darkblue}
      \bf
      \begin{tabular}{c}
        effective $(b-p)$-sphere
        \\
        transversal to $p$-brane
      \end{tabular}
    }
    }
  };

\draw[draw=lightgray, fill=lightgray]
  (-5,-4)
  rectangle
  (5,4);

\draw[gray, line width=2]
  (-5, 4)
  to
  (-5,-4);

\draw[gray, line width=2]
  (+5, 4)
  to
  (+5,-4);

\draw[dashed, fill=lightgray]
  (-5,-4)
    --
  (+5,-4)
    --
  (+5+1.5,-4-1.1)
    --
  (-5+1.5,-4-1.1);

\draw[dashed, fill=lightgray]
  (+5,-7pt)
    --
  (+5+1.5,-7pt-1.1cm)
    --
  (+5+1.5,-4cm-1.1cm)
    --
  (+5,-4);

\draw[lightgray, line width=2]
  (+5,-7pt)
    to
  (+5+1.5,-7pt-1.1cm);

\draw[dashed]
  (-5,-4)
  to
  (-5+1.5,-4-1.1);

\draw[gray, line width=2]
  (+5, -7pt)
  to
  (+5,-4);

\draw[white, line width=13]
  (-5.5,0)
  to
  (0,0);

\draw[gray, dashed, line width=1.5]
  (0,0)
  ellipse
  (1.2 and 3);

\draw[white, line width=13]
  (0,0)
  to
  (5.5,0);

\draw
  (.8,-4.6)
  node
  {
    \scalebox{.7}{
      \color{darkblue}
      \bf
      asymptotic $b$-brane worldvolume
      \color{black}
      $\mathbb{R}^{b,1}$
    }
  };

\draw
  (-4.5,-4.9)
  node
  {
    \scalebox{.7}{
      \rotatebox[origin=c]{-38}{
        \color{darkblue}
        \bf
        \def\arraystretch{.8}
        \begin{tabular}{c}
          probe $p$-brane
          \\
          worldvolume
          \color{black} $\mathbb{R}^{p,1}$
        \end{tabular}
      }
    }
  };

\draw[->,gray, dashed]
  (0,0)
  to
  node[very near end]
  {
    \rotatebox[origin=c]{43}
    {
      \scalebox{.7}{
        \color{darkblue}
        \bf
        \begin{tabular}{c}
          radial
          \\
          \phantom{A}
        \end{tabular}
      }
    }
  }
  (180+43:3.6);

\draw
  (2.9,-.03)
  node
  {\scalebox{.8}{
    \color{darkblue}
    \bf
    solitonic $b$-brane singularity
  }};

\begin{scope}[shift=(20:1.2 and 3), scale=(1)]
  \shade[ball color=gray!40, opacity=.9] (0,0) circle (.2);
  \draw (-.2,0) arc (180:360:.2 and .06);
  \draw[densely dotted] (.2,0) arc (0:180:.2 and .06);
\end{scope}

\begin{scope}[shift=(-30:1.2 and 3), scale=(1.14)]
  \shade[ball color=gray!40, opacity=.9] (0,0) circle (.2);
  \draw (-.2,0) arc (180:360:.2 and .06);
  \draw[densely dotted] (.2,0) arc (0:180:.2 and .06);
\end{scope}

\begin{scope}[shift=(50:1.2 and 3), scale=(1.2)]
  \shade[ball color=gray!40, opacity=.9] (0,0) circle (.2);
  \draw (-.2,0) arc (180:360:.2 and .06);
  \draw[densely dotted] (.2,0) arc (0:180:.2 and .06);
\end{scope}

\begin{scope}[shift=(200:1.2 and 3), scale=(2.5)]
  \shade[ball color=gray!40, opacity=.9] (0,0) circle (.2);
  \draw (-.2,0) arc (180:360:.2 and .06);
  \draw[densely dotted] (.2,0) arc (0:180:.2 and .06);
\end{scope}

\begin{scope}[scale=2.24]
\draw
  (187.6:1.2 and 3)
  node
  {
    \scalebox{.8}{
      \color{darkblue}
      \bf
      \def\arraystretch{.8}
      \begin{tabular}{c}
        probe $p$-brane
        \\
        polarization
        {\color{black}$\Sigma$}
      \end{tabular}
      \hspace{2pt}
    }
  };
\end{scope}

\begin{scope}[scale=1.9]
\draw
  (160:1.2 and 3)
  node
  {
    \scalebox{.8}{
      \color{darkblue}
      \bf
      \begin{tabular}{c}
        $(9-b)$-sphere
        \\
        around $b$-brane
      \end{tabular}
      \hspace{4pt}
    }
  };
\end{scope}

\draw[<->, dashed, gray, line width=1.3]
  (-4.8, 3.6)
  to
  node
  {
    \scalebox{.8}{
      \colorbox{lightgray}{
        \color{darkblue}
        \bf
        transversal $(b-p)$-space
      }
    }
  }
  (+4.8, 3.6);

\draw
  (6,3)
  node
  {
    \begin{tabular}{l}
      \scalebox{1.2}{$\infty$}
      \\
      \hspace{-.2cm}
      \scalebox{.8}{
        \color{darkblue}
        \bf
        \begin{tabular}{l}
          transversal
          \\
          infinity
        \end{tabular}
      }
    \end{tabular}
  };

\draw
  (-5.7,3.55)
  node
  {
    \scalebox{1.2}{$\infty$}
  };

\begin{scope}[shift=(270:1.2 and 3), scale=(1.8)]
  \shade[ball color=gray!40, opacity=.9] (0,0) circle (.2);
  \draw (-.2,0) arc (180:360:.2 and .06);
  \draw[densely dotted] (.2,0) arc (0:180:.2 and .06);
\end{scope}

\end{tikzpicture}
}
}

\end{tabular}

\vspace{-.2cm}

\noindent
Observe that these Cohomotopy charges
\eqref{CohomotopyChargeOnBlackBraneAsymptoticBoundary}
subsume low unstable {\it homotopy groups of spheres}:

\hspace{-.9cm}
\begin{tabular}{lll l}
  $\mathclap{\phantom{\vert^{\vert^{\vert}}}}$
  {\bf (i)}
  $\mathclap{\phantom{\vert_{\vert_{\vert}}}}$
  &
  $b = p$
  &
  i.e.: probe $p$-branes near their own horizon
  &
  $
    \Rightarrow
    \,
    \mbox{
    $
      \big(
        \mathrm{M}p @ \mathrm{M}p
      \big)
      \!
    $
    BndrCharges
    }
    \simeq
    \pi_{9-p}
    (
      S^4
    )
  $
  \\
  $\mathclap{\phantom{\vert^{\vert^{\vert}}}}$
  {\bf (ii)}
  $\mathclap{\phantom{\vert_{\vert_{\vert}}}}$
  &
  $b = 9$
  &
  i.e.: probe $p$-branes near an MO9-plane
  &
  $
    \Rightarrow
    \,
    \mbox{
    $
      \big(
        \mathrm{M}p @ \mathrm{M}9
      \big)
      \!
    $
    BndrCharges
    }
    \simeq
    \pi_{9-p}
    (
      S^4
    )
    \,\times\,
    \mathbb{Z}^{\mathrm{HW}}_2
  $
\end{tabular}

\vspace{1mm}
\begin{equation}
  \label{HomotopyGroupsOfThe4Sphere}
\mbox{
\begin{tabular}{|r||c|c|c|c|c|c|c|c|c|c|}
  \hline
  $
    \mathclap{\phantom{\vert^{\vert}}}
    p  =
    \mathclap{\phantom{\vert_{\vert}}}
  $
  &
  $0$
  &
  $1$
  &
  $2$
  &
  $3$
  &
  $4$
  &
  $5$
  &
  $6$
  &
  $7$
  &
  $8$
  &
  $9$
  \\
  \hline
  \hline
  $
    \mathclap{\phantom{\vert^{\vert^{\vert}}}}
    \pi_{9-p}
    \big(
      S^4
    \big)
    =
    \mathclap{\phantom{\vert^{\vert_{\vert_{\vert_{\vert}}}}}}
  $
  &
  $\mathbb{Z}_2^2$
  &
  $\mathbb{Z}_2^2$
  &
  $
    {
      \color{purple}
      \mathbb{Z}
    }
      \;\oplus\;
    \mathbb{Z}_{12}$
  &
  $\mathbb{Z}_2$
  &
  $\mathbb{Z}_2$
  &
  $
    {
      \color{purple}
      \mathbb{Z}
    }
  $
  &
  $0$
  &
  $0$
  &
  $0$
  &
  $0$
  \\
  \hline
\end{tabular}
}
\end{equation}
\vspace{-.1cm}

\label{M5BraneProbesAtMO9Planes}
\hspace{-.9cm}
\begin{tabular}{ll}

  \begin{minipage}[left]{6.2cm}

  {\bf M5-Brane probes at
  MO9-planes and the Hopf degree theorem}
  is the special case
  \fbox{$b = 9$ and $p = 5$}
  of \eqref{CohomotopyChargeOnBlackBraneAsymptoticBoundary}.
  This was discussed, under \hyperlink{HypothesisH}{Hypothesis H},
  in \cite{SS19a}\cite{BSS19}\cite{SS20c}
  in the generality of orbifold geometry and
  equivariant Cohomotopy (\cite{SS20b}):
  Here the free $\mathbb{Z}_2^{\mathrm{HW}}$-action
  identifies $S^{9 - b} = S^0$ with a point and leaves
  a transversal space which is an ADE-orbifold of
  $\mathbb{R}^{9-p} = \mathbb{R}^4$.

  In this transversal space the probe M5-branes appear,
  under the Pontrjagin isomorphism
  (\cref{PontrjaginIsomorphismForNonCompactManifolds}),
  as points
  (i.e. without polarization, as expected).

  \end{minipage}

  &

\raisebox{0pt}{
\begin{minipage}[left]{7cm}
\scalebox{.75}{
\begin{tikzpicture}[decoration=snake]

  \shade[right color=lightgray, left color=white]
    (3,-3)
      --
    (-1,-1)
      --
        (-1.21,1)
      --
    (2.3,3);

  \draw[dashed]
    (3,-3)
      --
    (-1,-1)
      --
    (-1.21,1)
      --
    (2.3,3)
      --
    (3,-3);

  \begin{scope}[rotate=(+8)]
  \draw[dashed]
    (1.5,-1)
    ellipse
    (.2 and .37);
  \draw
   (1.5,-1)
   to node{
     \rotatebox[origin=c]{7}{
     \scalebox{.7}{
     \color{orangeii}
     \bf
     \begin{tabular}{c}
      \;\;\;\;\;\;\;\;\;\;\;\;probe M5-brane
       \\
      \phantom{A}
     \end{tabular}
     }
     }
   }
   (-2.2,-1);
  \draw
   (1.5+1.2,-1)
   to
   (4,-1);
  \end{scope}

  \begin{scope}[shift={(-.2,1.4)}, scale=(.96)]
  \begin{scope}[rotate=(+8)]
  \draw[dashed]
    (1.5,-1)
    ellipse
    (.2 and .37);
  \draw
   (1.5,-1)
   to
   (-2.3,-1);
  \draw
   (1.5+1.35,-1)
   to
   (4.1,-1);
  \end{scope}
  \end{scope}

  \begin{scope}[shift={(-1,.5)}, scale=(.7)]
  \begin{scope}[rotate=(+8)]
  \draw[dashed]
    (1.5,-1)
    ellipse
    (.2 and .32);
  \draw
   (1.5,-1)
   to
   (-1.8,-1);
  \end{scope}
  \end{scope}

  \begin{scope}[shift={(9.6,.6)}, scale=(.9)]

  \begin{scope}[scale=1]

  \shade[ball color=gray!40, opacity=.6] (0,0) circle (2);
  \draw (-2,0) arc (180:360:2 and 0.6);
  \draw[dashed] (2,0) arc (0:180:2 and 0.6);

  \draw (-114:1.5) node
    {
      \large
      \color{blue}
      $S^4$
    };
  \draw (+90:2.2) node
    {
      \scalebox{.8}{
        $0$
      }
    };
  \draw (-90:2.2) node
    {
      $\infty$
    };

  \end{scope}

  \end{scope}

  \begin{scope}
  \clip
    (2,0)
    rectangle
    (8,4);
  \draw[line width=3.5, white]
    (3.3,.9) to[bend left=20] (9.5,2.38);
  \end{scope}
  \draw[line width=1, olive, arrows={[scale=1.2]|->[scale=2]}]
    (3.3,.9) to[bend left=20] (9.5,2.38);

  \begin{scope}
  \clip
    (2,0)
    rectangle
    (8,-4);
  \draw[line width=3.5, white]
    (2.92,-2.2) to[bend left=7] (9.5,-1.2);
  \end{scope}
  \draw[line width=1, olive, arrows={[scale=1.2]|->[scale=2]}]
    (2.92,-2.2) to[bend left=7] (9.5,-1.2);

 \draw
   (1,4)
   node
   {
     $
       \overset{
         \;\;\;
         \scalebox{1.1}{$
         \underset{
           \raisebox{3pt}{
             \tiny
             \color{darkblue}
             \bf
             \begin{tabular}{c}
               spacetime seen by fields
               vanishing at transversal infinity
             \end{tabular}
           }
         }{
           \scalebox{.8}{$
           X \,:=\,
           \mathbb{R}^{5,1}
           \times
           \overset{ \simeq \, S^4 }{
           \overbrace{
           \big(
             \mathbb{R}^{9-5}
             \times
             S^{9-9}
             /
             \mathbb{Z}^{\mathrm{HW}}_2
           \big)_{\mathrm{cpt}}
           }
           }
           $}
         }
         $}
       }{
         \overbrace{
           \phantom{------------------}
         }
       }
     $
   };

 \draw
   (5.6,4.1)
   node
   {
     \scalebox{1.5}{
     \xymatrix@C=30pt{
      \ar[rr]^-{\scalebox{.55}{$c$}}_-{
        \scalebox{.7}{
          \tiny
          \color{greenii}
          \bf
          \begin{tabular}{c}
            Cohomotopy charge
          \end{tabular}
        }
      }
      &&
     }
     }
   };

 \draw
   (9.6, 3.8)
   node
   {
     $
       \overset{
         \;
         \scalebox{1.1}{$
           \underset{
             \raisebox{3pt}{
               \tiny
               \color{darkblue}
               \bf
               classifying space for 4-Cohomotopy
             }
           }{
           \scalebox{.8}{$
             S^4
           $}
           }
           \mathclap{\phantom{\big(\mathbb{R}^2 \big)_{\mathrm{cpt}}}}
         $}
       }{
       \overbrace{
         \phantom{-----------}
       }
       }
     $
   };

\draw
  (6.3,-3.4)
  node
  {
    \scalebox{1}{$
  [ c ]
  \;\in\;
  \underset{
    \mathclap{
    \!\!\!\!\!\!\!\!\!\!\!\!\!\!
    \!\!\!\!\!\!\!\!\!\!\!\!\!\!
    \mbox{
      \tiny
      \color{darkblue}
      \bf
      \begin{tabular}{c}
        total flux = homotopy class
      \end{tabular}
    }
    }
  }{
  \big\{
    X
    \longrightarrow
    S^4
  \big\}_{
    \scalebox{.7}{$
      \!\!\big/\mathrm{hmpty}
    $}
  }
  }
  \;\simeq\;\;
  \underset{
    \mathclap{
    \mbox{
      \tiny
      \color{darkblue}
      \bf
      \begin{tabular}{c}
        \\
        charge
        \\
        lattice
      \end{tabular}
    }
    }
  }{
    \mathbb{Z}
  }
  $}
  };

\end{tikzpicture}
}
\end{minipage}
}

\end{tabular}

\vspace{1pt}

\noindent
Regarding here $n$-Cohomotopy of $n$-manifolds
as ordinary $n$-cohomology, via the Hopf degree theorem
(Ex. \ref{M5BraneChargeInOrdinaryCohomology} below),
this situation is a higher-dimensional
analogue of the vortex strings seen in superconductors
(p. \pageref{VortexStrings}).

\newpage

\noindent
{\bf M2-Brane polarization and the quaternionic-Hopf fibration.}
\label{M2BranePolarizationAndTheQuaternionicHopfFibration}
The special case \fbox{$b = 2$ and $p = 2$} of \eqref{SolitonicBraneSpacetime}
with $M^7 \,=\, S^7$
is that of the near-horizon geometry of black M2-branes
(\cite{DuffStelle91}\cite{DuffGibbonsTownsend95}\cite{Page83},
i.e. Freund-Rubin compactifications
\cite{FreundRubin80} of 11-dimensional supergravity on $S^7$)
with M2-brane probes.
Similarly, the case \fbox{ $b=9$ and $p = 2$ }
with $X \,\simeq\, S^7 \vee S^7 $ (by Prop. \ref{CompactificationOfProductSpace})
is that of
M2-branes probing the vicinity of an MO9-plane
in heterotic M-theory.
In both cases the set \eqref{HomotopyGroupsOfThe4Sphere}
of cohomotopical brane charges contains an integer summand,
which is
generated by the {\it quaternionic Hopf fibration}
$ [h_{\mathbb{H}}] \,=\, 1 \,\in\, \mathbb{Z} \,\subset\, \pi^4(S^7)$.
Since this $h_{\mathbb{H}}$ is an $\mathrm{SU}(2) \simeq S^3$-fiber bundle
over $S^4$, the brane worldvolume that corresponds to this
cohomotopical unit charge under the Pontrjagin isomorphism
\eqref{PontrjaginIsomorphismForNonCompactManifolds}
is a 6-manifold wrapped on a 3-sphere:

\vspace{-5mm}
\begin{equation}
  \label{M2BranesPTTheorem}
  \hspace{2.6cm}
  \xymatrix@C=22pt{
    \mathllap{
    \mbox{
      \tiny
      \color{darkblue}
      \bf
      \begin{tabular}{c}
        polarized M2-brane worldvolume /
        \\
        M5-brane wrapped on 3-sphere
      \end{tabular}
    }
    }
    \overset{
      \mathclap{
      \raisebox{6pt}{
        \tiny
        \color{orangeii}
        \bf
        \begin{tabular}{c}
          unit probe
          \\
          brane worldvolume
          \\
          under Hypothesis H
        \end{tabular}
      }
      }
    }{
      \mathbb{R}^{2,1} \times S^3
    }
    \;
    \ar@{^{(}->}[rr]
      _-{
        \;
        \mathrm{fib}(h_{\mathbb{H}})
        \;
      }
      ^-{
        \mathclap{
        \mbox{
          \tiny
          \color{greenii}
          \bf
          \begin{tabular}{c}
            Pontrjagin
            \\
            construction \eqref{PontrjaginIsomorphismForNonCompactManifolds}
            \\
            {\phantom{a}}
          \end{tabular}
        }
        }
      }
    \ar[d]
      ^-{
        \mathclap{\phantom{\vert^{\vert}}}
        \mathrm{id} \times h_{\mathbb{C}}
        \mathclap{\phantom{\vert_{\vert}}}
      }
    &
    {\phantom{AAAA}}
    &
    \mathbb{R}^{2,1}
    \times S^7
    \ar[rr]
      _-{\scalebox{0.7}{$
                 \mathrm{hmtpy}
$}        }^-{\simeq}
    \ar@/^1.5pc/[rrrr]
      |-{
        \mbox{
          \tiny
          \color{orangeii}
          \bf
          \begin{tabular}{c}
            unit M2-brane Page charge
            \\
            under Hypothesis H
          \end{tabular}
        }
      }
    \ar[d]
      |-{
        \mathclap{\phantom{\vert^{\vert}}}
        \mbox{
          \tiny
          \color{darkblue}
          \bf
          \begin{tabular}{c}
            M/IIA circle
            \\
            fibration
          \end{tabular}
        }
        \mathclap{\phantom{\vert_{\vert}}}
      }
    &
    {\phantom{AA}}
    &
    S^7
    \ar[rr]
      |-{\; h_{\mathbb{H}} \;}
      _-{
        \mbox{
          \tiny
          \color{greenii}
          \bf
          \begin{tabular}{c}
            {\phantom{a}}
            \\
            quaternionic Hopf fibration
          \end{tabular}
        }
      }
    \ar[d]
      |-{
        \mathllap{
          \mbox{
            \tiny
            \color{greenii}
            \bf
            \begin{tabular}{c}
              complex
              \\
              Hopf fibration
            \end{tabular}
          }
          \!\!\!\!
        }
        \mathclap{\phantom{\vert^{\vert}}}
        h_{\mathbb{C}}
        \mathclap{\phantom{\vert_{\vert}}}
      }
    &
    {\phantom{AAAAAA}}
    &
    S^4
    \ar@{=}[d]
    \\
    \mathllap{
    \mbox{
      \tiny
      \color{darkblue}
      \bf
      \begin{tabular}{c}
        polarized D2-brane worldvolume /
        \\
        D4-brane wrapped on 2-sphere
      \end{tabular}
    }
    }
    \mathbb{R}^{2,1}
    \times
    S^2
    \;
    \ar@{^{(}->}[rr]
      ^-{
        \;
        \mathrm{fib}(t_{\mathbb{H}})
        \;
      }
    &&
    \mathbb{R}^{2,1}
    \times
    \mathbb{C}P^3
    \ar[rr]
      ^-{
            \scalebox{0.7}{$
                 \mathrm{hmtpy}
$}         }_-{\simeq}
    \ar@/_1.5pc/[rrrr]
      |-{
        \mbox{
          \tiny
          \color{orangeii}
          \bf
          \begin{tabular}{c}
            unit D2-brane charge
            \\
            under Hypothesis H
          \end{tabular}
        }
      }
    &&
    \mathbb{C}P^3
    \ar[rr]
      |-{\;
        t_{\mathbb{H}}
      \;}
      ^-{
        \mbox{
          \tiny
          \color{greenii}
          \bf
          \begin{tabular}{c}
            Atiyah-Penrose fibration
            \\
            {\phantom{a}}
          \end{tabular}
        }
      }
    &&
    S^4
  }
\end{equation}

\noindent
Noticing that unit flux through the $S^7$ around a black
M2 must be M2-brane Page charge
(emphasized in \cite[(1.2)]{PTW15}, we discuss this below in \cref{M2BraneChargeAndTheHopfInvariant})
induced from non-trivial
but cohomologically trivialized 4-flux
(discussed below in \cref{M5ThreeFlux})
this cohomotopical analysis neatly
matches the M($2 \mapsto 5$)-brane polarization phenomenon
expected in the string theory literature
\cite{Bena00}\cite{BenaNudelman00}\cite{BenaWarner04}\cite{SMS08}\cite{GRRV08}\cite{BGKM14}
(review in \cite[\S 6]{BLMP13}),
notably including the expectation \cite{NPR09}\cite[\S 6.4.2]{BLMP13}
that the $S^3$-polarization of the M2-branes is,
under M/IIA-duality, fibered over
the $S^2$-polarization of D2-branes via the Hopf
map $h_{\mathbb{C}}$:  under
\hyperlink{HypothesisHOnHomotopicallyFlatSpacetimes}{Hypothesis H}
and via Pontrjagin's isomorphism \eqref{PontrjaginIsomorphismForNonCompactManifolds},
this is implied by Rem. \ref{CircleReductionOfCohomotopyChargeInR7}  below.
In particular, this means that the $S^3$s seen here in
$\mathbb{S}_3 \simeq (M\mathrm{Fr})_3$
do correspond to the fuzzy funnel/spheres
seen
via \hyperlink{HypothesisH}{Hypothesis H} in \cite{SS19b}.

We discuss further aspects of this situation in
\cref{M2BraneChargeAndTheHopfInvariant}, \cref{VicinityOfBlackM2BranesAndOrientedCohomology}
and \cref{GreenSchwarzMechanismAndComplexEOrientations}
below, see also the conclusion in Remark \ref{UnitM2BraneChargeUnderHypothesisH}.

\medskip

\noindent
{\bf D6/D8-Branes in Type I' and the May-Segal theorem.}
The special case
\fbox{$b = 9$ and $p = 6,8$}
of \eqref{CohomotopyChargeOnBlackBraneAsymptoticBoundary}
is that of probe 6-branes and probe 8-branes
in the vicinity of an MO9-plane,
hence of D6/D8-branes in the vicinity of an O8-plane in strongly coupled Type I'
string theory, as discussed in \cite{SS19b}:
By  \eqref{HomotopyGroupsOfThe4Sphere}
the Cohomotopy charges of these D6/D8-branes
as such vanish identically;
which is in line with the fact that only the $D(\leq 5)$-branes
are meant to have M-theoretic lifts to fundamental M-branes
(to M2/M5-branes). The nature of D6/D8-branes in M-theory
must be more subtle:
Indeed, beyond the mere Cohomotopy set
lies the full {\it Cohomotopy cocycle space}:

\vspace{-3mm}
$$
  \overset{
    \mathclap{
    \raisebox{3pt}{
      \tiny
      \color{darkblue}
      \bf
      \begin{tabular}{c}
        $n$-Cohomotopy
        \\
        cocycle space
      \end{tabular}
    }
    }
  }{
    {\widetilde \boldpi}{}^{\, n}
    \big(
      X
    \big)
  }
  \;:=\;
  \mathrm{Maps}^{\ast/\!}
  \big(
    X,
    S^n
  \big)
  \,,
  {\phantom{AAAA}}
  \overset{
    \mathclap{
    \raisebox{3pt}{
      \tiny
      \color{darkblue}
      \bf
      \begin{tabular}{c}
        $n$-Cohomotopy set
      \end{tabular}
    }
    }
  }{
    {\widetilde \pi}{}^{\, n}
    \big(
      X
    \big)
  }
  \;=\;
  \pi_0
  {\widetilde \boldpi}{}^{\, n}
  \big(
    X
  \big)
  \,,
$$
a higher homotopy type
of which the Cohomotopy classes
are only the connected components;
hence which, under
\hyperlink{HypothesisHOnHomotopicallyFlatSpacetimes}{\it Hypothesis H},
is a
{\it moduli stack of brane configurations}
with their gauge- and higher gauge-of-gauge
transformations,
of which the plain Cohomotopy charge
\eqref{CohomotopyChargeOnBlackBraneAsymptoticBoundary}
only captures the lowest gauge-equivalence classes.

\medskip
The analog of the Pontrjagin theorem \eqref{PontrjaginIsomorphismForNonCompactManifolds}
in this situation is the {\it May-Segal theorem},
which identifies these Cohomotopy cocycle spaces
(not just with single brane configurations but)
with the {\it moduli spaces of configurations} of D6/D8-branes
as per their positions in their transversal space
$\mathbb{R}^{9-p}$

\vspace{-.3cm}
\begin{equation}
  \label{MaySegalTheorem}
  \xymatrix@C=5em{
    \overset{
      \mathclap{
      \raisebox{6pt}{
        \tiny
        \color{darkblue}
        \bf
        \begin{tabular}{c}
          configuration space of
          \\
          points in $\mathbb{R}^{9-p}$
          carrying labels in $\mathbb{R}^{p-5}_{\scalebox{.6}{cpt}}$
        \end{tabular}
      }
      }
    }{
      \mathrm{Conf}
      \big(
        \mathbb{R}^{9-p}
        ;
        \mathbb{R}^{p-5}_{\scalebox{.6}{cpt}}
      \big)
    }
    \quad
    \ar@/^1pc/[rr]
      ^-{
        \;\;
        \mbox{
          \tiny
          \color{greenii}
          \bf
          assign Cohomotopy charge
        }
        \;\;
      }
    \ar@/_1pc/@{<-}[rr]
      _-{
        \;\;
        \mbox{
          \tiny
          \color{greenii}
          \bf
          find worldvolume of given charge
        }
        \;\;
      }
    \ar@{}[rr]
      |-{ \simeq_{{}_{\mathrm{hmtpy}}} }
    &
    {\phantom{AAAAAAAAAAA}}
    &
    \overset{
      \raisebox{3pt}{
        \tiny
        \color{darkblue}
        \bf
        \begin{tabular}{c}
          4-Cohomotopy
          cocycle space
        \end{tabular}
      }
    }{
      {\widetilde \boldpi}{}^4
      \big(
        \mathbb{R}^{9-p}_{\scalebox{.6}{cpt}}
      \big).
    }
     }
\end{equation}
In a negative-dimensional
parallel to the $(p \mapsto 5)$-brane polarization process
via Pontrjagin's theorem in \eqref{CohomotopyChargeOnBlackBraneAsymptoticBoundary},
the May-Segal theorem \eqref{MaySegalTheorem} asserts that
the points/branes in these configurations carry labels
in $\mathbb{R}^{p-5}_{\scalebox{.6}{cpt}}$,
which is just the modulus for an M5-brane
positioned {\it inside} the $p$-brane (possibly at infinity).

\medskip
For detailed discussion of
Hypothesis H in this low-codimension sector of
M-brane physics we refer to \cite{SS19b}.

\newpage

\noindent
{\bf Branes probing bulk spacetime and the suspension homomorphism.}
The unstable framed Cobordism sets
\eqref{PontrjaginEquivalenceBetweenUnstableFramedCobordismAndCohomotopy}
naturally appear in sequences where more and more ambient
space dimensions are made available for the bordisms
to propagate through (see \hyperlink{FigureS}{\it Figure S}):
\begin{equation}
  \label{SequenceOfSuspensioHomomrphismsOnCobordismSets}
  \hspace{-8mm}
  \raisebox{48pt}{
  \xymatrix@R=6pt@C=2.8em{
    \Sigma^{d-n}
    \subset
    M^d
    \;
    \ar@{^{(}->}[r]^-{
      \overset{
        \mathrlap{
        \raisebox{6pt}{
          \tiny
          \color{greenii}
          \bf
          make more ambient spatial dimensions available
          to bordisms
          $\longrightarrow$
        }
        }
      }{
        (\mathrm{id}, 0)
      }
    }
    &
    M^d \times \mathbb{R}^1
    \;
    \ar@{^{(}->}[r]^-{
      (\mathrm{id}, 0)
    }
    &
    M^d \times \mathbb{R}^2
    \; \ar@{^{(}..>}[r]
    &
    M^d \times \mathbb{R}^\infty
    \\
    \mathllap{
      \mbox{
        \tiny
        \color{darkblue}
        \bf
        \begin{tabular}{c}
          unstable
          \\
          framed
          \\
           Cobordism
        \end{tabular}
        \!\!\!\!\!
      }
    }
    \mathrm{Cob}^n_{\mathrm{Fr}}
    (
      M^d
    )
    \ar[r]
      ^-{\sigma}
    \ar[dd]
      |<<<{
        \mbox{
          \tiny
          \color{greenii}
          \bf
          \begin{tabular}{c}
            Cohomotopy charge map
          \end{tabular}
        }
      }
    &
    \mathrm{Cob}^{n+1}_{\mathrm{Fr}}
    \big(
      M^d \times \mathbb{R}^1
    \big)
    \ar[r]
      ^-{\sigma}
    \ar[dd]
      ^-{\simeq}
    &
    \mathrm{Cob}^{n+2}_{\mathrm{Fr}}
    \big(
      M^d \times \mathbb{R}^2
    \big)
    \ar[dd]
      ^-{\simeq}
    \ar@{..>}[r]
    &
    \widetilde {M \mathrm{Fr}}{}^n
    \big(
      M^d_{\scalebox{.6}{cpt}}
    \big)
    \mathrlap{
      \!\!\!\!\!
      \mbox{
        \tiny
        \color{darkblue}
        \bf
        \begin{tabular}{c}
          stable
          \\
          framed
          \\
          Cobordism
        \end{tabular}
      }
    }
    \ar@{=}[dd]
      |-{
        \mathclap{\phantom{\vert^{\vert}}}
        \mbox{
          \tiny
          \color{greenii}
          \bf
          stable Pontrjagin-Thom isom.
        }
        \mathclap{\phantom{\vert_{\vert}}}
      }
    \\
    \\
    \mathllap{
      \mbox{
        \tiny
        \color{darkblue}
        \bf
        \begin{tabular}{c}
          unstable
          \\
          Cohomotopy
        \end{tabular}
      }
      \;
    }
    {\widetilde \pi}^{n}
    \big(
      M^d_{\scalebox{.6}{cpt}}
    \big)
    \ar[r]
      ^-{ (-)\wedge S^1 }
      _-{
        \underset{
          \mathrlap{
          \raisebox{-6pt}{
            \tiny
            \color{greenii}
            \bf
            suspension homomorphism
            $\longrightarrow$
          }
          }
        }{
          {\phantom{\sigma}}
        }
      }
    &
    {\widetilde \pi}^{n+1}
    \big(
      (M^d \times \mathbb{R}^1)_{\scalebox{.6}{cpt}}
    \big)
    \ar[r]
      ^-{ (-)\wedge S^1 }
    &
    {\widetilde \pi}^{n+1}
    \big(
      (M^d \times \mathbb{R}^2)_{\scalebox{.6}{cpt}}
    \big)
    \ar@{..>}[r]
    &
    \widetilde {\mathbb{S}}{}^n
    \big(
      M^d_{\scalebox{.6}{cpt}}
    \big)
    \mathrlap{
      \;
      \mbox{
        \tiny
        \color{darkblue}
        \bf
        {\begin{tabular}{c}
          stable
          \\
          Cohomotopy
        \end{tabular}}
      }
    }
    }
  }
\end{equation}

\vspace{-.1cm}

\noindent
Seen under the Cohomotopy charge map \eqref{PontrjaginIsomorphismForNonCompactManifolds}, this
is equivalently the sequence of
{\it suspension homomorphisms} on reduced Cohomotopy sets,
given by forming the smash product $c \wedge S^1$ of
Cohomotopy cocycle maps $X \xrightarrow{\;c\;} S^n$ with
$\mathbb{R}^1_{\scalebox{.6}{cpt}} = S^1$:

\vspace{-.5cm}
\begin{equation}
  \label{SuspensionHomomorphismOnCohomotopySets}
  \hspace{6mm}
  \mathllap{
    \mbox{
      \tiny
      \color{greenii}
      \bf
      \begin{tabular}{c}
        suspension
        \\
        homomorphism
      \end{tabular}
    }
  }
  :
  \xymatrix@C=1.5em{
    {\widetilde \pi}{}^n
    (
      X^d
    )
    \ar@{=}[r]
    &
    \pi_0
    \mathrm{Maps}^{\ast/}\!
    \big(
      X^d, S^n
    \big)
    \ar[rr]^-{ (-) \wedge S^1 }
    &&
    \pi_0
    \mathrm{Maps}^{\ast/}\!
    \big(
      \Sigma X^d  , S^{n+1}
    \big)
    \ar@{=}[r]
    &
    {\widetilde \pi}{}^{n+1}
    (
      \Sigma X^d
    )\,.
  }
\end{equation}
\vspace{-.4cm}

\medskip

\hspace{-3mm}
\begin{tabular}{ll}
\hypertarget{FigureS}{}
\begin{minipage}[left]{6.5cm}

{\footnotesize
{\bf Figure S} -- {\bf The suspension homomorphism
on Cobordism}
$
  \mathrm{Cob}^n_{\mathrm{Fr}}
  \big(
    M^d
  \big)
  \xrightarrow{\sigma}
  \mathrm{Cob}^{n+1}_{\mathrm{Fr}}
  \big(
    M^d \times \mathbb{R}^1
  \big)
$
takes (normally framed) submanifolds
$\Sigma^{d-n} \,\subset\, M^d$ to their equivalence
classes under (normally framed) bordisms that may
explore an extra bulk dimension.
}
\end{minipage}

&

\begin{minipage}[left]{7cm}

\begin{tikzpicture}

  \draw[fill=blue, fill opacity=.3, draw opacity=0]
    (-2.5-.3,-.4)
      --
    (-2.5+.3,+.4)
      --
    (+4.5+.3,+.4)
      --
    (+4.5-.3,-.4);

  \draw[fill=blue, fill opacity=.1, draw opacity=0]
    (-2.5-.3,-.4+2.2)
    to
    (-2.5-.3,-.4)
    to
    (-2.5+.3,+.4)
    to
    (-2.5+.3,+.4+2.2)
    to
    (-2.5-.3,-.4+2.2);

 \draw
   (-2,-.1)
   node
   {
     \xymatrix@C=-3pt@R=-3pt{
       & {\;\;\;\;}&
       \\
       {\phantom{a}}
       \\
       \ar[rrr]^<<<<{\scalebox{.6}{$M^d$}} &&{\phantom{aa}}&
       \\
       & \ar[ruuu]
     }
   };

 \draw
   (-2.6,+.7)
   node
   {
     \xymatrix{
       \ar@{<-}[d]_-{ \mathbb{R}^{\!1} \!\! }
       \\
       &
     }
   };

  \begin{scope}[shift={(-1,0)}]
    \draw[line width=2]
      (-.5,0) arc (180+10:360+10:.5 and .12);
    \draw[line width=2, densely dotted]
      (-.5,0) arc (180+10:0+10:.5 and .12);
  \end{scope}

  \begin{scope}[shift={(1.5,0)}]
    \draw[line width=2]
      (-.5,0) arc (180+10:360+10:.5 and .12);
    \draw[line width=2, densely dotted]
      (-.5,0) arc (180+10:0+10:.5 and .12);
  \end{scope}

  \begin{scope}[shift={(3,0)}]
    \draw[line width=2]
      (-.5,0) arc (180+10:360+10:.5 and .12);
    \draw[line width=2, densely dotted]
      (-.5,0) arc (180+10:0+10:.5 and .12);
  \end{scope}

  \shade[ball color=darkgray!90!gray, draw opacity=0.0, fill opacity=.5]
    (-.5-1,0)
      arc (180:360:.5 and .12)
      .. controls (-1+.5-.3,1) and (1.5-.5+.3,1) .. (1.5-.5,0)
      arc (180:360:.5 and .12)
      .. controls (1.5+.5-.2,.4) and (3-.5+.2,.4) .. (3-.5,0)
      arc (180:360:.5 and .12)
      .. controls (+3+.5+.5,2.5) and (-1-.5-.5,2.5) .. (-.5-1,0);


\end{tikzpicture}

\end{minipage}

\end{tabular}

\medskip

\noindent
{\bf Holography and Freudenthal suspension theorem.}
Here \hyperlink{FigureS}{Figure S}
reflects the qualitative picture of

\noindent {\bf (a)} Polyakov's holographic principle,
where QCD quarks are boundaries of flux strings
that probe into a hidden bulk dimension;
which,

\noindent  {\bf (b)} in M-theoretic holography
translates to bulk $p=2$-branes with
boundaries constrained to asymptotic
$b = 5,9$-brane boundaries.
Indeed, the {\it Freudenthal suspension theorem} says that
the suspension homomorphisms \eqref{SuspensionHomomorphismOnCohomotopySets}
become {\it iso}morphisms after
a finite number of $k$ steps:
\vspace{-1mm}
$$
  \underset{
    \mathclap{
    \mbox{
      \tiny
      \color{darkblue}
      \bf
      \begin{tabular}{c}
        codimension
      \end{tabular}
    }
    }
  }{
    n + k
  }
  \;\; > \;\;
  \underset{
    \mathclap{
    \mbox{
      \tiny
      \color{darkblue}
      \bf
      \begin{tabular}{c}
        ambient
        \\
        dimension
      \end{tabular}
    }
    }
  }{
    (d + k)
  }/2
  +
  1
  \;\;\;\;\;\;
    \Rightarrow
  \;\;\;\;\;\;
  \xymatrix{
    \mathrm{Cob}^{n+k}_{\mathrm{Fr}}
    \big(
      X^d \times \mathbb{R}^k
    \big)
    \ar@{=}[r]^-{ \sigma }
    &
    \mathrm{Cob}^{n+k+1}_{\mathrm{Fr}}
    \big(
      X^d \times \mathbb{R}^{k+1}
    \big)
  }
$$

\vspace{-2mm}
\noindent from which stage on the sequence \eqref{SequenceOfSuspensioHomomrphismsOnCobordismSets}
of unstable framed Cobordism/Cohomotopy sets {\it stabilize}
to the {\it stable} framed Cobordism cohomology group,
equivalently the {\it} stable Cohomotopy group of
$M^d_{\scalebox{.6}{cpt}}$.
This implies
for the probe $p \geq 2$-brane charges
in black $b$-brane backgrounds,
from  \eqref{CohomotopyChargeOnBlackBraneAsymptoticBoundary} above,
that their charge stabilizes after
revealing $k = 1$ extra bulk dimensions, hence that;
under
\hyperlink{HypothesisHOnHomotopicallyFlatSpacetimes}{\it Hypothesis H}:

\vspace{1mm}
{\it Allowing $(p \geq 2)$-brane interactions to
probe the radial bulk direction $\mathbb{R}^1_{\mathrm{rad}}$
of the ambient black $b$-brane spacetime
abelianizes their Cohomotopy charges from unstable
Cohomotopy to stable Cohomotopy.}

\vspace{-.5cm}
\begin{equation}
  \label{BulkInteractionsAsStabilization}
  \hspace{1mm}
  \xymatrix@C=2.7em{
    \overset{
      \raisebox{3pt}{
        \tiny
        \color{darkblue}
        \bf
        \begin{tabular}{c}
          charges \eqref{CohomotopyChargeOnBlackBraneAsymptoticBoundary}
          of probe $p$-branes at
          \\
          black $b$-brane horizons
        \end{tabular}
      }
    }{
      \big(
        \mathrm{M}p @ \mathrm{M}b
      \big)
      \mathrm{BndrCharges}
    }
    \ar[rr]
      ^-{
        \mbox{
          \tiny
          \color{greenii}
          \bf
          \begin{tabular}{c}
            allow $p$-brane interactions
            \\
            to probe into the bulk
          \end{tabular}
        }
      }
    \ar@{=}[d]
      |-{
        \mathclap{\phantom{\vert^{\vert}}}
        \mbox{
          \tiny
          \color{greenii}
          \bf
          Hypothesis H
        }
        \mathclap{\phantom{\vert_{\vert}}}
      }
    &&
    \overset{
      \raisebox{3pt}{
        \tiny
        \color{darkblue}
        \bf
        \begin{tabular}{c}
          charges of probe $p$-branes in
          \\
          bulk around black $b$-branes
        \end{tabular}
      }
    }{
      \big(
        \mathrm{M}p @ \mathrm{M}b
      \big)
      \mathrm{BulkCharges}
    }
    \ar@{==}[d]
    \\
    \mathllap{
      \mbox{
        \tiny
        \color{darkblue}
        \bf
        \begin{tabular}{c}
          unstable
          \\
          Cohomotopy
        \end{tabular}
      }
    \!\!\!}
    {\widetilde \pi}{}^4
    \Big(
      \big(
        \mathbb{R}^{b -p}
        \times
        S^{9-b}
      \big)_{\scalebox{.6}{cpt}}
    \Big)
    \ar@{=}[d]
      |-{
        \mathclap{\phantom{\vert^{\vert}}}
        \mbox{
          \tiny
          \color{greenii}
          \bf
          Pontrjagin Thm.
        }
        \mathclap{\phantom{\vert_{\vert}}}
      }
    \ar@/^1.8pc/@<+4pt>@{-->}[rr]
      ^-{
        \;\;
        \mbox{
          \tiny
          \color{greenii}
          \bf
          stabilization (abelianization)
        }
        \;\;
      }
    &
    {\widetilde \pi}{}^{4 {\color{purple} + 1 }  }
    \Big(
      \big(
        \mathbb{R}^{b - p}
        \times
        S^{9-b}
        \times
        {\color{purple}
          \mathbb{R}^1_{\mathrm{rad}}
        }
      \big)_{\scalebox{.6}{cpt}}
    \Big)
    \ar@{=}[d]
      |-{
        \mathclap{\phantom{\vert^{\vert}}}
        \mbox{
          \tiny
          \color{greenii}
          \bf
          Pontrjagin Thm.
        }
        \mathclap{\phantom{\vert_{\vert}}}
      }
    \ar@{=}[r]
      ^-{ p \geq 2 }
      _-{
        \mbox{
          \tiny
          \color{greenii}
          \bf
          \begin{tabular}{c}
            Freudenthal
            \\
            Thm.
          \end{tabular}
        }
      }
    &
    \;\;
    {\widetilde {\mathbb{S}}}{}^4
    \Big(
      \big(
        \mathbb{R}^{b -p}
        \times
        S^{9-b}
      \big)_{\scalebox{.6}{cpt}}
    \Big) \!\!\!\!\!
    \mathrlap{
            \mbox{
        \tiny
        \color{darkblue}
        \bf
        \begin{tabular}{c}
          stable
          \\
          Cohomotopy
        \end{tabular}
      }
    }
    \ar@{=}[d]
      |-{
        \mathclap{\phantom{\vert^{\vert}}}
        \mbox{
          \tiny
          \color{greenii}
          \bf
          \begin{tabular}{c}
            Pontrjagin-Thom Isom.
          \end{tabular}
        }
        \mathclap{\phantom{\vert_{\vert}}}
      }
    \\
    \mathllap{
      \mbox{
        \tiny
        \color{darkblue}
        \bf
        \begin{tabular}{c}
          unstable
          \\
          framed
          \\
          Cobordism
        \end{tabular}
      }
      \!\!\!
    }
    \mathrm{Cob}^{4}_{\mathrm{Fr}}
    \big(
      \mathbb{R}^{b-p}
      \times
      S^{9-b}
    \big)
    \ar[r]^-{ \sigma }
      _-{
        \mbox{
          \tiny
          \color{greenii}
          \bf
          \begin{tabular}{c}
            allow cobordisms
            \\
            to probe into the bulk
          \end{tabular}
        }
      }
    &
    \mathrm{Cob}^{4 {\color{purple} +1 } }_{\mathrm{Fr}}
    \big(
      \mathbb{R}^{b - p}
      \times
      S^{9-b}
      \times
      {\color{purple}
        \mathbb{R}_{\mathrm{rad}}
      }
    \big)
    \ar@{==}[r]
    &
    {\widetilde {M \mathrm{Fr}}}{}^4
    \Big(
      \big(
        \mathbb{R}^{b -p}
        \times
        S^{9-b}
      \big)_{\scalebox{.6}{cpt}}
    \Big)
    \mathrlap{
      \!\!\!\!\!
      \mbox{
        \tiny
        \color{darkblue}
        \bf
        \begin{tabular}{c}
          stable
          \\
          framed
          \\
          Cobordism
        \end{tabular}
      }
    }
  }
\end{equation}

\newpage

\noindent Combining all this, we arrive at:
\label{MTheoryCompactifiedOnK3AndTheThirdStableStem}
\vspace{1mm}
\noindent
{\bf M-Theory compactified on K3 and the third stable stem.}
We saw in \eqref{M2BranesPTTheorem} that,
under
\hyperlink{HypothesisHOnHomotopicallyFlatSpacetimes}{\it Hypothesis H},
M2-branes probing their own
asymptotic horizon geometry $\mathbb{R}^{2,1} \times S^7$
appear polarized to M5-brane
worldvolumes $\mathbb{R}^{2,1} \times S^3$ wrapped on 3-spheres,
whose cohomotopical brane charges \eqref{PontrjaginIsomorphismForNonCompactManifolds}
are the non-torsion elements $n \in \mathbb{Z} \subset \pi_7(S^4)$.
But moreover,
only a quotient of these asymptotic brane charges is visible
inside the bulk spacetime, namely only their image
$n \,\mathrm{mod}\, 24$
in the third stable homotopy group of spheres
$\mathbb{Z}_{24} \simeq \mathbb{S}_3$
\eqref{FromUnstableToStable4CohomotopyOf7Sphere}.
Under the Pontrjagin isomorphism \eqref{PontrjaginIsomorphismForNonCompactManifolds}
this means that
in the vicinity of 24 probe branes
the bulk spacetime
$\mathbb{R}^{2,1} \times ( \mathbb{R}^8 \setminus \{0\})$
admits a compactification

\vspace{-6mm}
\begin{equation}
  \label{CompactificationOnY4}
  \hspace{-4mm}
  \underset{
    \mathclap{
    \raisebox{-3pt}{
      \tiny
      \color{darkblue}
      \bf
      \begin{tabular}{c}
        24-punctured
        \\
        K3-surface
      \end{tabular}
    }
    }
  }{
  \left.
  \def\arraystretch{.6}
  \begin{array}{l}
    {
      \color{orangeii}
      Y^4
    }
    =
    \\
    \mathrm{K}3 \!\setminus\! \sqcup_{{}_{24}} D^4
  \end{array}
  \!\!\!\!\!
  \right\}
  }
  \;\Rightarrow\!\!\!\!\!\!\!\!\!\!\!\!\!\!\!\!
  \xymatrix@R=20pt@C=20pt{
    \overset{
      \mathclap{
      \raisebox{3pt}{
        \tiny
        \color{orangeii}
        \bf
        \begin{tabular}{c}
          $Y^4$-compactification of
          \\
          spacetime near probe M-branes
        \end{tabular}
      }
      }
    }{
      \mathbb{R}^{5,1} \times \mathbb{R}^1 \times Y^4
    }
    \;
    \ar@{^{(}->}[rr]
      ^-{
        \mbox{
          \tiny
          \color{greenii}
          \bf
          Pontrjagin construction
        }
      }
    &&
    \overset{
      \mathclap{
      \raisebox{3pt}{
        \tiny
        \color{darkblue}
        \bf
        \begin{tabular}{c}
          bulk spacetime
          \\
          near black M2-branes
        \end{tabular}
      }
      }
    }{
      \mathbb{R}^{2,1} \times \mathbb{R}^8 \setminus \{0\}
    }
    \;
    \ar@{^{(}->}[r]
    &
    \mathbb{R}^{2,1}
      \!\!\times\!
    S^7
      \!\!\!\times\!\!
    (\mathbb{R}^1_{\mathrm{rad}})_{\scalebox{.6}{cpt}}
    \qquad
    \ar@/_1.8pc/[r]
      |-{
                \underset{
           \raisebox{-4pt}{
              \tiny
              \color{orangeii}
              \bf
              trivialized in bulk
            }
         }{
          \Sigma(24 \cdot [h_{\mathbb{H}}])
        }
            }
      ^>>>>>>>{{}^{{}^{\ }}}="t"
    \ar@/^1.8pc/[r]
      |-{ \;0\; }
      _>>>>>{\ }="s"
    &
    \Sigma S^4
    \\
    \mathclap{\phantom{\vert^{\vert^{\vert}}}}
    \underset{\{1,\cdots, 24\}}{\sqcup}
    \underset{
      \mathclap{
      \raisebox{+5pt}{
        \tiny
        \color{darkblue}
        \bf
        \begin{tabular}{c}
          probe M2-branes
          \\
          (polarized to M5s)
        \end{tabular}
      }
      }
    }{
    \big(
      \mathbb{R}^{2,1}
      \times
      S^3
    \big)
    }
    \times
    \underset{
      \mathrlap{
      \;\;\;\;\;\;
      \raisebox{-7pt}{
        \tiny
        \color{darkblue}
        \bf
        \begin{tabular}{l}
          their vicinity
          \\
          (tubular
          neighborhood)
        \end{tabular}
      }
      }
    }{
      \mathbb{R}^4
    }
    \;
    \ar@{^{(}->}[u]
      _-{
        \mathrm{bdry}
      }
    \ar@{^{(}->}[rr]^-{
      \mbox{
        \tiny
        \color{greenii}
        \bf
        Pontrjagin construction
      }
    }
    &{\phantom{AA}}&
    \mathclap{\phantom{\vert^{\vert^{\vert}}}}
    \underset{
      \mathclap{
      \raisebox{-3pt}{
        \tiny
        \color{darkblue}
        \bf
        \begin{tabular}{c}
          asymptotic boundary
          \\
          near black M2-branes
        \end{tabular}
      }
      }
    }{
      \mathbb{R}^{2,1} \times S^7
    }
    \ar@{^{(}->}[u]
    \ar[rr]
      _<<<<<<<<<<<<<<<<<<{24 \cdot [h_{\mathbb{H}}] }
      ^<<<<<<<<<<<<<<<<<<{
        \mbox{
          \tiny
          \color{greenii}
          \bf
          \begin{tabular}{c}
            24 units of
            \\
            cohomotopical brane charge
          \end{tabular}
        }
      }
    &&
    S^4
    \ar@{=>}
      _-{
        \mathclap{\phantom{\vert^{\vert}}}
        \color{orangeii}
        \vdash Y^4
        \mathclap{\phantom{\vert_{\vert}}}
      }
      "s"; "t"
  }
\end{equation}
on a 4-dimensional compact manifold $Y^4$ with 24 punctures
--
namely given by the
tubular neighborhood $\simeq Y^4 \times \mathbb{R}^4$
of a normally framed 4d cobordism between the 24 $S^3$s.
But that $Y^4$ is given, up to higher cobordism,
by the 24-punctured K3-manifold
(e.g. \cite[\S 2.6]{WangXu10}\cite{SP17}):
\vspace{-2mm}
$$
\hspace{-2mm}
  \raisebox{80pt}{
  \xymatrix@C=1pt@R=4em{
    &
    {\phantom{AAA}}
    &
    \scalebox{.8}{$
      \pi^4(S^7)
       =
      \pi_7(S^4)
    $}
    \ar@{=}[d]
    \ar@{}[r]|-{ \scalebox{.6}{$\ni$} }
    &
    \!\!\!\!\!\!\!\!\!
    \scalebox{.8}{$
      (n,0)
    $}
    \\
    &&
    \scalebox{.8}{$
      \mathrm{Cob}^4_{\mathrm{Fr}}
      \big(
        S^7
      \big)
    $}
    \ar@{}[r]|-{\scalebox{.6}{$\ni$}}
    &
    \!\!\!\!\!\!\!\!\!\!
    \scalebox{.8}{$
      n
        \cdot
      [S^3 = \mathrm{SU}(2)]
    $}
    \ar@{|->}[ddll]
    \\
    \pi^{4+1}\big(
      S^7
        \times
        (\mathbb{R}^1_{\mathrm{rad}})_{\scalebox{.6}{cpt}}
    \big)
    =
    \mathbb{S}_3
    \ar@{<-}[uurr]
      ^-{
        \mbox{
          \tiny
          \color{greenii}
          \bf
          \begin{tabular}{c}
            bulk interactions /
            \\
            stabilization
          \end{tabular}
        }
      }
      \ar@{=}[d]
      \ar@{}[r]|-{ \ni }
      &
      n &  \!\!\!\!\!\!\!\!\!\!\!\!\!\!\!\!\!\!\!\! \mathrm{mod} \;  24
      {\phantom{ \mathrm{K3} \setminus \cdot D^4 }\;\;\;\;\;\;\;}
      \\
      \mathrm{Cob}^{4+1}_{\mathrm{Fr}}
      \big(
        S^7 \times  \mathbb{R}^1_{\mathrm{rad}}
      \big)
      =
      {M\mathrm{Fr}}{}_3
      \;
      \ar@{}[r]|-{\ni}
      &
      \;
      n \cdot [S^3]
      &
      \mathrm{mod}\,
      \big(
        \mathrm{K3} \setminus 24\cdot D^4
      \big)
    }
  }
  \hspace{-.2cm}
  \qquad
  \raisebox{-120pt}{
  \scalebox{.8}{
\begin{tikzpicture}

  \begin{scope}[shift={(1.8,0)}]

  \draw[fill=lightgray, fill opacity=.7, draw opacity=0]
    (-4,-4)
      rectangle
    (2.2,4);

  \draw[fill=lightgray, fill opacity=.7, draw opacity=0]
    (-4-.05,-4)
      --
    (-4-.05,+4)
      --
    (-4-.05-1.6,+4-1.6)
      --
    (-4-.05-1.6,-4-1.6);

  \draw
    (-3.2, +3.5)
    node
    {$\mathbb{R}^8 \setminus \{0\}$};

  \end{scope}

  \draw[densely dotted, line width=.33]
    (-3,0) arc (180:0:3 and .7);

  \begin{scope}[rotate=(-90), line width=.33]
  \draw[densely dotted]
    (-3,0) arc (180:0:3 and .7);
  \end{scope}

  \begin{scope}[rotate=(6)]

  \shade[ball color=gray!60, opacity=2]
    (-3:3)
      -- (3:3)
      .. controls (4:1.1) and (11:1.2) ..
      (15-3:3)
      -- (15+3:3)
      .. controls (15+4:1.2) and (15+11:1.2)
      .. (15+15-3:3)
      -- (2*15+3:3)
      .. controls (2*15+4:1.2) and (2*15+11:1.2)
      .. (2*15+15-3:3)
      -- (3*15+3:3)
      .. controls (3*15+4:1.2) and (3*15+11:1.2)
      .. (3*15+15-3:3)
      -- (4*15+3:3)
      .. controls (4*15+4:1.2) and (4*15+11:1.2)
      .. (4*15+15-3:3)
      -- (5*15+3:3) .. controls
         (5*15+4:1.2) and
         (5*15+11:1.2)
      .. (5*15+15-3:3)
      -- (6*15+3:3) .. controls
         (6*15+4:1.2) and
         (6*15+11:1.2)
      .. (6*15+15-3:3)
      -- (7*15+3:3) .. controls
         (7*15+4:1.2) and
         (7*15+11:1.2)
      .. (7*15+15-3:3)
      -- (8*15+3:3) .. controls
         (8*15+4:1.2) and
         (8*15+11:1.2)
      .. (8*15+15-3:3)
      -- (9*15+3:3) .. controls
         (9*15+4:1.2) and
         (9*15+11:1.2)
      .. (9*15+15-3:3)
      -- (10*15+3:3) .. controls
         (10*15+4:1.2) and
         (10*15+11:1.2)
      .. (10*15+15-3:3)
      -- (11*15+3:3) .. controls
         (11*15+4:1.2) and
         (11*15+11:1.2)
      .. (11*15+15-3:3)
      -- (12*15+3:3) .. controls
         (12*15+4:1.2) and
         (12*15+11:1.2)
      .. (12*15+15-3:3)
      -- (13*15+3:3) .. controls
         (13*15+4:1.2) and
         (13*15+11:1.2)
      .. (13*15+15-3:3)
      -- (14*15+3:3) .. controls
         (14*15+4:1.2) and
         (14*15+11:1.2)
      .. (14*15+15-3:3)
      -- (15*15+3:3) .. controls
         (15*15+4:1.2) and
         (15*15+11:1.2)
      .. (15*15+15-3:3)
      -- (16*15+3:3) .. controls
         (16*15+4:1.2) and
         (16*15+11:1.2)
      .. (16*15+15-3:3)
      -- (17*15+3:3) .. controls
         (17*15+4:1.2) and
         (17*15+11:1.2)
      .. (17*15+15-3:3)
      -- (18*15+3:3) .. controls
         (18*15+4:1.2) and
         (18*15+11:1.2)
      .. (18*15+15-3:3)
      -- (19*15+3:3) .. controls
         (19*15+4:1.2) and
         (19*15+11:1.2)
      .. (19*15+15-3:3)
      -- (20*15+3:3) .. controls
         (20*15+4:1.2) and
         (20*15+11:1.2)
      .. (20*15+15-3:3)
      -- (21*15+3:3) .. controls
         (21*15+4:1.2) and
         (21*15+11:1.2)
      .. (21*15+15-3:3)
      -- (22*15+3:3) .. controls
         (22*15+4:1.2) and
         (22*15+11:1.2)
      .. (22*15+15-3:3)
      -- (23*15+3:3) .. controls
         (23*15+4:1.2) and
         (23*15+11:1.2)
      .. (23*15+15-3:3)
      ;

 \foreach \n in {1, ..., 3}
 {
   \draw
     (\n*15:3.25) node {\scalebox{.8}{\color{darkblue}$S^3$}};
 }

 \draw
   (-.3,.3)
   node
   {

    ${\scalebox{.6}{\bf\color{darkblue} K3}}
    \mathrlap{\raisebox{0pt}{\scalebox{.6}{$\setminus ( 24 \cdot D^4 )$}}}$
   };

  \end{scope}

  \begin{scope}
    \clip
      (-134.9-0.2:0.96) rectangle (-134.9+0.2:0.1);
    \draw[densely dotted]
      (-134.9-0.2:4.3) to (-134.9-0.2:.4);
    \draw[densely dotted]
      (-134.9-0.0:4.3) to (-134.9+0.0:.4);
    \draw[densely dotted]
      (-134.9+0.2:4.3) to (-134.9+0.2:.4);
  \end{scope}
  \begin{scope}[shift={(-134.9-0.2:.4)}]
    \draw[densely dotted, line width=1.5pt]
      (-120:.1) arc (-120:360-150:.1);
  \end{scope}

  \begin{scope}[rotate=(180)]
    \shade[ball color=green!40, opacity=.6] (0,0) circle (3.005);
  \end{scope}

  \begin{scope}[rotate=(-90)]
  \draw[line width=.33]
    (-3,0) arc (180:360:3 and .7);
  \end{scope}

  \begin{scope}[rotate=(6)]

  \foreach \n in {1, ..., 24}
  {
    \draw[line width=.33, fill=lightgray]
     (\n*15-3.1:3)
       arc (\n*15-3.1:\n*15+2.9:3)
     to[bend right=70] (\n*15-2.9:3);
  }

   \end{scope}

  \draw[line width=.33]
    (-3,0) arc (180:360:3 and .7);

  \begin{scope}
    \clip
      (-134.9:4.3) rectangle (-134.9:0.97);
    \draw
      (-134.9-0.2:4.3) to (-134.9-0.2:.2);
    \draw
      (-134.9+0.0:4.3) to (-134.9+0.0:.2);
    \draw
      (-134.9+0.2:4.3) to (-134.9+0.2:.2);
  \end{scope}

  \draw
    (-134.9:0.97)+(.23,-.2)
    node
    {\scalebox{.8}{\color{darkblue}$\mathbf{S}^7$}};

  \draw
    (-134.9:4.3)+(.42,0)
    node
    {\scalebox{.8}{\color{darkblue}$\mathbb{R}^1_{\mathrm{rad}}$}};

\end{tikzpicture}
  }
  }
$$

\noindent
In conclusion:

\vspace{2mm}

\noindent
\hspace{.04cm}
\fbox{
\begin{minipage}[left]{16.8cm}
{\it
In the bulk spacetime of M-theory near black M2-branes,
\hyperlink{HypothesisH}{\it Hypothesis H} implies that
the total charge of 24 probe M-branes
cancels out, their flux through the 24 encircling 3-spheres
being absorbed along the compact K3-fiber of
a spontaneous KK-compactification of spacetime near the probe branes. }
\end{minipage}
}

\medskip

Of course, K3-compactifications of M-theory have
received much attention, due to a famous argument that
they should be dual to heterotic string theory
compactified on $\mathbb{T}^2$ \cite[\S 6]{HullTownsend95}\cite[\S 4]{Witten95}.
However, there seems to have been no argument why
spontaneous K3-compactifications of M-theory should be dynamically
preferred in the first place, hence why the theory
would share the theorists' interest in these compactifications.
The above conclusion suggests that
\hyperlink{HypothesisHOnHomotopicallyFlatSpacetimes}{\it Hypothesis H}
provides this
missing aspect of M-theory: The 24-punctured K3 fiber arises
in M-theory near black M2-branes
as the framed cobordism which is,
under Pontrjagin's theorem,  the trivialization of 24 units
of cohomotopical M-brane that is implied in the bulk spacetime
by Freudenthal's theorem.

\newpage

\label{24BraneCancellationInWorldvolumeSection}
\noindent {\bf Tadpole cancellation of 24 $\mathrm{M}5_{\mathrm{HET}}$-branes
via GS mechanism on K3
and the Poincar{\'e}-Hopf theorem.}
While

\vspace{-.1cm}
\hspace{-.9cm}
\begin{tabular}{ll}

\begin{minipage}[left]{6cm}
we arrived at the above picture from consideration of
M2-branes polarized into M5-branes wrapped on $S^3$s,
it is, in the end, only the
measurement of 3-flux through the $S^3$s that signifies brane charges,
while the reconstruction of the brane worldvolumes sourcing these
charges depends on the spacetime perspective:
If we think of the above configuration
from a different perspective, where the
$\mathbb{R}^1$-factor in the top left of
\eqref{CompactificationOnY4} is interpreted as
going along a Ho{\v r}ava-Witten fiber, then
the same $S^3$s appear as encircling
$\mathrm{NS5}_{\mathrm{HET}}$-branes in heterotic
string theory on K3.
\end{minipage}

 &

 \hspace{1cm}

\begin{minipage}[left]{9cm}
\begin{tikzpicture}

\draw
  (0,0)
  node
  {
  \xymatrix@R=14pt@C=14pt{
    &
    &
    \rotatebox[origin=c]{+90}{
        \tiny
        \color{darkblue}
        \bf
        $\mathrm{NS}5_{\mathrm{HET}}$-brane
    }
    &
    \rotatebox[origin=c]{+90}{
        \tiny
        \color{darkblue}
        \bf
        and its horizon
    }
    \\
    \ar@{~>}@<-23pt>[d]_-{
      \mathclap{
        \!\!\!\!\!\!\!\!\!\!\!\!\!
        \rotatebox{+90}{$
            \mathclap{
              \mbox{
                \tiny
                HW compactification on $\mathbb{T}^3$
              }
            }
          $}
        }
      }
    &
    [0,1]
    \times
    \mathbb{T}^3
    \ar@{}[dd]|-{ \times }
    \ar@{<-^{)}}[r]
    &
    \;\mathbb{R}^3\;
    \ar@{}[d]|-{ \times }
    &
    &
    \mathrlap{
      \mbox{
        \tiny
        \color{orangeii}
        \bf
        transverse space to
      }
    }
    \\
    &
    &
    \mathbb{R}^{2,1}
    \ar@{}[r]|-{ \times }
    \ar@{=}[dl]
    &
    \mathclap{\phantom{\vert_{\vert}}}
    S^3
    \ar@{^{(}->}[d]
    &
    \mathrlap{
      \mbox{
        \tiny
        \color{orangeii}
        \bf
        M2 polarized into M5
      }
    }
    \\
    \mathllap{
      \mbox{
        \tiny
        \color{darkblue}
        \bf
        7d HET on K3
      }
    }
    &
    \mathbb{R}^{2,1}
    \ar@{}[rr]|-{ \times }
    &&
    \;\;
    \mathclap{
      \mathrm{K3} \setminus \sqcup D^4
    }
    \;\;
    \\
    &
    \rotatebox[origin=c]{+90}{
        \tiny
        \color{orangeii}
        \bf
        7d HW on $\mathbb{T}^3$
    }
    &
    \ar@{<~}[r]_-{
      \mbox{
        \tiny
        \bf
        KK-compactification on $\mathrm{K3}$
      }
    }
    &&
  }

  };

\draw[gray, fill=gray, fill opacity=.2, draw opacity=0]
  (-1.8,-1.7)
  rectangle
  (.36,1.5);

\draw[gray, fill=gray, fill opacity=.2, draw opacity=0]
  (-1.8,-1.7)
  rectangle
  (3,-.5);

\end{tikzpicture}

\end{minipage}

\end{tabular}

Indeed, heterotic string theory on K3 has been argued to require
(in the absence of gauge flux) precisely 24 NS5-branes
transverse the the K3, in order for their tadpoles
along the compact fiber to cancel:
In the traditional derivation
(\cite[p. 50]{Schwarz97}\cite[p. 30]{Johnson98}\cite[\S 1.1]{ChoiKobayashi19})
this follows from the
differential relation encoding the Green-Schwarz mechanism,
which requires that the $H_3$-flux density of the
$\mathrm{NS}5_{\mathrm{HET}}$-branes trivializes the
fractional Pontrjagin 4-form $\tfrac{1}{2}p_1(\nabla)$
of the compact space, which
for a complex surface like K3 is its Euler form $\rchi_4(\nabla)$:
\begin{equation}
  \label{GreenSchwarzMechanismOnK3}
  \underset{
    \mathclap{
    \raisebox{-3pt}{
      \tiny
      \color{darkblue}
      \bf
      \begin{tabular}{c}
        Green-Schwarz mechanism
        \\
        (for vanishing gauge flux $F_2$)
      \end{tabular}
    }
    }
  }{
  \underbrace{
    d H_3
      \,=\,
    \tfrac{1}{2}p_1(\nabla)
  }
  }
  \;
  \underset{
    \mathclap{
    \raisebox{-3pt}{
      \tiny
      \color{darkblue}
      \bf
      \begin{tabular}{c}
        on complex
        \\
        surface
      \end{tabular}
    }
    }
  }{
  \underbrace{
    \,=\,
    \rchi_4(\nabla)
  }
  }
  \;\in\;
  \Omega^4_{\mathrm{dR}}
  \big(
    \mathrm{K3} \setminus n \cdot D^4
  \big)
  \;\;\;\;\;
  \xRightarrow{
    \mbox{
      \tiny
      \color{greenii}
      \bf
      \begin{tabular}{c}
        Poincar{\'e}
        \\
        lemma
      \end{tabular}
    }
  }
  \;\;\;\;\;
  n \cdot
  \underset{
    \mathclap{
    \mbox{
      \tiny
      \color{darkblue}
      \bf
      \begin{tabular}{c}
        charge of
        \\
        enclosed $\mathrm{NS}5_{\mathrm{HET}}$
      \end{tabular}
    }
    }
  }{
  \underbrace{
    \int_{S^3} H^3
  }
  }
    \;=\;
  \underset{
    \mathclap{
    \mbox{
      \tiny
      \color{darkblue}
      \bf
      \begin{tabular}{c}
        Euler number of K3
      \end{tabular}
    }
    }
  }{
  \underbrace{
    \int_{\mathrm{K3}} \rchi_4(\nabla)
      \;=\;
    24
  }
  }
  \,.
\end{equation}
(We highlight in passing that making rigorous even
this traditional derivation in itself already requires appeal to Cohomotopy:
Namely one needs to prove that an $H_3$-form on
$\mathrm{K3}\setminus 24\cdot D^4$
may actually be constructed
with the required properties
(the differential relation $d H_3 = \rchi_4(\nabla)$
subject to the normalization $\int_{S^3} H^3 = 1$):
This follows by appeal to the Poincar{\'e}-Hopf theorem
to obtain a cocycle in the {\it J-twisted} 3-Cohomotopy
(as in \cite[\S 2.5]{FSS19b})
on $\mathrm{K3}\setminus 24 \cdot D^4$, whose de Rham image
under the cohomotopical character map \cite{FSS20c}
is then guaranteed
to satisfy \eqref{GreenSchwarzMechanismOnK3}
(by \cite[Prop. 2.5]{FSS19b}).

\medskip
The constraint that branes transverse to a K3
must appear in multiples of 24,
is also known for D7-branes in F-theory compactified on
elliptically fibered K3 \cite[p. 5]{Sen96}\cite[p. 5]{Vafa96}\cite[p. 6]{Lerche99}\cite{DouglasParkSchnell14}
in order for their total charges in the compact dimensions to cancel out
\cite[p. 34]{Denef08}.
Under T-duality this is expected to imply the analogous statement for
D6-branes in IIA-theory \cite[Footn. 2]{Vafa96}.
While the mechanism for D7-branes is superficially rather different
from the above argument \eqref{GreenSchwarzMechanismOnK3}
for NS5-branes, it has recently been argued that both
correspond to each other under suitable stringy dualities
(\cite[\S III]{BBLR18}, following analysis in \cite{AspinwallMorrison97}).

\medskip

\noindent {\bf $H_3$-Flux and Toda brackets in Cohomotopy.}
Hence it remains to see how a 3-flux $H_3$ satisfying
a Green-Schwarz condition \eqref{GreenSchwarzMechanismOnK3}
arises from first principles under
\hyperlink{HypothesisHOnHomotopicallyFlatSpacetimes}{\it Hypothesis H}.
This discussion occupies most of the following sections:

\noindent
In \cref{M5ThreeFlux} we see the 3-flux appear,
charge-quantized via the Adams e-invariant on the stable
homotopy groups of spheres.

\noindent
In \cref{TadpoleCancellationAndSUBordismWithBoundaries}
we see its geometric incarnation in the form of a Green-Schwarz
mechanism on cobordisms, namely as Conner-Floyd's formulation
of the e-invariant.

\noindent
In \cref{GreenSchwarzMechanismAndComplexEOrientations}
we see the more general Green-Schwarz mechanism with gauge flux
in the form of heterotic line bundles.

\noindent
In fact, in its usual incarnation in
ordinary cohomology we had derived
the $H_3$-flux and its Green-Schwarz mechanism
from Hypothesis H already previously in
\cite{FSS20b}\cite{SS20c} (see also \cite{FSS20a}).
Here we see how the $H_3$-flux is refined to
generalized cohomology, in line with the refinement
of $G_4$-flux to Cohomotopy theory.

\newpage

\subsection{M5-brane charge and the Adams d-Invariant}
\label{M5BraneChargeAndMltiplicativeCohomologyTheory}

\noindent {\bf Unit M5-brane charge and Unital cohomology.}
Given a Whitehead-generalized cohomology theory $E$ \eqref{WhiteheadGeneralizedCoohomology},
we ask that there is a notion of {\it unit M5-brane charge}
as measured in $E$-cohomology.
Since, under
\hyperlink{HypothesisHOnHomotopicallyFlatSpacetimes}{\it Hypothesis H},
the actual M5-brane charge
of a homotopically-flat spacetime $X$ (Remark \ref{FramedSpacetimes})
is measured by classes in Cohomotopy \eqref{CohomotopyInIntroduction},
hence by homotopy classes of maps to the 4-sphere,
this means that
there must be a certain unit element in the $E$-cohomology of the 4-sphere:
$[G^E_{4,\mathrm{unit}}] \,\in\, {\widetilde E}{}^4(S^4)$.

\medskip
A sufficient condition for this to exist is
that the cohomology theory is {\it multiplicative}
(Def. \ref{MultiplicativeCohomologyTheory}),
so that its coefficient ring $E_\bullet$
\eqref{WhiteheadGeneralizedCoohomology}
has the structure of a
$\mathbb{Z}$-graded-commutative ring,
and in particular has a multiplicative unit element $1^E \in E_0$.
The image under the 4-fold suspension
isomorphism \eqref{SuspensionIsomorphism}
of this unit $1^E$ serves the purpose:
\begin{equation}
  \label{4SuspendedUnitInECohomology}
    \mathllap{
      \mbox{
        \tiny
        \color{darkblue}
        \bf
        \begin{tabular}{c}
          multiplicative unit in
          \\
          $E$-cohomology ring...
        \end{tabular}
      }
    }
    1^E
    \,\in\,
    \pi_0(E)
    \;=\;
    {\widetilde E}{}^0(S^0)
    \xrightarrow
      [ \;\;\Sigma^4\;\; ]
      {\simeq}
    {\widetilde E}{}^4(S^4)
    \mathrlap{
      \mbox{
        \tiny
        \color{darkblue}
        \bf
        \begin{tabular}{c}
          ... suspended to canonical unit
          \\
          in $E$-cohomology of 4-sphere.
        \end{tabular}
      }
    }
\end{equation}
In order to bring out its interpretation as
unit M5-brane charge seen in $E$-cohomology, we denote this
$E$-unit in degree 4 as follows:
\begin{equation}
  \label{UnitG4Flux}
  \mathllap{
    \mbox{
      \tiny
      \color{darkblue}
      \bf
      \begin{tabular}{c}
        unit M5-brane charge
        \\
        seen in $E$-cohomology
      \end{tabular}
    }
    \!\!\!
  }
  \big[
  G^E_{4,\mathrm{unit}}
  \;:=\;
  \Sigma^4 (1^E)
  \;:\;
  \xymatrix{
    S^4
    \ar[r]
    &
    E\degree{4}
  }
  \big]
  \;\;
  \in
  \;
  {\widetilde E}{}^4(X)
  \,.
\end{equation}
In fact, for any multiplicative cohomology theory $E$,
the suspended unit elements $\Sigma^n(1^E)$ \eqref{4SuspendedUnitInECohomology}
are jointly the components of a {\it unique}
homomorphism of ring spectra (Def. \ref{MultiplicativeCohomologyTheory})
out of the
sphere spectrum $\mathbb{S}$ (Example \ref{ExamplesOfMultiplicativeCohomologyTheories}):
\begin{equation}
  \label{UnitMorphismOfSpectra}
  \xymatrix@C=6em@R=4pt{
    &
    \mathllap{
      \mbox{
        \tiny
        \color{darkblue}
        \bf
        sphere spectrum
      }
      \;\;
    }
    \mathbb{S}
    \ar[rr]
      |-{ \; e^E \; }
      ^-{
        \mbox{
          \tiny
          \begin{tabular}{c}
            \color{greenii}
            \bf
            unit of $E$-ring spectrum
            \\
            (unique multiplicative map from $\mathbb{S}$)
            \\
            {\phantom{a}}
          \end{tabular}
        }
      }
    &
    {\phantom{AAAA}}
    &
    E
    \mathrlap{
      \;
      \mbox{
        \tiny
        \color{darkblue}
        \bf
        any ring spectrum
      }
    }
    \\
    \Sigma^{\infty} S^n
    \ar@{}[r]|-{\simeq}
    &
    \Sigma^n \mathbb{S}
    \ar[rr]|-{ \;\Sigma^n (e^E)\; }
    &&
    \Sigma^n E
    \\
    &
    S^n
    \ar[rr]
      |-{ \;\Sigma^n (1^E)\; }
      _-{
        \mbox{
          \tiny
          \begin{tabular}{c}
            \\
            {
              \color{greenii}
              \bf
              suspensions of
              $1^E := [e^E] \in \pi_0(E)\;$
            }
            (canonical units in $\widetilde E(S^n)$)
          \end{tabular}
        }
      }
    &&
    \Omega^{\infty - n } E
    \ar@{}[r]|-{=:}
    &
    E\degree{n}
  }
\end{equation}
The corresponding cohomology operations are
the {\it Hurewicz-Boardman homomorphisms} in
$E$-cohomology  \cite[\S II.6]{Adams74}\cite{Hunton95}\cite{Arlettaz04}:
\begin{equation}
  \label{UnitCohomologyOperations}
  \xymatrix@R=-2pt{
    \mathllap{
      \mbox{
        \tiny
        \color{darkblue}
        \bf
        \begin{tabular}{c}
          stable
          \\
          Cohomotopy
        \end{tabular}
      }
    }
    \widetilde {\mathbb{S}}{}^n(X)
    \ar[rr]
      ^-{
        \mathclap{
        \mbox{
          \tiny
          \color{greenii}
          \bf
          \begin{tabular}{c}
            Hurewicz-Boardman homomorphism
            \\
            \phantom{A}
          \end{tabular}
        }
        }
      }
      |-{
        \;
        \beta^n_X
        \;
      }
    &&
    \widetilde E{}^n(X)
    \mathrlap{
      \mbox{
        \tiny
        \color{darkblue}
        \bf
        \begin{tabular}{c}
          $E$-cohomology
        \end{tabular}
      }
    }
    \\
    \big[
      X
        \xrightarrow{ c }
      \mathbb{S}\degree{n}
    \big]
    &\longmapsto&
    \big[
      X
        \xrightarrow{ c }
      \mathbb{S}\degree{n}
        \xrightarrow{ \scalebox{.6}{$\Sigma^n(1^E)$} }
      E\degree{n}
    \big]
    \,.
  }
\end{equation}

\medskip

\noindent {\bf General M5-brane charge and the d-Invariant.}
General M5-brane charge on
a homotopically flat spacetime $X$ (Remark \ref{FramedSpacetimes}) is,
by \hyperlink{HypothesisHOnHomotopicallyFlatSpacetimes}{\it Hypothesis H},
given by classes in the 4-Cohomotopy of $X$,
hence homotopy classes of maps
$X \xrightarrow{\;c\;}  S^4$.
For $E$ a multiplicative cohomology theory, the corresponding
charge as measured in the $E$-cohomology of $X$ is the base change
along this map of the
$E$-unit M5-brane charge \eqref{UnitG4Flux}:
\begin{equation}
  \label{M5BraneFluxInEtheoryByPullback}
  \underset{
    \mathclap{
    \raisebox{-5pt}{
      \tiny
      \color{darkblue}
      \bf
      \begin{tabular}{c}
        general M5-brane charge
        \\
        seen in $E$-cohomology
      \end{tabular}
    }
    }
  }{
    \big[
      G_4^E(c)
    \big]
  }
  \;:=\;
  \underset{
    \mathclap{
    \raisebox{-5pt}{
      \tiny
      \color{darkblue}
      \bf
      \begin{tabular}{c}
        pullback of unit M5-brane charge in $E$-cohomology
        \\
        along classifying map of its charge in Cohomotopy
      \end{tabular}
    }
    }
  }{
  \big[
    c^\ast
    \big(
      G^E_{4,\mathrm{unit}}
    \big)
  \big]
  \;=\;
  \big[
    X
      \xrightarrow{\;\;c\;\;}
    S^4
      \xrightarrow{\; \scalebox{0.6}{$\Sigma^4(1^E)$}\;}
    E\degree{4}
  \big]
  }
  \;\;\in\;\;
  \widetilde E^4(X)
  \,.
\end{equation}
This construction,
mapping Cohomotopy classes to $E$-cohomology classes,
is the {\it Adams d-invariant} $d_E(c)$  (Def. \ref{dInvariant})
of the Cohomotopy class $c$ seen in $E$-cohomology:
\vspace{-3mm}
\begin{equation}
  \label{dInvariantAsMapOnCohomotopy}
  \xymatrix@R=-18pt{
    \overset{
      \mathclap{
      \raisebox{6pt}{
        \tiny
        \color{darkblue}
        \bf
        Cohomotopy
      }
      }
    }{
      \widetilde \pi^4(X)
    }
    \ar[rr]^-{
      \overset{
        \mathclap{
        \raisebox{3pt}{
          \tiny
          \color{greenii}
          \bf
          d-invariant
        }
        }
      }{
        d^4_E
      }
    }
    &&
    \overset{
      \mathclap{
      \raisebox{6pt}{
        \tiny
        \color{darkblue}
        \bf
        $E$-Cohomology
      }
      }
    }{
      \widetilde E^4(X)
    }
    \\
    \underset{
      \raisebox{-3pt}{
        \tiny
        \begin{tabular}{c}
          \bf
          \color{darkblue}
          full M-brane charge
          \\
          (under Hypothesis H)
        \end{tabular}
      }
    }{
      \big[
        X
          \xrightarrow{\;\;c\;\;}
        S^4
     \big]
    }
    \ar@{}[rr]|-{ \longmapsto }
    &&
    \big[
      G_4^{E}(c)
    \big]
    \;:=\;
    \underset{
      \mathclap{
      \raisebox{-1pt}{
        \tiny
        \color{orangeii}
        \bf
        M-brane charge in $E$-cohomology
      }
      }
    }{
    \big[
      X
        \xrightarrow{\;\;c\;\;}
      S^4
        \xrightarrow{\; \scalebox{0.6}{$\Sigma^4(1^E)$}\;}
      E\degree{4}
    \big]
    \,.
    }
  }
\end{equation}

\newpage

\noindent {\bf Hypothesis H and Initiality of the sphere spectrum.}
Any homomorphism of multiplicative cohomology theories
$E \xrightarrow{\;\phi\;} F$ (Def. \ref{MultiplicativeCohomologyTheory})
(a multiplicative cohomology operation) induces a
natural comparison map between the corresponding M-brane charge quantization laws \eqref{dInvariantAsMapOnCohomotopy}.
In this system of multiplicative charge quantization laws, one is universal:
The sphere spectrum $\mathbb{S}$
is initial among homotopy-commutative ring spectra,
via the unit morphisms \eqref{UnitMorphismOfSpectra},
and so {\it stable Cohomotopy is initial} among multiplicative
cohomology theories --
we may say: among multiplicative charge quantization laws.
It seems suggestive that among
all imaginable charge quantization laws in physics,
M-theory would correspond to this exceptional one.
This is what
\hyperlink{HypothesisHOnHomotopicallyFlatSpacetimes}{\it Hypothesis H} asserts.

\medskip
Concretely, initiality of $\mathbb{S}$ means that
any choice $E$ of
M5-brane charge quantization \eqref{M5BraneFluxInEtheoryByPullback}
factors through
stable Cohomotopy $\mathbb{S}$
via the cohomotopical d-invariant
$d_{\mathbb{S}}$ \eqref{dInvariantAsMapOnCohomotopy}
followed by the $E$-Boardman homomorphism
$\beta_E$ \eqref{UnitCohomologyOperations}:
\begin{equation}
  \label{dInvariantAndBoardman}
  \xymatrix@R=-1pt{
    \widetilde \pi^4(X)
    \ar[rr]|-{
      \color{greenii}
      \;d^4_{\mathbb{S}}\;
    }
    \ar@/^1.8pc/[rrrr]|-{
      \color{greenii}
      \;d^4_E\;
    }
    &&
    \widetilde {\mathbb{S}}{}^4(X)
    \ar[rr]|{
      \;
      \color{purple}
      \beta^4_E
      \;
    }
    &&
    \widetilde E^4(X)
    \\
    [c]
    &\longmapsto&
    \big[
      G^{\mathbb{S}}_4\!(c)
    \big]
    &\longmapsto&
    \big[
      G^E_4\!(c)
    \big]
    \\
    \big[
      X \xrightarrow{c} S^4
    \big]
    &\longmapsto&
    \big[
      X
        \xrightarrow{c}
      S^4
        \xrightarrow{\scalebox{0.6}{$ \Sigma^4(1^{\mathbb{S}})$} }
      \mathbb{S}\degree{4}
    \big]
   &\longmapsto&
    \big[
      X
        \xrightarrow{c}
      S^4
        \xrightarrow{\scalebox{0.6}{$ \Sigma^4(1^E)$} }
      E\degree{4}
    \big]
    \,.
   }
\end{equation}

\noindent
Since the $G_4$-flux as seen in any $E$
is, thereby, a function of the $G_4$-flux seen in
stable Cohomotopy $E = \mathbb{S}$, we shall also write
\begin{equation}
  \label{GFluxes}
  G^{\mathbb{S}}_4\mathrm{Fluxes}
  \big(
    X
  \big)
  \;\coloneqq\;
  \widetilde {\mathbb{S}}{}^4(X)
\end{equation}
for the stable 4-Cohomotopy group of a spacetime $X$,
hence for the stable image of the unstable 4-Cohomotopy of $X$
(Remark \ref{DInvariantInStableCohomotopyIsStableClass}).
This notation serves to bring out the beginning of a
pattern that continues with the group
$H^E_{n-1}\mathrm{Fluxes}(X)$ in \cref{M5ThreeFlux}
below.

\medskip

While measurement of the 4-flux $G_4$ itself
requires only the unital structure of a multiplicative
cohomology theory, by \eqref{M5BraneFluxInEtheoryByPullback},
the product structure enters in describing the
{\it dual} flux $G_7$
(whose rational image appeared around \eqref{RationalCohomotopy}):

\medskip

\noindent {\bf Dual M5-brane flux and Multiplicative cohomology.}
Using the product structure (cup product)
in a given multiplicative cohomology theory $E$,
we may form, in particular,
the square of the M5-brane charge \eqref{M5BraneFluxInEtheoryByPullback}.
But the square of the {\it unit} M5-brane flux $G^E_{4,\mathrm{unit}}$
\eqref{UnitG4Flux} trivializes canonically
(by Prop. \ref{CanonicalTrivializationOfCupSquareOvernSphere} below)
via a homotopy/gauge transformation \eqref{NonabelianCohomologyInIntroduction},
which we denote\footnote{
  We include a ``$2$'' in the notation because, after
  rationalization (see Example \ref{DualFluxInOrdinaryCohomology}),
  this is the conversion factor to the conventional normalization
  of the 7-flux form $G_7$ \cite[(3)]{FSS19c}.
}
$- 2 G^E_{7,\mathrm{unit}}$, as shown on the left here:
\vspace{-2mm}
\begin{equation}
  \label{TrivializationOfCupSquareOfUnitM5BraneCharge}
  \raisebox{26pt}{
  \xymatrix@R=36pt@C=54pt{
    S^4
    \ar[rr]
      ^-{
          {\color{greenii}
            G^E_{4,\mathrm{unit}}
          }
          \;=\;
          \Sigma^4 (1^E)
        }
      _>>>>{\ }="s"
    \ar[d]
    &&
    E\degree{4}
    \ar[d]^-{ (-)^2 }
    \\
    \ast
    \ar[rr]_-{0}
      ^-<<<<{\ }="t"
    &&
    E\degree{8}
    \ar@{=>}
      _-{
        \mathclap{
        \rotatebox{25}{
          \scalebox{.6}{$
            \mathclap{
              \!\!\!\!\!\!\!\!\!\!\!\!\!\!\!\!\!\!\!\!\!\!\!
              d\,
              \big(
                {
                  \color{orangeii}
                  2 G^E_{7,\mathrm{unit}}
                }
              \big)
              =
              -
              {\color{greenii}
                G^E_{4,\mathrm{unit}}
              }
              \scalebox{.9}{$\cup\,$}
              {\color{greenii}
                G^E_{4,\mathrm{unit}}
              }
            }
          $}
        }
        }
      }
      "s"; "t"
  }
  }
  \qquad
  \Rightarrow
  \qquad
  \big(
    X \xrightarrow{\;c\;}S^4
  \big)
  \;\;
  \mapsto
  \;\;
  \raisebox{26pt}{
  \xymatrix@R=36pt@C=54pt{
    X
    \ar[rr]
      ^-{
          {
            \color{greenii}
            G^E_{4}(c)
          }
          \;=\;
          c^\ast G^E_{4,\mathrm{unit}}
        }
      _>>>>{\ }="s"
    \ar[d]
    &&
    E\degree{4}
    \ar[d]^-{ (-)^2 }
    \\
    \ast
    \ar[rr]_-{0}
      ^-<<<<{\ }="t"
    &&
    E\degree{8}
    \ar@{=>}
      _-{
        \mathclap{
        \rotatebox{25}{
          \scalebox{.6}{$
            \mathclap{
              \!\!\!\!\!\!\!\!\!\!\!\!\!\!\!\!\!\!
              \!\!\!\!\!\!\!\!\!\!\!\!\!\!\!\!\!\!
              d\,
              \big(
                {
                  \color{orangeii}
                  2 G^E_{7}(c)
                }
              \big)
              =
              -
              {\color{greenii}
                G^E_{4}(c)
              }
              \scalebox{.9}{$\cup\,$}
              {\color{greenii}
                G^E_{4}(c)
              }
            }
          $}
        }
        }
      }
      "s"; "t"
  }
  }
\end{equation}
By naturality of the cup square, this implies
that for given Cohomotopy charge $X \xrightarrow{\;c\;} S^4$,
the homotopy
\begin{equation}
  \label{G7FluxInECohomology}
  \mathllap{
    \mbox{
      \tiny
      \color{orangeii}
      \bf
      \begin{tabular}{c}
        dual M5-brane flux
        \\
        seen in $E$-cohomology
      \end{tabular}
    }
  }
  \!\!\!
  G^E_{7}(c)
  \;:=\;
  c^\ast
  \big(
    G^E_{7,\mathrm{unit}}
  \big)
\end{equation}
trivializes the
cup product square of the M5-brane charge measured in $E$-cohomology
\eqref{M5BraneFluxInEtheoryByPullback},
as shown on the right
of \eqref{TrivializationOfCupSquareOfUnitM5BraneCharge},
and thus is the dual\footnote{
  Once the geometric aspect of (super-)gravity is taken
  into account, its equations of motion force
  a differential form representative of the rational
  image of $G_7$ (see Example \ref{DualFluxInOrdinaryCohomology})
  to be {\it Hodge dual} to that of $G_4$, whence the terminology.
  Our consideration of the cohomological charge quantization
  of the pair $(G_4, G_7)$ without -- or rather before -- considering
  its geometric self-duality constraint is directly analogous
  to how the K-theoretic charge quantization of the set
  of RR-flux forms is considered before
  imposing the respective self-duality constraint.
}
flux analog of
\eqref{M5BraneFluxInEtheoryByPullback}.

\medskip

\begin{examples}[Examples of multiplicative cohomology theories]
  \label{ExamplesOfMultiplicativeCohomologyTheories}
Some examples of multiplicative
(Def. \ref{MultiplicativeCohomologyTheory}, except for the last items)
Whitehead-generalized cohomology theories \eqref{WhiteheadGeneralizedCoohomology},
with the notation we will use for them:

\begin{center}
\def\arraystretch{1.5}
\begin{tabular}{|ll|l|l|}
  \hline
  $E$
  &
  &
  \begin{minipage}[left]{7cm}
    {\bf Multiplicative generalized cohomology}
  \end{minipage}
  &
  \multirow{1}{*}{
  \begin{minipage}[left]{5cm}
    \cite{ConnerFloyd66}\cite{Adams74}\cite{Kochman96}\cite{TamakiKono06}
  \end{minipage}
  }
  \\
  \hline
  \hline
  $H R$
    & &
  \begin{minipage}[left]{7cm}
    Ordinary cohomology with $R$ coefficients
  \end{minipage}
  &
  \multirow{2}{*}{
  \begin{minipage}[left]{5cm}
    $\mathclap{\phantom{\vert^{\vert}}}$\cite[p. 243]{Eilenberg40}
    \\
    \cite[p. 520-521]{EML54b}
    \\
    \cite[\S 19]{Steenrod72}\cite[\S 22]{May99}
  \end{minipage}
  }
  \\
  \cline{1-2}
  $H^{\mathrm{ev}}\!R$
  &
    $
    \mathllap{
      :=
      \;
    }
    \underset{k}{\oplus}
    \Sigma^{2k} HR
    $
  &
  \begin{minipage}[left]{7cm}
    Even-periodic ordinary cohomology
  \end{minipage}
  &
  \\
  \hline
  ${K\mathrm{U}}$
    & &
  \begin{minipage}[left]{7cm}
    Complex topological K-theory
  \end{minipage}
  &
  \multirow{2}{*}{
    \begin{minipage}[left]{5cm}
      \cite{AtiyahHirzebruch61}\cite{Atiyah64}
      \\
      \cite{Karoubi78}
    \end{minipage}
  }
  \\
  \cline{1-2}
  ${K\mathrm{O}}$
    & &
  \begin{minipage}[left]{7cm}
    Orthogonal topological K-theory
  \end{minipage}
  &
  \\
  \hline
  $\mathbb{S}$
  &
  $\mathllap{\simeq \;\; } M \mathrm{Fr}$
  &
  \begin{minipage}[left]{7cm}
    Stable Cohomotopy /
    $\mathclap{\phantom{\vert^{\vert^{\vert^{\vert}}}}}$
    \\
    framed Cobordism
    $\mathclap{\phantom{\vert_{\vert_{\vert_{\vert}}}}}$
  \end{minipage}
  &
  \multirow{1}{*}{
  \begin{minipage}[left]{5cm}
    \cite[p. 204]{Adams74}\cite[[p. 1]{Rognes04}/
    \\
    \cite[p. 27]{ConnerFloyd66}\cite[\S IV, p. 40]{Stong68}
  \end{minipage}
  }
  \\
  \hline
  $M \! f$
  &
  &
  \begin{minipage}[left]{7cm}
    Cobordism cohomology
  \end{minipage}
  &
  \hspace{-2pt}
  \multirow{2}{*}{
    \begin{minipage}[left]{5cm}
      \cite{Atiyah61}\cite{Quillen71}
      \\
      \cite{Stong68}
    \end{minipage}
  }
  \\
  \cline{1-3}
  &
  $M \mathrm{O}$
  &
  \begin{minipage}[left]{7cm}
    Cobordism
  \end{minipage}
  &
  \\
  \cline{1-3}
  &
  $M \mathrm{U}$
  &
  \begin{minipage}[left]{7cm}
    Complex Cobordism
  \end{minipage}
  &
  \hspace{-2pt}
  \multirow{1}{*}{
    \begin{minipage}[left]{5cm}
      \cite[\S 12]{ConnerFloyd66}
      \\
      \cite[\S 4.4]{Kochman96}\cite[\S 6]{TamakiKono06}
    \end{minipage}
  }
  \\
  \cline{1-3}
  &
  $M \mathrm{SU}$
  &
  \begin{minipage}[left]{7cm}
    Special complex Cobordism
  \end{minipage}
  &
  \multirow{1}{*}{
  \begin{minipage}[left]{5cm}
    \cite{LLP17}\cite{CLP19}
  \end{minipage}
  }
  \\
  \hline
  \hline
  $\!\!M\!\!f\!/\mathbb{S}$
  &
  $
    \mathllap{
      =
    }
    \,M \!(f\!\!,\mathrm{Fr})
  $
  &
  \begin{minipage}[left]{7cm}
    Cobordism with framed boundaries
  \end{minipage}
  &
  \\
  \hline
  &
  $M \!(\mathrm{U},\mathrm{Fr})$
  &
  \begin{minipage}[left]{7cm}
    Complex Cobordism
    $\mathclap{\phantom{\vert^{\vert^{\vert}}}}$
    \\
    with framed boundaries
    $\mathclap{\phantom{\vert_{\vert_{\vert_{\vert}}}}}$
  \end{minipage}
  &
  \hspace{-2pt}
  \multirow{2}{*}{
    \begin{minipage}[left]{5cm}
      \cite[\S 16]{ConnerFloyd66}\cite[\S 6]{ConnerSmith69}\cite{Smith71}
      \\
      \cite{Laures00}
    \end{minipage}
  }
  \\
  \cline{1-3}
  &
  $M \!(S\mathrm{U},\mathrm{Fr})$
  &
  \begin{minipage}[left]{7cm}
    Special complex Cobordism
    $\mathclap{\phantom{\vert^{\vert^{\vert}}}}$
    \\
    with framed boundaries
    $\mathclap{\phantom{\vert_{\vert_{\vert_{\vert}}}}}$
   \end{minipage}
  &
  \\
  \hline
\end{tabular}
\end{center}

\end{examples}

\begin{example}[\bf Ordinary M5-brane charge and Hopf winding degree]
  \label{M5BraneChargeInOrdinaryCohomology}
  If $E = H R$ is ordinary cohomology,
  then, under the de Rham homomorphism,
  the unit $G_4$-flux \eqref{UnitG4Flux}
  is given by the volume form on the 4-sphere:
  $$
    G^{HR}_{4,\mathrm{unit}}
    \,\simeq\,
    \big[
      \mathrm{vol}_{S^4}
    \big]
    \,\in\,
    H^4( S^4; \mathbb{Z})
      \longrightarrow
    H^4( S^4; R)
    \,.
  $$
  Thus, if spacetime $X \,\simeq\, X^4$
  is homotopy equivalent to a connected closed manifold of dimension 4,
  the {\it Hopf degree theorem}
  (\cite[\S 9]{Pontrjagin55}, review in \cite[\S 7.5]{Kobin16})
  says that
  the assignment \eqref{dInvariantAsMapOnCohomotopy}
  is an isomorphism, where the d-invariant

\vspace{-.4cm}
\begin{equation}
  \label{WindingDegreeAsMapOnCohomotopy}
  \xymatrix@R=-10pt@C=4em{
    \overset{
      \mathclap{
      \raisebox{6pt}{
        \tiny
        \color{darkblue}
        \bf
        Cohomotopy
      }
      }
    }{
      \widetilde \pi^4\big(X^4\big)
    }
    \ar[rr]^-{
      \overset{
        \mathclap{
        \raisebox{3pt}{
          \tiny
          \color{greenii}
          \bf
          Hopf winding degree map
        }
        }
      }{
        \mathclap{\phantom{\vert^{\vert}}}
        \mathrm{deg}
      }
    }_-{ \simeq }
    &&
    \overset{
      \mathclap{
      \raisebox{6pt}{
        \tiny
        \color{darkblue}
        \bf
        ordinary cohomology
      }
      }
    }{
      \widetilde{HZ}{}^4(X^4)
      \;\simeq\;
      \mathbb{Z}
    }
    \\
    \big[
      X
        \xrightarrow{\;\;c\;\;}
      S^4
    \big]
    \ar@{}[rr]|-{ \longmapsto }
    &&
    \big[
      G_4^{HR}(c)
    \big]
    \;:=\;
    c^\ast
    \big[
      \mathrm{vol}_{S^4}
    \big]
  }
\end{equation}
  is the {\it Hopf winding degree} of the given
  map $X^4 \xrightarrow{c} S^4$
  (see also Example \ref{CohomotopyChargeOfPoints}).
  In this sense, the d-invariant \eqref{dInvariantAsMapOnCohomotopy}
  generalizes integer mapping degree to generalized cohomology, whence the name.

\medskip
  In the physics literature, this situation is familiar from
  Freund-Rubin compactifications of 11-dimensional supergravity
  on spacetimes locally of the topology
  $\mathbb{R}^{5,1} \times \mathbb{R}_{\mathrm{rad}} \times S^4$
  (\cite{FreundRubin80}), which are the near-horizon geometries
  of $n$ coincident black M5-branes (\cite{Gueven92}\cite[\S 2.1.2]{AFCS99}).
  Here the 4-flux $G_4$ is, on the $S^4$-factor, $n$ times the
  volume form. Under Hypothesis H, this is identified as the
  $d_{H\mathbb{R}}$-invariant \eqref{dInvariantAsMapOnCohomotopy}
  of
  the Cohomotopy cocycle
  $\big[\mathbb{R}^{5,1} \times \mathbb{R}_{\mathrm{rad}} \times S^4
  \,\simeq\, S^4
  \xrightarrow{\;c\;} S^4 \big]$
  with Hopf degree/winding
  number \eqref{WindingDegreeAsMapOnCohomotopy} equal to $n$.

\medskip
  (In the more general global case,
  as per Remark \ref{FramedSpacetimes},
  on spacetimes that are $S^4$-fibrations
  and measuring charge in tangentially twisted Cohomotopy,
  this statement finds a parametrized generalization reflecting
  M5-brane anomaly cancellation; this is discussed in \cite{SS20a}.)
\end{example}

\begin{example}[\bf Ordinary dual M5-brane flux and the Sullivan model of the 4-sphere]
  \label{DualFluxInOrdinaryCohomology}
  For $E = H \mathbb{R}$ BEING ordinary rational cohomology with real coefficients,
  as in Example \ref{UnitG4Flux} and under the fundamental theorem
  of rational homotopy theory, the topological equation of
  motion \eqref{TrivializationOfCupSquareOfUnitM5BraneCharge} is
  the differential relation
  \vspace{-1mm}
  $$
    d
    \big(
      2G^{\mathbb{R}}_{7,\mathrm{unit}}
    \big)
    \,=\,
    -
    G^{\mathbb{R}}_{4,\mathrm{unit}}
    \wedge
    G^{\mathbb{R}}_{4,\mathrm{unit}}
  $$

  \vspace{-1mm}
\noindent   in the Sullivan model for the 4-sphere
  (recalled as \cite[Ex. 3.68]{FSS20c}),
  and its image
  under the cohomotopical character map \cite[\S 5.3]{FSS20c}
  in the rational/de Rham cohomology of spacetime:
   \vspace{-1mm}
  $$
    d
    \big(
      2G^{\mathbb{R}}_{7}(c)
    \big)
    \,=\,
    -
    G^{\mathbb{R}}_{4}(c)
    \wedge
    G^{\mathbb{R}}_{4}(c)
    \,.
  $$

    \vspace{-1mm}
\noindent
  But this is the defining relation of the dual 7-flux form
  in 11-dimensional supergravity.
  The fact that this
  appears as the image of Cohomotopy theory as seen
  in ordinary rational cohomology is a key motivation
  for Hypothesis H
  \cite[\S 2.5]{Sati13}\cite{FSS15b}\cite{FSS16a}\cite[(57)]{FSS19a}\cite[Prop. 2.5]{FSS19b}.
\end{example}

\medskip

\noindent {\bf Flux degree and Adams operations.}
While flux densities are thought of as differential forms
(or their classes in De Rham cohomology and hence in ordinary rational cohomology)
of a given degree, such may not manifestly be provided by
a class in a Whitehead-generalized cohomology $E$
measuring M-brane charge \eqref{dInvariantAsMapOnCohomotopy}.
Extra structure on $E$ which does
allow to extract form degree under its character map
are {\it Adams operations}.
The Adams operations on topological K-theory
(\cite[\S 5]{Adams62}\cite[\S IV.7]{Karoubi78}, review in \cite[\S 11]{Wirthmueller12})
are a system of cohomology operations indexed by $k \in \mathbb{N}$
\vspace{-4mm}
\begin{eqnarray}
  \label{AdamsOperationsOnComplexKTheory}
 & \xymatrix{
    X
    \ar@{}[rr]|-{ \longmapsto }
    &&
    \widetilde {{K\mathrm{U}}}{}^0(X)
    \ar[rr]^-{ \psi^k }
    &&
    \widetilde {{K\mathrm{U}}}{}^0(X)
  }
\\[-5pt]
   \label{AdamsOperationsOnOrthogonalKTheory}
 & \xymatrix{
    X
    \ar@{}[rr]|-{ \longmapsto }
    &&
    \widetilde {{K\mathrm{O}}}{}^0(X)
    \ar[rr]^-{ \psi^k }
    &&
    \widetilde {{K\mathrm{O}}}{}^0(X)
    \,.
  }
\end{eqnarray}

\vspace{-2mm}
\noindent
These being cohomology operations means that they are
natural transformations of cohomology groups, hence
commute with pullback along any map $X \xrightarrow{\;f\;} Y$,
in that the following squares commute:
\vspace{-2mm}
\begin{equation}
  \label{AdamsOperationsCommuteWithPullback}
  \xymatrix@R=12pt{
    \widetilde {{K\mathrm{U}}}{}^0(X)
    \ar[rr]^-{ \psi^k }
    \ar@{<-}[d]_-{ f^\ast }
    &&
    \widetilde {{K\mathrm{U}}}{}^0(X)
    \ar@{<-}[d]^-{ f^\ast }
    \\
    \widetilde {{K\mathrm{U}}}{}^0(Y)
    \ar[rr]^-{ \psi^k }
    &&
    \widetilde {{K\mathrm{U}}}{}^0(Y)
  }
\end{equation}

\vspace{-3mm}
\noindent
Less subtle but otherwise analogous cohomology operations $\psi^k_H$
exist on ordinary even cohomology $H^{\mathrm{ev}}R$
(Example \ref{ExamplesOfMultiplicativeCohomologyTheories})
defined as multiplication by $k^r$ in any degree $2r$:
\vspace{-2mm}
$$
  \xymatrix@R=-1pt{
    \widetilde {H^{\mathrm{ev}}R}{}^0(X)
    \ar[rr]^-{ \psi_H^k }
    &&
    \widetilde {H^{\mathrm{ev}}R}{}^0(X)
    \\
    {\phantom{A}}
    \\
    \mathclap{\phantom{\vert^{\vert^{\vert^{\vert^{\vert}}}}}}
    \widetilde {HR}{}^{2r}(X)
    \ar@{^{(}->}[uu]
    \ar[rr]^-{  }
    &&
    \mathclap{\phantom{\vert^{\vert^{\vert^{\vert^{\vert}}}}}}
    \widetilde {HR}{}^{2r}(X)
    \ar@{^{(}->}[uu]
    \\
    [
      \alpha_{2r}
    ]
    \ar@{}[rr]|-{ \longmapsto }
    &&
    k^{r}
    \cdot
    [
      \alpha_{2r}
    ]
  }
$$

\vspace{-1mm}
\noindent
We have (\cite[Thm. 5.1 (vi)]{Adams62} \cite[Thm. V.3.27]{Karoubi78})
that the Chern character map is compatible
with the Adams operations, in that the following diagrams all commute
\vspace{-1mm}
\begin{equation}
  \label{ChernCharacterCompatibilityWithAdamsOperations}
  \raisebox{20pt}{
  \xymatrix@R=1em{
    \widetilde{{K\mathrm{U}}}{}^0(X)
    \ar[rr]^-{ \mathrm{ch} }
    \ar[d]_-{ \psi^k }
    &&
    \widetilde{H^{\mathrm{ev}}\mathbb{Q}}{}^0(X)
    \ar[d]^{ \psi^k_H }
    \\
    \widetilde{{K\mathrm{U}}}{}^0(X)
    \ar[rr]^-{ \mathrm{ch} }
    &&
    \widetilde{H^{\mathrm{ev}}\mathbb{Q}}{}^0(X)
    \,,
  }
  }
  \phantom{AAAAA}
  \raisebox{20pt}{
  \xymatrix@R=1em{
    \widetilde{{K\mathrm{O}}}{}^0(X)
    \ar[rr]^-{ \mathrm{ch} }
    \ar[d]_-{ \psi^k }
    &&
    \widetilde{H^{\mathrm{ev}}\mathbb{Q}}{}^0(X)
    \ar[d]^{ \psi^k_H }
    \\
    \widetilde{{K\mathrm{O}}}{}^0(X)
    \ar[rr]^-{ \mathrm{ch} }
    &&
    \widetilde{H^{\mathrm{ev}}\mathbb{Q}}{}^0(X)
  }
  }
\end{equation}

  \vspace{-1mm}
\noindent
This means that a class
$\big[V\big] \,\in\, \widetilde {{K\mathrm{U}}}{}^0(X)$
represents a single flux form in degree $2r$ if it is
an eigenvector of eigenvalue $k^{r}$ for the Adams operations
$\psi^k$:
\vspace{-2mm}
\begin{equation}
  \label{SingleDegreeChernCharacterClasses}
  \psi^k
  [
    V
  ]
  \,=\,
  k^r
  \cdot
  [
    V
  ]
  \;\;\;\;\;
   \Rightarrow
  \;\;\;\;\;
  \mathrm{ch}
  (
    [
      V
    ]
  )
  \;\in\;
  H^{2r}
  (
    X;\, \mathbb{Q}
  )
  \;\hookrightarrow\;
  \widetilde{H^{\mathrm{ev}}\mathbb{Q}}{}^0(X)
\end{equation}
For instance, on the $2r$-dimensional sphere,
where the Chern character of any reduced class
is necessary concentrated on classes in degree $2r$,
all reduced K-theory classes are eigenvectors
to eigenvalue $k^{r}$ of the
Adams operations $\psi^k$
(e.g. \cite[p. 45 (47 of 67)]{Wirthmueller12}):
\vspace{-2mm}
\begin{equation}
  \label{AdamsOperationsOverSpheres}
  \xymatrix@C=3em{
    \widetilde {{K\mathrm{U}}}{}^0
    (
      S^{2r}
    )
    \ar[rr]^-{
      \psi^k(-)
      \;=\;
      k^r \cdot(-)
    }
    &&
    \widetilde {{K\mathrm{U}}}{}^0
    (
      S^{2r}
    )
    \,.
  }
\end{equation}

\subsection{M5 brane $H_3$-Flux and the Toda brackets}
 \label{M5ThreeFlux}

\noindent {\bf Fluxless backgrounds and Vanishing d-invariant.}
A {\it fluxless} background
in M-theory is meant to be one
for which the $G_4$-flux vanishes, by which we shall mean that
it vanishes in cohomology (Remark \ref{FluxlessBackgrounds}).
Under
\hyperlink{HypothesisHOnHomotopicallyFlatSpacetimes}{\it Hypothesis H}
and seen in $E$-cohomology via \eqref{M5BraneFluxInEtheoryByPullback},
this means that the
$d_E$-invariant \eqref{dInvariantAsMapOnCohomotopy}
of the Cohomotopy charge vanishes:
\begin{equation}
  \label{Fluxlessness}
  \big[
    c
  \big]
  \,\in\,
  \widetilde \pi^4(X)
  \;\;
  \mbox{is $E$-fluxless}
  \;\;\;\;\;\;
  \Leftrightarrow
  \;\;\;\;\;\;
  \big[
    G^E_4\!(c)
  \big]
  \;= 0\;
  \,\in\,
  \widetilde E^4(X)
  \;\;\;\;\;\;
  \Leftrightarrow
  \;\;\;\;\;\;
  d^E_E(c)
  \;=\;
  0
  \,.
\end{equation}
Notice that the Cohomotopy class $[c]$ itself may be non-trivial
even if its $E$-flux vanishes, which is
(\cite[\S 3.8]{FSS19b}) a key source of
subtle effects implied by Hypothesis H
(we expand on this in
Example \ref{VanishingG4SFluxInVicinityOfM2Branes}
and \cref{TadpoleCancellationAndSUBordismWithBoundaries}
below).

\medskip

\noindent {\bf $H_3$-Flux and Trivializations of the d-invariant.}
When the $G_4^E(c)$-flux/$d_E(c)$-invariant \eqref{dInvariantAsMapOnCohomotopy}
vanishes,
there are $E$-gauge transformations \eqref{NonabelianCohomologyInIntroduction},
to be denoted $H^E_3\!(c)$,
which witness this vanishing:
\begin{equation}
  \label{HE3Homotopy}
  \big[
    G_4^E(c)
  \big]
  \,=\,
  0
  \qquad
    \Leftrightarrow
  \qquad
  \exists
  \quad
  \left(
  \raisebox{25pt}{
  \xymatrix@C=90pt@R=30pt{
    X
    \ar[d]_-{
      c
    }
    \ar[r]_>>{\ }="s"
    &
    \ast
    \ar[d]^-{0}
    \\
    S^4
    \ar[r]^-{\qquad
      G^E_{4,\mathrm{unit}}
    }^<<{\ }="t"
    &
    E\degree{4}
    \ar@{=>}_{
      \mathclap{
      \rotatebox[origin=c]{25}{
        \scalebox{.7}{
        $
          \mathclap{
            \!\!\!\!\!\!\!\!\!\!\!\!\!\!\!\!\!\!\!
            d \,
            {
              \color{orangeii}
              H^E_3\!(c)
            }
            \,=\,
            G_4^E(c)
          }
        $
      }}
      }
    } "s"; "t"
  }
  }
  \right)
  \,.
\end{equation}
These trivializing homotopies are in general not unique, even up to
homotopy. The difference
of homotopy classes of
any two such homotopies is a class in the 3rd $E$-cohomology of
$X$:
\begin{equation}
  \label{DifferenceOfHFluxes}
  \big[
    H^E_{3}\!(c)_a
     \,-\,
    H^E_{3}\!(c)_b
  \big]
  \;\;
    =
  \;\;
  \left[
  \raisebox{40pt}{
  \xymatrix{
    &&
    \ast
    \ar@/^.7pc/[drr]|-{ \;0\; }_<{\ }="s1"
    \\
    X
    \ar@/_.7pc/[drr]^>{\ }="t2"
    \ar@/^.7pc/[urr]
    \ar[rrrr]|-{ \;G^E_{4}(c)\; }
      _-{\ }="s2"
      ^-{\ }="t1"
    &&
    &&
    E\degree{4}
    \\
    &&
    \ast
    \ar@/_.7pc/[urr]|-{ \;0\; }
    \ar@{=>}
      _-{
        \mathclap{\phantom{\vert^{\vert^{\vert}}}}
        { \color{orangeii} H^E_{3}\!(c)_a}
        \mathclap{\phantom{\vert_{\vert}}}
      }
      "s1"; "t1"
    \ar@{=>}
      _-{
        \mathclap{\phantom{\vert^{\vert^{\vert}}}}
        - { \color{orangeii} H^E_{3}\!(c)_b}
        \mathclap{\phantom{\vert_{\vert}}}
      }
      "s2"; "t2"
  }
  }
  \right]
  \;\;\;\;\;
  \in
  \;
  \pi_1
  \mathrm{Maps}^{\ast/}
  \!
  \big(
    X
    \,,\,
    E\degree{4}
  \big)
  \;\simeq\;
  \widetilde E^3(X)
  \,,
\end{equation}
and every such class arises this way.
This means that for each Cohomotopy charge
$c$ with vanishing $G_4^E$-flux \eqref{Fluxlessness},
hence with trivial $d_E$-invariant,
the set $H^E_3\mathrm{Fluxes}(X,c)$ of possible
{\it choices of trivializations}
$\big[ H^E_3(c) \big]$ \eqref{HE3Homotopy}
carries the structure of a {\it torsor} (a principal bundle over the point)
for the third $E$-cohomology group of $X$ (Lemma \ref{TorsorOfNullHomotopies}):
 \vspace{-2mm}
\begin{equation}
  \label{3FluxesFormTorsorOverThirdECohomologyOfX}
  [c] \in \pi^4(X),
  \;
  \big[G_4^E(c)\big] = 0
  \;\;\;\;\;\;\;\;\;\;
    \Rightarrow
  \;\;\;\;\;\;\;\;\;\;
  H^E_3\mathrm{Fluxes}(X,c)
  \;\;
    \in
  \;\;
  \big(
    \widetilde E^3
    (
      X
    )
  \big)
  \mathrm{Torsors}
  \,.
\end{equation}

\vspace{-2mm}
\noindent
Noticing that the set of
$H^E_3\mbox{Fluxes}(X,c)$ depends only on the
stabilized Cohomotopy charge \eqref{StableCohomotopy}
$
  [c]
  \,\in\,
  \widetilde \pi^4(X)
  \xrightarrow{\;\Sigma^\infty\;}
  \widetilde {\mathbb{S}}{}^4(X)
$
we write (see Def. \ref{TrivializationsOfThedInvariant}):
\begin{equation}
  \label{EFluxes}
  \hspace{-2cm}
  \xymatrix{
    \mathllap{
      \mbox{
        \tiny
        \color{darkblue}
        \bf
        {\begin{tabular}{c}
          $H_3$-fluxes in $E$
          \\
          for $G_4$-fluxes in $\mathbb{S}$
        \end{tabular}}
      }
    }
    H^E_3\mathrm{Fluxes}(X)
    \ar[d]
    \mathrlap{
      \;
      \coloneqq
      \;
      \underset{
        [c] \in \widetilde {\mathbb{S}}{}^4(X)
      }{\bigsqcup}
      H^E_3\mathrm{Fluxes}(X,c)
    }
    \\
    \mathllap{
      \mbox{
        \tiny
        \color{darkblue}
        \bf
        {\begin{tabular}{c}
          $G_4$-fluxes in
          \\
          stable Cohomotopy
        \end{tabular}}
      }
    }
    G^E_4\mathrm{Fluxes}(X)
    \mathrlap{
      \;
      \coloneqq
      \;\;\;
      \widetilde {\mathbb{S}}{}^4(X)
    }
  }
\end{equation}
for the set of all such 3-fluxes as the stable Cohomotopy
class of the Cohomotopy charge $c$ varies, hence fibered
over the set of $G^E_4\mathrm{Fluxes}$ \eqref{GFluxes},
with fiber over $[c]$ empty if $\big[ G^E_4\!(c)\big] \neq 0$
and otherwise
an $\widetilde E^3(X)$-torsor \eqref{3FluxesFormTorsorOverThirdECohomologyOfX}.
This is the (discrete) {\it moduli space}
of (homotopy/gauge equivalence classes of)
choices of pairs of 4-flux and 3-flux.

\begin{example}[$H^{\mathbb{S}}_3$-Flux near M2-branes and the Order of the third stable stem]
  \label{VanishingG4SFluxInVicinityOfM2Branes}
  In the initial case that $E = \mathbb{S}$ is
  taken to be stable Cohomotopy itself
  \eqref{StableCohomotopy}
  and $X \simeq S^7$ is spacetime in the vicinity of
  M2-branes \eqref{M2BranesPTTheorem}, the unstable
  Cohomotopy charges lie in $\mathbb{Z} \times \mathbb{Z}_{12}$
  \eqref{HomotopyGroupsOfThe4Sphere},
  while their underlying 4-flux seen in
  $\mathbb{S}$-cohomology/stable Cohomotopy
  vanishes precisely on those such charges that
  in the first factor are multiples of $24$
  (as in \cite[(8)]{FSS19b}):
  \vspace{-2mm}
\begin{equation}
  \label{FromUnstableToStable4CohomotopyOf7Sphere}
  \hspace{-4mm}
  \xymatrix@R=2pt@C=2.8em{
    \mbox{
      \tiny
      \color{darkblue} \bf
      \begin{tabular}{c}
        $\phantom{a}$
        \\
        quaternionic
        \\
        Hopf fibration
      \end{tabular}
    }
    &
    \mathclap{
      [S^7 \overset{h_{\mathbb{H}}}{\to} S^4]
    }
    \ar@{}[d]|-{
      \mbox{
        \begin{rotate}{-90}
          $\!\!\!\!\!\mapsto$
        \end{rotate}
      }
    }
    \ar@{}[r]|>>>>{\in}
    &
    \overset
    {
      \mathclap{
      \mbox{
        \tiny
        \color{darkblue} \bf
        \begin{tabular}{c}
          non-abelian/unstable
          \\
          Cohomotopy group
          \\
          $\phantom{a}$
        \end{tabular}
      }
      }
    }
    {
      \widetilde \pi^4(S^7)
    }
    \ar@{}[d]|-{
      \mbox{
        \begin{rotate}{-90}
          $\!\!\!\simeq$
        \end{rotate}
      }
    }
    \ar[rrr]^-{
      \overset{
        \raisebox{2pt}{
          \tiny
          \color{orangeii} \bf
          \begin{tabular}{c}
            Boardman
            \\
            homomorphism
          \end{tabular}
        }
      }{
        \beta^4
      }
    }
    &&&
    \overset{
      \mathclap{
      \mbox{
        \tiny
        \color{darkblue} \bf
        \begin{tabular}{c}
          abelian/stable
          \\
          Cohomotopy group
          \\
          $\phantom{a}$
        \end{tabular}
      }
      }
    }{
      \;\widetilde{\mathbb{S}}^4(S^7)\;
    }
    \ar@{}[d]|-{
      \mbox{
        \begin{rotate}{-90}
          $\!\!\!\simeq$
        \end{rotate}
      }
    }
    \ar@{}[r]|<{\ni}
    &
    \mathclap{
      \;\;
      \Sigma^\infty[S^7 \overset{h_{\mathbb{H}}}{\to} S^4]
    }
    \ar@{}[d]|-{
      \!\!\!\!\!
      \mbox{
        \begin{rotate}{-90}
          $\!\!\!\!\!\mapsto$
        \end{rotate}
      }
    }
    &
    \mbox{
      \tiny
      \color{darkblue} \bf
      \begin{tabular}{c}
        stabilized
        \\
        quaternionic
        \\
        Hopf fibration
      \end{tabular}
    }
    \\
    \mbox{
      \tiny
      \color{darkblue} \bf
      \begin{tabular}{c}
        non-torsion
        \\
        generator
      \end{tabular}
    }
    &
    (1,0)
    \ar@{}[r]|>>>{ \in }
     &
    \mathbb{Z} \times \mathbb{Z}_{12}
    \ar[rrr]_-{ (n,a) \, \mapsto \,  (n \,\,\mathrm{mod}\,\, {\color{purple}24}) }
    &&&
    \mathbb{Z}_{\color{purple}24}
    \ar@{}[r]|<<<<{\ni}
    &
    1
    &
    \mbox{
      \tiny
      \color{darkblue} \bf
      \begin{tabular}{c}
        torsion
        \\
        generator
      \end{tabular}
    }
  }
\end{equation}
\end{example}

\noindent
Hence:

\vspace{2mm}

\noindent
\hspace{.04cm}
\fbox{
\begin{minipage}[left]{16.8cm}
{\it
$H^{\mathbb{S}}_3$-Flux near M2-branes for Cohomotopy charge
$c = n \cdot [h_{\mathbb{H}}]$
measured in $\mathbb{S}$-Cohomology
exists precisely if $n$
is a multiple of 24.}
Moreover, comparison with \eqref{CompactificationOnY4}
shows that:
{\it This cohomotopical $H^{\mathbb{S}}_3$-flux
witnesses the spontaneous compactification of M-theory on
K3 near probe M2-branes.}
\end{minipage}
}

\begin{example}[Ordinary and K-theoretic 3-flux]
  \label{H3FluxInKUCohomology}
  Let spacetime  \eqref{SolitonicBraneSpacetime}
  be homotopy-equivalent to any odd-dimensional
  sphere
  (such as $X \simeq S^7$ for an M2-brane background \eqref{CompactificationOnY4})
  and consider measuring M-brane charge \eqref{dInvariantAsMapOnCohomotopy}
  in
  either even integral cohomology
  or in complex K-theory (Notation \ref{ExamplesOfMultiplicativeCohomologyTheories}):
  $$
    X \,\simeq\, S^{2n+1}
    \;\;\;\;\;\;
    \mbox{and}
    \;\;\;\;\;\;
    E \,\in\,
    \big\{
      H^{\mathrm{ev}}\!\mathbb{Z}
      \,,\,
      {K\mathrm{U}}
    \big\}
    \,.
  $$
  Then, by the suspension isomorphism and by (Bott-)periodicity,
  we have:

  \noindent

  \begin{itemize}
  \vspace{-.2cm}
  \item[{\bf (a)}]
  $\widetilde E^4(X) \simeq 0$,
  and hence
  for {\it all} Cohomotopy
  charges $\big[X \xrightarrow{c} S^4 \big]$
  the $d_{{K\mathrm{U}}}$-invariant \eqref{dInvariantAsMapOnCohomotopy}
  vanishes,
  \\
  $\big[ G^{{K\mathrm{U}}}_4(c) \big] = 0$, \eqref{Fluxlessness}
  and an $H^{{K\mathrm{U}}}_3$-flux exists \eqref{HE3Homotopy};

  \vspace{-.2cm}
  \item[{\bf (b)}]
  $\widetilde E^3(X) \simeq \mathbb{Z}$,
  and hence
  the possible $H_3$-fluxes
  \eqref{3FluxesFormTorsorOverThirdECohomologyOfX}
  form a $\mathbb{Z}$-torsor.
  \end{itemize}
  \vspace{-.2cm}

  \noindent
  If we suggestively denote the
  $\mathbb{Z}$-action on this $\mathbb{Z}$-torsor as
  addition of multiples of a {\it unit of closed 3-flux}
  \begin{equation}
    \label{ZTorsorActionOnKU3Fluxes}
    \xymatrix@R=-10pt{
      E\mathrm{Fluxes}(X,c)
      \times
      \mathbb{Z}
      \ar[rr]
      &&
      E\mbox{Fluxes}(X,c)
      \\
      \big(
        \underset{
          \mathclap{
          \raisebox{-3pt}{
            \tiny
            \color{darkblue}
            \bf
            {\begin{tabular}{c}
              any
              \\
              3-flux
            \end{tabular}}
          }
          }
        }{
        \big[
          H^{E}_{3}\!(c)
        \big]
        }
        \,,
        \;
        n
      \big)
      \ar@{}[rr]|-{ \longmapsto }
      &&
      \big[
        H^{E}_{3}\!(c)
      \big]
      +
      n
       \cdot
      \underset{
        \mathclap{
        \raisebox{-3pt}{
          \tiny
          \color{darkblue}
          \bf
          {\begin{tabular}{c}
            unit of
            \\
            closed 3-flux
          \end{tabular}}
        }
        }
      }{
      \big[
        H^{E}_{3, \mathrm{unit}}
      \big]
      }
    }
  \end{equation}

  \vspace{-2mm}
\noindent
  then this means that for all $c$, a choice of
  {\it reference 3-flux}
  -- to be denoted $C^{{K\mathrm{U}}}_{3}\!(c)$ --
  is equivalently an
  isomorphism of $\mathbb{Z}$-torsors,
  specifically the one of the following form
  (a trivialization of $\mathbb{Z}$-torsors):
  \begin{equation}
    \label{TrivializationOfZTorsorOfKU3Fluxes}
    \xymatrix@R=-6pt{
      \mathbb{Z}
      \ar[rr]^-{
            \simeq
        }
          &&
      H^E_3\mathrm{Fluxes}(X,c)
      \\
      n
      \ar@{}[rr]|-{ \longmapsto }
      &&
      \underset{
        \mathclap{
        \raisebox{-3pt}{
          \tiny
          \color{darkblue}
          \bf
          C-field
        }
        }
      }{
        \big[
          C^{E}_{3}\!(c)
        \big]
      }
      \;+\;
      \underset{
        \mathclap{
        \raisebox{-3pt}{
          \tiny
          \color{darkblue}
          \bf
          closed 3-flux
        }
        }
      }{
      n
        \cdot
      \big[
        H^{E}_{3,\mathrm{unit}}
      \big]
      }
    }
  \end{equation}
\end{example}

\noindent
{\bf Observables on $H_3$-Flux and Toda brackets.}
A choice of $H^E_3$-flux \eqref{HE3Homotopy} by itself
does not give a cohomology class, and is hence not
an invariant observable.
By forming differences \eqref{DifferenceOfHFluxes} we
do get cohomology classes, and hence observables,
of pairs of $H_3$-fluxes, measuring one relative to another.
In order to make $H_3$-flux itself be observable
we need such a relative construction with the
reference point fixed by other means.
Since $H^E_3$-flux witnesses the vanishing of $G^E_4$-flux,
whose reference point is, in a sense, the unit 4-flux
$G^E_{4,\mathrm{unit}}$ \eqref{UnitG4Flux}, we obtain
such a reference point
in any  (non-abelian) cohomology theory
$A$ in which the $G_{4,\mathrm{unit}}$-flux has been trivialized,
regarded there through a given cohomology operation
$E^4 \xrightarrow{\;\phi\;} B A$:
\begin{equation}
  \label{TrivializationOfUnitG4FluxInATheory}
  \raisebox{20pt}{
  \xymatrix@C=25pt{
    S^4
    \ar[rr]
      ^-{ G^E_{4,\mathrm{unit}} }
      _>>{\ }="s"
    \ar[d]
      ^>>{\ }="t"
    &&
    E^4
    \ar[d]
      ^-{ \phi }
    \\
    \ast
    \ar[rr]
      _-{ 0 }
    &&
    B A
    \ar@{=>}
      "s"; "t"
      ^-{
        \mathclap{
          \rotatebox{27}{
            \scalebox{.6}{
              $
              \mathclap{
                d\,
                {\color{orangeii}
                  O^A
                }
                \;=\;
                - G^\phi_{4,\mathrm{unit}}
                \!\!\!\!\!\!\!\!\!\!\!\!\!\!
              }
              $
            }
          }
        }
      }
    }
  }
\end{equation}
Given this, we may measure $H^E_3$-flux \eqref{HE3Homotopy}
in $A$-cohomology,
by forming the pasting composite with \eqref{TrivializationOfUnitG4FluxInATheory}:
\begin{equation}
  \label{TodaBracketObservableOnH3Flux}
  \underset{
    \raisebox{-6pt}{
      \tiny
      \color{darkblue}
      \bf
      \begin{tabular}{c}
        Observable on $H^E_3$-flux
        \\
        relative to $O^A$
      \end{tabular}
    }
  }{
  \raisebox{35pt}{
  \xymatrix{
    X
    \ar@/^2.4pc/[ddrr]
      ^-{
        0
      }
      _-{\ }="s"
    \ar@/_2.4pc/[ddrr]
      _-{
        0
      }
      ^-{\ }="t"
    \\
    \\
    &&
    B A
    \ar@{=>}
      |-{
        \mathclap{\phantom{\vert^{\vert^{\vert}}}}
        \color{orangeii}
        O^A
        (
          H^E_3(c)
         )
        \mathclap{\phantom{\vert_{\vert_{\vert}}}}
      }
      "s"; "t"
  }
  }
  }
  \;\;\;
  \coloneqq
  \;\;\;
  \underset{
    \raisebox{-3pt}{
      \tiny
      \color{darkblue}
      \bf
      Refined Toda bracket
    }
  }{
  \raisebox{42pt}{
  \xymatrix@C=35pt{
    X
    \ar[rr]
      _>>{\ }="s1"
    \ar[d]
      _-{ c }
      ^>>{\ }="t1"
    &&
    \ast
    \ar[d]
      ^-{ 0 }
    \\
    S^4
    \ar[rr]
      |-{
        \;
        G^E_{4,\mathrm{unit}}
        \;
      }
      _>>{\ }="s"
    \ar[d]
      ^>>{\ }="t"
    &&
    E^4
    \ar[d]
      ^-{ \phi }
    \\
    \ast
    \ar[rr]
      _-{ 0 }
    &&
    B A
    \ar@{=>}
      "s1"; "t1"
      ^-{
        \mathclap{
          \rotatebox{17}{
            \scalebox{.6}{
              $
              \mathclap{
                d\,
                {\color{orangeii}
                  H^E_3(c)
                }
                \;=\;
                G^E_{4}(c)
                   \!\!\!\!\!\!\!\!\!\!\!\!\!\!\!
              }
              $
            }
          }
        }
        }
    \ar@{=>}
      "s"; "t"
      ^-{
        \mathclap{
          \rotatebox{20}{
            \scalebox{.6}{
              $
              \mathclap{
                d\,
                {\color{orangeii}
                  O^A
                }
                \;=\;
                - G^\phi_{4,\mathrm{unit}}
                \!\!\!\!\!\!\!\!\!\!\!\!\!\!\!
              }
              $
            }
          }
        }
      }
    }
  }
  }
  \;\;\;\;
  \begin{aligned}
    \in
    \;\;\;\;
    \;
    &
   \pi_1
    \mathrm{Maps}^{\ast/\!\!}
    (
      X, B A
    )
    \\
     =
    \;
    &
    \pi_0
    \mathrm{Maps}^{\ast/\!\!}
    (
      X, A
    )
    \\
    =
    \;
    &
    {\widetilde A}(X)
  \end{aligned}
\end{equation}

This construction
of {\it observables on trivializing fluxes}
hence this {\it secondary invariant}
is a refined {\it Toda bracket} (Def. \ref{RefinedTodaBracket})
of {(a)} the Cohomotopy charge $c$
with {(b)} its unit $G^E_{4,\mathrm{unit}}$ in $E$-cohomology
and {(c)} its observation through $\phi$,
as discussed in detail below in \cref{TodaBrackets}.

\medskip
\noindent
In the following we consider a sequence of
{\bf examples of Toda-bracket observables on $H^E_3$-flux}:

\vspace{-3mm}
\begin{enumerate}
  \setlength\itemsep{-3pt}
\item
  The stably {\it universal} $H^E_3$-observable
  $O^{E/\mathbb{S}}$
  takes values in the
  {\it Adams cofiber cohomology theory}
  $E/\mathbb{S}$ (Def. \ref{UnitCofiberCohomologyTheories})
  --
  discussed as
  Prop. \ref{H3FluxIsQuantizedInAdamsCofiberCohomologyTheory}
  below
  (a special case of Prop. \ref{EFluxesAreCocycleInCofiberTheory}
  further below).

\item
 The {\it refined Adams e-invariant}
 sees a charge lattice of
 rational values of $H^{K\mathrm{U}}_3$-fluxes
 --
 discussed in \cref{M5BraneC3FieldAndTheAdamseInvariant}.

\item
 The {\it refined Conner-Floyd e-invariant}
 measures the
 flux of $H_3$ through 3-spheres around branes
 transversal to local CY2-compactifications --
 discussed in
 (\cref{TadpoleCancellationAndSUBordismWithBoundaries}

\item
 The {\it Hopf invariant} is the $H^{H \mathbb{Z}}_3$-observable
 that measures the {\it Page charge} of M2-branes
 corresponding to the given $H_3/C_3$-field --
 discussed in \cref{M2BraneChargeAndTheHopfInvariant}.

\end{enumerate}

\vspace{-3mm}
\noindent
First, to better understand
observables on the moduli spaces \eqref{EFluxes} of $E$-fluxes,
we turn attention to the {\it classifying maps}
of the 3-flux:

\medskip

\noindent {\bf Classifying maps for 3-flux and Homotopy cofiber sequences.}
Given a pair of maps out of one domain space $X$,
we denote their {\it homotopy cofiber-product} or {\it homotopy pushout}
(of one map along the other) by the label ``(po)''\footnote{
Often we notationally suppress the homotopy (the double arrow)
filling this square. Conversely, {\it all}
cells of all diagrams
we display in this article are homotopy-commutative,
with coherent homotopies filling them, even if these are not
made notationally explicit.}
(e.g. \cite[Def. A.23]{FSS20c}), as shown on the left
in the following.
In particular, when $X \xrightarrow{\;c\;} B$
is any map, and $A = \ast$ is the point,
this is the  {\it homotopy cofiber} $C_c$, shown on the right:
\vspace{-2mm}
\begin{equation}
  \label{HomotopyPushoutAndHomotopyCofiberSpace}
  \raisebox{20pt}{
  \xymatrix@C=3em{
    X
    \ar[rr]_>>>>{\ }="s"
    \ar[d]
    &&
    A
    \ar[d]
    \\
    B
    \ar[rr]^<<<<{\ }="t"
    &&
    A \underset{X}{\sqcup}^h B
    \mathrlap{
      \!\!\!
      \mbox{
        \tiny
        \color{darkblue}
        \bf
        \begin{tabular}{c}
          homotopy
          \\
          pushout
        \end{tabular}
      }
    }
    \ar@{=>}
      _-{
        \mathclap{
          \rotatebox[origin=c]{30}{
            \scalebox{.52}{$
              \mathclap{
                \!\!\!\!\!\!\!\!\!\!\!\!\!\!\!\!\!\!\
                \mbox{
                  \color{orangeii}
                  \bf
                  universal homotopy
                }
              }
            $}
          }
        }
      }
      ^{ \mbox{\tiny\rm(po)} }
      "s"; "t"
  }
  }
  \phantom{AAAAAAAAAAAAA}
  \raisebox{20pt}{
  \xymatrix@C=37pt{
    X
    \ar[rr]_>>>>{\ }="s"
    \ar[d]_-c
    &&
    \ast
    \ar[d]
    \mathrlap{
      \!\!\!
      \mbox{
        \tiny
        \color{darkblue}
        \bf
        \begin{tabular}{c}
          point space
        \end{tabular}
      }
    }
    \\
    B
    \ar[rr]^<<<<{\ }="t"
    &&
    C_c
    \mathrlap{
      \!\!\!
      \mbox{
        \tiny
        \color{darkblue}
        \bf
        \begin{tabular}{c}
          homotopy
          \\
          cofiber space
        \end{tabular}
      }
    }
    \ar@{=>}
      _-{
        \mathclap{
          \rotatebox[origin=c]{29}{
            \scalebox{.52}{$
              \mathclap{
                \!\!\!\!\!\!\!\!\!\!\!\!\!\!\!
                \!\!\!\!\!
                \mbox{
                  \color{orangeii}
                  \bf
                  universal homotopy
                }
              }
            $}
          }
        }
      }
      ^{ \mbox{\tiny\rm(po)} }
      "s"; "t"
    }
  }
\end{equation}

\vspace{-3mm}
\noindent
These homotopies being universal means that
homotopies $\varphi$ under
the original pair of maps are {\it classified} by
maps $\vdash \varphi$ out of the homotopy pushout,
via factorization through the universal homotopy:
\vspace{-2mm}
\begin{equation}
  \label{HomotopyPushoutPropertyInIntroduction}
  \raisebox{20pt}{
  \xymatrix@R=1.5em{
    X
    \ar[rr]_>>>>{\ }="s"
    \ar[d]
    &&
    A
    \ar@[greenii][d]^-{
      \color{greenii}
      f
    }
    \\
    B
    \ar@[greenii][rr]_-{
      \color{greenii}
      g
    }^<<<<{\ }="t"
    &&
    \color{darkblue}
    C
    \ar@[orangeii]@{=>}|-{
      \color{orangeii}
      \mathclap{\phantom{\vert}}
      \;\;\varphi\;\;
    } "s"; "t"
  }
  }
  \;\;\;\;\;\;\;\;
    \simeq
  \;\;\;\;\;\;\;\;
  \raisebox{20pt}{
  \xymatrix@R=1.5em{
    X
    \ar[rr]_>>>>{\ }="s"
    \ar[d]
    &&
    A
    \ar[d]
    \ar@[greenii]@/^1pc/[ddr]^-{
      \color{greenii}
      f
    }
    \\
    B
    \ar[rr]^<<<<{\ }="t"
    \ar@[greenii]@/_1pc/[drrr]_-{
      \color{greenii}
      g
    }
    &&
    A \underset{X}{\sqcup}^h B
    \ar@[orangeii]@{-->}[dr]_<<<{
      \color{orangeii}
      \vdash \varphi
      \!\!\!\!\!
    }
    \\
    && &
    \color{darkblue}
    C
    \ar@{=>}^{ \mbox{\tiny (po)} } "s"; "t"
  }
  }
\end{equation}
Moreover, such homotopy pushout squares \eqref{HomotopyPushoutAndHomotopyCofiberSpace}
satisfy the {\it pasting law}, which says that if in a
``homotopy pasting diagram'', as shown in the diagram
\vspace{-2mm}
\begin{equation}
  \label{PastingLaw}
  \raisebox{42pt}{
  \xymatrix@C=37pt@R=14pt{
    {\;\phantom{X}\;}
    \ar[r]_>>{\ }="s1"
    \ar[d]^>>{\ }="t1"
    &
    {\;\phantom{X}\;}
    \ar[d]
    \\
    {\;\phantom{X}\;}
    \ar[r]_>>{\ }="s2"
    \ar[d]^>>{\ }="t2"
    &
    {\;\phantom{X}\;}
    \ar[d]
    \\
    {\;\phantom{X}\;}
    \ar[r]
    &
    {\;\phantom{X}\;}
    \ar@{=>}^-{
      \mbox{
        \tiny
        \rm
        (po)
      }
    }
       "s1"; "t1"
    \ar@{=>} "s2"; "t2"
  }
  }
  \end{equation}
 the top square is a homotopy pushout, then the bottom square is
so if and only if the total rectangle is.
In particular, this means
\vspace{-2mm}
\begin{equation}
  \label{PastingForIteratedHomotopyCofiber}
  \raisebox{35pt}{
  \xymatrix@R=1em{
    X
    \ar[rr]_->>{\ }="s1"
    \ar[d]_c^>>{\ }="t1"
    &&
    \ast
    \ar[d]
    \\
    Y
    \ar[rr]_->>{\ }="s2"
    \ar[d]^>>{\ }="t2"
    &&
    C_c
    \ar[d]
    \\
    \ast
    \ar[rr]
    &&
    \Sigma X
    \ar@{=>}^-{ \mbox{\tiny\rm(po)} }
      "s1"; "t1"
    \ar@{=>}^-{ \mbox{\tiny\rm(po)} }
      "s2"; "t2"
  }
  }
\end{equation}
that the homotopy cofiber of a
cohomotopy cofiber is the suspension of the original domain.
The resulting long
{\it homotopy cofiber sequences}
\vspace{-2mm}
\begin{equation}
\label{HomotopyCofiberSequence}
\xymatrix{
  X
    \ar[r] &
  Y
   \ar[r] &
  \Sigma X
    \ar[r] &
  \Sigma Y
    \ar[r] &
  \Sigma Z
    \ar[r] &
  \cdots
  }
\end{equation}

\vspace{-1mm}
\noindent
are sent by any Whitehead-generalized cohomology theory
$\widetilde E(-)$ to the corresponding long exact sequence
\eqref{LongExactSequenceInECohomology} in
$E$-cohomology:
\vspace{-3mm}
\begin{equation}
  \label{LongExactSequenceInECohomology}
  \hspace{-3mm}
  \xymatrix{
    \cdots
      \ar@{<-}[r]
      &
    \widetilde E^{n+1}(Z)
      \ar@{<-}[r]
     &
    \widetilde E^n(X)
      \ar@{<-}[r]
    &
    \widetilde E^n(Y)
      \ar@{<-}[r]
    &
    \widetilde E^n(Z)
      \ar@{<-}[r]
    &
    \widetilde E^{n-1}(X)
      \ar@{<-}[r]
    &
    \cdots
  }
\end{equation}

\vspace{-2mm}
\noindent
Applied to the 3-flux homotopies $H^E_{3}\!(c)$ from \eqref{HE3Homotopy},
the equivalence
\eqref{HomotopyPushoutPropertyInIntroduction} says that these are
classified by
$E$-cohomology classes $\big[\vdash H^E_3\!(c)\big]$ of the cofiber space
$C_c$ of the Cohomotopy charge $c$:
\begin{equation}
  \label{ClassifyingCohomologyClassFor3Flux}
  \big[
    \vdash H_3^E\!(c)
  \big]
  \;\,\in\,\;
  \widetilde E^4
  \big(
    C_c
  \big)
  \,,
  \phantom{AAAA}
  \mathrm{s.t.}
  \;\;\;\;
  q_c^\ast
  \big[
    \vdash H^E_3\!(c)
  \big]
  \;=\;
  \big[
    G^E_{4,\mathrm{unit}}
  \big]
  \,.
\end{equation}
\begin{equation}
  \label{FluxlessnessDiagram}
  \raisebox{24pt}{
  \xymatrix@C=100pt@R=30pt{
    X
    \ar[d]_-{
      c
    }
    \ar[r]_>>{\ }="s"
    &
    \ast
    \ar[d]^-{0}
    \\
    S^4
    \ar[r]_-{
      G^E_{4,\mathrm{unit}}
    }^<<{\ }="t"
    &
    E\degree{4}
    \ar@{=>}_{
      \mathclap{
      \rotatebox[origin=c]{22}{
        \scalebox{.7}{
        $
          \mathclap{
            \!\!\!\!\!\!\!\!\!\!\!\!\!\!\!\!
            d \,
            {
              \color{orangeii}
              H^E_3\!(c)
            }
            \,=\,
            G_4^E(c)
          }
        $
      }}
      }
    } "s"; "t"
  }
  }
  \;\;\;\;\;\;\;
    \simeq
  \;\;\;\;\;\;\;
  \raisebox{24pt}{
  \xymatrix@R=10pt@C=40pt{
    X
    \ar[rr]_>>>{\ }="s"
    \ar[dd]_c^>>{\ }="t"
    &&
    \ast
    \ar[dd]
    \ar@/^1pc/[dddr]^-{ \;0\; }
    \\
    \\
    S^4
    \ar[rr]|-{ \;q_c\; }
    \ar@/_1pc/[drrr]|-{
      \;\;
      G^E_{4,\mathrm{unit}}
      \;\;
    }
    &&
    C_c
    \ar@{-->}[dr]|-{
      \;\;
      \color{orangeii}
      \vdash H_3^E\!(c)
      \;\;
    }
    \\
    && &
    E\degree{4}
    \ar@{=>}
      ^-{ \mbox{\tiny(po)} }
      "s"; "t"
  }
  }
\end{equation}
Furthermore,  \eqref{PastingLaw} says that this is controlled
by long homotopy cofiber sequence of this form:
\vspace{-2mm}
\begin{equation}
  \label{HomotopyCofiberSequenceOfCohomotopyCharge}
  \xymatrix{
    X
    \ar[r]^-{c}
    &
    S^4
    \ar[r]^-{ q_c }
    &
    C_c
    \ar[r]^-{ p_c }
    &
    \Sigma X
    \ar[r]
    &
    \cdots
    \,,
  }
  \phantom{AAA}
  \mbox{e.g.}
  \phantom{AAA}
  \xymatrix{
    S^{n}
    \ar[r]^-{c}
    &
    S^4
    \ar[r]^-{ q_c }
    &
    C_c
    \ar[r]^-{ p_c }
    &
    S^{n+1}
    \ar[r]
    &
    \cdots
    \,.
  }
\end{equation}

\medskip

\noindent
{\bf $H^E_3$-Flux charge quantization and the Adams cofiber $E/\mathbb{S}$-theory.}
The {\it stably universal} way to fill the bottom square
in \eqref{TodaBracketObservableOnH3Flux},
and hence the
stably universal Toda-bracket observable on $H^E_3$-flux,
is given by the defining homotopy square of the
cohomology theory $E/\mathbb{S}$ which is the
homotopy cofiber of the unit map \eqref{UnitMorphismOfSpectra}
-- the {\it Adams cofiber cohomology theory}
(Def. \ref{UnitCofiberCohomologyTheories}):

\begin{prop}[$H^E_3$-Flux measured in unit cofiber cohomology]
  \label{H3FluxIsQuantizedInAdamsCofiberCohomologyTheory}
  Let spacetime $X \,\simeq\, S^{3 + d}$
  be homotopy equivalent to a sphere\footnote{
    This assumption is not necessary;
    we make it just for focus of exposition.
  }
  (\cref{MBraneWorldvolumesAndThePontrjaginConstruction})
  and
  measure M-brane charge in a multiplicative
  cohomology theory $E$ (\cref{M5BraneChargeAndMltiplicativeCohomologyTheory}).
  Then, the following stably universal Toda bracket observable \eqref{TodaBracketObservableOnH3Flux}
  on $H^E_3$
  \vspace{-1mm}
 \begin{equation}
  \label{MappingHFluxToClassInCofiberCohomology}
  H^E_{n-1}\mathrm{Fluxes}
  (
    S^{n+d-1}
  )
  \;\ni
  \!\!\!\!\!\!\!
  \raisebox{34pt}{
  \xymatrix{
    S^{n + d - 1}
    \ar[dd]
      |-{
        \mathclap{\phantom{\vert^{\vert}}}
        \color{greenii}
        c
        \mathclap{\phantom{\vert_{\vert}}}
      }
      ^>>>{\ }="t"
    \ar[rr]_>>>{\ }="s"
    &&
    \ast
    \ar[dd]
      |-{
        \mathclap{\phantom{\vert^{\vert}}}
        0
        \mathclap{\phantom{\vert_{\vert}}}
      }
    \\
    \\
    S^n
    \ar[rr]
      |-{
        \;
        \Sigma^n (1^E)
        \;
      }
    &&
    E\degree{n}
    \ar@{=>}
      |-{
        \mathclap{\phantom{\vert^{\vert^{\vert}}}}
        \color{orangeii}
        H^E_{n-1}\!(c)
        \mathclap{\phantom{\vert_{\vert_{\vert}}}}
      }
     "s"; "t"
    }
  }
  \;\;\;\;\;
    \longmapsto
  \!\!\!\!\!
  \raisebox{76pt}{
  \xymatrix@R=1.6em@C=3em{
   S^{n + d - 1}
   \ar[dd]
     |-{
       \mathclap{\phantom{\vert^{\vert}}}
       \color{greenii}
       c
       \mathclap{\phantom{\vert_{\vert}}}
     }
   \ar[rr]
   \ar@{}[ddrr]|-{
     \rotatebox[origin=c]{-45}{\color{orangeii}$\big\Downarrow$}
     \mbox{\tiny\rm (po)}
   }
   &&
   \ast
   \ar[dr]
   \ar[dd]
   \\
   && &
   \ast
   \ar[dd]|-{
     \mathclap{\phantom{\vert^{\vert}}}
     0
     \mathclap{\phantom{\vert_{\vert}}}
   }
   \\
   S^{n}
   \ar[dd]
   \ar[rr]|-{ \;q_c\; }
   \ar@{}[ddrr]|-{
     \rotatebox[origin=c]{-45}{$\big\Downarrow$}
     \mbox{\tiny\rm (po)}
   }
   \ar@/_.7pc/[drrr]
     |<<<<<{ \;\Sigma^{n} (1^{E})\;\; }
     |>>>>>>>>>>{ {\phantom{AA}} \atop {\phantom{AA}} }
   &&
   C_c
   \ar[dd]
     |>>>>>>{
       \mathclap{\phantom{\vert^{\vert}}}
       p_c
       \mathclap{\phantom{\vert_{\vert}}}
     }
   \ar@{-->}[dr]|-{
     \vdash
     {
       \color{orangeii}
       H^E_{n-1}\!(c)
     }
   }
   &&
   \\
   && &
   E\degree{n}
   \ar[dd]
     |-{
       \mathclap{\phantom{\vert^{\vert^{\vert}}}}
       i^E
       \mathclap{\phantom{\vert_{\vert}}}
     }
   \\
   \ast
   \ar@/_.7pc/[drrr]|-{ \;0\; }
   \ar[rr]
   &&
   S^{n + d}
   \ar@{-->}[dr]|-{
     \mathllap{\vdash}
     (
       {\color{greenii}
         G^{\mathbb{S}}_n\!(c)
       }
       \,,\,
       {\color{orangeii}
         H^E_{n-1}\!(c)
       }
     )
   }
   &&
   \\
   && &
   (E/\mathbb{S})\degree{n}
  }
  }
  \!\!\!\!\!\!\!\!\!
  \!\!\!\!\!\!\!\!\!
  \!\!\!\!\!\!\!\!\!
  \!\!\!\!\!\!\!\!\!
\end{equation}

\vspace{-3mm}
\noindent
  constitutes a lift from the set \eqref{EFluxes}
  of classes of possible $H^E_3$-fluxes on $X$
  to the degree $d$-cohomology ring \eqref{WhiteheadGeneralizedCoohomology}
  of the cofiber cohomology theory
  $E\!/\mathbb{S}$ \eqref{CofiberOfRingSpectrumUnitAndBoundaryHomomorphism}
  through its boundary map:
  \vspace{-2mm}
  \begin{equation}
    \raisebox{30pt}{
    \xymatrix@R=5pt{
      &
      \scalebox{0.7}{$
      \big(
        \big[
          {
            \color{greenii}
            c
          }
        \big],
        \big[
          {
            \color{orangeii}
            H^E_3\!(c)
          }
        \big]
      \big)
      $}
      \ar@{|->}[rr]
      &&
      \scalebox{0.7}{$
      \big[
      \vdash
      (
          {
            \color{greenii}
            G^E_4(c)
          }
        ,
          {
            \color{orangeii}
            H^E_3\!(c)
          }
      )
      \big]
      $}
      \\
      \scalebox{0.7}{$
      \big(
        \big[
          {
            \color{greenii}
            c
          }
        \big],
        \big[
          {
            \color{orangeii}
            H^E_3\!(c)
          }
        \big]
      \big)
      $}
    \ar@{|->}[dd]
    &
      H^E_3\mathrm{Fluxes}
      \big(
        S^{3 + d}
      \big)
      \ar@{-->}[rr]
      \ar[dd]
      &&
      \big(
        E \!/\mathbb{S}
      \big)_d
      \ar[dd]^{\partial}
      \\
      \\
      \scalebox{0.7}{$
        \big[
          {
            \color{greenii}
            G^{\mathbb{S}}_4\!(c)
          }
        \big]
        $}
        &
      G^E_4\mathrm{Fluxes}
      \big(
        S^{3+d}
      \big)
      \ar@{=}[rr]
      &&
      \mathbb{S}_{\mathrlap{d-1}}
    }
    }
  \end{equation}

\end{prop}

\newpage

We prove this as
Prop. \ref{EFluxesAreCocycleInCofiberTheory} below.

\begin{remark}[Observables on $H^E_3$ from multiplicative cohomology operations]
  \label{ObservablesOnH3FromMultiplicativeCohomologyOperations}
  One way to extract more specific information
  from
  the universal cofiber Toda-bracket observable
  (Def. \ref{H3FluxIsQuantizedInAdamsCofiberCohomologyTheory})
  is by postcomposition with
  a multiplicative cohomology operation
  $\xymatrix{E \ar[r]^-{\phi}  & F}$:
  Namely,
  any such canonically sits in a homotopy-commutative square
  of the form
  \begin{equation}
    \label{CofiberSquareOnMultiplicativeCohomologyOperation}
    \xymatrix{
      E
      \ar[rr]
        ^-{ \phi }
        _>>{\ }="s"
      \ar[d]
        _-{
          \mathclap{\phantom{\vert^{\vert}}}
          i^E
          \mathclap{\phantom{\vert_{\vert}}}
        }
        ^>>{\ }="t"
      &&
      F
      \ar[d]
        ^-{
          \mathclap{\phantom{\vert^{\vert}}}
          i^F
          \mathclap{\phantom{\vert_{\vert}}}
        }
      \\
      E/\mathbb{S}
      \ar[rr]
        ^-{
          \phi/\mathbb{S}
        }
      &&
      F/\mathbb{S}
      \ar@{=>}
        |-{
         \mathclap{\phantom{\vert^{\vert^{\vert}}}}
          \;\;
          i^\phi
          \;\;
          \mathclap{\phantom{\vert_{\vert_{\vert}}}}
        }
        "s"; "t"
    }
  \end{equation}
  whose pasting composite with the
  diagram \eqref{MappingHFluxToClassInCofiberCohomology}
  for the universal observable
  hence yields the following new Toda-bracket
  observable \eqref{TodaBracketObservableOnH3Flux}:
\begin{equation}
  \label{UniversalObserbableComposedWithMultiplicativeOperation}
  \underset{
    \raisebox{-6pt}{
      \tiny
      \color{darkblue}
      \bf
      \begin{tabular}{c}
        Observable on $H^E_3$-flux
        \\
        relative to
        $O^{E\!/\mathbb{S}}
          \scalebox{.6}{$\Box$}
        i^{\phi}$
      \end{tabular}
    }
  }{
  \raisebox{35pt}{
  \xymatrix{
    X
    \ar@/^2.4pc/[ddrr]
      ^-{
        0
      }
      _-{\ }="s"
    \ar@/_2.4pc/[ddrr]
      _-{
        0
      }
      ^-{\ }="t"
    \\
    \\
    &&
    F\!/\mathbb{S}
    \ar@{=>}
      |-{
        \mathclap{\phantom{\vert^{\vert^{\vert}}}}
        \color{orangeii}
        O^{F\!/\mathbb{S}}
        (
          H^E_3(c)
        )
        \mathclap{\phantom{\vert_{\vert_{\vert}}}}
      }
      "s"; "t"
  }
  }
  }
  \;\;\;
  \coloneqq
  \;\;\;
  \underset{
    \raisebox{-3pt}{
      \tiny
      \color{darkblue}
      \bf
      composite refined Toda bracket
    }
  }{
  \raisebox{60pt}{
  \xymatrix@C=6em{
    X
    \ar[rr]
      _>>{\ }="s1"
    \ar[d]
      _-{ c }
      ^>{\ }="t1"
    &&
    \ast
    \ar[d]
      ^-{ 0 }
    \\
    S^4
    \ar[rr]
      |-{
        \;
        G^E_{4,\mathrm{unit}}
        \;
      }
      _>>{\ }="s"
    \ar[d]
      ^>>{\ }="t"
    &&
    E^4
    \ar@/^.4pc/[dl]
      _-{
       \qquad  i^E
      }
      \ar[d]
        ^-{ \phi^4 }
        _-{\ }="s3"
    \\
    \ast
    \ar[r]
      ^-{\quad 0 }
    \ar@{=}[d]
    &
    E\!/\mathbb{S}
    \ar@/_1pc/[dr]
      _<<<<<<<<<<<<<<<<{
        (\phi/\mathbb{S})^4 \!\!\!\!\!
      }
      ^<<<<<{\ }="t3"
    &
    F^4
    \ar[d]
      ^-{
        (i^F)^4
      }
    \\
    \ast
    \ar[rr]
      ^<<<<<<<<<<<<<<<<{
       0
      }
    &
    &
    F\!/\mathbb{S}
    \ar@{=>}
      ^-{
        (i^\phi)^4
      }
      "s3"; "t3"
    \ar@{=>}
      "s1"; "t1"
      _<<<<<<<<<<<<<<<<<<<<<<<<<<<<<<<<<<<<<{
        \mathclap{
          \rotatebox{10}{
            \scalebox{.6}{
              $
              \mathclap{
                d\,
                {\color{orangeii}
                  H^E_3(c)
                }
                \;=\;
                G^E_{4}(c)
                            }
              $
            }
          }
        }
        }
    \ar@{=>}
      "s"; "t"
      _<<<<<<<<<<<<<<<<<<<<<<<<<<<<<<<<<<<<<<<<<<<{
        \mathclap{
          \rotatebox{10}{
            \scalebox{.6}{
              $
              \mathclap{
                d\,
                {\color{orangeii}
                  O^{E\!/\mathbb{S}}
                }
                \;=\;
                - G^{(i^E)}_{4,\mathrm{unit}}
                \!\!\!\!\!\!\!\!\!\!\!\!\!\!\!
              }
              $
            }
          }
        }
      }
    }
  }
  }
\end{equation}

\end{remark}

\newpage

\subsection{M5 brane $C_3$-Field and the Adams e-Invariant}
\label{M5BraneC3FieldAndTheAdamseInvariant}

\noindent {\bf $H^{{K\mathrm{U}}}_3$-Flux and the Adams $\mathrm{e}$-Invariant.}
Since the Chern character on complex K-theory $K \mathrm{U}$
and the Pontrjagin character on $K \mathrm{O}$

\vspace{-.6cm}
\begin{equation}
  \label{ChernCharacterAndPontrjaginCharacter}
  \xymatrix@R=14pt@C=6em{
    K \mathrm{O}
    \ar[d]
      _-{
        \mathrm{cplx}
      }
    \ar[dr]
      ^-{
        \overset{
          \mathrlap{
            \raisebox{4pt}{
              \tiny
              \color{greenii}
              \bf
              Pontrjagin character
            }
          }
        }{
          \mathrm{ph}
        }
      }
    \\
    K\mathrm{U}
    \ar[r]
      ^-{ \mathrm{ch} \;\;\;\;\; }
      _-{
        \mbox{
          \tiny
          \color{greenii}
          \bf
          Chern character
        }
      }
    &
    H^{\mathrm{ev}}\mathbb{Q}
  }
\end{equation}
\vspace{-.3cm}

\noindent
are multiplicative cohomology operations,
they induce, via Remark \ref{ObservablesOnH3FromMultiplicativeCohomologyOperations},
Toda-bracket observables
$O^{(H^{\mathrm{ev}}\mathbb{Q}\!/\mathbb{S})}$
\eqref{TodaBracketObservableOnH3Flux}
on $H^{K\mathrm{U}}_3$- and $H^{K\mathrm{O}}_3$-fluxes.
We call these observables the
{\it $\widehat{e}_{K\mathrm{U}}$-invariant}
and
{\it $\widehat{e}_{K\mathrm{O}}$-invariant}
(Def. \ref{LiftedEInvariantDiagrammatically} below)
since,
in the case that spacetime
\eqref{CohomotopyChargeOnBlackBraneAsymptoticBoundary}
is homotopy equivalent to an (odd-dimensional) sphere,
this observable
\vspace{-2mm}
$$
  \xymatrix@C=38pt{
    H^{K\mathrm{U}}_{2n-1}\mathrm{Fluxes}
    \big(
      S^{2(n+d)-1}
    \big)
    \ar@/^1.5pc/[rrrrr]
      ^-{
        {\widehat e}_{K\mathrm{U}}
      }
    \ar[rr]
      _-{
        O^{ K\mathrm{U}/\mathbb{S} }
      }
    &&
    \big(
      (K \mathrm{U})/\mathbb{S}
    \big)_{2d-1}
    \ar[rr]
      _-{
       \mathrm{ch}/\mathbb{S}
      }
    &&
    \big(
      (H^{\mathrm{ev}\!}\mathbb{Q})/\mathbb{S}
    \big)_{2d-1}
    \ar[r]_-{ \mathrm{spl}_0 }
    &
    \mathbb{Q}
  }
$$
\begin{equation}
  \label{DiagrammatichatecInvariant}
  \hspace{-.4cm}
  \raisebox{30pt}{
  \xymatrix@R=15pt@C=25pt{
    S^{2(n + d) - 1}
    \ar[dd]
      _-{
        \mathclap{\phantom{\vert^{\vert}}}
        \color{greenii}
        c
        \mathclap{\phantom{\vert_{\vert}}}
      }
      ^>>>{\ }="t"
    \ar[rr]_>>>{\ }="s"
    &&
    \ast
    \ar[dd]
      ^-{
        \mathclap{\phantom{\vert^{\vert}}}
        0
        \mathclap{\phantom{\vert_{\vert}}}
      }
    \\
    \\
    S^{2n}
    \ar[rr]
      _-{
        \;
        \Sigma^{2n} (1^{{K\mathrm{U}}})
        \;
      }
    &&
    {K\mathrm{U}}^{2n}
    \ar@{=>}
      |-{
        \mathclap{\phantom{\vert^{\vert^{\vert}}}}
        \color{orangeii}
        H^{{K\mathrm{U}}}_{2n-1}\!(c)
        \mathclap{\phantom{\vert_{\vert_{\vert}}}}
      }
     "s"; "t"
    }
  }
  \;\;\;\;\;
    \longmapsto
  \!\!\!\!\!
  \raisebox{96pt}{
  \xymatrix@R=1.5em@C=24pt{
   S^{2(n + d) - 1}
   \ar[dd]
     _-{
       \mathclap{\phantom{\vert^{\vert}}}
       \color{greenii}
       c
       \mathclap{\phantom{\vert_{\vert}}}
     }
   \ar[rr]
   \ar@{}[ddrr]|-{
     \rotatebox[origin=c]{-45}{\color{orangeii}$\big\Downarrow$}
     \mbox{\tiny (po)}
   }
   &&
   \ast
   \ar[dr]
   \ar[dd]
   \\
   && &
   \ast
   \ar[dd]^-{
     \mathclap{\phantom{\vert^{\vert}}}
     0
     \mathclap{\phantom{\vert_{\vert}}}
   }
   \\
   S^{2n}
   \ar[dd]
   \ar[rr]^-{ \;q_c\; }
   \ar@{}[ddrr]|-{
     \rotatebox[origin=c]{-45}{$\big\Downarrow$}
     \mbox{\tiny (po)}
   }
   \ar@/_.7pc/[drrr]
     |<<<<<{ \;\Sigma^{2n} (1^{K\mathrm{U}})\;\; }
     |>>>>>>{ {\phantom{AA}} \atop {\phantom{AA}} \;}
   &&
   C_c
   \ar[dd]
     _>>>>>>{
       \mathclap{\phantom{\vert^{\vert}}}
       p_c
       \mathclap{\phantom{\vert_{\vert}}}
     }
   \ar@{-->}[dr]|-{
     \vdash
     {
       \color{orangeii}
       H^{{K\mathrm{U}}}_{2n-1}\!(c)
     }
   }
   &&
   \\
   && &
   {K\mathrm{U}}^{2n}
   \ar[dd]
     ^-{
       \mathclap{\phantom{\vert^{\vert^{\vert}}}}
       i^{{K\mathrm{U}}}
       \mathclap{\phantom{\vert^{\vert}}}
     }
   \ar[dr]
     ^-{
       \;\mathrm{ch}\;
     }
   \\
   \ast
   \ar@/_.7pc/[drrr]_-{ \;0\; }
   \ar[rr]
   &&
   S^{2(n + d)}
   \ar@{-->}[dr]|-{
     \mathllap{\vdash}
     (
       {\color{greenii}
         G^{\mathbb{S}}_{2n}\!(c)
       }
       \,,\,
       {\color{orangeii}
         H^{{K\mathrm{U}}}_{2n-1}\!(c)
       }
     )
   }
   &&
   (H^{\mathrm{ev}}\mathbb{Q})^{2n}
   \ar[dd]
     ^-{
       \mathclap{\phantom{\vert^{\vert^{\vert}}}}
       i^{H^{\mathrm{ev}}\mathbb{Q}}
       \mathclap{\phantom{\vert^{\vert}}}
     }
   \\
   && &
   ({K\mathrm{U}}/\mathbb{S})^{2n}
   \ar[dr]
     _-{
       \;\mathrm{ch}/\mathbb{S}\;
     }
   \\
   && &&
   \big(
     (H^{\mathrm{ev}}\mathbb{Q})/\mathbb{S}
   \big)^{2n}
  }
  }
\end{equation}
coincides
with the lifts
through $\mathbb{Q} \xrightarrow{\;} \mathbb{Q}/\mathbb{Z}$
of the classical Adams $\mathrm{e}$-invariant
(Def. \ref{ClassicalConstructionOfAdamseCInvariant} below)
on Cohomotopy charges \eqref{CohomotopyChargeOnBlackBraneAsymptoticBoundary}
in the $(2n-1)$st stem:

\begin{prop}[\bf The $\widehat e_{K\mathrm{U}}$-invariant
 defines a charge lattice of $H^{{K\mathrm{U}}}_3$-flux]
 \label{ChargeLatticeOfHKU3Flux}
 For $X \simeq S^{2d + 3}$
 \eqref{H3FluxInKUCohomology},
 the ${\widehat e}_{K \mathrm{U}}$-invariant
 \eqref{DiagrammatichatecInvariant}
 embeds  the $\mathbb{Z}$-torsors of $H^{{K\mathrm{U}}}_3$-fluxes
(from Example \ref{H3FluxInKUCohomology}),
as $\mathbb{Z}$-sets, into the rational numbers $\mathbb{Q}$,
such that this covers the classical
Adams $\mathrm{e}$-invariant (Def. \ref{ClassicalConstructionOfAdamseCInvariant})
on the underlying Cohomotopy classes,
in that we have a commuting diagram as follows:
 \vspace{-2mm}
\begin{equation}
  \label{TheLiftedAdamseCInvariant}
  \hspace{-4mm}
  \raisebox{38pt}{
  \xymatrix@R=-10pt@C=2.8em{
    &
    \overset{
      \mathclap{
      \raisebox{6pt}{
        \tiny
        \color{darkblue}
        \bf
        \begin{tabular}{c}
          $\mathbb{Z}$-torsor
          of trivializations
          \\
          of d-invariant $G^{{K\mathrm{U}}}_4(c)$
        \end{tabular}
      }
      }
    }{
      H^{K\mathrm{U}}_3\mathrm{Fluxes}(X,c)
    }
    \;
   \ar@{^{(}->}[r]
    \ar@{}[ddr]|-{ \mbox{\tiny\rm(pb)} }
    \ar[dd]
      |-{
        \mbox{
          \tiny
          \color{greenii}
          \bf
          \begin{tabular}{c}
            forget
            \\
            3-flux
          \end{tabular}
        }
        \!\!
      }
    \;\;\;
    \ar@{^{(}->}@/_1.46pc/[rrr]
    &
    H^{K\mathrm{U}}_3\mathrm{Fluxes}(X)
    \ar[rr]^-{
      \overset{
        \mathclap{
        \raisebox{6pt}{
          \tiny
          \color{greenii}
          \bf
          \begin{tabular}{c}
            observable on $H^{K\mathrm{U}}_3$-fluxes
            \\
            induced by Pontrjagin character
          \end{tabular}
        }
        }
      }{
        \;\widehat e_{K\mathrm{U}}\;
      }
    }
    \ar[dd]
      |<<<<{ \phantom{AA} }
    &&
    \mathbb{Q}
    \ar[dd]
    \ar@{}[r]|-{\in}
    &
    \mathbb{Z}\mathrm{Sets}
    \ar[dd]|-{
      \mbox{
        \tiny
        \color{greenii}
        \bf
        \begin{tabular}{c}
          quotient by
          \\
          $\mathbb{Z}$-action
        \end{tabular}
      }
    }
    \\
    \underset{
      \mathclap{
      \raisebox{-3pt}{
        \tiny
        \color{darkblue}
        \bf
        \begin{tabular}{c}
          a Cohomotopy class
          \\
          with d-invariant trivialization
        \end{tabular}
      }
      }
    }{
    \Big\{\!
      \big(
      \overset{
        \mathclap{
        \rotatebox{25}{
          \rlap{
          \tiny
          \color{darkblue}
          \bf
          M5 charge in Cohomotopy
          }
        }
        }
      }
      [c]
      \,,\,
      \overset{
        \mathclap{
        \rotatebox{25}{
          \rlap{
          \tiny
          \color{darkblue}
          \bf
          3-flux in K-theory
          }
        }
        }
      }{
        H^{{K\mathrm{U}}}_3\!(c)
      }
      \big)
   \! \Big\}
    }
    \ar[dr]
    \;
    \ar@{^{(}->}[ur]
    \\
    &
    \big\{
      \underset{
        \mathclap{
        \raisebox{-3pt}{
          \tiny
          \color{darkblue}
          \bf
          \begin{tabular}{c}
            a Cohomotopy class
          \end{tabular}
        }
        }
      }{
        [c]
      }
    \big\}
    \;
    \ar@{^{(}->}[r]
    &
    \underset{
      \raisebox{-3pt}{
        \tiny
        \color{darkblue}
        \bf
        \begin{tabular}{c}
          Cohomotopy in
          \\
          $(4n-1)$st stem
        \end{tabular}
      }
    }{
      \pi^4\big( S^{4n + 3} \big)
    }
    \ar[rr]^-{
      \;e_{\mathbb{R}}\;
    }_-{
      \mathclap{
      \raisebox{-3pt}{
        \tiny
        \color{greenii}
        \bf
        {\begin{tabular}{c}
          Adams' ${\mathrm{e}_{\mathrm{Ad}}}$-invariant
        \end{tabular}}
      }
      }
    }
    &&
    \mathbb{Q}/\mathbb{Z}
    \ar@{}[r]|-{ \in }
    &
    \mathrm{Sets}
    \,.
  }
  }
\end{equation}

\end{prop}
\noindent
We prove this as Theorem \ref{DiagrammaticeCCoincidesWithClassicaleCInvariant}
below.

\newpage

\noindent
\noindent {\bf The $C_3$-field and vanishing of the Adams e-invariant.}
Prop. \ref{ChargeLatticeOfHKU3Flux}
says that the $\widehat e_{K\mathrm{U}}$-invariant
gives integer-spaced rational number-values to
$H_3$-fluxes measured in topological K-theory.
Hence {\it when} these rational numbers are all integers,
{\it then} the $\widehat e_{K\mathrm{U}}$-invariant
gives a trivialization of the
$\mathbb{Z}$-torsor of 3-fluxes \eqref{ZTorsorActionOnKU3Fluxes},
and hence a
{\it choice of $C_3$-field} \eqref{TrivializationOfZTorsorOfKU3Fluxes}
seen in K-theory.
But the diagrams
\eqref{TheLiftedAdamseCInvariant}
say that this is the case precisely when the
classical Adams e-invariant vanishes:
\begin{equation}
  \label{VanishingAdamsInvariantImpliesCFieldKOCase}
  \overset{
    \mathclap{
    \raisebox{3pt}{
      \tiny
      \color{darkblue}
      \bf
      \begin{tabular}{c}
        Adams $\mathrm{e}$-invariant vanishes
      \end{tabular}
    }
    }
  }{
    {\mathrm{e}_{\mathrm{Ad}}}(c)
    \,=\,
    0
    \;\;\in\;
    \mathbb{Q}/\mathbb{Z}
  }
  \phantom{AAAAA}
    \Longleftrightarrow
  \phantom{AAAAA}
    \overset{
      \mathclap{
      \raisebox{3pt}{
        \tiny
        \color{darkblue}
        \bf
        \begin{tabular}{c}
          ${\widehat e}_{K\mathrm{U}}$-observable
          realizes C-field in $K\mathrm{U}$-theory
        \end{tabular}
      }
      }
    }{
    \xymatrix@R=-6pt{
      \mathbb{Z}
      \ar@{<-}[rr]
        ^-{
          {\widehat e}_{K\mathrm{U}}
        }
        _-{ \simeq }
      &&
      H^{K\mathrm{U}}_3\mathrm{Fluxes}(X,c)
      \\
      n
      \ar@{}[rr]|-{
      \rotatebox[origin=0]{180}{
        $\longmapsto$
      }
      }
      &&
      \underset{
        \mathclap{
        \raisebox{-3pt}{
          \tiny
          \color{darkblue}
          \bf
          C-field
        }
        }
      }{\scalebox{.8}{$
        \big[
          C^{{K\mathrm{U}}}_{3}\!(c)
        \big]$}
      }
      \; \scalebox{.8}{$$}+\;
      \underset{
        \mathclap{
        \raisebox{-3pt}{
          \tiny
          \color{darkblue}
          \bf
          closed 3-flux
        }
        }
      }{ \scalebox{.8}{$
      n
        \cdot
      \big[
        H^{{K\mathrm{U}}}_{3,\mathrm{unit}}
      \big]$}
      }
    }
  }
\end{equation}

Concretely,
near M2-branes
$X \simeq S^7$ \eqref{M2BranesPTTheorem}
with Cohomotopy charge in the 3rd stem
being an integral multiple
$
  [c]
  \;=\;
  n \cdot [h_{\mathbb{H}}]
$
of the quaternionic Hopf fibration
(Example \ref{VanishingG4SFluxInVicinityOfM2Branes}),
the classical Adams invariant is (by Example \ref{AdamseInvariantOn3rdStableStem}):
\begin{equation}
  \label{ClassicalAdamsInvariantOn3rdStem}
  {\mathrm{e}_{\mathrm{Ad}}}( n \cdot h_{\mathbb{H}})
  \,=\,
  \left[
    \tfrac{n}{12}
  \right]
  \;\in\;
  \mathbb{Q}/\mathbb{Z}
  \,.
\end{equation}

\noindent
With
\eqref{VanishingAdamsInvariantImpliesCFieldKOCase}
this means, in conclusion:

\vspace{2mm}

\noindent
\hspace{.04cm}
\fbox{
\begin{minipage}[left]{16.8cm}
{\it
When measured in $K\mathrm{U}$-theory,
integral $H_3$-flux relative to a $C_3$-field
exists in the vicinity of M2-branes
precisely when the background Cohomotopy charge
\eqref{FromUnstableToStable4CohomotopyOf7Sphere}
is an integer multiple of $12$.}
\end{minipage}
}

\begin{remark} [Integrality conditions in various theories]
\label{From12To24}
{\bf (i)} The above condition for
integral $H_3$-flux and a $C_3$-field near M2-branes
to be visible in $K \mathrm{U}$-theory
is similar to but a little weaker
than the condition
that $H_3$-flux is seen at all in stable Cohomotopy $\mathbb{S}$.
The latter requires
background Cohomotopy charge in multiples of 24
(Example \ref{VanishingG4SFluxInVicinityOfM2Branes})
instead of just 12,
and is witnessed by a spontaneous KK-compactification
\eqref{CompactificationOnY4}
on $\mathrm{K3}$ (p. \pageref{M2BranesPTTheorem}),
with K3 regarded as a bare 4-manifold.

\noindent {\bf (ii)} We next see in \cref{TadpoleCancellationAndSUBordismWithBoundaries} below
that when the observation of $H_3$ in $K \mathrm{U}$-theory
is made through
$M S\mathrm{U} \to M \mathrm{U} \to K \mathrm{U}$,
then the $\mathrm{K}3$-compactification emerges again,
now with its $S\mathrm{U}$-structure.
\end{remark}

\medskip

\newpage

\subsection{$\mathrm{NS5}_{\mathrm{HET}}$-brane charge and Conner-Floyd's e-invariant}
 \label{TadpoleCancellationAndSUBordismWithBoundaries}

We had found in \cref{M5BraneC3FieldAndTheAdamseInvariant} that
the refined $\widehat e_{K \mathrm{U}}$-invariant
of a 4-Cohomotopy charge near M2-branes,
is a $\mathbb{Q}$-valued measure of the possible $H_3$-fluxes
on such a background ($\mathbb{Z}$-valued when the classical
Adams invariant vanishes).
But since this Cohomotopy charge corresponds, under the
Pontrjagin-construction,
to the bordism class of a 3-manifold, interpreted in
\cref{MBraneWorldvolumesAndThePontrjaginConstruction} as
the polarization of the worldvolume of probe M2-branes
into M5-branes on 3-manifolds, it ought to be the case that the
refined Adams e-invariant has an equivalent interpretation
as a Cobordism invariant measuring
actual $H_3$-flux through these 3-folds.

\medskip
In the form of a cobordism invariant, an M-theoretic interpretation
of the Adams invariant has been discussed in \cite{Sati14}.
Here we expand on this interpretation,
discussing how it connects to
\hyperlink{HypothesisH}{\it Hypothesis H} and how, in this
context, it encodes $H_3$-flux through 3-spheres
and the Green-Schwarz anomaly
cancellation of 24 5-branes transverse to a K3-surface.

\medskip

\noindent {\bf Extended brane worldvolumes
 and $M\mathrm{U}/\mathbb{S}$-Cobordism.}
In the case that $E = M \!f$ is Cobordism cohomology theory
(Example \ref{ExamplesOfMultiplicativeCohomologyTheories})
the Adams cofiber cohomology theory
$M \!f / \mathbb{S}$ \eqref{CofiberOfRingSpectrumUnitAndBoundaryHomomorphism}
is Cobordism cohomology
for $f$-structured manifolds with {\it framed boundaries}
(\cite[p. 97]{ConnerFloyd66}\cite[p. 102]{Stong68}):
\begin{equation}
  \label{ComplexRelativeCobordismRings}
  \mathclap{
    \raisebox{0pt}{
      \tiny
      \color{darkblue}
      \bf
      \begin{tabular}{c}
        Cobordism ring for
        \\
        $f$-structure with
        framed boundaries
      \end{tabular}
    }
    }
   \qquad \qquad  \quad
   \Omega^{f,\mathrm{Fr}}_\bullet
  \; :=\;
  \big(
    M\!f / \mathbb{S}
  \big)_\bullet
  \,,
  \phantom{AA}
  \mbox{e.g.:}
  \phantom{AA}
  \Omega^{\mathrm{U},\mathrm{Fr}}_\bullet
  \;:=\;
  \big(
    M \mathrm{U} / \mathbb{S}
  \big)_\bullet
  \,,
  \phantom{A}
  \Omega^{S\mathrm{U},\mathrm{Fr}}_\bullet
  \;:=\;
  \big(
    M S\mathrm{U} / \mathbb{S}
  \big)_\bullet
  \,.
\end{equation}
Therefore, Prop. \ref{H3FluxIsQuantizedInAdamsCofiberCohomologyTheory}
here says that $H_3$-fluxes measured in
$M \mathrm{U}$-theory
(\cref{M5ThreeFlux})
on spacetimes $X \,\simeq\, S^{3 + 2d}$
are seen via the universal Toda-bracket observable
$O^{M \mathrm{U}/\mathbb{S}}$
\eqref{MappingHFluxToClassInCofiberCohomology}
as bordism classes of $2d$-dimensional $\mathrm{U}$-manifolds
whose framed {\it boundaries}
are the (polarized) brane worldvolumes from \cref{MBraneWorldvolumesAndThePontrjaginConstruction}:
\vspace{-2mm}
\begin{equation}
  \label{IdentificationOfH3MUFluxesWithMUModSCobordism}
  \xymatrix@C=44pt@R=15pt{
    \mathllap{
      \mbox{
        \tiny
        \color{darkblue}
        \bf
        \begin{tabular}{c}
          $H_3$-fluxes seen in
          \\
          $\mathrm{U}$-Cobordism cohomology
        \end{tabular}
      }
    }
    H^{M \mathrm{U}}_3\mathrm{Fluxes}(X)
    \ar[rr]
      _-{
       O^{M \mathrm{U}/\mathbb{S}}
      }
      ^-{
        \mbox{
          \tiny
          \color{greenii}
          \bf
          universal Toda-bracket observable
        }
        \mathclap{\phantom{\vert_{\vert_{\vert}}}}
      }
    \ar[d]
    &&
    (M \mathrm{U}/\mathbb{S})_{2d}
    \ar[d]^{ \partial }
    \mathrlap{
      \!\!
      \mbox{
        \tiny
        \color{darkblue}
        \bf
        \begin{tabular}{c}
          extended complex
          \\
          brane worldvolumes
        \end{tabular}
      }
    }
    \\
    \mathllap{
      \mbox{
      \tiny
      \color{darkblue}
      \bf
      \begin{tabular}{c}
        underlying $G_4$-flux seen in
        \\
        stable Cohomotopy
      \end{tabular}
      }
    }
    G^{\mathbb{S}}_4\mathrm{Fluxes}(X)
    \ar[rr]
      ^-{ \simeq }
      _-{
        \mathclap{\phantom{\vert^{\vert^{\vert}}}}
        \mbox{
          \tiny
          \color{greenii}
          \bf
          Pontrjagin-Thom isomorphism
        }
      }
    &&
    (M\mathrm{Fr})_{2d\mathrlap{-1}}
    \mathrlap{
      \;\;\;
      \mbox{
        \tiny
        \color{darkblue}
        \bf
        \begin{tabular}{c}
          brane worldvolumes
          \\
          (compact part)
        \end{tabular}
      }
    }
  }
\end{equation}
$$
\vspace{-2mm}
\hspace{-1mm}
\begin{array}{ccccc}
  \Big(
  \big[
    {
      \color{greenii}
      c
    }
  \big],
  \big[
    {
      \color{orangeii}
      H^{M\mathrm{U}}_3\!(c)
    }
  \big]
  \Big)
  &
  \;\;\;
    \overset{
      O^{M S\mathrm{U}/\mathbb{S}}
    }{
      \longmapsto
    }
  \;\;\;
  &
  \raisebox{20pt}{
  \xymatrix@C=1.5em{
    S^{3 + 2d}
    \ar@/^1.7pc/[drr]_-{\ }="s"
    \ar@/_1.7pc/[drr]^-{\ }="t"
    \\
    &&
    \;
    (M\mathrm{U}\!/\mathbb{S})\degree{4}
    \ar@{=>}
      _{ }
      ^-{
        \!\!\!\!
        {\color{purple}M^{2d}_{\mathrm{U},\mathrm{Fr}}}
      }
    "s"; "t"
  }
  }
  &
  \;
  :=
  \;
  &
  \raisebox{56pt}{
  \xymatrix@R=1.2em@C=2.8em{
   S^{3 + 2d }
   \ar[dd]_-{
     \color{greenii}
     c
   }
   \ar[rr]
   \ar@{}[ddrr]|-{
     \rotatebox[origin=c]{-45}{\color{orangeii}$\big\Downarrow$}
     \mbox{\tiny (po)}
   }
   &&
   \ast
   \ar[dd]
   \ar[dr]
   \\
   &&
   &
   \ast
   \ar[dd]|-{
     \mathclap{\phantom{\vert^{\vert}}}
     0
     \mathclap{\phantom{\vert_{\vert}}}
   }
   \\
   S^{4}
   \ar[dd]
   \ar[rr]|-{ \;q_c\; }
   \ar@{}[ddrr]|-{
     \rotatebox[origin=c]{-45}{$\big\Downarrow$}
     \mbox{\tiny (po)}
   }
   \ar@/_.72pc/[drrr]
     |>>>>>>>>>{ \phantom{AA}  }
     |>>>>>{
       \;
       G^{M\mathrm{U}}_{4,\mathrm{unit}}
       \;
     }
   &&
   C_c
   \ar[dd]
   \ar@{-->}[dr]^-{
     \vdash
     {\color{orangeii}
       H^{M\mathrm{U}}_3\!(c)
     }
   }
   \\
   && &
   M\mathrm{U}\degree{4}
   \ar[dd]
   \\
   \ast
   \ar@/_.7pc/[drrr]|-{ \;0\; }
   \ar[rr]
   &&
   S^{4 + 2d}
   \ar@{-->}[dr]^-{
     \mathclap{\phantom{\vert^{\vert}}}
     \;
     \vdash
     {\color{purple}
       M^{2d}_{\mathrm{U},\mathrm{Fr}}
     }
     \mathclap{\phantom{\vert_{\vert}}}
   }
   &&
   \\
   && &
   (\mathrm{MU}\!/\mathbb{S})\degree{4}
   \,.
  }
  }
  \end{array}
$$

\medskip

\noindent
{\bf KK-Compactification on K3 and $S\mathrm{U}$-Cobordism.}
The analogous statement \eqref{IdentificationOfH3MUFluxesWithMUModSCobordism} holds for
measurement of M-brane charge in $E = M S\mathrm{U}$.
Specifically, near M2-brane horizons $X \simeq S^7$ \eqref{M2BranesPTTheorem} the $O^{M S\mathrm{U}/\mathbb{S}}$-observable
sees the $H^{M S\mathrm{U}}$-flux embodied as
(the cobordism class of)
a 4d $S\mathrm{U}$-manifold with 3d $\mathrm{Fr}$-boundaries,
which by the discussion on p. \pageref{24BraneCancellationInWorldvolumeSection}
we may dually understand as
Kaluza-Klein-compactification fibers punctured
by transversal heterotic 5-branes:
\begin{equation}
  \label{IdentificationOfH3MSUFluxesWithMSUModSCobordism}
  \xymatrix@C=44pt@R=15pt{
    \mathllap{
      \mbox{
        \tiny
        \color{darkblue}
        \bf
        \begin{tabular}{c}
          $H_3$-fluxes seen in
          \\
          $S\mathrm{U}$-Cobordism cohomology
          \\
          near black M2-branes
        \end{tabular}
      }
    }
    H^{M S\mathrm{U}}_3\mathrm{Fluxes}(S^7)
    \ar[rr]
      _-{
       O^{M S\mathrm{U}}
      }
      ^-{
        \mbox{
          \tiny
          \color{greenii}
          \bf
          universal Toda-bracket observable
        }
        \mathclap{\phantom{\vert_{\vert_{\vert}}}}
      }
    \ar[d]
    &&
    (M S\mathrm{U}/\mathbb{S})_{2d}
    \ar[d]^{ \partial }
    \mathrlap{
      \!\!
      \mbox{
        \tiny
        \color{darkblue}
        \bf
        \begin{tabular}{c}
          KK-compactification space
          \\
          punctured by
          \\
          transversal 5-branes
        \end{tabular}
      }
    }
    \\
    \mathllap{
      \mbox{
      \tiny
      \color{darkblue}
      \bf
      \begin{tabular}{c}
        underlying $G_4$-flux see in
        \\
        stable Cohomotopy
      \end{tabular}
      }
    }
    G^{\mathbb{S}}_4\mathrm{Fluxes}(S^7)
    \ar[rr]
      ^-{ \simeq }
      _-{
        \mathclap{\phantom{\vert^{\vert^{\vert}}}}
        \mbox{
          \tiny
          \color{greenii}
          \bf
          Pontrjagin-Thom isomorphism
        }
      }
    &&
    (M\mathrm{Fr})_{3}
    \mathrlap{
      \;\;\;\;
      \mbox{
        \tiny
        \color{darkblue}
        \bf
        \begin{tabular}{c}
          3-spheres around
          \\
          transversal 5-branes
        \end{tabular}
      }
    }
  }
\end{equation}
But now the
abelian group $(M S\mathrm{U}/\mathbb{S})_4$
is an extension of $(M \mathrm{Fr})_3$ by
$(M S\mathrm{U})_4$ (Prop. \ref{CofiberECohomologyAsExtensionOfStableCohomotopyByECohomology})
and  the non-torsion elements in $(M S\mathrm{U})_\bullet$
are generated by Calabi-Yau manifolds (Prop. \ref{NonTorsionSUCobordismRingGeneratedByCYs}).
Specifically, $(M S\mathrm{U})_4$ is spanned
by (the cobordism class of) the
$\mathrm{K3}$-surface,
regarded as an $S\mathrm{U}$-manifold
(Prop. \ref{K3SpansSUCobordismInDegree4}).
This then means that
the observable $O^{M S\mathrm{U}/\mathbb{S}}$ \eqref{MappingHFluxToClassInCofiberCohomology}
sees the cancellation of the stable Cohomotopy charge of
24 probe M-branes near black M2 \eqref{CompactificationOnY4}
witnessed by the appearance of a transversal $\mathrm{K3}$-surface,
as before in plain Cohomotopy (p. \pageref{CompactificationOnY4})
but now seen with its $S\mathrm{U}$-structure:

\vspace{-.7cm}

\begin{equation}
  \label{K3CompactificationWithItsSUStructureAppears}
  \hspace{-7mm}
  \raisebox{44pt}{
  \xymatrix@R=-8pt@C=3.8em{
    &
    &&
    \overset{
      \mathclap{
      \raisebox{6pt}{
        \tiny
        \color{darkblue}
        \bf
        \begin{tabular}{c}
          $H_3$-fluxes near black M2s
          \\
          measured in $M S\mathrm{U}$-cohomology
        \end{tabular}
      }
      }
    }{
      H^{M S\mathrm{U}}_3\mathrm{Fluxes}(S^7)
    }
    \ar[d]
      _-{ O^{M S\mathrm{U}/\mathbb{S}} }
      ^-{
        \!\!\!\!\!\!\!\!\!\!
        \mbox{
          \tiny
          \color{greenii}
          \bf
          \begin{tabular}{c}
            Toda observable
            \\
            {\color{black} \eqref{MappingHFluxToClassInCofiberCohomology}}
          \end{tabular}
        }
      }
    \ar[rr]
      ^-{
        \mbox{
          \tiny
          \color{greenii}
          \bf
          \begin{tabular}{c}
            underlying $G_4$-flux
            \\
            in stable Cohomotopy
          \end{tabular}
        }
      }
    &&
    G^{\mathbb{S}}_4\mathrm{Fluxes}(S^7)
    \ar@{=}[d]
      ^-{
        \mbox{
          \tiny
          \color{greenii}
          \bf
          PT-isomorphism
        }
      }
    \\
    0
    \ar[r]
    &
    \overset{
      \mathclap{
      \raisebox{9pt}{
        \tiny
        \color{darkblue}
        \bf
        \begin{tabular}{c}
          cobordism classes of
          \\
          closed 4d $S \mathrm{U}$-manifolds
        \end{tabular}
      }
      }
    }{
      (M S\mathrm{U})_4
    }
    \;
    \ar@{^{(}->}[rr]^-{ i }
    \ar@{=}[d]
      _-{
        \mbox{
          \tiny
          \color{greenii}
          \bf
          \begin{tabular}{c}
            $(M S\mathrm{U})_\bullet$ is generated
            \\
            by Calabi-Yau manifolds
            \\
            {\color{black}
              Prop.
              \ref{NonTorsionSUCobordismRingGeneratedByCYs},
              \ref{K3SpansSUCobordismInDegree4}
            }
          \end{tabular}
        }
        \!\!\!\!\!\!
      }
    &&
    \underset{
      \mathclap{
      \raisebox{-9pt}{
        \tiny
        \color{darkblue}
        \bf
        \begin{tabular}{c}
          cobordism classes of
          \\
          compact 4d $S\mathrm{U}$-manifolds
          \\
          with $\mathrm{Fr}$-boundaries
        \end{tabular}
      }
      }
    }{
      (M S\mathrm{U}/\mathbb{S})_4
    }
    \ar@{->>}[rr]^-{
      \partial
    }
    _-{
      \mbox{
        \tiny
        \color{greenii}
        \bf
        boundary map
      }
    }
    &&
    (M \mathrm{Fr})_3
    \ar[r]
    \ar@{=}[d]
    &
    0
    \\
    &
    \underset{
      \mathclap{
      \raisebox{-6pt}{
        \tiny
        \color{darkblue}
        \bf
        \begin{tabular}{c}
          copies of
          \\
          the $\mathrm{K3}$-surface
        \end{tabular}
      }
      }
    }{
      \mathbb{Z}
      \big\langle
        [\mathrm{K3}]
      \big\rangle
    }
    \;
    \ar@{^{(}->}[rr]
    &&
    \underset{
    \mathclap{
    \raisebox{-6pt}{
      \tiny
      \color{purple}
      \bf
      \begin{tabular}{c}
        the cancellation of 24 units of M-brane charge
        \\
        is seen by $O^{M S\mathrm{U}/\mathbb{S}}$ as the appearance
        \\
        of the $\mathrm{K3}$-surface, with its $S\mathrm{U}$-structure
      \end{tabular}
    }
    }
    }{
      \mathbb{Z}
      \big\langle
        [\mathrm{K3}]
      \big\rangle
      \,\rtimes\,
      \mathbb{Z}_{24}
    }
    \ar@{->>}[rr]
    &&
    \underset{
      \mathclap{
      \raisebox{-6pt}{
        \tiny
        \color{darkblue}
        \bf
        \begin{tabular}{c}
          in bulk around black M2-branes
          \\
          charge of 24 probe branes cancels out
        \end{tabular}
      }
      }
    }{
      \mathbb{Z}_{24}
    }
  }
  }
\end{equation}

\medskip
\noindent
{\bf D-brane charge and the Conner-Floyd e-invariant.}
Given a Kaluza-Klein-compactification on a complex curve
$M^{4}_{S\mathrm{U}}$ with vanishing first Chern class,
such as a K3-surface
\eqref{K3CompactificationWithItsSUStructureAppears},
the traditional expression for its intrinsic D-brane charge,
given (\cite[(1.1)]{MinasianMoore97}\cite[Cor. 8.5]{BMRS08})
as evaluation of the
{\it square root of the Todd class} \eqref{SquareRootOfToddClassOnSU6Folds}
on the fundamental class $[M^4]$,
reduces \eqref{ToddNumberOfSUCurve}
to half the {\it Todd number} (Example \ref{SquareRootOfToddClassOfSUManifolds}):
\begin{equation}
  \label{DBraneChargeOfComplexSUCurve}
  \mathllap{
    \mbox{
      \tiny
      \color{darkblue}
      \bf
      \begin{tabular}{c}
        intrinsic D-brane charge
        \\
        of spacetime on $M^4_{S\mathrm{U}}$
      \end{tabular}
    }
  }
  \sqrt{\hat A}\; [M^{4}_{S\mathrm{U}}]
  \overset{
    \mathclap{
    \raisebox{3pt}{
      \tiny
      \color{greenii}
      \bf
      $S\mathrm{U}$-structure
    }
    }
  }{
    \;\;\;\;\;
      =
    \;\;\;\;\;
  }
  \sqrt{\mathrm{Td}}\; [M^{4}_{S\mathrm{U}}]
  \overset{
    \mathclap{
    \raisebox{3pt}{
      \tiny
      \color{greenii}
      \bf
      $\mathrm{dim}_{\mathbb{R}} = 4$
    }
    }
  }{
  \;\;\;\;\;
    =
  \;\;\;\;\;
  }
  \tfrac{1}{2} \mathrm{Td}\; [M^{4}_{S\mathrm{U}}]
  \mathrlap{
    \mbox{
      \tiny
      \color{darkblue}
      \bf
      half the Todd number
    }
  }
  \end{equation}
But the Todd character on $S\mathrm{U}$-Cobordism
is (Prop. \ref{ToddCharaxcterOnSUManifoldsIsPontrjaginCharacterOnKOThomClass})
a multiplicative cohomology operation,
equal to the composite of the Conner-Floyd $K \mathrm{O}$-orientation
(Prop. \ref{ConnerFloydKOrientation})
with the Pontrjagin character. This implies (Prop. \ref{CharactersOnAdamsCofiberTheories}) that
the intrinsic D-brane charge \eqref{DBraneChargeOfComplexSUCurve}
is a Toda-bracket observable \eqref{TodaBracketObservableOnH3Flux}
on $H^{M S\mathrm{U}}_3$-fluxes, that comes (via Remark \ref{ObservablesOnH3FromMultiplicativeCohomologyOperations})
from the universal $M S\mathrm{U}$-observable \eqref{IdentificationOfH3MSUFluxesWithMSUModSCobordism}:

\vspace{-.7cm}

$$
  \xymatrix@C=32pt{
    \overset{
      \mathclap{
      \raisebox{6pt}{
        \tiny
        \color{darkblue}
        \bf
        \begin{tabular}{c}
          $H_3$-fluxes near black M2s
          \\
          measured in $M S\mathrm{U}$-theory
        \end{tabular}
      }
      }
    }{
      H^{M S\mathrm{U}}_3\mathrm{Fluxes}
      (
        S^7
      )
    }
    \ar[rr]
      |-{
        \;
        O^{M S\mathrm{U}/\mathbb{S}}
        \;
      }
      ^-{
        \mbox{
          \tiny
          \color{greenii}
          \bf
          \begin{tabular}{c}
            universally observed as an
            \\
            $S\mathrm{U}$-manifold with $\mathrm{Fr}$-bdry
            \\
            {\phantom{a}}
          \end{tabular}
        }
      }
    \ar@/_1.2pc/[rrrr]
      |-{
        \;
        O^{K \mathrm{O}/\mathbb{S}}
        \;
      }
      |>>>>>>>>>>>>>>>>>{
        \colorbox{white}{
          \hspace{-10pt}
          \tiny
          \color{greenii}
          \bf
          \begin{tabular}{c}
            observed as
            cofiber K-theory class
          \end{tabular}
          \hspace{-10pt}
        }
      }
    \ar@/_2pc/@<-4pt>[rrrrrrr]
      |-{
        \;
        {\widehat e}_{K\mathrm{U}}
        \,=\,
        {\widehat e}_{K\mathrm{O}}
        \;
      }
      _-{
        \mathclap{\phantom{\vert^{\vert}}}
        \mbox{
          \tiny
          \color{greenii}
          \bf
          refined e-invariant
        }
      }
    &&
    (M S\mathrm{U} / \mathbb{S})_4
    \ar[rr]
      |-{
        \;
        \sigma_{S\mathrm{U}}/\mathbb{S}
        \;
      }
    \ar@/^1.2pc/[rrrr]
      |-{
        \;
        \mathrm{Td}/\mathbb{S}
        \;
      }
    \ar@/^1.8pc/@<+5pt>[rrrrr]
      |-{
        \;
        \mathrm{Td}/\mathbb{S}
        \;
      }
      ^-{
        \mbox{
          \tiny
          \color{greenii}
          \bf
          Todd class of cplx mfds with frmd bdry
        }
        \mathclap{\phantom{\vert_{\vert}}}
      }
    &&
    (K \mathrm{O} / \mathbb{S})_4
    \ar[rr]
      |-{
        \;
        \mathrm{ph}/\mathbb{S}
        \;
      }
    &
    &
    \big(
      (H^{\mathrm{ev}}\mathbb{Q})/\mathbb{S}
    \big)_4
    \ar[r]
      |-{
        \;\mathrm{spl}_0\;
      }
    &
    \mathbb{Q}
  }
$$
\begin{equation}
  \label{ReproducingTheConnerFloydEInvariant}
\end{equation}

\vspace{-1.3cm}
$$
\hspace{-1mm}
  \big(
    {
    \color{greenii}
      G^{\mathbb{S}}_4(c)
    },
    {\color{orangeii}
      H^{M S\mathrm{U}}_3(c)
    }
  \big)
  \;
  \mapsto
  \!\!\!\!\!\!\!\!\!\!
  \raisebox{20pt}{
  \xymatrix@C=.6em{
    S^7
    \ar@/^1.7pc/[drr]_-{\ }="s"
    \ar@/_1.7pc/[drr]^-{\ }="t"
    \\
    &&
        \big(
      (H^{\mathrm{ev}}\mathbb{Q})/\mathbb{S}
    \big)^8
    \ar@{=>}
      ^-{
        \!\!\!\!\!
        \mathrm{Td}
        [
          {\color{purple} M^{4}_{S\mathrm{U},\mathrm{Fr}}}
        ]
      }
    "s"; "t"
  }
  }
  \!\!\!\!\!\!
  :=
  \!\!\!\!\!\!
  \raisebox{86pt}{
  \xymatrix@R=1.2em@C=1.8em{
   S^7
   \ar[dd]
     _-{
       \color{greenii}
       c
     }
   \ar[rr]
   \ar@{}[ddrr]|-{
     \rotatebox[origin=c]{-45}{\color{orangeii}$\big\Downarrow$}
     \mbox{\tiny (po)}
   }
   &&
   \ast
   \ar[dr]
   \ar[dd]
   \\
   && &
   \ast
   \ar[dd]^-{
         0
   }
   \\
   S^4
   \ar[dd]
   \ar[rr]
   \ar@{}[ddrr]|-{
     \rotatebox[origin=c]{-45}{$\big\Downarrow$}
     \mbox{\tiny (po)}
   }
   \ar@/_.7pc/[drrr]
     |<<<<<{ \;\Sigma^4 (1^{M\mathrm{U}})\;\; }
     |>>>>>>>>>{ {\phantom{AA}} \atop {\phantom{AA}} }
   &&
   C_c
   \ar[dd]
   \ar@{-->}[dr]^-{
     \vdash
     {\color{orangeii}
       H^{M S\mathrm{U}}_3(c)
     }
   }
   &&
   \\
   && &
   (M S\mathrm{U})^4
   \ar[dd]
   \ar[dr]_-{
     \;\sigma_{S\mathrm{U}}\;
   }
   \ar@/^1.7pc/[ddrr]^>>>>>>>>>>>{
     \;
     \mathclap{\phantom{\vert^{\vert}}}
     \mathrm{Td}
     \mathclap{\phantom{\vert_{\vert}}}
     \;
   }
   \\
   \ast
   \ar@/_.7pc/[drrr]|-{ \;0\; }
   \ar[rr]
   &&
   S^{2(n+d)}
   \ar@{-->}[dr]^-{
     \;
     \mathclap{\phantom{\vert^{\vert}}}
     \color{purple}M^{4}_{S\mathrm{U},\mathrm{Fr}}
     \mathclap{\phantom{\vert_{\vert}}}
   }
   &&
   {K\mathrm{O}}^4
   \ar[dr]_-{
     \;
     \mathrm{ph}
     \;
     \mathclap{\phantom{\vert_{\vert}}}
   }
   \ar[dd]|>>>>>>{ \phantom{\vert_{\vert}} }
   \\
   && &
   (M S\mathrm{U}/\mathbb{S})^4
   \ar@/^1.8pc/[ddrr]
     |>>>>>>>>>{
       \mathclap{\phantom{\vert^{\vert}}}
       \;\mathrm{Td}/\mathbb{S}\;
       \mathclap{\phantom{\vert_{\vert}}}
     }
   \ar[dr]_-{ \;\sigma_{\mathrm{U}}/\mathbb{S}\; }
   &&
   (H^{\mathrm{ev}}\mathbb{Q})^4
   \ar[dd]
   \\
   && &&
   ({K\mathrm{O}}/\mathbb{S})^4
   \ar[dr]_-{ \;\mathrm{ph}/\mathbb{S}\; }
   \\
   && && &
   \big(
     (H^{\mathrm{ev}}\mathbb{Q})/\mathbb{S}
   \big)^4
  }
  }
$$

\noindent
Here the top composite operation in \eqref{ReproducingTheConnerFloydEInvariant}
\begin{equation}
  \label{ConnerFloydeInvariantInDictionary}
  \xymatrix@R=1.5em{
    [H^{M S\mathrm{U}}_3(c)]
      \ar@{|->}[d]
      \ar@{|->}[r]
      &
    [M^{4}_{S\mathrm{U}, \mathrm{Fr}}]
      \ar@{|->}[r]
      &
    \mathrm{Td}[M^{4}_{S\mathrm{U}, \mathrm{Fr}}]
      \ar@{}[r]
        |-{ \in }
      \ar@{|->}[d]
      &
    \mathbb{Q}
    \ar[d]
    \\
    [c]
    \ar@{|->}[rr]
      ^-{
        \mathrm{e}_{\mathrm{CF}}
      }
      _-{
        \mbox{
          \tiny
          \color{greenii}
          \bf
          Conner-Floyd e-invariant
        }
      }
    &&
    \mathrm{Td}[M^{4}_{S\mathrm{U}, \mathrm{Fr}}]
    &
    \mathbb{Q}/\mathbb{Z}
  }
\end{equation}
is the construction of
Conner-Floyd's
{\it geometric} {\it cobordism-theoretic}
e-invariant \eqref{GeometricExpressionOfTheeInvariant}.
Our Toda-bracket-theoretic
construction \eqref{ReproducingTheConnerFloydEInvariant}
makes manifest
(with a glance at Prop. \ref{OnCofiberTheoryToddCharacterRestrictionToMSU})
that this is {\it equal} to the
refined $\widehat e_{K \mathrm{U}}$-invariant
\eqref{DiagrammatichatecInvariant}
on the induced $H^{K\mathrm{U}}_{3}$-fluxes,
as shown by the commuting diagram at the top of
\eqref{ReproducingTheConnerFloydEInvariant}.
But, by Prop. \ref{ChargeLatticeOfHKU3Flux}, this implies
that modulo integers both coincide with the
Adams e-invariant;
which is Conner-Floyd's classical theorem
(Prop. \ref{TheConnerFloydeInvariant}):
\begin{equation}
  \label{ConnerFloydeInvariantEqualsAdamseInvariantInDictionary}
  \raisebox{20pt}{
  \xymatrix@C=2pt@R=.6em{
    \overset{
      \mathclap{
      \raisebox{3pt}{
        \tiny
        \color{darkblue}
        \bf
        \begin{tabular}{c}
          universal Toda-bracket observable
          \\
          on $H_3$-flux measured in $M S\mathrm{U}$
        \end{tabular}
      }
      }
    }{
    {\widehat e}_{K\mathrm{U}}
    \big(
      H^{M S\mathrm{U}}_3(c)
    \big)
    }
    \quad
    \ar@{|->}[d]
    \ar@{=}[rr]
    &
    {\phantom{AA}}
    &
    \overset{
      \raisebox{3pt}{
        \mbox{
          \tiny
          \color{purple}
          \bf
          \begin{tabular}{c}
            actual $H_3$-flux
            \\
            through punctures
            \rlap{
              \color{black}
              \eqref{RationalToddNumberByGreenSchwarzMechanism}
            }
          \end{tabular}
        }
      }
    }{
      \mathrm{Td}
      \big[M^4_{S\mathrm{U},\mathrm{Fr}}\big]
    }
    \ar@{|->}[d]
    \ar@{}[rr]
      |-{\in}
    &\;\;&
    \mathbb{Q}
    \ar[d]
    \\
    \underset{
      \mathclap{
      \raisebox{-3pt}{
        \tiny
        \color{darkblue}
        \bf
        \begin{tabular}{c}
          Adams
          \\
          e-invariant
        \end{tabular}
      }
      }
    }{
      {\mathrm{e}_{\mathrm{Ad}}}(c)
    }
    \ar@{=}[rr]
    &&
    \underset{
      \mathclap{
      \raisebox{-3pt}{
        \tiny
        \color{darkblue}
        \bf
        \begin{tabular}{c}
          Conner-Floyd
          \\
          e-invariant
        \end{tabular}
      }
      }
    }{
      {\mathrm{e}_{\mathrm{CF}}}
      \big(
        \partial M^4_{S\mathrm{U}, \mathrm{Fr}}
      \big)
    }
    \ar@{}[rr]
      |-{\in}
    &\;\;&
    \mathbb{Q}/\mathbb{Z}
  }
  }
\end{equation}

\medskip

\noindent
{\bf Green-Schwarz mechanism for $\mathrm{NS}5$-Branes and Twisted 3-Cohomotopy.}
The identification \eqref{ConnerFloydeInvariantEqualsAdamseInvariantInDictionary}
reveals diffe-rential-geometric content underlying
the abstract assignment $\widehat e_{K\mathrm{U}}$
of rational charges to  $H^{K \mathrm{U}}_3$-fluxes
(from Prop. \ref{ChargeLatticeOfHKU3Flux}):
In the case that $M^{4}_{S\mathrm{U},\mathrm{Fr}}$ is a
a punctured KK-compactification space \eqref{IdentificationOfH3MSUFluxesWithMSUModSCobordism}
with non-empty framed boundary being a disjoint union of 3-spheres around
transversal 5-branes (p. \pageref{24BraneCancellationInWorldvolumeSection}),
it carries
(by Prop. \ref{RationalToddNumberAsBoundaryIntegral})
a differential 3-form $H_3$ whose
closure is twisted by the Euler-form
and whose boundary integral
equals $\tfrac{1}{12}$ of the the rational Todd number \eqref{ConnerFloydeInvariantEqualsAdamseInvariantInDictionary}
\vspace{-2mm}
\begin{equation}
  \label{RationalToddNumberByGreenSchwarzMechanism}
  \underset{
    \mathclap{
    \raisebox{-3pt}{
      \tiny
      \color{darkblue}
      \bf
      \begin{tabular}{c}
        KK-compactification
        \\
        space
      \end{tabular}
    }
    }
  }{
    M^4_{S\mathrm{U},\mathrm{Fr}}
  }
  \,,
  \quad
  \mbox{s.t.}
  \quad
  \partial
  M^4_{S\mathrm{U}}
  \;=
  \!\!
  \underset{
    \mathclap{
    \raisebox{-3pt}{
      \tiny
      \color{darkblue}
      \bf
      \begin{tabular}{c}
        spheres around
        \\
        transversal 5-branes
      \end{tabular}
    }
    }
  }{
    \underset{
      1 \leq k \leq n
    }{\sqcup} S_k^3
  }
  \;\;\;\;\;\;
    \Rightarrow
  \;\;\;\;\;\;
   \exists
   \,
   \underset{
     \mathclap{
     \raisebox{-3pt}{
       \tiny
       \color{darkblue}
       \bf
       \begin{tabular}{c}
         ordinary NS
         \\
         3-form flux
       \end{tabular}
     }
     }
   }{
      H_3
   }
   \in \Omega^3_{\mathrm{dR}}(M^4_{S\mathrm{U},\mathrm{Fr}})
    \,,
    \;\;\;\;\;
    \mbox{s.t.}
    \;\;
    \left\{
      \mathclap{\phantom{
        \begin{array}{l}
          A
          \\
          A
          \\
          A
        \end{array}
      }}
    \right.
    \begin{array}{l}
    \overset{
      \mathclap{
      \raisebox{+3pt}{
        \tiny
        \color{darkblue}
        \bf
        \begin{tabular}{c}
          Green-Schwarz Bianchi identity
          \\
          (for vanishing gauge flux)
        \end{tabular}
      }
      }
    }{
      d\,H_3 = \rchi_4(\nabla^{T M^4})
      \,,
    }
    \\
    \mathrm{Td}[M^4_{\mathrm{U}, \mathrm{Fr}}]
    \;=\;
    \tfrac{1}{12}
    \underset{1 \leq k \leq n}{\sum}
    \;
    \underset{
      \mathclap{
      \raisebox{-3pt}{
        \tiny
        \color{darkblue}
        \bf
        \begin{tabular}{c}
          ordinary charge of $k$th stack
          \\
          of transversal 5-branes
        \end{tabular}
      }
      }
    }{
      \underbrace{
        \int_{S^3_k} H_3
      }
    }\;.
    \end{array}
  \end{equation}
Notice that the proof of this implication,
in Prop. \ref{RationalToddNumberAsBoundaryIntegral} below,
proceeds (via Lemma \ref{RealH3FluxOnK3AwayFrom24BraneInsertionsExists})
by regarding $H_3$ as the Cohomotopical character of a
cocycle in J-twisted 3-Cohomotopy on $M^4$; which is
made possible by the Poincar{\'e}-Hopf theorem
(just as discussed for $H_7$-flux in \cite[\S 2.5]{FSS19b}).
This is in line with
the emergence,
from \hyperlink{HypothesisH}{\it Hypothesis H},
of
$H_3$-flux quantization in twisted 3-Cohomotopy
seen in \cite{FSS20a}\cite{SS20c}.

\vspace{2mm}
\noindent
In conclusion, so far:

\vspace{2mm}

\noindent
\hspace{.04cm}
\fbox{
\begin{minipage}[left]{16.8cm}
{\it
  When the Cohomotopy charge \eqref{FromUnstableToStable4CohomotopyOf7Sphere}
  near black M2-branes \eqref{M2BranesPTTheorem}
  is a multiple of 12 (in particular a multiple of 24),
  then the integer values of the $H^{K\mathrm{U}}_3$-flux
  that are
  abstractly observed by the ${\widehat e}_{K\mathrm{U}}$-invariant
  (Prop. \ref{ChargeLatticeOfHKU3Flux})
  equals ordinary $H^{H\mathbb{R}}_3$-flux
  (i.e. periods of a differential 3-form $H_3$)
  through the 3-spheres around
  transversal brane insertion points in the
  complex surface on
  which spacetime has spontaneously compactified
  via \eqref{CompactificationOnY4}.
}
\end{minipage}
}

\medskip

Specifically, when $M^4_{S\mathrm{U},\mathrm{Fr}}$
in \eqref{RationalToddNumberByGreenSchwarzMechanism}
is a punctured K3-surface \eqref{CompactificationOnY4},
now regarded with its $S\mathrm{U}$-structure, then
this boundary integral \eqref{RationalToddNumberByGreenSchwarzMechanism}
is (still by Prop. \ref{RationalToddNumberAsBoundaryIntegral}) the
Euler number of $\mathrm{K3}$:
\begin{equation}
  \label{BraneCharge24OnK3FromGreenSchwarzMechanism}
  M^4_{S\mathrm{U},\mathrm{Fr}}
    \;=\;
  \mathrm{K3}
    \setminus
   \underset{
     1 \leq k \leq n
   }{\sqcup}
  D_k^4
  \;\;\;\;\;
    \Rightarrow
  \;\;\;\;\;
  \underset{
    \mathclap{
    \raisebox{-3pt}{
      \tiny
      \color{darkblue}
      \bf
      \begin{tabular}{c}
        total charge of
        \\
        transversal 5-branes
      \end{tabular}
    }
    }
  }{
  \underbrace{
    \underset{1 \leq k \leq n}{\sum}
    \int_{S^3_k} H^3
  }
  }
  \;=\;
  24
  \,.
\end{equation}
\noindent

However, by comparison to the traditional string theory lore
(e.g. \cite[p. 50]{Schwarz97}), we see that:

\vspace{-.2cm}
\begin{itemize}

\vspace{-.2cm}
\item
\eqref{RationalToddNumberByGreenSchwarzMechanism}
is the Green-Schwarz mechanism for
the ordinary 3-flux $H_3$ of  heterotic NS5-brane sources
(as on p. \ref{24BraneCancellationInWorldvolumeSection})
transversal to the compactification space
(here in the case of vanishing gauge flux);

\vspace{-.2cm}
\item
\eqref{BraneCharge24OnK3FromGreenSchwarzMechanism} is
Green-Schwarz mechanism specified to K3-compactifications,
implying that transversal heterotic NS5-branes must
appear in multiples of 24 in order to cancel to
total flux in the compactification space
(heterotic tadpole cancellation).

\end{itemize}
\vspace{-.2cm}

\noindent
In further conclusion, we find the following correspondence,
under \hyperlink{HypothesisH}{\it Hypothesis H},
between
{(a)}
the Green-Schwarz mechanism for heterotic NS5-branes
transversal to K3-compactifications
and
{(b)}
the construction and properties of the
Conner-Floyd e-invariant:

\medskip

\hspace{-.6cm}
\begin{tabular}{|p{13em}|c|p{13.5em}||p{1.5em}|}
  \hline
  \begin{minipage}[center]{5.2cm}
    \bf
    $\mathclap{\phantom{\vert}}$
    GS anomaly cancellation in
    \\
    $\mathclap{\phantom{\vert_{\vert}}}$
    KK-compactification on K3
  \end{minipage}
  &
  \begin{minipage}[center]{4.6cm}
    \begin{center}
    $\longleftrightarrow$
    \end{center}
  \end{minipage}
  &
  \begin{minipage}[center]{5.2cm}
    \bf
    $\mathclap{\phantom{\vert^{\vert}}}$
    Vanishing of
    $\scalebox{.7}{$\tfrac{1}{2}$}\mathrm{e}_{\scalebox{.57}{$\mathrm{CF}$}}$-invariant
    on
    \\
    $\mathclap{\phantom{\vert_{\vert}}}$
    Cohomotopy charge
    $\xymatrix@C=18pt{S^7\!\! \ar[r]^<<<{\;c\;} & S^4}$
  \end{minipage}
  &
  \\
  \hline
  \hline
  \begin{minipage}[center]{5.2cm}
    $\mathclap{\phantom{\vert^{\vert}}}$
    Asymptotic boundary of $n$
    \\
    $\mathclap{\phantom{\vert_{\vert}}}$
    $p$-branes of $\mathrm{codim} = 4$
    \\
    $\mathclap{\phantom{\vert_{\vert}}}$
    carrying charges $\big\{Q_k \in \mathbb{n}\big\}_{k = 1}^n$
  \end{minipage}
  &
  \begin{minipage}[center]{4.6cm}
    \begin{center}
    $
      \mathbb{R}^{p,1}
      \times
      \Big(
        \underset{n}{\sqcup}
        \;
        S^3
      \Big)
    $
    \end{center}
  \end{minipage}
  &
  \begin{minipage}[center]{5.2cm}
    $\mathclap{\phantom{\vert^{\vert}}}$
    Class in framed Cobordism
    $
      \xymatrix@R=2pt{
        \left[
          \underoverset{k=1}{n}{\sqcup}
          S^3_{{}_{\!\!\mathrm{nfr} = Q_k}}
        \right]
        \ar@{}[d]|<{
          \scalebox{.8}{
          \rotatebox{-90}{
            $\in$
          }
          }
        }
          \ar@{}[r]|-{ \mapsto }
        &
        \big[
          S^7
            \xrightarrow{
             \scalebox{.6}{$\underoverset{i=k}{N}{\sum}
              Q_k
              \cdot h_{\mathbb{H}}
           $} }
          S^4
        \big]
        \ar@{}[d]|<<{
          \scalebox{.8}{
          \rotatebox{-90}{
            $\in$
          }
          }
        }
        \\
        (M\mathrm{Fr})_3
        \ar[r]^-{
          \raisebox{3pt}{\scalebox{.7}{$\mathrm{Src}$}}
        }_-{ \;\raisebox{2.5pt}{\scalebox{.7}{$\simeq$}}\; }
        &
        \mathbb{S}_3
      }
    $
  \end{minipage}
  &
  \begin{minipage}[left]{.7cm}
    \hspace{-.2cm}
    \cref{MBraneWorldvolumesAndThePontrjaginConstruction}
  \end{minipage}
  \\
  \hline
  \begin{minipage}[center]{5.2cm}
    $\mathclap{\phantom{\vert^{\vert}}}$
    Spacetime compactified on
    \\
    $\mathclap{\phantom{.}}$
    generalized CY manifold $Y^4$
    \\
    $\mathclap{\phantom{\vert_{\vert}}}$
    around $n$ branes of $\mathrm{codim}=4$
  \end{minipage}
  &
  \begin{minipage}[center]{4.6cm}
    \begin{center}
    $
      \mathbb{R}^{p,1}
      \times
      \Big(
        Y^4
        \setminus
        \underset{n}{\sqcup} \, D^4
      \Big)
    $
    \end{center}
  \end{minipage}
  &
  \begin{minipage}[center]{5.2cm}
    $\mathclap{\phantom{\vert^{\vert^{\vert^{\vert}}}}}$
    $
    \xymatrix@R=4pt@C=7pt{
      \mbox{Lift to }
      \;\;\;
      \big[
        Y^4
        \setminus
        \underset{n}{\sqcup} \, D^4
      \big]
      \ar@{}[r]|-{ \in }
      &
      (M S\mathrm{U}/\mathbb{S})_4
      \ar@<-10pt>[d]^-{ \partial }
      \\
      \mbox{of}
      \;\;
      \left[
        \underoverset{i = 1}{N}{\sqcup} S^3_{{}_{\!\!\mathrm{nfr}=Q_i}}
      \right]
      \in
      \mathbb{S}_3
      \ar@{}[r]|-{\simeq}
      &
      (M\mathrm{Fr})_3
      }
    $
  \end{minipage}
  &
  \multirow{2}{*}{
    \hspace{-12pt}
    \begin{tabular}{c}
    \phantom{A}
    \\
    \phantom{A}
    \\
    \eqref{ConnerFloydeInvariantEqualsAdamseInvariantInDictionary}
    \end{tabular}
  }
  \\
  \cline{1-3}
  \begin{minipage}[center]{5.2cm}
    D-brane charge
    \\
    in compact space
  \end{minipage}
  &
  \begin{minipage}[center]{4.8cm}
    \begin{center}
    $
    \begin{aligned}
      \mathclap{\phantom{\int^{\vert}}}
      \sqrt{
        \mathrm{Td}
      }
      \big[
        Y^4
          \setminus
        \underset{N}{\sqcup}
      \,
      D^4
      \big]
     &
     \mathclap{\phantom{\int}}
     =
      \tfrac{1}{2} \!\!\!
      \underset{\scalebox{.5}{$
        Y^4
          \setminus
        \underset{N}{\sqcup}
        \,
        D^4
      $}}
      {\int}
      \!\!\!
      \tfrac{1}{12}
      c_2(\nabla)
      \\
      &
      \;=\;
      \mathclap{\phantom{\vert_{\vert}}}
      \tfrac{1}{24}
      \underoverset{i=1}{N}{\sum}
      Q_i
      \mathclap{\phantom{\vert_{\vert_{\vert_{\vert_{\vert_{\vert_{\vert}}}}}}}}
    \end{aligned}
    $
    \end{center}
  \end{minipage}
  &
  \begin{minipage}[center]{5.2cm}
    Conner-Floyd bordism formula
    \\
    for
    $\scalebox{.7}{$\tfrac{1}{2}$}\mathrm{e}_{\scalebox{.57}{$\mathrm{Ad}$}}$-invariant
  \end{minipage}
  &
  \\
  \hline
  \begin{minipage}[center]{5.2cm}
    $\mathclap{\phantom{\vert^{\vert}}}$
    3-flux density on
    $Y^4 \setminus
      \underset{n}{\sqcup} D^4 $
    \\
    $\mathclap{\phantom{\vert_{\vert}}}$
    satisfying GS Bianchi identity
  \end{minipage}
  &
  \begin{minipage}[center]{4.6cm}
    \begin{center}
    $
      \begin{aligned}
        \mathclap{\phantom{\vert^{\vert^{\vert^{\vert}}}}}
        d \, H_3 & = \; \tfrac{1}{2}p_1(\nabla)
        \\
        &
        \underset{
          \mathclap{
          \raisebox{-3pt}{
            \tiny
            \color{black}
            (by $S\mathrm{U}$-structure)
          }
          }
        }{
          = c_2(\nabla) \,=\, \rchi_4(\nabla)
        }
        \mathclap{\phantom{\vert_{\vert_{\vert_{\vert}}}}}
      \end{aligned}
    $
    \end{center}
  \end{minipage}
  &
  \begin{minipage}[center]{5.2cm}
    J-twisted  3-cohomotopical
    \\
    character
  \end{minipage}
  &
  \begin{minipage}[left]{.7cm}
  \eqref{CharacterImageOfTwistedCohomotopyCocycleInEvenDimensions}
  \end{minipage}
  \\
  \hline
  \begin{minipage}[center]{5.2cm}
    $\mathclap{\phantom{\vert^{\vert}}}$
    Charge of $k$th stack of branes
  \end{minipage}
  &
  \begin{minipage}[center]{4.6cm}
    \begin{center}
    $
      \mathclap{\phantom{\vert^{\vert^{\vert^{\vert}}}}}
      \underset{
        S^3_k
        \mathclap{\phantom{\vert_{\vert_{\vert}}}}
      }{\int}
      H_3
      \;=\;
      \mathrm{deg}
      \big(
        h|_{S^3_k}
      \big)
    $
    \end{center}
  \end{minipage}
  &
  \begin{minipage}[center]{5.2cm}
    Poincar{\'e}-Hopf index
  \end{minipage}
  &
  \begin{minipage}[left]{.7cm}
    \eqref{PoincareHopfIndex}
  \end{minipage}
  \\
  \hline
  \begin{minipage}[center]{5.2cm}
    $\mathclap{\phantom{\vert^{\vert}}}$
    GS-anomaly-free 5-brane charge
  \end{minipage}
  &
  \begin{minipage}[center]{4.6cm}
    \begin{center}
    $
      \mathclap{\phantom{\int^{\vert^{\vert^{\vert^{\vert}}}}}}
      \underset{
        \mathclap{
          \underset{
            \mathclap{
              \scalebox{.4}{$
                1 \leq k \leq n
                \;\;
              $}
            }
          }{\sqcup}
          \;
          S^3_k
        }
      }{\int}
     \; H_3
      \;\overset{!}{=}\;
      \mathclap{\phantom{\vert_{\vert_{\vert}}}}
      \rchi_4[Y^4]
    $
    \end{center}
  \end{minipage}
  &
  \begin{minipage}[center]{5.2cm}
    $\mathclap{\phantom{\vert^{\vert^{\vert}}}}$
    Poincar{\'e}-Hopf theorem
    $\mathclap{\phantom{\vert_{\vert_{\vert}}}}$
  \end{minipage}
  &
  \begin{minipage}[left]{.7cm}
    \eqref{3FormTrivializingEulerFormOnPuncturedK3}
  \end{minipage}
  \\
  \hline
  \begin{minipage}[center]{5.2cm}
    $\mathclap{\phantom{\vert^{\vert}}}$
    The case of $Y^4 \,=\, \mathrm{K3}$
  \end{minipage}
  &
  \begin{minipage}[center]{4.6cm}
    \begin{center}
    $
      \mathclap{\phantom{\int^{\vert^{\vert^{\vert^{\vert}}}}}}
      \underset{
        \mathclap{
          \underset{
            \mathclap{
              \scalebox{.4}{$
                1 \leq k \leq n
                \;\;
              $}
            }
          }{\sqcup}
          \;
          S^3_k
        }
      }{\int}
     \; H_3
      \;\overset{!}{=}\;
      \mathclap{\phantom{\vert_{\vert_{\vert}}}}
      24
    $
    \end{center}
  \end{minipage}
  &
  \begin{minipage}[center]{5.4cm}
    Vanishing
    $\scalebox{.65}{$\tfrac{1}{2}$}\mathrm{e}_{\mathrm{Ad}}$-invariant,
    \\
    Vanishing d-invariant in $E = \mathbb{S}$
  \end{minipage}
  &
  \begin{minipage}[left]{.7cm}
    \eqref{ClassicalAdamsInvariantOn3rdStem}
    \\
    \eqref{FromUnstableToStable4CohomotopyOf7Sphere}
  \end{minipage}
  \\
  \hline
\end{tabular}

\medskip

\newpage

\subsection{M2-Brane Page charge and the Hopf invariant}
\label{M2BraneChargeAndTheHopfInvariant}

\noindent {\bf Page charge and the $E$-Steenrod-Whitehead integral formula.}
We consider again the case that spacetime
is homotopy equivalent to a 7-sphere, $X \,\simeq\, S^7$,
as in the vicinity of M2-branes \eqref{M2BranesPTTheorem},
hence with $H^E_3$-flux \eqref{HE3Homotopy} of the
form shown on the left here:
\begin{equation}
  \label{H3FluxOn7SphereSpacetime}
  \raisebox{24pt}{
  \xymatrix@C=9em@R=30pt{
    \mathllap{
      \mbox{
        \tiny
        \color{darkblue}
        \bf
        \begin{tabular}{c}
          near-horizon
          \\
          of M2-brane
        \end{tabular}
      }
      \!\!\!
    }
    S^7
    \ar[d]_-{
      \mathllap{
        \mbox{
          \tiny
          \color{greenii}
          \bf
          \begin{tabular}{c}
            general
            \\
            Cohomotopy charge
          \end{tabular}
        }
      }
      c
    }
    \ar[r]_>>{\ }="s"
    &
    \ast
    \ar[d]
    \\
    S^4
    \ar[r]^-{ \qquad
      G^E_{4,\mathrm{unit}}
    }^<<{\ }="t"
    &
    E^4
    \,.
    \ar@{=>}_{
      \mathclap{
      \rotatebox[origin=c]{25}{
        \scalebox{.7}{
        $
          \mathclap{
            \!\!\!\!\!\!\!\!\!\!\!\!\!\!\!\!\!\!\!\!\!\!
            d \,
            \big(
              {
                \color{orangeii}
                H^E_{3}(c)
              }
            \big)
              =
            G^E_{4}(c)
          }
        $
      }}
      }
    } "s"; "t"
  }
  }
  \phantom{AAAA}
  \raisebox{26pt}{
  \xymatrix@R=36pt@C=54pt{
    S^4
    \ar[rr]
      ^-{
          {\color{greenii}
            G^E_{4,\mathrm{unit}}
          }
          \;=\;
          \Sigma^4 (1^E)
        }
      _>>>>{\ }="s"
    \ar[d]
    &&
    E^4
    \ar[d]^-{ (-)^2 }
    \\
    \ast
    \ar[rr]
      ^-{0}
      ^-<<<<{\ }="t"
    &&
    E^8
    \ar@{=>}
      _-{
        \mathclap{
        \rotatebox{25}{
          \scalebox{.6}{$
            \mathclap{
              \!\!\!\!\!\!\!\!\!\!\!\!\!\!\!\!\!\!\!\!\!\!\!
              d\,
              \big(
                {
                  \color{orangeii}
                  2 G^E_{7,\mathrm{unit}}
                }
              \big)
              =
              -
              {\color{greenii}
                G^E_{4,\mathrm{unit}}
              }
              \scalebox{.9}{$\cup\,$}
              {\color{greenii}
                G^E_{4,\mathrm{unit}}
              }
            }
          $}
        }
        }
      }
      "s"; "t"
  }
  }
\end{equation}
for any Cohomotopy charge $c$,
measured in any multiplicative cohomology theory $E$ \eqref{dInvariantAsMapOnCohomotopy}.

\medskip
The diagram on the right of \eqref{H3FluxOn7SphereSpacetime}
recalls the general trivialization
\eqref{TrivializationOfCupSquareOfUnitM5BraneCharge}
of the cup square of the unit $G^E_{4,\mathrm{unit}}$-flux.
This induces the corresponding Toda-bracket observable
\eqref{TodaBracketObservableOnH3Flux}
on $H^E_3$, given as the
pasting diagram of the two diagrams in \eqref{H3FluxOn7SphereSpacetime}.
We now denote this Toda-bracket observable induced from the
trivialization of $(G_4)^2$ by
\vspace{-2mm}
$$
  24 \, N^E_{\mathrm{M2}}\!(c)
  \;\in\;
  \pi_1 \mathrm{Maps}
  \big(
    S^7,
    \,
    E\degree{8}
  \big)
  \;\simeq\;
  E_0
  \,,
$$

\vspace{-2mm}
\noindent where a formal normalization factor of 24 is included for
reasons discussed in Remark \ref{UnitM2BraneChargeUnderHypothesisH}.
Hence this element is, by definition,
given by the following pasting composite:

\vspace{-2mm}
\begin{equation}
  \label{EPageChargeDiagrammatic}
  \overset{
    \mathclap{
    \raisebox{6pt}{
      \tiny
      \color{darkblue}
      \bf
      \begin{tabular}{c}
        $E$-number of M2s =
        \\
        $E$-Hopf invariant
      \end{tabular}
    }
    }
  }{
    {\color{purple}
     24 \,N^E_{\!\mathrm{M2}}\!(c)
    }
  }
  \;\;\;\;\;\;\;\;\;\;
  :=
  \;\;\;\;\;\;\;\;\;\;
  \overset{
    \mathclap{
    \raisebox{5pt}{
      \tiny
      \color{darkblue}
      \bf
      \begin{tabular}{c}
        $E$-Page charge =
        \\
        $E$-Steenrod-Whitehead integral
      \end{tabular}
    }
    }
  }{
  \int_{S^{7}}
  \Big(
    {\color{orangeii}
      H^E_3\!(c)
    }
    \cup
    G^E_4\!\!(c)
    \;+\;
    {
    \color{greenii}
    2 G^E_7\!(c)
    }
  \Big)
  }
  \phantom{AAAAAA}
  \mbox{
    for:
    $
      \left\{
      \begin{aligned}
        d \,
        {
          \color{orangeii}
          H^E_3\!(c)
        }
        &
        = G^E_4\!\!(c)
        \\
        d \,
        {\color{greenii}
          \big(
            2G^E_7\!(c)
          \big)
        }
        &
        = - \big( G^E_4\!\!(c)\big)^2
      \end{aligned}
      \right.
    $
  }
\end{equation}
$$
  \raisebox{20pt}{
  \xymatrix{
    S^{7}
    \ar@/^2.3pc/[drr]_-{\ }="s"
    \ar@/_2.3pc/[drr]^-{\ }="t"
    \\
    &&
    E_{8}
    \ar@{=>}|-{
      \mathclap{\phantom{\vert^{\vert^{\vert}}}}
      \Sigma^{7} \!
      \big(
        {
        \color{purple}
        24 \cdot N^E_{\mathrm{M2}}\!(c)
        }
      \big)
      \mathclap{\phantom{\vert_{\vert}}}
    } "s"; "t"
  }
  }
  \;
  :=
  \;
  \raisebox{86pt}{
  \xymatrix{
   S^{7}
   \ar[dd]_-{ c }
   \ar[rr]
   \ar@{}[ddrr]|-{
     \rotatebox[origin=c]{-45}{\color{orangeii}$\big\Downarrow$}
     \mbox{\tiny (po)}
   }
   &&
   \ast
   \ar[dr]
   \ar[dd]
   \\
   && &
   \ast
   \ar[dd]^-{0}
   \\
   S^4
   \ar[dd]
   \ar[rr]|-{
     \;
     q_c
     \;
   }
   \ar@{}[ddrr]|-{
     \rotatebox[origin=c]{-45}{\color{greenii}$\big\Downarrow$}
     \mbox{\tiny (po)}
   }
   \ar@/_.7pc/[drrr]
     |<<<<<<{ \;G^E_{4,\mathrm{unit}}\; }
     |>>>>>>>{ \phantom{AAA}  }
   &&
   C_c
   \ar[dd]|>>>>>>>{
     \mathclap{\phantom{\vert^{\vert}}}
     p_c
     \mathclap{\phantom{\vert_{\vert}}}
   }
   \ar@{-->}[dr]|<<<<<<{
     \mathclap{\phantom{\vert^{\vert^{\vert}}}}
     \vdash
     {\color{orangeii}
       H^E_3\!(c)
     }
     \mathclap{\phantom{\vert_{\vert_{\vert}}}}
   }
   \\
   && &
   E\degree{4}
   \ar[dd]^-{ (-)^{2_{\cup}} }
   \\
   \ast
   \ar@/_.7pc/[drrr]|-{ \;0\; }
   \ar[rr]
   &&
   S^{8}
   \ar@{-->}[dr]|<<<<<<{
     \!\!\!\!\!\!
     \mathclap{\phantom{\vert^{\vert}}}
     \scalebox{.71}{$
       \vdash
       {\color{greenii}
         2 G^E_{7,\mathrm{unit}}
       }
      $}
     \mathclap{\phantom{\vert_{\vert}}}
   }
   \\
   && &
   E\degree{8}
  }
  }
  \;
  =
  \;
  \raisebox{72pt}{
  \xymatrix{
   S^{7}
   \ar[dd]_-{ c }
   \ar@[white][rr]
   &&
   \color{white}
   \ast
   \ar@[white][dr]
   \\
   && &
   \color{white}
   \ast
   \\
   S^4
   \ar[dd]
   \ar@[white][rr]
   \ar@/_.7pc/[drrr]
     |<<<<<<<{
       \;G^E_{4,\mathrm{unit}}\;
       \mathclap{\phantom{\vert_{\vert}}}
     }
     _>>>>>>>{\ }="s"
   &&
   \color{white}
   C_c
   \ar@[white]@{-->}[dr]^-{  }
   \\
   && &
   E\degree{4}
   \ar[dd]|-{
     \mathclap{\phantom{\vert^{\vert}}}
     (-)^{2_{\cup}}
     \mathclap{\phantom{\vert_{\vert}}}
   }
   \\
   \ast
   \ar@/_.7pc/[drrr]|-{ \;0\; }^<<<<<{\ }="t"
   \ar@[white][rr]
   &&
   \color{white}
   S^{8}
   \ar@[white]@{-->}[dr]|-{
     \color{white}
     \mathclap{\phantom{\vert^{\vert}}}
     \Sigma^{8} \kappa
     \mathclap{\phantom{\vert_{\vert}}}
   }
   \\
   && &
   E\degree{8}
   \ar@{=>}^-{
     \;\;\,
     \scalebox{.62}{\clap{
     \rotatebox[origin=c]{51}{\clap{$
       \;\;
       d\,
       {\color{greenii}
         (2 G^E_{7,\mathrm{unit}})
       }
       =
       -
       \big(
         G^E_{4,\mathrm{unit}}
       \big)^2
     $}}
     }
     }
   }
     "s"; "t"
  }
  }
  \hspace{-4cm}
  \raisebox{88pt}{
  \xymatrix{
   S^{7}
   \ar[dd]|-{
     \mathclap{\phantom{\vert^{\vert}}}
     c
     \mathclap{\phantom{\vert_{\vert}}}
   }
   \ar[rr]
   &&
   \ast
   \ar[dr]
   \\
   && &
   \ast
   \ar[dd]^-{0}_<<{\ }="s"
   \\
   S^4
   \ar@/_.7pc/[drrr]
     |<<<<<<<{ \;G^E_{4,\mathrm{unit}}\; }
   &&
   \\
   && &
   E\degree{4}
   \ar[dd]^-{ (-)^{2_{\cup}} }
   \\
   \color{white}
   \ast
   &&
   {\color{white}
   S^{8}}
   \\
   && &
   E\degree{8}
   \ar@{=>}
     _{
       \mathclap{
         \scalebox{.62}{
         \rotatebox[origin=c]{54}{\clap{$
           d
           {\color{orangeii}
             H^E_3\!(\hspace{-.5pt}c\hspace{-.5pt})
           }
           \hspace{-.5pt}=\hspace{-.5pt}
           G^E_{4,\mathrm{unit}}
         $}}
         }
       }
       \;
     }
     "s"; "s"+(-14,-19)
  }
  }
$$
Here

\noindent
{\bf (a)} on the right we highlight the two homotopies
\eqref{H3FluxOn7SphereSpacetime} and \eqref{TrivializationOfCupSquareOfUnitM5BraneCharge}
that are being composed,
while

\noindent
{\bf (b)} in the middle we are showing their classifying maps
via \eqref{HomotopyPushoutPropertyInIntroduction},
and {\bf (c)} on the left we highlight that the composite is a
single self-homotopy of the 0-cocycle on $S^7$ in $\widetilde E^8(-)$.

\medskip
This construction clearly exists for every multiplicative cohomology theory $E$
 (and it generalizes in the evident way to maps of the form
$S^{4n-1} \to S^{2n}$).
We observe now that it subsumes as special cases
all of the following classical constructions:

\newpage

\vspace{-.1cm}
\begin{itemize}

\vspace{-.2cm}
\item In ordinary cohomology $E = H \mathbb{Z}$...

\vspace{-.2cm}
\begin{itemize}

\vspace{-.2cm}
\item
  ... the middle diagram in \eqref{EPageChargeDiagrammatic}
  is the diagrammatic expression
  of the {\bf classical Hopf invariant} of  $c$:

  Namely, the homotopy-commutativity of the part

 \vspace{-.3cm}
\begin{minipage}[left]{14cm}
\begin{equation}
\hspace{-4mm}
  \raisebox{40pt}{
  \xymatrix@R=1pt@C=15pt{
    C_c
    \ar[drr]^-{
      \;
      \vdash
      {\color{orangeii}
        H^{H\mathbb{Z}}_3\!(c)
      }
      \;
    }
    \ar[dddd]_-{
      \mathclap{\phantom{\vert^{\vert}}}
      p_c
      \mathclap{\phantom{\vert_{\vert}}}
    }
    \\
    &&
    K(\mathbb{Z},4)
    \ar[dddd]^-{ (-)^{2_\cup} }
    \\
    \\
    \\
    S^{8}
    \ar[drr]_-{
      \;
      \vdash
      {\color{greenii}
        G^{H\mathbb{Z}}_7\!(c)
      }
      \;
    }
    \\
    && K(\mathbb{Z},8)
  }
  }
  \;\;\;\;\;
  \Leftrightarrow
  \;\;\;\;\;
  \begin{aligned}
  \big[
    \vdash
    {\color{orangeii}
      H^{H\mathbb{Z}}_3\!(c)
    }
  \big]^2
  &
  \;
  =
  \;
  p_c^\ast
  \big[
    \vdash
    {\color{greenii}
      G^{H\mathbb{Z}}_{7,\mathrm{unit}}
    }
  \big]
  \\
  &
  \;=\;
  p_c^\ast
  \Big(
  {\color{purple}
    h(c)
  }
  \cdot
  \Sigma^8\big( 1^{H \mathbb{Z}} \big)
  \Big)
  \;\;
  \in
  \;
  \widetilde{H\mathbb{Z}}{}^8(C_c)
  \end{aligned}
\end{equation}
\end{minipage}
\\
expresses that the cup square of the
degree-4 cohomology generator
$\big[ \vdash {\color{orangeii} H^{H\mathbb{Z}}_3 } \big]$
must be
{\it some} integer multiple ${\color{purple} h(c)}$
of the degree-8 cohomology generator, which in turn must
equal the pullback of that same multiple of the canonical generator
on $S^8$, the latter multiple being
$\big[ \vdash {\color{greenii} G^{H\mathbb{Z}}_7 } \big]$.
That multiple is the classical definition of the Hopf invariant $h(c)$
(e.g. \cite[p. 33]{MosherTangora86}).

But since, by the pasting law \eqref{PastingLaw},
the map $\vdash {\color{greenii} G^{H\mathbb{Z}}_7 }$
classifies also the total homotopy filling the total rectangle in
\eqref{EPageChargeDiagrammatic}, also that total rectangle
and hence the left hand side of \eqref{EPageChargeDiagrammatic} expresses
the classical Hopf invariant.

\item ... the right hand diagram in \eqref{EPageChargeDiagrammatic}
  is manifestly the diagrammatic expression of
  {\bf Steenrod's functional cup product}
  (\cite{Steenrod49}). If it happens
  on representatives that $G^{H\mathbb{Z}}_7\,(c) = 0$
  then this reduces, evidently,
  to {\bf Whitehead's integral formula}
  for the Hopf invariant (\cite{Whitehead47}, review in \cite[Prop. 7.22]{BT82}).
  That this specialization
  should not be necessary was already suggested in
  \cite[p. 17]{Haefliger78}; the more general formula
  found renewed attention in \cite[Ex. 1.9]{SinhaWalter08}.
  We gave a proof that the general formula \eqref{EPageChargeDiagrammatic}
  computes the Hopf invariant in \cite[Prop. 4.6]{FSS19c},
  using rational homotopy theory
  (as a special case of the generalization of this statement
  to classes in
  tangentially J-twisted Cohomotopy).
  The above pasting diagram
  decomposition is another proof.

\item ...under Hypothesis H --
whereby the terms
$G^E_4\!\!(c)$ \eqref{UnitG4Flux},
$G^E_7\!(c)$ \eqref{TrivializationOfCupSquareOfUnitM5BraneCharge},
$H^E_3\!(c)$ \eqref{HE3Homotopy},
in the
{\it Steenrod-Whitehead homotopy period integral formula
for the Hopf invariant} \eqref{EPageChargeDiagrammatic}
are interpreted as M-theory fluxes,
and using Examples
 \ref{M5BraneChargeInOrdinaryCohomology},
 \ref{DualFluxInOrdinaryCohomology},
 \ref{3FluxInOrdinaryCohomology}
for their incarnation in ordinary cohomology --
the formula \eqref{EPageChargeDiagrammatic}
manifestly expresses
\cite{FSS19c}
the
{\bf  Page charge} carried by M2-branes
surrounded by the 7-sphere \eqref{M2BranesPTTheorem}
according to
\cite[(8)]{Page83}\cite[(43)]{DuffStelle91} (review in \cite[(1.5.2)]{BLMP13}).

Just as the total electric flux through a 2-sphere around
a collection of electrons,
hence their total electric charge
(as on p. \pageref{VortexStrings})
is  proportional to their number,
with the proportionality constant being the unit of electric
charge, so the Page charge of M2-branes should be proportional
to their number $N_{\mathrm{M2}}$.
In \eqref{EPageChargeDiagrammatic} we declared that
proportionality constant, under Hypothesis H, to be 24,
due to the arguments listed in
Remark \ref{UnitM2BraneChargeUnderHypothesisH}.

\end{itemize}

\vspace{-.2cm}
\item In complex K-theory $E = {K\mathrm{U}}$...

\vspace{-.2cm}
\begin{itemize}

  \vspace{-.2cm}
  \item ... the middle diagram in \eqref{EPageChargeDiagrammatic}
  is the diagammatic expression
  of the {\bf K-theoretic Hopf invariant} of \cite{AdamsAtiyah66}:
  Namely, the homotopy-commutativity of

\vspace{-.3cm}
\begin{minipage}[left]{14cm}
\begin{equation}
\hspace{-4mm}
  \raisebox{40pt}{
  \xymatrix@R=1pt@C=15pt{
    C_c
    \ar[drr]^-{
      \;
      \vdash
      {\color{orangeii}
        H^{{K\mathrm{U}}}_3\!(c)
      }
      \;
    }
    \ar[dddd]_-{
      \mathclap{\phantom{\vert^{\vert}}}
      p_c
      \mathclap{\phantom{\vert_{\vert}}}
    }
    \\
    &&
    B \mathrm{U}
    \ar[dddd]^-{ (-)^{2_\cup} }
    \\
    \\
    \\
    S^{8}
    \ar[drr]_-{
      \;
      \vdash
      {\color{greenii}
        G^{{K\mathrm{U}}}_7\!(c)
      }
      \;
    }
    \\
    &&
    B \mathrm{U}
  }
  }
  \;\;\;\;\;
  \Leftrightarrow
  \;\;\;\;\;
  \begin{aligned}
  \big[
    \vdash
    {\color{orangeii}
      H^{{K\mathrm{U}}}_3\!(c)
    }
  \big]^2
  &
  \;
  =
  \;
  p_c^\ast
  \big[
    \vdash
    {\color{greenii}
      G^{{K\mathrm{U}}}_{7,\mathrm{unit}}
    }
  \big]
  \\
  &
  \;=\;
  p_c^\ast
  \Big(
  {\color{purple}
    h(c)
  }
  \cdot
  \Sigma^8\big( 1^{{K\mathrm{U}}} \big)
  \Big)
  \;\;
  \in
  \;
  \widetilde{{K\mathrm{U}}}{}^8(C_c)
  \end{aligned}
\end{equation}
\end{minipage}

\vspace{1mm}
\noindent now expresses
(proceeding just as in the discussion of the Adams $\mathrm{e}$-invariant
around \eqref{TheSESInKUForAdams})
the {\it choice} of a generator
$\big[ \vdash {\color{orangeii} H^{K\mathrm{U}}_3}(c) \big]$
lifting the canonical generator on $S^4$, whose cup square
is necessarily an integer multiple
${\color{purple} h_{{K\mathrm{U}}}(c) }$ of the
canonical generator
$\big[ \vdash {\color{greenii} H^{{K\mathrm{U}}}}_{7,\mathrm{unit}} \big]$
pulled back from $S^8$.
This multiple is the K-theoretic Hopf invariant
(review in \cite[p. 50]{Wirthmueller12}\cite[\S 26.1]{Quick14}).
\end{itemize}
\end{itemize}

\medskip

\subsection{M2-brane horizons and Ravenel $E$-orientations}
\label{VicinityOfBlackM2BranesAndOrientedCohomology}

In further specialization of the above discussion of the fluxless
(\cref{M5ThreeFlux}) vicinity of M2-branes \eqref{M2BranesPTTheorem}
we consider now the case
that $H^E_3$-flux \eqref{HE3Homotopy}
is chosen for the {\it unit} Cohomotopy charge
$[c] = [h_{\mathbb{H}}]$ \eqref{FromUnstableToStable4CohomotopyOf7Sphere}.

This case is of interest because it gives a {\it universal}
choice of $H^E_3$-flux \eqref{HE3Homotopy},
in that it implies $H^E_3\!(c)$-flux
for any integral Cohomotopy charge multiple
$c := n \cdot c_{\mathrm{unit}}$, by pullback along
$[n] \in \pi^7(S^7)$, hence by the pasting composite
shown on the right here:
\begin{equation}
  \label{Universal3FluxNearM2Branes}
  \hspace{-9mm}
  \mathllap{
    \mbox{
      \tiny
      \color{orangeii}
      \bf
      \begin{tabular}{c}
        3-flux for
        \\
        integer multiple
        \\
        Cohomotopy charge
      \end{tabular}
    }
    \!\!
  }
  \raisebox{29pt}{
  \xymatrix@R=40pt{
    S^7
    \ar[d]_-{
      \scalebox{.67}{$
        \def\arraystretch{.7}
        \begin{array}{l}
          c \,:=
          \\
          \; n \cdot h_{\mathbb{H}}
        \end{array}
      $}
    }
    ^>>{\ }="t"
    \ar[rr]_>>>{\ }="s"
    &&
    \ast
    \ar[d]^-{
      0
    }
    \\
    S^4
    \ar[rr]|-{ \;G^E_{4,\mathrm{unit}}\; }
    &&
    E\degree{4}
    \ar@{=>}^-{ H^E_3\!(c) } "s"; "t"
  }
  }
  \;\;\;\;\;\;\;\;
    :=
  \;\;\;\;\;\;\;\;
  \raisebox{29pt}{
  \xymatrix@R=40pt{
    S^7
    \ar[rr]_-{
      \;n\;
    }^-{
      \mbox{
        \tiny
        \color{greenii}
        \bf
        \begin{tabular}{c}
          multiple winding
          \\
          \phantom{-}
        \end{tabular}
      }
    }
    \ar[d]_-{
      \scalebox{.67}{$
        \def\arraystretch{.7}
        \begin{array}{l}
          c \,:=
          \\
          \; n \cdot h_{\mathrm{unit}}
        \end{array}
      $}
    }
    &&
    S^7
    \ar[d]_-{
      \scalebox{.7}{$
        \def\arraystretch{.8}
        \begin{array}{l}
          c_{\mathrm{unit}}
          \\
          \, := h_{\mathbb{H}}
        \end{array}
      $}
    }
    ^>>{\ }="t"
    \ar[rr]_>>>{\ }="s"
    &&
    \ast
    \ar[d]^-{
      0
    }
    \\
    S^4
    \ar@{=}[rr]
    &&
    S^4
    \ar[rr]|-{ \;G^E_{4,\mathrm{unit}}\; }
    &&
    E\degree{4}
    \ar@{=>}^-{ H^E_3\!(c_{\mathrm{unit}}) } "s"; "t"
  }
  }
  \mathrlap{
    \!\!
    \mbox{
      \tiny
      \color{orangeii}
      \bf
      \begin{tabular}{c}
        universal 3-flux for
        \\
        unit (quat. Hopf fib.)
        \\
        Cohomotopy charge
      \end{tabular}
    }
  }
\end{equation}

\medskip

\noindent {\bf Universal M2-brane backgrounds and Quaternionic orientation in $E$-Cohomology.}
In the vicinity of M2-branes, hence for $X \,\simeq\, S^7$
\eqref{M2BranesPTTheorem},
and given unit Cohomotopy charge
$[c] := 1 \cdot [h_{\mathbb{H}}]$
\eqref{FromUnstableToStable4CohomotopyOf7Sphere},
the corresponding
homotopy cofiber space \eqref{HomotopyPushoutAndHomotopyCofiberSpace}
is the quaternionic projective plane
$C_{h_{\mathbb{H}}} \simeq \mathbb{H}P^2$ (Remark \ref{CellStructureOfProjectiveSpaces}).
Therefore,
the choice of universal 3-flux $H^E_3$ \eqref{Universal3FluxNearM2Branes}
is equivalently \eqref{ClassifyingCohomologyClassFor3Flux}
the choice of a class
$$
  \big[ \scalebox{.8}{$\tfrac{1}{2}$}p_1^E \big]
  \;:=\;
  \big[ \vdash H^E_3(c_{\mathrm{unit}}) \big]
  \;\in\;
  E^4\big( \mathbb{H}P^2 \big)
$$
fitting into the diagram shown on the right here:
\begin{equation}
  \label{G4TrivializationOnS7IsQuaternionicOrientationToSecondStage}
  \overset{
    \mathclap{
    \raisebox{3pt}{
      \tiny
      \color{orangeii}
      \bf
      \begin{tabular}{c}
        universal 3-flux
        \\
        seen in $E$-cohomology
      \end{tabular}
    }
    }
  }{
  \raisebox{24pt}{
  \xymatrix@C=10em@R=30pt{
    \mathllap{
      \mbox{
        \tiny
        \color{darkblue}
        \bf
        \begin{tabular}{c}
          near-horizon
          \\
          of M2-brane
        \end{tabular}
      }
      \!\!\!
    }
    S^7
    \ar[d]|-{
      \mathllap{
        \mbox{
          \tiny
          \color{greenii}
          \bf
          \begin{tabular}{c}
            unit
            \\
            Cohomotopy charge
          \end{tabular}
        }
        \!\!\!\!
      }
      \mathclap{\phantom{\vert^{\vert}}}
      c_{\mathrm{unit}}
      \mathclap{\phantom{\vert_{\vert}}}
    }
    \ar[r]_>>{\ }="s"
    &
    \ast
    \ar[d]
    \\
    S^4
    \ar[r]_-{
      G^E_{4,\mathrm{unit}}
    }^<<{\ }="t"
    &
    E\degree{4}
    \ar@{=>}_{
      \mathclap{
      \rotatebox[origin=c]{22}{
        \scalebox{.7}{
        $
          \mathclap{
            \!\!\!\!\!\!\!\!\!\!\!\!\!\!\!\!\!\!\!\!\!\!
            d \,
            \big(
              {
                \color{orangeii}
                H^E_{3}\!(c_{\mathrm{unit}})
              }
            \big)
              =
            G^E_{4}(c_{\mathrm{unit}})~~~
          }
        $
      }}
      }
    } "s"; "t"
  }
  }
  }
  \;\;\;\;\;\;\;
  \simeq
  \;\;\;\;\;\;\;
  \overset{
    \mathclap{
    \raisebox{3pt}{
      \tiny
      \color{orangeii}
      \bf
      \begin{tabular}{c}
        quaternionic orientation
        \\
        in $E$-cohomology to 2nd stage
      \end{tabular}
    }
    }
  }{
  \raisebox{24pt}{
  \xymatrix@C=3em@R=10pt{
    S^7
    \ar[rr]_>>>{\ }="s"
    \ar[dd]_{h_{\mathbb{H}}}^>>{\ }="t"
    &&
    \ast
    \ar[dd]
    \ar@/^1pc/[dddr]^-{ \;0\; }
    \\
    \\
    S^4
    \ar[rr]|-{ \;q\; }
    \ar@/_1pc/[drrr]^-{ \Sigma^4 1 }
    &&
    \mathbb{H}P^2
    \ar@{-->}[dr]|-{
      \color{orangeii}
      \scalebox{.5}{$\tfrac{1}{2}$} p_1^{E}
    }
    \\
    && &
    E\degree{4}
    \ar@{=>}
      ^-{ \mbox{\tiny(po)} }
      "s"; "t"
  }
  }
  }
\end{equation}
But such a $\scalebox{.8}{$\tfrac{1}{2}$}p_1^E$
is equivalently the choice of a universal
{\it 10-dimensional quaternionic orientation} on $E$-cohomology,
(Def. \ref{QuaternionicOrientedCohomology}).

\medskip
More widely familiar than quaternionic oriented cohomology theory
is complex oriented cohomology theory (Def. \ref{ComplexOrientedCohomology}).
But in fact, ($4k+2$-dimensional) complex orientations $c^E_1$
{\it induce} ($4k + 2$-dimensional) quaternionic orientations
(Theorem \ref{FiniteRankQuaternionicOrientationFromFiniteRankComplexOrientation} below)
by taking the first $E$-Pontrjagin class of
a quaternionic vector bundle to be
the second Conner-Floyd $E$-Chern class $c^E_2$
(Prop. \ref{ConnerFloydChernClasses})
of the underlying complex vector bundle
\eqref{FiniteRankQuaternionicOrientationFromFiniteRankComplexOrientation}
\begin{equation}
  \tfrac{1}{2}p^E_1(-)
  \;:=\;
  c^E_2
  \big(
    (-)_{\mathbb{C}}
  \big)
  \,.
\end{equation}

\noindent
In conclusion:

\vspace{2mm}

\noindent
\hspace{.04cm}
\fbox{
\begin{minipage}[left]{16.8cm}
{\it
Under \hyperlink{HypothesisH}{Hypothesis H}, with M-brane charge
measured in $E$-cohomology \eqref{dInvariantAsMapOnCohomotopy},
the class of a universal 3-flux $H^E_3(c_{\mathrm{unit}})$ \eqref{Universal3FluxNearM2Branes}
near the horizon of M2-branes \eqref{M2BranesPTTheorem}
is equivalent to a choice of 10-dimensional quaternionic $E$-orientation,
such as is induced from a 10-dimensional complex $E$-orientation.}
\end{minipage}
}
\begin{equation}
\label{HE3FluxFromComplexOrientation}
\hspace{-6mm}
  \raisebox{-38pt}{
  \begin{tabular}{cccccc}
    \scalebox{.8}{
    \begin{tabular}{c}
      \color{darkblue}
      \bf
      $E$ is a complex oriented
      \\
      \color{darkblue}
      \bf
      multiplicative cohomology
      \\
      (in 10 dimensions)
      \\
      \phantom{a}
    \end{tabular}
    }
    &
    \raisebox{6pt}{
    $
      \overset{
        \mathclap{
        \raisebox{3pt}{
          \tiny
          Thm. \ref{FiniteRankQuaternionicOrientationFromFiniteRankComplexOrientation}
        }
        }
      }{
        \!
        \Longrightarrow
        \!
      }
    $
    }
    &
    \scalebox{.8}{
    \begin{tabular}{c}
      \color{darkblue}
      \bf
      $E$ is a quaternionic oriented
      \\
      \color{darkblue}
      \bf
      multiplicative cohomology
      \\
      (in 10 dimensions)
      \\
      \phantom{a}
    \end{tabular}
    }
    &
    \raisebox{6pt}{
    $
      \overset{
        \mathclap{
        \raisebox{3pt}{
          \tiny
          \eqref{G4TrivializationOnS7IsQuaternionicOrientationToSecondStage}
        }
        }
      }{
        \!\!\!
        \Longleftrightarrow
        \!\!\!
      }
    $
    }
    &
    \scalebox{.8}{
    \begin{tabular}{c}
      \color{darkblue}
      \bf
      $E$-cohomology detects
      \\
      \color{darkblue}
      \bf
      unit charges near M2-branes
      \\
      (under \hyperlink{HypothesisH}{Hypothesis H})
      \\
      \phantom{a}
    \end{tabular}
    }
    \\
    \raisebox{20pt}{
    \xymatrix@C=4em{
      \mathbb{C}P^1
      \mathclap{\phantom{\vert_{\vert}}}
      \ar@{^{(}->}[d]
      \ar[r]^-{
        \Sigma^2(1^E)
      }
      &
      E_2
      \\
      \mathbb{C}P^5
      \ar@{-->}[ur]_-{
        \color{orangeii}
        c_1^E
      }
    }
    }
    &&
    \raisebox{20pt}{
    \xymatrix@C=4em{
      \mathbb{H}P^1
      \mathclap{\phantom{\vert_{\vert}}}
      \ar@{^{(}->}[d]
      \ar[r]^-{
        \color{greenii}
        \Sigma^4(1^E)
      }
      &
      E\degree{4}
      \\
      \mathbb{H}P^2
      \ar@{-->}[ur]_-{
        \scalebox{.7}{$
        \begin{array}{l}
          \color{orangeii}
          \scalebox{.7}{$\tfrac{1}{2}$}p_1^E
          \\
          \;
          ( = c^E_2 )
        \end{array}
        $}
      }
    }
    }
    &&
  \raisebox{24pt}{
  \xymatrix@C=9em@R=30pt{
    S^7
    \ar[d]_-{
      c_{\mathrm{unit}}
    }
    \ar[r]_>>{\ }="s"
    &
    \ast
    \ar[d]
    \\
    S^4
    \ar[r]|-{
      \;
      \color{greenii}
      G^E_{4,\mathrm{unit}}
      \;
    }^<<{\ }="t"
    &
    E\degree{4}
    \ar@{==>}_{
      \mathclap{
      \rotatebox[origin=c]{23}{
        \scalebox{.7}{
        $
          \mathclap{
            \!\!\!\!\!\!\!\!\!\!\!\!\!\!\!\!\!
            \!\!\!\!\!\!\!\!\!\!\!\!\!\!\!\!\!
            d \,
            \big(
              {
                \color{orangeii}
                H^E_{3}\!(c_{\mathrm{unit}})
              }
            \big)
              =
            G^E_{4}\!(c_{\mathrm{unit}})
          }
        $
      }}
      }
    } "s"; "t"
  }
  }
  \end{tabular}
  }
\end{equation}

\medskip

\begin{remark}[Emergence of line bundle orientations in 10 dimensions]
  The 10-dimensional complex $E$-orientations
  appearing in \eqref{HE3FluxFromComplexOrientation}
  are {\it exactly} the data necessary to
  (fiberwise) orient general complex line bundles
  over 10-dimensional manifolds $X^{10}$
  (as discussed in \cref{Orientation}),
  hence are {\it exactly} the data needed to orient
  general heterotic line bundles in heterotic M-theory
  (discussed in \cref{GreenSchwarzMechanismAndComplexEOrientations})
  on globally hyperbolic
  11-dimensional spacetimes
   \footnote{Note that the heterotic theory viewed as a cobordism is studied in \cite{Sati-cob}.}
  of the form $\mathbb{R}^{0,1} \times X^{10}$.
  This seems remarkable, as the number 10
  {\it emerges} here,
  via Theorem \ref{FiniteRankQuaternionicOrientationFromFiniteRankComplexOrientation},
  from just
  {\hyperlink{HypothesisH}{\it Hypothesis H}} and
  the consideration of black codimension-8
  brane backgrounds in \eqref{Universal3FluxNearM2Branes}
  --
  see Remark \ref{RelationOfDimensionsForComplexAndQuaternionicOrientations}.
\end{remark}

\begin{example}[3-Flux in ordinary cohomology]
  \label{3FluxInOrdinaryCohomology}
  When $E = H R$ is ordinary cohomology,
  there is a unique quaternionic orientation
  (by Example \ref{ComplexOrientationInOrdinaryCohomology}),
  equivalently
  a unique choice of 3-flux $H^{H A}_{3,\mathrm{unit}}$
  (by \eqref{3FluxesFormTorsorOverThirdECohomologyOfX}
  and
  because
  $(\widetilde{HA})^3(S^7) \simeq \pi_4(HA) = 0$).
  In the case $A = \mathbb{R}$,
  via Example \ref{M5BraneChargeInOrdinaryCohomology}
  and under the fundamental theorem of rational homotopy theory
  (see \cite[Prop. 3.60]{FSS20c}) this corresponds to
  the degree-3 generator in the relative Sullivan model
  for the quaternionic Hopf fibration
  \cite[Prop. 3.20]{FSS19b} (\cite[Thm. 3.4]{FSS19c}).
\end{example}

 \newpage

\subsection{M5${}_{\mathrm{HET}}$-brane charge and Conner-Floyd classes}
\label{GreenSchwarzMechanismAndComplexEOrientations}

\noindent
{\bf The C-field in heterotic M-theory and the Twistor fibration.}
The (class of the) quaternionic Hopf fibration \eqref{M2BranesPTTheorem},
representing unit Cohomotopy charge in the vicinity of M2-brane horizons
\eqref{Universal3FluxNearM2Branes},
naturally factors through $\mathbb{C}P^3$,
via the complex Hopf fibration followed by the
Atiyah-Penrose {\it twistor fibration} $t_{\mathbb{H}}$
(Remark \ref{ComplexProjectiveSpacesOverQuaternionicProjectiveSpaces},
see \cite[\S 2]{FSS20b} for more pointers):
\begin{equation}
  \label{FactoringhHThroughtH}
  \xymatrix@C=3em{
    &&
    \mathbb{C}P^3
    \ar[d]
      ^-{
        t_{\mathbb{H}}
        \!\!\!
        \mathrlap{
          \mbox{
            \tiny
            \color{greenii}
            \bf
            \begin{tabular}{c}
              twistor
              \\
              fibration
            \end{tabular}
          }
        }
      }
    \\
    S^7
    \ar[urr]
      ^-{
        \mathllap{
          \mbox{
            \tiny
            \color{greenii}
            \bf
            \begin{tabular}{c}
              complex
              \\
              Hopf fibration
            \end{tabular}
          }
        }
        \!\!\!
        h_{\mathbb{C}}
      }
    \ar[rr]
      ^{
        h_{\mathbb{H}}
      }
      _{
        \mbox{
          \tiny
          \color{greenii}
          \bf
          \begin{tabular}{c}
            quaternionic
            \\
            Hopf fibration
          \end{tabular}
        }
      }
    &&
    S^4
    \,.
  }
\end{equation}
We may think also of $\mathbb{C}P^3$ as the classifying space
of a non-abelian cohomology theory \eqref{NonabelianCohomologyInIntroduction},
which is thus a close twistorial cousin of 4-Cohomotopy \eqref{CohomotopyInIntroduction},
called {\it twistorial Cohomotopy} in
\cite[\S 3.2]{FSS20b}\cite[Ex. 2.44]{FSS20c}\cite[Def. 2.48]{SS20c}:
\begin{equation}
  \label{TwistorialCohomotopy}
  \xymatrix{
    \mathllap{
      \mbox{
        \tiny
        \color{darkblue}
        \bf
        twistorial Cohomotopy
      }
    }
    \widetilde{\mathcal{T}}(X)
    \ar[d]
      _-{
        \mathllap{
          \mbox{
            \tiny
            \color{greenii}
            \bf
            \begin{tabular}{c}
              pushforward along
              \\
              twistor fibration
            \end{tabular}
          }
          \!\!
        }
        (t_{\mathbb{H}})_\ast
      }
    \ar@{}[r]
      |-{ := }
    &
    \pi_0 \mathrm{Maps}^{\ast/\!}
    \big(
       X
       \,,\,
       \mathbb{C}P^4
    \big)
    \\
    \mathllap{
      \mbox{
        \tiny
        \color{darkblue}
        \bf
        4-Cohomotopy
      }
    }
    {\widetilde \pi}{}^4(X)
    \ar@{}[r]
      |-{ := }
    &
    \pi_0 \mathrm{Maps}^{\ast/\!}
    \big(
       X
       \,,\,
       S^4
    \big)
    \,.
  }
\end{equation}
Evidence discussed in \cite{FSS20b}\cite{FSS20c}\cite{SS20c}
suggests that \hyperlink{HypothesisH}{\it Hypothesis H}
extends from M-theory to {\it heterotic M-theory}
(Ho{\v r}ava-Witten theory \cite{HoravaWitten95}\cite{Witten96}\cite{HoravaWitten96b}\cite{DOPW99}\cite{DOPW00}\cite{Ovrut02})
by lifting M-brane charges
from 4-Cohomotopy \eqref{CohomotopyInIntroduction}
to twistorial Cohomotopy \eqref{TwistorialCohomotopy}
(generally tangentially J-twisted, but here considered
on homotopically flat spacetimes and hence for vanishing twist,
see Remark \ref{FramedSpacetimes}).

Concretely, the lift of the Cohomotopical character map
\eqref{RationalCohomotopy} through
\eqref{TwistorialCohomotopy} turns out
\cite[Prop. 3.9]{FSS20b}\cite[Thm. 1.1]{SS20c}\cite[Ex. 5.23]{FSS20c}
to correspond to the appearance of the rational
second Chern class
$$
  c_2
  \big(
    \mathcal{L} \oplus \mathcal{L}^\ast
  \big)
  \;=\;
  c_1(\mathcal{L})
    \wedge
  c_1(\mathcal{L}^\ast)
  \;=\;
  -
  \big(
    c_1(\mathcal{L})
  \big)^2
  \;\;
    \in
  \;
  H^4(X; \mathbb{R})
$$
of a
$(\mathrm{U}(1) \hookrightarrow S\mathrm{U}(2) \hookrightarrow E_8)$-bundle
(a ``heterotic line bundle''
\cite{AGLP12}\cite{AGLP12}\cite{BBL17}\cite{FSS20c},
here of rank 2 as considered in \cite[\S 4.2]{ADO20a}\cite[\S 2.2]{ADO20b}):
\begin{equation}
  \label{HeteroticLineBundle}
  \xymatrix@R=5pt@C=3em{
    \overset{
      \mathclap{
      \raisebox{6pt}{
        \tiny
        \color{darkblue}
        \bf
        complex line bundle
      }
      }
    }{
      \mathcal{L}
    }
     \ar@(dl,dr)
    \ar@{|->}[rr]
      _-{
        \mbox{
          \tiny
          \bf
          \begin{tabular}{c}
            {
              \color{greenii}
              heterotic line bundle
            }
            \\
            of rank 2
          \end{tabular}
        }
      }
      &{\phantom{A}}&
    \qquad
    \overset{
      \mathclap{
      \raisebox{6pt}{
        \tiny
        \color{darkblue}
        \bf
        \begin{tabular}{c}
          rank-2 complex vector bundle
          \\
          with vanishing first Chern class
        \end{tabular}
      }
      }
    }{
      \mathcal{L} \oplus \mathcal{L}^\ast
    }
  \qquad   \ar@(dl,dr)
    \\
    \mathrm{U}(1)\;
   \ar@{^{(}->}[rr]
      _-{
        c \,\mapsto\, \mathrm{diag}(c,c^\ast)
      }
    &&
   \quad
    S\mathrm{U}(2)
    \mathrlap{
      \mbox{
        \tiny
        \color{greenii}
        \bf
        structure group
      }
    }
  }
\end{equation}
together with a 3-form $H^{H\mathbb{R}}_3$ satisfying
\begin{equation}
  \label{RationalHoravaWittenGreenSchwarzMechanism}
  \begin{aligned}
    d \, H^{H\mathbb{R}}_3
    & = \;
    G^{H\mathbb{R}}_4 - c_1(\mathcal{L}) \wedge c_1(\mathcal{L})
    \\
    & = \;
    G^{H\mathbb{R}}_4 + c_2\big( \mathcal{L} \oplus \mathcal{L}^\ast  \big).
  \end{aligned}
  \end{equation}
This  {\it Bianchi identity} is just that of the Ho{\v r}ava-Witten Green-Schwarz mechanism
in heterotic M-theory near MO9-planes \cite[(1.13)]{HoravaWitten96b}\cite[(3.9)]{DFM03}\cite[(161)]{SS20c},
specialized to rank-2 heterotic line bundles \eqref{HeteroticLineBundle}
and shown here for vanishing $\scalebox{.6}{$\tfrac{1}{2}$}p_1$
(following Remark \ref{FramedSpacetimes},
see \cite{FSS20b}\cite{SS20c}\cite{FSS20c}
for the general statement including
the summand of $\scalebox{.6}{$\tfrac{1}{2}$}p_1$).

\medskip

We now observe that the form of the M-theoretic Green-Schwarz mechanism
\eqref{RationalHoravaWittenGreenSchwarzMechanism}
appears, from {\hyperlink{HypothesisH}{\it Hypothesis}}, not just
in the rational approximation $E = H\mathbb{R}$, but is visible
for measurement \eqref{M5BraneFluxInEtheoryByPullback}
of $G_4$-flux in any multiplicative
generalized cohomology theory $E$:

\medskip

\noindent
{\bf Heterotic line bundles and the $E$-Whitney sum rule.}
Given $H^E_3(c)$-flux \eqref{HE3Homotopy}
for the unit 4-Cohomotopy charge
$c_{\mathrm{unit}} = [h_{\mathbb{H}}]$ near
the horizon of black M2-branes \eqref{Universal3FluxNearM2Branes},
which comes from a 10d complex $E$-orientation
\eqref{HE3FluxFromComplexOrientation},
we obtain a homotopy
\begin{equation}
  \label{HeteroticHE3Flux}
  \hspace{-8mm}
  \left.
  \mbox{
    \begin{tabular}{c}
      $H^E_3(c_{\mathrm{unit}})$-flux induced from
      \\
      10d complex $E$-orientation \eqref{HE3FluxFromComplexOrientation}
    \end{tabular}
  } \!\!\!\!\!\!
  \right\}
  \;\;\;
    \Rightarrow
  \;\;\;
  \raisebox{20pt}{
  \xymatrix@R=10pt@C=3em{
    &&
    \mathbb{C}P^3
    \ar[dd]
      _-{
        t_{\mathbb{H}}
      }
      ^>>>{\ }="t"
    \ar[rr]
      ^-{
        c^E_1 \cdot c^E_{1^\ast}
      }
      _>>>{\ }="s"
    &{\phantom{AAAA}}&
    E^4
    \ar@{=}[dd]
    \\
    \\
    S^7
    \ar[uurr]
      ^-{
        \overset{
        \mathllap{
        \mbox{
          \tiny
          \color{greenii}
          \bf
          \begin{tabular}{c}
            lift to
            \\
            twistorial Cohomotopy
          \end{tabular}
        }
        }
        }{
          t_{\mathbb{C}}
        }
      }
    \ar[rr]
      ^-{
        h_{\mathbb{H}}
      }
      _-{
        \mbox{
          \tiny
          \color{greenii}
          \bf
          unit Cohomotopy charge
        }
      }
    &&
    S^4
    \ar[rr]
      _-{
        G^E_{4,\mathrm{unit}}
      }
    &&
    E^4
    \ar@{=>}
      ^-{
        \mathclap{
          \rotatebox{12}{
            \scalebox{.6}{$
              \mathclap{
                \!
                d
                \,
                {\color{orangeii}
                H^E_{3,\mathrm{het}}
                }
                \,=\,
                G^E_{4,\mathrm{unit}}
                -
                c^E_1 \cdot c^E_{1^\ast}
              }
            $}
          }
        }
      }
      "s"; "t"
    }
  }
  \!\!\!
    {
      \mbox{
        \tiny
        \color{orangeii}
        \begin{tabular}{c}
          heterotic $H_3$-flux
          \\
          measured in
          \\
          $E$-cohomology
        \end{tabular}
      }
    }
\end{equation}
by co-restricting the given $H^E_3(c_{\mathrm{unit}})$-flux
from the quaternionic Hopf fibration $h_{\mathbb{H}}$
to the twistor fibration $t_{\mathbb{H}}$ \eqref{FactoringhHThroughtH}
as follows:
$$
  \begin{array}{ccc}
  \raisebox{45pt}{
  \xymatrix@C=38pt{
    S^7
    \ar[d]
      _<<<<{
        \mathclap{\phantom{\vert^{\vert}}}
        h_{\mathbb{C}}
        \mathclap{\phantom{\vert_{\vert}}}
      }
      ^>>>{\ }="t1"
    \ar[rr]
      _>>>>>>>>>>>>{\ }="s1"
    &
    {\phantom{
      \ast
    }}
    &
    \ast
    \ar[d]
    \\
    \mathbb{C}P^3
    \ar[d]
      _<<<<{
        \mathclap{\phantom{\vert^{\vert}}}
        t_{\mathbb{H}}
        \mathclap{\phantom{\vert_{\vert}}}
      }
    \;
    \ar[rr]
      ^-{
        \;
        c^E_1 \cdot c^E_{1^{\!\ast}}
        \;
      }
      _>>>>>{\ }="s"
    \ar[d]
    &
    {\phantom{
      \mathbb{C}P^5
    }}
    &
    E^4
    \ar@{=}[d]
    \\
    \mathbb{H}P^1
    \ar@/_1.4pc/[rr]
      ^<<<<<{\ }="t"
      |-{
        \;
        G^E_{4,\mathrm{unit}}
        \;
      }
    \;
    &
    {\phantom{
      \mathbb{H}P^2
    }}
    &
    E^4
    \ar@{=>}
      |-{
        \;\;
        \color{orangeii}
        H^E_{3,\mathrm{het}}
        \;\;
      }
      "s"; "t"
  }
  }
  &
  \;\;\;\;
  :=
  \;\;\;\;
  &
  \raisebox{45pt}{
  \xymatrix@C=6em{
    S^7
    \ar[d]
      _<<<<{
        \mathclap{\phantom{\vert^{\vert}}}
        h_{\mathbb{C}}
        \mathclap{\phantom{\vert_{\vert}}}
      }
      ^>>>{\ }="t1"
    \ar[r]
      _>>{\ }="s1"
    \ar@{}[dr]
      |-{
      }
    &
    \ast
    \ar[r]
    \ar[d]
    &
    \ast
    \ar[d]
    \\
    \mathbb{C}P^3
    \;
    \ar@{^{(}->}[r]
      _>>>{\ }="s2"
    \ar[d]
      _<<<<{
        \mathclap{\phantom{\vert^{\vert}}}
        t_{\mathbb{H}}
        \mathclap{\phantom{\vert_{\vert}}}
      }
      ^>>>{\ }="t2"
    &
    \mathbb{C}P^5
    \ar[d]
      _-{
        \scalebox{.5}{$
          \color{greenii}
          \def\arraystretch{.7}
          \begin{array}{c}
            \mathcal{L}
            \\
            \mapsdown
            \\
            \mathcal{L} \oplus \mathcal{L}^\ast
          \end{array}
        $}
      }
      ^>{\ }="t3"
    \ar[r]
      ^-{
        \;
        c^E_1
        \cdot
        c^E_{1^{\!\ast}}
        \;
      }
      _>>{\ }="s3"
    &
    E^4
    \ar@{=}[d]
    \\
    \mathbb{H}P^1
    \ar@/_1.4pc/[rr]
      |-{
        \;
        G^E_{4,\mathrm{unit}}
        \;
      }
    \;
    \ar@{^{(}->}[r]
    &
    \mathbb{H}P^2
    \ar[r]
      ^-{
       \;
       \color{orangeii}
       c^E_2
       \;
      }
    &
    E^4
    \ar@{=>}
      ^-{
        \mbox{
          \tiny
          \rm
          (po)
        }
      }
    "s1"; "t1"
    \ar@{=>}
      ^-{
        \mbox{
          \tiny
          \rm
          (po)
        }
      }
    "s2"; "t2"
    \ar@{=>}
      _-{
        \scalebox{.9}{
          \tiny
          WSR
        }
      }
    "s3"; "t3"
  }
  }
  \\
  &&
  \mathclap{\phantom{\vert^{\vert^{\vert^{\vert^{\vert^{\vert}}}}}}}
  \rotatebox{90}{\scalebox{1.1}{$=$}}
  \\
  \raisebox{35pt}{
  \xymatrix@C=38pt{
    S^7
    \ar[dd]
      _-{
        \mathclap{\phantom{\vert^{\vert}}}
        h_{\mathbb{H}}
        \mathclap{\phantom{\vert_{\vert}}}
      }
      ^>>>{\ }="t1"
    \ar[rr]
      _>>>>>>{\ }="s1"
    \ar@{}[dr]
      |-{
      }
    &
    {\phantom{\ast}}
    &
    \ast
    \ar[dd]
    \\
    {\phantom{\mathbb{C}^3}}
    \;
    \ar[d]
      ^>>>{\ }="t2"
    &
    {\phantom{\mathbb{C}^5}}
    &
    {\phantom{E^4}}
    \\
    S^4
    \ar@/_1.4pc/[rr]
      |-{
        \;
        G^E_{4,\mathrm{unit}}
        \;
      }
    &
    {\phantom{\mathbb{H}P^2}}
    &
    E^4
    \ar@{=>}
      ^-{
        \color{orangeii}
        H^E_{3}(c_{\mathrm{unit}})
      }
    "s1"; "t1"
  }
  }
  &
  \scalebox{1.1}{$=$}
  &
  \raisebox{35pt}{
  \xymatrix@C=6em{
    S^7
    \ar[dd]
      _-{
        \mathclap{\phantom{\vert^{\vert}}}
        h_{\mathbb{H}}
        \mathclap{\phantom{\vert_{\vert}}}
      }
      ^>>>{\ }="t1"
    \ar[r]
      _>>{\ }="s1"
    \ar@{}[dr]
      |-{
      }
    &
    \ast
    \ar[r]
    \ar[dd]
    &
    \ast
    \ar[dd]
    \\
    {\phantom{\mathbb{C}^3}}
    \;
    \ar[d]
      _<<<<{
        \mathclap{\phantom{\vert^{\vert}}}
        t_{\mathbb{H}}
        \mathclap{\phantom{\vert_{\vert}}}
      }
      ^>>>{\ }="t2"
    &
    {\phantom{\mathbb{C}^5}}
    &
    {\phantom{E^4}}
    \\
    \mathbb{H}P^1
    \ar@/_1.4pc/[rr]
      |-{
        \;
        G^E_{4,\mathrm{unit}}
        \;
      }
    \;
    \ar@{^{(}->}[r]
    &
    \mathbb{H}P^2
    \ar[r]
      ^-{
       \;
       {
       \color{orangeii}
         \scalebox{.6}{$\tfrac{1}{2}$}p^E_1
       }
       :=
       {
         \color{orangeii}
         c^E_2
       }
       \;
      }
    &
    E^4
    \ar@{=>}
      ^-{
        \mbox{
          \tiny
          \rm
          (po)
        }
      }
    "s1"; "t1"
  }
  }
  \end{array}
$$
Here, going counter-clockwise from the bottom left:

\vspace{-.15cm}
\begin{itemize}
  \vspace{-.2cm}
  \item
    The bottom left homotopy is the given $H^E_3$-flux
    \eqref{Universal3FluxNearM2Branes}.

  \vspace{-.2cm}
  \item
    The bottom right pasting decomposition is that through the
    corresponding cofiber space \eqref{FluxlessnessDiagram},
    which here is a 10d quaternionic $E$-orientation
    $\scalebox{.6}{$\tfrac{1}{2}$}p^E_1$
    \eqref{G4TrivializationOnS7IsQuaternionicOrientationToSecondStage}
    which, by assumption in \eqref{HeteroticHE3Flux},
    equals $c^E_2$ \eqref{HE3FluxFromComplexOrientation}.

  \vspace{-.2cm}
  \item The top right pasting decomposition is

  \vspace{-.2cm}
  \begin{itemize}

  \vspace{-.2cm}
  \item on the left the pasting law \eqref{PastingLaw}
  applied to the twistor factorization \eqref{FactoringhHThroughtH}
  with the induced appearance \eqref{CellStructureOfProjectiveSpace}
  of complex projective 5-space and the
  classifying map \eqref{TowerOfS2FibrationsConvergingTpBS1OverBSU2}
  for the assignment of heterotic line bundles \eqref{HeteroticLineBundle};

  \vspace{-.2cm}
  \item
    on the right the homotopy $\mathrm{WSR}$ exhibiting
    on classifying maps
    the {$E$-Whiteney sum rule} \eqref{WhitneySumRuleforEChernClasses}
    for evaluation of
    the second Conner-Floyd $E$-Chern class
    $c^E_2$ on these heterotic line bundles.
  \end{itemize}
  \vspace{-.2cm}

  \vspace{-.2cm}
  \item
   The top left square is obtained from this by
   forming horizontal pasting composites of these homotopy squares.
\end{itemize}

\noindent
In conclusion:

\vspace{2mm}

\noindent
\hspace{.04cm}
\fbox{
\begin{minipage}[left]{16.8cm}
{\it
  {\hyperlink{HypothesisH}{\it Hypothesis H}}
  implies not only
  the usual
  Ho{\v r}ava-Witten Green-Schwarz Bianchi identity
  \eqref{RationalHoravaWittenGreenSchwarzMechanism}
  in ordinary cohomology $E = H \mathbb{R}$
  (for rank-2 heterotic line bundles \eqref{HeteroticLineBundle}),
  but lifts \eqref{HeteroticHE3Flux}
  the structure of this relation
  to any multiplicative 10d complex-oriented cohomology
  theory $E$ in which \eqref{HE3FluxFromComplexOrientation}
  cohomotopical M-brane charge
  near black M2-branes is being measured \eqref{Universal3FluxNearM2Branes}.
}
\end{minipage}
}

\newpage

\section{The proofs}
\label{BordismAndStableHomotopyForMTheory}

Here we provide rigorous mathematical
background and proofs for the right hand side of the correspondence
that is
discussed in \cref{TheDictionary} and summarized in \cref{Conclusions}.

\medskip

While a fair bit of the following
background material is classical in algebraic topology,
it has not previously found its connection to
and application in mathematical physics and string theory.
Moreover, much of this classical material seems to
not have received modernized discussion before; for instance:

\medskip

{\bf (i)} Toda brackets are traditionally still discussed
in the style of Toda's original
article \cite{Toda62} from 1962. While
writing our diagrammatic re-formulation of Toda brackets
in \cref{TodaBrackets}, we discovered that this
perspective had been observed already by Hardie et al. in 1999
(see Remark \ref{ReferencesOnTodaBrackets} below);
but their treatment, in turn, does not seem to have received
attention nor application.

\vspace{1mm}
{\bf
(ii)} The essential reference for Conner-Floyd's
geometric e-invariant (\cref{RelativeCobordism})
(and related constructions such as the Conner-Floyd
K-orientation, Prop. \ref{ConnerFloydKOrientation} below)
has remained their original account \cite{ConnerFloyd66} from 1966.
We believe that our reformulation \eqref{ReproducingTheConnerFloydEInvariant}
via diagrammatic Toda-brackets makes usefully transparent
not only the general relation to observables on
trivializations of d-invariants (``H-fluxes'', in our correspondence
entry \cref{M5ThreeFlux})
but also the construction itself, notably the proof of
its equivalence to the classical Adams e-invariant
(Prop. \ref{TheConnerFloydeInvariant} below).

Moreover, the fact (Prop. \ref{RationalToddNumberAsBoundaryIntegral} below) that the Conner-Floyd construction is
really a special case of Chern-Simons-invariants computed via
``extended worldvolumes'' in the style of \cite{Witten83}
(relating it to the Green-Schwarz mechanism
in our dictionary entry \cref{TadpoleCancellationAndSUBordismWithBoundaries}),
seems to not have been touched upon before.

\vspace{1mm}
{\bf (iii)}
The bounded-dimensional generalization of
complex orientations on generalized $E$-cohomology
originates with \cite[\S 3]{Ravenel84},
we call them {\it Ravenel orientations} in \cref{Orientation}.
However, this reference, with most of its followups,
focuses on properties
of the associated {\it Ravenel spectra}.
The main reference with explicit discussion
of their role in bounded-dimensional complex orientation
remains Hopkins' thesis \cite[\S 2.1]{Hopkins84}.
Since any application of
complex oriented cohomology
to physics
(such as to D-brane charge and elliptic partition functions)
is necessarily bounded by spacetime dimension,
it is really these bounded-dimensional $E$-orientations
that matter here.

\medskip

Besides providing streamlined accounts of these
and related topics,
our main mathematical results here are the following two theorems,
crucial for the progression of the correspondence in
\cref{M5BraneC3FieldAndTheAdamseInvariant},
\cref{TadpoleCancellationAndSUBordismWithBoundaries}
and
\cref{VicinityOfBlackM2BranesAndOrientedCohomology},
\cref{GreenSchwarzMechanismAndComplexEOrientations},
respectively:

\vspace{1mm}
\noindent
{\bf (a)} Theorem \ref{DiagrammaticeCCoincidesWithClassicaleCInvariant},
confirms that the
Toda-bracket formulation of the
refined e-invariant (Def. \ref{LiftedEInvariantDiagrammatically})
does reproduce the classical Adams e-invariant;

\noindent
{\bf (b)} Theorem \ref{FiniteRankQuaternionicOrientationFromFiniteRankComplexOrientation},
shows how bounded-dimensional complex $E$-orientations
induce bounded-dimensional quaternionic $E$-orientations.

\medskip
While these statements are not surprising and
may be known to experts in the field,
they are not in the literature
(to the best of our knowledge),
and
there is some amount of fiddly detail involved in their proofs,
which we aim to spell out in detail. Hence we hope that this
self-contained and streamlined account will be useful.

\newpage

\subsection{Borsuk-Spanier Cohomotopy}
\label{BorsukSpanierCohomotopy}

We briefly set up basics
of homotopy theory to fix concepts and notation. For more details and
further pointers see \cite[\S A]{FSS20c}.

\vspace{2mm}
\noindent {\bf Categories.}
While we give a fairly self-contained account of
the more advanced notions in the following,
we make free use of elementary concepts in category theory and
homotopy theory: for joint introduction see
\cite{Riehl14}\cite{Richter20} and for concise review in
our context see \cite[\S A]{FSS20c}.
Here just our notation conventions:
\begin{notation}[Categories]
  \label{Categories}
  For $\mathcal{C}$ a category
  and for $X, A \,\in\, \mathcal{C}$ a pair of its objects, we write
  \vspace{-1mm}
  \begin{equation}
    \label{HomSet}
    \mathcal{C}(X,A) \;\in\; \mathrm{Sets}
  \end{equation}

    \vspace{-2mm}
\noindent  for the hom-set of morphisms in $\mathcal{C}$ from $X$ to $A$.
  This extends, of course, to a contravariant functor on $\mathcal{C}$
  in its first argument, which we denote as
    \vspace{-1mm}
  \begin{equation}
    \label{ContravariantHomFunctor}
    \mathcal{C}(-,A)
    \;:\;
    \xymatrix@R=-2pt{
      \mathcal{C}^{\mathrm{op}}
      \ar[r]
      &
      \mathrm{Sets}
      \\
      X
      \ar@{}[r]|-{\longmapsto}
      &
      \mathcal{C}(X,A)
      \,.
    }
  \end{equation}
\end{notation}
Elementary as this is, these hom-functors are of profound interest
in the case that $\mathcal{C}$ is the {\it homotopy category}
$\mathrm{Ho}\big(\mathbf{H}\big)$
of (the pointed objects in) a {\it model topos} ($\infty$-topos) $\mathbf{H}$
(see \cite[A.43]{FSS20c}\cite[\S 2.1]{SS20b} for more pointers), such as
is the case for the classical homotopy category
(Notation \ref{ClassicalHomotopyCategory} below): Then the
contravariant hom-functors \eqref{ContravariantHomFunctor}
are {\it non-abelian generalized cohomology theories}
\eqref{NonabelianCohomologyInIntroduction}, see
Def. \ref{ReducedNonAbelianCohomologyTheory} below.
\begin{notation}[Adjunctions]
  \label{Adjunctions}
  For $\mathcal{C}, \mathcal{D} \,\in\, \mathrm{Categories}$
  (Notation \ref{Categories}) a pair of functors
  $L : \xymatrix{\mathcal{D} \ar@{<->}[r] & \mathcal{C}} : R$
  is an {\it adjunction} with $L$ {\it left adjoint} and
  $R$ {\it right adjoint} if there is a natural isomorphism
      \vspace{-2mm}
  \begin{equation}
    \label{PairOfAdjointFunctors}
    \mathcal{C}
    \big(
      L(X)
      \,,\,
      A
    \big)
    \;\simeq\;
    \mathcal{D}
    \big(
      X
      \,,\,
      R(A)
    \big)
    \,,
    \;\;
    \;\;
    \mbox{to be denoted:}
    \;\;
    \;\;\;\;
    \;\;
    \xymatrix{
      \mathcal{C}
      \ar@<+6pt>@{<-}[rr]^-{L}_-{ \bot }
      \ar@<-6pt>[rr]_-{R}
      &&
      \mathcal{D}
      \,.
    }
  \end{equation}
\end{notation}

\medskip

\noindent {\bf The point at infinity.}
In applications to physics,
the role of {\it pointed} topological spaces,
so fundamental to (stable) homotopy theory, is to
encode the crucial concept of {\it fields vanishing at infinity}.
To bring this out, we recall some basics
of pointed topology
(e.g. \cite[p. 199]{Bredon93}; a good account is in \cite{Cutler20}).

\begin{notation}[Pointed topological spaces]
  \label{PointedTopologicalSpaces}
  We write
      \vspace{-2mm}
  \begin{equation}
    \label{CategoriesOfTopologicalSpaces}
    \mathrm{Spaces}
    \,,
    \mathrm{Spaces}^{\ast/}
    \;\;\;
    \in
    \;
    \mathrm{Categories}
  \end{equation}

      \vspace{-2mm}
\noindent
  for a convenient category of topological spaces
  (convenient in the technical sense of \cite{Steenrod67}),
  such as D-topological (i.e. Delta-generated) spaces
  (e.g. \cite[Def. 2.2]{SS20b}), and for its pointed version,
  respectively.
  The objects of the latter are spaces $X$
  equipped with a ``base''-point $\infty_X \in X$
  and the morphisms are continuous functions $f : X \to Y$
  respecting that choice: $f(\infty_X) = \infty_Y$.
  We refer to $\infty_X \in X$ as the {\it point at infinity}
  in $X$, because that is the role these points play in application
  to physics \eqref{OnePointCompactificationAndFunctionsOnIt}:
  \vspace{-2mm}
  $$
    \mathrm{Spaces}^{\ast/}
    \big(
      (X,\infty_X)
      \,,\,
      (Y,\infty_Y)
    \big)
    \;\;
    =
    \;\;
    \left\{
    \raisebox{14pt}{
    \xymatrix@R=9pt{
      X
      \ar[rr]
        ^-{
          \mbox{
            \tiny
            \color{greenii}
            \bf
            continuous function
          }
        }
        _-{
          f
        }
      &&
      Y
      \\
      \mathclap{\phantom{\vert}}
      \{\infty_X\}
      \ar@{^{(}->}[u]
      \ar[rr]
      &&
      \mathclap{\phantom{\vert}}
      \{\infty_Y\}
      \ar@{^{(}->}[u]
    }
    }
    \right\}
  $$
  The {\it convenience} of
  $\mathrm{Spaces}$ \eqref{CategoriesOfTopologicalSpaces}
  means, in particular, that this category is
  Cartesian closed, in that we have adjunctions \eqref{PairOfAdjointFunctors} of the following form:
  \vspace{-2mm}
  \begin{equation}
    \label{ProductHomAdjunctionForConvenientSpaces}
    \xymatrix@C=3em{
      \mathrm{Spaces}
      \ar@<+6pt>@{<-}[rr]^-{
        \overset{
          \mathclap{
          \raisebox{3pt}{
            \tiny
            \color{greenii}
            \bf
            product space
          }
          }
        }{
          X \times (-)
        }
      }_-{\bot}
      \ar@<-6pt>[rr]_-{
        \underset{
          \mathclap{
          \raisebox{-3pt}{
            \tiny
            \color{greenii}
            \bf
            mapping space
          }
          }
        }{
          \mathrm{Maps}(X,- )
        }
      }
      &&
      \;
      \mathrm{Spaces}
      \,,
      \;\;\;\;\;\;
      \mbox{for all $X \in \mathrm{Spaces}$}
      \,,
    }
  \end{equation}
  while $\mathrm{Spaces}^{\ast/}$ is
  closed symmetric monoidal, in that we have an adjunction
  \vspace{-2mm}
  \begin{equation}
    \label{SmashHomAdjunctionInPointedSpaces}
    \xymatrix@C=5em{
      \mathrm{Spaces}^{\ast/}
      \ar@<+6pt>@{<-}[rr]^-{
        \overset{
          \mathclap{
          \raisebox{3pt}{
            \tiny
            \color{greenii}
            \bf
            smash product space
          }
          }
        }{
          (X,\infty_X)
          \wedge
          (-)
        }
      }_-{\bot}
      \ar@<-6pt>[rr]_-{
        \underset{
          \mathclap{
          \raisebox{-3pt}{
            \tiny
            \color{greenii}
            \bf
            pointed mapping space
          }
          }
        }{
          \mathrm{Maps}^{\ast/}
          \big(
            (X,\infty_x), -
          \big)
        }
      }
      &&
      \;
      \mathrm{Spaces}^{\ast/}
      \,,
      \;\;\;\;\;\;
      \mbox{for all $(X,\infty_X) \in \mathrm{Spaces}^{\ast/}$}
      \,,
    }
  \end{equation}

  \vspace{-2mm}
  \noindent
  where $\mathrm{Maps}^{\ast/}\big( (X,\infty_X), (Y,\infty_Y)  \big)
  \,\subset\, \mathrm{Maps}(X,Y)$ is the subspace of maps that preserves
  the base points, and where

  $$
    \big(
      X, \infty_X
    \big)
    \wedge
    \big(
      Y, \infty_Y
    \big)
    \;\;\;:=\;\;\;
    \Bigg(
      \frac{
        X \times Y
      }{
        \{\infty_X\} \!\times\! Y
          \;\cup\;
        X \!\times\! \{\infty_Y\}
      }
      \,, \;
      \underset{
        \infty_{(X \wedge Y)}
      }{
      \underbrace{
      \big[
        \{\infty_X\} \!\times\! Y
          \;\cup\;
        X \!\times\! \{\infty_Y\}
      \big]
      }
      }
    \Bigg)
  $$
  is the {\it smash product}
  of pointed spaces: the quotient space
  of the product space
  where all those points that are at infinity in $X$ or in $Y$
  are identified with a single point at infinity.
  This yields the structure of a symmetric monoidal category,
  with tensor unit the 0-sphere $S^0 \,\coloneqq\, \ast \sqcup \infty$:
  \begin{equation}
    \label{SymmetricMonoidalCategoryOfPointedSpaces}
    \big(
      \PointedSpaces,
      S^0,
      \wedge
    \big)
    \;\;
    \in
    \;
    \mathrm{SymmetricMonoidalCategories}.
  \end{equation}
\end{notation}
\begin{notation}[Locally compact Hausdorff space with proper maps]
  \label{LCHSpaces}
  We write
  $
    \LCHSpaces
    \,\subset\,
    \Spaces
  $
  for the non-full subcategory of locally compact Hausdorff topological
  spaces with {\it proper} maps between them.
  We regard this as a symmetric monoidal category
  \begin{equation}
    \label{SymmetricMonoidalCategoryOfLCHSpaces}
    \big(
      \LCHSpaces,
      \ast,
      \times
    \big)
    \;\in\;
    \mathrm{SymmetricMonoidalCategories}
    \,.
  \end{equation}
  with tensor product
  given by forming topological product spaces
  (which, beware, is no longer the {\it cartesian} product
  in $\LCHSpaces$, due to the restriction of the morphisms to proper maps).
\end{notation}
\begin{prop}[Functorial one-point compactification {(e.g. \cite[Prop. 1.5]{Cutler20})}]
  \label{OnePointCompactification}
  On $\mathrm{LCHSpaces} \subset \mathrm{Spaces}$
  (Notation \ref{LCHSpaces})
  the construction of one-point compactifications \eqref{OnePointCompactificationAndFunctionsOnIt}
  extends to a functor:
  \vspace{-2mm}
  \begin{equation}
    \label{OnePointCompactificationFunctor}
    \xymatrix{
      \LCHSpaces
      \ar[rr]^-{ (-)_{\mathrm{cpt}} }
      &&
      \PointedSpaces
      \,.
    }
  \end{equation}
\end{prop}
\begin{example}[Adjoining infinity to a compact space]
  \label{AdjoiningInfinityToACompactSpace}
  If $X \,\in\, \mathrm{LCHSpaces}$ (Notation \ref{LCHSpaces}) is already compact,
  then its one-point compactification
  \eqref{OnePointCompactificationAndFunctionsOnIt}
  is given by disjoint union with the singleton space on the
  point at infinity, often denoted $X_+$:
  \vspace{-2mm}
  \begin{equation}
    \label{AdjoiningBasePointToACompactSpace}
    \mbox{
      $X \in \mathrm{Spaces}$ is compact
    }
    \;\;\;\;\;
    \Rightarrow
    \;\;\;\;\;
    X_{\mathrm{cpt}}
      \;=\;
    X_+
      \;:=\;
    X \sqcup \{\infty\}
    \,.
  \end{equation}

  \vspace{-2mm}
\noindent
  Hence:
  {\it In a compact space, a continuous trajectory
  does not escape to infinity}
  (unless it starts there, in which case it stays there).
\end{example}

\begin{example}[Spheres and Thom spaces as one-point compactifications]
  \label{SpheresAndThomSpacesAsOnePointCompactifications}
  $\,$

  \noindent
  {\bf (i)}  For $X = \mathbb{R}^n$ the $n$-dimensional Euclidean space,
  its one-point compactification
  \eqref{OnePointCompactificationAndFunctionsOnIt}
  is the $n$-sphere:
  \begin{equation}
    \label{PointedSpheres}
    \big( \mathbb{R}^n \big)_{\mathrm{cpt}}
    \;\simeq\;
    S^n
    \;\;\;
    \in
    \;
    \mathrm{Spaces}^{\ast/}
    \,.
  \end{equation}

  \vspace{-2mm}
  \noindent { \bf (ii)}  Notice, for $n = 0$,
  that the 0-sphere is the two-point space,
  by \eqref{AdjoiningBasePointToACompactSpace}:
   \vspace{-2mm}
  \begin{equation}
    \label{TheZeroSphere}
    S^0 \;\simeq\; \ast_{\mathrm{cpt}} \;\simeq\; \ast_+ \;\simeq\; \ast \sqcup \{\infty\}
    \,.
  \end{equation}

 \vspace{-2mm}
  \noindent { \bf (iii)}  More generally,
  for $\mathcal{V}_X \to X$ a real vector bundle,
  over a {\it compact} topological space,
  its one-point compactification
  is equivalently its  {\it Thom space}
  $\mathrm{Th}(\mathcal{V}_X)$,
  the quotient space of the unit disk bundle
  by the unit sphere bundle in $\mathcal{V}_X$
  (with respect to any choice of
  fiber metric).
  \begin{equation}
    \label{ThomSpaceAsQuotientOfDiskBundleBySphereBundle}
    \mbox{$X$ compact}
    \;\;\;
      \Rightarrow
    \;\;\;
    \big(
      \mathcal{V}_X
    \big)_{\mathrm{cpt}}
    \;\;
    \;
    \simeq
    \;
    \mathrm{Th}
    \big(
      \mathcal{V}_X
    \big)
    \;:=\;
    D\big( \mathcal{V}_X\big)
    /
    S\big( \mathcal{V}_X\big)
    \;\;\;
    \in
    \;
    \mathrm{Spaces}^{\ast/}
    \,.
  \end{equation}
\end{example}

\begin{prop}[One-point compactification is strong monoidal (e.g. {\cite[Prop. 1.6]{Cutler20}})]
  \label{CompactificationOfProductSpace}
  The one-point compactification functor \eqref{OnePointCompactificationFunctor}
  is strong monoidal with respect to
  \eqref{SymmetricMonoidalCategoryOfPointedSpaces}
  and
  \eqref{SymmetricMonoidalCategoryOfLCHSpaces}:
   \vspace{-2mm}
  $$
    (-)_{\mathrm{cpt}}
    \;:\;
    \xymatrix{
      \big(
        \LCHSpaces, \ast, \times
      \big)
      \ar[rr]
      &&
      \big(
        \PointedSpaces, S^0, \wedge
      \big)
    }
  $$

  \vspace{-2mm}
  \noindent
  in that there is a natural isomorphism between the
  one-point compactification of a product space and
  the smash product of the compactification of the two factors:
   \vspace{-1mm}
  \begin{equation}
    \label{CompactificationOfProductSpaceIsomorphicToSmashProduct}
    X,Y \,\in\, \LCHSpaces \,\subset\, \Spaces
    \;\;\;\;
      \Rightarrow
    \;\;\;\;
    \left\{
    \begin{aligned}
      \big(
        X \times Y
      \big)_{\mathrm{cpt}}
      &
      \;\simeq\;
      X_{\mathrm{cpt}} \wedge Y_{\mathrm{cpt}}
      \\
      \big(
        X \sqcup Y
      \big)_{\mathrm{cpt}}
      &
      \;\simeq\;
      X_{\mathrm{cpt}} \vee Y_{\mathrm{cpt}}
    \end{aligned}
    \right.
    \;\;
    \in
    \;
    \PointedSpaces
    \,.
  \end{equation}
\end{prop}

\medskip

\noindent {\bf Homotopy types.}
\begin{notation}[Classical homotopy category]
  \label{ClassicalHomotopyCategory}
  We write
   \vspace{-2mm}
  \begin{equation}
    \label{ClasicalHomotopyCategories}
    \xymatrix{
      \Spaces
      \ar[r]^-{ \gamma }
      &
      \HoSpaces
    }
    \,,\;
    \xymatrix{
      \PointedSpaces
      \ar[r]^-{ \gamma^{\ast/} }
      &
      \HoPointedSpaces
    }
    \;\;\;
    \in
    \;
    \mathrm{Categories}
  \end{equation}

   \vspace{-2mm}
 \noindent for the classical homotopy categories of
  (pointed) topological spaces
  (e.g. \cite[(338)]{FSS20c}), i.e.
  the localization of the categories of spaces
  (Def. \ref{PointedTopologicalSpaces})
  at the weak homotopy equivalences.
  We say that an object of
  \eqref{ClasicalHomotopyCategories} is a (pointed)
  {\it homotopy type}.
\end{notation}
\begin{example}[Pointed mapping spaces are homotopy fiber of evaluation map]
  \label{PointedMappingSpacesAreHomotopyFiberOfEvaluationMap}
  For $X, Y \,\in\, \HoPointedSpaces$
  (Notation \ref{ClassicalHomotopyCategory})
  their pointed mapping space \eqref{SmashHomAdjunctionInPointedSpaces}
  is the homotopy fiber (e.g. \cite[Def. A.23]{FSS20c})
  of the evaluation map out of the plain mapping space
  \eqref{ProductHomAdjunctionForConvenientSpaces}
  at the base-point:
  \vspace{-2mm}
  $$
    \xymatrix{
      \mathrm{Maps}^{\ast/}(X,Y)
      \ar[rr]^-{ F }
      &&
      \mathrm{Maps}(X,Y)
      \ar[rr]^-{ \mathrm{ev}_{\ast_X} }
      &&
      Y
    }
  $$

  \vspace{-2mm}
  \noindent
  Moreover,
  if $\pi_0(Y) = \ast$
  and $\pi_1(Y) = 1$, then the forgetful map
  $F$ is
  a weak homotopy equivalence (e.g. \cite[Prop. A4.2]{Hatcher01}).
\end{example}

\begin{example}[Homotopy groups as homs in the classical homotopy category]
  \label{HomotopyGroupsAsHomSetsOInTheClassicalHomotopyCategory}
  For $n \in \mathbb{N}$, with the $n$-sphere $S^n$
  \eqref{PointedSpheres} regarded as
  a pointed topological space (Def. \ref{ClassicalHomotopyCategory}),
  and for any $X \in \mathrm{Spaces}^{\ast/}$
  we have a natural identification
  of the $n$th homotopy group/set of $X$ (at its base-point)
  as the hom-set from $S^n$ to $X$ in the pointed homotopy category
  (Notation \ref{ClassicalHomotopyCategory}):
  \begin{equation}
    \label{HomotopyGroupsAsHoms}
    \begin{aligned}
    \pi_{n \geq 2}(X)
    & \simeq\;
    \HoPointedSpaces
    \big(
      S^n,\,
      X
    \big)
    \;\;\;\;
    \in
    \;
    \mathrm{AbelianGroups}
    \\
    \pi_1(X)
    & \simeq\;
    \HoPointedSpaces
    \big(
      S^1,\,
      X
    \big)
    \;\;\;\;
    \in
    \;
    \mathrm{Groups}
    \\
    \pi_0(X)
    &
    \simeq\;
    \HoPointedSpaces
    \big(
      S^0,\,
      X
    \big)
    \;\;\;\;
    \in
    \;
    \mathrm{Sets}
    \end{aligned}
  \end{equation}
\end{example}

\begin{example}[Suspension/Looping adjunction]
  \label{UnstableLoopingSuspensionAdjunction}
  The smash/hom-adjunction \eqref{SmashHomAdjunctionInPointedSpaces}
  descends to its derived adjunction
  (e.g. \cite[Prop. A.20]{SS20b}) on the
  classical homotopy category \eqref{ClasicalHomotopyCategories},
   \vspace{-2mm}
   \begin{equation}
    \xymatrix@C=5em{
      \HoPointedSpaces
      \ar@{<-}@<+6pt>[rr]^-{
        \overset{
          \raisebox{3pt}{
            \tiny
            \color{greenii}
            \bf
            derived smash product
          }
        }{
          \mathbb{L}
          \left(
            X
            \wedge
            (-)
          \right)
        }
      }_-{ \bot }
      \ar@<-6pt>[rr]_-{
        \underset{
          \mathclap{
          \raisebox{-3pt}{
            \tiny
            \color{greenii}
            \bf
            derived pointed mapping space
          }
          }
        }{
        \mathbb{R}
        \mathrm{Maps}^{\ast/}
        \!
        (
          X,-
        )
        }
      }
      &&
      \;
      \mathrm{Ho}
      \big(
        \mathrm{Spaces}^{\ast/}
      \big)
      \,,
    }
    \;\;\;\;\;\;
    \mbox{for all $X  \in \mathrm{Ho}\big(\mathrm{Spaces}^{\ast/}\big)$}
    \,.
  \end{equation}

   \vspace{-2mm}
  \noindent
  For $X$ pointed $n$-sphere \eqref{PointedSpheres}
  this is the derived {\it suspension/looping adjunction}:
   \vspace{-2mm}
  \begin{equation}
    \label{UnstableLoopingSuspensionAdjointFunctors}
    \xymatrix@C=3.5em{
      \mathrm{Ho}
      \big(
        \mathrm{Spaces}^{\ast/}
      \big)
      \ar@{<-}@<+6pt>[rr]^-{
        \overset{
          \raisebox{3pt}{
            \tiny
            \color{greenii}
            \bf
            reduced suspension
          }
        }{
          \Sigma (-)
          \,:=\,
          \mathbb{L}
          \big(
            S^1 \wedge (-)
          \big)
        }
      }_-{ \bot }
      \ar@<-6pt>[rr]_-{
        \underset{
          \raisebox{-3pt}{
            \tiny
            \color{greenii}
            \bf
            based loop space
          }
        }{
          \Omega(-)
          \,:=\,
          \mathbb{R}
          \mathrm{Maps}^{\ast/\!}
          (S^1,-)
        }
      }
      &&
      \;
      \mathrm{Ho}
      \big(
        \mathrm{Spaces}^{\ast/}
      \big)
      \,,
    }
    \;\;\;\;\;\;\;\;\;
    \xymatrix@C=3.5em{
      \mathrm{Ho}
      \big(
        \mathrm{Spaces}^{\ast/}
      \big)
      \ar@{<-}@<+6pt>[rr]^-{
        \overset{
          \raisebox{3pt}{
            \tiny
            \color{greenii}
            \bf
            reduced $n$th suspension
          }
        }{
          \Sigma^n (-)
          \,:=\,
          S^n \wedge (-)
        }
      }_-{ \bot }
      \ar@<-6pt>[rr]_-{
        \underset{
          \raisebox{-3pt}{
            \tiny
            \color{greenii}
            \bf
            based $n$th loop space
          }
        }{
          \Omega^n
          \,:=\,
          [S^n, -]
        }
      }
      &&
    \;  \mathrm{Ho}
      \big(
        \mathrm{Spaces}^{\ast/}
      \big)
      .
    }
  \end{equation}
\end{example}

Beware that we will {\it notationally suppress}
the derived-functor notation $\mathbb{L}(-)$, $\mathbb{R}(-)$.

\begin{notation}[Delooping]
  \label{Delooping}
  For $A \in \mathrm{Ho}(\mathrm{Spaces}^{\ast/})$
  (Notation \ref{ClassicalHomotopyCategory})
  and $n \in \mathbb{N}$, we say that the
  {\it $n$-fold delooping} of $A$ is (if it exists)
  the homotopy type
  $B^n A \,\in\, \mathrm{Ho}\big(\mathrm{Spaces}^{\ast/}\big)$
  such that:

  {\bf (i)} $\pi_{\bullet \leq n}\big( B^n A \big) \,\simeq\, \ast$
  (via Example \ref{HomotopyGroupsAsHomSetsOInTheClassicalHomotopyCategory});

  {\bf (ii)} $A \simeq \Omega^n B^n A$ (via Example \ref{AdjoiningBasePointToACompactSpace}).

\end{notation}
\begin{example}[Classifying spaces]
  \label{ClassifyingSpacesAsDeloopings}
  For $G$ a compact topological group, regarded as a pointed
  homotopy type $G \,\in\, \mathrm{Ho}(\mathrm{Spaces}^{\ast/})$
  (Notation \ref{ClassicalHomotopyCategory})
  via its neutral element,
  its delooping (Notation \ref{Delooping}) exists and is given
  by the classical classifying space $B G$.
  If $G$ is discrete this is also called the
  Eilenberg-MacLane space $K(G,1)$.
\end{example}

\begin{example}[Higher Eilenberg-MacLane spaces]
  \label{HigherEilenbergMacLaneSpaces}
  If the group in Example \ref{ClassifyingSpacesAsDeloopings}
  is a discrete abelian group $A$, then it admits
  arbitrary deloopings, called its higher
  Eilenberg-MacLane spaces, denoted:
   \vspace{-2mm}
  $$
    B^n A \;\simeq\; K(A,n)
    \,.
  $$
\end{example}

\medskip

\noindent {\bf Non-abelian cohomology.} As in \cite[Def. 2.1]{FSS20c}
 we say:
\begin{defn}[Reduced non-abelian cohomology, vanishing at infinity]
  \label{ReducedNonAbelianCohomologyTheory}
  For $A \,\in\, \mathrm{Ho}\big( \mathrm{Spaces}^{\ast/} \big)$
  (Notation \ref{ClassicalHomotopyCategory}):

   \noindent {\bf (i)}  We say that the
  functor on pointed spaces that it represents
   \vspace{-3mm}
  \begin{equation}
    \label{ReducedACohomologyAsFunctorOnPointedSpaces}
    \xymatrix@R=-3pt{
      \mathrm{Spaces}^{\ast/}
      \ar[r]^-{ \gamma }
      &
      \mathrm{Ho}
      \big(
        \mathrm{Spaces}^{\ast/}
      \big)^{\mathrm{op}}
      \ar[r]^-{
        \widetilde A
        (
          -
        )
      }
      &
      \mathrm{Sets}
      \\
      X \ar@{|->}[rr]
      &&
      \scalebox{0.9}{$     \mathrm{Ho}
      (\mathrm{Spaces}^{\ast/})
      (
        X,
        \,
        A
      )
      $}
    }
  \end{equation}

   \vspace{-2mm}
\noindent
  is the {\it non-abelian $A$-cohomology theory}
  in its {\it reduced} version, i.e. {\it vanishing it infinity}.

 \noindent {\bf (ii)}  This takes values in graded sets via looping \eqref{UnstableLoopingSuspensionAdjunction}:
  For $n \in \mathbb{N}$ we write
     \vspace{-2mm}
  $$
    \widetilde A^{-n}(-)
    \;:=\;
    \widetilde {
      \big(
        \Omega^n A
      \big)
    }
    (-)
    \;:=\;
    \mathrm{Ho}\big(\mathrm{Spaces}^{\ast/}\big)
    \big(
      -
      \,,\,
      \Omega^n A
    \big)
    \;\simeq\;
    \mathrm{Ho}\big(\mathrm{Spaces}^{\ast/}\big)
    \big(
      \Sigma^n(-)
      \,,\,
      A
    \big)
  $$

     \vspace{-2mm}
\noindent
  for reduced $A$-cohomology in negative degrees;
  and if there exist $n$-fold deloopings
  (Notation \ref{Delooping})
  $B^n A \,\in\, \mathrm{Spaces}^{\ast/}$, then we write
     \vspace{-1mm}
  $$
    \widetilde A^n(-)
    \;:=\;
    \widetilde{
      \big(
        B^n A
      \big)
    }
    (-)
    \;:=\;
    \mathrm{Ho}\big(\mathrm{Spaces}^{\ast/}\big)
    \big(
      -
      \,,\,
      B^n A
    \big)
  $$

     \vspace{-1mm}
\noindent
  for reduced $A$-cohomology in the respective positive degrees.
\end{defn}

\begin{example}[Ordinary non-abelian cohomology]
  \label{OrdinaryNonAbelianCohomology}
  For $A = G$ a compact topological group, with
  delooping given by its classifying space $B G$
  (Example \ref{ClassifyingSpacesAsDeloopings}),
  the resulting non-abelian cohomology in degree 1
  (Def. \ref{ReducedNonAbelianCohomologyTheory}) is
  the classical non-abelian $G$-cohomology which on
  compact spaces with a disjoint base point \eqref{AdjoiningBasePointToACompactSpace}
  classifies principal $G$-bundles:
  \begin{equation}
    \label{ClassicalNonabelianCohomologyInDegreeOne}
    \widetilde{(B G)}(X_+)
    \;\simeq\;
    H^1(X;\,G)
    \;\simeq\;
    G\mathrm{Bund}(X)_{/\sim}
    \;\;\;
    \in
    \;
    \mathrm{Sets}
    \,.
  \end{equation}

  Notice that in degree 0 the classical non-abelian cohomology
  of the point is the group of connected components of $G$,
  by \eqref{HomotopyGroupsAsHoms}:
  \begin{equation}
    \label{ClassicalNonabelianCohomologyOfPointInDegreeZero}
    \widetilde G(\ast_+)
    \;\simeq\;
    H^0(\ast;\, G)
    \;\simeq\;
    \pi_0(G)
    \;\;\;
    \in
    \;
    \mathrm{Groups}
    \,.
  \end{equation}
\end{example}

\begin{defn}[Reduced Cohomotopy]
  \label{UnstableCohomotopyVanishingAtInfinity}
  For $n \in \mathbb{N}$ and $A := S^n$
  the homotopy type of the pointed $n$-sphere \eqref{PointedSpheres},
  the corresponding reduced non-abelian cohomology theory
  (Def. \ref{ReducedNonAbelianCohomologyTheory}) is reduced
  {\it Cohomotopy theory} in degree $n$, denoted:
     \vspace{-1mm}
  \begin{equation}
    \label{UnstableCohomotopy}
    {\widetilde \pi}{}^n(-)
    \;:=\;
    \widetilde S^n(-)
    \;:=\;
    \mathrm{Ho}\big(\mathrm{Spaces}^{\ast/}\big)
    (
      -
      ,\,
      S^n
    )
    \,.
  \end{equation}
\end{defn}
\begin{remark}[Comparing reduced and unreduced Cohomotopy]
  \label{ComparingReducedAndUnreducedCohomotopy}
  By Example \ref{PointedMappingSpacesAreHomotopyFiberOfEvaluationMap},
  reduced and unreduced Cohomotopy sets differ at most in degree 0:
  For $X$ a topological space equipped with a point $\infty \in X$,
  the forgetful map
  from the reduced Cohomotopy of $(X,x)$
  -- hence the Cohomotopy of $X$ that {\it vanishes at infinity} --
  to the plain Cohomotopy of $X$
  is an isomorphism:

  \vspace{-.5cm}
  \begin{equation}
    \label{ReducedCohomotopyIsomorphicToUnreducedCohomotopy}
    \xymatrix{
      {\widetilde \pi}{}^{n \geq 1}
      \big(
        (X, \infty)
      \big)
      \ar[r]
        ^-{ \simeq }
      &
      \pi^{\geq 1}
      (
        X
      )
      \,.
    }
  \end{equation}
\end{remark}

\begin{example}[Cohomotopy of flat space vanishing at infinity is homotopy groups of spheres]
  By combining Example \ref{HomotopyGroupsAsHomSetsOInTheClassicalHomotopyCategory}
  with Example \ref{SpheresAndThomSpacesAsOnePointCompactifications}
  and Remark \ref{ComparingReducedAndUnreducedCohomotopy},
  we see that the Cohomotopy vanishing at infinity
  (Example \ref{UnstableCohomotopyVanishingAtInfinity})
  of Euclidean spaces is given by homotopy groups of spheres:
  \begin{equation}
    \label{CohomotopyVanishingAtInfinityOfEuclideanSpacesIfHomotopyGroupsOfSpheres}
    \mathllap{
      \mbox{
        \tiny
        \color{darkblue}
        \bf
        \begin{tabular}{c}
          reduced Cohomotopy
          \\
          of Euclidean spaces
          \\
          vanishing at infinity
        \end{tabular}
      }
    }
    {\widetilde \pi}{}^{n}
    \left(
      \big(
        \mathbb{R}^d
      \big)_{\mathrm{cpt}}
    \right)
    \;
    \underset{
      n \geq 1
    }{
      \simeq
    }
    \;
    \pi^{n}
    \left(
      \big(
        \mathbb{R}^d
      \big)_{\mathrm{cpt}}
    \right)
    \;\simeq\;
    \pi^n(
      S^d
    )
    \;=\;
    \pi_d
    (
      S^n
    )
    \mathrlap{
      \mbox{
        \tiny
        \color{darkblue}
        \bf
        \begin{tabular}{c}
          homotopy groups
          \\
          of spheres
        \end{tabular}
      }
    }
    \,.
  \end{equation}
\end{example}

\medskip

\subsection{Pontrjagin-Thom construction}

\noindent
{\bf Cohomotopy charge of Cobordism classes of normally-framed submanifolds.}
We briefly recall {\it Pontrjagin's theorem}
(Prop. \ref{PontrjaginIsomorphism} below)
identifying the $n$-Cohomotopy of a smooth manifold with the
cobordism classes of its normally framed submanifolds in codimension $n$.
Ponrjagin's construction
has come to be mostly known as the
{\it Pontrjagin-Thom construction},
after Thom considered the variant for
orientation structure and Lashof generalized both to
any choice of tangential structure
(see Remark \ref{AttributionsForPontrjaginTheorem} for historical references).
Review of the original Pontrjagin theorem
(Prop. \ref{PontrjaginIsomorphism})
may be found, apart from the original
exposition \cite{Pontrjagin55},
in \cite[\S B]{FreedUhlenbeck91}\cite[\S IX]{Kosinski93}
\cite[\S 7]{Milnor97}\cite[\S 16]{BestvinaKeenan02}\cite{Kervaire11}\cite[\S 1]{Sadykov13}\cite{Csepai20};
for illustration see \hyperlink{FigureD}{Figure D}.

\medskip

\begin{defn}[Normally framed cobordism {\cite[\S 6]{Pontrjagin55}\cite[\S IX.2]{Kosinski93}}]
  \label{NormallyFramedCobordism}
  Let $M^D$ be a smooth manifold.

  \noindent
  {\bf (i)}
  A {\it normally-framed submanifold}
  of $M^d$
  is a smooth submanifold
  $\Sigma^d$ of $M^D$
  equipped with a trivialization of its normal bundle
  $N \Sigma^d \coloneqq T_{\Sigma} M /_{\!\Sigma} T \Sigma $:

  \vspace{-.4cm}
  \begin{equation}
  \label{TrivializationOfNormalVectorBundle}
  \xymatrix@R=-4pt@C=4em{
    \overset{
      \mathclap{
      \raisebox{4pt}{
        \tiny
        \color{greenii}
        \bf
        smooth submanifold
        }
      }
    }{
      \Sigma^{d} \hookrightarrow X^D
    }
    \,,
    {\phantom{AAAAA}}
    \underset{
      \mathclap{
      \mbox{
        \tiny
        \color{darkblue}
        \bf
        \begin{tabular}{c}
          normal bundle
        \end{tabular}
      }
    }
    }{
      N\Sigma^d
    }
    \quad
    \ar[r]
      ^-{
        \mbox{\tiny \color{greenii} \bf normal framing}
        \mathclap{\phantom{\vert_{\vert}}}
      }
      |-{
        \;\simeq\;
      }
      _-{
        \mathclap{\phantom{\vert^{\vert}}}
        \mathrm{fr}
      }
    &
    \underset{
      \mbox{
        \tiny
        \color{darkblue}
        \bf
        \begin{tabular}{c}
          trivial vector bundle
          \\
          of rank $n \coloneqq D - d$
        \end{tabular}
      }
    }{
      \Sigma \times \mathbb{R}^{n}
    }
  }
\end{equation}
\vspace{-.5cm}

\noindent
{\bf (ii)} A {\it bordism}
\begin{equation}
  \label{BordismBetweenNormallyFramedSubmanifolds}
  \xymatrix{
    (\Sigma^d_0,\mathrm{fr}_0)
    \ar@{=>}[r]
      ^-{
        (\widehat \Sigma, \widehat{\mathrm{fr}})
      }
    &
    (\Sigma^d_1,\mathrm{fr}_1)
  }
\end{equation}
between a pair of {\it closed} (i.e. compact without boundary)
such normally-framed submanifolds \eqref{TrivializationOfNormalVectorBundle}
is a normally-framed submanifold
of the cylinder $M^D \times \mathbb{R}^1$
whose boundary at $i = 0,1 \in \mathbb{R}^1$ coincides
with $(\Sigma_i,\mathrm{fr}_i)$
(illustration on p. \pageref{AntiBranesAndNormalStructure}).

\noindent
The {\it relation} on normally framed submanifolds of
having a bordism \eqref{BordismBetweenNormallyFramedSubmanifolds}
between them is called {\it cobordism}. \footnote{
  In \cite[p. 42, Def. 3]{Pontrjagin55}
  cobordant submanifolds are called ``homologous'',
  thinking of Cobordism as a variant of singular homology.
  The modern term ``cobordant'' is due to \cite[p. 64]{Thom54}.
}

\noindent
{\bf (iii)} We denote the set of equivalence classes of the
cobordism relation \eqref{BordismBetweenNormallyFramedSubmanifolds}
between normally framed submanifolds of
codimension $n = D - d$ by:
\begin{equation}
  \label{CobordismSetNormallyFramedSubmanifolds}
  \mathrm{Cob}^n_{\mathrm{Fr}}
  \big(
    M^D
  \big)
  \;\coloneqq\;
    \left\{
    \!\!\!\!\!\!\!\!
    \mbox{
      \raisebox{2pt}{\footnotesize
      \begin{tabular}{c}
        Closed submanifolds $\Sigma^d \overset{}{\hookrightarrow} X^D$
        \\
        of dimension $d = D - n$
        \\
        and equipped with normal framing
      \end{tabular}
      }
    }
   \!\!\!\!\!\!\!\!\!
   \right\}_{\!\!\!\big/\mathrm{cobordism}}
   \,.
\end{equation}
\end{defn}

\begin{defn}[Cohomotopy charge sourced by submanifolds {\cite[p. 48]{Pontrjagin55}\cite[p. 179]{Kosinski93}}]
  \label{CohomotopyCharge}
  Let $\Sigma^d \hookrightarrow M^D$ be a smooth submanifold
  embedding.

  \noindent
  {\bf (i)}
  A {\it tubular neighborhood} of the submanifold
  is an extension
  $\exp_\Sigma$
  to an embedding of the normal bundle $N \Sigma$ into $M^D$.
  Its inverse $\exp^{-1}_\Sigma$, when
  extended to the one-point compactifications
  \eqref{OnePointCompactificationAndFunctionsOnIt}
  where it is known as the {\it Pontrjagin-collapse map}
  --
  regards
  all points in $M^D$ not inside this tubular neighborhood
  as being {\it at infinity}:
  \begin{equation}
    \label{TubularNeighborhoodAndPontrjaginCollapse}
    \raisebox{50pt}{
    \xymatrix@R=4pt@C=5em{
      &
      \;
      \Sigma
      \;
      \ar@{_{(}->}[dl]
        _-{
          \mbox{
            \tiny
            \color{darkblue}
            \bf
            submanifold
          }
          \;\;\;
        }
      \ar@{^{(}->}[dr]
      \\
      M^D
      \;
      \ar@{<-_{)}}[rr]
        ^-{
          \exp_\Sigma
          \;\;\;\;\;\;\;\;\;\;\;
        }
        _-{
          \mbox{
            \tiny
            \color{greenii}
            \bf
            \begin{tabular}{c}
              tubular
              neighborhood
            \end{tabular}
          }
        }
      &&
      \;\;
      N\Sigma^d
      \!\!\!
      \mathrlap{
        \mbox{
          \tiny
          \color{darkblue}
          \bf
          \begin{tabular}{c}
            normal
            \\
            bundle
          \end{tabular}
        }
      }
      \\
      \big(
        M^D
      \big)_{\mathrm{cpt}}
      \ar[rr]
        _-{
          \exp^{-1}_\Sigma
        }
        ^-{
          \mbox{
            \tiny
            \color{greenii}
            \bf
            Pontrjagin collapse
          }
        }
      &&
      \big( N \Sigma \big)_{\mathrm{cpt}}
      \\
      x
      \ar@{}[rr]
        |-{\longmapsto}
      &&
      \left\{
      \scalebox{0.75}{${
      \arraycolsep=2pt
      \begin{array}{lcl}
        \exp_{\Sigma}^{-1}(x)
        &\vert& x \in \mathrm{image}(\exp_\Sigma)
        \\
        \infty &\vert& \mbox{otherwise}
      \end{array}
      }$}
      \right.
    }
    }
  \end{equation}

  \noindent
  {\bf (ii)}
  Given, moreover,
  a normal framing $\mathrm{fr}$ on $\Sigma$
  (Def. \ref{NormallyFramedCobordism}),
  the {\it Cohomotopy charge} of
  the normally-framed submanifold $(\Sigma,\mathrm{fr})$
  is the homotopy class of the composite
  of the collapse \eqref{TubularNeighborhoodAndPontrjaginCollapse}
  with the normal framing \eqref{TrivializationOfNormalVectorBundle},
  regarded as a class
  in reduced $n$-Cohomotopy (Def. \ref{UnstableCohomotopyVanishingAtInfinity}):
  \begin{equation}
    \label{CohomotopyChargeConstruction}
    \hspace{-5mm}
    \Big[\!\!
    \xymatrix{
      (M^D)_{\mathrm{cpt}}
      \ar[rr]
        ^-{
          \exp^{-1}_\Sigma
        }
        \ar@/_1.6pc/[rrrrrrr]
          |-{
            \;
            \mbox{
              \tiny
              \color{greenii}
              \bf
              Cohomotopy charge
              of $(\Sigma, \mathrm{fr})$
            }
            \;
          }
      &&
      \big(
        N\Sigma
      \big)_{\mathrm{cpt}}
      \ar[rr]
        ^-{
          (\mathrm{fr})_{\mathrm{cpt}}
        }
        _-{
          \simeq
        }
      &&
      \big(
        \Sigma \times \mathbb{R}^n
      \big)_{\mathrm{cpt}}
      \ar[rr]
        ^-{
          (\mathrm{pr}_2)_{\mathrm{cpt}}
        }
      &&
      (\mathbb{R}^n)_{\mathrm{cpt}}
      \ar@{=}[r]
      &
      S^n
    } \!\!
    \Big]
    \;
    \in
    \;
    \widetilde{\pi}{}^n
    \Big(\!\!
      \big(
        M^D
      \big)_{\mathrm{cpt}}
    \Big)
    \,,
  \end{equation}

  \newpage

  \noindent where we used the functoriality of one-point compactification
  (Prop. \ref{OnePointCompactification})
  on proper maps (since $\Sigma$ is assumed to be closed, hence compact)
  and the fact \eqref{PointedSpheres}
  that the one-point compactification of
  Euclidean $n$-space is the $n$-sphere
  (Example \ref{SpheresAndThomSpacesAsOnePointCompactifications}).

  \noindent
  {\bf (iii)}
  This construction is clearly independent, up to
  homotopy of the resulting Cohomotopy charge cocycle
  \eqref{CohomotopyChargeConstruction},
  of {(a)} the choice $\exp_\Sigma$
  of tubular neighborhood \eqref{TubularNeighborhoodAndPontrjaginCollapse}
  and {(b)} the deformation of $(\Sigma,\mathrm{fr})$ along
  bordisms \eqref{BordismBetweenNormallyFramedSubmanifolds}, and
  hence constitutes a function on the
  Cobordism set \eqref{CobordismSetNormallyFramedSubmanifolds}:
  \begin{equation}
    \xymatrix{
      \mathllap{
        \mbox{
          \tiny
          \color{darkblue}
          \bf
          \begin{tabular}{c}
            normally-framed
            \\
            Cobordism
          \end{tabular}
        }
      }
      \mathrm{Cob}^n_{\mathrm{Fr}}
      \big( M^D \big)
      \ar[rr]
        ^-{
          \mbox{
            \tiny
            \color{greenii}
            \bf
            \begin{tabular}{c}
              Cohomotopy charge
              \\
              assignment
            \end{tabular}
          }
        }
      &&
      {\widetilde \pi}{}^n
      \Big(
        \big(
          M^D
        \big)_{\mathrm{cpt}}
      \Big)
      \mathrlap{
        \mbox{
          \tiny
          \color{darkblue}
          \bf
          \begin{tabular}{c}
            reduced
            \\
            Cohomotopy
          \end{tabular}
        }
      }
    }
  \end{equation}
\end{defn}

\begin{defn}[Submanifolds sourcing Cohomotopy charge {\cite[p. 44, Def. 4]{Pontrjagin55}}]
  \label{SubmanifoldsSourcingCohomotopyCharge}
  For $M^D$ a smooth manifold and
  $[c] \in {\widetilde \pi}{}^n\big( \big( M^D \big)_{\mathrm{cpt}} \big)$
  a class in the reduced $n$-Cohomotopy set
  (Def. \ref{UnstableCohomotopyVanishingAtInfinity})
  of its one-point compactification
  \eqref{OnePointCompactificationAndFunctionsOnIt},
  let $M^d \xrightarrow{c_{\mathrm{reg}}} S^n$
  be a representative for
  which $0 \in \big(\mathbb{R}^n\big)_{\mathrm{cpt}} = S^n$
  is a regular value.
  This means that the pre-image
  $\Sigma \coloneqq c^{-1}(0) \subset M^D$
  is a smooth submanifold, and that there is
  an open ball neighborhood
  $\mathbb{R}^n \simeq N\{0\} \subset \big(\mathbb{R}^n\big)_{\mathrm{cpt}} = S^n$
  of $0$
  whose pre-image is a tubular neighborhood
  \eqref{TubularNeighborhoodAndPontrjaginCollapse}
  of $\Sigma$,
  thereby equipped with a normal framing
  \eqref{TrivializationOfNormalVectorBundle},
  as exhibited by the following pasting composite of
  pullbacks of smooth manifolds:
  \begin{equation}
    \label{SourceManifoldFromCohomotopyCocycleViaPullbacks}
    \hspace{-1cm}
      \xymatrix@R.8em@C=2.5em{
        \mbox{
          \tiny
          \color{darkblue}
          \bf
          \begin{tabular}{c}
            submanifold
          \end{tabular}
        }
        &
        \Sigma
        \mathclap{\phantom{\vert_{\vert}}}
        \ar[rr]
        \ar@{}[ddrr]
          |-{
            \mbox{
              \tiny (pb)
            }
          }
        \ar@{^{(}->}[dd]
          _-{
            (\mathrm{id}, 0)
          }
        &{\phantom{AAAAAA}}&
        \{0\}
        \mathclap{\phantom{\vert_{\vert}}}
        \ar@{^{(}->}[dd]
        \\
        \\
        \ar@{}[dd]
          |-{
            \mbox{
              \tiny
              \color{greenii}
              \bf
              \begin{tabular}{c}
                normal
                \\
                framing
              \end{tabular}
            }
          }
        &
        \Sigma \times \mathbb{R}^n
        \ar[dd]_-{
          \mathrm{fr}^{-1}
        }^-{ \simeq }
        \ar@{}[ddrr]|-{ \mbox{ \tiny (pb)} }
        \ar[rr]
          |-{
            \;\mathrm{pr}_2\;
          }
        &&
        \mathbb{R}^n
        \ar[dd]^{ \simeq }
        \\
        \\
        \ar@{}[dd]
          |-{
            \mbox{
              \tiny
              \color{greenii}
              \bf
              \begin{tabular}{c}
                tubular
                \\
                neighborhood
              \end{tabular}
            }
          }
        &
        N\Sigma
        \mathclap{\phantom{\vert_{\vert}}}
        \ar@{}[ddrr]|-{ \mbox{ \tiny  (pb)} }
        \ar[rr]
        \ar@{^{(}->}[dd]
          _-{
            \exp_\Sigma
          }
        &&
        N\{0\}
        \mathclap{\phantom{\vert_{\vert}}}
        \ar@{^{(}->}[dd]
          ^-{
            \exp_{\{0\}}
          }
        \\
        \\
        \mbox{
          \tiny
          \color{darkblue}
          \bf
          \begin{tabular}{c}
            regularized
            \\
            Cohomotopy coccyle
          \end{tabular}
        }
        &
        M^D
        \ar[rr]|-{
          \;
          c_{\mathrm{reg}}
          \;
        }_-{\ }="s"
        \ar@/_2pc/[rr]_{ c }^-{\ }="t"
        &&
        S^n
        \ar@{=>}^{\;\simeq} "s"; "t"
      }
  \end{equation}

  It is clear that
  the normally framed submanifold $(\Sigma,\mathrm{fr})$
  obtained this way gets deformed along a bordism
  \eqref{BordismBetweenNormallyFramedSubmanifolds}
  when $c$ is deformed by a homotopy, so that this
  construction constitutes a function
  \begin{equation}
    \xymatrix{
      \mathllap{
        \mbox{
          \tiny
          \color{darkblue}
          \bf
          \begin{tabular}{c}
            normally-framed
            \\
            Cobordism
          \end{tabular}
        }
      }
      \mathrm{Cob}^n_{\mathrm{Fr}}
      \big( M^D \big)
      \ar@{<-}[rr]
        ^-{
          \mbox{
            \tiny
            \color{greenii}
            \bf
            \begin{tabular}{c}
            \end{tabular}
          }
        }
        ^-{
          \mathrm{Src}
        }
      &&
      {\widetilde \pi}{}^n
      \Big(
        \big(
          M^D
        \big)_{\mathrm{cpt}}
      \Big)
      \mathrlap{
        \mbox{
          \tiny
          \color{darkblue}
          \bf
          \begin{tabular}{c}
            reduced
            \\
            Cohomotopy
          \end{tabular}
        }
      }
    }
  \end{equation}
  to the Cobordism set \eqref{CobordismSetNormallyFramedSubmanifolds}.
\end{defn}

\begin{prop}[Pontrjagin's Theorem {\cite[p. 48, Thm. 10]{Pontrjagin55}\cite{Kosinski93}\cite[p. 50]{Milnor97}}]
  \label{PontrjaginIsomorphism}
  For $M^D$ a smooth manifold, the
  construction of

  {\bf (a)} Cohomotopy charge of
  normally framed submanifolds
  (Def. \ref{CohomotopyCharge}), and

  {\bf (b)} of submanifolds sourcing Cohomotopy charge
  (Def. \ref{SubmanifoldsSourcingCohomotopyCharge})

 \noindent  are inverse isomorphisms between

  {\bf (i)} the Cobordism set \eqref{CobordismSetNormallyFramedSubmanifolds}
  of normally framed submanifolds
  of codimension $n$, and

  {\bf (ii)} the reduced $n$-Cohomotopy (Def. \ref{UnstableCohomotopyVanishingAtInfinity})
  for all $0 \leq n \leq D$:
  \begin{equation}
    \label{ThePontrjaginIsomorphism}
    \xymatrix{
      \mathllap{
        \mbox{
          \tiny
          \color{darkblue}
          \bf
          \begin{tabular}{c}
            normally-framed
            \\
            Cobordism
          \end{tabular}
        }
      }
      \mathrm{Cob}^n_{\mathrm{Fr}}
      \big( M^D \big)
      \ar@<+6pt>[rr]
        ^-{
          \mbox{
            \tiny
            \color{greenii}
            \bf
            \begin{tabular}{c}
              Cohomotopy charge
              \\
              assignment
            \end{tabular}
          }
        }
        _-{
          \mathclap{\phantom{\vert}}
          \simeq
        }
      \ar@<-6pt>@{<-}[rr]
        ^-{
          \mbox{
            \tiny
            \color{greenii}
            \bf
            \begin{tabular}{c}
            \end{tabular}
          }
        }
        _-{
          \mathrm{Src}
        }
      &&
      \;\;
      {\widetilde \pi}{}^n
      \Big(
        \big(
          M^D
        \big)_{\mathrm{cpt}}
      \Big)
      \mathrlap{
        \mbox{
          \tiny
          \color{darkblue}
          \bf
          \begin{tabular}{c}
            reduced
            \\
            Cohomotopy
          \end{tabular}
        }
      }
    }
  \end{equation}
\end{prop}

\newpage
\begin{remark}[Attributions for Pontrjagin's theorem]
  \label{AttributionsForPontrjaginTheorem}
  The statement of Prop. \ref{PontrjaginIsomorphism}
  underlies already the announcement \cite{Pontrjagin36};
  in print it is made explicit in \cite{Pontrjagin38},
  briefly and for the case $M^D = S^D$.
  But it is only after Thom's variant of the construction
  (for normally-oriented instead of normally-framed submanifolds)
  appears much later \cite{Thom54} that
  Pontrjagin gives a comprehensive account of the construction
  \cite{Pontrjagin55}, still focusing on $M^D = S^D$.
  Later, most authors consider the theorem for the case
  that $M^D$ is any closed manifold (e.g. \cite[\S IX.5]{Kosinski93}).
  The further generalization in Prop. \ref{PontrjaginIsomorphism},
  where $M^D$ need not be closed, readily follows by the same
  kind of proof; it appears stated this way in, e.g.,
  \cite[p. 50]{Milnor97}\cite[p. 1]{Moran13}\cite[p. 12-13]{Csepai20}.
  The first statement of the
  construction in the generality of arbitrary
  tangential structure
  (``(B,f)-structure'' \cite[p. 258]{Lashof63}\cite[\S 1.4]{Kochman96})
  is \cite[Thm. C]{Lashof63}:
  This general theorem by Lashof is the actual {\it Pontrjagin-Thom theorem}
  in modern parlance.
\end{remark}

\begin{example}[Cohomotopy charge of signed points {(e.g. \cite[\S IX.4]{Kosinski93})}]
  \label{CohomotopyChargeOfPoints}
  A
  normally-framed closed submanifold (Def. \ref{NormallyFramedCobordism})
  in maximal codimension $n = D$
  of an orientable manifold $M^D$
  is
  a finite subset of points
  $$
    \Sigma^0
    \,=\,
    \underset{
      1 \leq k \leq n
    }{\sqcup}
    \{x_k\}
    \;\subset\;
    M^D
    \,,
    {\phantom{AAAA}}
    \mathrm{fr}^0
    \,=\,
    (\sigma_1, \cdots, \sigma_k)
    \;\in\;
    \{\pm 1\}^n
  $$

  \vspace{-2mm}
\noindent   each equipped with a signed
  {\it unit charge} $\sigma_k \in \{ \pm 1\}$
  (the orientation of open balls around each
  point relative to the ambient orientation).
  If $M^D$ is, moreover, connected, then
  any such configuration is cobordant,
  via pair creation/annihilation
  bordisms as illustrated in \eqref{AntiBranesAndNormalStructure},
  to a subset of points
  whose framing labels are either all $+$ or all $-$;
  whence the Cobordism classes \eqref{CobordismSetNormallyFramedSubmanifolds}
  are, manifestly, classified by the integers:
  \vspace{-3mm}
  $$
    \mbox{
      $M^D$
      is connected
      \&
      orientable
    }
    \;\;\;\Rightarrow\;\;\;
    \raisebox{20pt}{
    \xymatrix@R=1.2em{
      \mathrm{Cob}^D_{\mathrm{Fr}}
      (
        M^D
      )
      \ar@{=}[dr]
        _{
          \mbox{
            \tiny
            \color{greenii}
            \bf
            \begin{tabular}{c}
              sum of
              \\
              charges
            \end{tabular}
          }
        }
      \ar@{<->}[rr]
        \mbox{
          \tiny
          Prop. \ref{PontrjaginIsomorphism}
        }
      &&
      {\widetilde \pi}^D
      \Big(
        \big(
          M^D
        \big)_{\mathrm{cpt}}
      \Big).
      \ar@{=}[dl]
        ^-{
          \!\!\!\!
          \mbox{
            \tiny
            \color{greenii}
            \bf
            \begin{tabular}{c}
              Hopf
              \\
              winding degree
            \end{tabular}
          }
        }
      \\
      &
      \mathbb{Z}
    }
    }
  $$

  \vspace{-2mm}
\noindent   Under the Pontrjagin theorem (Prop. \ref{PontrjaginIsomorphism}),
  this says that the $D$-Cohomotopy of a connected and orientable
  $D$-manifold is given by {\it winding number}:
  which is the statement of
  {\it Hopf degree theorem} (see also Example \ref{M5BraneChargeInOrdinaryCohomology}).
  For more discussion of this situation in our context
  see \cite[\S 2.2]{SS19a}.
\end{example}

\begin{example}[Cohomotopy charge of 3-Sphere in $\mathbb{R}^7$]
  \label{CohomotopyChargeOf3SphereIn7Sphere}
  The normally-framed submanifold
  of $\mathbb{R}^7 \subset S^7$ which sources
  (Def. \ref{SubmanifoldsSourcingCohomotopyCharge})
  the 4-Cohomotopy charge represented by the quaternionic
  Hopf fibration $h_{\mathbb{H}}$
  is the 3-sphere $S^3 \,\simeq\, S\mathrm{U}(2)$
  equipped with the normal framing
  induced from its $S\mathrm{U}(2)$-invariant tangential framing:
  \vspace{-2mm}
  \begin{equation}
    \label{ThreeFoldSourcingTheQuaternionicHopfFibration}
    \raisebox{10pt}{
    \xymatrix@R=-2pt{
      \mathrm{Cob}^4_{\mathrm{Fr}}
      (
        \mathbb{R}^7
      )
      \ar@{<-}[rr]
        ^-{ \mathrm{Src} }
      &&
      {\widetilde \pi}{}^4
      (
        S^7
      )
      \\
      S^3_{{}_{\mathrm{nfr}=1}}
      \ar@{}[rr]
        |-{
          \rotatebox[origin=c]{180}{
          \scalebox{.7}{$
          \mathclap{
            \longmapsto
          }
          $}
          }
        }
      &&
      \big[
        S^7 \xrightarrow{h_{\mathbb{H}}} S^4
      \big]
      \,,
    }
    }
    \phantom{AAAA}
    \raisebox{13pt}{
    \xymatrix@R=1em@C=2.5em{
      S^3
      \ar[d]
      \ar[r]
      \ar@{}[dr]
        |-{
          \mbox{
            \tiny
            (pb)
          }
        }
      &
      S^7
      \ar@(ul,ur)^-{ S\mathrm{U}(2) }
      \ar[d]
        ^-{
          h_{\mathbb{H}}
          \mathrlap{
            \mbox{
              \tiny
              \color{greenii}
              \bf
              \begin{tabular}{c}
                quaternionic
                \\
                Hopf fibration
              \end{tabular}
            }
          }
        }
      \\
      \ast
      \ar[r]
      &
      S^4
    }
    }
  \end{equation}

  \vspace{-2mm}
  \noindent
  This follows from the pullback construction
  \eqref{SourceManifoldFromCohomotopyCocycleViaPullbacks}
  by the fact that the quaternionic Hopf fibration is an
  $S\mathrm{U}(2)$-principal bundle.
\end{example}
\begin{remark}[Circle-reduction of Cohomotopy charge in $\mathbb{R}^7$]
  \label{CircleReductionOfCohomotopyChargeInR7}
  The 7-sphere $S^7$,
  besides being a $S\mathrm{U}(2)$-principal bundle
  over $S^4$ via the quaternionic Hopf fibration,
  is also a $\mathrm{U}(1)$-principal bundle over
  complex projective 3-space $\mathbb{C}P^3$,
  via the complex Hopf fibration in 7d
  as reflected in the following pasting diagram
  of pullbacks of manifolds
  (the remaining factorization shown at the bottom
is through the Atiyah-Penrose twistor fibration $t_{\mathbb{H}}$,
see Remark \ref{ComplexProjectiveSpacesOverQuaternionicProjectiveSpaces},
and see \cite[\S 2]{FSS20b} for more pointers):
\vspace{-2mm}
$$
  \xymatrix@R=1.8em@C=5em{
    S^1
    \ar[d]
    \ar[r]
    \ar@{}[dr]|-{ \mbox{\tiny\rm(pb)} }
    &
    S^3
    \ar@(ul,ur)
      ^-{
        \scalebox{.7}{$\mathrm{U}(1)$}
      }
    \ar[d]
      _-{
        \mathclap{\phantom{\vert^{\vert}}}
        h_{\mathbb{C}}
        \mathclap{\phantom{\vert_{\vert}}}
      }
    \ar[r]
    \ar@{}[dr]|-{ \mbox{\tiny(pb)} }
    &
    S^7
    \ar@(ul,ur)
      ^-{
       \scalebox{.7}{$
         \mathrm{Sp}(1)
       $}
      }
    \ar[d]
      _-{
        \mathclap{\phantom{\vert^{\vert}}}
        h_{\mathbb{C}}
        \mathclap{\phantom{\vert_{\vert}}}
      }
    \ar@/^1.5pc/[dd]
      ^-{
        h_{\mathbb{H}}
      }
    &
    \ar@{}[d]
      |-{
        \mbox{
          \tiny
          \color{greenii}
          \bf
          \begin{tabular}{c}
            complex
            \\
            Hopf fibration
          \end{tabular}
        }
      }
    \ar@{}[dd]
      |-{
        \;\;\;\;\;\;\;
        \mbox{
          \tiny
          \color{greenii}
          \bf
          \begin{tabular}{c}
            quaternionic
            \\
            Hopf fibration
          \end{tabular}
        }
      }
    \\
    \ast
    \ar[r]
    &
    S^2
    \ar[d]
    \ar[r]
    \ar@{}[dr]|-{ \mbox{\tiny(pb)} }
    &
    \mathbb{C}P^3
    \ar[d]
      _-{
        \mathclap{\phantom{\vert^{\vert}}}
        t_{\mathbb{H}}
        \mathclap{\phantom{\vert_{\vert}}}
      }
    &
    \ar@{}[d]
      |-{
          \mbox{
            \tiny
            \color{greenii}
            \bf
            \begin{tabular}{c}
              Atiyah-Penrose
              \\
              twistor fibration
            \end{tabular}
          }
      }
    \\
    &
    \ast
    \ar[r]
    &
    S^4
    &
  }
$$
Under restriction to the $S\mathrm{U}(2) \simeq S^3$-fiber
of the quaternionic Hopf fibration,
this $\mathrm{U}(1)$-action becomes that of the
standard complex Hopf fibration in 3d.
This means that
{(a)} the present discussion
has an enhancement to $\mathrm{U}(1)$-equivariant cohomology
(as in \cite{SS20b}) where a circle-action is understood throughout,
where
{(b)}
the 3-sphere in Example \ref{CohomotopyChargeOf3SphereIn7Sphere}
appears equipped with its complex Hopf-fibration structure.
We will discuss this in more detail elsewhere.
\end{remark}

\medskip

\noindent
{\bf Product in Cohomotopy.}

\begin{prop}[Product of source manifolds from product in Cohomotopy {\cite[\S 6.1]{Kosinski93}}]
  \label{ProductInCohomotopy}
  Under Pontrjagin's theorem
  (Prop. \ref{PontrjaginIsomorphism}),
  composition of Cohomotopy classes
  \vspace{-1mm}
  $$
    \xymatrix@R=-2pt{
      {\widetilde \pi}{}^{n_1}\big( S^{\,n_2 + n_1} \big)
      \times
      {\widetilde \pi}{}^{\,n_2 + n_1}
      (
        X
      )
      \ar[r]
      &
      \pi^{n_1}
      (
        X
      )
      \\
      \big(
        [c_1]
        ,
        [c_2]
      \big)
      \ar[r]
      &
      [
        c_1 \circ c_2
      ]
    }
  $$

    \vspace{-2mm}
\noindent   corresponds
  to Cartesian product of
  normally-framed submanifolds,
  in that
  \vspace{-1mm}
  $$
    \mathrm{Src}
    \big(
      [c_1 \circ c_2]
    \big)
    \;\simeq\;
    \Big[
      \mathrm{Src}(c_2) \times \mathrm{Src}(c_1)
      \;\subset\;
      N \mathrm{Src}(c_2)
      \;\subset\;
      X
    \Big].
  $$
\end{prop}
\begin{proof}
  Using the pullback-description \eqref{SourceManifoldFromCohomotopyCocycleViaPullbacks},
  this follows by the pasting law \eqref{PastingLaw},
  applied to the following pasting diagram:
  $$
    \xymatrix@R=12pt@C=4em{
      \mathllap{
        \mbox{
          \tiny
          \color{darkblue}
          \bf
          \begin{tabular}{c}
            submanifold
            \\
            classif. by $c_1 \!\circ\! c_2$
          \end{tabular}
        }
      }
      \mathclap{\phantom{\vert_{\vert}}}
      \Sigma_2 \times \Sigma_1
      \ar@{^{(}->}[dd]
      \ar[rr]
      \ar@{}[ddrr]|-{\phantom{AAA}\mbox{\tiny\rm(pb)}}
      &&
      \mathclap{\phantom{\vert_{\vert}}}
      \overset{
        \mathclap{
        \raisebox{3pt}{
          \tiny
          \color{darkblue}
          \bf
          \begin{tabular}{c}
            submfd. class. by $c_1$
          \end{tabular}
        }
        }
      }{
        \Sigma_1 \times \{0\}
      }
      \ar[dd]
      \ar[rr]
      \ar@{}[ddrr]|-{ \mbox{\rm\tiny(pb)} }
      \ar@{^{(}->}[dd]
      &&
      \mathclap{\phantom{\vert_{\vert}}}
      \{0\}
      \ar@{^{(}->}[dd]
      \\
      \\
      \mathllap{
        \mbox{
          \tiny
          \color{darkblue}
          \bf
          \begin{tabular}{c}
            normal bundle of
            \\
            submfd. class. by $c_1 \!\circ\! c_2$
          \end{tabular}
        }
      }
      \mathclap{\phantom{\vert_{\vert}}}
      \Sigma_2 \times \Sigma_1
      \times \mathbb{R}^{n_1}
      \ar@{^{(}->}[d]
      \ar@{}[drr]|-{\mbox{\tiny\rm(pb)}}
      \ar[rr]
      &&
      \mathclap{\phantom{\vert_{\vert}}}
      \Sigma_1
        \times
      \mathbb{R}^{n_1}
      \ar@{^{(}->}[d]
      \ar[rr]|-{ \;\mathrm{pr}_2\; }
      \ar@{}[ddrr]|-{\mbox{\tiny\rm(pb)}}
      &&
      \mathclap{\phantom{\vert}}
      \mathbb{R}^{n_1}
      \ar@{^{(}->}[dd]
      \\
      \mathllap{
        \mbox{
          \tiny
          \color{darkblue}
          \bf
          \begin{tabular}{c}
            normal bundle of
            \\
            submfd. class. by $c_2$
          \end{tabular}
        }
      }
      \mathclap{\phantom{\vert}}
      \Sigma_2
        \times
      \mathbb{R}^{n_2 + n_1}
      \ar@{^{(}->}[d]
      \ar[rr]|-{ \;\mathrm{pr}_2\; }
      \ar@{}[drr]
        |-{
          \mbox{\tiny\rm(pb)}
          \mathclap{\phantom{\vert_{\vert_{\vert_{\vert_{\vert}}}}}}
        }
      &&
      \mathclap{\phantom{\vert}}
      \mathbb{R}^{n_2 + n_1}
      \ar@{^{(}->}[d]
      &&
      \\
      X
      \ar[rr]^-{
          (c_2)_{\mathrm{reg}}
        }
        _-{
          \mbox{
            \tiny
            \color{greenii}
            \bf
            regularized Cohomotopy cocycle
          }
        }
      &&
      S^{n_2 + n_1}
      \ar[rr]^-{
          (c_1)_{\mathrm{reg}}
        }
        _-{
          \mbox{
            \tiny
            \color{greenii}
            \bf
            regularized Cohomotopy cocycle
          }
        }
      &&
      S^{n_1}
    }
  $$
\end{proof}

As an example we have:

\begin{example}[3-Folds in $\mathbb{R}^7$ carrying integer Cohomotopy charge]
  \label{IntegerCharged3FoldsInR7}
  The normally framed submanifolds
  of $\mathbb{R}^7$ which source
  integer-valued
  Cohomotopy charge (Def. \ref{SubmanifoldsSourcingCohomotopyCharge})
  $N \in \mathbb{N} \hookrightarrow {\widetilde \pi}{}^4(S^7)$
  \eqref{FromUnstableToStable4CohomotopyOf7Sphere}
  are those bordant \eqref{BordismBetweenNormallyFramedSubmanifolds}
  to $N$ disjoint copies of the Lie-framed 3-sphere
  \eqref{ThreeFoldSourcingTheQuaternionicHopfFibration}:
  \vspace{-2mm}
  $$
    \xymatrix@R=-1pt{
      \mathbb{Z}
      \;
      \ar@{^{(}->}[r]
      &
      \mathbb{Z} \oplus \mathbb{Z}_{12}
      \;\simeq\;
      {\widetilde \pi}{}^4\big( S^7 \big)
      \ar[rr]^-{ \mathrm{Src} }
      &&
      \mathrm{Cob}^4_{\mathrm{Fr}}
      \big(
        \mathbb{R}^7
      \big)
      \\
      N
      \ar@{}[r]
        |-{
          \overset{
            \mbox{
              \tiny
              \eqref{Universal3FluxNearM2Branes}
            }
          }{
            \longmapsto
          }
        }
      &
      \big[
          S^7
          \xrightarrow{N}
          S^7
          \xrightarrow{h_{\mathbb{H}} }
          S^4
      \big]
      \ar@{}[rr]|-{ \longmapsto }
      &&
      \Big[\;
       \raisebox{2pt}{$ \underset{N}{\bigsqcup}
        \,
        S^3_{{}_{\mathrm{nfr}=1}}
     $} \Big]
    }
  $$
  With Prop. \ref{ProductInCohomotopy},
  this follows from Example \ref{CohomotopyChargeOf3SphereIn7Sphere}
  and Example \ref{CohomotopyChargeOfPoints}.
\end{example}

\newpage

\subsection{Adams d-invariant}

From the point of view of stable homotopy theory and
generalized cohomology theory, the
{\it d-invariant} of a map
(Def. \ref{dInvariant} below)
is an elementary concept whose interest is mainly in it being
the first in a sequence
of more interesting stable invariants, proceeding with the
{\it e-invariant} (\cref{TheAdamsEInvariant})
and the {\it f-invariant}.
However, from the broader perspective of unstable
homotopy theory and of non-abelian cohomology theory
(\cref{BorsukSpanierCohomotopy}),
passage to the d-invariant of a map is, conversely,
synonymous with {\it stabilization} \eqref{StabilizedSuspensionLoopingAdjunction}
and with evaluation in multiplicative cohomology.
It is in this form that the d-invariant appears in
\cref{M5BraneChargeAndMltiplicativeCohomologyTheory}.
Therefore, we begin by quickly reviewing some basics
of stable homotopy theory, streamlined towards
our applications.

\medskip

\noindent {\bf Stable homotopy theory.}
Non-abelian cohomology (Def. \ref{ReducedNonAbelianCohomologyTheory})
is so named since its cohomology groups are
in general not abelian,
as in \eqref{ClassicalNonabelianCohomologyOfPointInDegreeZero};
in fact they are in general just sets bare of group structure,
as in \eqref{ClassicalNonabelianCohomologyInDegreeOne}.
By Example \ref{HomotopyGroupsAsHomSetsOInTheClassicalHomotopyCategory}
this is related to the suspension/looping adjunction
(Example \ref{UnstableLoopingSuspensionAdjunction}) not being an
equivalence of categories,
hence of the classical homotopy category (Notation \ref{ClassicalHomotopyCategory})
{\it not being stable under looping};
for if it were then the
$n$th connective cover of the $n$th suspension
of any space would serve as its $n$-fold delooping
(Notation \ref{Delooping}).

\medskip
This situation reflects the rich nature of the classical
homotopy category: It is {\it non-linear} in a sense that
is made precise by {\it Goodwillie calculus} \cite{Goodwillie1990}.
Ultimately we are interested in exploring the full
non-linear structure of (co)homotopy theory.
But just as in ordinary calculus
and in ordinary perturbation theory, where
the first step towards fully understanding a non-linear object
is to understand its linear approximations,
so here the first step is to consider the homotopy-linear version
of the classical homotopy category.

\medskip
Since, by the above, this involves making it {\it stable under looping},
it is known as the {\it stable homotopy category}, for short:

\begin{notation}[Stable homotopy category]
  \label{StableHomotopyCategory}
  We write
  $
    \mathrm{Ho}
    \big(
      \mathrm{Spectra}
    \big)
 $
 for the {\it stable homotopy category}
 of spectra \eqref{AnOmegaSpectrum}
 (see \cite[(350)]{FSS20c} for pointers).
  By suitable (somewhat subtle) use of the
  adjunction \eqref{SmashHomAdjunctionInPointedSpaces},
  this is again closed symmetric monoidal,
  with unit the sphere spectrum $\mathbb{S}$:
  \begin{equation}
    \label{TheStableHomotopyCategory}
    \big(
      \mathrm{Ho}
      (
        \mathrm{Spectra}
      )
      ,\,
      \mathbb{S}
      ,\,
      \wedge
    \big)
    \;\;\;
    \in
    \;
    \mathrm{SymmetricMonoidalCategories}
     \end{equation}
  in that we have adjunctions
  \begin{equation}
    \label{SmashHomAdjunctionInStableHomotopyCategory}
    \xymatrix@C=5em{
      \mathrm{Ho}(\mathrm{Spectra})
      \ar@<+6pt>@{<-}[rr]^-{
        \overset{
          \mathclap{
          \raisebox{3pt}{
            \tiny
            \color{greenii}
            \bf
            smash product spectrum
          }
          }
        }{
          E
          \wedge
          (-)
        }
      }_-{\bot}
      \ar@<-6pt>[rr]_-{
        \underset{
          \mathclap{
          \raisebox{-3pt}{
            \tiny
            \color{greenii}
            \bf
            mapping spectrum
          }
          }
        }{
          [E,-]
        }
      }
      &&
      \;
      \mathrm{Ho}(\mathrm{Spectra})
      \,,
      \;\;\;\;\;\;
      \mbox{for all $E \in \mathrm{Ho}(\mathrm{Spectra})$}
      \,.
    }
  \end{equation}

\end{notation}

  Moreover, we have the {\it stabilization adjunction}
  (e.g. \cite[Ex. A.41]{FSS20c})
  between
  the classical homotopy category (Def. \ref{ClassicalHomotopyCategory})
  and the stable homotopy category (Def. \ref{StableHomotopyCategory}):
  \begin{equation}
    \label{StabilizationAdjunction}
    \xymatrix{
      \mathrm{Ho}
      \big(
        \mathrm{Spaces}^{\ast/}
      \big)
      \ar@{<-}@<+6pt>[rr]
        ^-{ \Sigma^\infty }
        _-{ \bot }
      \ar@<-6pt>[rr]
        _-{ \Omega^\infty }
      &&
      \;
      \mathrm{Ho}
      (
        \mathrm{Spectra}
      )
      \,,
    }
    \mathrlap{
    \phantom{AAA}
    \mathbb{S} = \Sigma^\infty S^0
    \,,
    }
  \end{equation}
  which stabilizes the suspension/looping adjunction
  (Example \ref{UnstableLoopingSuspensionAdjunction})
  in that the following diagram commutes
  (both that of right adjoint functors as well as,
  and equivalently, that of left adjoint functors)
  \begin{equation}
    \label{StabilizedSuspensionLoopingAdjunction}
    \raisebox{30pt}{
    \xymatrix@R=2em@C=5em{
      \mathrm{Ho}
      \big(
        \mathrm{Spaces}^{\ast/}
      \big)
      \ar@{<-}@<+6pt>[rr]^-{
        \overset{
          \raisebox{3pt}{
            \tiny
            \color{darkblue}
            \bf
            adjunction
          }
        }{
          \Sigma^n
        }
      }_-{ \bot }
      \ar@<-6pt>[rr]_-{ \Omega^n }
      \ar@<-6pt>[dd]_-{
        \mathllap{
          \mbox{
            \tiny
            \color{darkblue}
            \bf
            \begin{tabular}{c}
              stabilization
              \\
              adjunction
            \end{tabular}
          }
          \;\;
        }
        \Sigma^\infty
      }^-{ \dashv }
      \ar@{<-}@<+6pt>[dd]^-{ \Omega^\infty }
      &&
      \mathrm{Ho}
      \big(
        \mathrm{Spaces}^{\ast/}
      \big)
      \ar@<-6pt>[dd]_-{
        \Sigma^\infty
      }^-{ \dashv }
      \ar@{<-}@<+6pt>[dd]^-{ \Omega^\infty }
      \ar@{-->}@<-6pt>[ddll]|-{
        \mathclap{\phantom{\vert^{\vert^{\vert}}}}
        \;
        \Sigma^{\infty + n}
        \;
      }
      \ar@{<--}@<+6pt>[ddll]|-{
        \mathclap{\phantom{\vert^{\vert^{\vert}}}}
        \;
        \Omega^{\infty + n}
        \;
      }
      \\
      \\
      \mathrm{Ho}
      \big(
        \mathrm{Spectra}
      \big)
      \ar@{<-}@<+6pt>[rr]^-{ \Sigma^n }_-{
        {}_{{}_{\scalebox{.7}{$\simeq$}}}
      }
      \ar@<-6pt>[rr]_-{
        \underset{
          \raisebox{-3pt}{
            \tiny
            \color{darkblue}
            \bf
            adjoint equivalence
          }
        }{
          \Omega^n
        }
      }
      &&
      \mathrm{Ho}
      \big(
        \mathrm{Spectra}
      \big)
    }
    }
  \end{equation}

\begin{example}[Stable homotopy groups and Generalized cohomology of spheres]
 \label{StableHomotopyGroupsAndCohomologyOfSpheres}
 The hom-isomorphisms \eqref{PairOfAdjointFunctors}
 of the
 stabilization square of adjunctions \eqref{StabilizedSuspensionLoopingAdjunction}
 gives the identification
 of homotopy groups of spectra
 with the reduced generalized Cohomology
 of spheres:
 \eqref{WhiteheadGeneralizedCoohomology}:
$$
  \xymatrix@=12pt{
    &
    \mathllap{
      \mbox{
        \tiny
        \color{darkblue}
        \bf
        \begin{tabular}{c}
          homotopy groups
          \\
          of spectra
        \end{tabular}
      }
      \!\!
    }
    \pi_\bullet(E)
    \ar@{=}[d]
      ^-{
        \;
        \scalebox{.6}{$
          \forall \;\; k + \bullet \geq 0
        $}
      }
    \\
    \mathllap{
      \mbox{
        \tiny
        \color{darkblue}
        \bf
        \begin{tabular}{c}
          cohomology operations
          \\
          from stable Cohomotopy
        \end{tabular}
      }
      \!\!
    }
    \mathrm{Ho}
    (
      \mathrm{Spectra}
    )
    \big(
      \Sigma^k \mathbb{S}
      ,
      \Omega^{\bullet - k } E
    \big)
    \ar@{=}[r]
    &
    \mathrm{Ho}
    (
      \mathrm{Spectra}
    )
    \big(
      \Sigma^\infty S^{\bullet + k}
      ,
      \Sigma^k E
    \big)
    \ar@{=}[d]
    \\
    &
    \mathrm{Ho}
    (
      \mathrm{Spaces}^{\ast/}
    )
    \big(
      S^{\bullet + k}
      ,
      \Omega^{\infty - k} E
    \big)
    \ar@{=}[r]
    \ar@{=}[d]
    &
    {\widetilde E}^k
    \big(
      S^{\bullet + k}
    \big)
    \mathrlap{
      \mbox{
        \tiny
        \color{darkblue}
        \bf
        \begin{tabular}{c}
          reduced $E$-cohomology
          \\
          of spheres
        \end{tabular}
      }
      \!\!
    }
    \\
    &
    \pi_{\bullet + k}
    \big(
      E^k
    \big)
    \mathrlap{
      \mbox{
        \tiny
        \color{darkblue}
        \bf
        \begin{tabular}{c}
          homotopy groups of
          \\
          component spaces
        \end{tabular}
      }
      \!\!
    }
    &
  }
$$
\end{example}

\noindent
We will need the following deep fact about stable homotopy theory:

\begin{prop}[Fiber sequences of spectra coincide with cofiber sequences
(e.g. {\cite[\S III, Thm. 2.4]{LMS86}})]
  \label{InSpectraFiberSequencesAreCofiberSequences}
  $\,$

  \noindent
  A sequence
  \vspace{-2mm}
  $$
    \xymatrix{ \cdots \ar[r] &  E \ar[r] & F \ar[r] & G \ar[r] & \cdots }
  $$

\vspace{-1mm}
\noindent
  of spectra is a homotopy fiber sequence
  if and only if it is a homotopy cofiber sequence:
  \vspace{-2mm}
  \begin{equation}
    \label{HomotopyCartesianCoFiberSquaresOfSpectra}
    \raisebox{20pt}{
    \xymatrix@R=1.5em{
      E
      \ar[d]
        _-{
          \mathclap{\phantom{\vert^{\vert}}}
          \mathrm{fib}
          \mathclap{\phantom{\vert_{\vert}}}
        }
        ^>>>{\ }="t"
      \ar[rr]_>>>{\ }="s"
      &&
      \ast
      \ar[d]
      \\
      F
      \ar[rr]
      &&
      G
      \ar@{=>}
        ^-{
          \mbox{\tiny\rm(pb)}
        }
        "s"; "t"
    }
    }
    \;\;\;\;\;
    \Leftrightarrow
    \;\;\;\;\;
    \raisebox{20pt}{
    \xymatrix@R=1.5em{
      E
      \ar[d]^>>>{\ }="t"
      \ar[rr]_>>>{\ }="s"
      &&
      \ast
      \ar[d]
      \\
      F
      \ar[rr]
        _-{
          \;\mathrm{cofib}\;
        }
      &&
      G
      \ar@{=>}
        ^-{
          \mbox{\tiny\rm(po)}
        }
        "s"; "t"
    }
    }
  \end{equation}
\end{prop}
\noindent It follows
in particular that $\Omega^\infty$ \eqref{StabilizationAdjunction},
being a derived right adjoint
(see \cite[Prop. A.21]{FSS20c} for pointers),
takes homotopy co-fiber sequences
of spectra to homotopy fiber sequences of pointed spaces.

\medskip

\noindent {\bf Generalized cohomology.}

\begin{defn}[Whitehead-generalized cohomology theory]
  \label{WhiteheadGeneralizedCohomologyTheory}
  Given a stable homotopy type $E \in \mathrm{Ho}\big( \mathrm{Spectra}\big)$
  (Def. \ref{StableHomotopyCategory}), the
  corresponding reduced {\it generalized cohomology theory} in the sense of
  Whitehead is the functor
  \vspace{-2mm}
  $$
    \xymatrix@R=-3pt{
      \mathrm{Ho}
      \big(
        \mathrm{Spaces}^{\ast/}
      \big)^{\mathrm{op}}
      \ar[rr]^-{ \widetilde E^\bullet(-) }
      &&
      \mathbb{Z}\mathrm{GradedAbelianGroups}
      \\
      X
      \ar@{}[rr]|-{ \longmapsto }
      &&
    \scalebox{0.9}{$  \mathrm{Ho}
      (\mathrm{Spectra})
      \big(
        \Sigma^\infty X
        ,\,
        \Sigma^{\bullet} E
      \big)
      $}
      \,.
    }
  $$

  \noindent
  The {\it suspension isomorphism} in $E$-cohomology is the natural isomorphism
   \vspace{-2mm}
  \begin{equation}
    \label{SuspensionIsomorphism}
    \xymatrix@C=2em{
      \widetilde E^\bullet(X)
      \ar[r]^-{ \Sigma^n }_-{ \simeq }
      &
      \mathrm{Ho}
       (\mathrm{Spectra})
      \big(
        \Sigma^{\infty + n}  X
        ,\,
        \Sigma^{\bullet + n} E
      \big)
      \ar@{}[r]|-{ \simeq }
      &
      \mathrm{Ho}
       (\mathrm{Spectra})
      \big(
        \Sigma^{\infty } S^n \wedge X
        ,\,
        \Sigma^{\bullet + n} E
      \big)
      \ar@{}[r]|-{ \simeq }
      &
      \widetilde E^{\bullet + n}
      (
        S^n \wedge X
      )
    }.
  \end{equation}
\end{defn}

\medskip

\begin{defn}[Multiplicative cohomology theory
  {\cite[p. 23]{ConnerFloyd66}\cite[\S III.10]{Adams74}}]
  \label{MultiplicativeCohomologyTheory}
  A {\it homotopy-commutative ring spectrum} is
  a commutative monoid object $(E,1,\cdot)$ in the
  smash-monoidal stable homotopy category
  (Def. \ref{StableHomotopyCategory}):
  \begin{equation}
    \label{HomotopyCommutativeRingSpectrum}
    \big(
      E, 1, \cup
    \big)
    \;\in\;
    \mathrm{CommutativeMonoids}
    \big(
      \mathrm{Ho}
      (
        \mathrm{Spectra}
      )
      ,\,
      1,\,
      \wedge
    \big)
    \,.
  \end{equation}
  The Whitehead-generalized cohomology theory
  ${\widetilde E}(-)$
  \eqref{WhiteheadGeneralizedCoohomology}
  that is represented by a homotopy-commutative ring spectrum $E$
  \eqref{HomotopyCommutativeRingSpectrum} inherits cup-product
  operations that make it a {\it multiplicative cohomology theory}.

  In particular, the homotopy-ring structure
  \eqref{HomotopyCommutativeRingSpectrum} induces on the
  coefficient groups
  $E_\bullet$ \eqref{WhiteheadGeneralizedCoohomology}
  (the reduced E-cohomology
  of $S^0$, Example \ref{StableHomotopyGroupsAndCohomologyOfSpheres})
  the structure of a graded-commutative ring
  (\cite[\S III.10]{Adams74},
  review in \cite[\S 3]{Boardman95}\cite[\S C]{TamakiKono06}):
  \begin{equation}
    \label{CoefficientRingOfECohomologyTheory}
    \big(
      E_\bullet,
      1,
      \cdot
    \big)
    \;\;\;
    \in
    \;
    \mathrm{CommutativeMonoids}
    \big(
      \mathbb{Z}\mathrm{GradedAbelianGroups}
    \big).
  \end{equation}
\end{defn}

\medskip

\newpage

\noindent {\bf The d-invariant.}

\begin{defn}[d-invariant on Cohomotopy]
  \label{dInvariant}
  For $E$
  a multiplicative cohomology theory
  we say that the {\it d-invariant} of
  a Cohomotopy class
  $\big[ X \xrightarrow{\;c\;} S^n \big] \in \pi^n(X)$
  (for any $n \in \mathbb{N}$)
  seen in $E$-cohomology
  is the class of the pullback along $c$ of the
  $n$-fold suspended $E$-unit \eqref{UnitMorphismOfSpectra}:
\vspace{-2mm}
  $$
    d_E(c)
    \;:=\;
    \big[
      c^\ast \big( \Sigma^n 1 \big)
    \big]
    \;=\;
    \big[
      X
        \xrightarrow{c}
      S^n
        \xrightarrow{ \Sigma^n (1^E) }
      E\degree{n}
    \big]
    \;\;\in\;
    \widetilde E^n(X)
    \,.
  $$
\end{defn}

\begin{remark}[d-invariant for mapping spaces]
  More generally, for $X \xrightarrow{ f } Y$ a map
  with any codomain space, its
  {\it Adams $d_E$-invariant}
  is the element
  \vspace{-1mm}
  $$
    [f^\ast]
      \;\in\;
    \mathrm{Hom}_{E_{-\bullet}}
    \Big(
      \widetilde E^\bullet(Y)
      \,,\,
      \widetilde E^\bullet(X)
    \Big)
    .
  $$

\noindent   For the case $Y = S^n$ this reduces to
  Def. \ref{dInvariant}, since now the suspension
  isomorphism identifies
    \vspace{-1mm}
  $$
    \widetilde E^\bullet
    \big(
      \underset{
        Y
      }{
      \underbrace{
        S^n
      }
      }
    \big)
      \;\simeq\;
    E_{n-\bullet}
      \;\simeq\;
    E_{-\bullet}
    \big\langle
      \Sigma^n(1^E)
    \big\rangle
  $$

      \vspace{-1mm}
\noindent
  with the free $E_{-\bullet}$-module on
  a single generator in degree $n$. This follows {\cite[\S 3]{Adams66}}.
\end{remark}
The name ``d-invariant'' in Def. \ref{dInvariant}
alludes to its realization in ordinary cohomology:
\begin{example}[Hopf degree is d-invariant in ordinary cohomology]
  For $X$ a connected closed oriented manifold of dimension $n$,
  the d-invariant (Def. \ref{dInvariant}) of a map $\phi : X \to S^n$
  seen in ordinary integral cohomology $E = H\mathbb{Z}$
  is the {\it Hopf degree} in $H^n(X;\mathbb{Z}) \,\simeq\, \mathbb{Z}$.
\end{example}
In the other extreme:
\begin{example}[d-Invariant in stable Cohomotopy is stable class]
  \label{DInvariantInStableCohomotopyIsStableClass}
  The d-invariant in stable Cohomotopy $E = \mathbb{S}$
  is equivalently the stable class of a map:
  \vspace{-3mm}
  $$
    \begin{aligned}
      d_{\mathbb{S}}(c)
      &
      \;\simeq\;
      \big[
        X
          \xrightarrow{c}
        S^n
          \xrightarrow{ \Sigma^n (1^{\mathbb{S}}) }
        \mathbb{S}^n
        =
        \Omega^\infty \Sigma^\infty S^n
      \big]
      \\
      &
      \;\simeq\;
      \big[
        \Sigma^\infty X
          \xrightarrow{ \Sigma^\infty c }
        \Sigma^\infty S^n
      \big]
      \,.
    \end{aligned}
  $$
\end{example}

\medskip

\noindent
{\bf Trivializations of the d-invariant.}
The d-invariant is a ``primary'' homotopy invariant,
in that it is an invariant of the (stable) homotopy class
of a map of pointed spaces, hence of a {\it 1-morphism} in the
homotopy theory ($\infty$-category) of pointed homotopy types.
When this primary invariant vanishes,
then ``secondary'' homotopy invariants appear,
namely higher-homotopy invariants of (null) homotopies
witnessing this vanishing of the primary invariant, hence
invariants
of {\it 2-morphisms} in the $\infty$-category of pointed homotopy types:

\begin{defn}[Trivializations of the d-invariant]
  \label{TrivializationsOfThedInvariant}
  For $E$ a multiplicative cohomology theory
  and $n, d \in \mathbb{N}$,
  write
  \begin{equation}
    \label{SetOfTrivializationsOfdInvariant}
    \begin{aligned}
    H^E_{n-1}\mbox{Fluxes}\!\big( S^{n+d-1} \big)
    &
    \;\coloneqq\;
    \underset{
      [c] \in \mathbb{S}_{d-1}
    }{\sqcup}
    \pi_0
    \mathrm{Paths}_0^{c^\ast (1^E)}
    \Big(
    \mathrm{Maps}^{\ast/}\!
      \big(
        S^{n+d-1}
        \,,\,
        E^{n}
      \big)
    \Big)
    \\
    & \; = \;
    \underset{
      [
        {
          \color{greenii}
          c
        }
      ] \in \mathbb{S}_{d-1}
    }{\sqcup}
    \left\{
    \raisebox{27pt}{
    \xymatrix@C=9em@R=24pt{
      S^{n+d-1}
      \ar[rr]_>>{\ }="s"
      \ar[d]
        _-{
          \mathclap{\phantom{\vert^{\vert}}}
          \color{greenii}
          c
          \mathclap{\phantom{\vert_{\vert}}}
        }
      \ar[dr]|-{
        G^{\mathbb{S}}_{n}\!(c)
      }
      ^>>>>>{\ }="t"
      &&
      \ast
      \ar[d]
        ^-{
          \mathclap{\phantom{\vert^{\vert}}}
          0
          \mathclap{\phantom{\vert_{\vert}}}
        }
      \\
      S^{n}
      \ar@/_1.3pc/[rr]|-{ \;\Sigma^{n}(1^E)\; }
      \ar[r]|-{
        \;
        \Sigma^{n}
        (1^{\mathbb{S}})
        \;
      }
      &
      \mathbb{S}^{n}
      \ar[r]|-{
        \;
        (e^E)^{n}
        \;
      }
      &
      E^{n}
      \ar@{==>}
        |-{
          \;
          \color{orangeii}
          H^E_{n-1}\!(c)
          \;
        }
        "s"; "t"
    }
    }
    \right\}_{
      \mathrlap{
        \!\!\!\!\big/\scalebox{.6}{2-homotopy}
      }
    }
    \end{aligned}
  \end{equation}
  for the set of tuples consisting of the stable Cohomotopy
  class $\big[ G^{\mathbb{S}}_{n}\!(c) \big]$
  of a map $S^{n+d-1} \xrightarrow{\;c\;}S^{n}$
  and the 2-homotopy class
  $\big[ H^E_{n-1}\!(c) \big]$
  of a trivialization (if any) of its
  d-invariant $c^\ast(1^E)$
  \eqref{dInvariantAsMapOnCohomotopy}
  in $E$-cohomology.
  This set is canonically fibered over the underlying
  $G^{\mathbb{S}}_n\mathrm{Fluxes}(X)$
  \eqref{GFluxes}.
\end{defn}

\begin{remark}[Trivialization for vanishing classes]
In the special case
that $[c] = 0 \in \mathbb{S}_{d-1}$,
the trivialization $\big[ H^E_{n-1}\!(c) \big]$
in \eqref{SetOfTrivializationsOfdInvariant}
may be identified with a class in $E_{d-1}$.
For general $[c]$, however, the diagram
\eqref{SetOfTrivializationsOfdInvariant}
indicates that $\big[ H^E_{n-1}\!(c) \big]$ is like a
class in
{\it $G^{\mathbb{S}}_{n}\!(c)$-twisted $E$-cohomology}.
\end{remark}
We make this precise in \cref{TodaBrackets}
by understanding
$H^E_{n-1}$-fluxes as refined Toda brackets
(Def. \ref{RefinedTodaBracket}) of
{(1)}
the Cohomotopy charge
$[c]$,
with
{(2)} the unit $\Sigma^n(1^E)$,
and {(3)}
the canonical projection to the
classifying space of the
Adams cofiber cohomology theory
(Def. \ref{UnitCofiberCohomologyTheories}).

\medskip

\subsection{Toda brackets}
 \label{TodaBrackets}

The notion of {\it Toda brackets}
(\cite{Toda62}) is central to $M\mathrm{Fr}$-theory. \footnote{
  \cite[p. 17]{IsaksenWangXu20b}:
  ``Our philosophy is that the stable homotopy groups of spheres
  are not really understood until the Toda bracket structure is revealed.''}
Nevertheless, their definition is traditionally stated in
a somewhat intransparent way (Remark \ref{ReferencesOnTodaBrackets} below),
and their {\it indeterminancy} is often treated more as a
nuisance than as a feature. We use the occasion to bring out the
hidden elegance of the notion (Def. \ref{RefinedTodaBracket})
and highlight how the Toda bracket,
via its apparent ``indeterminancy'', is really
a {\it fibered moduli space} (albeit discrete):
the {\it Toda bundle} (Remark \ref{TheTodaBundle} below),
a special case of which are the moduli spaces of
$H^E_{3}$-fluxes for given $G^E_4$-fluxes \eqref{EFluxes};
this is Prop. \ref{EFluxesAreCocycleInCofiberTheory} below.

\medskip
Further below in \cref{TheAdamsEInvariant}, we see how the Adams e-invariant
and the Hopf invariant are special cases of ``refined'' Toda brackets,
namely of points in fibers of the Toda bundle.

\medskip

\noindent
{\bf Toda brackets.}

\begin{defn}[Zero-map and null-homotopy]
  \label{ZeroMapsAndNullHomotopy}
  Let $X, Y \in \mathrm{PointedTopologicalSpaces}$.

  \noindent
  {\bf (a)} The {\it zero map} or {\it zero morphism} between them
  is the unique function that is constant on the base point of $Y$.
  In particular, for $X = \ast$, the base point inclusion itself is
  the zero map, and we write
  \vspace{-2mm}
  \begin{equation}
    \label{ZeroMorphism}
    0
      :
    \xymatrix@C=16pt{
      X
      \ar[r]
      &
      \underset{
        \mathclap{
        \raisebox{-3pt}{
          \tiny
          \color{greenii}
          \bf
          zero morphism
        }
        }
      }{
        \ast
      }
      \ar[r]^-{ 0 }
      &
      Y
    }
    .
  \end{equation}

\vspace{-3mm}
  \noindent
  {\bf (b)} For any pointed map $X \xrightarrow{f} Y$ a
  {\it null-homotopy} is (if it exists) a pointed homotopy
  from (or to) the zero-map \eqref{ZeroMorphism}
  \vspace{-3mm}
  $$
    \xymatrix{
      X
      \ar@/^1.6pc/[rr]
        ^-{ \;0\; }
        _-{\ }="s"
      \ar@/_1.6pc/[rr]
        _-{ f }
        ^-{\ }="t"
      &&
      Y
      \ar@{=>}
        |-{
          \mathclap{\phantom{\vert^{\vert}}}
          \mbox{
            \tiny
            \color{orangeii}
            null-homotopy
          }
          \mathclap{\phantom{\vert_{\vert}}}
        }
        "s"; "t"
    }
    \;\;\;\;
      =
    \;\;\;
    \raisebox{37pt}{
    \xymatrix@R=7pt{
      \\
      &
      \ast
      \ar[dr]
        ^-{ 0 }
        _<{\ }="s"
      \\
      X
      \ar@/_1.6pc/[rr]
        _-{ \;f\; }
        ^-{\ }="t"
      \ar[ur]
      &&
      Y
      \ar@{=>}
        "s"; "t"
    }
    }
  $$
\end{defn}

\begin{lemma}[Torsor of null homotopies]
  \label{TorsorOfNullHomotopies}
  For $X \xrightarrow{f} Y$ a map of pointed topological spaces,
  its set of 2-homotopy classes of null-homotopies
  (Def. \ref{ZeroMapsAndNullHomotopy}) is either empty
  or is a torsor
  (a principal bundle over the point) for the group
  \begin{equation}
    \label{FundamentalGroupOfPointedMappingSpace}
    \underset{
      \mathllap{
      \simeq
      \,
      }
      \pi_0
      \mathrm{Maps}^{\ast/\!\!}
      \big(
        X, \Omega Y
      \big)
    }{
    \underbrace{
      \pi_1\mathrm{Maps}^{\ast/\!\!}
      \big(
        X, Y
      \big)
    }
    }
    \;\;
    =
    \;\;
    \left\{
      [g]
    \;
      \left\vert
    \;
    \xymatrix@C=12pt{
      X
      \ar@/^1pc/[rr]
        |-{ \;0\; }
        _-{\ }="s"
      \ar@/_1pc/[rr]
        |-{ \;0\; }
        ^-{\ }="t"
      &&
      Y
      \ar@{=>}
        ^{ g }
        "s"; "t"
    }
    \right.
    \right\}
  \end{equation}
  under its canonical action by composition of homotopies:
  \vspace{-2mm}
  $$
    \xymatrix{
      X
      \ar[rr]
        _-{\ }="s"
        |-{ \;0\; }
      \ar@/_1.7pc/[rr]
        |-{ \;f\; }
        ^-{\ }="t"
      &&
      Y
      \ar@{=>}
        ^-{\phi}
        "s"; "t"
    }
    \;\;\;\;
    \overset{ g }{\longmapsto}
    \;\;\;\;
    \xymatrix{
      X
      \ar@/^1.7pc/[rr]
        _-{\ }="s2"
        |-{ \;0\; }
      \ar[rr]
        ^-{\ }="t2"
        _-{\ }="s"
        |-{ \;0\; }
      \ar@/_1.7pc/[rr]
        |-{ \;f\; }
        ^-{\ }="t"
      &&
      Y
      \ar@{=>}
        ^-{\phi}
        "s"; "t"
      \ar@{=>}
        ^-{ g }
        "s2"; "t2"
    }
    \,.
  $$
\end{lemma}
\begin{proof}
  It is clear that this is an action. That this action is
  principal means that for evert pair $[\phi_1], [\phi_2]$
  of 2-homotopy classes of null-homotopies of $f$, there
  is a {\it unique} $[g] \coloneqq  [\phi_1]^{-1} [\phi_2] $
  which takes $[\phi_1]$ to $[\phi_2]$. This is clearly
  given by the 2-homotopy class of the composite of $\phi_2$ with
  the inverse of $\phi_1$:
  $$
    [\phi_1]^{-1}[\phi_2]
    \;\;
      =
    \;\;
    \left[
    \xymatrix{
      X
      \ar@/^1.7pc/[rr]
        _-{\ }="s2"
        |-{ \;0\; }
      \ar[rr]
        ^-{\ }="t2"
        _-{\ }="s"
        |-{ \;f\; }
      \ar@/_1.7pc/[rr]
        |-{ \;0\; }
        ^-{\ }="t"
      &&
      Y
      \ar@{=>}
        ^-{\phi_1^{-1}}
        "s"; "t"
      \ar@{=>}
        ^-{ \phi_2 }
        "s2"; "t2"
    }
    \right]
    \;\;\;
    \in
    \;\;
    \pi_1
    \mathrm{Maps}^{\ast/\!}
    \big(
      X, Y
    \big),
  $$
  establishing the statement.
\end{proof}

\newpage

\begin{defn}[Toda brackets]
  \label{RefinedTodaBracket}
  Given three consecutive maps
  of pointed spaces, equipped in pairs with null-homotopies
  (Def. \ref{ZeroMapsAndNullHomotopy})
$$
  \xymatrix{
    X_0
     \ar[rr]
       |-{ \;f_1\; }
       ^>{\ }="t1"
     \ar@/^2.4pc/[rrrr]
       ^-{ 0 }
       _-{\ }="s1"
    &&
    X_1
     \ar[rr]
       |-{ \;f_2\; }
     \ar@/_2.4pc/[rrrr]
       _-{ 0 }
       ^-{\ }="t2"
    &&
    X_2
     \ar[rr]
       |-{ \;f_3\; }
       _<{\ }="s2"
    &&
    X_3
    \,,
    \ar@{=>}
      ^-{
        \color{orangeii}
        \phi_1
      }
      "s1"; "t1"
    \ar@{=>}
      ^-{
        \color{orangeii}
        \phi_2
      }
      "s2"; "t2"
  }
$$
then the pasting composite of these null-homotopies
is a loop on the zero-map between $X_0$ and $X_3$,
$$
\hspace{-6cm}
  \raisebox{40pt}{
  \xymatrix{
    X_0
    \ar@/^2.5pc/[ddrr]
      ^-0
      _-{\ }="s"
    \ar@/_2.5pc/[ddrr]
      _-0
      ^-{\ }="t"
    \\
    \\
    &&
    X_3
    \ar@{=>}
      |-{
        \mathclap{\phantom{\vert^{\vert}}}
        \color{orangeii}
          ( \phi_2 \cdot f_1 )
          \circ
          (f_3 \cdot \phi_1)
        \mathclap{\phantom{\vert_{\vert}}}
      }
      "s"; "t"
  }
  }
  \;\;
    \coloneqq
  \;\;\;\;\;
  \raisebox{46pt}{
  \xymatrix{
    X_0
    \ar[rr]_>>{\ }="s1"
    \ar[d]
      _-{ f_1 }
      ^>>{\ }="t1"
    &&
    \ast
    \ar[d]^-{ 0 }
    \\
    X_1
    \ar[rr]
      |-{ \;f_2\; }
      _>>{\ }="s2"
    \ar[d]^>>{\ }="t2"
    &&
    X_2
    \ar[d]^-{ f_3 }
    \\
    \ast
    \ar[rr]
      _-{ 0 }
    &&
    X_3
    \ar@{=>}
      ^-{
        \color{orangeii}
        \phi_1
      }
      "s1"; "t1"
    \ar@{=>}
      ^-{
        \color{orangeii}
        \phi_2
      }
      "s2"; "t2"
  }
  }
  \;\;\;
  \begin{aligned}
    \in
    \;\;\;\;\;\;
   &
    \pi_1
    \mathrm{Maps}^{\ast/\!\!}
    (
      X_0, X_3
    )
    \\
    =
    \;
    &
    \pi_0
    \mathrm{Maps}^{\ast/\!\!}
    (
      X_0, \Omega X_3
    )
    \\
    =
    \;
    &
    \pi_0
    \mathrm{Maps}^{\ast/\!\!}
    (
      \Sigma X_0, X_3
    )
  \end{aligned}
$$
$$
\hspace{3.2cm}
    =
  \;\;\;
  \raisebox{45pt}{
  \xymatrix@R=12pt@C=4.5em{
    X_0
     \ar@/^.7pc/[rr]
    \ar@{-->}[r]
       _-{
        \scalebox{.58}{$
          \color{orangeii}
          \langle
            f_1,f_2, f_3
          \rangle_{(\phi_1,\phi_2)}
          \!\!\!\!\!
        $}
      }
    \ar[dd]
      _-{
        f_1
      }
    &
    \Omega X_3
    \ar[r]
      _>>>>>>>{\ }="s1"
    \ar[dd]
      ^>>>>>{\ }="t1"
    &
    \ast
    \ar[dd]
      ^-{ 0 }
    \\
    \\
    X_1
    \ar@/^.7pc/[rr]
      ^<<<<{
        \;f_2\;
      }
      |-{ \phantom{AAA} }
    \ar@{-->}[r]
      _ -{
        \color{orangeii}
        \vdash
        \phi_2
      }
    \ar[dd]
    &
    F_{f_3}
    \ar[r]
      _>>>>>>>{\ }="s2"
    \ar[dd]
      ^>>>>>{\ }="t2"
    &
    X_2
    \ar[dd]
      ^-{ f_3 }
    \\
    \\
    \ast
    \ar@{=}[r]
    \ar@/^.7pc/[rr]
      ^<<<<{
        \;0\;
      }
      |-{ \phantom{AAA} }
    &
    \ast
    \ar[r]
      _-{ 0 }
    &
    X_3
    \ar@{=>}
      ^-{
        \mbox{
          \tiny
          (pb)
        }
      }
      "s1"; "t1"
    \ar@{=>}
      ^-{
        \mbox{
          \tiny
          (pb)
        }
      }
      "s2"; "t2"
  }
  }
  \;\;\;\;\;
    =
  \;\;\;\;\;
  \raisebox{45pt}{
  \xymatrix@R=12pt@C=4.9em{
    X_0
    \ar@/_.7pc/[rr]
      |-{ \phantom{AAA} }
    \ar[r]
    \ar[dd]
      _-{
        f_1
      }
    &
    \ast
    \ar@{=}[r]
    \ar[dd]
      _>>>>>>>{\ }="s1"
    \ar@{}[ddr]
    &
    \ast
    \ar[dd]
      ^-{ 0 }
    \\
    \\
    X_1
    \ar@/_.7pc/[rr]
      |-{ \phantom{AAA} }
      _>>>>{
        \;f_2\;
      }
    \ar[r]
      ^>>>>>>>{\ }="t1"
    \ar[dd]
    &
    C_{f_1}
    \ar@{-->}[r]
      ^-{
        \color{orangeii}
        \vdash
        \phi_1
      }
    \ar[dd]
      _>>>>>>>{\ }="s2"
    \ar@{}[ddr]
    &
    X_2
    \ar[dd]
      ^-{ f_3 }
    \\
    \\
    \ast
    \ar[r]
      ^>>>>>>>{\ }="t2"
    \ar@/_.7pc/[rr]
      _<<<<{
        \;0\;
      }
    &
    \Sigma X_0
    \ar@{-->}[r]
       ^-{
        \scalebox{.58}{$
          \color{orangeii}
          \!\!
          \langle
            f_1,f_2, f_3
          \rangle_{(\phi_1,\phi_2)}
        $}
      }
    &
    X_3
    \ar@{=>}
      _-{
        \mbox{
          \tiny
          (po)
        }
      }
      "s1"; "t1"
    \ar@{=>}
      _-{
        \mbox{
          \tiny
          (po)
        }
      }
      "s2"; "t2"
  }
  }
$$
$$
  \hspace{9.5cm}
  \;\;\;\;\;\;\;
    =
  \;\;\;\;\;\;\;
  \raisebox{45pt}{
  \xymatrix@R=12pt@C=4.9em{
    X_0
    \;
    \ar@/_.7pc/[rr]
      |-{ \phantom{AAA} }
    \ar@{^{(}->}[r]
    \ar[dd]
      _-{
        f_1
      }
    \ar@{}[ddr]
      |-{
        \mbox{
          \tiny
          (po)
        }
      }
    &
    C_{X_0}
    \ar[r]
    \ar[dd]
    \ar@{}[ddr]
    &
    \ast
    \ar[dd]
      ^-{ 0 }
    \\
    \\
    X_1
    \mathclap{\phantom{\vert_{\vert}}}
    \;
    \ar@/_.7pc/[rr]
      _>>>>{
        \;f_2\;
      }
      |-{ \phantom{AAA} }
    \ar@{^{(}->}[r]
    \ar@{^{(}->}[dd]
    \ar@{}[ddr]
      |-{
        \mbox{
          \tiny(po)
        }
      }
    &
    C_{f_1}
    \ar@{-->}[r]
      ^-{
        \color{orangeii}
        \vdash \phi_1
      }
    \ar[dd]
    \ar@{}[ddr]
    &
    X_2
    \ar[dd]
      ^-{ f_3 }
    \\
    \\
    C_{X_1}
    \ar[r]
    \ar@/_.7pc/[rr]
      _<<<<{
        \;0\;
      }
    &
    \Sigma X_0
    \ar@{-->}[r]
       ^-{
        \scalebox{.58}{$
          \color{orangeii}
          \!\!
          \langle
            f_1,f_2, f_3
          \rangle_{(\phi_1,\phi_2)}
        $}
      }
    &
    X_3
    \,,
  }
  }
$$
which is classified, via the universal property
of the homotopy (co)fiber \eqref{HomotopyPushoutPropertyInIntroduction},
by a map from the suspension of $X_0$
or equivalently to the loop space of $X_3$.

\noindent {\bf (i)} The homotopy class of this classifying map
\begin{equation}
  \label{TheRefinedTodaBracket}
  {
    \color{orangeii}
    \langle
      f_1, f_2, f_3
    \rangle_{(\phi_1, \phi_2)}
  }
  \;\;\;
    \coloneqq
  \;\;\;
  \Big[
    \vdash
    \big(
      ( \phi_2 \cdot f_1 )
      \circ
      (f_3 \cdot \phi_1)
    \big)
  \Big]
  \;\;\;
  \begin{array}{rcl}
   \in
   &&
   \pi_0\mathrm{Maps}^{\ast/\!\!}
   (
     \Sigma X_0, X_3
   )
   \\
   &
  \quad  =
   &
      \pi_0\mathrm{Maps}^{\ast/\!\!}
   (
     X_0, \Omega X_3
   )
  \end{array}
\end{equation}
is the
{\it refined Toda bracket} of the triple of maps $(f_1, f_2, f_3)$,
depending on the chosen null-homotopies.

\noindent {\bf (ii)}  The {\it plain Toda bracket} is
the {\it set} of all refined Toda brackets \eqref{TheRefinedTodaBracket}
obtained as the choice of the pair $(\phi_1, \phi_2)$ of null-homotopies
is varied:
\begin{equation}
  \label{ThePlainTodaBracket}
  \langle
    f_1,
    f_2,
    f_3
  \rangle
  \;\coloneqq\;
  \left\{
    \langle
      f_1,
      f_2,
      f_3
    \rangle_{(\phi_1, \phi_2)}
    \left\vert
      \begin{array}{rcl}
        0
          &\overset{\phi_1}{\Rightarrow}&
        f_2 \circ f_1\,,
        \\
        f_3 \circ f_2
          &
          \underset{\phi_2}{\Rightarrow}
          &
          0
        \end{array}
    \right.
  \right\}
  \;\;
  \begin{array}{rcl}
   \subset
   &&
   \pi_0\mathrm{Maps}^{\ast/\!\!}
   (
     \Sigma X_0, X_3
   )
   \\
   &
   \quad =
   &
   \pi_0\mathrm{Maps}^{\ast/\!\!}
   (
     X_0, \Omega X_3
   )\;.
  \end{array}
\end{equation}
\end{defn}
\begin{remark}[Perspectives on the Toda bracket]
  \label{ReferencesOnTodaBrackets}
  {\bf (i)} The traditional way to state the definition of the
  Toda bracket \cite{Toda62} is essentially the last diagram
  shown in Def. \ref{RefinedTodaBracket},
  where the homotopy cofibers are modeled by
  topological cone constructions and where the Toda bracket
  appears as a ``consecutive extension of functions over cones'',
  these being the dashed morphisms in our diagram.

  \noindent
  {\bf (ii)} The more intrinsic description of the Toda bracket
  as a homotopy-coherent pasting composite,
  which we amplify at the beginning of the sequence of
  diagrams in Def. \ref{RefinedTodaBracket},
  is made more explicit in
  \cite[(0.2), (0.3)]{HardieKampsKieboom99}\cite[\S 3]{HardieMarcumOda01}\cite[(2.2)]{HardieKampsMarcum02}.

{
\noindent
{\bf (iii)} The Toda bracket in homotopy is the analogue of
the {\it Massey product} in cohomology (being Eckman-Hilton duals),
where composition of maps corresponds to cup products
and null-homotopies to coboundaries in cohomology.
The corresponding diagrammatic formulation of refined Massey products
is discussed in \cite[\S 3.2]{GS-Massey}.
}
\end{remark}

The perspective of the Toda bracket as the homotopy-pasting composite
makes immediate the characterization of the ``indeterminancy''
in the plain Toda bracket:
\begin{prop}[Toda bracket is torsor over Toda group]
  \label{TodaBracketIsTorsorOverTodaGroup}
  The plain Toda bracket \eqref{ThePlainTodaBracket}
  is either the empty set or is a torsor
  (a principal bundle over the point) for the
  direct product group
  $$
    \begin{aligned}
      \mathrm{TodaGroup}(X_0, X_1, X_2, X_3)
      & \coloneqq\;
      \pi_1 \mathrm{Maps}^{\ast/\!\!}
      (
        X_0, X_2
      )
      \times
      \pi_1 \mathrm{Maps}^{\ast/\!\!}
      (
        X_1, X_3
      )^{\mathrm{op}}
      \\
      & =
      \pi_0 \mathrm{Maps}^{\ast/\!\!}
      (
        X_0, \Omega X_2
      )
      \times
      \pi_0 \mathrm{Maps}^{\ast/\!\!}
      (
        X_1, \Omega  X_3
      )^{\mathrm{op}}
    \end{aligned}
  $$
  acting via composition of homotopies:
$$
  \langle
    f_1, f_2, f_3
  \rangle_{
    (
      {
        \color{orangeii}
        \phi_1
      }
      ,
      {
        \color{orangeii}
        \phi_2
      }
    )
  }
  \;\;=\;\;
  \raisebox{43pt}{
  \xymatrix{
    X_0
    \ar[drr]
      _<<<<<<<{\ }="s1"
      |<<<<<<<<{
        \;0\;
      }
    \ar[d]
      _-{ f_1 }
      ^>{\ }="t1"
    &&
    \\
    X_1
    \ar[rr]
      |-{ \;f_2\; }
      _>>>{\ }="s2"
    \ar[drr]
      ^>>>>>>>>>>{\ }="t2"
      |>>>>>>>>>>{
        \;0\;
      }
    &&
    X_2
    \ar[d]^-{ f_3 }
    \\
    &&
    X_3
    \ar@{=>}
      ^-{
        \color{orangeii}
        \phi_1
      }
      "s1"; "t1"
    \ar@{=>}
      ^-{
        \color{orangeii}
        \phi_2
      }
      "s2"; "t2"
  }
  }
  \;\;\;
  \longmapsto
  \;\;\;
  \raisebox{43pt}{
  \xymatrix{
    X_0
    \ar[drr]
      _<<<<<<<{\ }="s1"
      |<<<<<<<<{
        \;0\;
      }
      ^<<<<<<<<{\ }="tt1"
    \ar@/^2.2pc/[drr]
      |-{
        \;0\;
      }
      _<<<<<<<<{\ }="ss1"
    \ar[d]
      _-{ f_1 }
      ^>{\ }="t1"
    &&
    \\
    X_1
    \ar@/_2.2pc/[drr]
      |-{
        \;0\;
      }
      ^>>>>>>>>{\ }="tt2"
    \ar[rr]
      |-{ \;f_2\; }
      _>>>{\ }="s2"
    \ar[drr]
      ^>>>>>>>>>>{\ }="t2"
      _>>>>>>>>>>{\ }="ss2"
      |>>>>>>>>>>{
        \;0\;
      }
    &&
    X_2
    \ar[d]^-{ f_3 }
    \\
    &&
    X_3
    \ar@{=>}
      ^-{
        \color{orangeii}
        \phi_1
      }
      "s1"; "t1"
    \ar@{=>}
      ^-{
        \color{orangeii}
        \phi_2
      }
      "s2"; "t2"
    \ar@{=>}
      ^-{
        \color{greenii}
        g_1
      }
      "ss1"; "tt1"
    \ar@{=>}
      ^-{
        \color{greenii}
        g_2
      }
      "ss2"; "tt2"
  }
  }
  \;\;
    =
  \langle
    f_1, f_2, f_3
  \rangle_{
    (
      {
        \color{orangeii}
        \phi_1
      }
        \circ
      {
        \color{greenii}
        g_1
      }
      ,
      \;
      {
        \color{greenii}
        g_2
      }
      \circ
      {
        \color{orangeii}
        \phi_2
      }
    )
  }
  \;\;
$$
\end{prop}
\begin{proof}
  This follows as in Lemma \ref{TorsorOfNullHomotopies}.
\end{proof}

\begin{remark}[The Toda bundle]
 \label{TheTodaBundle}
In conclusion, for a fixed sequence $(X_0, X_1, X_2, X_3)$ of pointed spaces,
the Toda bracket in Def. \ref{RefinedTodaBracket}
is {\it not a function but a bundle}, which
{\it instead of values has fibers},
namely the plain Toda brackets
\eqref{ThePlainTodaBracket},
whose elements are the
refined Toda blackets \eqref{TheRefinedTodaBracket}:
\begin{equation}
  \label{TodaBundle}
  \raisebox{40pt}{
  \xymatrix@C=14pt@R=7pt{
    \overset{
      \mathclap{
      \raisebox{3pt}{
        \tiny
        \bf
        \begin{tabular}{c}
          \color{darkblue}
          element of fiber of Toda bundle:
          \\
          \color{greenii}
          a refined Toda bracket
        \end{tabular}
      }
      }
    }{
      \langle
        f_1, f_2, f_3
      \rangle_{(\phi_1,\phi_2)}
    }
    \ar@{}[r]|-{ \in }
    &
    \overset{
      \mathclap{
      \raisebox{3pt}{
        \tiny
        \bf
        \begin{tabular}{c}
          \color{darkblue}
          fiber of Toda bundle:
          \\
          \color{greenii}
          a Toda bracket
        \end{tabular}
      }
      }
    }{
      \langle
        f_1, f_2, f_3
      \rangle
    }
    \quad
    \ar@{}[ddrr]
      |-{ \mbox{\tiny(pb)} }
    \;
    \ar@{^{(}->}[rr]
    \ar[dd]
    &&
    \qquad
    \underset{
      \raisebox{3pt}{
        \hspace{-80pt}
        \tiny
        \bf
        \begin{tabular}{c}
          \color{darkblue}
          Toda bundle:
          \\
          \color{greenii}
          all Toda brackets
        \end{tabular}
      }
    }{
      \mathrm{TodaBundle}(X_0, \cdots\!, X_3)
    }
    \ar[dd]
    \ar@(ul,ur)@<-30pt>
      ^-{
        \overset{
          \mathclap{
          \raisebox{+6pt}{
            \tiny
            \bf
            \begin{tabular}{c}
              \color{darkblue}
              structure group of Toda bundle:
              \\
              \color{greenii}
              indeterminacy of Toda brackets
            \end{tabular}
          }
          }
        }{
          \mathrm{TodaGroup}(X_0,\, \cdots,  X_3)
        }
      }
    \\
    \\
    &
    \underset{
      \mathclap{
      \raisebox{-9pt}{
        \tiny
        \bf
        \begin{tabular}{c}
          \color{darkblue}
          point in base space:
          \\
          \color{greenii}
          a triple of arguments for Toda bracket
        \end{tabular}
      }
      }
    }{
      \big\{
        \big(
          [f_1], [f_2], [f_3]
        \big)
      \big\}
    }
    \quad
    \ar@{^{(}->}[rr]
    &&
    \quad
    \underset{
      \mathclap{
      \raisebox{-3pt}{
        \tiny
        \bf
        \begin{tabular}{c}
          \color{darkblue}
          base space of Toda bundle:
          \\
          \color{greenii}
          possible arguments of Toda brackets
        \end{tabular}
      }
      }
    }{
      \underset{i \in \{0,1,2\}}{\prod}
      \pi_0\mathrm{Maps}^{\ast/\!\!}
      (X_i,X_{i+1})
    }
  }
  }
\end{equation}
Away from the empty fibers, this bundle is
principal for the Toda group (by Prop. \ref{TodaBracketIsTorsorOverTodaGroup}).
\end{remark}

\medskip

\noindent
{\bf Adams cofiber cohomology.}
Since a choice of trivialization $H^E_{n-1}(c)$ of a d-invariant
$d_E(c) = G^E_{n}(c)$
(Def. \ref{TrivializationsOfThedInvariant})
gives ``half'' of the arguments of a refined Toda bracket
(Def. \ref{RefinedTodaBracket}),
namely
$$
\langle c, \Sigma^n(1^E), - \rangle_{(H^E_{n-1}(c), -)  }\;,
$$
it is natural to consider the {\it universal} completion
of the remaining two arguments.
This is provided by the boundary map in the
{\it Adams cofiber cohomology theory} $E/\mathbb{S}$:

\begin{defn}[Unit cofiber cohomology]
  \label{UnitCofiberCohomologyTheories}
For $E$ any multiplicative cohomology theory,
the homotopy cofiber \eqref{HomotopyPushoutAndHomotopyCofiberSpace}
of the its unit morphism \eqref{UnitMorphismOfSpectra}
deserves to be denoted $E/\mathbb{S}$
(but is
often abbreviated $\Sigma \overline{E}$,
following \cite[Thm 15.1, p. 319]{Adams74},
or just $\overline{E}$, as in \cite[Cor. 5.3]{Hopkins99}).
By the pasting law \eqref{PastingLaw}
this comes equipped with a cohomology operation $\partial^E$
from $E\!/\mathbb{S}$-cohomology to stable Cohomotopy
in one degree higher:
\vspace{-2mm}
\begin{equation}
  \label{CofiberOfRingSpectrumUnitAndBoundaryHomomorphism}
  \raisebox{25pt}{
  \xymatrix@C=40pt@R=-1pt{
    \overset{
      \mathclap{
      \raisebox{3pt}{
        \tiny
        \color{darkblue}
        \bf
        \begin{tabular}{c}
          stable
          \\
          Cohomotopy
        \end{tabular}
      }
      }
    }{
      \mathbb{S}
    }
    \ar[dd]
      ^>>{\ }="t"
    \ar[rr]
      |-{
        \;e^E\;
      }
      ^-{
        \mbox{
          \tiny
          \color{greenii}
          \bf
          \begin{tabular}{c}
            unit
            \\
            \phantom{a}
          \end{tabular}
        }
      }
      _>>{\ }="s"
    &&
    \overset{
      \mathclap{
      \raisebox{4pt}{
        \tiny
        \color{darkblue}
        \bf
        \begin{tabular}{c}
          $E$-cohomology
        \end{tabular}
      }
      }
    }{
      E
    }
    \ar[dd]
      _-{
        \mathclap{\phantom{\vert^{\vert^{\vert}}}}
        i^E
        \mathclap{\phantom{\vert_{\vert}}}
      }
      ^>>{\ }="t2"
    \ar[rr]
      ^-{
      }
      _>>{\ }="s2"
    &&
    \ast
    \ar[dd]
    \\
    {\phantom{A}}
    \\
    \ast
    \ar[rr]
    &&
    \underset{
      \mathclap{
      \raisebox{-4pt}{
        \tiny
        \color{darkblue}
        \bf
        \begin{tabular}{c}
          $E$-unit cofiber
          \\
          cohomology
          \\
          theory
        \end{tabular}
      }
      }
    }{
      E/\mathbb{S}
    }
    \ar[rr]
      |-{ \;\partial^E\; }
      _-{
        \mbox{
          \tiny
          \color{greenii}
          \bf
          \begin{tabular}{c}
            \phantom{-}
            \\
            boundary
            \\
            operation
          \end{tabular}
        }
      }
    &&
    \underset{
      \mathclap{
      \raisebox{-6pt}{
        \tiny
        \color{darkblue}
        \bf
        \begin{tabular}{c}
          shifted stable
          \\
          Cohomotopy
        \end{tabular}
      }
      }
    }{
      \Sigma \mathbb{S}
    }
    \,.
    \ar@{=>}^-{ \mbox{\tiny\rm(po)} }
      "s"; "t"
    \ar@{=>}^-{ \mbox{\tiny\rm(po)} }
      "s2"; "t2"
  }
  }
  \;\;\;\;\;\;\;\;\;\;
  \in
  \;
  \mathrm{Spectra}\;.
\end{equation}

By Prop. \ref{InSpectraFiberSequencesAreCofiberSequences},
this means that
for any $n \in \mathbb{N}$ we have the following homotopy
fiber sequence on component spaces:
\begin{equation}
  \label{BoundaryMapForCofiberTheoryOnComponentSpaces}
  \raisebox{40pt}{
  \xymatrix@R=10pt{
    (E/\mathbb{S})^{n-1}
    \ar[dd]
      _-{
        (\partial^E)^{n-1}
      }
      ^>>{\ }="t2"
    \ar[rr]
      _>>{\ }="s2"
    &&
    \ast
    \ar[dd]
    \\
    \\
    S^n
    \ar[rr]
      |-{
        \;
        \Sigma^n(1^E)
        \;
      }
      _>>>{\ }="s"
    \ar[dd]
      ^>>>{\ }="t"
    &&
    E^n
    \ar[dd]
      ^-{
        (i^E)^n
      }
    \\
    \\
    \ast
    \ar[rr]
    &&
    (E/\mathbb{S})^n
    \ar@{=>}
      ^-{
        \mbox{
          \tiny
          (pb)
        }
      }
      "s"; "t"
    \ar@{=>}
      ^-{
        \mbox{
          \tiny
          (pb)
        }
      }
      "s2"; "t2"
  }
  }
\end{equation}
\end{defn}

\vspace{-2mm}
\noindent
Proposition \ref{EFluxesAreCocycleInCofiberTheory} below
shows that this {\it cofiber theory} $E\!/\mathbb{S}$
is
the cohomology theory that classifies $H^E_3$-fluxes \eqref{HE3Homotopy}.

\begin{lemma}[Cofiber $E$-cohomology as extension of stable Cohomotopy by $E$-cohomology]
  \label{CofiberECohomologyAsExtensionOfStableCohomotopyByECohomology}
  For $E$ a multiplicative cohomology theory and
  $X$ a space, assume that the
  $E$-Boardman homomorphism
  $\widetilde {\mathbb{S}}{}^\bullet(X) \xrightarrow{ \;\beta^\bullet_X\; }
  \widetilde E^\bullet(X)$
  \eqref{UnitCohomologyOperations}
  is zero in degrees $n$ and $n + 1$
  -- for instance in that $X \simeq S^{k \geq 2}$,
  $n = 0$ and the
  groups $\widetilde E^0(S^k) = \pi_k(E)$ have no torsion --
  then the cohomology operations $i^E$, $\partial^E$ in
  \eqref{CofiberOfRingSpectrumUnitAndBoundaryHomomorphism}
  form a short exact sequence of cohomology groups:
  \vspace{-2mm}
  \begin{equation}
    \label{ShortExactSequenceForCofiberCohomologyTheory}
    \xymatrix@C=12pt{
      0
      \ar[r]
      &
      \widetilde E
      \big(
        X
      \big)
      \;
      \ar@{^{(}->}[rr]^-{ i }
      &&
      (
      \widetilde
      {
        E/\mathbb{S}
      }
      ){}^n(X)
      \ar@{->>}[rr]^-{ \partial }
      &&
      \widetilde {\mathbb{S}}{}^{n+1}(X)
      \ar[r]
      &
      0
    }.
  \end{equation}
\end{lemma}
\noindent
(This is in generalization of \cite[p. 102]{Stong68},
which in turn follows \cite[Thm. 16.2]{ConnerFloyd66}.)
\begin{proof}
  Generally,
  the long cofiber sequence of cohomology theories
  \eqref{CofiberOfRingSpectrumUnitAndBoundaryHomomorphism}
  induces a long exact sequence of cohomology groups
  (e.g. \cite[p. 197]{Adams74}):
  \vspace{-2mm}
  $$
    \xymatrix@C=12pt{
      \cdots
      \ar[r]
      &
      \widetilde {\mathbb{S}}{}^{n}(X)
      \ar[rr]^-{ 1^n_X }
      &&
      \widetilde E^n(X)
      \ar[rr]^-{ i^n_X }
      &&
      (\widetilde{E/\mathbb{S}})^n(X)
      \ar[rr]^-{ \partial^n_X }
      &&
      \widetilde S^{n+1}(X)
      \ar[rr]^-{ 1^{n+1}_X }
      &&
      \widetilde {E}^{n+1}(X)
      \ar[r]
      &
      \cdots
    }
  $$

  \vspace{-2mm}
\noindent
  Under the given assumption the two outermost morphisms
  shown are zero, and hence the sequence truncates as claimed.
\end{proof}

\begin{defn}[Induced cohomology operations on cofiber cohomology]
  \label{InducedCohomologyOperationsOnCofiberCohomology}
  Let $E \xrightarrow{ \;\phi\; } F$ be a
  multiplicative cohomology operation, so that in particular
  it preserves the units, witnessed
  by a homotopy-commutative square on the
  left here:
  $$
  \xymatrix@R=1.5em{
    \mathbb{S}
    \ar[rr]^-{ \;1^{E}\; }
    \ar@{=}[d]
    &&
    E
    \ar[d]
      _-{
        \mathclap{\phantom{\vert^{\vert}}}
        \phi
        \mathclap{\phantom{\vert_{\vert}}}
      }
    \ar[rr]
    &&
    E/\mathbb{S}
    \ar[rr]^-{ \;\partial^E\; }
    \ar[d]
      _-{
        \mathclap{\phantom{\vert^{\vert}}}
        \phi/\mathbb{S}
        \mathclap{\phantom{\vert_{\vert}}}
      }
    &&
    \Sigma \mathbb{S}
    \ar@{=}[d]
    \\
    \mathbb{S}
    \ar[rr]^-{ \;1^F\; }
    &&
    F
    \ar[rr]
    &&
    F/\mathbb{S}
    \ar[rr]^-{ \;\partial^F\; }
    &&
    \Sigma\mathbb{S}
    \,.
  }
$$
Then passing to homotopy cofibers yields
the induced cohomology operation $\phi/\mathbb{S}$ on cofiber theories
(Def. \ref{UnitCofiberCohomologyTheories}).
\end{defn}

\begin{example}[Chern character on cofiber of K-theory]
  \label{ChernCharacterOnCofiberOfKTheory}
  The Chern character on $K \mathrm{U}$
  and the Pontrjagin character on $K \mathrm{O}$
  are multiplicative, hence descend to operations
  on cofiber K-theories
  via Def. \ref{InducedCohomologyOperationsOnCofiberCohomology}:
  \vspace{-1mm}
  $$
    \xymatrix@C=4em@R=1.4em{
      K\mathrm{O}
      \ar[rr]
        ^-{
          \;
          \mathrm{ph}
          \;
        }
      \ar[dd]
        _-{
          \mathclap{\phantom{\vert^{\vert}}}
          \mathrm{cplx}
          \mathclap{\phantom{\vert_{\vert}}}
        }
      \ar[dr]
        _-{
          i^{K\mathrm{O}}
        }
      &&
      H^{\mathrm{ev}\!}\mathbb{Q}
      \ar[dr]
        ^-{
          i^{H^{\mathrm{ev}\!}\mathbb{Q}}
        }
      \ar[dd]
        |-{
          \phantom{ {A \atop A} }
        }
      \\
      &
      K \mathrm{O}/\mathbb{S}
      \ar[rr]
        ^<<<<<<<<<{
          \;
          \mathrm{ph}/\mathbb{S}
          \;
        }
      \ar[dd]
        _>>>>>>>>>>>>>>{
          \mathclap{\phantom{\vert^{\vert}}}
          \mathrm{cplx}/\mathbb{S}
          \mathclap{\phantom{\vert_{\vert}}}
        }
      &&
      (H^{\mathrm{ev}}\mathbb{Q})/\mathbb{S}
      \ar@{=}[dd]
      \\
      K\mathrm{U}
      \ar[dr]
        _-{
          i^{K\mathrm{U}}
        }
      \ar[rr]
        |-{ \phantom{AA} }
        ^>>>>>>>>{
          \;\mathrm{ch}\;
        }
      &&
      H^{\mathrm{ev}\!}\mathbb{Q}
      \ar[dr]
        ^-{
          i^{H^{\mathrm{ev}\!}\mathbb{Q}}
        }
      \\
      &
      K \mathrm{U}/\mathbb{S}
      \ar[rr]
        ^-{
                    \mathrm{ch}/\mathbb{S}
          \quad
        }
      &&
      (H^{\mathrm{ev}}\mathbb{Q})/\mathbb{S}
    }
  $$
\end{example}

\begin{remark}[Canonical splitting of cofiber of periodic rational cohomology]
  \label{CanonicalSplittingOfCofiberOfPeriodicRationalCohomology}
For $n,d \in \mathbb{N}$ with $d \geq 1$,

\noindent {\bf (i)} we have a canonically split short exact sequence of the form
\vspace{-2mm}
\begin{equation}
  \label{CanonicalSplittingOfCofiberTheoryOfEvenPeriodicOrdinaryOverSphere}
  \raisebox{20pt}{
  \xymatrix@C=12pt@R=10pt{
    0
    \ar[r]
    &
    \big(
      \widetilde {H^{\mathrm{ev}}\mathbb{Q}}
    \big)
    {}^{2n}
    \big(
      S^{2(n+d)}
    \big)
    \ar[rr]
    \ar@{=}[d]
    &&
    \big(
      \widetilde {(H^{\mathrm{ev}}\mathbb{Q})/\mathbb{S}}
    \big)
    {}^{2n}
    \big(
      S^{2(n+d)}
    \big)
    \ar[rr]
    \ar[d]
      ^-{
            \simeq
        }
      _{\mathrm{spl}_0}
    &&
    \widetilde {\mathbb{S}}
    {}^{2n+1}
    \big(
      S^{2(n+d)}
    \big)
    \ar[r]
    \ar@{=}[d]
    &
    0
    \\
    0
    \ar[r]
    &
    \mathbb{Q}
    \;
    \ar@{^{(}->}[rr]
    &&
    \mathbb{Q}
    \oplus
    \mathbb{S}_{2d-1}
    \ar@{->>}[rr]
    &&
    \mathbb{S}_{2d-1}
    \ar[r]
    &
    0
    \,.
  }
  }
\end{equation}
Here {\bf (a)} the short exact sequence at the top follows with
Lemma \ref{CofiberECohomologyAsExtensionOfStableCohomotopyByECohomology},
using that the stable stems $\mathbb{S}_{\bullet}$
are finite in positive degrees
(Serre's finiteness theorem), hence in particular pure torsion,
hence admit only the 0-morphism into $\mathbb{Q}$.
Moreover {\bf (b)} the
canonical splitting $\mathrm{spl}_0$ shown
is that induced from the inclusion $H\mathbb{Q} \hookrightarrow H^{\mathrm{ev}}\mathbb{Q}$
(which is multiplicative, hence in particular preserves the units):
\vspace{-2mm}
$$
  \xymatrix@C=12pt@R=10pt{
    0
    \ar[r]
    &
    \big(
      \widetilde {H\mathbb{Q}}
    \big)
    {}^{2n}
    \big(
      S^{2(n+d)}
    \big)
    \ar[rr]
    \ar@{=}[d]
    &&
    \big(
      \widetilde {(H\mathbb{Q})/\mathbb{S}}
    \big)
    {}^{2n}
    \big(
      S^{2(n+d)}
    \big)
    \ar[rr]
    \ar[d]^-{
          \simeq
    }
    &&
    \widetilde {\mathbb{S}}
    {}^{2n+1}
    \big(
      S^{2(n+d)}
    \big)
    \ar[r]
    \ar@{=}[d]
    &
    0
    \\
    0
    \ar[r]
    &
    0
    \;
    \ar@{^{(}->}[rr]
    &&
    \mathbb{S}_{2d-1}
    \ar@{=}[rr]
    &&
    \mathbb{S}_{2d-1}
    \ar[r]
    &
    0
    \,,
  }
$$
where on the left we used again the assumption that $d \neq 0$.

\noindent {\bf (ii)} We will use this canonical splitting together with the
canonical splitting of the even rational cohomology
of a cofiber space
$
  S^{2(n+d)-1}
    \xrightarrow{ \;c\; }
  S^{2n}
    \xrightarrow{ \;\; }
  C_c
$
given by the splitting into degrees $2(n+d)$ and $2n$
\vspace{-2mm}
\begin{equation}
  \label{CanonicalSplittingOfEvenRationalCohomologyOfCofiberSpace}
  \raisebox{20pt}{
  \xymatrix@C=12pt@R=10pt{
    0
    \ar[r]
    &
    \big(
      \widetilde {H^{\mathrm{ev}}\mathbb{Q}}
    \big)
    {}^{2n}
    \big(
      S^{2(n+d)}
    \big)
    \ar[rr]
    \ar@{=}[d]
    &&
    \widetilde {(H^{\mathrm{ev}}\mathbb{Q})}
    {}^{2n}
    \big(
      C_c
    \big)
    \ar[rr]
    \ar[d]
      ^-{
            \simeq
      }
      _{\mathrm{spl}_0}
    &&
    \widetilde {(H^{\mathrm{ev}}\mathbb{Q})}
    {}^{2n}
    \big(
      S^{2n}
    \big)
    \ar[r]
    \ar@{=}[d]
    &
    0
    \\
    0
    \ar[r]
    &
    \mathbb{Q}
    \;
    \ar@{^{(}->}[rr]
    &&
    \mathbb{Q}
    \oplus
    \mathbb{Q}
    \ar@{->>}[rr]
    &&
    \mathbb{Q}
    \ar[r]
    &
    0
    \,,
  }
  }
\end{equation}
which is induced in the same fashion from the inclusion
$H\mathbb{Q} \hookrightarrow H^{\mathrm{ev}}\mathbb{Q}$.

\noindent {\bf (iii)} Both of these splittings are hence characterized by the fact that
their corresponding retraction is
{\it projection onto rational cohomology in degree $2(n+d)$}.

\end{remark}

\medskip

\noindent
{\bf $H^E_{n-1}$-fluxes as Toda brackets.}

\begin{prop}[Trivializations of d-invariant
  lift to classes in cofiber theory via refined Toda bracket]
  \label{EFluxesAreCocycleInCofiberTheory}
  For $E$ a multiplicative cohomology theory and
  $n,d \in \mathbb{N}$

  \noindent
  {\bf (i)}
  there exists a lift from the set
  \eqref{SetOfTrivializationsOfdInvariant}
  of trivializations of the
  $d_E$-invariant of maps $S^{n+d - 1} \to S^n$
  through the boundary map $\partial^E$ \eqref{CofiberOfRingSpectrumUnitAndBoundaryHomomorphism}
  on coefficient groups \eqref{WhiteheadGeneralizedCoohomology}
  of the unit cofiber theory
  $E\!/\mathbb{S}$ \eqref{CofiberOfRingSpectrumUnitAndBoundaryHomomorphism}:

  \vspace{-.4cm}
  \begin{equation}
    \label{IdentificationOfFluxesWithCofiberCohomologyClasses}
    \raisebox{32pt}{
    \xymatrix@R=.2em{
    &
    \overset{
      \mathclap{
      \raisebox{4pt}{
        \tiny
        \color{darkblue}
        \bf
        trivialization of a $d_E$-invariant
      }
      }
    }{
    \scalebox{.7}{$
    \Big(
      \big[
        {
        \color{greenii}
        G^{\mathbb{S}}_n\!(c)
        }
      \big]
      \,,\,
      \big[
        {
        \color{orangeii}
        H^E_{n-1}\!(c)
        }
      \big]
     \Big)
     $}
     }
     \ar@{|->}[rr]
    &&
    \overset{
      \mathclap{
      \raisebox{4pt}{
        \tiny
        \color{darkblue}
        \bf
        refined Toda bracket
      }
      }
    }{
    \scalebox{.7}{$
    \big\langle
      {
      \color{greenii}
      c
      },
      {
        \Sigma^n(1^E)
      },
      (i^E)^n
    \big\rangle_{
      \big(
        {
          \color{orangeii}
          H^E_{n-1}(c)
        },\,
        \mathrm{(pb)}
      \big)
    }
    $}
    }
    \\
    \scalebox{0.7}{$
     \Big(
       \big[
         {
         \color{greenii}
         G^{\mathbb{S}}_n\!(c)
         }
       \big]
       \,,\,
       \big[
         {
         \color{orangeii}
         H^E_{n-1}\!(c)
         }
       \big]
      \Big)
     $}
    \ar@{|->}[dd]
    &
     \overset{ \mathllap{
        \mbox{
          \tiny
          \color{darkblue}
          \bf
          \begin{tabular}{c}
          \end{tabular}
        }
      }}{
      H^E_{n-1}\mbox{\rm{Fluxes}}\!
      \big(
        S^{n + d - 1}
      \big)
      }
      \ar@{-->}[rr]
        ^-{
          O^{ E/\mathbb{S} }
        }
      \ar[dd]
      &&
      \big(
        E \!/\mathbb{S}
      \big)_{d}
      \ar[dd]^{\partial}
      \\
      \\
      \scalebox{0.7}{$
        \big[
          {
            \color{greenii}
            G^{\mathbb{S}}_n\!(c)
          }
        \big]
      $}
      &
      G^{\mathbb{S}}_n\mathrm{Fluxes}
      \big(
        S^{n+d-1}
      \big)
      \ar[rr]^-{
        \simeq        }
      &&
      \mathbb{S}_{d-1}
    }
    }
  \end{equation}

  \noindent
 {\bf (ii)} given by sending
 an $H^E_{n-1}$-flux
 to its refined Toda bracket
 (Def. \ref{RefinedTodaBracket})
 with
 the universal null homotopy
 \eqref{BoundaryMapForCofiberTheoryOnComponentSpaces},
 denoted ``(pb)'' in \eqref{IdentificationOfFluxesWithCofiberCohomologyClasses},
 as made fully explicit in diagram
 \eqref{MappingHFluxToClassInCofiberCohomology}.
\end{prop}

\newpage

\begin{proof}
Regarding {\bf (i)}:
Since homotopy cofiber sequences of spectra are also
homotopy fiber sequences (Prop. \ref{InSpectraFiberSequencesAreCofiberSequences}),
the universal property of the defining cofiber sequence in
\eqref{CofiberOfRingSpectrumUnitAndBoundaryHomomorphism}
says that the
homotopy diagram in the definition \eqref{SetOfTrivializationsOfdInvariant},
when equivalently seen under the
stabilization adjunction \eqref{StabilizedSuspensionLoopingAdjunction},
factors uniquely, up to homotopy of cones, as shown here:
\vspace{-2mm}
\begin{equation}
  \label{TrivializationOfdInvariantFactoredInSpectra}
  \raisebox{24pt}{
  \xymatrix@R=14pt{
    \Sigma^{n+d-1}\mathbb{S}
    \ar[rr]_>>>{\ }="s"
    \ar[dd]^>>>{\ }="t"
      _-{
        \mathclap{\phantom{\vert^{\vert}}}
        {
          \color{greenii}
          \Sigma^\infty c
        }
        \mathclap{\phantom{\vert_{\vert}}}
      }
    &&
    \ast
    \ar[dd]
      ^-{
        \mathclap{\phantom{\vert^{\vert}}}
        0
        \mathclap{\phantom{\vert_{\vert}}}
      }
    \\
    \\
    \Sigma^n \mathbb{S}
    \ar[rr]
      _-{
        \; \Sigma^n(e^E) \;
      }
    &&
    \Sigma^n E
    \ar@{==>}
      |-{
        \mathclap{\phantom{\vert^{\vert^{\vert}}}}
        {
          \color{orangeii}
          H^E_{n-1}\!(c)
        }
        \mathclap{\phantom{\vert_{\vert_{\vert}}}}
      }
      "s"; "t"
  }
  }
  \;\;
  \simeq
  \;\;
  \raisebox{40pt}{
  \xymatrix@R=16pt@C=3.5em{
    \Sigma^{n+d-1}\mathbb{S}
    \ar@/_1.2pc/[dddrr]
      _-{
        \mathclap{\phantom{\vert^{\vert^{\vert}}}}
        \Sigma^\infty
        \color{greenii}
        c
        \mathclap{\phantom{\vert_{\vert_{\vert}}}}
      }
      ^-{\ }="t2"
    \ar@/^1.2pc/[rrrrd]
    \ar@{-->}[drr]
      |-{
        \scalebox{.7}{$
        \mathclap{\phantom{\vert^{\vert^{\vert}}}}
        \;\;
        \vdash
        {
          \big(
            {
              \color{greenii}
              G^E_n(c)
            }
            ,
            {
              \color{orangeii}
              H^E_{n-1}\!(c)
            }
          \big)
        }
        \mathclap{\phantom{\vert_{\vert_{\vert}}}}
        $}
      }
      _>>>{\ }="s2"
    \\
    &&
    \Sigma^{n-1}
    (\mathbb{S}/E)
    \ar[rr]_>>>{\ }="s"
    \ar[dd]^>>>{\ }="t"
      _-{
        \mathclap{\phantom{\vert^{\vert}}}
        \partial
        \mathclap{\phantom{\vert_{\vert}}}
      }
    &&
    \ast
    \ar[dd]
      ^-{
        \mathclap{\phantom{\vert^{\vert}}}
        0
        \mathclap{\phantom{\vert_{\vert}}}
      }
    \\
    \\
    &&
    \Sigma^n \mathbb{S}
    \ar[rr]
      |-{
        \;\Sigma^n(e^E)\;
      }
    &&
    \Sigma^n E
    \ar@{=>}
      ^-{ \mbox{\tiny\color{orangeii}(pb)} }
    "s"; "t"
    \ar@{=>}
    "s2"; "t2"
  }
  }
\end{equation}
Now the dashed morphism on the right of \eqref{TrivializationOfdInvariantFactoredInSpectra}
represents an element in $(E/\mathbb{S})_d$ and the
homotopy-commutativity of the left triangle on the right
shows that this bijection
$\big[ H^E_{n-1}\!(c)\big] \mapsto
\big[ \vdash H^E_{n-1}\!(c)\big]$
makes the square in \eqref{IdentificationOfFluxesWithCofiberCohomologyClasses}
commute.

\medskip

\noindent
Regarding {\bf (ii)}:
Let
$
  \big[
    S^{n + d - 1}
    \xrightarrow{\;c\;}
    S^{n}
  \big]
  \;\in\;
   \pi^n
   \big(
     S^{n+d-1}
   \big)
$
be a given class in Cohomotopy
and consider the following construction of a
homotopy pasting diagram in $\mathrm{Spectra}$, all of whose cells are
homotopy pushouts:
\begin{equation}
  \label{ConstructionOfMdFromC}
  \raisebox{40pt}{
  \xymatrix@R=9pt@C=5.5em{
    \Sigma^\infty S^{n + d - 1}
    \ar[dd]_-{
      \Sigma^\infty
      {\color{greenii}
        c
      }
    }
    \ar[rr]
    &&
    \ast
    \ar[dd]
    \\
    \\
    \Sigma^\infty S^{n}
    \ar[dd]
    \ar[rr]|-{
    }
    \ar@/_1.4pc/[rrr]
      |<<<<<<<<<<<<<<<<<<<{
        \; \Sigma^{n}(e^{E}) \;
      }
    &&
    \Sigma^\infty C_c
    \ar[dd]|<<<<{ \phantom{A} }
    \ar@{-->}[r]
    \ar@{}[r]
    |-{
      \scalebox{.7}{$
        \begin{array}{c}
          \vdash
          {\color{orangeii}
            H^E_{n-1}\!(c)
          }
          \\
          {\phantom{a}}
        \end{array}
      $}
    }
    \ar@/^2pc/[rr]
    &
    \Sigma^n E
    \ar[r]
    \ar[dd]
    &
    \ast
    \ar[dd]
    \\
    \\
    \ast
    \ar[rr]
    &&
    \Sigma^{\infty\scalebox{.6}{$+1$}} S^{n + d - 1}
    \ar[r]^-{
      \color{purple}
      M^{d}
    }
    \ar@/_1.2pc/[rr]|-{
      \;
      \Sigma^{\infty\scalebox{.5}{$+1$}}
      {\color{greenii}
        c
      }
      \;
    }
    &
    \Sigma^{n}(E\!/\mathbb{S})
    \ar[r]^{ \partial }
    &
    \Sigma^{\infty\scalebox{.6}{$+1$}} S^{n}
  }
  }
\end{equation}
For given $H^E_{n-1}\!(c)$,
this diagram is constructed as follows
(where we say ``square'' for
any {\it single} cell and ``rectangle'' for the pasting composite
of any adjacent {\it pair} of them):

\begin{itemize}

\vspace{-.2cm}
\item
The two squares on the left are the stabilization of the
homotopy pushout squares defining the cofiber space $C_c$
\eqref{ClassifyingCohomologyClassFor3Flux}
and the suspension of $S^{n + d - 1}$
\eqref{PastingForIteratedHomotopyCofiber}.

\vspace{-.2cm}
\item
The bottom left rectangle
(with $\Sigma^n(e^E)$ at its top)
is the homotopy pushout \eqref{CofiberOfRingSpectrumUnitAndBoundaryHomomorphism}
defining $\Sigma^n(E\!/\mathbb{S})$.

\vspace{-.2cm}
\item
The classifying map
$\vdash H^E_{n-1}(c)$
for the given $(n-1)$-flux
\eqref{ClassifyingCohomologyClassFor3Flux},
shown as a dashed arrow,
completes a co-cone under
the bottom left square. Thus the
map ${\color{purple}M^d}$ forming the
bottom middle square is uniquely implied by the
homotopy pushout property \eqref{HomotopyPushoutPropertyInIntroduction}
of the bottom left square.
Moreover, the pasting law \eqref{PastingLaw}
implies that this
bottom middle square is itself homotopy cartesian.

\vspace{-.2cm}
\item
The bottom right square is the
homotopy pushout \eqref{CofiberOfRingSpectrumUnitAndBoundaryHomomorphism}
defining $\partial$.

\vspace{-.2cm}
\item
By the pasting law \eqref{PastingLaw}
it follows that also the
bottom right rectangle is homotopy cartesian,
hence that, after the two squares on the left, it
exhibits the third step in the long homotopy cofiber sequence
\eqref{HomotopyCofiberSequence}
of $\Sigma^\infty c$.
This means that its total bottom
morphism is $\Sigma^{\infty + 1} c$, and hence
that $\partial \big[ M^d \big] = [c]$.
\end{itemize}
\vspace{-.1cm}

In conclusion,
these construction steps yield a map
map
${\color{orangeii}H^E_{n-1}\!(c)}
  \mapsto
  {
  \color{purple}
    M^d
  }
$ over the
stable class of $c$.

We claim that this assignment is bijective over
any $\Sigma^\infty c$:
To that end, assume conversely that $M^d$ is given,
and with it the diagram \eqref{ConstructionOfMdFromC}
except for the dashed arrow.
But since the bottom right square is
a homotopy pushout \eqref{CofiberOfRingSpectrumUnitAndBoundaryHomomorphism}
it is also a homotopy pullback (by Prop. \ref{InSpectraFiberSequencesAreCofiberSequences}),
whence  a dashed morphism is uniquely implied.
By its uniqueness, this reverse assignment
$M^{2d} \mapsto H^E_{n-1}\!(c)$ must be
the inverse of the previous construction.

Finally, comparison of the diagrams
\eqref{MappingHFluxToClassInCofiberCohomology} and
\eqref{ConstructionOfMdFromC} shows that
$M^d$ corresponds to the refined Toda bracket in question
under the hom-isomorphism of the stabilization adjunction
\eqref{StabilizationAdjunction}
$$
  \big[
  {
    \color{purple}
    M^d
  }
  \big]
  \;\leftrightarrow\;
  \big[
    \vdash
    (
      {
        \color{greenii}
        G^E_n(c)
      },
      {
        \color{orangeii}
        H^E_{n-1}(c)
      }
    )
  \big]
  \;=\;
    \big\langle
      {
      \color{greenii}
      c
      },
      {
        \Sigma^n(1^E)
      },
      (i^E)^n
    \big\rangle_{
      \big(
        {
          \color{orangeii}
          H^E_{n-1}(c)
        },
        \mathrm{(pb)}
      \big)
    }
  \,.
$$
\vspace{-.4cm}
\end{proof}

\begin{remark}[Higher homotopies]
  We may improve the lift \eqref{IdentificationOfFluxesWithCofiberCohomologyClasses}
  in Prop. \ref{EFluxesAreCocycleInCofiberTheory}
  to a bijection by retaining the
  2-homotopy class of the
  the left homotopy on the right of \eqref{TrivializationOfdInvariantFactoredInSpectra}
  (using the homotopy-universal property of $\infty$-limits,
  e.g. \cite[Def. 1.2.13.4, Thm. 4.2.4.1]{Lurie06}).
  For example, in the extreme case when
  $E = \mathbb{S}$ we have $E/\mathbb{S} \simeq 0$,
  so that there is no information in the lift
  \eqref{IdentificationOfFluxesWithCofiberCohomologyClasses}
  itself,
  while {\it all} the information about the $H_{n-1}$-flux
  is solely in the cone homotopies
  on the right of\eqref{TrivializationOfdInvariantFactoredInSpectra},
  these being the very homotopies that define $H^{\mathbb{S}}_{n-1}$
  in Def. \ref{TrivializationsOfThedInvariant},
  Example \ref{VanishingG4SFluxInVicinityOfM2Branes}.
\end{remark}

\medskip

\noindent {\bf $H^E_3$-Fluxes and extraordinary flat differential Cohomotopy.}

\begin{remark}[$H^E_3$-Fluxes as extraordinary flat differential Cohomotopy]
  \label{H3FluxesAsExtraordinaryFlatDifferentialCohomotopy}
Proposition \ref{EFluxesAreCocycleInCofiberTheory} implies that the set of
$H^E_3\mathrm{Fluxes}(X)$ is a variant
of the flat stable {\it differential Cohomotopy} of $X$:

\noindent {\bf (i)} Recall from
\cite[Ex. 434]{FSS20c}
that {\it differential stable Cohomotopy}
in degree 4 (just for definiteness)
is the cohomology classified by the homotopy pullback
\begin{equation}
  \label{Differential4Cohomotopy}
  \raisebox{28pt}{
  \xymatrix@C=4em{
    \overset{
      \mathclap{
      \raisebox{4pt}{
        \tiny
        \color{darkblue}
        \bf
        \begin{tabular}{c}
          classifying space for
          \\
          \color{purple}
          differential
          stable 4-Cohomotopy
        \end{tabular}
      }
      }
    }{
      {\mathbb{S}}{}^4_{\mathrm{conn}}
    }
    \ar[d]^>>{\ }="t"
    \ar[rr]_>>>{\ }="s"
    &&
    \overset{
      \mathclap{
      \raisebox{4pt}{
        \tiny
        \color{darkblue}
        \bf
        \begin{tabular}{c}
          classifying stack for
          \\
          differential 4-forms
        \end{tabular}
      }
      }
    }{
      \Omega_{\mathrm{dR}}^4(-)_{\mathrm{cl}}
    }
    \ar[d]
      ^-{
        \!\!\!\!\!\!\!
        \mbox{
          \tiny
          \color{greenii}
          \bf
          \begin{tabular}{c}
            de Rham
            \\
            theorem
          \end{tabular}
        }
      }
    \\
    \underset{
      \mathclap{
      \raisebox{-4pt}{
        \tiny
        \color{darkblue}
        \bf
        \begin{tabular}{c}
          classifying space for
          \\
          stable 4-Cohomotopy
        \end{tabular}
      }
      }
    }{
      \mathbb{S}^4
    }
    \ar[rr]
      |-{
        \;
          \mathrm{ch}_{\mathbb{S}^4}
          \,=\,
          \Sigma^4 ( e^{H\mathbb{R}})
        \;
      }
      ^-{
        \mbox{
          \tiny
          \color{greenii}
          \bf
          \begin{tabular}{c}
            Chern-Dold character map
          \end{tabular}
        }
        \mathclap{\phantom{\vert_{\vert}}}
      }
      _-{
        \mathclap{\phantom{\vert^{\vert}}}
        \mbox{
          \tiny
          \begin{tabular}{c}
            here:
            {
              \color{greenii}
              \bf
              unit map
            }
          \end{tabular}
        }
      }
    &{\phantom{AAAAAAAA}}&
    \underset{
      \mathclap{
      \raisebox{-4pt}{
        \tiny
        \color{darkblue}
        \bf
        \begin{tabular}{c}
          classifying space for
          \\
          ordinary real 4-cohomology
        \end{tabular}
      }
      }
    }{
      (H \mathbb{R})^4
    }
    \ar@{=>}^-{ \mbox{\tiny\rm(pb)} }
      "s"; "t"
  }
  }
\end{equation}
formed in the
homotopy theory of smooth $\infty$-stacks,
where the smooth classifying space
$\Omega^4_{\mathrm{dR}}(-)_{\mathrm{cl}}$
of closed differential 4-forms exists and
serves to capture the $G_4$-flux as a differential form.

\noindent {\bf (ii)} While smooth $\infty$-stacks are beyond the scope
of our discussion here, this technicality goes
away as we focus on {\it flat} differential Cohomotopy
where the 4-flux form vanishes, as befits the fluxless
backgrounds that we are focusing on in \cref{TheDictionary}.
In that case the above diagram \eqref{Differential4Cohomotopy}
reduces to the one shown on the left here:
\begin{equation}
  \label{FlatDifferential4Cohomotopy}
\quad
  \raisebox{28pt}{
  \xymatrix{
    \overset{
      \mathclap{
      \raisebox{4pt}{
        \tiny
        \color{darkblue}
        \bf
        \begin{tabular}{c}
          classifying space for
          \\
          { \color{purple}
            flat}
            \\
          { \color{purple}  differential
            stable 4-Cohomotopy
          }
        \end{tabular}
      }
      }
    }{
      \mathbb{S}{}^4_{\mathrm{flat}}
    }
    \ar[d]^>>{\ }="t"
    \ar[rr]_>>>{\ }="s"
    &&
    \ast
    \ar[d]
      ^-{
        \mathclap{\phantom{\vert^\vert}}
        0
        \mathclap{\phantom{\vert_\vert}}
      }
    \\
    \underset{
      \mathclap{
      \raisebox{-4pt}{
        \tiny
        \color{darkblue}
        \bf
        \begin{tabular}{c}
          classifying space for
          \\
          stable 4-Cohomotopy
        \end{tabular}
      }
      }
    }{
      \mathbb{S}^4
    }
    \ar[rr]
      |-{
        \;
          \mathrm{ch}_{\mathbb{S}^4}
          \,=\,
          \Sigma^4 ( e^{H\mathbb{R}})
        \;
      }
      _-{
        \mathclap{\phantom{\vert^{\vert}}}
        \mbox{
          \tiny
          \begin{tabular}{c}
            {
              \color{greenii}
              \bf
              $H\mathbb{R}$-unit map
            }
          \end{tabular}
        }
      }
    &{\phantom{AAAAAAA}}&
    \underset{
      \mathclap{
      \raisebox{-4pt}{
        \tiny
        \color{darkblue}
        \bf
        \begin{tabular}{c}
          classifying space for
          \\
          ordinary real 4-cohomology
        \end{tabular}
      }
      }
    }{
      (H \mathbb{R})^4
    }
    \ar@{=>}^-{ \mbox{\tiny\rm(pb)} }
      "s"; "t"
  }
  }
  \phantom{A}
  \mbox{
    \begin{tabular}{c}
      special case of:
    \end{tabular}
  }
  \phantom{AA}
  \raisebox{28pt}{
  \xymatrix{
    \overset{
      \mathclap{
      \raisebox{4pt}{
        \tiny
        \color{darkblue}
        \bf
        \begin{tabular}{c}
          classifying space for
          \\
          \color{purple}
          $H^E_3$-fluxes
        \end{tabular}
      }
      }
    }{
      (E/\mathbb{S})^3
    }
    \ar[d]^>>{\ }="t"
    \ar[rr]_>>>{\ }="s"
    &&
    \ast
    \ar[d]
      ^-{
        \mathclap{\phantom{\vert^\vert}}
        0
        \mathclap{\phantom{\vert_\vert}}
      }
    \\
    \underset{
      \mathclap{
      \raisebox{-4pt}{
        \tiny
        \color{darkblue}
        \bf
        \begin{tabular}{c}
          classifying space for
          \\
          stable 4-Cohomotopy
        \end{tabular}
      }
      }
    }{
      \mathbb{S}^4
    }
    \ar[rr]
      |-{
        \;
          \Sigma^4 ( e^{E})
        \;
      }
      _-{
        \mathclap{\phantom{\vert^{\vert}}}
        \mbox{
          \tiny
          \begin{tabular}{c}
            {
              \color{greenii}
              \bf
              $E$-unit map
            }
          \end{tabular}
        }
      }
    &{\phantom{AAAAAAA}}&
    \underset{
      \mathclap{
      \raisebox{-4pt}{
        \tiny
        \color{darkblue}
        \bf
        \begin{tabular}{c}
          classifying space for
          \\
          degree-4 $E$-cohomology
        \end{tabular}
      }
      }
    }{
      E^4
    }
    \mathrlap{\,.}
    \ar@{=>}^-{ \mbox{\tiny\rm(pb)} }
      "s"; "t"
  }
  }
\end{equation}
\noindent {\bf (iii)} While in general the Chern-Dold character on
a generalized cohomology theory $A$ is the rationalization
map \cite[\S 4.1]{FSS20c}, given on spectra as the smash product with the
unit $e^{H \mathbb{R}}$ \eqref{UnitMorphismOfSpectra}
of the Eilenberg-MacLane spectrum
\vspace{-2mm}
$$
  \mathrm{ch}_A
  \;:\;
  \xymatrix{
    A
    \ar[r]^-{ \;\simeq\; }
    &
    A \wedge \mathbb{S}
    \ar[rr]
      ^-{
        \mathrm{id}_A \wedge e^{H \mathbb{R}}
      }
    &&
    A \wedge H \mathbb{R}
    \,,
  }
$$
here for $A = \mathbb{S}$ this is reduces to the
unit map itself, as shown in \eqref{FlatDifferential4Cohomotopy}.
But this means
(by Def. \ref{UnitCofiberCohomologyTheories} with Prop. \ref{InSpectraFiberSequencesAreCofiberSequences})
that the classifying spectrum for flat
differential Cohomotopy is the unit cofiber spectrum
\eqref{CofiberOfRingSpectrumUnitAndBoundaryHomomorphism}
of
$H\mathbb{R}$, as highlighted on the right of \eqref{FlatDifferential4Cohomotopy}.

\noindent {\bf (iv)} It follows by  Prop. \ref{EFluxesAreCocycleInCofiberTheory} that
flat differential Cohomotopy
is exactly the theory of $H^{H\mathbb{R}}_3$-fluxes
\eqref{EFluxes} in the sense of Def. \ref{TrivializationsOfThedInvariant}:
$$
  {\mathbb{S}}{}^4_{\mathrm{flat}}
  \;\;\;\;
    \simeq
  \;\;\;\;
  \big(
    (H\mathbb{R})/\mathbb{S}
  \big)^3
  \,,
  {\phantom{AAAAAAA}}
  \overset{
    \mathclap{
    \raisebox{4pt}{
      \tiny
      \color{darkblue}
      \bf
      \begin{tabular}{c}
        flat differential
        \\
        4-Cohomotopy of $X$
      \end{tabular}
    }
    }
  }{
    \widetilde {\mathbb{S}}{}^4_{\mathrm{flat}}
    (X)
  }
  \;\;\;\simeq\;\;\;
  H^{H\mathbb{R}}_3\mathrm{Fluxes}(X)\;.
$$
\noindent {\bf (v)} Conversely, this means that for
Whitehead-generalized cohomology theories $E$,
the
$H^{E}_3\mathrm{Fluxes}$
\eqref{EFluxes} from Def. \ref{TrivializationsOfThedInvariant}:
constitute a generalization of flat differential Cohomotopy
where the trivialization of the $G_4$-flux by the $H_3$-flux
happens not in  ordinary real cohomology, but in
a Whitehead-generalized (``extraordinary'') cohomology theory.
\end{remark}

\medskip

\subsection{Adams e-invariant}
\label{TheAdamsEInvariant}

While the Adams e-inavariant of a map
(recalled below as Def. \ref{ClassicalConstructionOfAdamseCInvariant})
exists whenever the d-invariant vanishes,
its classical constructions proceeds
through a refined quantity $\widehat e$
which is an invariant of
the given map {\it equipped with a choice of trivialization}
of its d-invariant (Def. \ref{TrivializationsOfThedInvariant}).
Since it is these choices of trivializations that are identified with $H$-fluxes in
\cref{M5ThreeFlux}, we now re-cast the
construction of the e-invariant from the more
abstract perspective of Toda brackets (Def. \ref{RefinedTodaBracket})
that makes this
refined homotopy-dependence explicit
(Def. \ref{LiftedEInvariantDiagrammatically},
Theorem \ref{DiagrammaticeCCoincidesWithClassicaleCInvariant}).
We find that this perspective renders key properties of the e-invariant
transparently manifest, notably it reduces
Conner-Floyd's cobordism interpretation
to an immediate corollary (discussed in \cref{RelativeCobordism} below).

\medskip

\noindent {\bf The refined e-invariant as a Toda-bracket.}

\begin{defn}[The $\widehat e_{{K\mathrm{U}}}$-invariant]
 \label{LiftedEInvariantDiagrammatically}
We define the
{\it  ${\widehat e}_{{K \mathrm{U}}}$-invariant}
to be the composite of
\begin{itemize}
\vspace{-.2cm}
\item[{\bf (i)}]
the refined Toda-bracket
$O^{K\mathrm{U}/\mathbb{S}}$ \eqref{IdentificationOfFluxesWithCofiberCohomologyClasses}
from Prop. \ref{EFluxesAreCocycleInCofiberTheory}
for $E = {K\mathrm{O}}$,

\vspace{-.2cm}
\item[{\bf (ii)}]
with the cofiber Chern character from Example \ref{ChernCharacterOnCofiberOfKTheory},
yielding the refined Toda-bracket \eqref{DiagrammatichatecInvariant},

\vspace{-.2cm}
\item[{\bf (iii)}]
regarded as a rational number under
the canonical splitting from Remark \ref{CanonicalSplittingOfCofiberOfPeriodicRationalCohomology}:
\end{itemize}
\begin{equation}
  \label{FunctionDiagrammatichatecInvariant}
  \!\!\!\!\!\!\!\!\!\!\!\!\!\!\!
  \xymatrix@C=50pt@R=2pt{
    H^{K\mathrm{U}}_{2n-1}\mbox{Fluxes}
    \big(
      S^{2(n+d)-1}
    \big)
    \ar[r]
      _-{
        O^{K\mathrm{U}/\mathbb{S}}
      }
      \ar@/^1.2pc/[rr]
        ^-{
          O^{(H^{\mathrm{ev}\!}\mathbb{Q})/\mathbb{S}}
        }
    &
    ({K\mathrm{U}}/\mathbb{S})_{2d}
    \ar[r]_-{
      \;
      \mathrm{ch}/\mathbb{S}
      \;
    }
    &
    \big((H^{\mathrm{ev}}\mathbb{Q})/\mathbb{S}\big)_{2d}
    \ar[r]^-{
          \simeq
        }
      _-{ \mathrm{spl}_0 }
    &
    \mathbb{S}_{2(n+d)-1}
    \oplus
    \mathbb{Q}\;.
    \\
  \scalebox{0.8}{$  \Big(
      [{\color{greenii}c}]
      ,
      \big[
        {\color{orangeii}
        H^{{K\mathrm{U}}}_{2n-1}\!(c)
        }
      \big]
    \Big)
    $}
    \ar@{|->}[rrr]
    &
    &&
    \scalebox{0.8}{$  \Big(
      [{\color{greenii}c}]
      ,
      {\color{orangeii}
        \widehat e_{{K\mathrm{U}}}(c)
      }
    \Big)
    $}
  }
\end{equation}
\end{defn}

\begin{prop}[Refined $\widehat e_{{K\mathrm{U}}}$-invariant]
  \label{TheDiagrammaticeKUInvariant}
  The refined $\widehat e_{{K\mathrm{U}}}$-invariant
  (from Def.  \ref{LiftedEInvariantDiagrammatically})
  on
  $H^{{K\mathrm{U}}}_{n-1}\mathrm{Fluxes}\big(S^{2(n+d)-1}\big)$
  descends
  to a $\mathbb{Q}/\mathbb{Z}$-valued function
  $e_{{K\mathrm{U}}}$
  on
  $G^{\mathbb{S}}_{n}\mathrm{Fluxes}\big(S^{2(n+d)-1}\big)$
  \eqref{GFluxes}:
  \vspace{-2mm}
  \begin{equation}
    \label{DescendingTheDiagrammaticeKUInvariant}
    \raisebox{24pt}{
    \xymatrix@R=1.4em{
      H^{{K\mathrm{U}}}_{n-1}
      \mathrm{Fluxes}
      \big(
        S^{2(n+d)-1}
      \big)
      \ar[d]
      \ar[rr]^-{
          \;
          \widehat e_{{K\mathrm{U}}}
          \;
        }
      &&
      \mathbb{Q}
      \ar[d]
      \\
      G^{\mathbb{S}}_n\mathrm{Fluxes}
      \big(
        S^{2(n+d)-1}
      \big)
      \ar[rr]
        ^-{
          \;e_{{K\mathrm{U}}}\;
        }
      &&
      \mathbb{Q}/\mathbb{Z}
      \,.
    }
    }
  \end{equation}
\end{prop}
\begin{proof}
Using the same three ingredients that enter Def. \ref{LiftedEInvariantDiagrammatically}, we obtain the
following commuting diagram of abelian groups,
where the middle vertical composite is
$\widehat e_{{K\mathrm{U}}}$:
$$
  \xymatrix@R=18pt{
    \mathbb{Z}
    \;
    \ar[rr]
    \ar@{=}[d]
    &&
    H^{{K\mathrm{U}}}_{2n-1}\mbox{Fluxes}
    \big(
      S^{2(n+d)}
    \big)
    \ar[rr]
    \ar@{=}[d]
    &&
    \widetilde {\mathbb{S}}{}^{2n+1}
    \big(
      S^{2(n+d)}
    \big)
    \ar@{=}[d]
    \\
    \widetilde { {K\mathrm{U}} }{}^{2n}
    \big(
      S^{2(n+d)}
    \big)
    \ar[rr]
    \ar@{^{(}->}[d]_-{
         \mathrm{ch}
      }
    &&
    \big(\widetilde { ({K\mathrm{U}}/\mathbb{S}) }\big){}^{2n}
    \big(
      S^{2(n+d)}
    \big)
    \ar[d]
      _-{
               \mathrm{ch}/\mathbb{S}
              }
    \ar[rr]
    &&
    \widetilde {\mathbb{S}}{}^{2n+1}
    \big(
      S^{2(n+d)}
    \big)
    \ar@{=}[d]
    \\
    \big(
      \widetilde {H^{\mathrm{ev}}\mathbb{Q}}
    \big){}^{2n}
    \big(
      S^{2(n+d)}
    \big)
    \ar@{=}[d]
    \ar[rr]
    &&
    \big(
      \widetilde { (H^{\mathrm{ev}}\mathbb{Q})/\mathbb{S} }
    \big){}^{2n}
    \big(
      S^{2(n+d)}
    \big)
    \ar[rr]
    \ar[d]
      ^-{
          \simeq
      }_-{\mathrm{spl}_0}
    &&
    \widetilde {\mathbb{S}}{}^{2n+1}
    \big(
      S^{2(n+d)}
    \big)
    \ar@{=}[d]
    \\
    \mathbb{Q}
    \;
    \ar@{^{(}->}[rr]
    &&
    \mathbb{Q}
    \oplus
    \mathbb{S}_{2d-1}
    \ar@{->>}[rr]
    &&
    \mathbb{S}_{2d-1}
  }
$$
Now
{\bf (a)} Lemma \ref{CofiberECohomologyAsExtensionOfStableCohomotopyByECohomology}
shows that the middle horizontal rows are exact;
and
{\bf (b)} the left vertical morphism is the canonical injection
(since the Chern character preserves the unit).
Therefore (a) any two choices of $H^{{K\mathrm{U}}}_{n-1}\!(c)$
differ by an integer in the top row, and (b) this translates
to an integer difference as we pass down to the bottom row.
\end{proof}

\medskip

\noindent {\bf Recovering the classical e-invariant.}
We prove (Theorem \ref{DiagrammaticeCCoincidesWithClassicaleCInvariant} below) that the diagrammatically defined
e-invariant $\widehat e_{{K\mathrm{U}}}$
from Prop. \ref{TheDiagrammaticeKUInvariant}
coincides with Adams's classical ${\mathrm{e}_{\mathrm{Ad}}}$-invariant.
First we recall the classical construction
(\cite[\S 3,7]{Adams66}, review in \cite[\S 29]{Quick14}):

\begin{defn}[Classical construction of Adams $\mathrm{e}$-invariant]
 \label{ClassicalConstructionOfAdamseCInvariant}
For $n,d \in \mathbb{N}$, and
$\big[ S^{2(n+d)-1 } \xrightarrow{c} S^{2n} \big]
\in\pi^{2n}\big(S^{2(n+d)-1} \big)$,
consider the following construction.
Since
${K\mathrm{U}}_{2(n+d)-1} = 0$
(so that the $d_{{K\mathrm{U}}}$-invariant of all such $c$ vanishes)
the long exact sequence in ${K\mathrm{U}}$-cohomology
along the cofiber sequence of $c$
\vspace{-2mm}
$$
  \xymatrix{
    \cdots
    \ar@{<-}[r]
    &
    \underset{
      = \, 0
    }{
    \underbrace{
      \widetilde{{K\mathrm{U}}}{}^{0}
      \big(
        S^{2(n+d)-1}
      \big)
    }
    }
    \ar@{<-}[r]^-{ c^\ast }
    &
    \widetilde{{K\mathrm{U}}}{}^0
    \big( S^{2d} \big)
    \ar@{<-}[r]^-{
      q_c^\ast
    }
    &
    \widetilde{{K\mathrm{U}}}{}^0
    \big(
      C_c
    \big)
    \ar@{<-}[r]^-{
      p_c^\ast
    }
    &
    \widetilde{{K\mathrm{U}}}{}^0
    \big(
      S^{2(n+d)}
    \big)
    \ar@{<-}[r]^-{  }
    &
    \underset{
      = \, 0
    }{
    \underbrace{
      \widetilde{{K\mathrm{U}}}{}^{-1}
       \big(
         S^{2n}
       \big)
    }
    }
    \ar@{<-}[r]
    &
    \cdots
  }
$$

\vspace{-2mm}
\noindent truncates to a short exact sequence, which we may
identify as follows:
\vspace{-2mm}
\begin{equation}
  \label{TheSESInKUForAdams}
  \raisebox{30pt}{
  \xymatrix@R=10pt@C=20pt{
    0
      \ar@{<-}[r]
    &
    \widetilde {{K\mathrm{U}}}{}^0
    \big(
      S^{2n}
    \big)
    \ar@{<-}[rr]^-{
      q_c^\ast
    }
    \ar@{=}[d]
    &&
    \widetilde {{K\mathrm{U}}}{}^0
    \big(
      C_c
    \big)
    \ar@{=}[d]
    \ar@{<-}[rr]^-{
      p_c^\ast
    }
    &&
    \widetilde {{K\mathrm{U}}}{}^0
    \big(
      S^{2(n+d)}
    \big)
    \ar@{<-}[r]
    \ar@{=}[d]
    &
    0
    \\
    0
      \ar@{<-}[r]
    &
    \;\mathbb{Z}\;
    \ar@{=}[d]
    \ar@{<<-}[rr]^-{ }
    &&
    \mathbb{Z}
    \oplus
    \mathbb{Z}
    \ar@{<-^{)}}[rr]^-{ }
    \ar@{=}[d]
    &&
    \;
    \mathbb{Z}
    \ar@{=}[d]
    \;
    \ar@{<-}[r]
    &
    0
    \\
    &
    \Big\langle
      \big[
        G^{{K\mathrm{U}}}_{2n,\mathrm{unit}}
      \big]
    \Big\rangle
    &&
    \Big\langle
      \big[
        \vdash
        C^{{K\mathrm{U}}}_{2n-1}\!(c)
      \big]
      \,
      ,
      \,
      \big[
        \vdash
        H^{{K\mathrm{U}}}_{2n-1,\mathrm{unit}}
      \big]
    \Big\rangle
    &&
    \Big\langle
      \big[
        \vdash
        H^{{K\mathrm{U}}}_{2n-1,\mathrm{unit}}
      \big]
    \Big\rangle
    \\
    &
    \big[
      G^{{K\mathrm{U}}}_{2n,\mathrm{unit}}
    \big]
    \ar@{|->}[rr]^-{ \mathrm{spl}^{{K\mathrm{U}}} }
    &&
    \big[
      \vdash
      C^{{K\mathrm{U}}}_{2n-1}\!(c)
    \big]
    {\phantom{
      \big[
        \vdash
        H^{{K\mathrm{U}}}_{2n-1,\mathrm{unit}}
      \big]
    }}
    \;\;\;
  }
  }
\end{equation}

\vspace{-2mm}
\noindent
Here we have observed that the two outer groups are the integers,
by the suspension isomorphism and Bott-periodicity.
Such short exact sequences split, and any choice
of splitting $\mathrm{spl}$ identifies the middle group as a
direct sum $\mathbb{Z} \oplus \mathbb{Z}$, whose
canonical generators we have indicated by suggestive symbols.

\medskip

Now, using
that the Adams operations $\psi^k$ \eqref{AdamsOperationsOnComplexKTheory}
over the $(2r)$-sphere are given \eqref{AdamsOperationsOverSpheres}
by multiplication with $k^{r}$,
and that they commute with pullback \eqref{AdamsOperationsCommuteWithPullback},
it follows from \eqref{TheSESInKUForAdams} that
 \vspace{-3mm}
\begin{equation}
  \label{ActionOfAdamsOperationsOnSplitKTheoryOfCofiberSpace}
  \xymatrix@R=-2pt{
    \widetilde {{K\mathrm{U}}}{}^0
    \big(
      C_c
    \big)
    \ar[rr]^{ \psi^k }
    &&
    \widetilde {{K\mathrm{U}}}{}^0
    \big(
      C_c
    \big)
    \\
    \big[
      \vdash
      C^{{K\mathrm{U}}}_{2n-1}\!(c)
    \big]
    \ar@{}[rr]|-{\longmapsto}
    &&
    k^{n}
    \cdot
    \big[
      \vdash
      C^{{K\mathrm{U}}}_{2n-1}\!(c)
    \big]
    \,+\,
    {\color{orangeii}
      \mu_k(c)
    }
    \cdot
    \big[
      \vdash
      H^{{K\mathrm{U}}}_{2n-1,\mathrm{unit}}
    \big]
    \\
    \big[
      \vdash
      H^{{K\mathrm{U}}}_{2n-1,\mathrm{unit}}
    \big]
    \ar@{}[rr]|-{\hspace{-1.67cm}\longmapsto}
    &&
    k^{n+d}
    \cdot
    \big[
      \vdash
      H^{{K\mathrm{U}}}_{2n-1, \mathrm{unit}}
    \big]
  }
\end{equation}
for {\it some} assignment
\begin{equation}
  \label{ThePrecursorOfTheeCInvariant}
  (c,k)
  \;\longmapsto\;
  {\color{orangeii}
    \mu_k(c)
  }
  \;\in\;
  \mathbb{Z}
  \,.
\end{equation}
Then the {\it classical Adams $\mathrm{e}$-invariant}
of $c$ is
\begin{equation}
  \label{AdamseCInvariant}
  \overset{
    \mathclap{
    \raisebox{4pt}{
      \tiny
      \color{darkblue}
      \bf
      Adams $\mathrm{e}$-invariant
    }
    }
  }{
    {\mathrm{e}_{\mathrm{Ad}}}(c)
  }
  \;:=\;
  \left[
    \widehat {\mathrm{e}_{\mathrm{Ad}}}(c)
  \right]_{\mathrm{mod}\,\mathbb{Z}}
  \;\;
  \in
  \mathbb{Q}/\mathbb{Z}
  \,,
  \phantom{AA}
  \mbox{for}
  \;\;\;\;
  \overset{
    \mathclap{
    \raisebox{4pt}{
      \tiny
      \color{darkblue}
      \bf
      \begin{tabular}{c}
        refined
        \\
        Adams $\mathrm{e}$-invariant
      \end{tabular}
    }
    }
  }{
    \widehat {\mathrm{e}_{\mathrm{Ad}}}(c)
  }
  \;:=\;
  \frac{
    {\color{orangeii}
      \mu_k(c)
    }
  }{
    k^{n}
    \big(
      k^d - 1
    \big)
  }
  \;\;
  \in
  \;
  \mathbb{Q}
  \,.
\end{equation}
It is readily checked that this is well-defined in that it is...

\begin{itemize}

\item
...independent of the choice of $k \geq 1$ -- this
follows by imposing the commutativity of the Adams operations
$\psi^{k_1}\circ \psi^{k_2} \,=\, \psi^{k_2} \circ \psi^{k_1}$
on the transformation law \eqref{ActionOfAdamsOperationsOnSplitKTheoryOfCofiberSpace};

\item
...independent of the choice of
splitting of \eqref{TheSESInKUForAdams} in that
$$
  \big[
    \vdash
    C^{{K\mathrm{U}}}_{2n-1}\!(c)
  \big]
  \;\longmapsto\;
  \big[
    \vdash
    C^{{K\mathrm{U}}}_{2n-1}\!(c)
  \big]
  \,+\,
  n\cdot
  \big[
    \vdash
    H^{{K\mathrm{U}}}_{2n-1,\mathrm{unit}}
  \big]
  \;\;\;\;\;\;\;\;\;
  \Rightarrow
  \;\;\;\;\;\;\;\;\;
  \widehat {\mathrm{e}_{\mathrm{Ad}}}
  \;\longmapsto\;
  \widehat {\mathrm{e}_{\mathrm{Ad}}}
  \,+\,
  n
  \,.
$$
This follows by direct
inspection of \eqref{ActionOfAdamsOperationsOnSplitKTheoryOfCofiberSpace}.
\end{itemize}

\noindent Alternatively, the number $\widehat {\mathrm{e}_{\mathrm{Ad}}}$
may be defined without reference to the Adams operations.
Since the Chern character map $\mathrm{ch}$
respects \eqref{ChernCharacterCompatibilityWithAdamsOperations}
the Adams operations,
the effect of \eqref{ActionOfAdamsOperationsOnSplitKTheoryOfCofiberSpace}
on Chern characters in
$\widetilde {H^{\mathrm{ev}}\mathbb{Q}}{}^0(C_c)$
is given, in matrix notation, by:
$$
  \psi^k
  \,
  \left(
  \!\!\!
  \def\arraystretch{1.4}
  \begin{array}{l}
    \mathrm{ch}
    \big[
      \vdash
      C^{{K\mathrm{U}}}_{2n-1}(c)
    \big]
    \\
    \mathrm{ch}
    \big[
      \vdash
      H^{{K\mathrm{U}}}_{2n-1,\mathrm{unit}}
    \big]
  \end{array}
  \!\!\!
  \right)
  \;\;\;
    =
  \;\;\;
  \left( \!\!\!
  \def\arraystretch{1.4}
  \begin{array}{cc}
    k^n
      &
    {\color{orangeii}
      \widehat {\mathrm{e}_{\mathrm{Ad}}}(c)
    }
    \,\cdot\,
    k^n \big( k^d  - 1\big)
    \\
    0
    &
    k^{n + d}
  \end{array}
   \!\!\!  \right)
  \;\cdot\;
  \left(
  \!\!\!
  \def\arraystretch{1.4}
  \begin{array}{l}
    \mathrm{ch}
    \big[
      \vdash
      C^{{K\mathrm{U}}}_{2n-1}(c)
    \big]
    \\
    \mathrm{ch}
    \big[
      \vdash
      H^{{K\mathrm{U}}}_{2n-1,\mathrm{unit}}
    \big]
  \end{array}
  \!\!\!
  \right)
  \,.
$$
This admits diagonalization, with eigenvectors
\vspace{-2mm}
\begin{equation}
  \label{EigenvectorForRationalAdamsOperations}
  \def\arraystretch{1.5}
  \begin{array}{rcll}
    \mathrm{ch}
    \big[
      \vdash
      C^{{K\mathrm{U}}}_{2n-1}\!(c)
      -
      {\color{orangeii}
        \widehat {\mathrm{e}_{\mathrm{Ad}}}(c)
      }
      \cdot
      \vdash
      H^{{K\mathrm{U}}}_{3,\mathrm{unit}}
    \big]
    &
    \longmapsto
    &
    k^n
    \cdot
    &
    \mathrm{ch}
    \big[
      \vdash
      C^{{K\mathrm{U}}}_{2n-1}\!(c)
      -
      {\color{orangeii}
        \widehat {\mathrm{e}_{\mathrm{Ad}}}(c)
      }
      \cdot
      \vdash
      H^{{K\mathrm{U}}}_{2n-1,\mathrm{unit}}
    \big]\;,
    \\
    \mathrm{ch}
    \big[
      \vdash
      H^{{K\mathrm{U}}}_{2n-1,\mathrm{unit}}
    \big]
    &
    \longmapsto
    &
    k^{n + d}
    \cdot
    &
    \mathrm{ch}
    \big[
      \vdash
      H^{{K\mathrm{U}}}_{2n+1,\mathrm{unit}}
    \big]\;.
  \end{array}
\end{equation}
Since the component of the Chern character transforming
with $k^r$ is \eqref{SingleDegreeChernCharacterClasses}
exactly the component of cohomological degree $2r$, this means that
the refined e-invariant $\widehat {\mathrm{e}_{\mathrm{Ad}}}(c)$
\eqref{AdamseCInvariant}
is equivalently the degree-$2(n+d)$-component
of the Chern character of the choice of splitting \eqref{TheSESInKUForAdams}:
\vspace{-4mm}
\begin{equation}
  \label{AdamseCInvariantAsTopDegreeComponentOfChernCharacter}
  \hspace{-4mm}
  \raisebox{31pt}{
  \xymatrix@R=8pt@C=16pt{
    \widetilde {{K\mathrm{U}}}{}^{2n}
    \big(
      S^{2n}
    \big)
    \ar[rr]
      _-{ \;\mathrm{spl}^{{K\mathrm{U}}} \circ \, i_r \; }
      ^-{
        \mbox{
          \tiny
          \color{greenii}
          \bf
          \begin{tabular}{c}
            choose lift
            \\
            \phantom{a}
          \end{tabular}
        }
      }
    &&
    \widetilde {{K\mathrm{U}}}{}^{2n}
    \big(
      C_c
    \big)
    \ar[rr]
      _-{ \;\mathrm{ch}\; }
      ^-{
        \mbox{
          \tiny
          \color{greenii}
          \bf
          \begin{tabular}{c}
            Chern character
            \\
            \phantom{a}
          \end{tabular}
        }
      }
    &&
    \widetilde {H^{\mathrm{ev}}\mathbb{Q}}{}^{2n}
    \big(
      C_c
    \big)
    \ar[rr]
      _-{  \;p_l \, \circ  \,\mathrm{spl}^{H^{\mathrm{ev}}\mathbb{Q}}_0\; }
      ^-{
        \mathclap{
        \mbox{
          \tiny
          \color{greenii}
          \bf
          \begin{tabular}{c}
            projection onto
            \\
            degree $2(n+d)$
                 \end{tabular}
        }
        }
      }
    &&
    \widetilde {
      H\mathbb{Q}
    }{}^{2(n+d)}(C_c)
    \ar@{=}[r]
    &
    \mathbb{Q}
    \\
    \big[
      G^{{K\mathrm{U}}}_{2n,\mathrm{unit}}
    \big]
    \ar@{|->}[rr]
    &&
    \big[
      \vdash C^{{K\mathrm{U}}}_{2n-1}\!(c)
    \big]
    \ar@{|->}[rr]
    &&
    \mathrm{ch}
    \big[
      \vdash C^{{K\mathrm{U}}}_{2n-1}\!(c)
    \big]
    \ar@{|->}[rr]
      _-{ \mbox{\tiny \eqref{EigenvectorForRationalAdamsOperations}} }
    &&
    {\color{orangeii}
      \widehat {\mathrm{e}_{\mathrm{Ad}}}(c)
    }
    \cdot
    \mathrm{ch}
    \big[
      \vdash H^{{K\mathrm{U}}}_{2n-1,\mathrm{unit}}
    \big]
    \ar@{|->}[r]
    &
    {\color{orangeii}
      \widehat {\mathrm{e}_{\mathrm{Ad}}}(c)
    }
  }
  }
\end{equation}
\end{defn}

\begin{example}[Adams e-invariant on 3rd stable stem]
 \label{AdamseInvariantOn3rdStableStem}
On the third stem
$\mathbb{S}_3 \;\simeq\; \mathbb{Z}_{24} \coloneqq \mathbb{Z}/24$
\eqref{FromUnstableToStable4CohomotopyOf7Sphere},
the
classical Adams e-invariant (Def. \ref{ClassicalConstructionOfAdamseCInvariant})
is:
\vspace{-2mm}
$$
  \xymatrix@C=7pt@R=1.5em{
    \mathbb{S}_\bullet
    \ar[d]^-{
      e_{\color{blue}\mathbb{R}}
    }_-{\simeq}
    \ar@{}[r]|-{\ni}
    &
    \big[
      h_{\mathbb{H}}
    \big]
    \ar@{}[dd]|-{ \mapsdown }
    \\
    \mathclap{\phantom{\vert_{\vert}}}
    \mathbb{Z}_{24}
    \ar@{^{(}->}[d]
    \\
    \mathbb{Q}/\mathbb{Z}
    \ar@{}[r]|-{\ni}
    &
    \big[
      \tfrac{1}{24}
    \big]
  }
  \phantom{AAAAAAAAAAA}
  \xymatrix@C=7pt@R=1.5em{
    \mathbb{S}_\bullet
    \ar@{->>}[d]^-{
      e_{\color{blue}\mathbb{C}}
    }_-{\simeq}
    \ar@{}[r]|-{\ni}
    &
    \big[
      h_{\mathbb{H}}
    \big]
    \ar@{}[dd]|-{ \mapsdown }
    \\
    \mathclap{\phantom{\vert_{\vert}}}
    \mathbb{Z}_{12}
    \ar@{^{(}->}[d]
    \\
    \mathbb{Q}/\mathbb{Z}
    \ar@{}[r]|-{\ni}
    &
    \big[
      \tfrac{1}{12}
    \big]
  }
$$
\end{example}
\noindent
This is due to  \cite[Prop. 7.14, Ex. 7.17]{Adams66}.

\begin{remark}
  Observe that both the
  classical construction of the ${\mathrm{e}_{\mathrm{Ad}}}$-invariant in the form
  \eqref{AdamseCInvariantAsTopDegreeComponentOfChernCharacter}
  as well as our diagrammatic construction in
  Def. \ref{LiftedEInvariantDiagrammatically}
  produce a number by reference to one of the
  {\it canonical splittings} of Remark \ref{CanonicalSplittingOfCofiberOfPeriodicRationalCohomology},
  given by projection onto the rational cohomology in degree
  $2(n+2)$.
\end{remark}

Hence to relate the two constructions we need to
relate these two canonical splittings:
\begin{lemma}[Compatibility of the two canonical splittings]
  \label{CompatibilityOfTheTwoCanonicalSplittings}
  Given non-trivial
  $\big[ S^{2(n+d)-1} \xrightarrow{\;c\;} S^{2n}\big]
  \,\in \, \mathbb{S}_{2d-1}$
  for any $n,d \in \mathbb{N}$ with $d \geq 1$, we have
  a commuting diagram

  \vspace{-3mm}
  $$
    \raisebox{30pt}{
    \xymatrix@C=10pt@R=6pt{
      &&
      \big(
        \widetilde { (H^{\mathrm{ev}}\mathbb{Q})/\mathbb{S} }
      \big)^{2n}
      \big(
        S^{2(n+d)}
      \big)
      \ar[dd]
        _-{
          \mathclap{\phantom{\vert^{\vert}}}
          p^\ast
          \mathclap{\phantom{\vert_{\vert}}}
        }
      \ar@{=}[dr]^-{ \;\mathrm{spl}_0\; }
      \\
      &&
      &
      \mathbb{Q} \oplus \mathbb{S}_{2d-1}
      \ar[dd]^-{ \binom{\mathrm{id}~ \cdot}{\,0~\cdot}
      }
      \\
      \big(
        \widetilde { H^{\mathrm{ev}}\mathbb{Q} }
      \big)^{2n}
      \big(
        C_c
      \big)
      \ar@{=}[dr]_-{ \;\mathrm{spl}_0\; }
      \ar[rr]^-{ \;i_\ast\; }
      &&
      \big(
        \widetilde { (H^{\mathrm{ev}}\mathbb{Q})/\mathbb{S} }
      \big)^{2n}
      \big(
        C_c
      \big)
      \ar@{=}[dr]
      \\
      &
      \mathbb{Q} \oplus \mathbb{Q}
      \ar[rr]|-{\;\; \binom{\mathrm{id}~ \cdot}{\,0~\cdot} \;\;
        }
      &&
      \mathbb{Q} \oplus
      \big(
        \mathbb{Q}/\mathbb{Z}
        \oplus'
        \mathbb{S}_{2d-1}/\mathbb{Z}
      \big)
      \,,
    }
    }
  $$
  \vspace{-5mm}
\end{lemma}
\noindent
Here:

\vspace{-.1cm}
\begin{itemize}

  \vspace{-.2cm}
  \item
  the rear morphisms are pushforward along
  the projection $E \xrightarrow{i} E/\mathbb{S}$ \eqref{CofiberOfRingSpectrumUnitAndBoundaryHomomorphism}
  and pullback along $C_c \xrightarrow{p} S^{2(n+d)}$,
  respectively;

  \vspace{-.2cm}
  \item
  $\mathrm{spl}_0$ denotes the canonical isomorphisms
   from Remark \ref{CanonicalSplittingOfCofiberOfPeriodicRationalCohomology};

  \vspace{-.2cm}
  \item
  the matrices act on row vectors
  by multiplication from the right;

\vspace{-.2cm}
\item
$\mathbb{S}_{d-1}
  \,\coloneqq\,
  \mathrm{cof}
  \big(
    \mathbb{Z}
      \xrightarrow{c}
    \mathbb{S}_{d-1}
  \big)
$;

\vspace{-.2cm}
\item the symbol $\oplus'$ denotes some
possibly non-trivial extension,
left undetermined, of
$\mathbb{Q}/\mathbb{Z}$ by
$\mathbb{S}_{d-1}/\mathbb{Z}$
and vice versa.

\end{itemize}
\begin{proof}
Consider the diagram
\vspace{-3mm}
$$
  \pi_0
  \mathrm{Maps}
\scalebox{1}{$
  \left(
  \raisebox{48pt}{
    \xymatrix@R=1pt{
      \vdots
\\
      \\
      S^{2(n+d)}
         \ar@{->}[uu]
      \ar@{<-}[dd]
      \\
      \\
      C_c
      \ar@{<-}[dd]
      \ar@{}[r]|-{ , }
      &
    (H^{\mathrm{ev}}\mathbb{Q})^{2n}
    \ar[r]
    &
    \big(
      (H^{\mathrm{ev}}\mathbb{Q})
      /
      \mathbb{S}
    \big)^{2n}
    \ar[r]
    &
    \mathbb{S}^{2n-1}
      \\
      \\
      S^{2n}
      \\
      \\
      \vdots
      \ar@{->}[uu]
    }
  }
  \right)
  $}
$$
all of whose rows and columns are exact
(since $\mathrm{Maps}^{\ast/}$ sends both homotopy cofiber sequences
in the first argument as well as homotopy fiber sequences in the second
argument to homotopy fiber sequences, and
using that these induce long exact sequences on homotopy groups).

By definition of reduced cohomology groups, this
diagram is equal to the following (in the sector shown):
$$
\hspace{-1mm}
  \xymatrix@C=14pt@R=12pt{
    \widetilde {\mathbb{S}}{}^{2n}
    \big(
      S^{2n+1}
    \big)
    \ar[d]
    \ar[r]
    &
    \widetilde { (H^{\mathrm{ev}}\mathbb{Q}) }{}^{2n}
    \big(
      S^{2n+1}
    \big)
    \ar[r]
    \ar[d]
    &
    (\widetilde {(H^{\mathrm{ev}}\mathbb{Q})/\mathbb{S}}){}^{2n}
    \big(
      S^{2n+1}
    \big)
    \ar[d]
    \ar[r]
    &
    \widetilde {\mathbb{S}}{}^{2n+1}
    \big(
      S^{2n+1}
    \big)
    \ar[d]
    \\
    \widetilde {\mathbb{S}}{}^{2n}
    \big(
      S^{2(n+d)}
    \big)
    \ar[r]
    \ar[d]
    &
    \widetilde {(H^{\mathrm{ev}}\mathbb{Q})}{}^{2n}
    \big(
      S^{2(n+d)}
    \big)
    \;
    \ar[r]
    \ar[d]
    &
    (\widetilde {(H^{\mathrm{ev}}\mathbb{Q})/\mathbb{S}}){}^{2n}
    \big(
      S^{2(n+d)}
    \big)
    \ar[r]
    \ar[d]
    &
    \widetilde {\mathbb{S}}{}^{2n+1}
    \big(
      S^{2(n+d)}
    \big)
    \ar[d]
    \ar[r]
    &
    \widetilde {(H^{\mathrm{ev}}\mathbb{Q})}{}^{2n+1}
    \big(
      S^{2(n+d)}
    \big)
    \ar[d]
    \\
    \widetilde {\mathbb{S}}{}^{2n}
    \big(
      C_c
    \big)
    \ar[d]
    \ar[r]
    &
    \widetilde {(H^{\mathrm{ev}}\mathbb{Q})}{}^{2n}
    \big(
      C_c
    \big)
    \;
    \ar[r]
    \ar[d]
    &
    (\widetilde {(H^{\mathrm{ev}}\mathbb{Q})/\mathbb{S}}){}^{2n}
    \big(
      C_c
    \big)
    \ar[r]
    \ar[d]
    &
    \widetilde {\mathbb{S}}{}^{2n+1}
    \big(
      C_c
    \big)
    \ar[d]
    \ar[r]
    &
    \widetilde {(H^{\mathrm{ev}}\mathbb{Q})}{}^{2n+1}
    \big(
      C_c
    \big)
    \\
    \widetilde {\mathbb{S}}{}^{2n}
    \big(
      S^{2n}
    \big)
    \ar[d]
    \ar[r]
    &
    \widetilde {(H^{\mathrm{ev}}\mathbb{Q})}{}^{2n}
    \big(
      S^{2n}
    \big)
    \ar[r]
    \ar[d]
    &
    (\widetilde {(H^{\mathrm{ev}}\mathbb{Q})/\mathbb{S}}){}^{2n}
    \big(
      S^{2n}
    \big)
    \ar[r]
    &
    \widetilde {\mathbb{S}}{}^{2n+1}
    \big(
      S^{2n}
    \big)
    \\
    \widetilde {\mathbb{S}}{}^{2n}
    \big(
      S^{2(n+d)-1}
    \big)
    \ar[r]
    &
    \widetilde {(H^{\mathrm{ev}}\mathbb{Q})}{}^{2n}
    \big(
      S^{2(n+d)-1}
    \big)
  }
$$
Evaluating  all the cohomology groups on spheres yields:
\vspace{-2mm}
$$
\hspace{-3mm}
  \xymatrix@C=16pt@R=14pt{
    {\phantom{
    \widetilde {\mathbb{S}}{}^{2n}
    \big(
      S^{2n+1}
    \big)
    }}
    \hspace{-23pt}
    \mathclap{
     \color{blue}
     \mathbb{S}_{1} \oplus 0
    }
    \hspace{+23pt}
    \ar[d]
    \ar[r]
    &
    {\phantom{
    \widetilde { (H^{\mathrm{ev}}\mathbb{Q}) }{}^{2n}
    \big(
      S^{2n+1}
    \big)
    }}
    \hspace{-38pt}
    \mathclap{
      \color{blue}
      0
    }
    \hspace{+38pt}
    \ar[r]
    \ar[d]
    &
    (\widetilde {(H^{\mathrm{ev}}\mathbb{Q})/\mathbb{S}}){}^{2n}
    \big(
      S^{2n+1}
    \big)
    \ar[d]
    \ar[r]
    &
    {\phantom{
    \widetilde {\mathbb{S}}{}^{2n+1}
    \big(
      S^{2n+1}
    \big)
    }}
    \hspace{-30pt}
    \mathclap{
      \color{blue}
      0 \oplus \mathbb{Z}
    }
    \hspace{+30pt}
    \ar[d]
      _-{
        \mathclap{\phantom{\vert^{\vert}}}
        \color{blue}
        c
        \mathclap{\phantom{\vert_{\vert}}}
      }
    \\
    {\phantom{
    \widetilde {\mathbb{S}}{}^{2n}
    \big(
      S^{2(n+d)}
    \big)
    }}
    \hspace{-32pt}
    \mathclap{
      \color{blue}
      \mathbb{S}_{2d} \oplus 0
    }
    \hspace{+32pt}
    \ar[r]^-{ \color{blue}\;0\; }
    \ar[d]
    &
    {\phantom{
    \widetilde {(H^{\mathrm{ev}}\mathbb{Q})}{}^{2n}
    \big(
      S^{2(n+d)}
    \big)
    }}
    \hspace{-41pt}
    \mathclap{
      \color{blue}
      \mathbb{Q} \oplus 0
    }
    \hspace{+41pt}
    \;
    \ar[r]
    \ar[d]
    &
    (\widetilde {(H^{\mathrm{ev}}\mathbb{Q})/\mathbb{S}}){}^{2n}
    \big(
      S^{2(n+d)}
    \big)
    \ar[r]
    \ar[d]
    &
    {\phantom{
    \widetilde {\mathbb{S}}{}^{2n+1}
    \big(
      S^{2(n+d)}
    \big)}}
    \hspace{-30pt}
    \mathclap{
      \color{blue}
      0 \oplus \mathbb{S}_{2d-1}
    }
    \hspace{+30pt}
    \ar[d]
    \ar[r]
    &
    {\phantom{
    \widetilde {(H^{\mathrm{ev}}\mathbb{Q})}{}^{2n+1}
    \big(
      S^{2(n+d)}
    \big)
    }}
    \hspace{-47pt}
    \mathclap{
      \color{blue}
      0
    }
    \hspace{+47pt}
    \ar[d]
    \\
    \widetilde {\mathbb{S}}{}^{2n}
    \big(
      C_c
    \big)
    \ar[d]
    \ar[r]
    &
    \widetilde {(H^{\mathrm{ev}}\mathbb{Q})}{}^{2n}
    \big(
      C_c
    \big)
    \;
    \ar[r]
    \ar[d]
    &
    (\widetilde {(H^{\mathrm{ev}}\mathbb{Q})/\mathbb{S}}){}^{2n}
    \big(
      C_c
    \big)
    \ar[r]
    \ar[d]
    &
    \widetilde {\mathbb{S}}{}^{2n+1}
    \big(
      C_c
    \big)
    \ar[d]
    \ar[r]
    &
    \widetilde {(H^{\mathrm{ev}}\mathbb{Q})}{}^{2n+1}
    \big(
      C_c
    \big)
    \\
    {\phantom{
    \widetilde {\mathbb{S}}{}^{2n}
    \big(
      S^{2n}
    \big)
    }}
    \hspace{-19pt}
    \mathclap{
      \color{blue}
      0 \oplus \mathbb{Z}
    }
    \hspace{+19pt}
    \ar[d]
      _-{
        \mathclap{\phantom{\vert^\vert}}
        \color{blue}
        c
        \mathclap{\phantom{\vert_\vert}}
      }
    \ar[r]
      ^-{
        \color{blue}
         1 \mapsto 1
      }
    &
    {\phantom{
    \widetilde {(H^{\mathrm{ev}}\mathbb{Q})}{}^{2n}
    \big(
      S^{2n}
    \big)
    }}
    \hspace{-32pt}
    \mathclap{
      \color{blue}
      0 \oplus \mathbb{Q}
    }
    \hspace{+32pt}
    \ar[r]
    \ar[d]
    &
    (\widetilde {(H^{\mathrm{ev}}\mathbb{Q})/\mathbb{S}}){}^{2n}
    \big(
      S^{2n}
    \big)
    \ar[r]
    &
    {\phantom{
    \widetilde {\mathbb{S}}{}^{2n+1}
    \big(
      S^{2n}
    \big)
    }}
    \hspace{-25pt}
    \mathclap{
      \color{blue}
      0
    }
    \hspace{+25pt}
    \\
    {\phantom{
    \widetilde {\mathbb{S}}{}^{2n}
    \big(
      S^{2(n+d)-1}
    \big)
    }}
    \hspace{-24pt}
    \mathclap{
      \color{blue}
      0 \oplus \mathbb{S}_{2d-1}
    }
    \hspace{+24pt}
    \ar[r]
    &
    {\phantom{
    \widetilde {(H^{\mathrm{ev}}\mathbb{Q})}{}^{2n}
    \big(
      S^{2(n+d)-1}
    \big)
    }}
    \hspace{-47pt}
    \mathclap{
      \color{blue}
      0
    }
    \hspace{+47pt}
  }
$$
From this we recognize various split exact sequences,
using that $\mathrm{Ext}^1(-,\mathbb{Q}) = 0$
and
$\mathrm{Ext}^1(\mathbb{Z},-) = 0$:
\vspace{-2mm}
$$
  \xymatrix@C=16pt@R=14pt{
    {\phantom{
    \widetilde {\mathbb{S}}{}^{2n}
    \big(
      S^{2n+1}
    \big)
    }}
    \hspace{-23pt}
    \mathclap{
     \mathbb{S}_{1} \oplus 0
    }
    \hspace{+23pt}
    \ar[d]
    \ar[r]
    &
    {\phantom{
    \widetilde { (H^{\mathrm{ev}}\mathbb{Q}) }{}^{2n}
    \big(
      S^{2n+1}
    \big)
    }}
    \hspace{-38pt}
    \mathclap{
      0
    }
    \hspace{+38pt}
    \ar[r]
    \ar[d]
    &
    {\phantom{
    (\widetilde {(H^{\mathrm{ev}}\mathbb{Q})/\mathbb{S}}){}^{2n}
    \big(
      S^{2n+1}
    \big)
    }}
    \hspace{-48pt}
    \mathclap{
      \color{blue}
      0 \oplus \mathbb{Z}
    }
    \hspace{+48pt}
    \ar[d]
    \ar[r]
    &
    {\phantom{
    \widetilde {\mathbb{S}}{}^{2n+1}
    \big(
      S^{2n+1}
    \big)
    }}
    \hspace{-30pt}
    \mathclap{
      0 \oplus \mathbb{Z}
    }
    \hspace{+30pt}
    \ar[d]
      _-{
        \mathclap{\phantom{\vert^{\vert}}}
        c
        \mathclap{\phantom{\vert_{\vert}}}
      }
    \\
    {\phantom{
    \widetilde {\mathbb{S}}{}^{2n}
    \big(
      S^{2(n+d)}
    \big)
    }}
    \hspace{-32pt}
    \mathclap{
      \mathbb{S}_{2d} \oplus 0
    }
    \hspace{+32pt}
    \ar[r]^-{ \;0\; }
    \ar[d]
    &
    {\phantom{
    \widetilde {(H^{\mathrm{ev}}\mathbb{Q})}{}^{2n}
    \big(
      S^{2(n+d)}
    \big)
    }}
    \hspace{-41pt}
    \mathclap{
      \mathbb{Q} \oplus 0
    }
    \hspace{+41pt}
    \;
    \ar[r]
    \ar[d]
    &
    {\phantom{
    (\widetilde {(H^{\mathrm{ev}}\mathbb{Q})/\mathbb{S}}){}^{2n}
    \big(
      S^{2(n+d)}
    \big)
    }}
    \hspace{-45pt}
    \mathclap{
      \color{purple}
      \mathbb{Q}
        \oplus
      \mathbb{S}_{2d-1}
    }
    \hspace{+45pt}
    \ar[r]
    \ar[d]
    &
    {\phantom{
    \widetilde {\mathbb{S}}{}^{2n+1}
    \big(
      S^{2(n+d)}
    \big)}}
    \hspace{-30pt}
    \mathclap{
      0 \oplus \mathbb{S}_{2d-1}
    }
    \hspace{+30pt}
    \ar[d]
    \ar[r]
    &
    {\phantom{
    \widetilde {(H^{\mathrm{ev}}\mathbb{Q})}{}^{2n+1}
    \big(
      S^{2(n+d)}
    \big)
    }}
    \hspace{-47pt}
    \mathclap{
      0
    }
    \hspace{+47pt}
    \ar[d]
    \\
      \color{blue}
      (\mathbb{S}_{2d}/\mathbb{S}_1)
      \oplus
      \mathbb{Z}
    \ar[d]_-{
      \color{blue}
      \mathclap{\phantom{\vert^{\vert}}}
      0 \,\oplus\, \mathrm{ord}(c)
      \mathclap{\phantom{\vert_{\vert}}}
    }
    \ar[r]
    &
    {\phantom{
    \widetilde {(H^{\mathrm{ev}}\mathbb{Q})}{}^{2n}
    \big(
      C_c
    \big)
    }}
    \hspace{-30pt}
    \mathclap{
      \color{purple}
      \mathbb{Q}
        \oplus
      \mathbb{Q}
    }
    \hspace{+30pt}
    \;
    \ar[r]
    \ar[d]
    &
    (\widetilde {(H^{\mathrm{ev}}\mathbb{Q})/\mathbb{S}}){}^{2n}
    \big(
      C_c
    \big)
    \ar[r]
    \ar[d]
    &
    \color{blue}
    0 \oplus
    \mathbb{S}_{2d-1}/\mathbb{Z}
    \ar[d]
    \ar[r]
    &
    {\phantom{
    \widetilde {(H^{\mathrm{ev}}\mathbb{Q})}{}^{2n+1}
    \big(
      C_c
    \big)
    }}
    \hspace{-36pt}
    \mathclap{
      \color{blue}
      0
    }
    \hspace{+36pt}
    \\
    {\phantom{
    \widetilde {\mathbb{S}}{}^{2n}
    \big(
      S^{2n}
    \big)
    }}
    \hspace{-19pt}
    \mathclap{
      0 \oplus \mathbb{Z}
    }
    \hspace{+19pt}
    \ar[d]
      _-{
        \mathclap{\phantom{\vert^\vert}}
        c
        \mathclap{\phantom{\vert_\vert}}
      }
    \ar[r]
      ^-{
        \; 1 \mapsto 1\;
      }
    &
    {\phantom{
    \widetilde {(H^{\mathrm{ev}}\mathbb{Q})}{}^{2n}
    \big(
      S^{2n}
    \big)
    }}
    \hspace{-32pt}
    \mathclap{
      0 \oplus \mathbb{Q}
    }
    \hspace{+32pt}
    \ar[r]
    \ar[d]
    &
    {\phantom{
    (\widetilde {(H^{\mathrm{ev}}\mathbb{Q})/\mathbb{S}}){}^{2n}
    \big(
      S^{2n}
    \big)
    }}
    \hspace{-43pt}
    \mathclap{
      \color{blue}
      0 \oplus \mathbb{Q}/\mathbb{Z}
    }
    \hspace{+43pt}
    \ar[r]
    &
    {\phantom{
    \widetilde {\mathbb{S}}{}^{2n+1}
    \big(
      S^{2n}
    \big)
    }}
    \hspace{-25pt}
    \mathclap{
      0
    }
    \hspace{+25pt}
    \\
    {\phantom{
    \widetilde {\mathbb{S}}{}^{2n}
    \big(
      S^{2(n+d)-1}
    \big)
    }}
    \hspace{-24pt}
    \mathclap{
      0 \oplus \mathbb{S}_{2d-1}
    }
    \hspace{+24pt}
    \ar[r]
    &
    {\phantom{
    \widetilde {(H^{\mathrm{ev}}\mathbb{Q})}{}^{2n}
    \big(
      S^{2(n+d)-1}
    \big)
    }}
    \hspace{-47pt}
    \mathclap{
      0
    }
    \hspace{+47pt}
  }
$$
Here the splittings shown in purple
we may {\it choose to be the canonical ones}
from Remark \ref{CanonicalSplittingOfCofiberOfPeriodicRationalCohomology}.

\noindent
Further from this, and using

\vspace{-.1cm}
\begin{enumerate}

\vspace{-.2cm}
\item
again that $\mathrm{Ext}^1(-,\mathbb{Q}) = 0$;

\vspace{-.2cm}
\item
that
$\mathrm{Ext}^1(-, A \oplus B )
  \simeq
  \mathrm{Ext}^1(-,A) \oplus \mathrm{Ext}^1(-,B)$;

\vspace{-.2cm}
\item commutativity of the middle square with the two
 purple entries

\end{enumerate}
\vspace{-.2cm}

\noindent
the remaining entry and the maps into it must be as claimed:
\vspace{-2mm}
$$
  \xymatrix@C=16pt@R=16pt{
    {\phantom{
    \widetilde {\mathbb{S}}{}^{2n}
    \big(
      S^{2n+1}
    \big)
    }}
    \hspace{-23pt}
    \mathclap{
     \mathbb{S}_{1} \oplus 0
    }
    \hspace{+23pt}
    \ar[d]
    \ar[r]
    &
    {\phantom{
    \widetilde { (H^{\mathrm{ev}}\mathbb{Q}) }{}^{2n}
    \big(
      S^{2n+1}
    \big)
    }}
    \hspace{-38pt}
    \mathclap{
      0
    }
    \hspace{+38pt}
    \ar[r]
    \ar[d]
    &
    {\phantom{
    (\widetilde {(H^{\mathrm{ev}}\mathbb{Q})/\mathbb{S}}){}^{2n}
    \big(
      S^{2n+1}
    \big)
    }}
    \hspace{-48pt}
    \mathclap{
      0 \oplus \mathbb{Z}
    }
    \hspace{+48pt}
    \ar[d]
    \ar[r]
    &
    {\phantom{
    \widetilde {\mathbb{S}}{}^{2n+1}
    \big(
      S^{2n+1}
    \big)
    }}
    \hspace{-30pt}
    \mathclap{
      0 \oplus \mathbb{Z}
    }
    \hspace{+30pt}
    \ar[d]
      _-{
        \mathclap{\phantom{\vert^{\vert}}}
        c
        \mathclap{\phantom{\vert_{\vert}}}
      }
    \\
    {\phantom{
    \widetilde {\mathbb{S}}{}^{2n}
    \big(
      S^{2(n+d)}
    \big)
    }}
    \hspace{-32pt}
    \mathclap{
      \mathbb{S}_{2d} \oplus 0
    }
    \hspace{+32pt}
    \ar[r]^-{ \;0\; }
    \ar[d]
    &
    {\phantom{
    \widetilde {(H^{\mathrm{ev}}\mathbb{Q})}{}^{2n}
    \big(
      S^{2(n+d)}
    \big)
    }}
    \hspace{-41pt}
    \mathclap{
      \mathbb{Q} \oplus 0
    }
    \hspace{+41pt}
    \;
    \ar[r]
    \ar[d]
    &
    {\phantom{
    (\widetilde {(H^{\mathrm{ev}}\mathbb{Q})/\mathbb{S}}){}^{2n}
    \big(
      S^{2(n+d)}
    \big)
    }}
    \hspace{-45pt}
    \mathclap{
      \mathbb{Q} \oplus \mathbb{S}_{2d-1}
    }
    \hspace{+45pt}
    \ar[r]
    \ar[d]
      _-{
         \binom{\mathrm{id}~ \cdot}{\,0~\cdot}      }
    &
    {\phantom{
    \widetilde {\mathbb{S}}{}^{2n+1}
    \big(
      S^{2(n+d)}
    \big)}}
    \hspace{-30pt}
    \mathclap{
      0 \oplus \mathbb{S}_{2d-1}
    }
    \hspace{+30pt}
    \ar[d]
    \ar[r]
    &
    {\phantom{
    \widetilde {(H^{\mathrm{ev}}\mathbb{Q})}{}^{2n+1}
    \big(
      S^{2(n+d)}
    \big)
    }}
    \hspace{-47pt}
    \mathclap{
      0
    }
    \hspace{+47pt}
    \ar[d]
    \\
      (\mathbb{S}_{2d}/\mathbb{S}_1)
      \oplus
      \mathbb{Z}
    \ar[d]_-{
      \mathclap{\phantom{\vert^{\vert}}}
      0 \,\oplus\, \mathrm{ord}(c)
      \mathclap{\phantom{\vert_{\vert}}}
    }
    \ar[r]
    &
    {\phantom{
    \widetilde {(H^{\mathrm{ev}}\mathbb{Q})}{}^{2n}
    \big(
      C_c
    \big)
    }}
    \hspace{-30pt}
    \mathclap{
      \mathbb{Q} \oplus \mathbb{Q}
    }
    \hspace{+30pt}
    \;
    \ar[r]
      ^-{
        \binom{\mathrm{id}~ \cdot}{\,0~\cdot}      }
    \ar[d]
    &
      \color{purple}
      \mathbb{Q}
      \oplus
      \!
      \big(
        \mathbb{Q}/\mathbb{Z}
        \oplus'
        \mathbb{S}_{2d-1}/\mathbb{Z}
      \big)
    \ar[r]
    \ar[d]
    &
    0 \oplus
    \mathbb{S}_{2d-1}\!/\mathbb{Z}
    \ar[d]
    \ar[r]
    &
    {\phantom{
    \widetilde {(H^{\mathrm{ev}}\mathbb{Q})}{}^{2n+1}
    \big(
      C_c
    \big)
    }}
    \hspace{-36pt}
    \mathclap{
      0
    }
    \hspace{+36pt}
    \\
    {\phantom{
    \widetilde {\mathbb{S}}{}^{2n}
    \big(
      S^{2n}
    \big)
    }}
    \hspace{-19pt}
    \mathclap{
      0 \oplus \mathbb{Z}
    }
    \hspace{+19pt}
    \ar[d]
      _-{
        \mathclap{\phantom{\vert^\vert}}
        c
        \mathclap{\phantom{\vert_\vert}}
      }
    \ar[r]
      ^-{
        \; 1 \mapsto 1\;
      }
    &
    {\phantom{
    \widetilde {(H^{\mathrm{ev}}\mathbb{Q})}{}^{2n}
    \big(
      S^{2n}
    \big)
    }}
    \hspace{-32pt}
    \mathclap{
      0 \oplus \mathbb{Q}
    }
    \hspace{+32pt}
    \ar[r]
    \ar[d]
    &
    {\phantom{
    (\widetilde {(H^{\mathrm{ev}}\mathbb{Q})/\mathbb{S}}){}^{2n}
    \big(
      S^{2n}
    \big)
    }}
    \hspace{-43pt}
    \mathclap{
      0 \oplus \mathbb{Q}/\mathbb{Z}
    }
    \hspace{+43pt}
    \ar[r]
    &
    {\phantom{
    \widetilde {\mathbb{S}}{}^{2n+1}
    \big(
      S^{2n}
    \big)
    }}
    \hspace{-25pt}
    \mathclap{
      0
    }
    \hspace{+25pt}
    \\
    {\phantom{
    \widetilde {\mathbb{S}}{}^{2n}
    \big(
      S^{2(n+d)-1}
    \big)
    }}
    \hspace{-24pt}
    \mathclap{
      0 \oplus \mathbb{S}_{2d-1}
    }
    \hspace{+24pt}
    \ar[r]
    &
    {\phantom{
    \widetilde {(H^{\mathrm{ev}}\mathbb{Q})}{}^{2n}
    \big(
      S^{2(n+d)-1}
    \big)
    }}
    \hspace{-47pt}
    \mathclap{
      0
    }
    \hspace{+47pt}
  }
$$
\end{proof}

Now we may conclude:

\begin{theorem}[Identification of $e$-invariants]
  \label{DiagrammaticeCCoincidesWithClassicaleCInvariant}
  The refined $\widehat e_{{K\mathrm{U}}}$-invariant
  from Def. \ref{LiftedEInvariantDiagrammatically}
  coincides with the quantity $\widehat {\mathrm{e}_{\mathrm{Ad}}}$
  in \eqref{AdamseCInvariant}:
  \begin{equation}
    \label{EqualityOfRefinedeCInvariants}
    \widehat e_{{K\mathrm{U}}}(-,-)
    \;=\;
    \widehat {\mathrm{e}_{\mathrm{Ad}}}(-,-)
    \,.
  \end{equation}
  In particular, the diagrammatic $e_{{K\mathrm{U}}}$-invariant
  from Prop. \ref{TheDiagrammaticeKUInvariant}
  coincides with the classical Adams $\mathrm{e}$ invariant
  (Def. \ref{ClassicalConstructionOfAdamseCInvariant}):
  $$
    e_{{K\mathrm{U}}}(-)
    \;=\;
    {\mathrm{e}_{\mathrm{Ad}}}(-)
    \,.
  $$
\end{theorem}
\begin{proof}
  The homotopy-commuting rectangle in the
  bottom right part of the defining diagram
  \eqref{DiagrammatichatecInvariant}
  says that
  $$
    p^\ast
    \Big(
      \widehat e_{{K\mathrm{U}}}(c)
      ,
      [c]
    \Big)
    \;=\;
    i_\ast
    \Big(
      \mathrm{ch}
      \big[
        \vdash C^{{K\mathrm{U}}}_{2n-1}(c)
      \big]
    \Big)
    \,.
  $$
  By Lemma \ref{CompatibilityOfTheTwoCanonicalSplittings}
  this means that the image of both sides
  along their canonical retractions
  (Remark \ref{CanonicalSplittingOfCofiberOfPeriodicRationalCohomology})
  onto degree=$2(n+d)$ rational cohomology $\simeq \mathbb{Q}$
  coincide. But by definitions
  \eqref{FunctionDiagrammatichatecInvariant} and
  \eqref{AdamseCInvariantAsTopDegreeComponentOfChernCharacter}
  this is the statement of equality \eqref{EqualityOfRefinedeCInvariants}.
\end{proof}

\medskip

\subsection{Conner-Floyd e-invariant}
 \label{RelativeCobordism}

\noindent
{\bf Calabi-Yau manifolds and $S\mathrm{U}$-Cobordism.}
The special unitary Cobordism ring is tightly related to the
complex geometry of Calabi-Yau manifolds:

\begin{prop}[Torsion in the $S\mathrm{U}$-Cobordism ring {\cite[Thm. 5.8b, Thm. 5.11a]{CLP19}}]
  \label{TorsionInTheSUCobordismRing}
  The Cobordism ring $(M S\mathrm{U})_\bullet$
  of $S\mathrm{U}$-manifolds
  has only 2-torsion, and that is concentrated in degrees
  $1,2 \,\mathrm{mod}\, 8$.
\end{prop}

\begin{prop}[Non-torsion $S\mathrm{U}$-Cobordism ring
is generated by Calabi-Yau manifolds {\cite[Thm. 2.4]{LLP17}}]
  \label{NonTorsionSUCobordismRingGeneratedByCYs}
  Away from its 2-torsion (Prop. \ref{TorsionInTheSUCobordismRing})
  the $S\mathrm{U}$-cobordism ring is multiplicatively
  generated by Calabi-Yau manifolds, in that every element
  in $(M S\mathrm{U})_\bullet[\scalebox{.6}{$\tfrac{1}{2}$}]$
  is equal to a polynomial over
  $\mathbb{Q}$
  of $S\mathrm{U}$-Cobordism classes of Calabi-Yau manifolds.
\end{prop}

In particular:
\begin{prop}[K3 spans $S\mathrm{U}$-Cobordism in degree 4 {\cite[Ex. 3.1]{LLP17}\cite[Thm. 13.5a]{CLP19}}]
  \label{K3SpansSUCobordismInDegree4}
  The $S\mathrm{U}$-Cobordism ring in degree 4
  consists precisely of integer multiples of the class
  $[\mathrm{K3}]$
  of any non-toroidal K3-surface:
  $$
    (M S\mathrm{U})_4
    \;\simeq\;
    \mathbb{Z}
    \big\langle
      [\mathrm{K3}]
    \big\rangle
    \,.
  $$
\end{prop}

\medskip

\noindent
{\bf The Conner-Floyd K-orientation.}

\begin{prop}[Conner-Floyd K-orientation {\cite[\S 5, p. 29]{ConnerFloyd66}}]
  \label{ConnerFloydKOrientation}
  There is a homotopy-commutative diagram
  \begin{equation}
   \label{ConnerFloydOrientationCompatibilityDiagram}
   \raisebox{20pt}{
   \xymatrix@R=1.5em{
     M S\mathrm{U}
     \ar[rr]^-{ \sigma_{S\mathrm{U}} }
     \ar[d]
     &&
     K \mathrm{O}
     \ar[d]
     \\
     M \mathrm{U}
     \ar[rr]
       ^-{ \sigma_{\mathrm{U}} }
     &&
     K \mathrm{U}
   }
   }
  \end{equation}
  of homotopy-commutative ring spectra,
  where the vertical morphism are the canonical ones.
\end{prop}
\begin{remark}
  The Conner-Floyd K-orientation of Prop. \ref{ConnerFloydKOrientation}
  is directly analogous to the more widely known
  {\it Atiyah-Bott-Shapiro orientation} \cite{AtiyahBottShapiro64}
  \begin{equation}
   \raisebox{20pt}{
   \xymatrix@R=1em{
     M \mathrm{Spin}
     \ar[rr]^-{ \sigma_{\mathrm{Spin}} }
     &&
     K \mathrm{O}
     \\
     M \mathrm{Spin}^c
     \ar[rr]
       ^-{ \sigma_{\mathrm{Spin}^c} }
     &&
     K \mathrm{U}\;.
   }
   }
  \end{equation}
(All these horizontal maps are ``orientations''
in the sense of Example \ref{UniversalOrientationsOnMUAndMSp}.)
\end{remark}

\medskip

\noindent {\bf The rational Todd class.}
\begin{defn}[Todd class]
  \label{ToddClass}
  For $M^{\leq 6}_{\mathrm{U}}$ a $\mathrm{U}$-manifold
  of real dimension $\leq 6$,
  its {\it Todd class} is the following rational combination
  of Chern classes of the given stable
  $\mathrm{U}$-structured tangent bundle:
  \begin{equation}
    \label{ToddClassPolynomialUpToDegree6}
    \mathrm{Td}
    \big(
      M^{\leq 6}_{\mathrm{U}}
    \big)
    \;\coloneqq\;
    1
      +
    \tfrac{1}{2}
    c_1
      +
    \tfrac{1}{12}
    \big(
      c_1^2 + c_2
    \big)
    +
    \tfrac{1}{24} c_1 c_2
    \;\;\;
    \in
    \;
    H^\bullet
    \big(
      M^{\leq 6}_{\mathrm{U}};
      \mathbb{Q}
    \big)
    \,,
  \end{equation}
  where we abbreviate
  $c_i
    \coloneqq
   c_i
   \big(
     T_{\mathrm{st}}X^{\leq 6}_{\mathrm{U}}
   \big)
  $.
  \footnote{
  Of course the Todd class is defined
  more abstractly and in all dimensions (see e.g. \cite{SUZ});
  but we restrict attention as above for focus of the exposition.
  }
  For closed $M_{\mathrm{U}}$,
  their {\it Todd number}, i.e.
  the evaluation
  \begin{equation}
    \label{ToddNumber}
    \mathrm{Td}[M_{\mathrm{U}}]
    \;\coloneqq\;
    \mathrm{Td}(M_{\mathrm{U}})[M]
    \;\;\;
    \in
    \; \mathbb{Q}
  \end{equation}
  of the Todd class of the stable tangent bundle
  on the
  fundamental homology class $[M]$ of
  the underlying manifold,
  is a cobordism invariant and hence constitutes a
  function on the complex cobordism ring,
  which happens to be an integer (Prop. \ref{QuantizationOfToddNumbers}):
  $$
    \mathrm{Td}
    \;\colon\;
    (M \mathrm{U})_\bullet
    \longrightarrow
    \mathbb{Z}
      \longhookrightarrow
    \mathbb{Q}
    \,.
  $$
\end{defn}
\begin{example}[Square root of Todd class of $S\mathrm{U}$-manifolds]
  \label{SquareRootOfToddClassOfSUManifolds}
  On $S\mathrm{U}$-manifolds
  $
    [M^{\leq 6}_{S \mathrm{U}}]
      \in
    (M S\mathrm{U})_{\leq 6}
      \xrightarrow{\;}
    (M \mathrm{U})_{\leq 6}
    \,,
  $
  where the first Chern class vanishes
  $c_1(M_{S\mathrm{U}} ) = 0$,
  the Todd class \eqref{ToddClassPolynomialUpToDegree6}
  reduces to
  \begin{equation}
    \label{ToddClassOnSU6Folds}
    \mathrm{Td}
    (M^{\leq 6}_{S \mathrm{U}})
    \;=\;
    1 + \tfrac{1}{12} c_2
    \; ,
  \end{equation}
  and hence the {\it square root} of the Todd class
  (in the sense of formal power series of Chern classes)
  on these manifolds is
  \begin{equation}
    \label{SquareRootOfToddClassOnSU6Folds}
    \begin{aligned}
      \sqrt{
        \mathrm{Td}
      }
      \big(
        M^{\leq 6}_{S \mathrm{U}}
      \big)
      & = \;
      1
        +
      \tfrac{1}{2}
      \big(
        \tfrac{1}{12} c_2
      \big)
      +
      \underset{
        = 0
      }{
      \underbrace{
      -
      \tfrac{1}{8}
      \big(
        \tfrac{1}{12}
        c_2
      \big)^2
      +
      \tfrac{1}{16}
      \big(
        \tfrac{1}{12}
        c_2
      \big)^3
      +
      \cdots
      }
      }
      \\
      & = \;
      1 +
      \tfrac{1}{24} c_2
      \;\;\;\;\;\;\;
      \in
      \;
      H^\bullet
      \big(
        X^{\leq 6}_{S\mathrm{U}}
        ;
        \mathbb{Q}
      \big)
      \,.
    \end{aligned}
  \end{equation}
  Yet more specifically,
  if $M^{\leq 6}_{S\mathrm{U}} = M^4_{S\mathrm{U}}$
  is a {\it complex curve} with vanishing first Chern class,
  such as a Calabi-Yau 2-fold/K3-surface,
  then the second Chern class
  equals the Euler class
  (being the top Chern class),
  \begin{equation}
    \label{SecondChernClassIsEulerClassOnComplexCurves}
    c_2(M^4_{\mathrm{U}}) \;=\; \rchi_4(M^4)
    \;\;\;
    \in
    \;
    H^4(M^4; \mathbb{Q})
    \,,
  \end{equation}
  so that \eqref{SquareRootOfToddClassOnSU6Folds}
  reduces further to
  \begin{equation}
    \label{SquareRootOfToddClassOfK3}
    \sqrt{\mathrm{Td}}
    \big(
      M^4_{S\mathrm{U}}
    \big)
    \;=\;
    1
      +
    \tfrac{1}{24} \rchi_4
    \;\;\;\;\;
    \in
    H^\bullet
    \big(
      X;
      \mathbb{Q}
    \big)
    \,.
  \end{equation}
  Finally, in terms of
  characteristic numbers \eqref{ToddNumber},
  i.e. after evaluation
  on the fundamental homology class $[M^4]$,
  this means that
  the evaluation of the square root of the Todd class is
  half the evaluation of the Todd class itself:
  \begin{equation}
    \label{ToddNumberOfSUCurve}
    \sqrt{\mathrm{Td}}\, [M^4]
    \;=\;
    \tfrac{1}{24}\rchi_4[M^4]
    \;=\;
    \tfrac{1}{2}
    \mathrm{Td}(M^4_{S\mathrm{U}})[M^4]
    \;\;\;
    \in
    \;
    \mathbb{Z}
      \hookrightarrow
    \mathbb{Q}
    \,.
  \end{equation}
\end{example}

\begin{prop}[Quantization of Todd numbers]
  \label{QuantizationOfToddNumbers}
  The Todd number \eqref{ToddNumber}
  of any closed $\mathrm{U}$-manifold is integer,
  and that of an $S\mathrm{U}$-manifold of
  dimension $4 \;\mathrm{mod}\; 8$ is
  moreover divisible by 2.
\end{prop}

\begin{remark}[Rational Todd numbers via Chern-Weil theory]
  \label{ToddNumbersViaChernWeilTheory}
  By Chern-Weil theory (review in \cite{FSS20c})
  the Todd number (Def. \ref{ToddClass}) of a
  smooth closed
  $\mathrm{U}$-manifold $M_{\mathrm{U}}$ may equivalently be computed
  as the integral of a differential form:
  Choosing any unitary connection
  $\nabla \coloneqq \nabla^{T_{\mathrm{st}}M_{\mathrm{U}}}$
  on the stable tangent bundle $T_{\mathrm{st}} M_{\mathrm{U}}$
  and
  evaluating its curvature 2-form $F_\nabla$
  in the invariant polynomials
  $
    \langle \cdots \rangle
      \;\colon
    \mathrm{Sym}^\bullet(\mathfrak{u})
      \xrightarrow{\;\;}
    \mathbb{R}
  $
  on the unitary Lie algebra $\mathfrak{u}$
  yields the {\it Chern forms} $c_{2k}(\nabla)$
  and hence the {\it Todd form} $\mathrm{Td}(\nabla)$,
  by inserting the Chern forms in place of the Chern classes
  in the defining polynomial expression \eqref{ToddClassPolynomialUpToDegree6}:
  $$
    \nabla \coloneqq \nabla^{T_{\mathrm{st}}M_{\mathrm{U}}}
    \,,
    \phantom{AA}
    c_k(\nabla)
    \;\coloneqq\;
    \langle
      \underset{
        \mbox{
          \tiny
          $k$ factors
        }
      }{
      \underbrace{
        F_\nabla \wedge \cdots \wedge F_\nabla
      }
      }
    \rangle
    \;\in\;
    \Omega^{2i}_{\mathrm{dR}}
    \big(
      M_{\mathrm{U}}
    \big)
    \,,
    \phantom{AAA}
    \mathrm{Td}(\nabla)
    \;\coloneqq\;
    \mathrm{Td}(c_i(\nabla))
    \;\in\;
    \Omega^{\bullet}_{\mathrm{dR}}
    \big(
      M_{\mathrm{U}}
    \big)
    \,.
  $$
  Then the Todd number \eqref{ToddNumber}
  is equal to the integral of this
  differential Todd form over the manifold:
  \begin{equation}
    \label{RationalToddNumberAsIntegralOfCharacteristicForm}
    \mathrm{Td}[M_{\mathrm{U}}]
    \;=\;
    \int_{M_{\mathrm{U}}} \mathrm{Td}(\nabla^{T M})
    \,.
  \end{equation}
  An advantage of this differential re-formulation of the Todd number
  is that it generalizes to
  (compact) $\mathrm{U}$-manifolds
  {\it with boundaries}. Specifically for $\mathrm{U}$-manifolds
  $M_{\mathrm{U},\mathrm{Fr}}$
  with framed boundaries
  it yields again an invariant,
  now generally taking rational values:
  \begin{equation}
    \label{RationalToddNumberOnUManifoldsWithFramedBoundaries}
    \xymatrix@R=2pt{
      \mathllap{
        \mbox{
          \tiny
          \color{darkblue}
          \bf
          \begin{tabular}{c}
            Cobordism ring of
            \\
            $\mathrm{U}$-manifolds
            \\
            with $\mathrm{Fr}$-boundaries
          \end{tabular}
        }
        \!\!\!
      }
      (M \mathrm{U}/\mathbb{S})_\bullet
      \ar[rr]
        ^-{ \mathrm{Td} }
        _-{
          \mbox{
            \tiny
            \color{greenii}
            \bf
            \begin{tabular}{c}
              rational Todd number
            \end{tabular}
          }
        }
      &&
      \mathbb{Q}
      \\
      [M_{\mathrm{U},\mathrm{Fr}}]
      \ar[rr]
      &&
      \int_{M}
      \mathrm{Td}(
        \nabla^{
          T_{\mathrm{st}}
          M_{\mathrm{U},\mathrm{Fr}}
        }
      )
      \,.
    }
  \end{equation}
\end{remark}

This differential-geometric formulation of the
Todd class allows it to be further expressed by
boundary data alone:

\begin{lemma}[Euler class of 4-manifolds with boundary via twisted 3-forms]
  \label{RealH3FluxOnK3AwayFrom24BraneInsertionsExists}
Let $M^4$ be a closed manifold
of real dimension 4, and let
$$
  \underset{
    1 \leq k \leq n
  }{\sqcup}
  D_k^4
    \hookrightarrow
  M^4
$$
be an embedding of a {\it positive} number $n \,\geq 1\, \in \mathbb{N}$
of disjoint open balls.
Then,
for any choice of tangent connection  $\nabla$
on the complement manifold
$M^4 \setminus (\underset{\scalebox{.55}{$1 \leq k \leq n$}}{\sqcup} D_k^4)$
with boundary
$\underset{\scalebox{.55}{$1 \leq k \leq n$}}{\sqcup} S^3_k$,
there
exists a differential 3-form $H_3$ which

\noindent {\bf (a)} trivializes the Euler-form in de Rham cohomology,
and

\noindent {\bf (b)} integrates over the boundary 3-spheres to the Euler number
of $M^4$:
\begin{equation}
  \label{3FormTrivializingEulerFormOnPuncturedK3}
  \exists
  \;
  H_3
    \,\in\,
  \Omega^3_{\mathrm{dR}}
  \big(
    M^4 \setminus (\sqcup_{n} D^4)
  \big)
  \;\;\;\;\;
  \mbox{\rm s.t.}
  \;\;\;\;\;
  \left\{
  \begin{aligned}
    &
    d\,H_3 \;=\; \rchi_4(\nabla)
    \\
    &
    \int_{
      \sqcup_{n} S^3
    }
    H_3
    \;=\;
    \rchi_4[M^4]
    \,.
  \end{aligned}
  \right.
\end{equation}
\end{lemma}

\begin{proof}
We use the standard argument of choosing any vector field with isolated zeros on $M^4$ and then deforming
that continuously to move all of these into the open balls.
With this, the complement manifold
$M^4 \setminus \big( \sqcup_n D^4 \big)$
carries a nowhere-vanishing smooth vector field $v$
and hence, after re-scaling, a section $h$
of the 3-sphere fiber bundle associated with the tangent bundle
\begin{equation}
  \label{SectionOf3SphereBundleOnComplementOfOpenBallsIn4Manifold}
  v
    \,\in\,
  \Gamma_{M^4}
  \Big(
    T
    \big(
      M^4
        \setminus
      \sqcup_n
      D^4
    \big)
  \Big)
  \,,
  {\phantom{AAA}}
  h
  \,\coloneqq\,
  v/\left\vert v\right\vert
    \;\in\;
  \Gamma_{M^4}
  \Big(
    S\big(
      T
      (
        M^4
          \setminus
        \sqcup_n
        D^4
      )
    \big)
  \Big)
  \,.
\end{equation}
After choosing trivializations of
the tangent bundle, and hence its 3-sphere bundle,
on an open neighborhood of each $D_k^4$, this section
restricts over each boundary component to a map
$h|_{S_k^3} \,:\, S^3 \to S^3$, with Hopf winding degree
\begin{equation}
  \label{PoincareHopfIndex}
  \mbox{
    \tiny
    \color{darkblue}
    \bf
    \begin{tabular}{c}
      Poincar{\'e}-Hopf
      index
      \\
      of $h$ at $x_k$
    \end{tabular}
  }
  \mathrm{deg}\big(h|_{S_k^3}\big)
  \;=\;
  \int_{S^3_k}
  \big(
    h_{S^3_k}^\ast \mathrm{vol}_{S^3}
  \big)
  \mbox{
    \tiny
    \color{darkblue}
    \bf
    \begin{tabular}{c}
      Hopf degree
      \\
      of $h|_{S^3_k}$
    \end{tabular}
  }
\end{equation}
By the Poincar{\'e}-Hopf theorem
(e.g. \cite[Sec. 15.2]{DNF85},
see \cite[(83)]{FSS19b} for review in our context),
the sum of these degrees is the Euler characteristic of $M^4$:
\begin{equation}
  \label{PHTheoremFormula}
  \underset{
    1 \leq k \leq n
  }{\sum}
  \mathrm{deg}\big(h|_{S_k^3}\big)
  \;=\;
  \rchi_4[M^4]
  \,.
\end{equation}
Now, the section $h$ \eqref{SectionOf3SphereBundleOnComplementOfOpenBallsIn4Manifold}
constitutes
a cocycle in the $\tau \coloneqq T M^4$-twisted
4-Cohomotopy of $M^4 \setminus \sqcup_n D^4$
(\cite[Def. 2.1]{FSS19b});
and by \cite[Prop. 2.5 (i)]{FSS19b},
the image of this cocycle
under the twisted cohomotopical character map (\cite{FSS20c})
$$
  \xymatrix@R=3pt{
    \overset{
      \mathclap{
      \raisebox{3pt}{
        \tiny
        \color{darkblue}
        \bf
        \begin{tabular}{c}
          J-twisted
          \\
          3-Cohomotopy
        \end{tabular}
      }
      }
    }{
      \pi^{\tau}\big( M^4 \big)
    }
    \ar[rr]
      _-{ \mathrm{ch}^\tau_\pi }
      ^-{
        \mbox{
          \tiny
          \color{greenii}
          \bf
          twisted 3-cohomotopical character map
        }
      }
    &
    {\phantom{AAAA}}
    &
    \qquad
    \overset{
      \mathclap{
      \raisebox{3pt}{
        \tiny
        \color{darkblue}
        \bf
        \begin{tabular}{c}
          twisted non-abelian
          \\
          de Rham cohomology
        \end{tabular}
      }
      }
    }{
      H^{\tau_{\mathrm{dR}}}_{\mathrm{dR}}
    }
    \big(
      M^4;
      \mathfrak{l}S^3
    \big)
    \ar@{=}[d]
    \\
    &&
    \Big\{
      \left.
      H_3 \,\in\,
      \Omega^3_{\mathrm{dR}}
      \big(
        M^4
      \big)
      \;\right\vert\;
      d\,H_3 = \rchi_4(\nabla^{T M^4})
    \Big\}
  }
$$
\vspace{-.2cm}
\begin{equation}
  \label{CharacterImageOfTwistedCohomotopyCocycleInEvenDimensions}
  \hspace{-2.5cm}
  \raisebox{20pt}{
  \xymatrix@C=3em@R=10pt{
    & S(T M^4)
    \ar[d]
    \\
    X
    \ar@{-->}[ur]
      ^-{h}
    \ar@{=}[r]
    &
    X
  }
  }
  {\phantom{AAAAA}}
  \longmapsto
  {\phantom{AAAAA}}
  \raisebox{22pt}{
  \xymatrix@C=3em@R=10pt{
    & \widehat H_3
    \\
    H_3
    \coloneqq
    h^\ast \widehat H_3
    \ar@{<-|}[ur]
      ^-{h^\ast}
    &
  }
  }
\end{equation}
is a differential form $H_3$ satisfying
$d\, H_3 \,=\, \rchi_4(\nabla)$.
Moreover, this $H_3$ is \cite[(45)]{FSS19b}
the pullback along $h$ of a fiber-wise unit volume form
$\widehat H_3$
on the spherical fibration. This implies the claim.

More concretely:
There exists (e.g. \cite[\S 6.6, Thm. 6.1]{Walschap04})
a differential 3-form $\widehat H_3$ on the
total space of the 3-sphere bundle of the tangent bundle of
$M^4$, which
{(a)} trivializes the pullback of the Euler-form along the
bundle projection $p$ and {(b)} restricts to the
unit volume form on each 3-sphere fiber $S^3_x$:
\begin{equation}
  \label{3FormTrivializingEulerFormOnPuncturedK3}
  \exists
  \;
  \widehat H_3
    \,\in\,
  \Omega^3_{\mathrm{dR}}
  \bigg(
    S
    \Big(
      T
      \big(
        M^4 \setminus (\sqcup_{n} D^4)
    \Big)
  \bigg)
  \;\;\;\;\;
  \mbox{\rm s.t.}
  \;\;\;\;\;
  \left\{
  \begin{aligned}
    &
    d\,\widehat H_3 \;=\; p^\ast \rchi_4(\nabla)
    \\
    &
    \int_{
      S_x^3
    }
    \widehat H_3
    \;=\;
    1
    \,.
  \end{aligned}
  \right.
\end{equation}
Therefore, the pullback form
$$
  H_3 \;\coloneqq\; h^\ast \widehat H_3
$$
satisfies
$$
  d H_3
    \;=\;
  d h^\ast \widehat H_3
    \;=\;
  h^\ast d \widehat H_3
    \;=\;
  h^\ast p^\ast \rchi_4(\nabla)
    \;=\;
  \rchi_4(\nabla)
$$
and
$$
  \int_{\sqcup_n S^3}
  H_3
  \;=\;
  \int_{\sqcup_n S^3}
  h^\ast \widehat H_3
  \;=\;
  \underset{
    1 \leq k \leq n
  }{\sum}
    \int_{S_k^3}
    \big(
      h|_{S^3_k}
    \big)^\ast \mathrm{vol}_{S^3}
  \;=\;
  \underset{
    1 \leq k \leq n
  }{\sum}
    \mathrm{deg}
    \big(
      h|_{S^3_k}
    \big)
  \;=\;
  \rchi_4[M^4]
  \,,
$$
where in the last step we used \eqref{PHTheoremFormula}.
\end{proof}

In conclusion we highlight the following special case:
\begin{prop}[Rational Todd number of punctured complex curve as boundary integral]
  \label{RationalToddNumberAsBoundaryIntegral}
  Let
  $$
    M^4_{S\mathrm{U},\mathrm{Fr}}
    \;\;\;\;
    \mbox{\rm s.t.}
    \;\;\;\;
    \partial M^4
    \;=\;
    \underset{
      1 \leq k \leq n
    }{\sqcup}
    S^3_k
  $$
  be a
  real 4-dimensional
  $S \mathrm{U}$-manifold whose
  non-empty framed boundary is a disjoint union of 3-spheres.
  Then its rational Todd number \eqref{RationalToddNumberOnUManifoldsWithFramedBoundaries}
  equals $\tfrac{1}{12}$th of the boundary integral of a differential
  3-form $H_3$ whose de Rham differential
  equals the Euler form:
  \begin{equation}
    \label{RationalToddNumberAsBoundaryIntegration}
    \exists \,
    H_3
    \,\in\,
    \Omega^3_{\mathrm{dR}}
    \big(
      M^4
    \big)
    \,,
    \;\;\;\;
    \mbox{\rm s.t.}
    \;\;\;\;\;
    d\, H_3 \;=\; \rchi_4(\nabla^{T M^4_{\mathrm{U},\mathrm{Fr}}})
    \;\;\;
    \mbox{\rm and}
    \;\;\;
    \mathrm{Td}\,[M^4_{\mathrm{U},\mathrm{Fr}}]
    \;=\;
    \tfrac{1}{12}
    \underset{
      1 \leq k \leq n
    }{\sum}
    \int_{ S^3_k }
    H^3
    \,.
  \end{equation}
  Moreover, if $M^4_{S\mathrm{U},\mathrm{Fr}}$
  is the complement of $n \geq 1$ disjoint open balls
  in a closed $S\mathrm{U}$-manifold $M^4_{S\mathrm{U}}$, then
  this equals $\tfrac{1}{12}$th of the Euler number of that closed manifold:
  \begin{equation}
    \label{RationalToddNumberOfOpen4BallComplementIsEulerNumberOfClosedManifold}
    M^4_{S\mathrm{U},\mathrm{Fr}}
    \;=\;
    M^4_{S\mathrm{U}}
      \setminus
    \underset{
      1 \leq k \leq n
    }{\sqcup}
    \;\;\;\;\;\;\;\;\;
    \Rightarrow
    \;\;\;\;\;\;\;\;\;
    \mathrm{Td}\,[M^4_{S\mathrm{U},\mathrm{Fr}}]
    \;=\;
    \rchi_4[M^4_{S\mathrm{U}}]
    \,.
  \end{equation}
\end{prop}
\begin{proof}
  The existence of the differential 3-form
  and the final statement \eqref{RationalToddNumberOfOpen4BallComplementIsEulerNumberOfClosedManifold}
  follows by
  Lemma  \ref{RealH3FluxOnK3AwayFrom24BraneInsertionsExists}.
  To see \eqref{RationalToddNumberAsBoundaryIntegration}
  we compute as follows:
  $$
  \begin{aligned}
    \mathrm{Td}\,[M^4_{\mathrm{U},\mathrm{Fr}}]
    & =
    \tfrac{1}{12} c_2[M^4_{\mathrm{U},\mathrm{Fr}}]
    \\
    & =
    \tfrac{1}{12} \int_{M^4} c_2(\nabla^{T M^4_{\mathrm{U},\mathrm{Fr}}})
    \\
    & =
    \tfrac{1}{12} \int_{M^4} \rchi_4(\nabla^{T M^4_{\mathrm{U},\mathrm{Fr}}})
    \\
    & =
    \tfrac{1}{12} \int_{M^4} d\, H_3
    \\
    & =
    \tfrac{1}{12}
    \int_{\partial M^4} H^3
    \,.
  \end{aligned}
  $$
  Here the first line is \eqref{ToddClassOnSU6Folds},
  the second line uses \eqref{RationalToddNumberAsIntegralOfCharacteristicForm},
  the third line is \eqref{SecondChernClassIsEulerClassOnComplexCurves},
  the fourth line is item (a) in \eqref{3FormTrivializingEulerFormOnPuncturedK3}
  and the last line is the Stokes theorem.
\end{proof}

\medskip

\noindent
{\bf The Todd character.}
\begin{prop}
[Todd character on $\mathrm{U}$-manifolds is Chern character of $K\mathrm{U}$-Thom class]
\label{RationalToddCharacterIsChernCharacterOfThomClass}
The composite of the $\sigma_{\mathrm{U}}$-orientation
of $K\mathrm{U}$ \eqref{ConnerFloydOrientationCompatibilityDiagram}
with the Chern character is the Todd character
\begin{equation}
  \label{CompositeOfSigmaUWithChernCharacterIsToddGenus}
  \xymatrix{
    M\mathrm{U}
    \ar[r]
      ^-{ \sigma_{\mathrm{U}} }
    \ar@/_1pc/[rr]
      _-{ \mathrm{Td} }
    &
    K\mathrm{U}
    \ar[r]
      ^-{ \mathrm{ch} }
    &
    H^{\mathrm{ev}}\!\mathbb{Q}
    \,,
  }
\end{equation}
i.e. the morphism of homotopy commutative ring spectra
whose  induced morphism of coefficient rings
is the Todd number (Def. \ref{ToddClass})
\begin{equation}
  \label{TheToddGenus}
  \xymatrix@R=4pt{
    (M \mathrm{U})_\bullet
    \ar[r]^-{ \mathrm{Td}_\bullet }
    &
    (K \mathrm{U})_\bullet
    \,\simeq\,
    \mathbb{Z}[\beta_2]
    \;
    \ar@{^{(}->}[r]
    &
    \mathbb{Q}[\beta_2]
    \\
    [M_{\mathrm{U}}^{2d}]
    \ar@{|->}[rr]
    &&
    \mathrm{Td}
    [M_{\mathrm{U}}^{2d}]
    \cdot
    \beta_2^d
    \,,
  }
\end{equation}
\end{prop}
\noindent
This goes back to \cite[\S 6]{ConnerFloyd66}
and is formulated explicitly as above in \cite[p. 303]{Smith73}
(review in \cite[\S 2.3.2]{Spiegel13}).
In components, \eqref{CompositeOfSigmaUWithChernCharacterIsToddGenus}
encodes the fact that the Todd class
of a complex vector bundle in
rational cohomology equals,
under the Thom isomorphism in complex K-theory,
the Chern character of a $K\mathrm{U}$-Thom class;
in which form the statement is given in
\cite[\S V, Thm. 4.4]{Karoubi78}.

\medskip
We record the following two consequences:
\begin{prop}
[Todd character on $S\mathrm{U}$-manifolds is
Pontrjagin character of $K\mathrm{O}$-Thom class]
  \label{ToddCharaxcterOnSUManifoldsIsPontrjaginCharacterOnKOThomClass}
  After restriction
  from $\mathrm{U}$-manifolds to $S\mathrm{U}$-manifolds,
  the rational Todd character
  factors as the Conner-Floyd K-orientation $\sigma_{S\mathrm{U}}$
  \eqref{ConnerFloydKOrientation}
  on
  $K\mathrm{O}$ followed by the Pontrjagin character:
  \vspace{-2mm}
 \begin{equation}
  \label{CompositeOfSigmaSUWithPontrjaginCharacterIsToddGenus}
  \xymatrix@C=3.5em{
    M S\mathrm{U}
    \ar[r]
      ^-{ \sigma_{S\mathrm{U}} }
    \ar@/_1pc/[rr]
      _-{ \mathrm{Td} }
    &
    K\mathrm{O}
    \ar[r]
      ^-{ \mathrm{ph} }
    &
    H^{\mathrm{ev}}\!\mathbb{Q}
    \,.
  }
\end{equation}
\end{prop}
\begin{proof}
  Observe that we have the following homotopy-commutative
  pasting diagram:
    \vspace{-2mm}
  \begin{equation}
   \label{ToddCharacterRestrictionToMSU}
   \xymatrix@R=1.5em{
     M S\mathrm{U}
     \ar[rr]^-{ \sigma_{S\mathrm{U}} }
     \ar[d]
     &&
     K \mathrm{O}
     \ar[rr]
       ^-{
         \mathrm{ph}
       }
     \ar[d]
     &&
     H^{\mathrm{ev}}\!\mathbb{Q}
     \ar@{=}[d]
     \\
     M \mathrm{U}
     \ar@/_1pc/[rrrr]
       _-{ \mathrm{Td} }
     \ar[rr]
       ^-{ \sigma_{\mathrm{U}} }
     &&
     K \mathrm{U}
     \ar[rr]
       ^-{
         \mathrm{ch}
       }
     &&
     H^{\mathrm{ev}}\!\mathbb{Q}
   }
  \end{equation}

    \vspace{-2mm}
\noindent  Here the left square is from Prop. \ref{ConnerFloydKOrientation},
  the bottom triangle is Prop. \ref{RationalToddCharacterIsChernCharacterOfThomClass},
  while
  the right square is the defining
  relation \eqref{ChernCharacterAndPontrjaginCharacter}
  between the Chern- and Pontrjagin character.
  Thus the homotopy-commutativity of the total outer diagram
  implies the claim.
\end{proof}
\begin{prop}[Chern-, Pontrjagin- and Todd-character on Adams cofiber theories]
  \label{CharactersOnAdamsCofiberTheories}
  All of
  {(a)} the  Conner-Floyd K-orientations
  (Prop. \ref{ConnerFloydKOrientation}),
  {(b)}
  the Chern- and Pontrjagin-character \eqref{ChernCharacterAndPontrjaginCharacter}
  and {(c)} the Todd character \eqref{TheToddGenus}
  descend to the Adams cofiber theories
  (Def. \ref{UnitCofiberCohomologyTheories})
  where they form the following homotopy-commutative diagram:
  \begin{equation}
   \label{OnCofiberTheoryToddCharacterRestrictionToMSU}
   \raisebox{20pt}{
   \xymatrix@R=1.7em{
     (M S\mathrm{U})/\mathbb{S}
     \ar[rr]^-{ \sigma_{S\mathrm{U}}/\mathbb{S} }
     \ar[d]
     &&
     (K \mathrm{O})/\mathbb{S}
     \ar[rr]
       ^-{
         \mathrm{ph}/\mathbb{S}
       }
     \ar[d]
     &&
     (H^{\mathrm{ev}}\!\mathbb{Q})/\mathbb{S}
     \ar@{=}[d]
     \\
     (M \mathrm{U})/\mathbb{S}
     \ar@/_1pc/[rrrr]
       _-{ \mathrm{Td}/\mathbb{S} }
     \ar[rr]
       ^-{ \sigma_{\mathrm{U}}/\mathbb{S} }
     &&
     (K \mathrm{U})/\mathbb{S}
     \ar[rr]
       ^-{
         \mathrm{ch}/\mathbb{S}
       }
     &&
     (H^{\mathrm{ev}}\!\mathbb{Q})/\mathbb{S}
   }
   }
  \end{equation}
\end{prop}
\begin{proof}
This follows
with Def. \ref{InducedCohomologyOperationsOnCofiberCohomology}
from \eqref{ToddCharacterRestrictionToMSU},
once we know that all maps involved are multiplicative:
For the Chern- and Pontrjagin character this is
Example \ref{ChernCharacterOnCofiberOfKTheory},
while for the K-orientations $\sigma_{\mathrm{U}}$
and $\sigma_{S\mathrm{U}}$ this is
Prop. \ref{ConnerFloydKOrientation}.
\end{proof}

\medskip

\noindent
{\bf The Conner-Floyd e-invariant.}

\begin{prop}[The Conner-Floyd e-invariant {\cite[Thm. 16.2]{ConnerFloyd66}}]
 \label{TheConnerFloydeInvariant}
The Adams $\mathrm{e}_{\mathrm{Ad}}$-invariant (Def. \ref{ClassicalConstructionOfAdamseCInvariant})
on stable Cohomotopy elements
$[c] \,\in\, \mathbb{S}_\bullet$
is expressed geometrically as the rational
Todd class \eqref{RationalToddNumberOnUManifoldsWithFramedBoundaries},
modulo integers, of any compact $\mathrm{U}$-manifold
$M^{\bullet+1}_{\mathrm{U},\mathrm{Fr}}$ whose
framed boundary is the manifold corresponding to $[c]$
under the Pontrjagin-Thom isomorphism
\begin{equation}
  \label{GeometricExpressionOfTheeInvariant}
  \underset{
    \mathclap{
    \raisebox{-3pt}{
      \tiny
      \color{darkblue}
      \bf
      \begin{tabular}{c}
        Adams'
        \\
        e-invariant
      \end{tabular}
    }
    }
  }{
    \mathrm{e}_{\mathrm{Ad}}
    (
      c
    )
  }
  \;=\;
  \underset{
      \mathclap{
      \mbox{
        \tiny
        \color{darkblue}
        \bf
        \begin{tabular}{c}
          Conner-Floyd's
          \\
          e-invariant
        \end{tabular}
      }
      }
  }{
    \mathrm{e}_{\mathrm{CF}}
    \big(
      M^d_{\mathrm{Fr}}
    \big)
  }
  \;\coloneqq\;
  \mathrm{Td}
  \big[
    M^{d+1}_{\mathrm{U},\mathrm{Fr}}
  \big]
  \,,
  \phantom{AA}
  \mbox{
    \rm
    for
    $
      \partial M^{d+1}_{\mathrm{U},\mathrm{Fr}}
      \,=\,
      M^{d}_{\mathrm{Fr}}
      \xleftrightarrow
        [
          \mbox{
            \tiny
            \color{greenii}
            \bf
            PT-iso
          }
        ]
        {\;}
      [c]
    $.
  }
\end{equation}
\end{prop}
In more detail, consider the following diagram:
\begin{equation}
 \label{ConnerFloydeInvariantShortExactSequence}
 \hspace{-8mm}
  \xymatrix@R=5pt@C=3em{
    0
    \ar[r]
    &
    \overset{
      \mathclap{
      \raisebox{6pt}{
        \tiny
        \color{darkblue}
        \bf
        \begin{tabular}{c}
          cobordism classes of
          \\
          compact $\mathrm{U}$-manifolds
          \\
          without boundary
        \end{tabular}
      }
      }
    }{
      \;
      (M \mathrm{U})_{\bullet + 1}
      \;
    }
    \ar[dd]_-{
      \mathclap{\phantom{\int^{\vert}_{\vert}}}
      \mathrm{Td}
    }
    \ar@{^{(}->}[rr]
    &
    \ar@{}[dd]|-{
      \mathclap{
        \mbox{
          \tiny
          \color{greenii}
          \bf
          \begin{tabular}{c}
            Todd class
          \end{tabular}
        }
      }
    }
    &
    \overset{
      \mathclap{
      \raisebox{6pt}{
        \tiny
        \color{darkblue}
        \bf
        \begin{tabular}{c}
          cobordism classes of
          \\
          compact $\mathrm{U}$-manifolds
          \\
          with framed boundary
        \end{tabular}
      }
      }
    }{
    \;
    (M \mathrm{U}/\mathbb{S})_{\bullet + 1}
    \;
    }
    \ar@{->>}[rr]^-{
      \;\partial\;
    }_-{
        \raisebox{-3pt}{
          \tiny
          \color{greenii}
          \bf
          boundary map
        }
    }
    \ar[dd]_-{
      \mathclap{\phantom{\int^{\vert}_{\vert}}}
      \mathrm{Td}
    }
    &&
    \overset{
      \mathclap{
      \raisebox{3pt}{
        \tiny
        \color{darkblue}
        \bf
        \begin{tabular}{c}
          cobordism classes of
          \\
          framed manifolds
        \end{tabular}
      }
      }
    }{
      \;
      (M \mathrm{Fr})_\bullet
      \;
    }
    \ar@{<->}[rr]^-{
      \scalebox{.7}{
      \raisebox{3.7pt}{$
        \simeq
      $}
      }
    }_-{
      \mbox{
        \tiny
        \color{greenii}
        \bf
        \begin{tabular}{c}
          Pontrjagin-Thom
          \\
          isomorphism
        \end{tabular}
      }
    }
    \ar@{-->}[dd]
      _-{
        \mathrm{e}_{\mathrm{CF}}
      }
    &&
    \overset{
      \mathclap{
      \raisebox{6pt}{
        \tiny
        \color{darkblue}
        \bf
        \begin{tabular}{c}
          stable Cohomotopy
          \\
          ground ring
        \end{tabular}
      }
      }
    }{
      \mathbb{S}_\bullet
    }
    \ar[dd]_-{
      e_{\mathrm{Ad}}
    }^-{
      \mathrlap{
        \!\!\!\!\!\!\!\!
        \mbox{
          \tiny
          \color{greenii}
          \bf
          \begin{tabular}{c}
            Adams
            \\
            e-invariant
          \end{tabular}
        }
      }
    }
    \\
    \\
    0 \ar[r]
    &
    \;\mathbb{Z}\;
    \ar@{^{(}->}[rr]
    &&
    \mathbb{Q}
    \ar[rr]
    &&
    \mathbb{Q}/\mathbb{Z}
    \ar@{=}[rr]
    &&
    \mathbb{Q}/\mathbb{Z}
    \,,
  }
\end{equation}
where the top row is exact, by Lemma \ref{CofiberECohomologyAsExtensionOfStableCohomotopyByECohomology},
(so that, in particular, the boundary map is surjective and
hence does admit the lifts assumed in \eqref{GeometricExpressionOfTheeInvariant})
while the two left vertical morphisms are
from Remark \ref{ToddNumbersViaChernWeilTheory},
thus inducing the dashed morphism $\mathrm{e}_{\mathrm{CF}}$.
The claim is that the square on the right commutes.

\medskip
This is, at its heart, a consequence of the
fact that the Todd character is the Chern character of any
Thom class (Prop. \ref{RationalToddCharacterIsChernCharacterOfThomClass}),
as made fully manifest by our diagrammatic construction
\eqref{ReproducingTheConnerFloydEInvariant}.

\begin{remark}[Conner-Floyd e-invariant on $S\mathrm{U}$-manifolds {\cite[p. 104]{ConnerFloyd66}}]
Since the Todd number on $S\mathrm{U}$-manifolds
of real dimension $4 \;\mathrm{mod}\; 8$
is
divisible by 2 (Prop. \ref{QuantizationOfToddNumbers}),
it follows that in this situation also
the Conner-Floyd e-invariant (Def. \ref{TheConnerFloydeInvariant})
is divisible by 2.

\noindent {\bf (i)} Hence if follows that \eqref{ConnerFloydeInvariantShortExactSequence}
refines to the following diagram:
\vspace{-2mm}
\begin{equation}
\label{FractionalConnerFloydeInvariant}
\hspace{-8mm}
  \xymatrix@R=5pt@C=2.7em{
    0
    \ar[r]
    &
    \overset{
      \mathclap{
      \raisebox{6pt}{
        \tiny
        \color{darkblue}
        \bf
        \begin{tabular}{c}
          cobordism classes of
          \\
          compact $S\mathrm{U}$ manifolds
          \\
          without boundary
        \end{tabular}
      }
      }
    }{
      \;
      (M S\mathrm{U})_{8\bullet + 4}
      \;
    }
    \ar[dd]_-{
      \mathclap{\phantom{\int^{\vert}_{\vert}}}
      \scalebox{.6}{$\tfrac{1}{2}$}
      \mathrm{Td}
    }
    \ar@{^{(}->}[rr]
    &
    \ar@{}[dd]|-{
      \mathclap{
        \mbox{
          \tiny
          \color{greenii}
          \bf
          \begin{tabular}{c}
            half Todd class
          \end{tabular}
        }
      }
    }
    &
    \overset{
      \mathclap{
      \raisebox{6pt}{
        \tiny
        \color{darkblue}
        \bf
        \begin{tabular}{c}
          cobordism classes of
          \\
          compact $S\mathrm{U}$-manifolds
          \\
          with framed boundary
        \end{tabular}
      }
      }
    }{
    \;
    (M \mathrm{U}/\mathbb{S})_{8\bullet + 4}
    \;
    }
    \ar@{->>}[rr]^-{
      \;\partial\;
    }_-{
        \raisebox{-3pt}{
          \tiny
          \color{greenii}
          \bf
          boundary map
        }
    }
    \ar[dd]_-{
      \mathclap{\phantom{\int^{\vert}_{\vert}}}
      \scalebox{.6}{$\tfrac{1}{2}$}
      \mathrm{Td}
    }
    &&
    \overset{
      \mathclap{
      \raisebox{3pt}{
        \tiny
        \color{darkblue}
        \bf
        \begin{tabular}{c}
          cobordism classes of
          \\
          framed manifolds
        \end{tabular}
      }
      }
    }{
      \;
      (M \mathrm{Fr})_{8\bullet + 3}
      \;
    }
    \ar@{<->}[rr]^-{
      \scalebox{.7}{
      \raisebox{3.7pt}{$
        \simeq
      $}
      }
    }_-{
      \mbox{
        \tiny
        \color{greenii}
        \bf
        \begin{tabular}{c}
          Pontrjagin-Thom
          \\
          isomorphism
        \end{tabular}
      }
    }
    \ar@{-->}[dd]
      _-{
        \scalebox{.6}{$\tfrac{1}{2}$}
        \mathrm{e}_{\mathrm{CF}}
      }
    &&
    \overset{
      \mathclap{
      \raisebox{6pt}{
        \tiny
        \color{darkblue}
        \bf
        \begin{tabular}{c}
          stable Cohomotopy
          \\
          ground ring
        \end{tabular}
      }
      }
    }{
      \mathbb{S}_{8\bullet + 3}
    }
    \ar[dd]_-{
      \scalebox{.6}{$\tfrac{1}{2}$}
      e_{\mathrm{Ad}}
    }^-{
      \mathrlap{
        \!\!\!\!\!\!\!\!
        \mbox{
          \tiny
          \color{greenii}
          \bf
          \begin{tabular}{c}
            Adams
            \\
            $e_{\mathbb{R}}$-invariant
          \end{tabular}
        }
      }
    }
    \\
    \\
    0 \ar[r]
    &
    \;\mathbb{Z}\;
    \ar@{^{(}->}[rr]
    &&
    \mathbb{Q}
    \ar[rr]
    &&
    \mathbb{Q}/\mathbb{Z}
    \ar@{=}[rr]
    &&
    \mathbb{Q}/\mathbb{Z}
    \,.
  }
\end{equation}

\noindent {\bf (ii)} The divided Adams e-invariant
on the right coincides with the invariant called
$\mathrm{e}_{{}_{\mathbb{R}}}$ in \cite{Adams66},
which is defined in terms of Adams operation in $K\mathrm{O}$
by the same formula \eqref{AdamseCInvariant}
that defines $\mathrm{e}_{\mathrm{Ad}}$ by Adams operations on
$K\mathrm{U}$.

\noindent {\bf (iii)} Beware that the alternative formula
\eqref{AdamseCInvariantAsTopDegreeComponentOfChernCharacter}
for $\mathrm{e}_{\mathrm{Ad}}$ in terms of the Chern character
does {\it not} have a direct analogue that computes
$e_{\mathbb{R}}$: Replacing the Chern character on $K\mathrm{U}$
with the Pontrjagin character on $K\mathrm{O}$
still computes $e_{\mathrm{Ad}} = 2 e_{{}_{\mathbb{R}}}$.
Since it is the Chern character formula
\eqref{AdamseCInvariantAsTopDegreeComponentOfChernCharacter}
for the Adams e-invariant that is captured
(according to Theorem \ref{DiagrammaticeCCoincidesWithClassicaleCInvariant})
by the Toda bracket observable
$\widehat e_{K\mathrm{U}}$
(Def. \ref{LiftedEInvariantDiagrammatically})
the analogous Toda-bracket observable
$\widehat e_{K\mathrm{O}}$
built from the Pontrjagin character on $K\mathrm{O}$
still lifts $\mathrm{e}_{\mathrm{Ad}}$ and not
$\mathrm{e}_{{}_{\mathbb{R}}}$.
\end{remark}

\newpage

\subsection{Hopf invariant}

We give an abstract discussion of the
{\it Hopf invariant} in generalized cohomology,
streamlined to make manifest its nature as a
refined Toda-bracket (Def. \ref{RefinedTodaBracket}),
on the same footing as
the refined Adames e-invariant
\cref{TheAdamsEInvariant}
and the refined Conner-Floyd e-invariant \cref{RelativeCobordism},
all unified by the notion of observables on
trivializations
(``$H_{n-1}$-fluxes'') of the d-invariant in \cref{M5ThreeFlux}.

\medskip

\noindent {\bf The cup-square cohomology operation and its trivialization.}

\begin{defn}[Cup square cohomology operation]
 \label{CupSquareCohomologyOperation}
For $E$ a multiplicative cohomology theory
(Def. \ref{MultiplicativeCohomologyTheory})
and $n \in \mathbb{N}$,
we say that the {\it cup-square operation}
in $E$-cohomology of degree $n$
is the unstable cohomology operation
\vspace{-2mm}
$$
  \big[
    E\degree{n}
      \xrightarrow{ (-)^2 }
    E\degree{2n}
  \big]
  \;\in\;
  \widetilde E{}^{2n}
  \big(
    E\degree{n}
  \big)
$$

\vspace{-2mm}
\noindent which represents (via the Yoneda lemma)
over each $X \in \PointedHomotopyTypes$
the operation
\vspace{-4mm}
$$
\xymatrix{
  \big[
    X,
    (-)^2
  \big]
  \;:\;
  \big[
    X,
    E\degree{n}
  \big]
  \,=\,
  \widetilde E{}^{n}(X)
  \ar[rr]^-{
    \Delta_{\widetilde E{}^{n}(X)}
  }
  &&
  \widetilde E{}^{n}(X)
  \times
  \widetilde E{}^{n}(X)
  \ar[r]^-{
    \cup^E_X
  }
  &
  \widetilde E{}^{2n}(X)
  \,=\,
  \big[
    X,
    E\degree{2n}
  \big]}
  .
$$
\end{defn}
\noindent
Of course, this exists more generally
for $k$th cup powers for any $k \in \mathbb{N}$.
Review for the case of $E = {K\mathrm{U}}$
is in \cite[p. 44]{Wirthmueller12}.

\begin{lemma}[Connective covers of ring spectra]
  \label{ConnectiveCoversOfRingSpectra}
  For $E$ a ring spectrum with $E_\infty$-structure,
  its connective cover $E \langle 0 \rangle$
  inherits the structure of a ring spectrum such that the
  coreflection
  $
    E\langle 0 \rangle
      \xrightarrow{ \;\epsilon_E\; }
    E
  $
  is a homomorphism of ring spectra.
\end{lemma}

\noindent
This is due to
\cite[Prop. VII 4.3]{May77}\cite[Prop. 7.1.3.13]{Lurie17}.

\begin{prop}[Canonical trivialization of cup square over $n$-sphere]
  \label{CanonicalTrivializationOfCupSquareOvernSphere}
For $E$ a multiplicative cohomology theory
(Def. \ref{MultiplicativeCohomologyTheory})
and $n \in \mathbb{N}$ with $n \geq 1$,
the cup square of every element
$[c] \in \widetilde E{}^n\big( S^n \big)$
is trivial; and if $n \geq 2$ then it is canonically trivialized.
\end{prop}
\begin{proof}
  By the suspension isomorphism, $c$ is the
  $n$-fold suspension of an element in $E_0$,
  and hence is represented by a cocycle
  that factors through the connective cover
  cohomology theory $E\langle 0  \rangle$.
  But since connective covers of ring spectra
  are equipped with compatible ring spectrum structure
  (by Lemma \ref{ConnectiveCoversOfRingSpectra})
  so does its cup square, in that we have
  a solid diagram as follows, with the vertical
  morphisms on the right representing the
  cup square cohomology operation via
  Def. \ref{CupSquareCohomologyOperation}
  \vspace{-2mm}
  \begin{equation}
  \label{CanonicalTrivializationOfCupSquareOnnSphere}
  \raisebox{22pt}{
  \xymatrix{
    S^n
    \ar[rr]
      ^-{ \;c\; }
      _>>>{\ }="s"
    \ar[d]^>>>{\ }="t"
    &&
    E\degree{n}
    \ar[d]
      ^-{ (-)^2 }
    \\
    \ast
    \ar[rr]
    &&
    E\degree{2n}
    \ar@{=>}
      "s"; "t"
  }
  }
  \qquad
    \coloneqq
  \qquad
  \raisebox{22pt}{
  \xymatrix@C=4em{
  S^n
  \ar[r]
    _>>>{\ }="s"
  \ar[d]
    ^>>>{\ }="t"
  &
  (
    E\langle0\rangle
  )\degree{n}
  \ar[d]
    _-{ (-)^2 }
  \ar[r]
    ^-{ \;\
        (
          \epsilon_E
        )\degree{n}
    \; }
  &
  E\degree{n}
  \ar[d]
    ^-{ (-)^2 }
  \\
  \ast
  \ar[r]
  &
  E\degree{2n}
  \ar[r]
    ^-{ \;
      ( \epsilon_E )\degree{2n}
    \;}
  &
  (
    E
  )\degree{2n}.
  \ar@{==>}
    ^-{ \exists ! }
    "s"; "t"
  }
  }
  \end{equation}

    \vspace{-2mm}
\noindent  But, since $(E\langle 0\rangle)\degree{2n}$
  is $2n$-connective
  (because the stable homotopy groups
    $
      \pi_k
      \big(
        E\langle 0 \rangle
      \big)
      \,=\,
      \pi_{k + n}\big( (E\langle 0 \rangle)\degree{n} \big)
    $
    of a connective spectrum vanish for $k < 0$, by definition),
    we have
   (using with $n \geq 1$ that $n < 2n$)
  $
    \pi_n
    \big(
      (E\langle 0\rangle)\degree{2n}
    \big)
    \;=\;
    0
    \,,
  $
  meaning that there exists a dashed homotopy, as shown.
  If $n \geq 2$ then also the next higher homotopy
  group is trivial and the dashed homotopy exists
  uniquely, thus inducing a canonical homotopy
  filling the full diagram.
\end{proof}

\newpage

\noindent {\bf The $E$-Hopf invariant.}

\begin{defn}[$E$-Hopf invariant]
 \label{RefinedHopfInvariant}
For $E$ a multiplicative cohomology theory
(Def. \ref{MultiplicativeCohomologyTheory})
and $n \in \mathbb{N}$ with $n \geq 1$
we say that the
{\it refined $E$-Hopf invariant} is the function
$
    H^E_{n-1}
    \mathrm{Fluxes}
    (
      S^{2n-1}
    )
    \xrightarrow{ \kappa_E }
    E_0
  $,
which sends any $H^{E}_{n-1}$-flux
(Def. \ref{TrivializationsOfThedInvariant}, \eqref{FluxlessnessDiagram})
\vspace{-2mm}
$$
  \raisebox{10pt}{
  \xymatrix@R=15pt{
    S^{2n-1}
    \ar[d]
      _-{
        \mathclap{\phantom{\vert^{\vert}}}
        \themap
        \mathclap{\phantom{\vert_{\vert}}}
      }
      ^>>>{\ }="t"
    \ar[rr]_>>>{\ }="s"
    &&
    \ast
    \ar[d]
    \\
    S^n
    \ar[rr]
      |-{ \; \Sigma^n(1^E) \; }
    &&
    E\degree{n}
    \ar@{=>}
      ^-{
        \color{orangeii}
        h_{n-1}
      }
      "s"; "t"
    }
  }
$$

\vspace{-1mm}
\noindent
to the class
in
$
  \pi_1
  \mathrm{Maps}^{\ast/}
  \big( S^{2n-1}, E\degree{2n} \big)
  \,\simeq\,
  \pi_0 \mathrm{Maps}^{\ast/}
  \big( S^{2n}, E\degree{2n} \big)
  \,\simeq\,
  E_0
$
of its pasting composite with the canonical trivialization of the
cup square from Prop. \ref{CanonicalTrivializationOfCupSquareOvernSphere}:
\vspace{-2mm}
\begin{equation}
  \label{HopfInvariantDiagrammatic}
  \overset{
    \mathclap{
    \raisebox{6pt}{
      \tiny
      \color{darkblue}
      \bf
      $E$-Hopf invariant
    }
    }
  }{
    {\color{purple}
    \kappa_E
    }
  }
  \;\;\;\;\;\;\;\;\;\;
  :=
  \;\;\;\;\;\;\;\;\;\;
  \overset{
    \mathclap{
    \raisebox{5pt}{
      \tiny
      \color{darkblue}
      \bf
      \begin{tabular}{c}
        homotopy Whitehead integral
        {\color{black}$\,$/}
        \\
        functional cup product in $E$
      \end{tabular}
    }
    }
  }{
  \int_{S^{2n-1}}
  \Big(
    {
    \color{orangeii}
    h_{n-1} \cup \themap^\ast \omega_n
    }
    \;+\;
    {
    \color{greenii}
    \themap^\ast \omega_{2n-1}
    }
  \Big)
  }
  \phantom{AAAAAA}
  \mbox{
    for:
    $
      \left\{
      \begin{aligned}
        \color{orangeii}
        d \,h_{n-1}
        &
        \color{orangeii}
        = \themap^\ast \omega_n
        \\
        \color{greenii}
        d \,\omega_{2n-1}
        &
        \color{greenii}
        = - \omega_{n} \cup \omega_n
      \end{aligned}
      \right.
    $
  }
\end{equation}
$$
  \raisebox{20pt}{
  \xymatrix{
    S^{2n-1}
    \ar@/^1.7pc/[drr]_-{\ }="s"
    \ar@/_1.7pc/[drr]^-{\ }="t"
    \\
    &&
    E\degree{2n}
    \ar@{=>}
      |-{
      \Sigma^{2n-1} \! \color{purple}\kappa_E
      }
      "s"; "t"
  }
  }
  \;
  :=
  \;
  \raisebox{86pt}{
  \xymatrix{
   S^{2n-1}
   \ar[dd]
     _-{
           \themap
           }
   \ar[rr]
   \ar@{}[ddrr]
     |-{
     \rotatebox[origin=c]{-45}{\color{orangeii}$\big\Downarrow$}
     \mbox{\tiny (po)}
   }
   &&
   \ast
   \ar[dr]
   \ar[dd]
   \\
   && &
   \ast
   \ar[dd]
     |-{
       \mathclap{\phantom{\vert^{\vert}}}
       0
       \mathclap{\phantom{\vert_{\vert}}}
     }
   \\
   S^n
   \ar[dd]
   \ar[rr]
   \ar@{}[ddrr]|-{
     \rotatebox[origin=c]{-45}{\color{greenii}$\big\Downarrow$}
     \mbox{\tiny (po)}
   }
   \ar@/_.7pc/[drrr]
     |<<<<<<{
       \;\;\; \Sigma^n (1^E)\;\;\;\;
     }
     |>>>>>>>{
       {\phantom{AA}} \atop {\phantom{AA}}
     }
   &&
   C_c
   \ar[dd]
   \ar@{-->}[dr]
     |-{
       \mathclap{\phantom{\vert^{\vert}}}
       \vdash
       \color{orangeii}
       h_{n-1}
       \mathclap{\phantom{\vert_{\vert}}}
     }
   \\
   && &
   E\degree{n}
   \ar[dd]^-{ (-)^2 }
   \\
   \ast
   \ar@/_.7pc/[drrr]_-{ \;0\; }
   \ar[rr]
   &&
   S^{2n}
   \ar@{-->}[dr]|-{
     \mathclap{\phantom{\vert^{\vert}}}
     \Sigma^{2n} \color{purple}\kappa
     \mathclap{\phantom{\vert_{\vert}}}
   }
   \\
   && &
   E\degree{2n}
  }
  }
  \;
  =
  \;
  \raisebox{72pt}{
  \xymatrix{
   S^{2n-1}
   \ar[dd]
     _-{
             \themap
           }
   \ar@[white][rr]
   &&
   \color{white}
   \ast
   \ar@[white][dr]
   \\
   && &
   \color{white}
   \ast
   \\
   S^n
   \ar[dd]
   \ar@[white][rr]
   \ar@/_.7pc/[drrr]
     |<<<<<<<{
       \;\Sigma^n (1^E)
       \mathrlap{ \;\;=:\; \omega_n }
       \;
       \mathclap{\phantom{\vert_{\vert}}}
     }
     _>>>>>>>{\ }="s"
   &&
   \color{white}
   C_c
   \ar@[white]@{-->}[dr]
   \\
   && &
   E\degree{n}
   \ar[dd]|-{
     \mathclap{\phantom{\vert^{\vert}}}
     (-)^{2}
     \mathclap{\phantom{\vert_{\vert}}}
   }
   \\
   \ast
   \ar@/_.7pc/[drrr]_-{ \;0\; }^<<<<<{\ }="t"
   \ar@[white][rr]
   &&
   \color{white}
   S^{2n}
   \ar@[white]@{-->}[dr]|-{
     \color{white}
     \mathclap{\phantom{\vert^{\vert}}}
     \Sigma^{2n} \kappa
     \mathclap{\phantom{\vert_{\vert}}}
   }
   \\
   && &
   E\degree{2n}
   \ar@{=>}^-{
     \color{greenii}
     \!\!\!\!\!\!\!\!
     \scalebox{.7}{$
     \mathclap{
     \rotatebox[origin=c]{46}{$
       d \omega_{\,2n\raisebox{-.4pt}{\scalebox{.6}{$-1$}}}
       \,=\,
       - \omega_n \scalebox{.8}{$\cup$} \omega_n
       \!\!\!\!\!\!\!\!\!\!\!\!\!\!\!\!\!
       \!\!\!\!\!\!\!\!\!\!\!\!\!\!\!\!\!
       \!\!\!\!
     $}
     }
     $}
   }
     "s"; "t"
  }
  }
  \hspace{-4.5cm}
  \raisebox{88pt}{
  \xymatrix{
   S^{2n-1}
   \ar[dd]_-{
     \themap
   }
   \ar[rr]
   \ar@{}[ddrr]|-{
     \rotatebox[origin=c]{-45}{$\big\Downarrow$}
     \mbox{\tiny (po)}
   }
   &&
   \ast
   \ar[dr]
   \ar[dd]
   \\
   && &
   \ast
   \ar[dd]^-{0}_<<<<{\ }="s"
   \\
   S^n
   \ar[rr]
   \ar@/_.7pc/[drrr]
     |<<<<<<<{ \;\;\; \Sigma^n (1^E) \;\;\; }
   &&
   C_f
   \ar@[white][dd]
   \ar@{-->}[dr]^-{ c }
   \\
   && &
   E\degree{n}
   \ar[dd]^-{ (-)^{2_{\cup}} }
   \\
   \color{white}
   \ast
   &&
   {\color{white}
   S^{2n}}
   \\
   && &
   E\degree{2n}
   \ar@{=>}
     _{
       \color{orangeii}
       \mathclap{
         \scalebox{.69}{
         \rotatebox[origin=c]{46}{$
           d h_{n\raisebox{-.4pt}{\scalebox{.6}{$-1$}}}
           = \! \themap^\ast \omega_n
           \!\!\!\!
         $}
         }
         \!\!\!\!\!\!\!\!\!\!\!\!\!\!\!\!\!\!\!
       }
     }
     |>>>>>>>>>{ {\phantom{AA}} \atop {\phantom{AA}} }
     "s"; "s"+(-17,-17)
  }
  }
$$
\end{defn}
The two ways to read this diagram yield
{\it two component definitions of the $E$-Hopf invariant}:

\vspace{1mm}
\noindent   {\bf (1)} {\bf Hopf-Adams/Atiyah-style definition}.
    On the left in \eqref{HopfInvariantDiagrammatic},
  the choice of classifying
  map $\vdash h_{n-1}$ is equivalently
  a choice of extension of the pullback
  of the canonical generator from $S^n$;
  and the homotopy-commuting square
  repeated on the right
  says that the cup square of this lift is
  an $E_0$-multiple $\kappa$ of the canonical generator
  on $S^{2n}$.
  \vspace{-2mm}
  $$
    \raisebox{35pt}{
    \xymatrix@R=-1pt{
      C_c
      \ar@{-->}[rrd]
        ^-{
          \vdash
          \color{orangeii}
          h_{n-1}
        }
      \ar[ddd]
        _-{
          \mathclap{\phantom{\vert^{\vert}}}
          p_c
          \mathclap{\phantom{\vert_{\vert}}}
        }
      \\
      &&
      E\degree{n}
      \ar[ddd]^{ (-)^2 }
      \\
      \\
      S^{2n}
      \ar[drr]
        _-{
          \;\;
            \Sigma^{2n}
            \color{purple}
            \kappa
          \;\;
        }
      \\
      &&
      E\degree{2n}
    }
    }
    \qquad \Leftrightarrow \qquad
\big[ \vdash h_{n-1}\big]^2
      \,=\,
           \kappa
        \cdot
      p_c^\ast \big[\Sigma^{2n}(1^E)\big].
  $$

  \vspace{-2mm}
\noindent  For $E = H \mathbb{F}_2$
  this gives Hopf's original definition
  (\cite[p. 33]{MosherTangora86}),
  while for $E = {K\mathrm{U}}$
  this gives Adams-Atiyah's K-theoretic definition of the Hopf invariant
  \cite{AdamsAtiyah66}
  (review in \cite[p. 50]{Wirthmueller12}).

\vspace{1mm}
   \noindent
   {\bf (2)} {\bf Whitehead-Steenrod-Haefliger-type definition.}
   On the right in
   \eqref{HopfInvariantDiagrammatic}
   we see,
   in the case that $E = H\mathbb{R}$ is ordinary real cohomology,
   and translating to differential forms via
   the fundamental theorem of rational homotopy theory,
   we get the
   classical homotopy Whitehead-Steenrod integral formula
   \cite{Whitehead47} (review in \cite[Prop. 7.22]{BT82})
   \cite{Steenrod49}\cite[p. 17]{Haefliger78}
   (review in \cite[Ex. 1.9]{SinhaWalter08})
   as discussed in our context in
  \cite[Prop. 4.6]{FSS19c}\cite[Prop. 3.20]{FSS19b}.
  (The symbols in \eqref{HopfInvariantDiagrammatic}
  are chosen such as to match the notation
  used in \cite{FSS19b}\cite{FSS19c} in this rational case.)

\medskip

Hence the {\it proof} that these two formulations
of the classical Hopf invariant are equivalent
is trivialized by the diagrammatic Definition \ref{RefinedHopfInvariant},
as is its generalization to coefficients in
any multiplicative cohomology theory $E$.

\newpage

\subsection{Ravenel orientations}
\label{Orientation}

\noindent
{\bf Finite-dimensional $E$-orientations.}
The traditional term {\it complex-oriented $E$-cohomology}
(e.g. \cite{Hopkins99}), reviewed in a moment (the formal
definition is Def. \ref{ComplexOrientedCohomology} below)
is short for {\it universal orientation of fibers
of complex vector bundles in the Whitehead-generalized
cohomology theory $E$}; analogously there is real- and
quaternionic oriented cohomology (Def. \ref{QuaternionicOrientedCohomology} below).
Here ``universal'' means that a fiber-wise orientation is
chosen for {\it all} vector bundles
over {\it all} (paracompact) topological spaces at once,
and compatibly so under pullback --
which means to choose an orientation on the universal
vector bundles over the corresponding
infinite-dimensional (as cell complexes) classifying spaces:
$$
  \phantom{AAAAAAAAA}
  \raisebox{24pt}{
  \xymatrix@R=1.5em{
    \mathllap{
      \mbox{
        \tiny
        \color{darkblue}
        \bf
        \begin{tabular}{c}
          {\it any} $G$-vector bundle
        \end{tabular}
      }
    }
    \mathcal{V}_X
    \ar[d]
    \ar[rr]
    \ar@{}[drr]|-{ \mbox{\tiny(pb)} }
    &&
    \mathcal{V}_{B G}
    \ar[d]
    \mathrlap{
      \!\!\!\!\!
      \mbox{
        \tiny
        \color{darkblue}
        \bf
        \begin{tabular}{c}
          universal
          \\
          $G$-vector bundle
        \end{tabular}
      }
    }
    \\
    \mathllap{
      \mbox{
        \tiny
        \color{darkblue}
        \bf
        \begin{tabular}{c}
          {\it any}
          \\
          topological space
        \end{tabular}
      }
    }
    X
    \ar[rr]
      ^-{
        \;\;
        \vdash \mathcal{V}_X
        \;\;
      }
      _-{
        \mathclap{\phantom{\vert^{\vert^{\vert}}}}
        \mbox{
          \tiny
          \color{greenii}
          \bf
          classifying map
        }
      }
    &&
    B G
    \mathrlap{
      \!\!\!\!\!
      \mbox{
        \tiny
        \color{darkblue}
        \bf
        \begin{tabular}{c}
          classifying space
          \\
          for $G$-vector bundles
          \\
          over any space
        \end{tabular}
      }
    }
  }
  }
    \phantom{AAAAAAAAAAAAAA}
    \mbox{
      $E$-orientation of
      $\mathcal{V}_{B G}$
    }
    \;\;
    \Rightarrow
    \;\;
    \underset{\mathcal{V}_X}{\forall}
    \mbox{
      $E$-orientation of
      $\mathcal{V}_{X}$.
    }
$$

However, in applications, such as to physics,
there is often an
upper bound both on the dimension of relevant base spaces
(spacetimes) as well as on the rank of the vector bundle
(field content),
so that the standard notion of oriented
$E$-cohomology involves an infinite tower of redundant data.
Therefore we turn attention to
{\it finite-dimensional $E$-orientations}, namely to
$E$-orientations universally chosen (only)
on vector bundles of rank $\leq r + 1$
over $\leq d$-dimensional base spaces
and those pulled back from these, for some fixed
upper bound on $d + 2r$ (for complex vector bundles)
or on $d + 4r$ (for quaternionic vector bundles).
For complex vector bundles this
finite-dimensional notion of
$E$-orientations appears briefly in
\cite[\S 1.2]{Hopkins84}\cite[\S. 6.5]{Ravenel86},
but does not seem to have found attention in bundle theory
(beyond formal investigation into the associated Thom spectra,
now known as {\it Ravenel's spectra}
\cite[\S 3]{Ravenel84}\cite[\S 6.5]{Ravenel86}\cite[Thm. 3]{DHJ88}\cite{Beardsley19}).

\medskip
In particular, our discussion in \cref{VicinityOfBlackM2BranesAndOrientedCohomology}
shows (with Lemma \ref{UniversalFiniteDimensionalLineBundles} below)
that {\it Hypothesis H} leads to the appearance of
$E$-orientations on 10-dimensional quaternionic line bundles,
in this sense, namely of quaternionic line bundles
whose classifying map factors through the 11-skeleton
of the full classifying space:
$$
\hspace{-2cm}
  \xymatrix@C=4em@R=7pt{
    &
    &
    &
    \mathllap{
      \mbox{
        \tiny
        \color{darkblue}
        \bf
        \begin{tabular}{c}
          classifying space for $\mathbb{H}$-line bundles
          \\
          over 10-dimensional spaces
        \end{tabular}
      }
      \!\!\!\!
    }
    \mathbb{H}P^2
    \mathclap{\phantom{\vert_{\vert}}}
    \ar@{^{(}->}[dd]
      ^-{
        \!\!
        \mathrlap{
        \mbox{
          \tiny
          {
          \color{greenii}
          \bf
          11-skeleton
          }
        }
        }
      }
    \\
    \\
    \underset{
      \mathclap{
      \raisebox{-3pt}{
        \tiny
        \color{darkblue}
        \bf
        \begin{tabular}{c}
          spacetime
        \end{tabular}
      }
      }
    }{
      \mathbb{R}^{0,1}
      \times
      X^{10}
    }
    \ar@{->>}[r]
    &
    \;\;
    \underset{
      \mathclap{
      \mbox{
        \tiny
        \color{darkblue}
        \bf
        \begin{tabular}{c}
          10-dim
          space
        \end{tabular}
      }
      }
    }{
      X^{10}
    }
    \;\;
    \ar@{-->}[uurr]
    \ar[rr]
      ^-{
        \quad
        \vdash \mathcal{V}_{X^{10}}
           }
      _-{
        \mathclap{\phantom{\vert^{\vert^{\vert^{\vert}}}}}
        \mathclap{
        \mbox{
          \tiny
          \color{greenii}
          \bf
          \begin{tabular}{c}
            classifying
            \\
            map
          \end{tabular}
        }
        }
      }
    &
    &
    \mathbb{H}P^\infty
    \mathrlap{
      \;
      =
      \;
      B \mathrm{Sp}(1)\;.
    }
  }
$$

While quaternionic $E$-orientations have not found much
attention yet by themselves, we know that they
subsume the widely-studied complex $E$-orientations, in that
every complex $E$-orientation induces a quaternionic
$E$-orientation
(see Prop. \ref{QuaternionicOrientationFromComplexOrientation} below).
Here we generalize this relation to finite-dimensional
orientations and observe
(Theorem \ref{FiniteRankQuaternionicOrientationFromFiniteRankComplexOrientation} below)
that $(4n+2)$-dimensional complex
orientations induce $4n+2$-dimensional quaternionic orientations
(which is not as immediate as it may seem from the coincidence of
the two dimensions; see Remark \ref{RelationOfDimensionsForComplexAndQuaternionicOrientations}).

\medskip
The traditional terminology of ``oriented cohomology theory''
makes use of a number of tacit identifications,
beginning with the standard definition
of ``complex $E$-orientation'' (see Def. \ref{ComplexOrientedCohomology} below) being all phrased in terms
of a first $E$-Chern class with a superficially ad-hoc condition
on it, and no mentioning of any notion of orientation.
Therefore, we first give a brief exposition of the
relevant concept formation
(the oriented reader may want to skip ahead):

\medskip

\noindent {\bf Orientations of Euclidean space.}
  An ordinary {\it orientation} of a
  differentiable manifold $Y$ is
  traditionally defined to be the choice of a volume form
  (a nowhere vanishing top degree differential form)
  up to pointwise positive rescaling.
  Assuming that $Y$ is closed, hence compact,
  we may equivalently remember just the class of this
  form in de Rham cohomology, up to positive rescaling:
  \vspace{-2mm}
  $$
    \mathrm{vol}_Y
    \;\in\;
    \xymatrix@C=3em{
      \underset{
        \raisebox{-3pt}{
          \tiny
          \color{darkblue}
          \bf
          \begin{tabular}{c}
            differential
            forms
          \end{tabular}
        }
      }{
        \Omega_{\mathrm{dR}}^{\mathrm{dim}(Y)}(Y)
      }
      \ar[rr]
        _-{
          \;[-]\;
        }
        ^-{
          \mathclap{
          \mbox{
            \tiny
            \color{greenii}
            \bf
            \begin{tabular}{c}
              pass to total volume
            \end{tabular}
          }
          }
        }
      &&
      \quad
      \underset{
        \mathclap{
        \raisebox{-3pt}{
          \tiny
          \color{darkblue}
          \bf
          de Rham cohomology
        }
        }
      }{
        H_{\mathrm{dR}}^{\mathrm{dim}(Y)}(Y)
      }
    }.
  $$

  \vspace{-2mm}
\noindent
  If $Y$ were already oriented otherwise, we could ask that
  $[\mathrm{vol}_Y]$ be rescaled to
  a {\it unit} $\pm 1 \,\in\, \mathbb{Z} \subset \mathbb{R}$,
  {\it with respect to} the
  reference orientation. That number $\pm 1 \in \mathbb{Z}$ would
  be the choice of {\it relative orientation}.
  But since we want to
  define that reference orientation in the first place,
  the normalization demand on it is, conversely, that it be
  {\bf (a)} integral and {\bf (b)} minimally so, such that
  any other integral class is a unique integral multiple.
  This means to demand that

\noindent  {\bf (a)} the orienting volume class lifts to integral cohomology:
\vspace{-2mm}
  $$
    \underset{
      \mathclap{
      \raisebox{-3pt}{
        \tiny
        \color{darkblue}
        \bf
        \begin{tabular}{c}
          integral
          \\
          volume class
        \end{tabular}
      }
      }
    }{
      [\mathrm{vol}_Y]
    }
    \;\in\;
    \xymatrix@C=2em{
      \underset{
        \mathclap{
        \raisebox{-2pt}{
          \tiny
          \color{darkblue}
          \bf
          integral cohomology
        }
        }
      }{
        H^{\mathrm{dim}(Y)}
        \big(
          Y; \mathbb{Z}
        \big)
      }
      \ar[rr]^-{
        \mbox{
          \tiny
          \color{greenii}
          \bf
          \begin{tabular}{c}
            extension
            \\
            of scalars
          \end{tabular}
        }
      }
      &&
      \underset{
        \mathclap{
        \raisebox{-2pt}{
          \tiny
          \color{darkblue}
          \bf
          real cohomology
        }
        }
      }{
        H^{\mathrm{dim}(Y)}
        \big(
          Y; \mathbb{R}
        \big)
      }
      \ar[rr]
        ^-{
          \mbox{
            \tiny
            \color{greenii}
            \bf
            \begin{tabular}{c}
              de Rham
              \\
              theorem
            \end{tabular}
          }
        }
      _-{
        \simeq
      }
      &&
      \underset{
        \mathclap{
        \raisebox{-2pt}{
          \tiny
          \color{darkblue}
          \bf
          de Rham cohomology
        }
        }
      }{
        H_{\mathrm{dR}}^{\mathrm{dim}(Y)}
        \big(
          Y
        \big)
      }
    }
  $$
  \noindent

\vspace{-2mm}
\noindent  {\bf (b)} every other integral cohomology $n$-class of $Y$ is obtained
  by external cup product of $[\mathrm{vol}_Y]$ with
  an integral class on the point:
  \begin{equation}
    \label{OrdinaryOrientationClassAsGeneratorInIntegralCohomology}
    \underset{
      \mathclap{
      \raisebox{-3pt}{
        \tiny
        \color{darkblue}
        \bf
        \begin{tabular}{c}
          top
          integral
          cohomology
        \end{tabular}
      }
      }
    }{
      H^{\mathrm{dim}(Y)}\big(Y;\, \mathbb{Z}\big)
    }
     \;\simeq\;
    \underset{
      \mathclap{
      \raisebox{-3pt}{
        \tiny
        \color{darkblue}
        \bf
        \begin{tabular}{c}
          generated
          \\
          under
          cupping
        \end{tabular}
      }
      }
    }{
      H^0\big(\ast;\, \mathbb{Z}\big)
    }
    \big\langle
      \underset{
        \mathrlap{
        \;
        \raisebox{-3pt}{
          \tiny
          \color{darkblue}
          \bf
          \begin{tabular}{c}
            from normalized
            \\
            volume class
          \end{tabular}
        }
        }
      }{
        [\mathrm{vol}_Y]
      }
    \big\rangle
    \;\;
    \in
    \;
    H^0\big(\ast;\, \mathbb{Z}\big)
    \,
    \mathrm{Modules}.
  \;
  \end{equation}
  If $Y$ is not compact, we apply this logic to
  those volume forms that {\it vanish at infinity} on $Y$
  -- their classes are naturally
  identified with classes as above,
  defined on the one-point compactification space
  $X \coloneqq Y_{\mathrm{cpt}}$.

\medskip
  In the special case that $Y = V^n$ is a Euclidean space
  of finite dimension $n$,
  its one-point compactification
  $( V^n)_{\mathrm{cpt}} = S^n$ is the sphere
  of that dimension.
  Since the {\it reduced}
  integral cohomology of the $n$-sphere
  is all concentrated in degree
  $n$, and since
  the integral cohomology of the point is
  concentrated in degree 0,
  the above condition
  \eqref{OrdinaryOrientationClassAsGeneratorInIntegralCohomology}
  on an orienting volume class on $V^n$
  may equivalently be stated in terms of the full graded
  cohomology ring as
  \begin{equation}
    \label{OrientingThenSphereInIntegralCohomology}
    \underset{
      \mathclap{
      \raisebox{-3pt}{
        \tiny
        \color{darkblue}
        \bf
        \begin{tabular}{c}
          reduced cohomology
        \end{tabular}
      }
      }
    }{
      {\widetilde H}^\bullet
      \big(
        (V^n)^{\!\mathrm{cpt}}; \mathbb{Z}
      \big)
    }
    \;\simeq\;
    \underset{
      \mathclap{
      \raisebox{-3pt}{
        \tiny
        \color{darkblue}
        \bf
        \begin{tabular}{c}
          generated
          over
          \\
          ground ring
        \end{tabular}
      }
      \;
      }
    }{
      H^\bullet(\ast;\mathbb{Z})
    }
    \big\langle
      \underset{
        \mathrlap{
        \;
        \raisebox{-3pt}{
          \tiny
          \color{darkblue}
          \bf
          \begin{tabular}{c}
            from normalized
            \\
            volume class
          \end{tabular}
        }
        }
      }{
        [\mathrm{vol}_{S^n}]
      }
    \big\rangle
    \;\;\;
    \in
    \;
    H^\bullet(\ast;\mathbb{Z})
    \,
    \mathrm{Modules}
    \,.
  \end{equation}
  Any two choices of $[\mathrm{vol}_Y]$ that satisfy this
  condition \eqref{OrientingThenSphereInIntegralCohomology}
  differ by multiplication by a {\it unit in the graded ground ring},
  which here is either of
  $\pm 1 \in \mathbb{Z}
    \simeq
   H^0(\ast;\mathbb{Z})
    \simeq
   H^\bullet(\ast; \mathbb{Z})
   \,.
  $
  While \eqref{OrientingThenSphereInIntegralCohomology}
  is a heavy way of speaking about the 2-element
  set of choices of $[\mathrm{vol}_{S^n}]$,
  it has the advantage that it
  readily generalizes to any
  Whitehead-generalized multiplicative cohomology theory $E$:

\medskip
  In direct generalization of \eqref{OrientingThenSphereInIntegralCohomology},
  one says that an orientation of
  the Euclidean space $V^n$ in $E$-cohomology
  is the choice of a class
  $[\mathrm{vol}_{S^n}^E] \in {\widetilde E}{}^\bullet(S^n)$
  which is a {\it generator} of the reduced cohomology as
  a module over the graded ground ring:
  \begin{equation}
    \label{OrientingThenSphereInGeneralizedCohomology}
    \underset{
      \mathclap{
      \raisebox{-3pt}{
        \tiny
        \color{darkblue}
        \bf
        \begin{tabular}{c}
          reduced
          \\
          $E$-cohomology
        \end{tabular}
      }
      }
    }{
      {\widetilde E}^\bullet(S^n)
    }
    \;\;\simeq\;\;
    \underset{
      \mathclap{
      \raisebox{-3pt}{
        \tiny
        \color{darkblue}
        \bf
        \begin{tabular}{c}
          generated
          over
          \\
          $E$-ground ring
        \end{tabular}
      }
      \;\;\;
      }
    }{
      E^\bullet(\ast)
    }
    \big\langle
      \;
      \underset{
        \mathrlap{
        \;\;\;
        \raisebox{-3pt}{
          \tiny
          \color{darkblue}
          \bf
          \begin{tabular}{c}
            from normalized
            \\
            $E$-volume class
          \end{tabular}
        }
        }
      }{
        [\mathrm{vol}^E_{S^n}]
      }
      \;
    \big\rangle
    \;\;\;
    \in
    \;
    E^\bullet(\ast)
    \,
    \mathrm{Modules}
    \,.
  \end{equation}

  Now any two choices of
  $E$-orientations, hence of classes $[\mathrm{vol}^E_{S^n}]$ that solve
  \eqref{OrientingThenSphereInGeneralizedCohomology},
  differ by multiplication
  with a unit (a multiplicatively invertible element)
  in the ground ring $E_{- \bullet} \coloneqq E^\bullet(\ast)$.
  This may be a set of choices much larger than the two orientations
  possible in ordinary cohomology. In particular,
  the characteristic property of Whitehead-generalized cohomology,
  namely that it allows the violation of the {\it dimension axiom}
  satisfied by ordinary cohomology, means that relative orientations
  may have non-vanishing degree.

\medskip
  On the other hand, there is one bit of extra information
  already provided with a
  Whitehead-generalized cohomology theory $E$, namely a choice of
  natural
  {\it suspension isomorphisms}
   $
    \xymatrix{
      \widetilde E{}^{\bullet}(X)
      \ar[r]_-{\simeq}
        ^-{\sigma^E}
      &
      \widetilde E{}^{\bullet+n}(\Sigma^n X)
    }
  $\!.
  For multiplicative theories
  this choice implies a canonical choice of $E$-orientation
  of $V^n$, namely that given by the image under
  $n$-fold suspension of the
  {\it canonical unit element} $1^E$
  in the ground ring:
  \vspace{-2mm}
  \begin{equation}
    \label{CanonicalEOrientationOnEuclideanSpace}
    \xymatrix@R=-6pt{
      E^0(\ast)
      \ar[r]
        _-{
          \simeq
        }
      &
        \widetilde E{}^{0}(S^0)
      \ar[r]
        _-{
          \simeq
        }
        ^-{\sigma^E}
      &
      \widetilde E{}^{n}(S^n)
      \\
      \underset{
        \mathrlap{
        \;\;\;\;\;\;\;\;\;\;\;\;\;\;\;\;
        \raisebox{-8pt}{
          \tiny
          \color{darkblue}
          \bf
          \begin{tabular}{l}
            suspension isomorphism in
            \\
            multiplicative $E$-cohomology induces...
          \end{tabular}
        }
        }
      }{
        1^E
      }
      \ar@{|->}[rr]
      &&
      \underset{
        \mathrlap{
        \raisebox{-3pt}{
          \tiny
          \color{darkblue}
          \bf
          \begin{tabular}{c}
            ...canonical $E$-orientation class
            \\
            of Euclidean $n$-space/the $n$-sphere
          \end{tabular}
        }
        }
      }{
        [\mathrm{vol}^E_{S^n}]
        \coloneqq
        \Sigma^n(1^E)
      }
    }
  \end{equation}

\medskip
\noindent {\bf Orientation of parametrized Euclidean spaces.}
Given not just one, but a {\it bundle} of Euclidean spaces
over a parameter space $X$, hence a $V^n$-fiber bundle
$\mathcal{V}_X \xrightarrow{\;\;} X$, then a choice
of $E$-orientations of each fiber $\mathcal{V}_{\{x\}} \simeq V^n$, continuously varying in the
parameter $x \in X$, is naturally defined
to be a single class in the one-point compactification of the bundle
-- its {\it Thom space}
 $\mathrm{Th}\big( \mathcal{V}_X \big) = \big( \mathcal{V}_X \big)_{\mathrm{cpt}}$
 if $X$ is compact, which we shall assume --
 whose restriction to any fiber $\mathcal{V}_{\{x\}}$
 is an $E$-orientation \eqref{OrientingThenSphereInGeneralizedCohomology}
 of that Euclidean space:
 \vspace{-2mm}
\begin{equation}
  \label{FiberwiseOrientationOfBundleOfEuclideanSpaces}
  \xymatrix@R=3pt@C=2em{
    \mathcal{V}_X
    \ar@{<-^{)}}[rr]
      ^-{ \;i_x\; }
    &&
    \;
    \mathcal{V}_{\{x\}}
    \ar@{<-}[r]
      ^-{
        \simeq
      }
    &
    V^n
    \\
    \widetilde E{}^\bullet
 \big(         \big(
        \mathcal{V}_X
      \big)_{\mathrm{cpt}}
            \big)
     \ar[rrr]
      ^-{
        \;
        (i_x)^\ast
        \;
      }
    &&&
    \widetilde E{}^\bullet
    \big(
      (V^n)_{\mathrm{cpt}}
    \big)
    \\
    \mathllap{
      \mbox{
        \tiny
        \begin{tabular}{c}
          \color{darkblue}
          \bf
          $X$-parametrized $E$-orientation class
          \\
          = ``Thom class'' in $E$-cohomology
        \end{tabular}
      }
    }
    [\mathrm{vol}^E_{\mathcal{V}_X}]
    \ar@{|->}[rrr]
      ^-{
        \mbox{
          \tiny
          \color{greenii}
          \bf
          restriction to fiber
        }
      }
    && &
    [\mathrm{vol}^E_{S^n}]
    \mathrlap{
      \mbox{
        \tiny
        \color{darkblue}
        \bf
        \begin{tabular}{c}
          $E$-orientation class
          \\
          of Euclidean $n$-space/the $n$-sphere
        \end{tabular}
      }
    }
  }
\end{equation}

\vspace{-2mm}
\noindent Notice here, by the homotopy invariance of
Whitehead-generalized cohomology theory,
that the fiber restrictions to any pair of
points $x,y \in X$ in the same connected component
are isomorphic
\vspace{-2mm}
$$
  [x] = [y] \,\in\, \pi_0(X)
  \phantom{AAAA}
    \Rightarrow
  \phantom{AAAA}
  \xymatrix{
    {\widetilde E}^\bullet(\mathcal{V}_{\{y\}})
    \ar[r]
      ^-{
        \simeq
      }
    &
    {\widetilde E}^\bullet(\mathcal{V}_{\{x\}})
    \,,
  }
$$

\vspace{-2mm}
\noindent
and {\it uniquely} isomorphic if $X$ is simply connected,
$\pi_1(X) = 1$.

\medskip
Therefore, over connected
and simply-connected base spaces $X$,
it is sufficient to ask that $[\mathrm{vol}^E_{\mathcal{V}_X}]$
restricts to an $E$-orientation class on any one fiber;
and without restriction of generality, this may be taken
to be the canonical $E$-orientation \eqref{CanonicalEOrientationOnEuclideanSpace}
induced by the suspension isomorphism provided with $E$.

\medskip

\noindent {\bf Orientation of parametrized vector spaces.}
If we equip Euclidean space $V^n$ with the structure of
a $\mathbb{K}$-vector space, and consider
parametrizations
that respect this linear structure, then we are dealing,
of course, with fiberwise $E$-orientation of
$\mathbb{K}$-{\it vector bundles} $\mathcal{V}_X$.
A key point of the theory of $E$-orientations is that, in this
case, the orientation classes/Thom classes
\eqref{FiberwiseOrientationOfBundleOfEuclideanSpaces}
on $\mathcal{V}_X$
pull back, along the zero-section, to {\it characteristic classes}
on the base, and in fact bijectively so, if the resulting system of
characteristic classes is characterized appropriately:

\medskip
In the case where $E = H\mathbb{Z}$
is ordinary integral cohomology and $\mathbb{K} = \mathbb{C}$
is the complex numbers, this is just the classical fact
that top-degree {\it Chern classes} are equivalently the
pullback of any orientation class/Thom class along the zero-section.
The generalization of this statement to Whitehead-generalized
cohomology theories $E$ is the theory of
{\it Conner-Floyd $E$-chern classes}
(see Prop. \ref{ConnerFloydChernClasses} below):
$$
  \xymatrix@R=8pt{
    \mathllap{
      \mbox{
        \tiny
        \color{darkblue}
        \bf
        \begin{tabular}{c}
          complex
          \\
          vector bundle
        \end{tabular}
      }
    }
    \mathcal{V}_X
    \ar[dd]
    &&
    \widetilde E{}^\bullet
    \big(
      \mathcal{V}_X
    \big)
    \ar[dd]
      _-{
        \mbox{
          \tiny
          \color{greenii}
          \bf
          \begin{tabular}{c}
            pullback
            \\
            along 0-section
          \end{tabular}
        }
      }
    \ar@{}[r]
      |-{
       \ni
      }
    &
    \big[\mathrm{vol}^E_{\mathcal{V}_X}\big]
    \mathrlap{
      \mbox{
        \tiny
        \begin{tabular}{c}
          \color{darkblue}
          \bf
          $X$-parametrized $E$-orientation class
          \\
          = ``Thom class'' in $E$-cohomology
        \end{tabular}
      }
    }
    \ar@{<->}[dd]
    \\
    \\
    X
    &&
    E{}^\bullet
    \big(
      X
    \big)
    \ar@{}[r]
      |-{ \ni }
    &
    c^E_{\mathrm{rnk}(\mathcal{V}_X)}
    \mathrlap{
      \mbox{
        \tiny
        \color{darkblue}
        \bf
        \begin{tabular}{c}
          top-degree
          \\
          Conner-Floyd $E$-Chern class
        \end{tabular}
      }
    }
  }
$$

\vspace{-2mm}
\noindent
It is this equivalence between
{\bf (a)} universal fiberwise $E$-orientations of complex vector bundles
and
{\bf (b)} universal $E$-Chern classes
that makes
``complex oriented $E$-cohomology theory''
often look more like ``Conner-Floyd $E$-Chern class theory''.
It is from the latter perspective that
the notion naturally emerges from {\it Hypothesis H} in
\cref{VicinityOfBlackM2BranesAndOrientedCohomology}.

\medskip

We now turn to formal discussion of these matters.

\medskip

\noindent {\bf Projective spaces.} To set up notation
and to be fully explicit about some fine-print needed later on,
we briefly recall projective spaces and their
tautological line bundles over
topological skew-fields, i.e. including the
non-commutative quaternionic case, which requires
some care and leads to the all-important effect of
Prop. \ref{MapsFromCPTpHPConvergeToSkewDiagonalMap} below.

\begin{notation}[Ground fields]
 \label{GroundFields}
We write $\mathbb{K} \in \{\mathbb{R}, \mathbb{C}, \mathbb{H}\}$
for the topological (skew-)fields of real numbers, complex numbers
or quaternions, respectively.

\noindent {\bf (i)} We write
$$
  \mathbb{K}^\times
  \;\coloneqq\;
  \big(\mathbb{K}, \cdot \big) \setminus \{0\}
  \;\;\;
  \in
  \;
  \mathrm{Groups}
$$
for their multiplicative groups (their {\it groups of units}).

\noindent {\bf (ii)} As topological groups these are homotopy equivalent to
the groups of unit-norm elements
$$
  S(\mathbb{H})
  \;\coloneqq\;
  \big\{
    v \in \mathbb{H}
    \,\big\vert\,
    v \cdot v^\ast = 1
  \big\}
  \;\;\;
  \in
  \;
  \mathrm{Groups}
  \,,
$$
which, specifically, are these compact Lie groups:
\vspace{-2mm}
$$
  S(\mathbb{R})
  \;=\;
  \mathbb{Z}/2
  \,,
  \phantom{AAA}
  S(\mathbb{C})
  \;=\;
  \mathrm{U}(1)
  \,,
  \phantom{AAA}
  S(\mathbb{H})
  \;=\;
  \mathrm{Sp}(1)
    \,\simeq\,
  \mathrm{SU}(2)
  \,.
$$

\vspace{-2mm}
\noindent {\bf (iii)} In particular, their classifying spaces are
weakly homotopy equivalent:
\vspace{-2mm}
\begin{equation}
  \label{ClassifyingSpaceOfKtimesIsThatOfSK}
  B \mathbb{K}^\times
  \;\simeq\;
  B\big( S(\mathbb{K}) \big)
  \,.
\end{equation}
\end{notation}

\begin{notation}[Projective spaces and their tautological line bundles]
  \label{ProjectiveSpaces}
  For $\mathbb{K} \,\in\, \{\mathbb{R}, \mathbb{C}, \mathbb{H}\}$
  and $n \in \mathbb{N}$ we have:

  {\bf (i)}
   the {\it $\mathbb{K}$-projective space}:

  \vspace{-.4cm}
  \begin{equation}
    \label{KProjeciveSpace}
    \begin{aligned}
    \mathbb{K}P^n
      \,\coloneqq\,
    P \big(\mathbb{K}^{n+1}\big)
    & \coloneqq\,
    \big(
      \mathbb{K}^{n+1}
      \setminus
      \{0\}
    \big)
    \big/
    \mathbb{K}^\times
    \end{aligned}
  \end{equation}

\hspace{-.4cm}
\begin{tabular}{l|l}
  {\bf (ii)}
   its {\it tautological $\mathbb{K}$-line bundle}:
  &
  {\bf (iii)}
  its {\it dual tautological $\mathbb{K}$-line bundle}:
  \\
  \begin{minipage}[left]{8cm}
  \begin{equation}
    \label{TheTautologicalKLineBundle}
    \xymatrix@C=8pt@R=16pt{
      \mathcal{L}_{{}_{\mathbb{K}P^n}}
      \ar[d]
      \ar@{}[r]|-{ := }
      &
      \frac{
        \big(
          \mathbb{K}P^{n+1}
          \setminus
          \{0\}
        \big)
          \times
        \mathbb{K}^\ast
      }
      {
        \mathclap{\phantom{\vert^{\vert}}}
        \mathbb{K}^\times
      }
      \ar[d]^-{ [v,\,t] \mapsto [v] }
      \ar[rr]^-{
        [v,\,t]
        \;\mapsto\;
        v \cdot t
      }
      &
      {\phantom{AAAA}}
      &
      \mathbb{K}^{n+1}
      \\
      \mathbb{K}P^n
      \ar@{}[r]|-{ = }
      &
      \frac{
        \big(
          \mathbb{K}P^{n+1}
          \setminus
          \{0\}
        \big)
          \times
        \ast
      }
      {
        \mathclap{\phantom{\vert^{\vert}}}
        {\mathbb{K}}^\times
      }
    }
  \end{equation}
  \end{minipage}
  \hspace{.2cm}
  &
  \hspace{.2cm}
  \begin{minipage}[left]{8cm}
 \begin{equation}
    \label{TheDualTautologicalKLineBundle}
    \xymatrix@C=8pt@R=16pt{
      \mathcal{L}^\ast_{{}_{\mathbb{K}P^n}}
      \ar[d]
      \ar@{}[r]|-{ := }
      &
      \frac{
        \big(
          \mathbb{K}P^{n+1}
          \setminus
          \{0\}
        \big)
          \times
        \mathbb{K}
      }
      {
        \mathclap{\phantom{\vert^{\vert}}}
        \mathbb{K}^\times
      }
      \ar[d]^-{ [v,\,t] \mapsto [v] }
      \ar[rr]^-{
        [v,\,t]
        \;\mapsto\;
        [
          (v,\, t)
        ]
      }
      &
      {\phantom{AAAA}}
      &
      \mathbb{K}P^{n+1}
      \\
      \mathbb{K}P^n
      \ar@{}[r]|-{ = }
      &
      \frac{
      \big(
        \mathbb{K}P^{n+1}
        \setminus
        \{0\}
      \big)
        \times
      \ast
      }
      {
        \mathclap{\phantom{\vert^{\vert}}}
        \mathbb{K}^\times
      }
    }
  \end{equation}
  \end{minipage}
\end{tabular}

\noindent Here:

\vspace{-.05cm}
\begin{enumerate}[{\bf (i)}]

\vspace{-.2cm}
\item we write elements of $\mathbb{K}^{r+1}$
  as lists $w = (v, v_{r+1} ) = (v_1, \cdots, v_r, v_{r+1})$;

\vspace{-.3cm}
\item
  \raisebox{-6pt}{
  \hspace{-.3cm}
  \begin{tabular}{ll}
    \begin{minipage}[left]{8cm}
      we regard $\mathbb{K}^{r}$ as equipped with
      the right $\mathbb{K}^\times$-action by multiplication from
      the {\it right}
    \end{minipage}
  &
  \begin{minipage}[left]{8cm}
  \begin{equation}
    \label{RightActionOnKVectorSpace}
    \raisebox{12pt}{
    \xymatrix@R=-3pt{
      \mathbb{K}^r \times \mathbb{K}^\times
      \ar[r]
      &
      \mathbb{K}^r
      \\
      \big(
        (v_1, \cdots, v_r)
        ,\,
        q
      \big)
      \ar@{}[r]|-{\longmapsto}
      &
      (v_1 \!\cdot\! q ,\, \cdots , v_r \!\cdot\! q )
    }
    }
  \end{equation}
  \end{minipage}
  \end{tabular}
  }

\vspace{-.2cm}
\item
  \raisebox{-6pt}{
  \hspace{-.3cm}
  \begin{tabular}{ll}
    \begin{minipage}[left]{8cm}
      in contrast,
      $\mathbb{K}^\ast$ denotes $\mathbb{K}$ equipped with the
      right
      $\mathbb{K}^\times$-action by
      {\it inverse multiplication} from the {\it left};
    \end{minipage}
  &
  \begin{minipage}[left]{8cm}
  \begin{equation}
    \label{RightActionOnKVectorSpace}
    \raisebox{12pt}{
    \xymatrix@R=-3pt{
      \mathbb{K}^\ast \times \mathbb{K}^\times
      \ar[r]
      &
      \mathbb{K}^\ast
      \\
      \big(
        t
        ,\,
        q
      \big)
      \ar@{}[r]|-{\longmapsto}
      &
      q^{-1} \!\cdot\! t
    }
    }
  \end{equation}
  \end{minipage}
  \end{tabular}
  }

\vspace{-.2cm}
\item
  $
    \frac{
      (-)\times(-)
    }
    {
      \mathclap{\phantom{\vert^{\vert}}}
      \mathbb{K}^\times
    }
  $
  denotes the quotient
space of a product of right $k^\times$-spaces
by their diagonal action, and $[-]$ denotes its elements
as equivalence classes of those of the original space
(so $[(v_1, \cdots, v_n)] = [v_1 : \cdots : v_n]$ ).

\end{enumerate}
\vspace{-.1cm}

\noindent These quotient spaces become $\mathbb{K}$-vector bundles
(line bundles) with respect to the remaining left actions:
For $\mathcal{L}^\ast_{\mathbb{K}P^n}$
\eqref{TheDualTautologicalKLineBundle}
this is by left multiplication on $\mathbb{K}$,
but for $\mathcal{L}_{\mathbb{K}P^n}$
\eqref{TheTautologicalKLineBundle} this is by
{\it conjugate} multiplication from the {\it right}:
\begin{equation}
  \label{LeftActionOnTautologicalLineBundle}
  \raisebox{12pt}{
  \xymatrix@R=-2pt{
    \mathbb{K}^\times
    \times
    \mathcal{L}_{{}_{\mathbb{K}P^n}}
    \ar[rr]
    &&
    \mathcal{L}_{{}_{\mathbb{K}P^n}}
    \\
    \big(
      q, \,
      [v, t]
    \big)
    \ar[rr]
    &&
    [ v,\, t \!\cdot\! q^\ast ]
  }
  }
  \phantom{AAAAA}
  \raisebox{12pt}{
  \xymatrix@R=-2pt{
    \mathbb{K}^\times
    \times
    \mathcal{L}^\ast_{{}_{\mathbb{K}P^n}}
    \ar[rr]
    &&
    \mathcal{L}^\ast_{{}_{\mathbb{K}P^n}}
    \\
    \big(
      q, \,
      [v, t]
    \big)
    \ar[rr]
    &&
    [ v,\, q \!\cdot\!  t ]
    \,.
  }
  }
\end{equation}

While the horizontal map in \eqref{TheTautologicalKLineBundle}
exhibits the tautological line bundle as the ``blow-up''
of the origin in $\mathbb{K}^{n+1}$, the horizontal map in
\eqref{TheDualTautologicalKLineBundle} embeds the
dual tautological line bundle into $\mathbb{K}P^{n+1}$,
as the complement of the single remaining point
$[(v,\infty)] \coloneqq [(0,1)]$.
Hence the embedding \eqref{TheDualTautologicalKLineBundle}
extends over this point to become a homeomorphism
between the Thom space (Example \ref{SpheresAndThomSpacesAsOnePointCompactifications})
of the dual tautological line bundle
and the next projective space
\vspace{-2mm}
\begin{equation}
  \label{ThomSpaceOfDualTautologicalLineBundleIsNextProjectiveSpace}
  \xymatrix@C=3em{
    \underset{
      \mathclap{
      \raisebox{-6pt}{
        \tiny
        \color{darkblue}
        \bf
        \begin{tabular}{c}
          projective $n$-space
        \end{tabular}
      }
      }
    }{
      \mathbb{K}P^n
    }
    \;
    \ar@{^{(}->}[rr]
      ^-{
        [v]
        \,\mapsto\,
        [(v,0)]
        \mathclap{\phantom{\vert_{\vert}}}
      }
      _-{
        \mathclap{\phantom{\vert^{\vert^{\vert}}}}
        \mbox{
          \tiny
          \color{greenii}
          \bf
          \begin{tabular}{c}
            zero-section
          \end{tabular}
        }
      }
    &{\phantom{AAAAA}}&
    \underset{
      \mathclap{
      \raisebox{-3pt}{
        \tiny
        \color{darkblue}
        \bf
        \begin{tabular}{c}
          Thom space of
          \\
          tautological line bundle
          \\
          over projective $n$-space
        \end{tabular}
      }
      }
    }{
    \mathrm{Th}
    \left(
      \mathcal{L}^\ast_{{}_{\mathbb{K}P^n}}
    \right)
    }
    \ar[rrr]
      _-{
        \underset{
          \raisebox{-3pt}{
            \tiny
            \color{greenii}
            \bf
            homeomorphism
          }
        }{
          \simeq
        }
      }
      ^-{
      [v,\, t]
      \,\mapsto\,
      \left\{
      \scalebox{.7}{$
        {\begin{array}{lcl}
          {[(0,1)]}
          &\vert& t = \infty
          \\
          {[(v,\,t)]}
          &\vert&
          \mbox{else}
        \end{array}}
      $}
      \right.
    }
    &{\phantom{AAAAA}}&&
    \underset{
      \mathclap{
      \raisebox{-6pt}{
        \tiny
        \color{darkblue}
        \bf
        \begin{tabular}{c}
          projective $(n+1)$-space
        \end{tabular}
      }
      }
    }{
      \mathbb{K}P^{n+1}
    }
  }
\end{equation}

\vspace{-2mm}
\noindent (compare to \cite[\S III, Lemma 3.8]{TamakiKono06}, where
the dualization $\mathcal{L}^\ast$ is actually implicit in the use of an inner product).

  \noindent
  The colimit (in topological spaces)
  over the sequence of inclusions
  of projective spaces
  (the coordinate ordering convention here is crucial for \eqref{DiagramOfProjectiveSpacesThomSpacesAndZeroSections} below):
  \vspace{-5mm}
  \begin{equation}
    \label{InclusionsOfProjectiveSpacesByAdjoiningZeroFromTheLeft}
    \underset{
      \longrightarrow
    }{\lim}
    \left(
    \raisebox{32pt}{
    \xymatrix@C=15pt@R=-1pt{
      &
    \scalebox{0.7}{$  [v,\,t]$}
      \ar@{}[r]|-{\longmapsto}
      &
      \scalebox{0.7}{$  [(0,v),\, t]$}
      \\
      \cdots
      \;
      \ar@{^{(}->}[r]
      &
     \;\; \mathcal{L}^\ast_{\mathbb{K}P^n}
      \;\;
      \ar[dd]
      \ar@{^{(}->}[r]
      &
      \;\;
      \mathcal{L}^\ast_{\mathbb{K}P^n}
      \;\;
      \ar[dd]
      \ar@{^{(}->}[r]
      &
      \cdots
      \\
      {\phantom{A}}
      \\
      \cdots
      \;
      \ar@{^{(}->}[r]
      &
      \;\; \mathbb{K}P^n \;\ar@{^{(}->}[r] & \mathbb{K}P^{n+1}
     \;\; \ar@{^{(}->}[r]
      &
      \cdots
      \\
      &
        \scalebox{0.7}{$[v]$}
      \ar@{}[r]|-{\longmapsto}
      &
       \scalebox{0.7}{$  [(0,\,v)]$}
    }
    }
    \right)
    \;\;\;
      \simeq
    \;\;\;
    \raisebox{58pt}{
    \xymatrix@C=9pt@R=4pt{
      &
      {\phantom{
        [v,\,t]
      }}
      &
      \\
      \mathcal{L}^\ast_{\mathbb{K}P^\infty}
      \ar[dd]
      \ar@{=}[r]
      &
      \overset{
        \mathllap{
        \raisebox{6pt}{
          \tiny
          \color{darkblue}
          \bf
          \begin{tabular}{c}
            universal
            $\mathbb{K}$-line bundle
          \end{tabular}
        }
        }
      }{
        E \mathbb{K}^\times
          \!
          \underset{
            \scalebox{.6}{$\mathbb{K}^{\mathrlap{\times}}$}
          }{\times}
        \mathbb{K}
      }
      \ar[dd]
      \\
      {\phantom{A}}
      \\
      \mathbb{K}P^\infty
      \ar@{=}[r]
      &
      B \mathbb{K}^\times
      \\
      &
      {\phantom{
        [v]
      }}
    }
    }
  \end{equation}

    \vspace{-7mm}
\noindent  yields the {\it universal $\mathbb{K}$-line bundle}
  over the
  infinite projective spaces,
  homotopy equivalent to the classifying spaces
  \eqref{ClassifyingSpaceOfKtimesIsThatOfSK}:
  \vspace{0mm}
  \begin{equation}
    \label{InfiniteProjectiveSpaces}
    \mathbb{K}P^\infty
    \;\simeq\;
    B\big( S(\mathbb{K})\big)
    \,,
    \;\;
    i.e.:
    \phantom{AAAAA}
    \mathbb{R}P^\infty
    \,\simeq\,
    B \mathbb{Z}_2
    \,,
    \phantom{AAA}
    \mathbb{C}P^\infty
    \,\simeq\,
    B \mathrm{U}(1)
    \,,
    \phantom{AAA}
    \mathbb{H}P^\infty
    \,\simeq\,
    B \mathrm{SU}(2)
    \,.
  \end{equation}

  In summary, there is the following commuting diagram
  of
  sequences of projective spaces,
  their dual tautological line bundles
  and their Thom spaces, whose colimit
  is the {\it universal $\mathbb{K}$-line bundle}
  over the classifying space:
\begin{equation}
  \label{DiagramOfProjectiveSpacesThomSpacesAndZeroSections}
  \phantom{S^{\mathrm{dim}_{{}_{\mathbb{R}}}\!\!(\mathbb{K})}}
  \raisebox{48pt}{
  \xymatrix@C=20pt@R=8pt{
    \mathllap{
      S^{\mathrm{dim}_{{}_{\mathbb{R}}}\!\!(\mathbb{K})}
      =
      \;\;
    }
    \mathbb{K}P^1
    \ar@{^{(}->}[r]
    &
    \mathbb{K}P^2
    \ar@{=}[d]
    \ar@{^{(}->}[r]
    &
    \;\;
    \cdots
    \;\;
    \ar@{^{(}->}[r]
    &
    \mathbb{K}P^{n+1}
    \ar@{=}[d]
    \;
    \ar@{^{(}->}[rr]
      |-{
        \;
        \scalebox{.76}{
        \raisebox{-5pt}{$
          \cdots
        $}
        }
      \;}
    &
    &
    \mathbb{K}P^{\infty}
    \ar@{=}[d]
    \\
    &
    \mathrm{Th}
    \big(
      \mathcal{L}^\ast_{\mathbb{K}P^1}
    \big)
    \ar@{^{(}->}[r]
    &
    \;\;
    \cdots
    \;\;
    \ar@{^{(}->}[r]
    &
    \mathrm{Th}
    \big(
      \mathcal{L}^\ast_{\mathbb{K}P^n}
    \big)
    \;
    \ar@{^{(}->}[rr]
      |-{
        \;
        \scalebox{.76}{
        \raisebox{-5pt}{$
          \cdots
        $}
        }
      \;}
    &
    &
    \mathrm{Th}
    \big(
      \mathcal{L}^\ast_{\mathbb{K}P^\infty}
    \big)
    \ar@{=}[r]
    &
    \mathrm{Th}
    \Big(
      E \mathbb{K}^\times
      \!
      \underset{
        \scalebox{.6}{$\mathbb{K}^{\mathrlap{\times}}$}
      }{\times}
      \mathbb{K}
    \Big)
    \\
    \\
    &
    \mathllap{
      S^{\mathrm{dim}_{{}_{\mathbb{R}}}\!\!(\mathbb{K})}
      =
      \;\;
    }
    \mathbb{K}P^1
    \ar@{^{(}->}[r]
    \ar[uu]
     ^-{\scalebox{0.8}{$
       0_{{}_{
         \mathrm{Th}
         (
           \mathcal{L}^\ast_{\mathbb{K}P^1}
         )
       }}
       \mathclap{\phantom{\vert_{\vert_{\vert_{\vert}}}}}
       $}
     }
    &
    \;\;\cdots\;\;
    \ar@{^{(}->}[r]
    &
    \mathbb{K}P^{n}
    \ar[uu]
     ^-{\scalebox{0.8}{$
       0_{{}_{
         \mathrm{Th}
         (
           \mathcal{L}^\ast_{\mathbb{K}P^{\,n}}
         )
       }}
       \mathclap{\phantom{\vert_{\vert_{\vert_{\vert}}}}}
       $}
     }
    \;
    \ar@{^{(}->}[rr]
      ^-{
        \;
        \scalebox{.76}{
        \raisebox{-5pt}{$
          \cdots
        $}
        }
      \;}
    &
    &
    \mathbb{K}P^{\infty}
    \ar[uu]
     ^-{\scalebox{0.8}{$
       0_{{}_{
         \mathrm{Th}
         (
           \mathcal{L}^\ast_{\mathbb{K}P^{\, \infty}}
         )
       }}
       \mathclap{\phantom{\vert_{\vert_{\vert_{\vert}}}}}
       $}
     }
     \ar@{=}[r]
     &
     B \mathbb{K}^\times
     \ar[uu]
     ^-{
           \simeq
           }
           }
        }
\end{equation}

Notice that, in the above conventions,
the horizontal inclusions \eqref{InclusionsOfProjectiveSpacesByAdjoiningZeroFromTheLeft}
are by adjoining a zero-coordinate to the left of the list,
while the vertical inclusions, being the zero-sections
under the identification \eqref{ThomSpaceOfDualTautologicalLineBundleIsNextProjectiveSpace},
are by adjoining a zero-coordinate to the right of the list,
so that the squares in \eqref{DiagramOfProjectiveSpacesThomSpacesAndZeroSections}
indeed commute.
\end{notation}
\begin{lemma}[Cell structure of projective spaces]
  \label{CellStructureOfProjectiveSpaces}
  For $n \in \mathbb{N}$, the projective spaces
  (Notation \ref{ProjectiveSpaces}) are equivalently
  quotient spaces of $k$-spheres
  \begin{equation}
    \label{ProjectiveSpaceAsQuotientOfSpheres}
    \mathbb{R}P^n
    \;\simeq\;
    S^n / \mathbb{Z}_2
    \,,
    \;\;\;\;\;\;\;\;\;
    \mathbb{C}P^n
    \;\simeq\;
    S^{2n+1} / \mathrm{U}(1)
    \,,
    \;\;\;\;\;\;\;\;\;
    \mathbb{H}P^n
    \;\simeq\;
    S^{4n+3} / \mathrm{Sp}(1)
    \,,
  \end{equation}
  and are related by homotopy pushouts of the following form (cell attachments):
  \begin{equation}
    \label{CellStructureOfProjectiveSpace}
    \raisebox{20pt}{
    \xymatrix@C=32pt@R=11pt{
      S^n
      \ar[r]
      \ar[d]
      \ar@{}[dr]|-{ \mbox{\tiny\rm(po)} }
      &
      \mathbb{R}P^n
      \ar[d]
      \\
      \ast
      \ar[r]
      &
      \mathbb{R}P^{n+1}
    }
    }
    \phantom{AAA}
    \raisebox{20pt}{
    \xymatrix@C=32pt@R=11pt{
      S^{2n+1}
      \ar[r]
      \ar[d]
      \ar@{}[dr]|-{ \mbox{\tiny\rm(po)} }
      &
      \mathbb{C}P^n
      \ar[d]
      \\
      \ast
      \ar[r]
      &
      \mathbb{C}P^{n+1}
    }
    }
    \phantom{AAA}
    \raisebox{20pt}{
    \xymatrix@C=32pt@R=11pt{
      S^{4n+3}
      \ar[r]
      \ar[d]
      \ar@{}[dr]|-{ \mbox{\tiny\rm(po)} }
      &
      \mathbb{H}P^n
      \ar[d]
      \\
      \ast
      \ar[r]
      &
      \mathbb{H}P^{n+1}
      \,,
    }
    }
  \end{equation}
  where the top morphisms are the quotient projections
  from \eqref{ProjectiveSpaceAsQuotientOfSpheres}.
\end{lemma}
As indicated on the far right of
\eqref{DiagramOfProjectiveSpacesThomSpacesAndZeroSections}, we have the following basic fact, of crucial importance in orientation-theory  (e.g. \cite[\S I, Ex. 2.1]{Adams74}\cite[Lem. 2.6.5]{Kochman96}):
\begin{prop}[Zero-section into Thom space of universal $\mathbb{K}$-line
  bundle is weak equivalence]
  \label{ZeroSectionIntoThomSpaceOfUniversalLineBundleIsEquivalence}
  The 0-section into the Thom space of the universal
  $\mathbb{K}$-line bundle, on the far right of \eqref{DiagramOfProjectiveSpacesThomSpacesAndZeroSections},
  is a weak homotopy equivalence:
  $$
    \xymatrix{
      \mathllap{
        \mbox{
          \tiny
          \color{darkblue}
          \bf
          classifying space
        }
      }
      \;
      B \mathbb{K}^\times
      \ar[rr]
        ^-{ \simeq }
        _-{
        \mbox{
          \tiny
          \color{greenii}
          \bf
          zero-section
        }
      }
      &&
      \mathrm{Th}
      \big(
         \mathcal{L}^\ast_{\mathbb{K}P^\infty}
      \big)
      \mathrlap{
        \mbox{
          \tiny
          \color{darkblue}
          \bf
          \begin{tabular}{c}
            Thom space of
            \\
            universal $\mathbb{K}$-line bundle
          \end{tabular}
        }
      }
    }
  $$
\end{prop}
\noindent
\begin{proof}
  Observing that, in the present case, the sphere bundle
  is contractible:
  $$
    S\big( \mathcal{L}^\ast_{\mathbb{K}P^\infty} \big)
    \;\simeq\;
    S\big(
      E (S(\mathbb{K}))
        \underset{
          \mathclap{
            \scalebox{.6}{$S(\mathbb{K})$}
          }
        }{\times}
      \mathbb{K}
    \big)
    \;\simeq\;
      E (S(\mathbb{K}))
        \underset{
          \mathclap{
            \scalebox{.6}{$S(\mathbb{K})$}
          }
        }{\times}
      S(\mathbb{K})
    \;\simeq\;
    E (S(\mathbb{K}))
    \;\simeq\;
    \ast
    \,,
  $$
  and that the zero-section of the unit disk bundle is,
  manifestly and generally,  a homotopy equivalence
  $X \xrightarrow{\simeq} D(\mathcal{V}_X) $, we have
  the solid part of the following commuting diagram:
  $$
    \xymatrix@C=3em{
      \ast
      \;
      \ar@{^{(}->}[rr]
      \ar[d]
        _-{
             \simeq
        }
      &&
      D\big( \mathcal{L}^\ast_{\mathbb{K}P^\infty} \big)
      \ar[rr]
      \ar@{=}[d]
      &&
      D\big( \mathcal{L}^\ast_{\mathbb{K}P^\infty} \big)
      \ar@{-->}[d]
      \ar@{<-}[r]^-{ \;0\; }
      &
      B \mathbb{K}^\times
      \ar[dl]
      \\
      S\big( \mathcal{L}^\ast_{\mathbb{K}P^\infty} \big)
      \;
      \ar@{^{(}->}[rr]
        _-{ \in \mathrm{Cofibrations} }
      &&
      D\big( \mathcal{L}^\ast_{\mathbb{K}P^\infty} \big)
      \ar[rr]
        ^-{ \mbox{\tiny cofiber} }
        _-{ \mbox{\tiny $\Rightarrow$ homotopy cofiber} }
      &&
      \mathrm{Th}
      \big(
        \mathcal{L}^\ast_{\mathbb{K}P^\infty}
      \big)
    }
  $$
 Now observing that the inclusion of the sphere bundle
 into the disk bundle is a relative cell complex inclusion,
 (using that our projective spaces are CW-complexes,
 by Lemma \ref{CellStructureOfProjectiveSpaces}),
 so that its cofiber is in fact a model for its
 homotopy cofiber,
  the claim follows
  by the respect of homotopy cofibers for weak equivalences
  (see. e.g., \cite[Ex. A.24]{FSS20c}).
\end{proof}

\noindent
In parametrized generalization of Notation \ref{ProjectiveSpaces}
we have:
\begin{notation}[Projective bundles]
  \label{ProjectiveBundles}
  For $\mathbb{K} \,\in\, \{\mathbb{R}, \mathbb{C}, \mathbb{H}\}$
  and
  $\mathcal{V}_{{}_X} \,\in\, \mathbb{K}\mathrm{VectorBundles}_{/X}$,
  we have the {\it projective bundle}
  \begin{equation}
    \label{ProjectiveBundle}
    P
    \big(
      \mathcal{V}_{{}_X}
    \big)
    \;:=\;
    \Big(
      \mathcal{L}_{{}_X}
      \setminus
      \overset{
        \mathclap{
        \mbox{
          \tiny
          \color{darkblue}
          \bf
          0-section
        }
        }
      }{
        \overbrace{
          X \times \{0\}
        }
      }
    \Big)
    /
    \mathbb{K}^\times
    \,,
  \end{equation}
  with respect to the evident projection map to $X$.
  This is a fiber bundle with typical fiber
  $\mathbb{K}P^{r-1}$, where
  $r := \mathrm{rnk}_{\mathbb{K}}\big( \mathbb{V}_{{}_X}\big)$
\end{notation}

\medskip

\noindent {\bf Relating complex-projective to quaternionic-projective spaces.}
Fixing an orthonormal basis for the quaternions
we have an induced fixed star-algebra inclusion
of the complex numbers, and hence a real-linear identification
of the quaternions with two copies of the complex numbers:
\begin{equation}
  \label{IdentifyingHWithCPlusCj}
  \raisebox{15pt}{
  \xymatrix@R=7pt{
    \mathbb{R}
    \big\langle
      1,\, \mathrm{i}
    \big\rangle
    \ar@{=}[d]
    \;
    \ar@{^{(}->}[rr]
    &&
    \mathbb{R}
    \big\langle
      1,\,
      \mathrm{i}
      ,\,
      \mathrm{j}
      ,\,
      \mathrm{k}
    \big\rangle
    \ar@{=}[d]
    \\
    \mathbb{C}
    \;
    \ar@{^{(}->}[rr]
    &&
    \mathbb{H}
  }
  }
  \;\;\;\;\;\;\;\;
  \Rightarrow
  \;\;\;\;\;\;\;\;
  \mathbb{H}
  \;\;
  \simeq_{{}_{\mathbb{R}}}
  \;\;
  \mathbb{C}
  \;\oplus\;
  \mathrm{j}
    \cdot
  \mathbb{C}
\end{equation}

With this choice of ordering (the factor of $\mathrm{j}$ being on the
left) the multiplication action of
$
  \xymatrix@C=6pt{
    \mathbb{C}
    \;
    \ar@{^{(}->}[r]
    &
    \mathbb{H}
  }
$
on $\mathbb{H}$ leads to the following $\mathrm{U}(1)$-module structures
(since $z \cdot \mathrm{j} = \mathrm{j} \cdot z^\ast$
for $z \in \mathbb{C} \hookrightarrow \mathbb{H}$ via \eqref{IdentifyingHWithCPlusCj}):
\begin{equation}
  \label{IdentificationOfHAsLeftRightCircleGroupModule}
  \mathbb{H}
  \;\simeq_{{}_{\mathbb{C}}}\;
  \mathbb{C} \oplus \mathbb{C}^\ast
  \;
  \in
  \;
  \mathrm{U}(1)\mathrm{LeftModules}
  \,,
  \phantom{AAAAA}
  \mathbb{H}
  \;\simeq_{{}_{\mathbb{C}}}\;
  \mathbb{C} \oplus \mathbb{C}
  \;\;
  \in
  \;
  \mathrm{U}(1)\mathrm{RightModules}
  \,,
\end{equation}
where the right action is the canonical one while the
{\it left} action is the sum of the canonical
one and its dual. In particular the quotient
of $\mathbb{H}^\times$ by the right $\mathbb{C}^\times$ action,
under \eqref{IdentifyingHWithCPlusCj}, is
$
  \mathbb{H}^\times / \mathbb{C}^\times
  \;=\;
  \mathbb{C}P^1
  \,.
$
Hence:
\begin{remark}[Complex projective spaces over Quaternionic-projective spaces]
\label{ComplexProjectiveSpacesOverQuaternionicProjectiveSpaces}
For each $n \in \mathbb{N}$, quotienting along the
inclusion \eqref{IdentifyingHWithCPlusCj}
yields $S^2$-fibration
of complex projective spaces over quaternionic projective spaces
(Notation \ref{ProjectiveSpaces}) of this form:
\begin{equation}
  \label{S2FibrationOfComplexProjectiveSpaceOverQuaternionicProjective}
  \raisebox{40pt}{
  \xymatrix@R=5pt@C=3pt{
    S^2
    \ar@{}[rr]|-{=}
    \ar[dr]
    &&
    \mathbb{H}^\times / \mathbb{C}^\times
    \ar[dr]
    \\
    &
    \mathbb{C}P^{2n + 1}
    \ar@{}[r]|-{=}
    \ar[dd]
    &&
    \;\;\;
    \mathclap{
    \phantom{\vert^{\vert^{\vert^{\vert}}}}
    \big(
      \mathbb{C}^{2n + 2} \!\setminus\! \{0\}
    \big)
      \big/
    \mathbb{C}^\times
    }
    \;\;\;
    \ar[dd]
    \ar@{}[rr]|>>{\ni\;}
    &&
    v \cdot \mathbb{C}^\times
    \ar@{|->}@<-10pt>[dd]
    \\
    \\
    &
    \mathbb{H}P^n
    \ar@{}[r]|-{=}
    &&
    \;\;\;
    \mathclap{
    \big(
      \mathbb{H}^{n + 1} \!\setminus\! \{0\}
    \big)
      \big/
    \mathbb{H}^\times
    }
    \;\;\;
    \ar@{}[rr]|>>>{\ni}
    &{\phantom{AAAAA}}&
    v \cdot \mathbb{H}^\times
  }
  }
\end{equation}

\noindent For $n = 1$ this is also known as the
{\it twistor fibration} $t_{\mathbb{H}}$, see \cite[\S 2]{FSS20b} for pointers.
As $n$ varies, these form a sequence of $S^2$-fibrations:
\begin{equation}
  \label{TowerOfS2FibrationsConvergingTpBS1OverBSU2}
  \hspace{-5mm}
  \raisebox{22pt}{
  \xymatrix@C=1.8em{
    \mathbb{C}P^1
    \;\ar@{^{(}->}[r]
    \ar[d]
    &
    \mathbb{C}P^2
    \;\ar@{^{(}->}[r]
    &
    \mathbb{C}P^3
    \ar[d]
      ^-{
        \mathclap{\phantom{\vert^\vert}}
        t_{\mathbb{H}}
        \mathclap{\phantom{\vert_\vert}}
      }
    \;\ar@{^{(}->}[r]
    &
    \mathbb{C}P^4
    \;\ar@{^{(}->}[r]
    &
    \mathbb{C}P^5
    \;\ar@{^{(}->}[r]
    \ar[d]
    &
    \mathbb{C}P^6
    \;\ar@{^{(}->}[r]
    &
    \mathbb{C}P^7
    \ar[d]
    \;\ar@{^{(}.>}[rr]
    &
    &
    \mathbb{C}P^\infty
    \ar@{}[r]|-{\simeq}
    \ar[d]
      _-{
        \scalebox{.7}{$
          \def\arraystretch{.75}
          \begin{array}{c}
            \mathcal{L}
            \\
            \mapsdown
            \\
            \mathcal{L} \oplus \mathcal{L}^\ast
          \end{array}
        $}
        \!\!
      }
    &
    B \mathrm{U}(1)
    \ar[d]
      |<<<<<{
        \mathclap{\phantom{\vert^{\vert}}}
        B
        (
          z \,\mapsto \mathrm{diag}(z,z^\ast)
        )
        \mathclap{\phantom{\vert_{\vert}}}
      }
    \\
    \mathbb{H}P^0
    \;\ar@{^{(}->}[rr]
    &&
    \mathbb{H}P^1
    \;\ar@{^{(}->}[rr]
    &&
    \mathbb{H}P^2
    \;\ar@{^{(}->}[rr]
    &&
    \mathbb{H}P^3
    \;\ar@{^{(}.>}[rr]
    &
    &
    \mathbb{H}P^\infty
    \ar@{}[r]|-{\simeq}
    &
    B \mathrm{SU}(2)
  }
  }
\end{equation}

\end{remark}

\begin{prop}[Maps from $\mathbb{C}P$ to $\mathbb{H}P$ converge to skew diagonal map]
  \label{MapsFromCPTpHPConvergeToSkewDiagonalMap}
  In the colimit $n \to \infty$,
  the morphism of classifying spaces on the right of
  \eqref{TowerOfS2FibrationsConvergingTpBS1OverBSU2}
  is  that induced from the group inclusion
  \raisebox{2.3pt}{
  \xymatrix@C=2em{
    \mathrm{U}(1)\;
    \ar@{^{(}->}[rr]
      ^-{
        \; z \,\mapsto\, \mathrm{diag}(z,z^\ast) \;
      }
    &{\phantom{AAA}}&
    \mathrm{SU}(2)
  }},
  as indicated.
\end{prop}
\begin{proof}
  The claim is equivalent to the statement that the
  complex vector bundle underlying the
  dual tautological quaternionic line bundle
  \eqref{TheTautologicalKLineBundle}
  pulls back to
  the Whitney sum of the
  tautological complex line bundle \eqref{TheTautologicalKLineBundle}
  with its dual \eqref{TheDualTautologicalKLineBundle},
  hence that we have a pullback diagram of spaces
  \vspace{-3mm}
  $$
    \raisebox{20pt}{
    \xymatrix@R=4pt@C=4em{
      \mathcal{L}^\ast_{\mathbb{C}P^n}
      \oplus_{{}_{\mathbb{C}P^n}}
      \mathcal{L}_{\mathbb{C}P^n}
      \ar[dd]
      \ar[rr]
        ^-{
        \mbox{
          \tiny
          \begin{tabular}{c}
            fiberwise
            \\
            left $\mathbb{C}$-linear isomorphism
          \end{tabular}
        }
      }
      \ar@{}[ddrr]|-{\mbox{\tiny(pb)}}
      &&
      \mathcal{L}^\ast_{\mathbb{H}P^n}
      \ar[dd]
      \\
      {\phantom{AA}}
      \\
      \mathbb{C}P^n
      \ar[rr]
      &&
      \mathbb{H}P^n
      \\
      v \cdot \mathbb{C}^\times
      \ar@{}[rr]|-{\longmapsto}
      &&
      v \cdot \mathbb{H}^\times
    }
    }
    \phantom{AAA}
      \Leftrightarrow
    \phantom{AAA}
      \underset{v}{\forall}
    \raisebox{22pt}{
    \xymatrix@C=3em{
      \mathbb{C}^\ast
        \oplus
      \mathbb{C}
      \ar[d]
      \ar[rr]
        ^-{
        \mbox{
          \tiny
          \begin{tabular}{c}
            left $\mathbb{C}$-linear isomorphism
          \end{tabular}
        }
      }
      &&
      \mathbb{H}
      \ar[d]
      \\
      \big\{
        v \cdot \mathbb{C}^\times
      \big\}
      \ar[rr]
      &&
      \big\{
        v \cdot \mathbb{H}^\times
      \big\}
    }
    }
  $$

    \vspace{-2mm}
\noindent  such that the top morphism is fiberwise complex linear
  for the {\it left} $\mathbb{C}$-actions \eqref{LeftActionOnTautologicalLineBundle}.
  On the typical fiber this means equivalently  that
  $\mathbb{H}
    \,\simeq_{{}_{\mathbb{C}}}\,
    \mathbb{C}
      \oplus
    \mathbb{C}^\ast
  $, which is the case by \eqref{IdentificationOfHAsLeftRightCircleGroupModule}.
\end{proof}

\medskip

\noindent
{\bf Bounded-dimensional complex and quaternionic orientation in $E$-cohomology.}

\begin{lemma}[Bounded-dimensional universal $\mathbb{K}$-line bundles]
  \label{UniversalFiniteDimensionalLineBundles}
  Let $n \in \mathbb{N}$.

  \noindent
  {\bf (a) } The dual tautological $\mathbb{C}$-line bundle
  $\mathcal{L}^\ast_{\mathbb{C}P^n}$
  \eqref{TheTautologicalKLineBundle} over
  $\mathbb{C}P^n$ is universal for complex line bundles
  over $d \leq 2n$-dimensional manifolds $X^{2n}$
  (more generally: over $\leq 2n$-dimensional cell complexes)
  in that every such has a classifying map
  that factors through $\mathbb{C}P^n \hookrightarrow
  \mathbb{C}P^\infty = B \mathrm{U}(1)$ and is
  hence isomorphic to a pullback of
  $\mathcal{L}^\ast_{\mathbb{C}P^n}$:
  \vspace{-2.5mm}

  $$
    \xymatrix{
      \mathllap{
        \mbox{
          \tiny
          \color{darkblue}
          \bf
          \begin{tabular}{c}
            complex line bundle
          \end{tabular}
        }
      }
      \mathcal{V}^\ast_X
      \ar[d]
      \ar[r]
      \ar@{}[dr]
        |-{\mbox{\tiny\rm(pb)}}
      &
      \mathcal{L}^\ast_{\mathbb{C}P^n}
      \ar[d]
      \ar[rr]
      \ar@{}[drr]
        |-{\mbox{\tiny\rm(pb)}}
      &&
      E \mathrm{U}(1)
        \underset{\mathrm{U}(1)}{\times}
      \mathbb{C}
      \ar[d]
      \\
      \mathllap{
        \mbox{
          \tiny
          \color{darkblue}
          \bf
          \begin{tabular}{c}
            base space of
            \\
            bounded dimension
            $\leq 2n$
          \end{tabular}
        }
      }
      X^{2n}
      \ar@/_1.3pc/[rrr]
        |-{
          \;
          \vdash \mathcal{V}_X
          \;
        }
      \ar@{-->}[r]
      &
      \mathbb{C}P^n
      \;
      \ar@{^{(}->}[r]
      &
      \mathbb{C}P^\infty
      \ar[r]
        ^-{
          \simeq
        }
      &
      B \mathrm{U}(1)
      \,.
    }
  $$

  \vspace{-.5mm}
  \noindent  and any two such factorizations are homotopic to each other.

  \noindent
  {\bf (b) } The dual tautological $\mathbb{H}$-line bundle
  $\mathcal{L}^\ast_{\mathbb{H}P^n}$
  \eqref{TheTautologicalKLineBundle} over
  $\mathbb{H}P^n$ is universal for quaternionic line bundles
  over $d \leq 4n+2$-dimensional manifolds $X^{4n+2}$
  (more generally: over $\leq 4n+2$-dimensional cell complexes)
  in that every such has a classifying map
  that factors through $\mathbb{H}P^n \hookrightarrow
  \mathbb{H}P^\infty = B \mathrm{Sp}(1)$ and is
  hence isomorphic to a pullback of
  $\mathcal{L}^\ast_{\mathbb{H}P^n}$:
  $$
    \xymatrix{
      \mathllap{
        \mbox{
          \tiny
          \color{darkblue}
          \bf
           {\begin{tabular}{c}
            quaternionic line bundle
           \end{tabular}}
        }
      }
      \mathcal{V}^\ast_X
      \ar[d]
      \ar[r]
      \ar@{}[dr]
        |-{\mbox{\tiny\rm(pb)}}
      &
      \mathcal{L}^\ast_{\mathbb{H}P^n}
      \ar[d]
      \ar[rr]
      \ar@{}[drr]
        |-{\mbox{\tiny\rm(pb)}}
      &&
      E \mathrm{Sp}(1)
        \underset{\mathrm{Sp}(1)}{\times}
      \mathbb{H}
      \ar[d]
      \\
      \mathllap{
        \mbox{
          \tiny
          \color{darkblue}
          \bf
          {\begin{tabular}{c}
            base space of
            \\
            bounded dimension $leq 4n+2$
          \end{tabular}}
        }
      }
      X^{4n+2}
      \ar@/_1.3pc/[rrr]
        |-{
          \;
          \vdash \mathcal{V}_X
          \;
        }
      \ar@{-->}[r]
      &
      \mathbb{H}P^n
      \;
      \ar@{^{(}->}[r]
      &
      \mathbb{H}P^\infty
      \ar[r]
        ^-{
          \simeq
        }
      &
      B \mathrm{Sp}(1)
      \,.
    }
  $$
  and any two such factorizations are homotopic to each other.
\end{lemma}
This goes back to \cite[p. 45, 67]{Chern51}.
For our purposes it will be useful to see this
as a transparent consequence of cellular approximation within
the familiar infinite classifying space:
\begin{proof}
  On the one hand, the topological space
  underlying a real $d$-dimensional manifold $X$ admits the structure of a
  $d$-dimensional CW-complex
  (e.g.  \cite[Ex. A.37]{FSS20c}).
  On the other hand, the sequence of
  finite dimensional projective spaces
  \eqref{DiagramOfProjectiveSpacesThomSpacesAndZeroSections}
  form the stages of a CW-complex structure on
  infinite projective space, by Lemma \ref{CellStructureOfProjectiveSpaces}.
  Therefore
  the cellular approximation theorem applies
  \cite[p. 404]{Spanier66}
  and says that any classifying map
  $X \xrightarrow{\;\;} \mathbb{K}P^\infty$
  factors, up to homotopy,
  through the $d$-skeleton of $\mathbb{K}P^\infty$,
  while any homotopy between such classifying maps factors
  still through its $d+1$-skeleton,
  hence through
  largest projective space of real dimension $\leq d + 1$.
\end{proof}

\begin{defn}[Bounded-dimensional complex orientation in $E$-cohomology]
 \label{ComplexOrientedCohomology}
 For $E$ be a multiplicative cohomology theory
 (Def. \ref{MultiplicativeCohomologyTheory}),
 and $n \in \mathbb{N}_+ \sqcup \{\infty\}$,
 a {\it universal $2n$-dimensional complex $E$-orientation}
 is a class
 $c_1^E \in E^2\big( \mathbb{C}P^n \big)$
 (to be called the
 {\it universal first $E$-Chern class} in dimension $2n$)
  whose restriction to $\mathbb{C}P^1$ is the suspended unit,
  hence such that we have a homotopy-commutative diagram of this
  form:
\vspace{-.3cm}
\begin{equation}
  \label{ComplexOrientationAsExtension}
  \raisebox{20pt}{
  \xymatrix@R=5pt@C=1em{
    \mathclap{\phantom{\vert_{\vert}}}
    \mathbb{C}P^1
    \ar@{^{(}->}[dd]
    \ar@{}[r]|-{\simeq}
    &
    S^2
    \ar[rrr]^-{
      \Sigma^2 (1^E)
    }
    &&&
    E\degree{2}
    \\
    \\
    \mathbb{C}P^n
    \ar@{-->}[uurrrr]_-{
      c_1^E
      \mathrlap{
        \!\!
        \mbox{
          \tiny
          \color{greenii}
          \bf
          \begin{tabular}{c}
            complex
            orientation
            \\
            to degree $n$
          \end{tabular}
        }
      }
    }
  }
  }
\end{equation}
\end{defn}
\noindent For $n = \infty$, this definition is classical
(\cite[Thm. 7.6]{ConnerFloyd66}\cite[\S II.2]{Adams74}\cite[\S 4.1]{Ravenel86}\cite[\S. 4.3]{Kochman96}\cite[\S 3.2]{TamakiKono06}\cite[\S 4]{Lurie10}).
For finite $n$, this has been considered
in \cite[\S 1.2]{Hopkins84}\cite[Def. 6.5.2]{Ravenel86}.
Analogously:
\begin{defn}[Bounded-dimensional quaternionic oriented $E$-cohomology]
 \label{QuaternionicOrientedCohomology}
 For $E$ be a multiplicative cohomology theory,
 and $n \in \mathbb{N} \sqcup \{\infty\}$,
 a {\it universal $(4n+2)$-dimensional quaternionic $E$-orientation}
 is a choice of a class
 $\scalebox{.8}{$\tfrac{1}{2}$}p^E_1 \in E^4\big( \mathbb{H}P^n \big)$
 (to be called the
 {\it universal first fractional $E$-Pontrjagin class}\footnote{
 In \cite{Baker92} the ``$\scalebox{.8}{$\tfrac{1}{2}$}$'' is omitted from the
 notation. But since this class does give the generator of
 $E^\bullet( \mathbb{H}P^\infty ) \,\simeq\, E^\bullet( B \mathrm{Spin}(3) )$
 we include that factor, for compatibility with the
 standard notation in the case that $E = H\mathbb{Z}$ is ordinary
 cohomology.
} in dimension $4n+2$)
whose restriction to $\mathbb{H}P^1$ is the suspended unit,
hence such that we have a homotopy-commutative diagram
of this form:

\vspace{-.3cm}
\begin{equation}
  \label{QuaternionicOrientationAsExtension}
  \raisebox{20pt}{
  \xymatrix@R=5pt@C=1em{
    \mathclap{\phantom{\vert_{\vert}}}
    \mathbb{H}P^1
    \ar@{^{(}->}[dd]
    \ar@{}[r]|-{\simeq}
    &
    S^4
    \ar[rrr]^-{
      \Sigma^4 (1^E)
    }
    &&&
    E\degree{4}
    \\
    \\
    \mathbb{H}P^n
    \ar@{-->}[uurrrr]_-{
      \scalebox{.7}{$\tfrac{1}{2}$} p_1^E
      \mathrlap{
        \!\!
        \mbox{
          \tiny
          \color{greenii}
          \bf
          \begin{tabular}{c}
            quaternionic
            orientation
            \\
            to degree $n$
          \end{tabular}
        }
      }
    }
  }
  }
\end{equation}
\end{defn}
\noindent For $n = \infty$, this
definition appears explicitly in \cite[Expl. 2.2.5]{Laughton08},
almost explicitly in \cite[Thm. 7.5]{ConnerFloyd66}\cite[\S 3.9]{TamakiKono06},
and somewhat implicitly in \cite[p. 2]{Baker92}.
The general Def. \ref{QuaternionicOrientedCohomology}
in finite dimension seems not to have received attention.

\begin{prop}[Complex-oriented cohomology of complex projective spaces]
  \label{ComplexOrientedCohomologyOfComplexProjectiveSpaces}
  Given a $2n$-dimensional complex $E$-orientation $c_1^E$
  (Def. \ref{ComplexOrientedCohomology}),
  we have for all $k \leq n$ an identification of the
  (un-reduced) $E$-cohomology ring of
  $\mathbb{C}P^k$ as
  the quotient polynomial algebra
  over the coefficient ring $E^\bullet(\ast)$
  generated by the
  $2n$-dimensional first $E$-Chern class
  \eqref{ComplexOrientationAsExtension}
  and quotiented to make its $(k+1)$st power vanish:
  \begin{equation}
    \label{ComplexOrientedECohomologyOfComplexProjectiveSpace}
    E^\bullet
    \big(
      \mathbb{C}P^k
    \big)
    \;\simeq\;
    E_\bullet
    \big[
      c_1^E
    \big]
    \big/
      \big(
        c_1^E
      \big)^{k+1}.
  \end{equation}
\end{prop}
\noindent This is classical,
e.g. \cite[Prop. 4.3.2(b)]{Kochman96}\cite[\S 4, Example 8]{Lurie10}.
The vanishing of $[c^E_1]^{k+1}$ follows as in
Prop. \ref{CanonicalTrivializationOfCupSquareOvernSphere}
(which gives the case $k = 1$).

\begin{remark}[First $E$-Chern classes of bounded-dimensional complex line bundles]
  \label{FirstEChernClassOfComplexLineBundles}
  Since $\mathbb{C}P^n$
  \eqref{InfiniteProjectiveSpaces}
  is a classifying space for complex line bundles
  over $d \leq 2n$-dimensional base spaces
  (Lemma \ref{UniversalFiniteDimensionalLineBundles}),
  a choice of
  $2n$-dimensional $E$-orientation (Def. \ref{ComplexOrientedCohomology})
  induces assignment to any
  $\mathcal{L}^\ast_X \,\in\, \mathbb{C}\mathrm{LineBundles}_{/X}$
  of an $E$-cohomology classs
  \begin{equation}
    \label{FirstEChernClassOfComplexLineBundle}
    c_1^E
    \big(
      \mathcal{L}^\ast_X
    \big)
    \;:=\;
    \Big[
      X
        \xrightarrow{ \vdash \mathcal{L}^\ast_X }
      \mathbb{C}P^n
        \xrightarrow{ c_1^E }
      E_2
    \Big]
    \;\in\;
    E^2(X)
    \,,
  \end{equation}
  to be called the {\it first $E$-Chern class}
  of the complex line bundle.
  In particular,
  the generating
  class in \eqref{ComplexOrientedECohomologyOfComplexProjectiveSpace}
  is that of the dual tautological line bundle \eqref{TheTautologicalKLineBundle},
  classified by the identity map:
  $c^E_1 = c^E_1\big( \mathcal{L}^\ast_{\mathbb{C}P^k} \big)
  \in E^2\big( \mathbb{C}P^k\big)$.
\end{remark}

\begin{example}[$2n$-Dimensional complex orientations in ordinary cohomology]
  \label{ComplexOrientationInOrdinaryCohomology}
  For $E = H R$ ordinary cohomology over a ring $R$,
  $(HR)_\bullet \,\simeq\, \mathbb{Z}$
  is concentrated in degree=0.
  Therefore, Prop. \ref{ComplexOrientedCohomologyOfComplexProjectiveSpaces}
  implies that the
  restriction morphisms
  $
    H^2\big( \mathbb{C}P^n; R \big)
      \xrightarrow{ \simeq }
    H^2\big( \mathbb{C}P^1; R \big)
      \;\simeq\;
    \mathbb{Z}
  $
  are isomorphisms for all $n \in \mathbb{N}_+ \sqcup \{\infty\}$.
  This means that there is a {\it unique} stage=$n$
  complex $H R$-orientation (Def. \ref{ComplexOrientedCohomology}).
  For this
  the $HR$-Chern class from Remark \ref{FirstEChernClassOfComplexLineBundles}
  is the ordinary Chern class
  $$
    c_1^{H R}
    \;=\;
    c_1
    \,.
  $$

  Analogously, there is a unique quaternionic orientation
  on ordinary cohomology.
\end{example}

\begin{example}[Universal orientations on $M\mathrm{U}$ and $M\mathrm{Sp}$]
  \label{UniversalOrientationsOnMUAndMSp}
  The weak equivalences from Prop. \ref{ZeroSectionIntoThomSpaceOfUniversalLineBundleIsEquivalence},
  regarded as maps
  \vspace{-2mm}
  $$
    \xymatrix{
      \mathbb{C}P^1
      \ar[rr]
        ^-{
          (c^{M\mathrm{U}}_1)^{\mathrm{univ}}
        }
        _-{
          \simeq
        }
      &&
      (M \mathrm{U})^2
    }
    \,,
    {\phantom{AAA}}
    \xymatrix{
      \mathbb{H}P^1
      \ar[rr]
        ^-{
          (\scalebox{.5}{$\tfrac{1}{2}$}p^{M\mathrm{Sp}}_1)^{\mathrm{univ}}
        }
        _-{
          \simeq
        }
      &&
      (M \mathrm{Sp})^4
    }
  $$
  constitute an unbounded
  complex orientation (Def. \ref{ComplexOrientedCohomology})
  of $M\mathrm{U}$
  and an unbounded quaternionic orientation
  (Def. \ref{QuaternionicOrientedCohomology})
  of $M\mathrm{Sp}$,
  respectively.

  In fact these orientations are {\it initial} among all
  complex/quaternionic orientations:
  Given a multiplicative cohomology theory $E$,
  there is a bijection between complex/quaternionic orientations
  on $E$ and homotopy-classes of homomomorphisms
  of homotopy-commutative ring spectra \eqref{HomotopyCommutativeRingSpectrum}
  from $M\mathrm{U}$
  (e.g. {\cite[\S 4, Lem. 4.1.13]{Ravenel86}\cite[\S 6, Thm. 8]{Lurie10}})
  or from $M\mathrm{Sp}$ (e.g.  {\cite[Ex. 2.2.9]{Laughton08}}),
  respectively:
  \begin{equation}
    \label{OrientationsBijectionsToMultiplicativeMaps}
    \underset{
      \mathclap{
      \raisebox{-3pt}{
        \tiny
        \color{darkblue}
        \bf
        \begin{tabular}{c}
          complex orientation
        \end{tabular}
      }
      }
    }{
    \big[
      \mathbb{C}P^\infty
      \xrightarrow{\;\;c^E_1\;\;}
      E^2
    \big]
    }
    \;\;
    \leftrightarrow
    \;\;
    \underset{
      \mathclap{
      \raisebox{-3pt}{
        \tiny
        \color{darkblue}
        \bf
        \begin{tabular}{c}
          homotopy-multiplicative
          \\
          map of ring spectra
        \end{tabular}
      }
      }
    }{
    \big[
      M\mathrm{U}
        \xrightarrow{\;\mathrm{mult}\;}
      E
    \big]
    }
    \,,
    {\phantom{AAAAAA}}
    \underset{
      \mathclap{
      \raisebox{-3pt}{
        \tiny
        \color{darkblue}
        \bf
        \begin{tabular}{c}
          quaternionic orientation
        \end{tabular}
      }
      }
    }{
    \big[
      \mathbb{H}P^\infty
      \xrightarrow{\;\;\scalebox{.5}{$\tfrac{1}{2}$}p^E_1\;\;}
      E^2
    \big]
    }
    \;\;
    \leftrightarrow
    \;\;
    \underset{
      \mathclap{
      \raisebox{-3pt}{
        \tiny
        \color{darkblue}
        \bf
        \begin{tabular}{c}
          homotopy-multiplicative
          \\
          map of ring spectra
        \end{tabular}
      }
      }
    }{
    \big[
      M\mathrm{Sp}
        \xrightarrow{\;\mathrm{mult}\;}
      E
    \big]
    \,.
    }
  \end{equation}
\end{example}

\medskip

\newpage

\subsection{Conner-Floyd classes}

We recall the construction of Conner-Floyd $E$-Chern classes
in complex-oriented cohomology, working out the minimal
data needed to construct them in the case of
bounded-dimensional complex orientations
(Prop. \ref{ConnerFloydChernClasses} below).
Then we use this to prove that the second
Conner-Floyd $E$-Chern class of a $4k+2$-dimensional
complex $E$-orientation induces
(the first fractional $E$-Pontrjagin class of)
a $4k+2$-dimenional quaternionic $E$-orientation
(Theorem \ref{FiniteRankQuaternionicOrientationFromFiniteRankComplexOrientation}
below).

\medskip

\begin{prop}[Bounded-dimensional Conner-Floyd $E$-Chern classes]
  \label{ConnerFloydChernClasses}
  For $r \leq n \in \mathbb{N}$, let
  $c_1^E$ be a $2n$-dimensional complex orientation
  in $E$-cohomology (Def. \ref{ComplexOrientedCohomology})
  and $X$ a CW-complex of bounded dimension,
  \vspace{-2mm}
  \begin{equation}
    \label{BoundOnDimensionAndRank}
    \mathrm{dim}(X) = 2(n-r)
    \,,
    \phantom{AAAA}
    \mathrm{rnk}_{\mathbb{C}}
    \big(
      \mathcal{V}_X
    \big)
    \;\leq\;
    r + 1
    \,.
  \end{equation}

\vspace{-2mm}
\noindent
Then there is an assignment to each
  $\mathcal{V}_X \,\in\, \mathbb{C}\mathrm{VectorBundles}_{/X}$
  with bounded rank \eqref{BoundOnDimensionAndRank}
  of an $E$-cohomology classes
  \vspace{-2mm}
  $$
    \Big\{
      c^E_n
      \big(
        \mathcal{V}^\ast_X
      \big)
      \;\in\;
      E^{2n}
      (
        X
      )
    \Big\}_{n \in \mathbb{N}}
  $$

  \vspace{-2mm}
\noindent
  such that the {\it total class}
  $$
    c^E
    \;\coloneqq\;
    1 + c_1^E + c_2^E + \cdots
  $$
  satisfies the same kind of conditions that
  characterize the usual Chern classes in ordinary cohomology:

  \vspace{-1mm}
  \begin{enumerate}[{\bf (i)}]
   \setlength\itemsep{1pt}
    \item For $X \xrightarrow{ \;f\; } Y$ a map, we have:

    \vspace{-1.1cm}
    \begin{equation}
      \label{CFClassesAreNatural}
      c^E\big( f^\ast\mathcal{V}^\ast_X \big)
      \,=\,
      f^\ast \big( c^E\big( \mathcal{V}^\ast_X \big) \big)\,.
    \end{equation}

    \item
      For $\mathcal{L}_X$ a line bundle we have:

      \vspace{-1.1cm}
      \begin{equation}
        \label{CFClassesExtendGivenFirstChernClass}
        c^E\big( \mathcal{L}^\ast_X \big)
          \,=\,
        1
          +
        \underset{
          \mathclap{
          \mbox{
            \tiny
            via \eqref{FirstEChernClassOfComplexLineBundle}
          }
          }
        }{
          \underbrace
          {
            c_1^E\big( \mathcal{L}^\ast_X \big)
          }
        }.
      \end{equation}
    \item The Whitney sum rule holds:

      \vspace{-1.2cm}
      \begin{equation}
        \label{WhitneySumRuleforEChernClasses}
        c^E ( \mathcal{V}^\ast_X \oplus_X \mathcal{W}^\ast_X)
        \;=\;
        c^E ( \mathcal{V}^\ast_X )
         \cdot
        c^E( \mathcal{W}^\ast_X )
        \,.
      \end{equation}
    \item The top degree is bounded by the rank:

      \vspace{-1.3cm}
      \begin{equation}
        \label{CFClassesHaveMaximalDegreeTheRankOfTheBundle}
        c^E\big( \mathcal{V}^\ast_X \big)
          \,=\,
        \sum_{k = 0}^{\mathrm{rnk}(\mathcal{V}_X)}
        c_k^E\big( \mathcal{V}^\ast_X \big)\,.
      \end{equation}
  \end{enumerate}
\end{prop}
\noindent
\begin{proof}
  Without the bound on dimension and rank, this is
  \cite[Thm. 7.6]{ConnerFloyd66}, see also \cite[I.4]{Adams74}.
  We just need to recall the construction of these
  {\it Conner-Floyd Chern classes} with attention to the
  required dimension in each step.
  So consider the pullback of the
  dual vector bundle
  $\mathcal{V}^\ast_X$ to
  the total space of the projective bundle
  $P\big( \mathcal {V}_X \big)$ \eqref{ProjectiveBundle}.
  This  pullback splits
  as a direct sum of:

  \noindent
  {\bf (a)} the dual tautological bundle
  $\mathcal{L}^\ast
    \coloneqq
  \mathcal{L}^\ast_{P(\mathcal{V}_X)}$ ,
  which is fiber-wise the dual tautological bundle
  \eqref{TheDualTautologicalKLineBundle} over $\mathbb{C}P^r$,

  \noindent
  {\bf (b)} a remaining orthogonal vector bundle
  $\mathcal{V}^\perp
    \coloneqq
    \big\{
      \left.
  \!\!      ( \eta_x, [v_x] )
      \,\in\,
      \mathcal{V}^\ast_X \times_{X} P(\mathcal{V}_X)
      \,
      \right\vert
      \,
        \eta_x(v_x) \,=\, 0
      \,
    \big\}
  $
  of rank $r$,

  \medskip
  \noindent
  as shown at the top of the following homotopy-commutative diagram:
  \vspace{-2mm}
  \begin{equation}
    \label{SplittingOfVectorBundleOverItsProjectiveBundle}
    \hspace{-.6cm}
    \raisebox{40pt}{
    \xymatrix@C=3.7em@R=16pt{
      &
      \mathcal{L}^\ast
        \oplus
      \mathcal{V}^{\perp}
      \ar[r]
      \ar[d]
      \ar@{}[dr]|-{\mbox{\tiny\rm(pb)}}
      &
      \mathcal{V}^\ast_X
      \ar[d]
      \\
      \mathbb{C}P^{r}
      \ar[r]^-{
          \;
          \mathrm{fib}(\pi)
          \;
        }
      &
      P\big( \mathcal{V}_X \big)
      \ar[r]|-{ \;\pi\; }
      \ar@{-->}[dr]
        _-{
         (\vdash \mathcal{L}^\ast,
          \,
          \vdash \mathcal{V}^\perp)
        \quad}
      &
      X^{ 2(n-r) }
      \ar[rr]
        ^-{
          \;
          \vdash \mathcal{V}^\ast_X
          \;
        }
      &&
      B \mathrm{U}(r+1)\;.
      \\
      & &
      {
             \mathbb{C}P^{2n}
            \times
         B \mathrm{U}(r)
      }
      \ar[r]
      &
      {
             B \mathrm{U}(1)
        \times
        B \mathrm{U}(r)
      }
      \ar[ur]
    }
    }
  \end{equation}

  \vspace{-2mm}

 \noindent
  Shown at the bottom
  of the diagram \eqref{SplittingOfVectorBundleOverItsProjectiveBundle}
  is a
  classifying map for this direct sum bundle, where we noticed that
  \begin{equation}
    \label{DimensionOfProjectiveBundle}
    \mathrm{dim}_{\mathbb{C}}
    \big(
      P
      \big(
        \mathcal{V}_X
      \big)
    \big)
    \;=\;
    \mathrm{dim}(X)
    +
    2
    \big(
    \mathrm{rnk}_{\mathbb{C}}
    \big(
      \mathcal{V}_X
    \big)
    -1
    \big)
    \;=\;
    2(n - r)
    +
    2 r
    \;=\;
    2 n\,,
  \end{equation}

\vspace{-2mm}
\noindent  and then used Lemma \ref{UniversalFiniteDimensionalLineBundles}
  to factor the classifying map
  $
    P(\mathcal{V}_X)
      \xrightarrow{ \vdash \mathcal{L}^\ast }
    B \mathrm{U}(1)
  $
  of $\mathcal{L}^\ast$
  through $\mathbb{C}P^{2n}$, up to homotopy.

  \noindent
  Via this factorization,
  we obtain the class \eqref{FirstEChernClassOfComplexLineBundle}
  \vspace{-2mm}
  $$
    c_1^E
    \big(
      \mathcal{L}^\ast
    \big)
    \;\coloneqq\;
    \big[
      P(\mathcal{V}_X)
        \xrightarrow{ \vdash \mathcal{L}^\ast }
      \mathbb{C}P^{2n}
        \xrightarrow{ c^E_1 }
      E^2
    \big]
    \;\in\;
    E^2
    \big(
      P(\mathcal{V}_X)
    \big),
  $$
  whose pullback along $\mathrm{fib}(\pi)$,
  yields the give first $E$-Chern class $c^E_1$
  (by Remark \ref{FirstEChernClassOfComplexLineBundles})
  and hence, by Prop. \ref{ComplexOrientedCohomologyOfComplexProjectiveSpaces},
  the polynomial generator
  of the $E$-cohomology ring
  \eqref{ComplexOrientedECohomologyOfComplexProjectiveSpace}.
  Therefore, the Leray-Hirsch theorem
  in its $E$-cohomology version
  (\cite[Thm. 7.4]{ConnerFloyd66}\cite[Thm. 3.1]{TamakiKono06})
  applies and says that

  \vspace{-2mm}
  \begin{equation}
    \label{ECohomologyOfProjectiveBundle}
    E^\bullet
    \big(
      P
      \big(
        \mathcal{V}_X
      \big)
    \big)
    \;\;\simeq\;\;
    E^\bullet(X)
    \Big\langle
      1, \,
      c_1^E(\mathcal{L})
      ,\,
      \big(
        c_1^E(\mathcal{L})
      \big)^2
      ,\,
      \cdots
      ,\,
      \big(
        c_1^E(\mathcal{L})
      \big)^n
    \Big\rangle
    \;\;\;
    \in
    \;
    E^\bullet(X)\mathrm{Modules}
    \,.
  \end{equation}
  Now, by the required properties
  \eqref{CFClassesAreNatural},
  \eqref{CFClassesExtendGivenFirstChernClass},
  \eqref{WhitneySumRuleforEChernClasses}
  of the $E$-Chern classes,
  they need to satisfy
  $$
    \pi^\ast
    \big(
      c^E(
        \mathcal{V}^\ast_X
      )
    \big)
    \;=\;
    c^E
    (
      \mathcal{V}^\perp
    )
    \cdot
    \big(
      1
        +
      c_1^E
      \big(
        \mathcal{L}^\ast
      \big)
    \big)
    \;\;\;\;
    \in \
    \;
    E^\bullet\big( P\big( \mathcal{V}_X\big)  \big)
    \,,
  $$
  hence equivalently:
  \begin{equation}
    \label{CFClassOfSplitVectorBundle}
    \begin{aligned}
      c^E
      \big(
        \mathcal{V}^\perp
      \big)
      &
      \; = \;
      \pi^\ast
      \big(
        c^E
        \big(
          \mathcal{V}^\ast_X
        \big)
      \big)
      \cdot
      \big(
        1
          +
        c_1^E
        \big(
          \mathcal{L}^\ast
        \big)
      \big)^{-1}
      \\
      & := \;
      \pi^\ast
      \big(
        c^E
        \big(
          \mathcal{V}^\ast_X
        \big)
      \big)
      \cdot
      \underset{k}{\sum}
      (-1)^k
      \big(
        c_1^E
        \big(
          \mathcal{L}^\ast
        \big)
      \big)^k
      \,.
    \end{aligned}
  \end{equation}
  But since
  $\mathrm{rnk}\big( \mathcal{V}^\perp_X\big) = r < r + 1$,
  the last condition \eqref{CFClassesHaveMaximalDegreeTheRankOfTheBundle}
  implies that \eqref{CFClassOfSplitVectorBundle}
  reduces in degree $2(r + 1)$ to
  \begin{equation}
    \label{DefiningEquationForConnerFloydClasses}
    \hspace{-2mm}
    \begin{aligned}
      0
      &
       =
      \pi^\ast
      \big(
        c^E_{r+1}
        \big(
          \mathcal{V}^\ast_X
        \big)
      \big)
      -
      \pi^\ast
      \big(
        c^E_{r}
        \big(
          \mathcal{V}^\ast_X
        \big)
      \big)
      \cdot
      c^E_1
      \big(
        \mathcal{L}^\ast
      \big)
      +
      \cdots
      +
      (-1)^r
      \pi^\ast
      \big(
        c^E_{1}
        \big(
          \mathcal{V}_X^\ast
        \big)
      \big)
      \cdot
      \big(
        c^E_1
        \big(
          \mathcal{L}^\ast
        \big)
      \big)^r
      \\
      &
       \phantom{=}\;
      +
      (-1)^{r+1}
      \big(
        c^E_1
        \big(
          \mathcal{L}^\ast
        \big)
      \big)^{r+1}
      \,.
    \end{aligned}
  \end{equation}

  \vspace{-2mm}
  \noindent
  By \eqref{ECohomologyOfProjectiveBundle},
  this equation \eqref{DefiningEquationForConnerFloydClasses}
  has a unique solution for classes
  $
    c_k^E( \mathcal{V}^\ast_X )
    \in
    E^{\bullet}
    (
      X
    )
   $.
  These are the Conner-Floyd $E$-Chern classes
  of $\mathcal{V}^\ast_X$
  for the given complex orientation $c^E_1$,
  usually defined for $c^E_1$ given on $\mathbb{C}P^\infty$,
  but here seen to depend only on
  $c^E_1 \in E^2(\mathbb{C}P^n)$,
  under the bounds \eqref{BoundOnDimensionAndRank}.
\end{proof}

\medskip

\noindent {\bf Quaternionic orientations induced from complex orientations.}
\begin{prop}[Complex orientation induces quaternionic orientation]
  \label{QuaternionicOrientationFromComplexOrientation}
  Given a complex $E$-orientation $c_1^E$
  (Def. \ref{ComplexOrientedCohomology}, i.e. unbounded,
  $\infty$-dimensional),
  the
  second Conner-Floyd $E$-Chern class (Prop. \ref{ConnerFloydChernClasses})
  defines a quaternionic $E$-orientation (Def. \ref{QuaternionicOrientedCohomology})
  by setting:
\vspace{-2mm}
\begin{equation}
  \label{QuaternionicOrientationInducedFromComplexOrientation}
  \hspace{-1.3cm}
  \scalebox{.8}{$\tfrac{1}{2}$}p_1^E
  \;\coloneqq\;
  c_2^E
  \big(
    (-)_{\mathbb{C}}
  \big)
  \;\;\;\;\;
    \mbox{
      i.e.:
    }
  \;\;\;\;\;
  \raisebox{20pt}{
  \xymatrix@R=6pt@C=2.5em{
    \mathclap{\phantom{\vert_{\vert}}}
    \mathbb{H}P^1
    \ar@{^{(}->}[dd]
    \ar@{-->}[rr]^-{ \Sigma^4 (1^E) }
    &&
    E\degree{4}
    \\
    \\
    \mathbb{H}P^\infty
    \ar@{}[r]|-{ \simeq }
    \ar@{->}[uurr]|-{
      \;\;
      \scalebox{.7}{$\tfrac{1}{2}$}p_1^E
      \;\;
    }
    &
    B \mathrm{SU}(2)
    \;
    \ar@{^{(}->}[r]
    &
    B \mathrm{U}(2)
    \ar[uu]_-{
      c_2^E
      \mathrlap{
        \mbox{
          \tiny
          \color{greenii}
          \bf
          \begin{tabular}{c}
            quaternionic orientation
            \\
            from complex orientation
          \end{tabular}
        }
      }
    }
    }
  }
\end{equation}
\end{prop}
\begin{proof}
  We need to show that the top triangle in
  \eqref{QuaternionicOrientationInducedFromComplexOrientation}
  homotopy-commutes, hence that the restriction of
  $c^E_2$ along
  $S^4 \,\simeq\, \mathbb{H}P^1 \hookrightarrow \mathbb{H}P^\infty
  \,\simeq\, B \mathrm{SU}(2)$ is the
  suspended unit class $\Sigma^4(1^E)$.

  But the $n$th Conner-Floyd Chern class $c^E_n$
  (induced by the given choice of $c^E_1$)
  is the pullback of an $E$-Thom class
  $
    \big[
    \mathrm{vol}^E_{
      E \mathrm{U}(n) \underset{\mathrm{U}(n)}{\times} \mathbb{C}^n
    }
    \big]
  $
  along the 0-section of the
  universal complex vector bundle over $B \mathrm{U}(n)$
  (e.g. \cite[p. 61]{TamakiKono06}).
  For the case $n = 2$ and restricted along
  $
    \mathbb{H}P^\infty
      \,\simeq\,
    B \mathrm{SU}(2)
      \xrightarrow{\;}
    B \mathrm{U}(2)
  $,
  this means that $c^E_2$
  is the restriction along the 0-section of
  an $E$-Thom class
  $
    \big[
      \mathrm{vol}^E_{ \mathcal{L}^\ast_{\mathbb{H}P^\infty} }
    \big]
  $
  on the universal $\mathbb{H}$-line bundle.

  Now the weak equivalence from
  Prop. \ref{ZeroSectionIntoThomSpaceOfUniversalLineBundleIsEquivalence}
  in the diagram \eqref{DiagramOfProjectiveSpacesThomSpacesAndZeroSections}
  means that the restriction of this Thom class
  $
    \big[
      \mathrm{vol}^E_{\mathcal{L}^\ast_{\mathbb{H}P^\infty}}
    \big]
  $
  to the base fiber
  --
    which is $\Sigma^4(1^E)$ by definition of Thom classes
  --
  equals the restriction of $c^E_2$ along
  $\mathbb{H}P^1 \hookrightarrow \mathbb{H}P^\infty$:
\begin{equation}
  \label{ProofThatc2IsAp1}
  \hspace{-.4cm}
  \raisebox{14pt}{
  \xymatrix@C=7pt@R=1.2em{
    \widetilde E {}^\bullet
    \big(
      S^4
    \big)
    \ar@{=}[d]
    \ar@{<-}[rr]^-{
      \mbox{
        \tiny
        \color{greenii}
        \bf
        restriction to
      }
    }
    _-{
      \mbox{
        \tiny
        \color{greenii}
        \bf
       fiber
      }
    }
    &
    {\phantom{AAAA}}
    &
    \widetilde E {}^\bullet
    \big(
      \mathrm{Th}
      (
        \mathcal{L}^\ast_{\mathbb{H}P^\infty}
      )
    \big)
    \ar@{=}[r]
    \ar[d]
     ^-{
       \simeq
     }
     &
    \widetilde E {}^\bullet
    \big(
      \mathrm{Th}
      (
        E \mathrm{SU}(2)
          \underset{
            \mathclap{
              \scalebox{.5}{$\mathrm{SU}(2)$}
            }
          }{\times}
        \mathbb{C}^2
      )
    \big)
    \ar[d]
     ^-{
       \simeq
     }
    &
    \Sigma^4(1^E)
    \ar@{<-|}[rr]
    \ar@{|->}[d]
    &
      {\phantom{A}}
    &
    \big[
      \mathrm{vol}^E_{\mathcal{L}^\ast_{\mathbb{H}P^1}}
    \big]
    \ar@{=}[r]
    &
    \big[
      \mathrm{vol}^E_{
        E\mathrm{SU}(2)
          \underset{
            \mathclap{
              \scalebox{.4}{$\mathrm{SU}(2)$}
            }
          }{\times}
        \mathbb{C}^2
      }
    \big]
    \ar@{<-|}[d]
    \\
    \widetilde E^{\bullet}
    \big(
      S^4
    \big)
    \ar@{<-}[rr]^-{
      \mbox{
        \tiny
        \color{greenii}
        \bf
        restriction to
      }
    }
    _-{
      \mbox{
        \tiny
        \color{greenii}
        \bf
         $\mathbb{H}P^1 \subset \mathbb{H}P^\infty$
      }
    }
    &&
    \widetilde E {}^{\bullet}
    \big(
      \mathbb{H}P^\infty
    \big)
    \ar@{=}[r]
    &
    \widetilde E {}^{\bullet}
    \big(
      B \mathrm{SU}(2)
    \big)
    &
    \Sigma^4(1^E)
    \ar@{<--|}[rr]
    &&
    \scalebox{.7}{$\tfrac{1}{2}$}p^E_1
    \ar@{|->}[r]
    &
    c^E_2
  }
  }
\end{equation}
\end{proof}

This situation generalizes to finite-stage orientations:

\begin{theorem}[Finite-dimensional quaternionic orientation from finite-dimensional complex orientation]
  \label{FiniteRankQuaternionicOrientationFromFiniteRankComplexOrientation}
  For $k \in \mathbb{N}$,
  given a
  $2(2k + 1) = 4k + 2$-dimensional
  complex $E$-orientation $c_1^E$
  (Def. \ref{ComplexOrientedCohomology}),
  the second
  $E$-Chern class (Prop. \ref{ConnerFloydChernClasses})
  defines a $4k + 2$-dimensional quaternionic orientation
  (Def. \ref{QuaternionicOrientedCohomology}) by setting
\vspace{-2mm}
\begin{equation}
  \label{BoundedDimensionalQuaternionicOrientationInducedFromComplexOrientation}
  \hspace{-1.3cm}
  \scalebox{.8}{$\tfrac{1}{2}$}p_1^E
  \;\coloneqq\;
  c_2^E
  \big(
    (-)_{\mathbb{C}}
  \big)
  \;\;\;\;\;
    \mbox{
      i.e.:
    }
  \;\;\;\;\;
  \raisebox{20pt}{
  \xymatrix@R=6pt@C=2.5em{
    \mathclap{\phantom{\vert_{\vert}}}
    \mathbb{H}P^1
    \ar@{^{(}->}[dd]
    \ar@{-->}[rr]^-{ \Sigma^4 (1^E) }
    &&
    E\degree{4}
    \\
    \\
    \mathbb{H}P^n
    \ar@{}[r]|-{ \simeq }
    \ar@{->}[uurr]|-{
      \;\;
      \scalebox{.7}{$\tfrac{1}{2}$}p_1^E
      \;\;
    }
    &
    \mathrm{sk}_{4n} B \mathrm{SU}(2)
    \;
    \ar@{^{(}->}[r]
    &
    \mathrm{sk}_{4n}
    B \mathrm{U}(2)
    \ar[uu]_-{
      c_2^E
      \mathrlap{
        \mbox{
          \tiny
          \color{greenii}
          \bf
          \begin{tabular}{c}
            finite-dimensional
            \\
            quaternionic orientation
            \\
            from finite-dimensional
            \\
            complex orientation
          \end{tabular}
        }
      }
    }
    }
  }
\end{equation}
\end{theorem}
\begin{proof}
  By Prop. \ref{QuaternionicOrientationFromComplexOrientation}
  the statement is true for complex orientations
  at sufficiently large dimension. We need to
  see in which dimension the complex orientation needs
  to be defined in order to provide the
  quaternionic orientation at the required dimension.
  In order to run the proof \eqref{ProofThatc2IsAp1}
  with $\mathbb{H}P^\infty$ replaced by
  $\mathbb{H}P^k$,
  we need the $E$-Chern class $c^E_2$ to be defined
  on the complex vector bundle of rank $r + 1 = 2$
  which underlies a quaternionic line bundle
  on $\mathbb{H}P^k$.
  By Prop. \ref{ConnerFloydChernClasses}
  and using that $\mathrm{dim}\big( \mathbb{H}P^k \big)
  = 2(2k)$
  this is given by
  a complex orientation of dimension
  $ 2( 2 k + r ) = 2(2 k + 1) = 4k + 2$.
\end{proof}

\begin{remark}[Relation of dimensions for complex- and quaternionic orientations]
\label{RelationOfDimensionsForComplexAndQuaternionicOrientations}
In summary, the matching of the dimensions
for complex- and quaternionic orientations
in Theorem \ref{FiniteRankQuaternionicOrientationFromFiniteRankComplexOrientation}
comes about from the interaction of two slightly subtle effects:

\noindent
{\bf (i)}
The construction of the {\it second} $E$-Chern class
on $\mathbb{H}P^k$ requires the {\it first} $E$-Chern
class to be defined in dimension
\vspace{-2mm}
$$
  \mathrm{dim}_{\mathbb{R}}
  \big(
    P_{\mathbb{C}}
    \big(
      \mathcal{L}^\ast_{\mathbb{H}P^k}
    \big)
  \big)
  \;=\;
  4k + 2
  \;=\;
  \mathrm{dim}
  \big(
    \mathbb{C}P^{2k+1}
  \big)
$$

\vspace{-1mm}
\noindent (by \eqref{DimensionOfProjectiveBundle} in the proof of
Prop. \ref{ConnerFloydChernClasses}).

\noindent
{\bf (ii)} The finite-dimensional projective spaces
$\mathbb{H}P^k$ and $\mathbb{C}P^{2k+1}$
are universal for
$\mathbb{H}$- and $\mathbb{C}$-line bundles,
respectively,
over spaces of coinciding dimension
\vspace{-2mm}
$$
  \mathrm{dim}_{\mathbb{R}}
  \big(
    \mathbb{H}P^{k+1}
  \big)
  - 2
  \;=\;
  4k + 2
  \;=\;
  \mathrm{dim}_{\mathbb{R}}
  \big(
    \mathbb{C}P^{(2k+1)+1}
  \big)
  -
  2
$$

\vspace{-2mm}
\noindent
(by Lemma \ref{UniversalFiniteDimensionalLineBundles}).

\end{remark}

\begin{example}[Orientation and near horizons of M2-branes]
  By Theorem \ref{FiniteRankQuaternionicOrientationFromFiniteRankComplexOrientation},
  a universal orientation of complex line bundles
  over base spaces of dimension 10
  induces a universal orientation of
  quaternionic line bundles over 10-dimensional base
  space, which
  equivalently
  is
  \eqref{G4TrivializationOnS7IsQuaternionicOrientationToSecondStage}
  a
  choice of universal $H^E_3$-flux near M2-brane horizons
  \eqref{Universal3FluxNearM2Branes}:
  \vspace{-2mm}
  $$
    \raisebox{22pt}{
    \xymatrix{
      \mathbb{C}P^1
      \ar@{^{(}->}[d]
      \ar[rr]
        ^-{
          \;\Sigma^2(1^E)\;
        }
      &&
      E^2
      \\
      \mathbb{C}P^5
      \ar@{-->}[urr]
        _-{ \;c^E_1\; }
    }
    }
    \;\;\;\;
    \Rightarrow
    \;\;\;\;
    \raisebox{30pt}{
    \xymatrix@R=1.2em
    {
      S^7
      \ar[dd]
        _-{
          \mathclap{\phantom{\vert^{\vert}}}
          h_{\mathbb{H}}
          \mathclap{\phantom{\vert_{\vert}}}
        }
        ^>>>{\ }="t"
      \ar[rr]_>>>{\ }="s"
      &&
      \ast
      \ar[dd]
      \\
      \\
      S^4
      \ar[rr]
        ^-{
          \;
          \Sigma^4(1^E)
          \;
        }
      &&
      E^4
      \ar@{=>}
        |{ \; H^E_3 \; }
        "s"; "t"
    }
    }
  $$
\end{example}

\newpage

\section{Conclusions}
\label{Conclusions}

\noindent
{\bf M-Brane charge quantization in Cobordism cohomology.}
Since the Pontrjagin theorem \eqref{PontrjaginIsomorphismForNonCompactManifolds}
identifies Cohomotopy theory
with framed Cobordism cohomology, the statement of
\hyperlink{HypothesisH}{\it Hypothesis H} is
equivalently:

\medskip

{\it M-Brane charge is quantized in framed Cobordism}.

\medskip

\noindent
We have seen in \cref{MBraneWorldvolumesAndThePontrjaginConstruction} that this is not just an abstract
mathematical identification,
but that Cobordism representatives of
brane charges in Cohomotopy
are naturally identified with
polarized brane worldvolumes \eqref{InterpretingProbeBraneWorldvolumes},
and their trivializations with
brane-punctured
Kaluza-Klein compactification of spacetimes
$\mathbb{R}^{D,1}$,
near the probe brane charges (p. \ref{MTheoryCompactifiedOnK3AndTheThirdStableStem}).
This is a consequence of how the {\it unstable} Pontrjagin-theorem
concerns
{(a) actual embedded submanifolds} $\Sigma^{D-n}$
equipped with
{(b) normal framings and hence Cartesian-product neighborhoods}
$\Sigma^{D - n} \times \mathbb{R}^n$,
as befits a KK-compactification on $\Sigma^{D-n}$
(near $\Sigma^{D-n}$).

\medskip

But since Cohomotopy/framed Cobordism
are, stably, the {\it initial} multiplicative cohomology theory
\eqref{dInvariantAndBoardman},
every other multiplicative cohomology serves to
approximate brane charge quantization in Cohomotopy.

Specifically, existence of universal $H_3$-fluxes near M2-branes
forces \eqref{G4TrivializationOnS7IsQuaternionicOrientationToSecondStage}
brane charge observation through
quaternionic-orientated --
in particular \eqref{HE3FluxFromComplexOrientation}, complex-orientated --
multiplicative cohomology theories.
But initial among oriented cohomology theories
are again Cobordism cohomology theories, namely
\eqref{OrientationsBijectionsToMultiplicativeMaps}
quaternionic Coordism $M\mathrm{Sp}$
and
complex Cobordism $M\mathrm{U}$, respectively.
And indeed, one may recover observations of charges in Cohomotopy,
hence in framed Cobordism,
from their observations in unitary Cobordism --
this is the statement of the Adams-Novikov spectral sequence
(p. \pageref{MBraneChargeInComplexOrientedCohomology}):

\medskip

{\it M-Brane charge quantization may be observed in complex Cobordism.}

\medskip

\noindent
Which is to say that, with Hypothesis H, M-brane charge is
not only {\it fundamentally} in framed Cobordism,
but, in the presence of black M2-brane sources,
M-brane charges are canonically approximated/perceived through
other flavors of Cobordism cohomology $M\!f$, such as
quaternionic and unitary cobordism,
but also, for instance, special unitary Cobordism  $M S\mathrm{U}$ \eqref{K3CompactificationWithItsSUStructureAppears},
(which connects with Calabi-Yau compactifications, Prop. \ref{NonTorsionSUCobordismRingGeneratedByCYs}).

\medskip

\noindent
In summary and in the form of a broad slogan\footnote{
The idea that brane charge should be quantized/vanish in
Cobordism cohomology may be compared to the
discussion in \cite[\S 5.2, 2nd paragraph]{McNamaraVafa17}},
Hypothesis H implies that:

\medskip

{\it M-brane charge quantization is seen in Cobordism theories.}

\medskip

\noindent
hence that

\medskip

{\it M-theory is controlled by $M\!f$-theory.}

\medskip

\noindent
This is a direct analogue -- a refinement -- of
the traditional conjecture that D-brane charge is quantized in
topological K-theory (p. \ref{DBraneChargeInKTheory}).
Indeed, if M-brane charge is in
Cobordism theory, then perceiving it through the lens of
K-orientations \eqref{ConnerFloydOrientationCompatibilityDiagram}
means to {\it approximate} it by K-theory.

\medskip

One place where brane charge quantization laws become
important is in discussing tadpole cancellations
(see \cite{SS19}), hence
the condition that total brane charge on compact spaces
is required to vanish: It is the charge quantization
law in generalized cohomology which determines what it actually
means for brane charges to vanish.
For example, the cancelling of charges of
24 transverse NS5/D7-branes
in M/F theory on K3 is faithfully reflected by
vanishing in the framed Cobordism group
$M\mathrm{Fr}_{7-4} \simeq \mathbb{S}_3 \simeq \mathbb{Z}/24$
\eqref{CompactificationOnY4}, Rem. \ref{UnitM2BraneChargeUnderHypothesisH}.

\medskip
These charge-less or {\it cohomologically fluxless} backgrounds
(Rem. \ref{FramedSpacetimes})
are witnessed by $H_3$-fluxes \eqref{HE3Homotopy},
which encode {\it how} the
total charge is cancelling,
observed through refined e-invariants \eqref{DiagrammatichatecInvariant}.
Seen in Cobordism cohomology,
such a brane charge cancellation is
represented by an actual cobordism between the vicinities
of the source branes.
For instance, the Conner-Floyd e-invariant \eqref{IdentificationOfH3MSUFluxesWithMSUModSCobordism}
manifestly sees the Green-Schwarz mechanism on
K3-surfaces with transversal NS5-brane punctures \eqref{RationalToddNumberByGreenSchwarzMechanism}.

\medskip

We close by tabulating, on pp. \pageref{CorrespondenceTable},
the correspondences between
M-theory and $M\!f$-theory obtained in
\cref{TheDictionary}.

\newpage

\noindent
{\bf A correspondence of concepts between M/F-Theory and $M \mathrm{Fr}$-Theory.}
\label{CorrespondenceTable}
The following tables survey the correspondence,
that we have obtained in \cref{TheDictionary},
between
phenomena expected in M-theory and made precise under
\hyperlink{HypothesisH}{Hypothesis H},
and
definitions/theorems in stable homotopy theory and
generalized cohomology, specifically revolving around
Cobordism cohomology $M \mathrm{Fr}$, $M S\mathrm{U}$ and
topological K-theory:

\medskip

{\hypertarget{TableD}{}}
\hspace{-.7cm}
{\small
\def\arraystretch{2.2}
\begin{tabular}{|p{13.5em}|c|p{13.5em}||p{1.5em}|}
  \hline
  \begin{minipage}[center]{5.2cm}
    \bf
    Non-perturbative physics
  \end{minipage}
  &
  \begin{minipage}[center]{5.65cm}
    \begin{center}
    $\longleftrightarrow$
    \end{center}
  \end{minipage}
  &
  \begin{minipage}[center]{5.2cm}
    \bf
    Algebraic Topology
  \end{minipage}
  &
  \begin{minipage}[left]{2cm}
    \hspace{-4pt}
    \bf \cref{FullMBraneChargeAndUnstableCohomotopyTheory}
  \end{minipage}
  \\
  \hline
  \hline
  \begin{minipage}[center]{5.2cm}
    Fields vanishing at $\infty$
  \end{minipage}
  &
  \begin{minipage}[center]{5.65cm}
    $$
      \LCHSpaces
      \xrightarrow{
        (-)_{\mathrm{cpt}}
      }
      \PointedSpaces
    $$
  \end{minipage}
  &
  \begin{minipage}[center]{5.2cm}
    (One-point compactification to)
    \\
    Pointed topological spaces
    $\phantom{\mathclap{\vert_{\vert}}}$
  \end{minipage}
  &
  \begin{minipage}[left]{2cm}
    \eqref{OnePointCompactificationAndFunctionsOnIt}
  \end{minipage}
  \\
  \hline
  \begin{minipage}[center]{5.2cm}
    Quantized charges
    \\
    in spacetime $X$ including $\infty$
  \end{minipage}
  &
  \begin{minipage}[center]{5.65cm}
    \begin{center}
    $
      \mathclap{
      \big[
        X
          \!\xrightarrow{c}\!
        A
        \!
      \big]
      \,\in\,
      \widetilde A
      \big(
        X
      \big)
      \,:\!=\,
      \pi_0
      \mathrm{Maps}^{\ast/}
      \!\!
      \big(
        X, A
      \big)
      }
    $
    \end{center}
  \end{minipage}
  &
  \begin{minipage}[center]{5.2cm}
    Reduced non-abelian
    \\
    generalized cohomology
    $\mathclap{\phantom{\vert_{\vert}}}$
  \end{minipage}
  &
  \begin{minipage}[left]{2cm}
    \eqref{NonabelianCohomologyInIntroduction}
  \end{minipage}
  \\
  \hline
\end{tabular}
\def\arraystretch{1}
}

\medskip

\hspace{-.7cm}
{\small
\def\arraystretch{2.2}
\begin{tabular}{|p{13.5em}|c|p{13.5em}||p{1.5em}|}
  \hline
  \begin{minipage}[center]{5.2cm}
    \bf
    M/F-Theory
  \end{minipage}
  &
  \begin{minipage}[center]{5.65cm}
    \begin{center}
    $
      \overset{
        \scalebox{.8}{
          \color{black}
          \hyperlink{HypothesisH}{\bf Hypothesis H}
        }
        \mathclap{\phantom{\vert_{\vert}}}
      }{
        \longleftrightarrow
      }
      \mathclap{\phantom{\vert_{\vert_{\vert_{\vert_{\vert}}}}}}
    $
    \end{center}
  \end{minipage}
  &
  \begin{minipage}[center]{5.2cm}
    \bf
    $M\mathrm{Fr}$-Theory
  \end{minipage}
  &
  \begin{minipage}[left]{2cm}
    \hspace{-4pt}
    \bf \cref{MBraneWorldvolumesAndThePontrjaginConstruction}
  \end{minipage}
  \\
  \hline
  \hline
  \begin{minipage}[center]{5.2cm}
    Full M-brane charge
    \\
    on parallelizable spacetime
    $\mathclap{\phantom{\vert_{\vert_{\vert}}}}$
  \end{minipage}
  &
  \begin{minipage}[center]{5.65cm}
    \begin{center}
      $
        \big[
          \,
          \raisebox{-1pt}{$
            X
              \xrightarrow{\;c\;}
            S^4$
          }
        \big]
        \,\in\,
        {\widetilde \pi}{}^4(X)
      $
    \end{center}
  \end{minipage}
  &
  \begin{minipage}[center]{5.2cm}
    Unstable
    \\
    Cohomotopy theory
  \end{minipage}
  &
  \begin{minipage}[left]{2cm}
    \eqref{ChargeQuantizationIn4CohomotopyOnHomotopicallyFlatSpacetime}
  \end{minipage}
  \\
  \hline
  \begin{minipage}[center]{5.2cm}
    Polarized worldvolume of
    \\
    branes sourcing the charge
  \end{minipage}
  &
  \begin{minipage}[center]{5.65cm}
    \begin{center}
    $
      \big[
      c^{-1}
      \big(
        0_{\mathrm{reg}}
      \big)
      \,\subset\,
      X
      \big]
      \,\in\,
      \mathrm{Cob}^{4}_{\mathrm{Fr}}
      (X)
    $
    \end{center}
  \end{minipage}
  &
  \begin{minipage}[center]{5.2cm}
    Unstable $\mathrm{Fr}$-Cobordism
    \\
    via Pontrjagin construction
    $\mathclap{\phantom{\vert_{\vert}}}$
  \end{minipage}
  &
  \begin{minipage}[left]{2cm}
    \eqref{PontrjaginIsomorphismForNonCompactManifolds}
  \end{minipage}
  \\
  \hline
  \begin{minipage}[left]{5.2cm}
      Charge of probe $p$-branes
      \\
      near black $b$-branes
      $\mathclap{\phantom{\vert_{\vert_{\vert}}}}$
  \end{minipage}
  &
  \begin{minipage}[center]{5.65cm}
    \begin{center}
      $
        \big[
          \big(
            \mathbb{R}^{b-p}
            \times
            S^{9-b}
          \big)_{\mathrm{cpt}}
          \xrightarrow{\;c\;}
          S^4
       \big]
      $
    \end{center}
  \end{minipage}
  &
  \begin{minipage}[center]{5.2cm}
    Unstable
    \\
    Homotopy groups of spheres
    $\mathclap{\phantom{\vert_{\vert}}}$
  \end{minipage}
  &
  \begin{minipage}[left]{2cm}
    \hspace{-5pt}
    \eqref{CohomotopyVanishingAtInfinityOfEuclideanSpacesIfHomotopyGroupsOfSpheres}
    \\
    \eqref{HomotopyGroupsOfThe4Sphere}
  \end{minipage}
  \\
  \hline
  \hline
  \begin{minipage}[center]{5.2cm}
    Charge of probe $p$-branes
    \\
    in bulk around black $b$-branes
    $\mathclap{\phantom{\vert_{\vert_{\vert}}}}$
  \end{minipage}
  &
  \begin{minipage}[center]{5.65cm}
    \begin{center}
      $
        \big[
          \big(
            \mathbb{R}^{b-p}
            \times
            S^{9-b}
          \big)_{\mathrm{cpt}}
          \xrightarrow{\;c\;}
          S^4
       \big]
       \wedge
       (\mathbb{R}^1_{\mathrm{rad}})_{\mathrm{cpt}}
      $
    \end{center}
  \end{minipage}
  &
  \begin{minipage}[center]{5.2cm}
    1-Stable
    \\
    Homotopy groups of spheres
    $\mathclap{\phantom{\vert_{\vert}}}$
  \end{minipage}
  &
  \begin{minipage}[left]{2cm}
    \eqref{BulkInteractionsAsStabilization}
  \end{minipage}
  \\
  \hline
  \begin{minipage}[center]{5.2cm}
    Linearized M-brane charge
    \\
    on parallelizable spacetime
    $\mathclap{\phantom{\vert_{\vert}}}$
  \end{minipage}
  &
  \begin{minipage}[center]{5.65cm}
    \begin{center}
      $
        \big[
          \raisebox{-2pt}{$\Sigma^\infty X
           \xrightarrow{\Sigma^\infty c }
          \Sigma^\infty S^4$}
        \big]^{\phantom{A}}
        \,\in\,
        \widetilde {\mathbb{S}}^4(X)
      $
    \end{center}
  \end{minipage}
  &
  \begin{minipage}[center]{5.2cm}
    Stable
    \\
    Cohomotopy theory
    $\mathclap{\phantom{\vert_{\vert}}}$
  \end{minipage}
  &
  \begin{minipage}[left]{2cm}
    \eqref{dInvariantAndBoardman}
  \end{minipage}
  \\
  \hline
  \begin{minipage}[center]{5.2cm}
    Polarized worldvolume class of
    \\
    branes sourcing linear charge
    $\mathclap{\phantom{\vert_{\vert}}}$
  \end{minipage}
  &
  \begin{minipage}[center]{5.65cm}
    \begin{center}
    $
      \mathrm{Src}(\Sigma^\infty c)
      \,
      \in
      \,
      {\widetilde {M\mathrm{Fr}}}{}^{4}(X)
    $
    \end{center}
  \end{minipage}
  &
  \begin{minipage}[center]{5.2cm}
    Stable $\mathrm{Fr}$-Cobordism cohomology
    \\
    via Pontrjagin-Thom isomorphism
    $\mathclap{\phantom{\vert_{\vert}}}$
  \end{minipage}
  &
  \begin{minipage}[left]{2cm}
    \eqref{BulkInteractionsAsStabilization}
  \end{minipage}
  \\
  \hline
\end{tabular}
\def\arraystretch{1}
}

\medskip

\hspace{-.7cm}
{\small
\def\arraystretch{2.2}
\begin{tabular}{|p{13.5em}|c|p{13.5em}||p{1.5em}|}
  \hline
  \begin{minipage}[center]{5.2cm}
    \bf
    M/F-Theory
  \end{minipage}
  &
  \begin{minipage}[center]{5.65cm}
    \begin{center}
    $
      \overset{
        \scalebox{.8}{
          \color{black}
          \hyperlink{HypothesisH}{\bf Hypothesis H}
        }
        \mathclap{\phantom{\vert_{\vert}}}
      }{
        \longleftrightarrow
      }
      \mathclap{\phantom{\vert_{\vert_{\vert_{\vert_{\vert}}}}}}
    $
    \end{center}
  \end{minipage}
  &
  \begin{minipage}[center]{5.2cm}
    \bf
    $M\mathrm{Fr}$-Theory
  \end{minipage}
  &
  \begin{minipage}[left]{2cm}
    \hspace{-4pt}
    \bf \cref{M5BraneChargeAndMltiplicativeCohomologyTheory}
  \end{minipage}
  \\
  \hline
  \hline
  \begin{minipage}[center]{5.2cm}
    Unit M5-brane charge
  \end{minipage}
  &
  \begin{minipage}[center]{5.65cm}
  \begin{center}
    $
      G^E_{4,\mathrm{unit}}
      \;:=\;
      \Sigma^4 (1^E)
      \,\in\,
      \widetilde E^4\big( S^4 \big)
    $
  \end{center}
  \end{minipage}
  &
  \begin{minipage}[center]{5.2cm}
    Unital cohomology theory
  \end{minipage}
  &
  \eqref{4SuspendedUnitInECohomology}
  \\
  \hline
  \begin{minipage}[center]{5.2cm}
    Total M5-brane charge
    \\
    in given units of charge
    $\mathclap{\phantom{\vert_{\vert}}}$
  \end{minipage}
  &
  \begin{minipage}[center]{5.65cm}
    \begin{center}
    $
      [G_4^E(c)]
      \;:=\;
      [c^\ast G^E_{4,\mathrm{unit}}]
      \;\in\;
      {\widetilde E}{}^4(X)
    $
    \end{center}
  \end{minipage}
  &
  \begin{minipage}[center]{5.2cm}
    Adams d-invariant
  \end{minipage}
  &
  \begin{minipage}[left]{2cm}
    \eqref{dInvariantAsMapOnCohomotopy}
  \end{minipage}
  \\
  \hline
  \begin{minipage}[center]{5.2cm}
    Dual M5-brane flux and
    \\
    Topological equation of motion
    $\mathclap{\phantom{\vert_{\vert}}}$
  \end{minipage}
  &
  \begin{minipage}[center]{5.65cm}
  \begin{center}
    $
      \xymatrix@C=17pt{
        G^E_{4}
         \cdot
        G^E_{4}
        \ar@{=>}[r]
          ^-{
            2G^E_{7}
          }
        &
        0
      }
      \in
      P
      \mathrm{Maps}^{\ast/\!\!}
      \big(
        X, E^8
      \big)
    $
  \end{center}
  \end{minipage}
  &
  \begin{minipage}[center]{5.2cm}
    Multiplicative
    cohomology theory
  \end{minipage}
  &
  \begin{minipage}[left]{2cm}
    \eqref{TrivializationOfCupSquareOfUnitM5BraneCharge}
  \end{minipage}
  \\
  \hline
  \begin{minipage}[center]{5.2cm}
    Computing M-brane charge from
    \\
    its measurement in multipl. units
    $\mathclap{\phantom{\vert_{\vert}}}$
  \end{minipage}
  &
  \begin{minipage}[center]{5.5cm}
    \begin{center}
      $
        \mathrm{Ext}^{\bullet}_{\widetilde E^\bullet(E)}
        \big(
          \widetilde E^\bullet(S^4)
          ,\,
          \widetilde E^\bullet(X)
        \big)
        \,\Rightarrow\,
        \widetilde {\mathbb{S}}^{4}(X)^{\wedge_E}
      $
    \end{center}
  \end{minipage}
  &
  \begin{minipage}[center]{5.2cm}
    Adams-Novikov spectral sequence
  \end{minipage}
  &
  \begin{minipage}[left]{2cm}
    p. \pageref{MBraneChargeInComplexOrientedCohomology}
  \end{minipage}
  \\
  \hline
\end{tabular}
\def\arraystretch{1}
}

\medskip

\medskip

\hspace{-.7cm}
{\small
\def\arraystretch{2.2}
\begin{tabular}{|p{13.5em}|c|p{13.5em}||p{1.5em}|}
  \hline
  \begin{minipage}[center]{5.2cm}
    \bf
    M/F-Theory
  \end{minipage}
  &
  \begin{minipage}[center]{5.65cm}
    \begin{center}
    $
      \overset{
        \scalebox{.8}{
          \color{black}
          \hyperlink{HypothesisH}{\bf Hypothesis H}
        }
        \mathclap{\phantom{\vert_{\vert}}}
      }{
        \longleftrightarrow
      }
      \mathclap{\phantom{\vert_{\vert_{\vert_{\vert_{\vert}}}}}}
    $
    \end{center}
  \end{minipage}
  &
  \begin{minipage}[center]{5.2cm}
    \bf
    $M\mathrm{Fr}$-Theory
  \end{minipage}
  &
  \begin{minipage}[left]{2cm}
    \hspace{-4pt}
    \bf \cref{M5ThreeFlux}
  \end{minipage}
  \\
  \hline
  \begin{minipage}[center]{5.2cm}
    $H_3$-Flux
    in given units
  \end{minipage}
  &
  \begin{minipage}[center]{5.65cm}
  \begin{center}
    $
      \xymatrix@C=20pt{
        0
        \ar@{=>}[r]
          ^-{\scalebox{0.8}{$
            H^E_3(c)
      $}    }
        &
        G^E_4(c)
      }
      \in
      P
      \mathrm{Maps}^{\ast/\!\!}
      \big(
        X, E^4
      \big)
    $
  \end{center}
  \end{minipage}
  &
  \begin{minipage}[center]{5.2cm}
    Trivialization of d-invariant
  \end{minipage}
  &
  \begin{minipage}[left]{2cm}
    \eqref{HE3Homotopy}

    \noindent
    \hspace{-5pt}
    \eqref{SetOfTrivializationsOfdInvariant}
  \end{minipage}
  \\
  \hline
  \begin{minipage}[center]{5.2cm}
    Observable on $H_3$-flux
  \end{minipage}
  &
  \begin{minipage}[center]{5.65cm}
    \begin{center}
      $
        O^{(-)}(H^E_3(c))
        \coloneqq
        \big\langle
          c, \, G^E_{4,\mathrm{unit}}, -
        \big\rangle_{(H^E_3(c), - )}
      $
    \end{center}
  \end{minipage}
  &
  \begin{minipage}[center]{5.2cm}
    Refined Toda bracket
  \end{minipage}
  &
  \begin{minipage}[left]{2cm}
    \eqref{TodaBracketObservableOnH3Flux}
  \end{minipage}
  \\
  \hline
  \begin{minipage}[center]{5.2cm}
    Universal
    \\
    observable on $H^E_3$-flux
    $\mathclap{\phantom{\vert_{\vert}}}$
  \end{minipage}
  &
  \begin{minipage}[center]{5.65cm}
    \begin{center}
      $
        \!\!\!\!\!
        O^{E/\mathbb{S}}(H^E_3)
        \coloneqq
        \big\langle
          c, \, G^E_{4,\scalebox{.6}{$\mathrm{unit}$}},
          \scalebox{.8}{$\mathrm{cof}$}(e^E)
        \big\rangle_{(H^E_3(c), \mathrm{po} )}
        \!\!\!\!\!
      $
    \end{center}
  \end{minipage}
  &
  \begin{minipage}[center]{5.2cm}
    Refined Toda bracket in
    \\
    Adams cofiber cohomology
  \end{minipage}
  &
  \begin{minipage}[left]{2cm}
    \eqref{MappingHFluxToClassInCofiberCohomology}
  \end{minipage}
  \\
  \hline
  \begin{minipage}[center]{5.2cm}
    Observation of $H^E_3$-flux
    \\
    through multiplicative operation
    $\mathclap{\phantom{\vert_{\vert}}}$
  \end{minipage}
  &
  \begin{minipage}[center]{5.65cm}
    \begin{center}
      $
        (E \xrightarrow{\phi} F)
        \;\rightsquigarrow\;
        O^{F/\mathbb{S}}
        \,\coloneqq\,
        \phi/\mathbb{S}
        \circ
        O^{E/\mathbb{S}}
      $
    \end{center}
  \end{minipage}
  &
  \begin{minipage}[center]{5.2cm}
    Box operation
    on Toda brackets
  \end{minipage}
  &
  \begin{minipage}[left]{2cm}
    \eqref{CofiberSquareOnMultiplicativeCohomologyOperation}
  \end{minipage}
  \\
  \hline
\end{tabular}
\def\arraystretch{1}
}

\medskip

\hspace{-.7cm}
{\small
\def\arraystretch{2.2}
\begin{tabular}{|p{13.5em}|c|p{13.5em}||p{1.5em}|}
  \hline
  \begin{minipage}[center]{5.2cm}
    \bf
    M/F-Theory
  \end{minipage}
  &
  \begin{minipage}[center]{5.65cm}
    \begin{center}
    $
      \overset{
        \scalebox{.8}{
          \color{black}
          \hyperlink{HypothesisH}{\bf Hypothesis H}
        }
        \mathclap{\phantom{\vert_{\vert}}}
      }{
        \longleftrightarrow
      }
      \mathclap{\phantom{\vert_{\vert_{\vert_{\vert_{\vert}}}}}}
    $
    \end{center}
  \end{minipage}
  &
  \begin{minipage}[center]{5.2cm}
    \bf
    $M\mathrm{Fr}$-Theory
  \end{minipage}
  &
  \begin{minipage}[left]{2cm}
    \hspace{-4pt}
    \bf \cref{M5BraneC3FieldAndTheAdamseInvariant}
  \end{minipage}
  \\
  \hline
  \hline
  \begin{minipage}[center]{5.2cm}
    Observation of $H^{K\mathrm{U}}_3$-flux
    \\
    through Chern character
    $\mathclap{\phantom{\vert_{\vert}}}$
  \end{minipage}
  &
  \begin{minipage}[center]{5.65cm}
    \begin{center}
    $
    \begin{aligned}
      &
      \widehat {\mathrm{e}}_{K\mathrm{U}}
      \big(
        H^{K\mathrm{U}}_3
      \big)
      \coloneqq
      \mathrm{spl}_0
      \circ
      \mathrm{ch}/\mathbb{S}
      \circ
      O^{K\mathrm{U}/\mathbb{S}}(H^{K\mathrm{U}}_3)
    \end{aligned}
    $
    \end{center}
  \end{minipage}
  &
  \begin{minipage}[center]{5.2cm}
    Refined Adams e-invariant
  \end{minipage}
  &
  \begin{minipage}[left]{2cm}
    \eqref{DiagrammatichatecInvariant}
  \end{minipage}
  \\
  \hline
  \begin{minipage}[center]{5.2cm}
    Obstruction to observation
    \\
    of integral $H_3$-flux
    $\mathclap{\phantom{\vert_{\vert}}}$
  \end{minipage}
  &
  \begin{minipage}[center]{5.65cm}
    \begin{center}
    $
      \mathrm{e}_{\mathrm{Ad}}(c)
      =
      {\widehat {\mathrm{e}}}_{K\mathrm{U}}
      \big(
        H^{K\mathrm{U}}_3)(c)
      \big)
      \,\mathrm{mod}\,
      \mathbb{Z}
    $
    \end{center}
  \end{minipage}
  &
  \begin{minipage}[center]{5.2cm}
    Classical Adams e-invariant
  \end{minipage}
  &
  \begin{minipage}[left]{2cm}
    \eqref{TheLiftedAdamseCInvariant}
  \end{minipage}
  \\
  \hline
  \begin{minipage}[center]{5.2cm}
    Existence of $C_3$-field
  \end{minipage}
  &
  \begin{minipage}[center]{5.65cm}
    \begin{center}
    $
      \begin{aligned}
        \mathclap{\phantom{\vert^{\vert^{\vert}}}}
        &
        [H^{K\mathrm{U}}_3(c)]
        =
        [C^{K\mathrm{U}}_3(c)]
        +
        n \cdot
        [H^{K \mathrm{U}}_{3,\mathrm{unit}}]
        \\
        &
        \Leftrightarrow
        \;
        \mathrm{e}_{\mathrm{Ad}}(c)
        \,=\,
        0
        \mathclap{\phantom{\vert_{\vert}}}
      \end{aligned}
    $
    \end{center}
  \end{minipage}
  &
  \begin{minipage}[center]{5.2cm}
    Vanishing of
    \\
    classical Adams e-invariant
  \end{minipage}
  &
  \begin{minipage}[left]{2cm}
    \eqref{VanishingAdamsInvariantImpliesCFieldKOCase}
  \end{minipage}
  \\
  \hline
\end{tabular}
\def\arraystretch{1}
}

\medskip

\medskip
\hspace{-.7cm}
{\small
\def\arraystretch{2.2}
\begin{tabular}{|p{13.5em}|c|p{13.5em}||p{1.5em}|}
  \hline
  \begin{minipage}[center]{5.2cm}
    \bf
    M/F-Theory
  \end{minipage}
  &
  \begin{minipage}[center]{5.65cm}
    \begin{center}
    $
      \overset{
        \scalebox{.8}{
          \color{black}
          \hyperlink{HypothesisH}{\bf Hypothesis H}
        }
        \mathclap{\phantom{\vert_{\vert}}}
      }{
        \longleftrightarrow
      }
      \mathclap{\phantom{\vert_{\vert_{\vert_{\vert_{\vert}}}}}}
    $
    \end{center}
  \end{minipage}
  &
  \begin{minipage}[center]{5.2cm}
    \bf
    $M\mathrm{Fr}$-Theory
  \end{minipage}
  &
  \begin{minipage}[left]{2cm}
    \hspace{-4pt}
    \bf \cref{TadpoleCancellationAndSUBordismWithBoundaries}
  \end{minipage}
  \\
  \hline
  \hline
  \begin{minipage}[center]{5.2cm}
    Spontaneous KK-compactification
    \\
    on K3 with 24 transversal branes
    $\mathclap{\phantom{\vert_{\vert_{\vert}}}}$
  \end{minipage}
  &
  \begin{minipage}[center]{5.65cm}
    \begin{center}
    $
      \!\!\!\!\!\!
      \xymatrix@C=28pt{
        24 \!\cdot\!
        [S^3_{\mathrm{fr}=1}]
        \ar@{=>}[r]
          ^-{ \scalebox{0.7}{$
            \mathrm{K3}
              \setminus
            \sqcup_{\scalebox{.7}{$24$}}
            D^4
          $}
          }
                  &
        0
      }
      \!\!
      \in
      P \mathrm{Maps}^{\ast/\!\!\!}
      \big(
        S^7\!\!\!\!,
        M\mathrm{Fr}^4
        \!
      \big)
      \!\!\!\!\!\!
    $
    \end{center}
  \end{minipage}
  &
  \begin{minipage}[center]{5.2cm}
    24-punctured K3 witnesses
    \\
    3rd stable stem $M \mathrm{Fr}_3 = \mathbb{Z}/24$
  \end{minipage}
  &
  \begin{minipage}[left]{2cm}
    \eqref{CompactificationOnY4}
  \end{minipage}
  \\
  \hline
  \begin{minipage}[center]{5.2cm}
    Observation of ordinary $H_3$-flux
    \\
    sourced by transversal branes
    $\mathclap{\phantom{\vert_{\vert_{\vert}}}}$
  \end{minipage}
  &
  \begin{minipage}[center]{5.65cm}
    \begin{center}
    $
      \widehat e_{K\mathrm{U}}
      \big(
        H^{M S\mathrm{U}}_3(n h_{\mathbb{H}})
      \big)
      \;=\;
      \mathrm{Td}
      \big[
        M^4_{S\mathrm{U}}
          \setminus
        \sqcup_{\scalebox{.5}{$n$}}
        D^4
      \big]
    $
    \end{center}
  \end{minipage}
  &
  \begin{minipage}[center]{5.2cm}
    Conner-Floyd's e-invariant
    \\
    via Todd character
    $\mathclap{\phantom{\vert_{\vert}}}$
  \end{minipage}
  &
  \begin{minipage}[left]{2cm}
    \eqref{CompactificationOnY4}
  \end{minipage}
  \\
  \hline
  \begin{minipage}[center]{5.2cm}
    Green-Schwarz mechanism for
    \\
    transversal 5-branes
    $\mathclap{\phantom{\vert_{\vert_{\vert}}}}$
  \end{minipage}
  &
  \begin{minipage}[center]{5.65cm}
    \begin{center}
    $
      d\,H_3 = \rchi_4(\nabla)
      ,
      \;\;\;
      \,
      \widehat{e}\,
      \big(
        H^{M S\mathrm{U}}_3
      \big)
      =
      \tfrac{1}{12}
      \int_{S^3_k} H_3
    $
    \end{center}
  \end{minipage}
  &
  \begin{minipage}[center]{5.2cm}
    Conner-Floyd's e-invariant
    \\
    via J-twisted 3-Cohomotopy
    $\mathclap{\phantom{\vert_{\vert}}}$
  \end{minipage}
  &
  \begin{minipage}[left]{2cm}
    \eqref{RationalToddNumberByGreenSchwarzMechanism}
  \end{minipage}
  \\
  \hline
\end{tabular}
\def\arraystretch{1}
}

\medskip

\hspace{-.7cm}
{\small
\def\arraystretch{2.2}
\begin{tabular}{|p{13.5em}|c|p{13.5em}||p{1.5em}|}
  \hline
  \begin{minipage}[center]{5.2cm}
    \bf
    M/F-Theory
  \end{minipage}
  &
  \begin{minipage}[center]{5.65cm}
    \begin{center}
    $
      \overset{
        \scalebox{.8}{
          \color{black}
          \hyperlink{HypothesisH}{\bf Hypothesis H}
        }
        \mathclap{\phantom{\vert_{\vert}}}
      }{
        \longleftrightarrow
      }
      \mathclap{\phantom{\vert_{\vert_{\vert_{\vert_{\vert}}}}}}
    $
    \end{center}
  \end{minipage}
  &
  \begin{minipage}[center]{5.2cm}
    \bf
    $M\mathrm{Fr}$-Theory
  \end{minipage}
  &
  \begin{minipage}[left]{2cm}
    \hspace{-4pt}
    \bf \cref{M2BraneChargeAndTheHopfInvariant}
  \end{minipage}
  \\
  \hline
  \hline
  \begin{minipage}[center]{5.2cm}
    M2-Brane Page charge
  \end{minipage}
  &
  \begin{minipage}[center]{5.65cm}
    \begin{center}
    $
      N^E_{\mathrm{M2}}(c)
      \;=\;
      \tfrac{1}{24}
      H
      \big(
        c
      \big)
    $
    \end{center}
  \end{minipage}
  &
  \begin{minipage}[center]{5.2cm}
    Hopf invariant
  \end{minipage}
  &
  \multirow{2}{*}{
    \begin{minipage}[left]{2cm}
      \hspace{-5pt}
      \eqref{EPageChargeDiagrammatic}
    \end{minipage}
  }
  \\
  \cline{1-3}
  \begin{minipage}[center]{5.2cm}
    M2-Brane Page charge as
    \\
    observable on $H_3$-flux
    $\mathclap{\phantom{\vert_{\vert}}}$
  \end{minipage}
  &
  \begin{minipage}[center]{5.65cm}
    \begin{center}
    $
      N^E_{\mathrm{M2}}(c)
      \;=\;
      \tfrac{1}{12}
      \int_{S^7}
      \big(
        \tfrac{1}{2}
        H_3 \wedge G_4
        +
        G_7
      \big)
    $
    \end{center}
  \end{minipage}
  &
  \begin{minipage}[center]{5.2cm}
    Whitehead integral formula/
    \\
    Steenrod functional cup product
    $\mathclap{\phantom{\vert_{\vert}}}$
  \end{minipage}
  &
  \begin{minipage}[left]{2cm}
  \end{minipage}
  \\
  \hline
\end{tabular}
\def\arraystretch{1}
}

\medskip

\hspace{-.7cm}
{\small
\def\arraystretch{2.2}
\begin{tabular}{|p{13.5em}|c|p{13.5em}||p{1.5em}|}
  \hline
  \begin{minipage}[center]{5.2cm}
    \bf
    M/F-Theory
  \end{minipage}
  &
  \begin{minipage}[center]{5.65cm}
    \begin{center}
    $
      \overset{
        \scalebox{.8}{
          \color{black}
          \hyperlink{HypothesisH}{\bf Hypothesis H}
        }
        \mathclap{\phantom{\vert_{\vert}}}
      }{
        \longleftrightarrow
      }
      \mathclap{\phantom{\vert_{\vert_{\vert_{\vert_{\vert}}}}}}
    $
    \end{center}
  \end{minipage}
  &
  \begin{minipage}[center]{5.2cm}
    \bf
    $M\mathrm{Fr}$-Theory
  \end{minipage}
  &
  \begin{minipage}[left]{2cm}
    \hspace{-4pt}
    \bf \cref{VicinityOfBlackM2BranesAndOrientedCohomology}

    \noindent
    \hspace{-4pt}
    \bf \cref{GreenSchwarzMechanismAndComplexEOrientations}
  \end{minipage}
  \\
  \hline
  \hline
  \begin{minipage}[center]{5.2cm}
    ${\phantom{A}}$
    \\
    $H_3$-flux in given units
    \\
    universal near black M2-branes
  \end{minipage}
  &
  \begin{minipage}[center]{5.65cm}
    \begin{center}
    $
      \xymatrix@C=20pt{
        0
        \ar@{=>}[r]
          ^-{\scalebox{0.7}{$
            H^E_3(h_{\mathbb{H}})
            $}
          }
        &
        G^E_4(h_{\mathbb{H}})
      }
      \in
      P
      \mathrm{Maps}^{\ast/\!\!}
      \big(
        S^7, E^4
      \big)
    $
    \end{center}
  \end{minipage}
  &
  \begin{minipage}[center]{5.2cm}
    10d quaternionic orientation
    \\
    in $E$-cohomology
  \end{minipage}
  &
  \begin{minipage}[left]{2cm}
    \eqref{G4TrivializationOnS7IsQuaternionicOrientationToSecondStage}
  \end{minipage}
  \\
  \cline{2-4}
  &
  \begin{minipage}[center]{5.65cm}
    \begin{center}
    $
      \xymatrix@C=20pt{
        c^E_2
        \ar@{=>}[r]
          ^-{
            H^E_{3,\mathrm{het}}
          }
        &
        G^E_{4,\mathrm{unit}}
      }
      \in
      P
      \mathrm{Maps}^{\ast/\!\!}
      \big(
        S^4, E^4
      \big)
    $
    \end{center}
  \end{minipage}
  &
  \begin{minipage}[center]{5.2cm}
    10d complex orientation
    \\
    in $E$-cohomology
  \end{minipage}
  &
  \begin{minipage}[left]{2cm}
    \eqref{G4TrivializationOnS7IsQuaternionicOrientationToSecondStage}
  \end{minipage}
  \\
  \hline
  \begin{minipage}[center]{5.2cm}
    \hspace{-.2cm}
    \scalebox{.93}{
    Ho{\v r}ava-Witten Green-Schwarz mech-
    }
    \\
    anism for heterotic line bundles
  \end{minipage}
  &
  \begin{minipage}[center]{5.65cm}
    \begin{center}
    $
      \!\!\!\!\!
      \xymatrix@C=14pt{
        c^E_1\cdot c^E_{1^\ast}
        \ar@{=>}[r]
          ^-{
            H^E_{3,\mathrm{het}}
          }
        &
        G^E_{4}(t_{\mathbb{H}})
      }
      \in
      P
      \mathrm{Maps}^{\ast/\!\!}
      \big(
        S^4, E^4
      \big)
      \!\!\!\!\!
    $
    \end{center}
  \end{minipage}
  &
  \begin{minipage}[center]{5.2cm}
    Factorization through
    \\
    twistor fibration
  \end{minipage}
  &
  \begin{minipage}[left]{2cm}
    \eqref{G4TrivializationOnS7IsQuaternionicOrientationToSecondStage}
  \end{minipage}
  \\
  \hline
\end{tabular}
\def\arraystretch{1}
}

\newpage

\noindent

\begin{remark}[{\bf Unit M2-brane charge and the Order of the third stable stem.}]
\label{UnitM2BraneChargeUnderHypothesisH}
A crucial subtlety of
{\hyperlink{HypothesisH}{\it Hypothesis H}}
is that it
unifies M2/M5-brane charge in a single
non-abelian cohomology theory, in that
the Borsuk-Spanier Cohomotopy $\pi^4(X)$ \eqref{CohomotopyInIntroduction}

$\bullet$ quantizes {\it both}
$G_4$-flux {\it and} its dual $G_7$-flux
\eqref{ChargeQuantizationIn4CohomotopyOnHomotopicallyFlatSpacetime};
hence

$\bullet$ measures {\it both}
M5-brane charge (p. \pageref{M5BraneProbesAtMO9Planes})
{\it and} M2-brane Page charge (\cref{M2BraneChargeAndTheHopfInvariant}).

\vspace{-3mm}
$$
  \xymatrix@R=-8pt@C=6em{
    &
    &
    S^7
    \ar[d]
      ^-{
        h_{\mathbb{H}}
        \mathrlap{
          \mbox{
            \tiny
            \color{greenii}
            \bf
            \begin{tabular}{c}
              quaternionic
              \\
              Hopf fibration
            \end{tabular}
          }
        }
      }
    \\
    \overset{
      \mathclap{
      \raisebox{6pt}{
        \tiny
        \color{darkblue}
        \bf
        \begin{tabular}{c}
          probe M5-brane wordvolume
          \\
          given by $\mathrm{Src}(c)$
        \end{tabular}
      }
      }
    }{
      \overbrace{
        \mathbb{R}^{2,1} \times S^3
      }
    }
    \times \mathbb{R}_+
    \;
    \ar@{^{(}->}[r]
    &
    \;
    \underset{
      \mathclap{
      \raisebox{-3pt}{
        \tiny
        \color{darkblue}
        \bf
        \begin{tabular}{c}
          vicinity of
          \\
          black M2-branes
        \end{tabular}
      }
      }
    }{
    \underbrace{
    \mathbb{R}^{2,1}
    \times
    S^7
    \times \mathbb{R}^+
    }
    }
    \ar[ur]
      _>>>>>>>>{
        N_{{}_{\mathrm{M2}}}
      }
      ^-{\qquad \quad
        \mathllap{
          \raisebox{-2pt}{
            \tiny
            \color{greenii}
            \bf
            \begin{tabular}{c}
              M2-brane Page charge
              \\
              in 7-Cohomotopy
            \end{tabular}
          }
        }
      }
      \ar@{-->}[r]
      _-{
        \raisebox{-5pt}{
          \tiny
          \color{greenii}
          \bf
          \begin{tabular}{c}
            induced
            M5-brane charge
            \\
            in 4-Cohomotopy
          \end{tabular}
      }
    }
    ^-{
      c
    }
    &
    S^4
  }
$$

For example, it is this interplay which,
under \hyperlink{HypothesisH}{\it Hypothesis H} and
the Pontrjagin isomorphism \eqref{PontrjaginIsomorphismForNonCompactManifolds},
gives rise to the phenomenon of M2-branes polarizing into M5-branes
as discussed on p. \pageref{M2BranePolarizationAndTheQuaternionicHopfFibration}.
But already ordinary polarization effects in
electromagnetic fields change the number of charged particles
per unit flux, at least locally.

\medskip

\noindent
Therefore, a key subtlety in identifying the physics
content of \hyperlink{HypothesisH}{Hypothesis H}
is to disentangle these two kinds of brane charges:
According to \cite[p. 13, \S 3.8]{FSS19b},
near the horizon of black M2-branes \eqref{M2BranesPTTheorem}
we are to regard the
{\color{purple}24}th multiple
$24 \cdot c_{\mathrm{unit}} \,=\, 24 \cdot [h_{\mathbb{H}}]$
\eqref{FromUnstableToStable4CohomotopyOf7Sphere}
of the non-torsion unit Cohomotopy charge
\eqref{Universal3FluxNearM2Branes}
as the unit of charge of a single black M2-brane,
where ${\color{purple} 24} = \left\vert\mathbb{S}_3\right\vert$
arises as the order of the third stable stem
\eqref{FromUnstableToStable4CohomotopyOf7Sphere}.

\medskip

Here in this article we have found further perspectives
that support this conclusion:

\medskip

{\small
\hspace{-.7cm}
\begin{tabular}{|c|c|c|}
  \hline
  \multicolumn{2}{|c|}{
  $\mathclap{\phantom{\vert^{\vert^{\vert}}}}$
  {\bf
    Rationale for identifying:
  }
  $
    \;\;\;
    \overset{
      \mathclap{
      \mbox{
        \tiny
        \color{darkblue}
        \begin{tabular}{c}
          \bf
          number of black M2-branes/
          \\
          {\bf M2-brane Page charge }
          {\color{black} \cref{M2BraneChargeAndTheHopfInvariant}}
        \end{tabular}
      }
      }
    }{
    \overbrace{
    \scalebox{1.2}{$
      N_{\mathrm{M2}}
    $}
    \Big(
    \,
      \underset{
        \mathclap{
        \mbox{
          \tiny
          \color{darkblue}
          \begin{tabular}{c}
   \bf                non-torsion
            \\
     \bf       Cohomotopy charge
          \end{tabular}
        }
        }
      }{
      \underbrace{
        \scalebox{1.2}{$
          {
            \color{greenii}
            n
          }
            \cdot
          h_{\mathbb{H}}
        $}
      }
      }
    \,
    \Big)
    }
    }
    \scalebox{1.2}{$
      \;
      \coloneqq
      \;
      {
        \color{greenii}
        n
      }
    $}
    \big/
    \underset{
      \mathllap{=}
      \left\vert
      \mathbb{S}_3
      \right\vert
    }{
    \underbrace{
    {
      \color{purple}
      \scalebox{1.2}{$
        24
      $}
    }
    }
    }
  $
  $\mathclap{\phantom{\vert_{\vert_{\vert}}}}$
  }
  &
  \bf
  References
  \\
  \hline
  \hline
  \begin{minipage}[left]{4cm}
    \bf
    Existence of $H_3$-flux and
    \\
    M2-brane Page charge
    \\
    in stable Cohomotopy $\mathbb{S}$
    \\
    and hence in the bulk
  \end{minipage}
  &
  \begin{minipage}[left]{9.2cm}
    $\mathclap{\phantom{\vert^{\vert^{\vert}}}}$
    With Cohomotopy charges seen in the bulk
    spacetime and hence
    \eqref{BulkInteractionsAsStabilization}
    in the initial multiplicative cohomology theory
    $E = \mathbb{S}$ (stable Cohomotopy),
    $H_3$-fluxes near M2-branes --
    and with them the very notion of M2-brane charge
    \eqref{EPageChargeDiagrammatic}
    --
    exist precisely for $n$ a multiple of 24.
    $\mathclap{\phantom{\vert_{\vert_{\vert}}}}$
  \end{minipage}
  &
  \begin{minipage}[left]{3.3cm}
    \cite[(32) \& \S 3.8]{FSS19b},
    \\
    \cref{M5ThreeFlux}, \cref{M2BraneChargeAndTheHopfInvariant}
  \end{minipage}
  \\
  \hline
  \begin{minipage}[left]{4cm}
    \bf
    Expected number of
    \\
    24 NS5/D7 branes in
    \\
    $\mathrm{M}_{\mathrm{HET}}$/F theory on K3
  \end{minipage}
  &
  \begin{minipage}[left]{9.2cm}
    $\mathclap{\phantom{\vert^{\vert^{\vert}}}}$
    Under the Pontrjagin-isomorphism,
    the multiple $n = 24$
    corresponds to a spontaneous Kaluza-Klein compactification
    of M-theory spacetime on a K3-fiber, with 24 transversal
    5-branes, dual to the 24 D7-branes thought to be required
    in F-theory on elliptically fibered K3.
    $\mathclap{\phantom{\vert_{\vert_{\vert}}}}$
  \end{minipage}
  &
  \begin{minipage}[left]{3.3cm}
    \cref{MBraneWorldvolumesAndThePontrjaginConstruction}
  \end{minipage}
  \\
  \hline
  \begin{minipage}[left]{4cm}
    \bf
    Expected coefficient in
    \\
    tadpole cancellation for
    \\
    M-theory on 8-manifolds
  \end{minipage}
  &
  \begin{minipage}[left]{9.2cm}
    $\mathclap{\phantom{\vert^{\vert^{\vert}}}}$
    In the generalization to M-brane charge on
    curved spacetimes measured in tangentially
    J-twisted Cohomotopy
    (discussed in \cite{FSS19b}\cite{FSS19c})
    the proportionality factor $1/24$ between
    $N_{M2}$ and the Page charge
    makes Hypothesis H imply the expected
    tadpole cancellation formula for M2-branes on
    8-manifolds.
    $\mathclap{\phantom{\vert_{\vert_{\vert}}}}$
  \end{minipage}
  &
  \begin{minipage}[left]{3.3cm}
    \cite[Prop. 3.22]{FSS19b}
  \end{minipage}
  \\
  \hline
  \begin{minipage}[left]{4cm}
    \bf
    Existence of $C_3$-field
    \\
    and integral $H_3$-flux
    \\
    in $K\mathrm{U}$- and $K\mathrm{O}$-theory
  \end{minipage}
  &
  \begin{minipage}[left]{9.2cm}
    $\mathclap{\phantom{\vert^{\vert^{\vert}}}}$
    With $H_3$-flux measured in $K\mathrm{U}$-cohomology,
    the universal stable observable $O^{M S\mathrm{U}}$
    of 3-flux, hence the $\widehat e_{K \mathrm{U}}$-invariant,
    sees a well-defined $C_3$-field and with it integrally
    quantized $H_3$-flux precisely when $n$ is a multiple of 12.
    $\mathclap{\phantom{\vert_{\vert_{\vert}}}}$
  \end{minipage}
  &
  \begin{minipage}[left]{3.3cm}
    \cref{M5BraneC3FieldAndTheAdamseInvariant}
  \end{minipage}
  \\
  \hline
  \begin{minipage}[left]{4cm}
    \bf
    Emergence of expected
    \\
    GS-mechanism behind 24
    \\
    5-branes transverse to $\mathrm{K3}$
  \end{minipage}
  &
  \begin{minipage}[left]{9.2cm}
    $\mathclap{\phantom{\vert^{\vert^{\vert}}}}$
    For $n$ a multiple of 24
    the observable on $H^{M S\mathrm{U}}_3$-flux
    induced by the Todd character
    witnesses the above appearance of 24 5-branes transversal
    to a K3-fiber explicitly as the solution to the
    Green-Schwarz anomaly cancellation condition
    on this compact space.
    $\mathclap{\phantom{\vert_{\vert_{\vert}}}}$
  \end{minipage}
  &
  \begin{minipage}[left]{3.3cm}
    \cref{TadpoleCancellationAndSUBordismWithBoundaries}
  \end{minipage}
  \\
  \hline
\end{tabular}
}
\end{remark}

\vspace{1cm}
\noindent  Hisham Sati, {\it Mathematics, Division of Science, New York University Abu Dhabi, UAE.}
\\
{\tt hsati@nyu.edu}
\\
\\
\noindent  Urs Schreiber, {\it Mathematics, Division of Science, New York University Abu Dhabi, UAE}
\\
{\tt us13@nyu.edu}

\end{document}